%% file: main.tex
\documentclass[11pt,twoside,urlcolor=blue, linkcolor=blue]{mitthesis} 
\usepackage{lgrind}
\usepackage{cmap}
\usepackage[T1]{fontenc}
\newcommand{\cblue}[1]{\textcolor{black}{#1}}
\newcommand{\cred}[1]{\textcolor{black}{#1}}
\newcommand{\e}{\hspace{1pt}\mathrm{e}}
\newcommand{\tp}{\text{p}}
\newcommand{\AMG}{\mathrm{AMG}}
\newcommand{\MCG}{\mathrm{MCG}}
\newcommand{\CP}{\text{CP}}
\newcommand{\middlewave}[1]{\raisebox{0.5em}{\uwave{\hspace{#1}}}}  
\newcommand{\E}{\mathop{\mathrm{E}}}
\newcommand{\A}{\mathop{\mathrm{A}}}
\newcommand{\B}{\mathop{\mathrm{B}}}

\newcommand{\bfW}{\mathbf{W}}
\newcommand{\bfV}{\mathbf{V}}
\newcommand{\figref}[1]{Fig.\,\ref{#1}}

\usepackage{graphicx}
\usepackage{amsmath}
\usepackage{amstext}
\usepackage{amssymb}
\usepackage{mathrsfs}
\usepackage{amsfonts}
\usepackage{amsbsy} 
\usepackage{dcolumn}
\usepackage{bm}
\usepackage{multirow}

\usepackage{booktabs}

\usepackage{color} 
\definecolor{lgray}{gray}{0.9}
\usepackage{linegoal}
\usepackage{titlesec}
\usepackage[normalem]{ulem}
\newcommand{\tr}{\mathop{\mathrm{tr}}}

\usepackage[colorlinks,citecolor=blue]{hyperref}
\definecolor{red}{rgb}{1,0,0}
\definecolor{blue}{rgb}{0,0,1}
\definecolor{dblue}{rgb}{0,0,0.4}
\definecolor{green}{rgb}{0,1,0}
\definecolor{black}{rgb}{0,0,0}
\definecolor{white}{rgb}{1,1,1}

\definecolor{brn}{rgb}{.8,.4,.0}
\definecolor{redo}{rgb}{1,.5,.0}
\definecolor{ddgrn}{rgb}{0,0.4,0}
\definecolor{dgrn}{rgb}{0,0.55,0}
\definecolor{dbl}{rgb}{0,0,0.5}

\usepackage[bbgreekl]{mathbbol}
\usepackage{amscd}

\newcommand{\Z}{\mathbb{Z}}

\newcommand{\R}{\mathbb{R}}

\newcommand{\bbn}{\mathbb{Z}}
\newcommand{\add}[1]{#1}
\newcommand{\re}{\mathrm{e}}
\newcommand{\iden}{I}
\newcommand{\ov}[1]{\overline{#1}}

\usepackage{textcomp}

\renewcommand{\t}[1]{\tilde{#1}} 
\newcommand{\ii}{\hspace{1pt}\mathrm{i}\hspace{1pt}}
\newcommand{\dd}{\hspace{1pt}\mathrm{d}}
\newcommand{\<}{\langle} 
\renewcommand{\>}{\rangle}

\newcommand{\eqn}[1]{eqn.~(\ref{#1})}

\newcommand{\Tr}{{\rm Tr}} 
 
\renewcommand{\Im}{{\rm Im}} 
\renewcommand{\Re}{{\rm Re}}

\newcommand{\prt}{\partial} 
 
\newcommand{\sgn}{{\rm sgn}}

\newcommand{\ie}{{\it i.e.~}}

\newcommand{\cC}{ {\cal C} } 
\newcommand{\cD}{ {\cal D} }

\newcommand{\cH}{ {\cal H} } 
\newcommand{\cK}{ {\cal K} }

\newcommand{\cN}{ {\cal N} } 
 
\newcommand{\cS}{ {\cal S} } 
\newcommand{\cT}{ {\cal T} } 
\newcommand{\cV}{ {\cal V} }

\newcommand{\cM}{ {\cal M} } 
\newcommand{\cW}{ {\cal W} } 

\newcommand{\tA}{ {\mathrm{A}} }

\newcommand{\al}{\alpha} 
\newcommand{\bt}{\beta} 
\newcommand{\del}{\delta}

\newcommand{\om}{\omega} 
 
\renewcommand{\th}{\theta} 
 
\newcommand{\si}{\sigma}

\newcommand{\bpm}{\begin{pmatrix}}
\newcommand{\epm}{\end{pmatrix}}
\newcommand{\bmm}{\begin{matrix}}
\newcommand{\emm}{\end{matrix}}

\newcommand\Ointint{\begingroup \displaystyle \unitlength 1pt
\int\mkern-7mu\begin{picture}(0,3)\put(0,3){\oval(10,8)}\end{picture}
\mkern-8mu\int\endgroup}

\newcommand{\bea}{\begin{eqnarray}}
\newcommand{\eea}{\end{eqnarray}}
\newcommand{\be}{\begin{equation}}
\newcommand{\ee}{\end{equation}}

\newcommand{\frm}[1]{\begin{center}\fbox{\parbox{5.5in}{\parindent=0pt #1}}\end{center}}

\newcommand{\tD}{\text{D}}
\newcommand{\tU}{\text{U}}
\newcommand{\sfS}{\mathsf{S}}
\newcommand{\sfT}{\mathsf{T}}
\newcommand{\sfC}{\mathsf{C}}
\newcommand{\sfc}{\mathsf{c}}

\newcommand{\sfP}{\mathsf{P}}
\newcommand{\sfR}{\mathsf{R}}
\newcommand{\ti}{\text{i}}

\newcommand{\GSD}{\text{GSD}}
\newcommand{\sfO}{\mathsf{O}}
\newcommand{\tI}{\text{I}}
\newcommand{\tII}{\text{II}}
\newcommand{\tIII}{\text{III}}
\newcommand{\tIV}{\text{IV}}
\newcommand{\tV}{\text{V}}
\usepackage{comment}

\newcommand{\sfU}{\mathsf{U}}

\newcommand{\diag}{\mathop{\mathrm{diag}}}
\newcommand{\rank}{\mathop{\mathrm{rk}}}

\renewcommand{\t}[1]{\tilde{#1}} 
\newcommand{\ep}{\hspace{1pt}\mathrm{e}}
\renewcommand{\>}{\rangle} 
\newcommand{\SU}{\mathrm{SU}}
\newcommand{\U}{\mathrm{U}}

\makeatletter
\usepackage{tikz}
\usetikzlibrary{arrows,matrix,calc,scopes,decorations.markings}
\newcommand{\foursimplex}[5]{
\begin{tikzpicture}[scale=1,baseline]
\coordinate (a) at (0,-0.1);
\coordinate (b) at (0.82,-0.79);
\coordinate (c) at (2.5,0);
\coordinate (d) at (1.25,1.5);
\coordinate (e) at ($ (a) ! 0.32 ! (c) $);
\coordinate (f) at ($ (a) ! 0.55 ! (c) $);
\coordinate (g) at ($ (a) ! 0.45 ! (c) $);
\coordinate (v4) at (2.25,1.6);
\draw (a) -- (e);
\draw (b) -- (d);
\draw (f) -- (c);
\draw (a) -- (b) -- (c) -- (d) -- cycle;
\draw[->,>=latex, line width=0.75pt] (a) -- ($ (a) ! 0.5 ! (b) $);
\draw[->,>=latex, line width=0.75pt] (b) -- ($ (b) ! 0.5 ! (c) $);
\draw[->,>=latex, line width=0.75pt] (c) -- ($ (c) ! 0.5 ! (d) $);
\draw[-,>=latex, line width=0.75pt] (g) -- (f);
\draw[>-,>=latex, line width=0.75pt] (f) -- (c);
\draw[->,>=latex, line width=0.75pt] (b) -- ($ (b) ! 0.5 ! (d) $);
\draw[->,>=latex, line width=0.75pt] (a) -- ($ (a) ! 0.5 ! (d) $);
\draw[->,>=latex, dashed, line width=0.7pt] (a) -- ($ (a) ! 0.5 ! (v4) $);
\draw[-,=latex, dashed, line width=0.7pt] ($ (a) ! 0.5 ! (v4) $) -- (v4);
\draw[->,>=latex, dashed, line width=0.4pt] (b) -- ($ (b) ! 0.5 ! (v4) $);
\draw[-,=latex, dashed, line width=0.4pt] ($ (b) ! 0.5 ! (v4) $) -- (v4);
\draw[->,>=latex, dashed, line width=0.4pt] (c) -- ($ (c) ! 0.5 ! (v4) $);
\draw[-,=latex, dashed, line width=0.4pt] ($ (c) ! 0.5 ! (v4) $) -- (v4);
\draw[->,>=latex, dashed, line width=0.4pt] (d) -- ($ (d) ! 0.5 ! (v4) $);
\draw[-,=latex, dashed, line width=0.4pt] ($ (d) ! 0.5 ! (v4) $) -- (v4);
\node[left] at(a) {\scalebox{0.75}{$#1$}};
\node[below] at(b) {\scalebox{0.75}{$#2$}};
\node[right] at(c) {\scalebox{0.75}{$#3$}};
\node[above] at(d) {\scalebox{0.75}{$#4$}};
\node[right] at(v4) {\scalebox{0.75}{$#5$}};
\node[below] at ($ (a) ! 0.4 ! (b) $) {\scalebox{0.75}{$g_{#1#2}$}};
\node[below] at ($ (b) ! 0.65 ! (c) $) {\scalebox{0.75}{$g_{#2#3}$}};
\node[right] at ($ (c) ! 0.5 ! (d) $) {\scalebox{0.75}{$g_{#3#4}$}};
\node[above] at ($ (d) ! 0.5 ! (v4) $) {\scalebox{0.75}{$g_{#4#5}$}};
\end{tikzpicture}
}

\newcommand{\SLTWOtransformSLeft}{
\begin{tikzpicture}[scale=0.8,baseline]
\coordinate (p1) at (0,0);
\coordinate (p12) at (0,0.4);
\coordinate (p2) at (0,1);
\coordinate (p13) at (.4,0);
\coordinate (p3) at (1,0);
\coordinate (p24) at (.4,1);
\coordinate (p34) at (1,0.4);
\coordinate (p4) at (1,1);
\coordinate (p14) at (.4,.4);
\coordinate (p23) at (.4,.6);
\draw [->,>=latex,line width=0.15pt] (-.5,0) -- (1.5,0);
\draw [->,>=latex,line width=0.15pt] (0,-0.5) -- (0,1.5);
\draw [-<,>=latex,line width=0.4pt] (p2) -- (p12);
\draw [-<,>=latex,line width=0.4pt] (p4) -- (p34);
\draw [-,>=latex,line width=0.4pt] (p34) -- (p3);
\draw [-<,>=latex,line width=0.4pt] (p3) -- (p13);
\draw [-,>=latex,line width=0.4pt] (p13) -- (p1);
\draw [-<,>=latex,line width=0.4pt] (p4) -- (p24);
\draw [-,>=latex,line width=0.4pt] (p24) -- (p2);
\draw [->,>=latex,line width=0.4pt] (p3) -- (p23);
\draw [-,>=latex,line width=0.4pt] (p23) -- (p2);
\node at (-0.15,-0.15) {1};
\node at (-0.15,1.15) {3};
\node at (1.15,-0.185) {2};
\node at (1.15,1.15) {4};
\node at (1.85,0.02) {$x$};
\node at (0,1.75) {$y$};
\node at (.5,-0.185) {$g_x$};
\node at (-0.25,.6) {$g_y$};
\end{tikzpicture}
}

\newcommand{\SLTWOtransformSRight}{
\begin{tikzpicture}[scale=0.8,baseline]
\coordinate (p1) at (0,0);
\coordinate (p2) at (-1,0);
\coordinate (p12) at (-.4,0);
\coordinate (p3) at (0,1);
\coordinate (p13) at (0,.4);
\coordinate (p34) at (-.4,1);
\coordinate (p4) at (-1,1);
\coordinate (p24) at (-1,.4);
\coordinate (p23) at (-.45,.55);
\draw [->,>=latex,line width=0.15pt] (-1.5,0) -- (1,0);
\draw [->,>=latex,line width=0.15pt] (0,-0.5) -- (0,1.5);
\draw [-<,>=latex,line width=0.4pt] (p2) -- (p12);
\draw [-,>=latex,line width=0.4pt] (p12) -- (p1);
\draw [-<,>=latex,line width=0.4pt] (p4) -- (p34);
\draw [-,>=latex,line width=0.4pt] (p34) -- (p3);
\draw [-<,>=latex,line width=0.4pt] (p3) -- (p13);
\draw [-,>=latex,line width=0.4pt] (p13) -- (p1);
\draw [-<,>=latex,line width=0.4pt] (p4) -- (p24);
\draw [-,>=latex,line width=0.4pt] (p24) -- (p2);
\draw [-<,>=latex,line width=0.4pt] (p2) -- (p23);
\draw [-,>=latex,line width=0.4pt] (p23) -- (p3);
\node at (0.15,-0.15) {1'};
\node at (-1.15,-0.15) {3'};
\node at (0.15,1.15) {2'};
\node at (-1.15,1.15) {4'};
\node at (1.35,0.02) {$x$};
\node at (0,1.75) {$y$};
\end{tikzpicture}
}

\newcommand{\SLTWOtransformTL}{
\begin{tikzpicture}[scale=0.8,baseline]
\coordinate (p1) at (0,0);
\coordinate (p12) at (0,0.4);
\coordinate (p2) at (0,1);
\coordinate (p13) at (.4,0);
\coordinate (p3) at (1,0);
\coordinate (p24) at (.4,1);
\coordinate (p34) at (1,0.4);
\coordinate (p4) at (1,1);
\coordinate (p14) at (.4,.4);
\coordinate (p23) at (.4,.6);
\coordinate (p0) at (-1,1);
\coordinate (p02) at (-.7,1);
\coordinate (p-1) at (-1,0);
\coordinate (p0-1) at (-1, .4);
\coordinate (p1-1) at (-.7, 0);
\draw [->,>=latex,line width=0.15pt] (-.1,0) -- (1.5,0);
\draw [->,>=latex,line width=0.15pt] (0,-0.5) -- (0,1.5);
\draw [-<,>=latex,line width=0.4pt] (p2) -- (p12);
\draw [-<,>=latex,line width=0.4pt] (p4) -- (p34);
\draw [-,>=latex,line width=0.4pt] (p34) -- (p3);
\draw [-<,>=latex,line width=0.4pt] (p3) -- (p13);
\draw [-,>=latex,line width=0.4pt] (p13) -- (p1);
\draw [-<,>=latex,line width=0.4pt] (p4) -- (p24);
\draw [-,>=latex,line width=0.4pt] (p24) -- (p2);
\draw [->,>=latex,line width=0.4pt] (p3) -- (p23);
\draw [-,>=latex,line width=0.4pt] (p23) -- (p2);
\draw [-<,>=latex, dotted, line width=0.4pt] (p2) -- (p02);
\draw [-,>=latex, dotted, line width=0.4pt] (p02) -- (p0);
\draw [-<,>=latex, dotted, line width=0.4pt] (p0) -- (p0-1);
\draw [-,>=latex, dotted, line width=0.4pt] (p0-1) -- (p-1);
\draw [-<,>=latex, dotted, line width=0.4pt] (p1) -- (p1-1);
\draw [-,>=latex, dotted, line width=0.4pt] (p1-1) -- (p-1);
\coordinate (p01) at (-.6, .6);
\draw [->,>=latex, dotted, line width=0.4pt] (p1) -- (p01);
\draw [-,>=latex, dotted, line width=0.4pt] (p01) -- (p0);
\coordinate (p6) at (2,1);
\coordinate (p46) at (1.4,1);
\draw [-<,>=latex, dotted, line width=0.4pt] (p6) -- (p46);
\draw [-,>=latex, dotted, line width=0.4pt] (p46) -- (p4);
\coordinate (p5) at (2,0);
\coordinate (p56) at (2,.4);
\draw [-<,>=latex, dotted, line width=0.4pt] (p6) -- (p56);
\draw [-,>=latex, dotted, line width=0.4pt] (p56) -- (p5);
\coordinate (p45) at (1.4,0.6);
\draw [->,>=latex, dotted, line width=0.4pt] (p5) -- (p45);
\draw [-,>=latex, dotted, line width=0.4pt] (p45) -- (p4);
\node at (-0.15,-0.15) {1};
\node at (-0.15,1.15) {3}; 
\node at (1.15,-0.185) {2}; 
\node at (1.15,1.15) {4};
\node at (-1.15,1.15) {$2^*$}; 
\node at (1.85,0.02) {$x$};
\node at (0,1.75) {$y$};
\node at (.5,-0.185) {$g_x$};
\node at (-0.25,.6) {$g_y$};
\end{tikzpicture}
}

\newcommand{\SLTWOtransformTR}{
\begin{tikzpicture}[scale=0.8,baseline]
\coordinate (p1) at (0,0);
\coordinate (p12) at (0.45,0.45);
\coordinate (p2) at (1,1);
\coordinate (p3) at (1,0);
\coordinate (p13) at (.45,0);
\coordinate (p14) at (.9,.45);
\coordinate (p24) at (1.35,1);
\coordinate (p34) at (1.425,.425);
\coordinate (p4) at (2,1);
\coordinate (p23) at (1., .65);
\coordinate (p0) at (0,1);
\coordinate (p02) at (0.4,1);
\draw [->,>=latex,line width=0.15pt] (-0.5,0) -- (1.5,0);
\draw [->,>=latex,line width=0.15pt] (0,-0.5) -- (0,1.5);
\draw [-<,>=latex,line width=0.4pt] (p2) -- (p12);
\draw [-,>=latex,line width=0.4pt] (p12) -- (p1);
\draw [-<,>=latex,line width=0.4pt] (p4) -- (p34);
\draw [-,>=latex,line width=0.4pt] (p34) -- (p3);
\draw [-<,>=latex,line width=0.4pt] (p3) -- (p13);
\draw [-,>=latex,line width=0.4pt] (p13) -- (p1);
\draw [-<,>=latex,line width=0.4pt] (p4) -- (p24);
\draw [-,>=latex,line width=0.4pt] (p24) -- (p2);
\draw [->,>=latex,line width=0.4pt] (p3) -- (p23);
\draw [-,>=latex,line width=0.4pt] (p23) -- (p2);
\draw [-<,>=latex, dotted, line width=0.4pt] (p2) -- (p02);
\draw [-,>=latex, dotted, line width=0.4pt] (p02) -- (p0);
\coordinate (p01) at (0,.65);
\draw [->,>=latex, dotted, line width=0.4pt] (p1) -- (p01);
\draw [-,>=latex, dotted, line width=0.4pt] (p01) -- (p0);
\coordinate (p5) at (2,0);
\coordinate (p45) at (2,.65);
\draw [->,>=latex, dotted, line width=0.4pt] (p5) -- (p45);
\draw [-,>=latex, dotted, line width=0.4pt] (p45) -- (p4);
\node at (-0.175,-0.18) {1'};
\node at (0.95,1.25) {3'};
\node at (1.15,-0.18) {2'};
\node at (2.15,1.25) {4'};
\node at (-0.3,1.25) {$2^*$'}; 
\node at (1.85,0.02) {$x$};
\node at (0,1.75) {$y$};
\end{tikzpicture}
}

\newcommand{\SLTHREEtransformSLeft}{
\begin{tikzpicture}[scale=0.8,baseline]
\coordinate (p1) at (0,0);
\coordinate (p15) at (0,0.5);
\coordinate (p5) at (0,1);
\coordinate (p12) at (.5,0);
\coordinate (p2) at (1,0);
\coordinate (p56) at (.5,1);
\coordinate (p6) at (1,1);
\coordinate (p26) at (1,.5);
\coordinate (p4) at (1+.5,0.35);
\coordinate (p8) at (1+.5,1.35); 
\coordinate (p3) at (.5,0.35);
\coordinate (p34) at (1.,0.35);
\coordinate (p13) at (.25,0.175); 
\coordinate (p24) at (1.25,0.175); 
\coordinate (p7) at (.5,1.35);
\coordinate (p37) at (.5,0.75); 
\coordinate (p57) at (.25,1.175); 
\coordinate (p78) at (1,1.35);
\coordinate (p68) at (1.25,1.175);
\coordinate (p8) at (1.5,1.35);
\coordinate (p48) at (1.5,.75);
\draw [->,>=latex,line width=0.15pt] (-.5,0) -- (2.1,0); 
\draw [->,>=latex,line width=0.15pt] (0,-0.5) -- (0,1.5); 
\draw [->,>=latex,line width=0.15pt] (0,0) -- (1.8,1.26);  
\draw [>-,>=latex,line width=0.4pt] (p12) -- (p2);
\draw [-,>=latex,line width=0.4pt] (p2) -- (p26);
\draw [>-,>=latex,line width=0.4pt] (p26) -- (p6);
\draw [-,>=latex,line width=0.4pt] (p56) -- (p5);
\draw [>-,>=latex,line width=0.4pt] (p56) -- (p6);
\draw [-,>=latex,line width=0.4pt] (p1) -- (p15);
\draw [>-,>=latex,line width=0.4pt] (p15) -- (p5);
\draw [-,>=latex,line width=0.4pt] (p2) -- (p24);
\draw [>-,>=latex,line width=0.4pt] (p24) -- (p4);
\draw [-,>=latex,line width=0.4pt] (p3) -- (p34);
\draw [>-,>=latex,line width=0.4pt] (p34) -- (p4);
\draw [-,>=latex,line width=0.4pt] (p4) -- (p8);
\draw [-,>=latex,line width=0.4pt] (p1) -- (p13);
\draw [>-,>=latex,line width=0.4pt] (p13) -- (p3);
\draw [-,>=latex,line width=0.4pt] (p3) -- (p37);
\draw [>-,>=latex,line width=0.4pt] (p37) -- (p7);
\draw [-,>=latex,line width=0.4pt] (p5) -- (p57);
\draw [>-,>=latex,line width=0.4pt] (p57) -- (p7);
\draw [-,>=latex,line width=0.4pt] (p7) -- (p78);
\draw [>-,>=latex,line width=0.4pt] (p78) -- (p8);
\draw [-,>=latex,line width=0.4pt] (p6) -- (p68);
\draw [>-,>=latex,line width=0.4pt] (p68) -- (p8);
\draw [-,>=latex,line width=0.4pt] (p4) -- (p48);
\draw [>-,>=latex,line width=0.4pt] (p48) -- (p8);
\node at (-0.15,-0.15) {1};
\node at (-0.15,1.15) {5};
\node at (1.,-0.185) {2};
\node at (.3,0.45) {3};
\node at (1.65,0.45) {4};
\node at (1.,1.15) {6};
\node at (.3,1.45) {7};
\node at (1.65,1.45) {8};
\node at (2.25,0.02) {$x$};
\node at (2,1.4) {$y$};
\node at (0,1.75) {$z$};
\node at (.5,-0.185) {$g_x$};
\node at (1.6,0.125) {$g_y$};
\node at (-0.25,.6) {$g_z$};
\end{tikzpicture}
}

\newcommand{\SLTHREEtransformSR}{
\begin{tikzpicture}[scale=0.8,baseline]
\coordinate (p1) at (0,0);
\coordinate (p15) at (0,0.5);
\coordinate (p5) at (0,1);
\coordinate (p12) at (.5,0);
\coordinate (p2) at (1,0);
\coordinate (p56) at (.5,1);
\coordinate (p6) at (1,1);
\coordinate (p26) at (1,.5);
\coordinate (p4) at (1+.5,0.35);
\coordinate (p8) at (1+.5,1.35); 
\coordinate (p3) at (.5,0.35);
\coordinate (p34) at (1.,0.35);
\coordinate (p13) at (.25,0.175); 
\coordinate (p24) at (1.25,0.175); 
\coordinate (p7) at (.5,1.35);
\coordinate (p37) at (.5,0.75); 
\coordinate (p57) at (.25,1.175); 
\coordinate (p78) at (1,1.35);
\coordinate (p68) at (1.25,1.175);
\coordinate (p8) at (1.5,1.35);
\coordinate (p48) at (1.5,.75);
\draw [->,>=latex,line width=0.15pt] (-.5,0) -- (1.8,0); 
\draw [->,>=latex,line width=0.15pt] (0,-0.5) -- (0,1.5); 
\draw [->,>=latex,line width=0.15pt] (0,0) -- (1.8,1.26);  
\draw [>-,>=latex,line width=0.4pt] (p12) -- (p2);
\draw [-,>=latex,line width=0.4pt] (p2) -- (p26);
\draw [>-,>=latex,line width=0.4pt] (p26) -- (p6);
\draw [-,>=latex,line width=0.4pt] (p56) -- (p5);
\draw [>-,>=latex,line width=0.4pt] (p56) -- (p6);
\draw [-,>=latex,line width=0.4pt] (p1) -- (p15);
\draw [>-,>=latex,line width=0.4pt] (p15) -- (p5);
\draw [-,>=latex,line width=0.4pt] (p2) -- (p24);
\draw [>-,>=latex,line width=0.4pt] (p24) -- (p4);
\draw [-,>=latex,line width=0.4pt] (p3) -- (p34);
\draw [>-,>=latex,line width=0.4pt] (p34) -- (p4);
\draw [-,>=latex,line width=0.4pt] (p4) -- (p8);
\draw [-,>=latex,line width=0.4pt] (p1) -- (p13);
\draw [>-,>=latex,line width=0.4pt] (p13) -- (p3);
\draw [-,>=latex,line width=0.4pt] (p3) -- (p37);
\draw [>-,>=latex,line width=0.4pt] (p37) -- (p7);
\draw [-,>=latex,line width=0.4pt] (p5) -- (p57);
\draw [>-,>=latex,line width=0.4pt] (p57) -- (p7);
\draw [-,>=latex,line width=0.4pt] (p7) -- (p78);
\draw [>-,>=latex,line width=0.4pt] (p78) -- (p8);
\draw [-,>=latex,line width=0.4pt] (p6) -- (p68);
\draw [>-,>=latex,line width=0.4pt] (p68) -- (p8);
\draw [-,>=latex,line width=0.4pt] (p4) -- (p48);
\draw [>-,>=latex,line width=0.4pt] (p48) -- (p8);
\node at (-0.15,-0.15) {1'};
\node at (-0.15,1.15) {3'};
\node at (1.,-0.185) {5'};
\node at (.3,0.45) {2'};
\node at (1.65,0.45) {6'};
\node at (1.,1.15) {7'};
\node at (.3,1.45) {4'};
\node at (1.65,1.45) {8'};
\node at (2.05,0.02) {$x$};
\node at (2,1.4) {$y$};
\node at (0,1.75) {$z$};
\node at (.5,-0.185) {};
\node at (-0.25,.6) {};
\end{tikzpicture}
}

\newcommand{\CocycleTriangleTWO}[5]{
\begin{tikzpicture}[scale=0.9,baseline]
\coordinate (c) at (0,0);
\coordinate (d) at (0,1.1);
\coordinate (a) at (210:1.1);
\coordinate (b) at (330:1.1);

\draw (a) node[left] {\scalebox{0.85}{$#1$}} -- (b) node[right] {\scalebox{0.85}{$#2$}} -- (c) node[below] {\scalebox{0.85}{$#3$}} -- (d) node[above] {\scalebox{0.85}{$#4$}} -- (a) -- (c);
\draw (b) -- (d);
\draw[->,>=latex,line width=0.65pt] (a) -- ($ (b) ! 0.5 ! (a) $); 
\draw[-<,>=latex,line width=0.65pt] (a) -- ($ (a) ! 0.5 ! (d) $);
\draw[-<,>=latex,line width=0.65pt] (b) -- ($ (b) ! 0.5 ! (d) $);

\ifnum #5=1 {
\draw[-<,>=latex,line width=0.65pt] (b) -- ($ (b) ! 0.5 ! (c) $);
\draw[->,>=latex,line width=0.65pt] (c) -- ($ (c) ! 0.5 ! (d) $);
\draw[-<,>=latex,line width=0.65pt] (a) -- ($ (a) ! 0.5 ! (c) $);}
\fi
\ifnum #5=2 {
\draw[-<,>=latex,line width=0.65pt] (b) -- ($ (b) ! 0.5 ! (c) $);
\draw[-<,>=latex,line width=0.65pt] (c) -- ($ (c) ! 0.5 ! (d) $);
\draw[-<,>=latex,line width=0.65pt] (a) -- ($ (a) ! 0.5 ! (c) $);}
\fi
\ifnum #5=3 {
\draw[->,>=latex,line width=0.65pt] (b) -- ($ (b) ! 0.5 ! (c) $);
\draw[-<,>=latex,line width=0.65pt] (c) -- ($ (c) ! 0.5 ! (d) $);
\draw[-<,>=latex,line width=0.65pt] (a) -- ($ (a) ! 0.5 ! (c) $);}
\fi
\ifnum #5=4 {
\draw[->,>=latex,line width=0.65pt] (b) -- ($ (b) ! 0.5 ! (c) $);
\draw[-<,>=latex,line width=0.65pt] (c) -- ($ (c) ! 0.5 ! (d) $);
\draw[->,>=latex,line width=0.65pt] (a) -- ($ (a) ! 0.5 ! (c) $);}
\fi
\end{tikzpicture}
}

\newcommand{\CocycleTriangle}[5]{
\begin{tikzpicture}[scale=0.9,baseline]
\coordinate (c) at (0,0);
\coordinate (d) at (0,1.1);
\coordinate (a) at (210:1.1);
\coordinate (b) at (330:1.1);

\draw (a) node[left] {\scalebox{0.85}{$#1$}} -- (b) node[right] {\scalebox{0.85}{$#2$}} -- (c) node[below] {\scalebox{0.85}{$#3$}} -- (d) node[above] {\scalebox{0.85}{$#4$}} -- (a) -- (c);
\draw (b) -- (d);
\draw[-<,>=latex,line width=0.65pt] (a) -- ($ (a) ! 0.5 ! (b) $);
\draw[-<,>=latex,line width=0.65pt] (a) -- ($ (a) ! 0.5 ! (d) $);
\draw[-<,>=latex,line width=0.65pt] (b) -- ($ (b) ! 0.5 ! (d) $);

\ifnum #5=1 {
\draw[-<,>=latex,line width=0.65pt] (b) -- ($ (b) ! 0.5 ! (c) $);
\draw[->,>=latex,line width=0.65pt] (c) -- ($ (c) ! 0.5 ! (d) $);
\draw[-<,>=latex,line width=0.65pt] (a) -- ($ (a) ! 0.5 ! (c) $);}
\fi
\ifnum #5=2 {
\draw[-<,>=latex,line width=0.65pt] (b) -- ($ (b) ! 0.5 ! (c) $);
\draw[-<,>=latex,line width=0.65pt] (c) -- ($ (c) ! 0.5 ! (d) $);
\draw[-<,>=latex,line width=0.65pt] (a) -- ($ (a) ! 0.5 ! (c) $);}
\fi
\ifnum #5=3 {
\draw[->,>=latex,line width=0.65pt] (b) -- ($ (b) ! 0.5 ! (c) $);
\draw[-<,>=latex,line width=0.65pt] (c) -- ($ (c) ! 0.5 ! (d) $);
\draw[-<,>=latex,line width=0.65pt] (a) -- ($ (a) ! 0.5 ! (c) $);}
\fi
\ifnum #5=4 {
\draw[->,>=latex,line width=0.65pt] (b) -- ($ (b) ! 0.5 ! (c) $);
\draw[-<,>=latex,line width=0.65pt] (c) -- ($ (c) ! 0.5 ! (d) $);
\draw[->,>=latex,line width=0.65pt] (a) -- ($ (a) ! 0.5 ! (c) $);}
\fi
\end{tikzpicture}
}

\newcommand{\tetrahedraTWO}[4]{
\begin{tikzpicture}[scale=1,baseline]
\coordinate (a) at (0,-0.1);
\coordinate (b) at (0.82,-0.79);
\coordinate (c) at (2.5,0);
\coordinate (d) at (1.25,1.5);
\coordinate (e) at ($ (a) ! 0.32 ! (c) $);
\coordinate (f) at ($ (a) ! 0.525 ! (c) $);
\draw (a) -- (e);
\draw (b) -- (d);
\draw (f) -- (c);
\draw (a) -- (b) -- (c) -- (d) -- cycle;
\draw[->,>=latex, line width=0.75pt] (a) -- ($ (a) ! 0.5 ! (b) $);
\draw[-<,>=latex, line width=0.75pt] (b) -- ($ (b) ! 0.5 ! (c) $);
\draw[-<,>=latex, line width=0.75pt] (c) -- ($ (c) ! 0.5 ! (d) $);
\draw[<-,>=latex, line width=0.75pt] (f) -- (c);
\draw[-<,>=latex, line width=0.75pt] (b) -- ($ (b) ! 0.5 ! (d) $);
\draw[-<,>=latex, line width=0.75pt] (a) -- ($ (a) ! 0.5 ! (d) $);

\node[left] at(a) {\scalebox{0.7}{$#1$}};
\node[below] at(b) {\scalebox{0.7}{$#2$}};
\node[right] at(c) {\scalebox{0.7}{$#3$}};
\node[above] at(d) {\scalebox{0.7}{$#4$}};

\node[below] at ($ (a) ! 0.4 ! (b) $) {\scalebox{0.7}{$#2{#1}^{-1}$}};
\node[below] at ($ (b) ! 0.65 ! (c) $) {\scalebox{0.7}{$#2{#3}^{-1}$}};
\node[right] at ($ (c) ! 0.5 ! (d) $) {\scalebox{0.7}{$#3{#4}^{-1}$}};
\node[below] at (f) {\scalebox{0.7}{$#1{#3}^{-1}$}};
\node[right] at ($ (b) ! 0.49 ! (d) $) {\scalebox{0.7}{$#2{#4}^{-1}$}};
\node[left] at ($ (a) ! 0.51 ! (d) $) {\scalebox{0.7}{$#1{#4}^{-1}$}};
\end{tikzpicture}
}

\newcommand{\tetrahedra}[4]{
\begin{tikzpicture}[scale=1,baseline]
\coordinate (a) at (0,-0.1);
\coordinate (b) at (0.82,-0.79);
\coordinate (c) at (2.5,0);
\coordinate (d) at (1.25,1.5);
\coordinate (e) at ($ (a) ! 0.32 ! (c) $);
\coordinate (f) at ($ (a) ! 0.525 ! (c) $);
\draw (a) -- (e);
\draw (b) -- (d);
\draw (f) -- (c);
\draw (a) -- (b) -- (c) -- (d) -- cycle;
\draw[-<,>=latex, line width=0.75pt] (a) -- ($ (a) ! 0.5 ! (b) $);
\draw[-<,>=latex, line width=0.75pt] (b) -- ($ (b) ! 0.5 ! (c) $);
\draw[-<,>=latex, line width=0.75pt] (c) -- ($ (c) ! 0.5 ! (d) $);
\draw[<-,>=latex, line width=0.75pt] (f) -- (c);
\draw[-<,>=latex, line width=0.75pt] (b) -- ($ (b) ! 0.5 ! (d) $);
\draw[-<,>=latex, line width=0.75pt] (a) -- ($ (a) ! 0.5 ! (d) $);

\node[left] at(a) {\scalebox{0.7}{$#1$}};
\node[below] at(b) {\scalebox{0.7}{$#2$}};
\node[right] at(c) {\scalebox{0.7}{$#3$}};
\node[above] at(d) {\scalebox{0.7}{$#4$}};

\node[below] at ($ (a) ! 0.4 ! (b) $) {\scalebox{0.7}{$#1{#2}^{-1}$}};
\node[below] at ($ (b) ! 0.65 ! (c) $) {\scalebox{0.7}{$#2{#3}^{-1}$}};
\node[right] at ($ (c) ! 0.5 ! (d) $) {\scalebox{0.7}{$#3{#4}^{-1}$}};
\node[below] at (f) {\scalebox{0.7}{$#1{#3}^{-1}$}};
\node[right] at ($ (b) ! 0.49 ! (d) $) {\scalebox{0.7}{$#2{#4}^{-1}$}};
\node[left] at ($ (a) ! 0.51 ! (d) $) {\scalebox{0.7}{$#1{#4}^{-1}$}};
\end{tikzpicture}
}

\newcommand*{\DashedArrow}[1][]{\mathbin{\tikz [baseline=-0.25ex,-latex, dashed,#1] \draw [#1] (0pt,0.5ex) -- (1.3em,0.5ex);}}%

\newcommand{\CaRep}{
\begin{tikzpicture}[scale=0.8,baseline]
\coordinate (p1) at (0,0);
\coordinate (p15) at (0,0.5);
\coordinate (p5) at (0,1);
\coordinate (p12) at (.5,0);
\coordinate (p2) at (1,0);
\coordinate (p56) at (.5,1);
\coordinate (p6) at (1,1);
\coordinate (p26) at (1,.5);
\coordinate (p4) at (1+.5,0.35);
\coordinate (p8) at (1+.5,1.35); 
\coordinate (p3) at (.5,0.35);
\coordinate (p34) at (1.,0.35);
\coordinate (p13) at (.25,0.175); 
\coordinate (p24) at (1.25,0.175); 
\coordinate (p7) at (.5,1.35);
\coordinate (p37) at (.5,0.75); 
\coordinate (p57) at (.25,1.175); 
\coordinate (p78) at (1,1.35);
\coordinate (p68) at (1.25,1.175);
\coordinate (p8) at (1.5,1.35);
\coordinate (p48) at (1.5,.75);
\coordinate (p14) at (0.75,0.175);
\coordinate (p58) at (0.75, 1.175);
\draw [->,>=latex,line width=0.15pt] (-.5,0) -- (1.8,0); 
\draw [-,>=latex, line width=0.15pt] (0,-0.5) -- (0,0); 
\draw [->,>=latex,line width=0.15pt] (0,1) -- (0,1.65); 
\draw [->,>=latex,line width=0.15pt] (.5,0.35) -- (1.8,1.26);  
\draw [>-,>=latex,line width=0.4pt] (p12) -- (p2);
\draw [-,>=latex, dashed, line width=0.4pt] (p2) -- (p26);
\draw [>-,>=latex, dashed, line width=0.4pt] (p26) -- (p6);
\draw [-,>=latex,line width=0.4pt] (p56) -- (p5);
\draw [>-,>=latex,line width=0.4pt] (p56) -- (p6);
\draw [-,>=latex, dashed, line width=0.4pt] (p1) -- (p15);
\draw [>-,>=latex, dashed, line width=0.4pt] (p15) -- (p5);
\draw [-,>=latex, line width=0.4pt] (p2) -- (p24);
\draw [>-,>=latex, line width=0.4pt] (p24) -- (p4);
\draw [-,>=latex, dotted, line width=0.4pt] (p3) -- (p34);
\draw [-,>=latex, dotted, line width=0.4pt] (p34) -- (p4);
\draw [-,>=latex, dotted, line width=0.4pt] (p1) -- (p13);
\draw [-,>=latex, dotted, line width=0.4pt] (p13) -- (p3);
\draw [-,dotted,>=latex,line width=0.4pt] (p3) -- (p37);
\draw [-, dotted, >=latex,line width=0.4pt] (p37) -- (p7);
\draw [-,>=latex, dotted, line width=0.4pt] (p5) -- (p57);
\draw [-,>=latex, dotted, line width=0.4pt] (p57) -- (p7);
\draw [-,>=latex, dotted, line width=0.4pt] (p7) -- (p78);
\draw [-,>=latex, dotted, line width=0.4pt] (p78) -- (p8);
\draw [-,>=latex,line width=0.4pt] (p6) -- (p68);
\draw [>-,>=latex,line width=0.4pt] (p68) -- (p8);
\draw [-,>=latex, dashed, line width=0.4pt] (p4) -- (p48);
\draw [>-,>=latex, dashed, line width=0.4pt] (p48) -- (p8);
\draw [-,>=latex, line width=0.4pt] (p1) -- (p14);
\draw [>-,>=latex, line width=0.4pt] (p14) -- (p4);
\draw [-,>=latex, line width=0.4pt] (p5) -- (p58);
\draw [>-,>=latex, line width=0.4pt] (p58) -- (p8);
\node at (-0.15,-0.15) {1};
\node at (-0.175,1.15) {1'};
\node at (1.,-0.185) {2};
\node at (.3,0.45) {3};
\node at (1.65,0.45) {4};
\node at (1.,1.15) {2'};
\node at (.3,1.45) {3'};
\node at (1.65,1.45) {4'};
\node at (2.05,0.02) {$x$};
\node at (2,1.4) {$y$};
\node at (0,1.8) {$t$};
\node at (-0.25,.6) {$c$};   
\node at (1.6,0.125) {$a$}; 
\node at (.5,-0.185) {$b$};  
\end{tikzpicture}
}

\newcommand{\CabRepL}{
\begin{tikzpicture}[scale=0.8,baseline]
\coordinate (p1) at (0,0);
\coordinate (p15) at (0,0.5);
\coordinate (p5) at (0,1);
\coordinate (p12) at (.5,0);
\coordinate (p2) at (1,0);
\coordinate (p56) at (.5,1);
\coordinate (p6) at (1,1);
\coordinate (p26) at (1,.5);
\coordinate (p4) at (1+.5,0.35);
\coordinate (p8) at (1+.5,1.35); 
\coordinate (p3) at (.5,0.35);
\coordinate (p34) at (1.,0.35);
\coordinate (p13) at (.25,0.175); 
\coordinate (p24) at (1.25,0.175); 
\coordinate (p7) at (.5,1.35);
\coordinate (p37) at (.5,0.75); 
\coordinate (p57) at (.25,1.175); 
\coordinate (p78) at (1,1.35);
\coordinate (p68) at (1.25,1.175);
\coordinate (p8) at (1.5,1.35);
\coordinate (p48) at (1.5,.75);
\coordinate (p14) at (0.75,0.175);
\coordinate (p58) at (0.75, 1.175);
\draw [->,>=latex,line width=0.15pt] (-.5,0) -- (1.8,0); 
\draw [->,>=latex,line width=0.15pt] (0,-0.5) -- (0,1.65); 
\draw [->,>=latex,line width=0.15pt] (.5,0.35) -- (1.8,1.26);  
\draw [>-,>=latex,line width=0.4pt] (p12) -- (p2);
\draw [-,>=latex,line width=0.4pt] (p2) -- (p26);
\draw [>-,>=latex,line width=0.4pt] (p26) -- (p6);
\draw [-,>=latex,line width=0.4pt] (p56) -- (p5);
\draw [>-,>=latex,line width=0.4pt] (p56) -- (p6);
\draw [-,>=latex,line width=0.4pt] (p1) -- (p15);
\draw [>-,>=latex,line width=0.4pt] (p15) -- (p5);
\draw [-,>=latex, line width=0.4pt] (p2) -- (p24);
\draw [>-,>=latex, line width=0.4pt] (p24) -- (p4);
\draw [-,>=latex, dotted, line width=0.4pt] (p3) -- (p34);
\draw [-,>=latex, dotted, line width=0.4pt] (p34) -- (p4);
\draw [-,>=latex,line width=0.4pt] (p4) -- (p8);
\draw [-,>=latex, dotted, line width=0.4pt] (p1) -- (p13);
\draw [-,>=latex, dotted, line width=0.4pt] (p13) -- (p3);
\draw [-,dotted,>=latex,line width=0.4pt] (p3) -- (p37);
\draw [-, dotted, >=latex,line width=0.4pt] (p37) -- (p7);
\draw [-,>=latex, dotted, line width=0.4pt] (p5) -- (p57);
\draw [-,>=latex, dotted, line width=0.4pt] (p57) -- (p7);
\draw [-,>=latex, dotted, line width=0.4pt] (p7) -- (p78);
\draw [-,>=latex, dotted, line width=0.4pt] (p78) -- (p8);
\draw [-,>=latex,line width=0.4pt] (p6) -- (p68);
\draw [>-,>=latex,line width=0.4pt] (p68) -- (p8);
\draw [-,>=latex,line width=0.4pt] (p4) -- (p48);
\draw [>-,>=latex,line width=0.4pt] (p48) -- (p8);
\draw [-,>=latex, line width=0.4pt] (p1) -- (p14);
\draw [>-,>=latex, line width=0.4pt] (p14) -- (p4);
\draw [-,>=latex, line width=0.4pt] (p5) -- (p58);
\draw [>-,>=latex, line width=0.4pt] (p58) -- (p8);
\node at (-0.15,-0.15) {1};
\node at (-0.175,1.15) {5};
\node at (1.,-0.185) {2};
\node at (.3,0.45) {3};
\node at (1.65,0.45) {4};
\node at (1.,1.15) {6};
\node at (.3,1.45) {7};
\node at (1.65,1.45) {8};
\node at (2.05,0.02) {$x$};
\node at (2,1.4) {$y$};
\node at (0,1.8) {$z$};
\node at (1.75,.9) {$b$};   
\node at (1.5,0.16) {$a$}; 
\node at (.5,-0.185) {$c$};  
\end{tikzpicture}
}

\newcommand{\CabRepR}{
\begin{tikzpicture}[scale=0.8,baseline]
\coordinate (p1) at (0,0);
\coordinate (p15) at (0,0.5);
\coordinate (p5) at (0,1);
\coordinate (p12) at (.5,0);
\coordinate (p2) at (1,0);
\coordinate (p56) at (.5,1);
\coordinate (p6) at (1,1);
\coordinate (p26) at (1,.5);
\coordinate (p4) at (1+.5,0.35);
\coordinate (p8) at (1+.5,1.35); 
\coordinate (p3) at (.5,0.35);
\coordinate (p34) at (1.,0.35);
\coordinate (p13) at (.25,0.175); 
\coordinate (p24) at (1.25,0.175); 
\coordinate (p7) at (.5,1.35);
\coordinate (p37) at (.5,0.75); 
\coordinate (p57) at (.25,1.175); 
\coordinate (p78) at (1,1.35);
\coordinate (p68) at (1.25,1.175);
\coordinate (p8) at (1.5,1.35);
\coordinate (p48) at (1.5,.75);
\coordinate (p14) at (0.75,0.175);
\coordinate (p58) at (0.75, 1.175);
\draw [->,>=latex,line width=0.15pt] (-.5,0) -- (1.8,0); 
\draw [->,>=latex,line width=0.15pt] (0,-0.5) -- (0,1.65); 
\draw [->,>=latex,line width=0.15pt] (.5,0.35) -- (1.8,1.26);  
\draw [>-,>=latex,line width=0.4pt] (p12) -- (p2);
\draw [-,>=latex,line width=0.4pt] (p2) -- (p26);
\draw [>-,>=latex,line width=0.4pt] (p26) -- (p6);
\draw [-,>=latex,line width=0.4pt] (p56) -- (p5);
\draw [>-,>=latex,line width=0.4pt] (p56) -- (p6);
\draw [-,>=latex,line width=0.4pt] (p1) -- (p15);
\draw [>-,>=latex,line width=0.4pt] (p15) -- (p5);
\draw [-,>=latex, line width=0.4pt] (p2) -- (p24);
\draw [>-,>=latex, line width=0.4pt] (p24) -- (p4);
\draw [-,>=latex, dotted, line width=0.4pt] (p3) -- (p34);
\draw [-,>=latex, dotted, line width=0.4pt] (p34) -- (p4);
\draw [-,>=latex,line width=0.4pt] (p4) -- (p8);
\draw [-,>=latex, dotted, line width=0.4pt] (p1) -- (p13);
\draw [-,>=latex, dotted, line width=0.4pt] (p13) -- (p3);
\draw [-,dotted,>=latex,line width=0.4pt] (p3) -- (p37);
\draw [-, dotted, >=latex,line width=0.4pt] (p37) -- (p7);
\draw [-,>=latex, dotted, line width=0.4pt] (p5) -- (p57);
\draw [-,>=latex, dotted, line width=0.4pt] (p57) -- (p7);
\draw [-,>=latex, dotted, line width=0.4pt] (p7) -- (p78);
\draw [-,>=latex, dotted, line width=0.4pt] (p78) -- (p8);
\draw [-,>=latex,line width=0.4pt] (p6) -- (p68);
\draw [>-,>=latex,line width=0.4pt] (p68) -- (p8);
\draw [-,>=latex,line width=0.4pt] (p4) -- (p48);
\draw [>-,>=latex,line width=0.4pt] (p48) -- (p8);
\draw [-,>=latex, line width=0.4pt] (p1) -- (p14);
\draw [>-,>=latex, line width=0.4pt] (p14) -- (p4);
\draw [-,>=latex, line width=0.4pt] (p5) -- (p58);
\draw [>-,>=latex, line width=0.4pt] (p58) -- (p8);
\node at (-0.15,-0.15) {1'};
\node at (-0.175,1.15) {5'};
\node at (1.,-0.185) {2'};
\node at (.3,0.45) {3'};
\node at (1.65,0.45) {4'};
\node at (1.,1.15) {6'};
\node at (.3,1.45) {7'};
\node at (1.65,1.45) {8'};
\node at (2.05,0.02) {$x$};
\node at (2,1.4) {$y$};
\node at (0,1.8) {$z$};
\end{tikzpicture}
}

\newcommand{\TQDthreeD}[6]{
\begin{tikzpicture}[scale=1,baseline]
\coordinate (a) at (0,-0.1);
\coordinate (b) at (0.82,-0.79);
\coordinate (c) at (2.5,0);
\coordinate (d) at (1.25,1.5);
\coordinate (e) at ($ (a) ! 0.32 ! (c) $);
\coordinate (f) at ($ (a) ! 0.65 ! (c) $);
\coordinate (g) at ($ (a) ! 0.55 ! (c) $);
\coordinate (v4) at (2.25,1.6);
\coordinate (center) at (1.2,0.1525);
\draw (a) -- (e);
\draw (b) -- (d);
\draw (f) -- (c);
\draw (a) -- (b) -- (c) -- (d) -- cycle;
\draw[-,>=latex, line width=0.4pt] (a) -- ($ (a) ! 0.5 ! (b) $);
\draw[-,>=latex, line width=0.4pt] (b) -- ($ (b) ! 0.5 ! (c) $);
\draw[-,>=latex, line width=0.4pt] (c) -- ($ (c) ! 0.5 ! (d) $);
\draw[-,>=latex, line width=0.4pt] (g) -- (f);
\draw[-,>=latex, line width=0.4pt] (f) -- (c);
\draw[-,>=latex, line width=0.4pt] (b) -- ($ (b) ! 0.5 ! (d) $);
\draw[-,>=latex, line width=0.4pt] (a) -- ($ (a) ! 0.5 ! (d) $);
\draw[-,>=latex, dashed, line width=0.4pt] (a) -- ($ (a) ! 0.5 ! (v4) $);
\draw[-,=latex, dashed, line width=0.4pt] ($ (a) ! 0.5 ! (v4) $) -- (v4);
\draw[-,>=latex, dashed, line width=0.4pt] (b) -- ($ (b) ! 0.5 ! (v4) $);
\draw[-,=latex, dashed, line width=0.4pt] ($ (b) ! 0.5 ! (v4) $) -- (v4);
\draw[-,>=latex, dashed, line width=0.4pt] (c) -- ($ (c) ! 0.5 ! (v4) $);
\draw[-,=latex, dashed, line width=0.4pt] ($ (c) ! 0.5 ! (v4) $) -- (v4);
\draw[-,>=latex, dashed, line width=0.4pt] (d) -- ($ (d) ! 0.5 ! (v4) $);
\draw[-,=latex, dashed, line width=0.4pt] ($ (d) ! 0.5 ! (v4) $) -- (v4);
\draw[-,=latex, dashed, line width=0.4pt] ($ (d) ! 0.5 ! (v4) $) -- (v4);
\draw[-,>=latex, line width=0.4pt] (a) -- (center);
\draw[-,>=latex, line width=0.4pt] (b) -- (center);
\draw[-,>=latex, line width=0.4pt] (c) -- (center);
\draw[-,>=latex, line width=0.4pt] (d) -- (center);
\draw[-,>=latex, dashed, line width=0.4pt] (v4) -- (center);
\node[left] at(a) {\scalebox{0.75}{$#1$}};
\node[below] at(b) {\scalebox{0.75}{$#2$}};
\node[right] at(c) {\scalebox{0.75}{$#3$}};
\node[above] at(d) {\scalebox{0.75}{$#4$}};
\node[right] at(v4) {\scalebox{0.75}{$#5$}};
\node at (1.35,0.275) {\scalebox{0.75}{$#6$}};
\end{tikzpicture}
}

\newcommand{\CbThreeRepL}{
\begin{tikzpicture}[scale=0.8,baseline]
\coordinate (p1) at (0,0);
\coordinate (p15) at (0,0.5);
\coordinate (p5) at (0,1);
\coordinate (p12) at (.5,0);
\coordinate (p2) at (1,0);
\coordinate (p56) at (.5,1);
\coordinate (p6) at (1,1);
\coordinate (p26) at (1,.5);
\coordinate (p4) at (1+.5,0.35);
\coordinate (p8) at (1+.5,1.35); 
\coordinate (p3) at (.5,0.35);
\coordinate (p34) at (1.,0.35);
\coordinate (p13) at (.25,0.175); 
\coordinate (p24) at (1.25,0.175); 
\coordinate (p7) at (.5,1.35);
\coordinate (p37) at (.5,0.75); 
\coordinate (p57) at (.25,1.175); 
\coordinate (p78) at (1,1.35);
\coordinate (p68) at (1.25,1.175);
\coordinate (p8) at (1.5,1.35);
\coordinate (p48) at (1.5,.75);
\coordinate (p14) at (0.75,0.175);
\coordinate (p58) at (0.75, 1.175);
\coordinate (p28) at (1.25,0.675);
\coordinate (p18) at (0.75,0.675);
\draw [->,>=latex,line width=0.15pt] (-.5,0) -- (1.8,0); 
\draw [-,>=latex,line width=0.15pt] (0,-0.5) -- (0,0); 
\draw [-,>=latex, dotted, line width=0.15pt] (0,0) -- (0,1); 
\draw [->,>=latex,line width=0.15pt] (0,1) -- (0,1.65); 
\draw [-,>=latex, dotted, line width=0.15pt] (.5,0.35) -- (1.5,1.05);  
\draw [->,>=latex,line width=0.15pt] (1.5,1.05) -- (1.8,1.26);  
\draw [>-,>=latex,line width=0.4pt] (p12) -- (p2);
\draw [-,>=latex, dotted, line width=0.4pt] (p2) -- (p26);
\draw [-,>=latex, dotted, line width=0.4pt] (p26) -- (p6);
\draw [-,>=latex, dotted, line width=0.4pt] (p56) -- (p5);
\draw [-,>=latex, dotted, line width=0.4pt] (p56) -- (p6);
\draw [-,>=latex, line width=0.4pt] (p2) -- (p24);
\draw [>-,>=latex, line width=0.4pt] (p24) -- (p4);
\draw [-,>=latex, dotted, line width=0.4pt] (p3) -- (p34);
\draw [-,>=latex, dotted, line width=0.4pt] (p34) -- (p4);
\draw [-,>=latex,line width=0.4pt] (p4) -- (p8);
\draw [-,>=latex, dotted, line width=0.4pt] (p1) -- (p13);
\draw [-,>=latex, dotted, line width=0.4pt] (p13) -- (p3);
\draw [-,dotted,>=latex,line width=0.4pt] (p3) -- (p37);
\draw [-, dotted, >=latex,line width=0.4pt] (p37) -- (p7);
\draw [-,>=latex, dotted, line width=0.4pt] (p5) -- (p57);
\draw [-,>=latex, dotted, line width=0.4pt] (p57) -- (p7);
\draw [-,>=latex, dotted, line width=0.4pt] (p7) -- (p78);
\draw [-,>=latex, dotted, line width=0.4pt] (p78) -- (p8);
\draw [-,>=latex, dotted,  line width=0.4pt] (p6) -- (p68);
\draw [-,>=latex, dotted, line width=0.4pt] (p68) -- (p8);
\draw [-,>=latex, dotted, line width=0.4pt] (p4) -- (p48);
\draw [>-,>=latex, dotted, line width=0.4pt] (p48) -- (p8);
\draw [-,>=latex, line width=0.4pt] (p1) -- (p14);
\draw [>-,>=latex, line width=0.4pt] (p14) -- (p4);
\draw [-,>=latex, dotted, line width=0.4pt] (p5) -- (p58);
\draw [-,>=latex, dotted, line width=0.4pt] (p58) -- (p8);
\draw [->,>=latex, line width=0.4pt] (p2) -- (p28);
\draw [-,>=latex, line width=0.4pt] (p28) -- (p8);
\draw [->,>=latex, line width=0.4pt] (p1) -- (p18);
\draw [-,>=latex, line width=0.4pt] (p18) -- (p8);
\node at (-0.15,-0.15) {1};
\node at (1.,-0.185) {2};
\node at (1.65,0.45) {4};
\node at (1.65,1.45) {8};
\node at (2.05,0.02) {$x$};
\node at (2,1.4) {$y$};
\node at (0,1.8) {$z$};
\node at (1.75,.9) {$b$};   
\node at (1.5,0.16) {$a$}; 
\node at (.5,-0.185) {$c$};  
\end{tikzpicture}
}

\newcommand{\CbThreeRepR}{
\begin{tikzpicture}[scale=0.8,baseline]
\coordinate (p1) at (0,0);
\coordinate (p15) at (0,0.5);
\coordinate (p5) at (0,1);
\coordinate (p12) at (.5,0);
\coordinate (p2) at (1,0);
\coordinate (p56) at (.5,1);
\coordinate (p6) at (1,1);
\coordinate (p26) at (1,.5);
\coordinate (p4) at (1+.5,0.35);
\coordinate (p8) at (1+.5,1.35); 
\coordinate (p3) at (.5,0.35);
\coordinate (p34) at (1.,0.35);
\coordinate (p13) at (.25,0.175); 
\coordinate (p24) at (1.25,0.175); 
\coordinate (p7) at (.5,1.35);
\coordinate (p37) at (.5,0.75); 
\coordinate (p57) at (.25,1.175); 
\coordinate (p78) at (1,1.35);
\coordinate (p68) at (1.25,1.175);
\coordinate (p8) at (1.5,1.35);
\coordinate (p48) at (1.5,.75);
\coordinate (p14) at (0.75,0.175);
\coordinate (p58) at (0.75, 1.175);
\coordinate (p28) at (1.25,0.675);
\coordinate (p18) at (0.75,0.675);
\draw [->,>=latex,line width=0.15pt] (-.5,0) -- (1.8,0); 
\draw [-,>=latex,line width=0.15pt] (0,-0.5) -- (0,0); 
\draw [-,>=latex, dotted, line width=0.15pt] (0,0) -- (0,1); 
\draw [->,>=latex,line width=0.15pt] (0,1) -- (0,1.65); 
\draw [-,>=latex, dotted, line width=0.15pt] (.5,0.35) -- (1.5,1.05);  
\draw [->,>=latex,line width=0.15pt] (1.5,1.05) -- (1.8,1.26);  
\draw [>-,>=latex,line width=0.4pt] (p12) -- (p2);
\draw [-,>=latex, dotted, line width=0.4pt] (p2) -- (p26);
\draw [-,>=latex, dotted, line width=0.4pt] (p26) -- (p6);
\draw [-,>=latex, dotted, line width=0.4pt] (p56) -- (p5);
\draw [-,>=latex, dotted, line width=0.4pt] (p56) -- (p6);
\draw [-,>=latex, line width=0.4pt] (p2) -- (p24);
\draw [>-,>=latex, line width=0.4pt] (p24) -- (p4);
\draw [-,>=latex, dotted, line width=0.4pt] (p3) -- (p34);
\draw [-,>=latex, dotted, line width=0.4pt] (p34) -- (p4);
\draw [-,>=latex,line width=0.4pt] (p4) -- (p8);
\draw [-,>=latex, dotted, line width=0.4pt] (p1) -- (p13);
\draw [-,>=latex, dotted, line width=0.4pt] (p13) -- (p3);
\draw [-,dotted,>=latex,line width=0.4pt] (p3) -- (p37);
\draw [-, dotted, >=latex,line width=0.4pt] (p37) -- (p7);
\draw [-,>=latex, dotted, line width=0.4pt] (p5) -- (p57);
\draw [-,>=latex, dotted, line width=0.4pt] (p57) -- (p7);
\draw [-,>=latex, dotted, line width=0.4pt] (p7) -- (p78);
\draw [-,>=latex, dotted, line width=0.4pt] (p78) -- (p8);
\draw [-,>=latex, dotted,  line width=0.4pt] (p6) -- (p68);
\draw [-,>=latex, dotted, line width=0.4pt] (p68) -- (p8);
\draw [-,>=latex, dotted, line width=0.4pt] (p4) -- (p48);
\draw [>-,>=latex, dotted, line width=0.4pt] (p48) -- (p8);
\draw [-,>=latex, line width=0.4pt] (p1) -- (p14);
\draw [>-,>=latex, line width=0.4pt] (p14) -- (p4);
\draw [-,>=latex, dotted, line width=0.4pt] (p5) -- (p58);
\draw [-,>=latex, dotted, line width=0.4pt] (p58) -- (p8);
\draw [->,>=latex, line width=0.4pt] (p2) -- (p28);
\draw [-,>=latex, line width=0.4pt] (p28) -- (p8);
\draw [->,>=latex, line width=0.4pt] (p1) -- (p18);
\draw [-,>=latex, line width=0.4pt] (p18) -- (p8);
\node at (-0.15,-0.15) {1'};
\node at (1.,-0.185) {2'};
\node at (1.65,0.45) {4'};
\node at (1.65,1.45) {8'};
\node at (2.05,0.02) {$x$};
\node at (2,1.4) {$y$};
\node at (0,1.8) {$z$};
\end{tikzpicture}
}

\begin{document}

\include{cover}
\pagestyle{plain}
\include{contents}

\include{chap1} 
\include{G-phase} 
\include{chap2} 
\include{chap3} 
\include{chap4} 
\include{QS_stSurgery} 
\include{Conclusion} 

\appendix
\include{biblio}
\end{document}

%% file: cover.tex
%
%
%
%
%
%
%
%
%
%
%

\title{\fontsize{18}{20}\selectfont{Aspects of Symmetry, Topology and Anomalies in Quantum Matter}}

\author{Juven Chun-Fan Wang}
\department{Department of Physics}

\degree{Doctor of Philosophy} 

\degreemonth{June}
\degreeyear{2015}
\thesisdate{May 18, 2015}


\supervisor{Xiao-Gang Wen}{Cecil and Ida Green Professor of Physics}

\chairman{Nergis Mavalvala}{Associate Department Head for Education and Professor of Physics}

\maketitle



\cleardoublepage
\setcounter{savepage}{\thepage}
\begin{abstractpage}
\input{abstract}

\end{abstractpage}


\cleardoublepage

\section*{Acknowledgments}



The singularity of human civilization's development will happen when the interstellar aliens or Artificial Intelligence arrive and surpass human beings, as warned 
by S.~Hawking and R.~Kurzweil.
The singularity of my PhD program had happened fortunately when I started to work with my supervisor, Prof. Xiao-Gang Wen.
The marvelous thing is that no matter what I am working on, with some good ideas to share 
or struggling with no clues, Xiao-Gang always provides great insights and comments.
He is like a brilliant careful detective finding the secret hidden floor in a fully dark museum. 
There are always more lights, new hints or directions after our discussions.
Maybe the truth is that he actually leads us in building new floors and creates the beautiful shining garden next to it.

Due to my background and education in diverse fields including condensed matter theory, high energy physics and some interests in statistical and bio-physics, there are many
people I should 
acknowledge.
Below I would like to acknowledge 
institutes and affiliated people at 
MIT, Perimeter Institute for Theoretical Physics (visited for 2 years), Harvard University, IAS Tsinghua University (visited in summer 2013) and National Taiwan University, where I have benefited 
during my stay and visit.
I feel deep appreciation for having had Hong Liu and Liang Fu as my thesis committee professors, and 
also to John McGreevy for being a thesis committee professor
at an earlier stage.
I must thank Patrick Lee and Senthil Todadri for teaching me Theory of Solids 8.511-512 and Stat Mech 8.333-334, 8.513, and for many lunch discussions with CMT faculty/postdoc members.
I enjoyed my time at MIT building 4, 6C, 8 and Green Center for Physics.
I also deeply thank administrative professors and staff: Krishna Rajagopal, Catherine Modica, Lesley Keaney, Maria Brennan, 
Crystal Nurazura Young, Sean Robinson, Margaret O'Meara, Charles Suggs and Scott Morley.
There are many other MIT professors who taught me, or taught a class together where I was a TA, or had insightful discussions with me, to name a few:
Allan Adams, William Detmold, Peter Dourmashkin, Jeremy England, 
Edward Farhi, Jeffrey Goldstone, Jeff Gore, Roman Jackiw, Robert Jaffe, Pablo Jarillo-Herrero, Mehran Kardar, Wolfgang Ketterle, Young Lee, Leonid Levitov, Iain Stewart, Jesse Thaler, Frank Wilczek, Barton Zwiebach, and Martin Zwierlein. 
I would like to thank them all together.
I am honored to have Professors Clifford Taubes, 
Eugene Demler, Anton Kapustin, John Preskill, Subir Sachdev, Nathan Seiberg, Edward Witten, Shing-Tau Yau and Anthony Zee, etc, interacting with me or 
teaching me some great sciences. 
My two-year stay at Perimeter Institute has been wonderful, special thanks to Dmitry Abanin, Ganapathy Baskaran, John Berlinsky, Davide Gaiotto, Sung-Sik Lee, Rob Myers, and Guifre Vidal, 
and also
to administrators Debbie Guenther and Diana Goncalves. 
I also learn from Xiao-Gang's fellow and IASTU alumni: Xie Chen, Zhencheng Gu, Michael Levin, Zheng-Xin Liu, Ying Ran, Joel Moore, Brian Swingle, Maissam Barkeshli, Cenke Xu and Claudio Chamon. 
It has been wonderful to have graduate students, postdocs and professors around: special thanks to Tian Lan,  Huan He, Janet Hung, Luiz Santos, Peng Ye, Yidun Wan where we worked projects together;
also thanks to Andrea Allais, Guy Bunin, Kuang-Ting Chen, Ching-Kai Chiu, Lukasz Cincio, Zhehao Dai, Ethan Dyer, Peng Gao, Tarun Grover, Yingfei Gu,
Tim Hsieh, Charles Hsu, 
Nabil Iqbal,
Wenjie Ji, 
Chao-Ming Jian,
Shenghan Jiang,
Wing-Ho Ko, Jaehoon Lee, Hai Lin, Fangzhou Liu, Yuan-Ming Lu, 
Raghu Mahajan, Jia-Wei Mei,
Max Metlitski, 
Efstathia Milaraki, 
Heidar Moradi, David Mross, Adam Nahum, Rahul Nandkishore, Yusuke Nishida, Andrew Potter, Silviu Pufu, Yang Qi, Maksym Serbyn, Shu-Heng Shao, Inti Sodemann, Justin Song, Evelyn Tang, David Vegh, 
Chong Wang, Zhong Wang, William Witczak-Krempa, Nai-Chang Yeh, Beni Yoshida, Yizhuang You, Nicole Yunger Halpern, and more.
MIT ROCSA and Harvard Taiwan Student Association deserve my appreciation.
Professors from Taiwan deserve my gratitude: Jiunn-Wei Chen, Pei-Ming Ho, Feng-Li Lin, NTU Center for Condensed Matter Science and more.
%
%
%
It is 100\% sure that there are people that I have forgotten to list here but to whom I am equally grateful.
I wish to thank them all together anonymously.



Last but not least, 
I would like to thank my family, my mom Killy S.C., my sister Lauren Y.C., and my father C.C. for their support. 

Finally, to plan for the future, I hope we can utilize our knowledge and experiences 
and build up essential new better kinds of science and technology,
and therefore try to save human civilization from the potential danger caused by the invasions of interstellar aliens or AI. \\

%% file: abstract.tex
%
%
%

To understand the new physics and richness of quantum many-body system phenomena is one of the stimuli 
driving the condensed matter community forward. 
Importantly, the new insights and solutions for condensed matter theory
sometimes come from the developed and developing knowledge of high energy theory, mathematical and particle physics, which is also true the other way around: 
%
Condensed matter physics has been providing 
crucial hints and playgrounds for the fundamental laws of high energy physics. 
In this thesis, we explore the aspects of symmetry, topology and anomalies in quantum matter with entanglement from both  
condensed matter and high energy theory viewpoints.
The focus of our research is on the gapped many-body quantum systems including symmetry-protected topological states (SPTs) and
topologically ordered states (TOs). 
We first explore the 
ground state structures of SPTs and TOs: 
the former can be symmetry twisted and the latter has robust degeneracy.
The Berry phases generated by transporting and overlapping ground state sectors 
potentially provide universal topological invariants that fully characterize the SPTs and TOs.
This framework 
provides us the aspects of symmetry and topology.
We establish a field theory representation of SPT invariants in any dimension to uncover group cohomology classification and beyond ---
the former for SPTs with gapless boundary gauge anomalies, the latter for SPTs with mixed gauge-gravity anomalies.
We study topological orders in 3+1 dimensions such as Dijkgraaf-Witten models, 
which support multi-string braiding statistics; the resulting patterns may be analyzed by the mathematical theory of knots and links.
We explore the aspects of surface anomalies of bulk gapped states from the bulk-edge correspondence:
The gauge anomalies of SPTs shed light on the construction of bosonic anomalies 
including Goldstone-Wilczek type,
and also guide us to design a non-perturbative lattice model 
regularizing the low-energy chiral fermion/gauge theory towards the Standard Model 
while overcoming the 
Nielsen-Ninomiya fermion-doubling problem without relying on Ginsparg-Wilson fermions. 
We conclude by utilizing aspects of 
both quantum mechanical topology and spacetime topology to
derive new formulas analogous to Verlinde's via geometric-topology surgery. 
This provides new insights 
for higher dimensional topological states of matter.

%% file: contents.tex
\tableofcontents
\listoffigures
\listoftables

%% file: chap1.tex
\chapter{Introduction} \label{chap:Intro}


\section{Background: Emergence, Reductionism and Many-Body Quantum Physics}

Emergence and reductionism respectively represent the hearts of condensed matter physics (CMP) and high energy physics (HEP).
Emergence describes remarkable collective phenomena. For example, given a set of elastic springs, initial conditions and their laws of interactions, 
intriguing properties such as resonances occur in a mattress formed by springs at the macroscopic level. 
Emergence emphasizes 
that certain many-body collective modes 
cannot be easily extrapolated from a few-body behavior. 
CMP, with emergence at its heart, 
asserts that
 to discover 
the collective laws of many-body quantum matters 
from the given individual basic components (such as qubits, quantum rotors, spins, bosons or fermions) 
requires as much creative effort as discovering the fundamental laws of a single component. 
In short, it is about P.W. Anderson's ``more is different\cite{Anderson}.''
Reductionism, on the other hand, posits that we can ``see the world in a grain of sand,'' as proclaimed by William Blake.
For example, 
reductionism assumes that colliding massive objects like nuclei and then observing 
what-and-how fragments result from that collision 
determine the fundamental laws among the basic elements.
The core distinction of CMP and HEP 
leads to 
differing working concepts and values. \\[-4.5mm]


However, 
past physics history suggests that the combined knowledge can enrich our understanding.
For example, the Anderson-Higgs mechanism describing the origin of all particles' mass in HEP is rooted in 
Anderson's CMP theory on plasma oscillation and superconductivity. 
The macro-scale inflation theory of the universe of 
HEP cosmology is 
inspired by the micro-scale supercooling phase transition of CMP. 
There are many more successful and remarkable examples, 
such as those examples concerning the interplay of fractionalization, quantum anomalies and topological non-perturbative aspects of quantum systems
such as quantum Hall states (see  \cite{ColemanA, Farhi:1982vt, Treiman:1986ep, wilczek1990fractional, Jackiw:1995be} 
for an overview), which we will gradually explore later.
In this thesis, we explore the aspects of symmetry, topology and anomalies in quantum matter from the intertwining viewpoints of CMP and HEP.

Why do we study quantum matter? Because the quantum matter not only resides in an outer space (the spacetime we are familiar with at the classical level),
but also resides in an inner but gigantic larger space, the Hilbert space of quantum systems.
The Hilbert space $\cH$ of a many-body quantum system is enormously huge. For a number of $N$ spin-$\frac{1}{2}$ particles, the
dimension of $\cH$  grows exponentially as $\dim(\cH)=2^N$. Yet we have not taken into account other degrees of freedom, like orbitals, charges and interactions, etc. 
So for a merely 1-mole of atoms with a tiny weight of a few grams, its dimension is $\dim(\cH) > 2^{6.02 \times 10^{23}}$!
Hence, studying the structure of 
Hilbert space may potentially guide us to systematically explore 
many mathematical structures both ones we have imagined and ones we have not yet imagined, and explore 
the possible old and new emergent principles hidden in all branches of physics, including CMP, HEP and even astro- or cosmology physics.

In particular, we will take a modest step, focusing on the many-body quantum systems at zero temperature where there are unique or degenerated bulk ground states 
well-separated from energetic excitation with finite energy gaps, while the surfaces of these states exhibit quantum anomalies.
These \emph{phases of matter} are termed symmetry-protected topological states (SPTs) and topologically ordered states (TOs).

\subsection{Landau symmetry-breaking orders, quantum orders, SPTs and topological orders: Classification and characterization} 
\label{Sec:ClassChara}

What are the \emph{phases of matter} (or the states of matter)? 
Phases of matter are the collective behaviors of many-body systems described by some macroscopic scale of parameters.
The important concepts to characterize the ``phases'' notions are \emph{universality}, \emph{phase transitions}, and \emph{fixed points} \cite{WilsonRGRMP}.
The universality class means that a large class of systems can exhibit the same or similar behavior even though their microscopic degrees of freedom 
can be very different. 
By tuning macroscopic scale of parameters such as temperature, doping or pressure,
the phase can encounter phase transitions. 
Particularly at the low energy and long wave length limit,
the universal behavior can be controlled by the fixed points of the phase diagram.
Some fixed points sit inside the mid of a phase region, some fixed points are critical points at the phase transition boundaries.
A powerful theory of universality class should describe the behavior of phases and phase transitions of matter.

Lev Landau established one such a powerful framework in 1930s known as Landau symmetry-breaking theory \cite{Landau1}.
It can describe many phases and phase transitions with \emph{symmetry-breaking orders},
including crystal or periodic charge ordering (breaking the continuous translational symmetry), 
ferromagnets (breaking the spin rotational symmetry) and superfluids (breaking the U(1) symmetry of bosonic phase rotation).
Landau symmetry-breaking theory still can be captured by the semi-classical Ginzburg-Landau theory \cite{GL5064,LL58}.

However, it is now known that there are certain \emph{orders} at zero temperature beyond the semi-classical Landau's symmetry-breaking orders.
The new kind of order is referred as \emph{quantum order} \cite{wen2004quantum}, where the quantum-many body behavior exhibits new phenomena 
without the necessity of classical analogy. 
The full scope of quantum order containing gapless or gapped excitations is too rich to be properly examined in my thesis. 
I will focus on the gapped quantum order: including SPTs \cite{XieSPT4,XieSPT5} and topological orders (TOs) \cite{W9039,Wen1210.1281}.
The first few examples of TOs are integer quantum Hall states (IQHs) discovered in 1980 \cite{Klitzing} and fractional quantum Hall states (FQHs) in 
1982 \cite{TsuiSG8259,Laughlin8395}. Quantum Hall states and TOs are exotic because
they are \emph{not distinguished by symmetry-breaking, local order parameters, or the long-range correlation}. 
These new kinds of orders require a new paradigm going beyond the old paradigm of Landau's theory.

{\bf Classification and characterization of quantum phases of matter}: So what exactly are SPTs and TOs?
SPTs and TOs are quantum phases of matter with bulk insulating gaps while the surfaces are anomalous (such as gapless edge modes) which cannot exist in its own dimensions.
One important strategy to guide us understand or even define the phases of matter is
doing the \emph{classification and characterization}.
By doing classification, we are counting the number of distinct states (of SPTs and TOs) and giving them a proper label and a name.
For example, giving the spacetime dimension and the symmetry group, etc; can we determined how many phases there are?
By doing characterization, we are listing their properties by physical observables. How can we potentially measure them in the experiments?

  Below we organize the key features of SPTs and TOs first, in Table \ref{table:SPTsTOs}. 
  There are a few important concepts
for physical measurement we need to introduce:
(i) ground state degeneracy, (ii) entanglement and (topological) entanglement entropy,
(iii) fractionalized charge and fractional statistics.

\begin{center}
\begin{table}[!h]
\begin{tabular}{l}
\hline
\hline
\cred{Symmetry-Protected Topological states (SPTs)} $\&$ \cblue{Topological Orders (TOs)}:\\
\hline
\hline
 {\bf \cred{Short}/\cblue{Long}} ranged entangled states at zero temperature.\\
$\leftrightarrow$ {\bf\cred{Yes}/\cblue{No}} deformed to trivial product states by local unitary transformations.\\
 {\bf \cred{No}/\cblue{Likely}} nontrivial topological entanglement entropy.\\
 {\bf \cred{No}/\cblue{Likely}} bulk fractionalized charge $\to$ \color{red} edge \color{black} may have fractionalized charges.\\
 {\bf \cred{No}/\cblue{Likely}} bulk anyonic statistics $\to$ \color{red} edge \color{black} may have degeneracy.\\
{\bf \cred{No}/\cblue{Yes (Likely)}} spatial topology-dependent {\bf GSD}.\\
{\bf \cred{No}/\cblue{Yes (Likely)} non-Abelian Berry's phases} on coupling const moduli space\\
\hline %
\hline %
\end{tabular}
\caption{Some properties of SPTs and TOs.}
\label{table:SPTsTOs}
\end{table}
\end{center}

\begin{figure}[!h]
\centering
\includegraphics[height=3in]{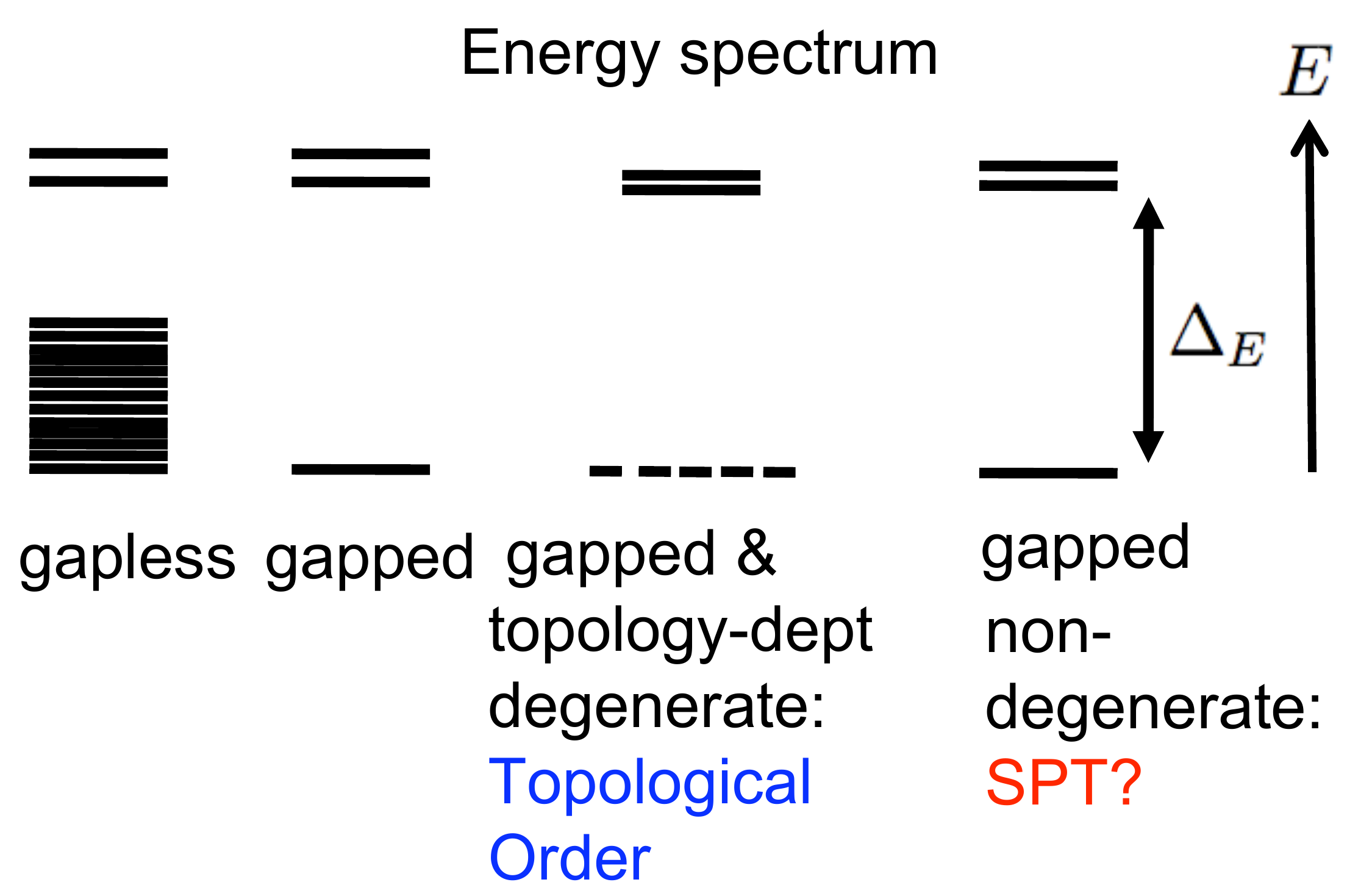}
\caption{Quantum matter: The energy spectra of gapless states, topological orders and symmetry-protected topological states (SPTs).
}
\label{1_energy_gap}
\end{figure}

The {\bf ground state degeneracy} (GSD) counts the number of linear independent 
ground states $| \psi \rangle$ on a topology-dependent manifold (such as a $d$-sphere $S^d$ or
a $d$-torus $T^d$) by solving the
Schr\"odinger equation: $H | \psi \rangle = E_{gd} | \psi \rangle$ with the
ground state energy $E_{gd}$. The possibility of the energy spectrum is
shown in Fig.\ref{1_energy_gap}.
In the infinite volume limit (thermodynamic limit at zero temperature limit), 
the gapless phase has continuous energy spectrum from the ground states.
The gapped phase has finite $\Delta_E$ in the energy spectrum $E$.
Topological order has robust GSD where the number usually depends on the system-topology (the exception can be chiral topological orders such as 
the integer quantum Hall state with $\nu=1$ filling-fraction or the E$_8$ bosonic quantum Hall state \cite{Kitaevhoneycomb}, they have GSD=1).
For example, a filling-fraction $\frac{1}{3}$-FQH state of Laughlin type, has a $3$ or $3^g$ fold degeneracy on a 2-torus or a genus $g$-Riemman surface respectively.
On the other hand, SPTs has a unique ground state independent of the spatial topology. 
In this sense, by measuring GSDs, TOs are potentially easier to be ``distinguished and detected'' than SPTs.

{\bf Entanglement} describes how a system of (quantum) states are correlated between subsystems, say $A$ and $B$; and describes how the system \emph{cannot} be described independently in the form of a pure product state
$| \Psi \rangle=| \psi_A \rangle \otimes | \psi_B \rangle \otimes \dots$.
Even though the full Hilbert space $\cH_{AB}$ is the tensor product form $\cH_{AB} =\cH_A \otimes \cH_B$, and the full basis of $\cH_{AB}$ can be spanned by
product states $\{ | u\rangle_A \otimes | v\rangle_B \}$, but the generic state would be:
$
| \psi \rangle =\sum_{\{ u,v\}} c_{u,v} | u\rangle_A \otimes | v\rangle_B
$
more general than a pure product state.
Entanglement entropy quantifies the entanglement by measuring how the subsystems are entangled with each other (see an introduction in \cite{preskill1998lecture,{pachos2012introduction}}).
Von Neumann entropy is defined 
by: $\mathcal{S}(\rho_A)=  -\text{Tr}[\rho_A\text{log}\rho_A] = \mathcal{S}(\rho_B)$
where $\rho_{A}=\mathrm{Tr}_B(\rho_{AB})$ and $\rho_{B}=\mathrm{Tr}_A(\rho_{AB})$,
here $\rho_{AB}$ is the density matrix, and $\rho_{AB}= |\Psi\rangle\langle\Psi|$ with the eigenstate sector $|\Psi\rangle$.  
More generally, one can define Renyi entropy $\mathcal{S}_\alpha (\rho_A) = \frac{1}{1-\alpha} \text{log} \text{Tr} (\rho^\alpha) = \mathcal{S}_\alpha(\rho_B)$,
where $\alpha \to 1$ then the Renyi entropy becomes the Von Neumann entropy. 
For the gapped 2+1D topological orders,
\footnote{The spacetime dimensionality definition used throughout the thesis is that $d+1$D means $d$-spatial and 1 temporal dimensions,
and $d$D means $d$-spatial dimensions.}
the von Neumann entropy $S_A=\alpha |\partial A| -\gamma+\dots$.
The $\partial A$ part is due to the area law,
where the possible contribution to the entanglement between two regions $A$ and $B$ should come from the 
regions near the boundary of $A$, namely $\partial A$. 
The $\dots$ term tend to be infinitesimal as $|\partial A| \to \infty$. 
Topological entanglement entropy (TEE)\cite{KitaevPreskill,LW0605} 
is the universal part captured by 
$
\gamma=\log \cD = \log (\sqrt{\sum_i d_i^2})
$
where $\cD$ is the total quantum dimension and $d_i$ is the quantum dimension for each particle labeled by $i$.
Basically the quantum dimension $d_i$ dictates the physical observables GSD of topological orders.
The quantum dimension characterizes the dimension growth of the Hilbert space 
 when an additional particle $i$ is inserted. This Hilbert space is named as the fusion Hilbert space
 $\cV(\cM)$ with a spatial manifold $\cM$.
For example, by putting $n$ anyons on a sphere, $\GSD=\dim(\cV(\cM)) \propto (d_i)^n$.
 We shall explain more the meaning of $\cD$ and $d_i$ later in the chapters.

{\bf Fractionalized charge and fractional statistics} are introduced in a review of selected papers in \cite{Jackiw:1995be}, \cite{wilczek1990fractional}.
Due to interactions, the emergent quasi-excitations of the system can have fractionalized charge and fractional statistics respect to the original
unit charge. To define fractional statistics as a meaningful measurable quantities in many-body systems, it requires the adiabatic braiding process between quasi-excitations
in a \emph{gapped phase at zero temperature}.
The wavefunction of the whole system will obtain a $e^{\ii \theta}$ phase with a fractional of $2 \pi$ value of $\theta$.
The excitation with fractional statistics is called \emph{anyon}.

The first known experimental example exhibits all exotic phenomena of 
(i) spatial topology-dependent GSD, (ii) entanglement and TEE,
(iii) fractionalized charge and fractional statistics, is the
FQHs with $\nu=1/3$-filling fraction discovered in 1982 \cite{TsuiSG8259,Laughlin8395}.
FQHs is a truly topologically ordered state. Some of other topological orders and SPTs \emph{may not} have all these nontrivial properties.
We should summarize them below.






For {\bf  SPTs}:
\begin{itemize}
\item Gapped-bulk {\bf short} ranged entangled states (SREs). 
\item No topological entanglement entropy. 
\item No bulk fractionalized charge $\to$ \color{black} edge \color{black} may carry fractionalized charge  
\item No bulk anyonic statistics (GSD = 1) $\to$ \color{black} gapped edge \color{black} may have degeneracy. 
\item The \color{black} bulk \color{black} realizes the symmetry with a global symmetry group $G$ \color{black} \emph{onsite} \color{black} (here we exclude the non-onsite space symmetry such as spatial translation or point group symmetry, etc; the time reveal symmetry can still be defined as an anti-unitary on-site symmetry).
The symmetry-operator is \emph{onsite}, if it has the form:
$
U(g)= \otimes _j
U_j(g), \ \ \ g \in G$.  
It can be written as the tensor product structure of $U_j(g)$ acting on each site $j$.
The \color{black}boundary \color{black} realizes the symmetry $G$ \color{black}non-onsite\color{black}, exhibiting one of the following:  
 (1) gapless edge modes, or 
 (2) {GSD} 
from symmetry breaking gapped boundary, or
 (3) {GSD} from the gapped surface topological order on the boundary.
\end{itemize}
%
%
For intrinsic {\bf topological orders}:
\begin{itemize}
\item Gapped-bulk {\bf long} ranged entangled states (LREs).  
\item Robust gapless edge states without the symmetry protection.
\item (usually) with topological entanglement entropy. 
\item (usually) Bulk fractionalized charge.  
\item (usually) Bulk fractionalized statistics. 
\item (1) Spatial topology-dependent {\bf GSD}.
\item (2) {\bf  Non-Abelian (Berry) geometric structure} on the Hamiltonian's coupling constant moduli space. 
\end{itemize}
Some remarks follow: The ``usually'' quoted above is to exclude some exceptional cases such as IQHs and E$_8$ QH states.
The {short} ranged entangled (SREs) and long ranged entangled states (LREs) are distinguished by
the \emph{local unitary} (LU) \emph{transformation}. SREs can be deformed to a trivial direct product state in the real space under
the LU transformation; SREs is distinguished from a trivial product state on each site only
if there is some symmetry-protection so that along the path connecting the state 
to a trivial product state breaks the symmetry.
LREs on the other hand cannot be connected to a trivial product state via LU transformation even if we remove all the symmetries. 
Thus the LU transformation is an important concept which guide us to classify the distinct states in SPTs and TOs by determining 
whether two states are connected via LU transformations.

{\bf The essences of orders}: Apart from the summary on physical comparison of Table \ref{table:SPTsTOs}, 
we comment that for there are microscopic or field theories,  
trying to capture \emph{the essences} of Landau's symmetry-breaking order, 
TOs and SPTs. For example, {\bf Landau's theory and Bardeen-Cooper-Schrieffer (BCS) theory} is powerful for understanding symmetry-breaking orders,
but the essence of symmetry-breaking order is indeed emphasized later as \emph{the long range correlation, symmetry-breaking local order parameters 
and the long-range order} [see C.N.Yang's review on the (off-diagonal) long range order\cite{CNYangRMP}].
Then, there are {\bf Laughlin's theory} for FQHs of topological orders, 
and there are topological quantum field theory ({\bf TQFT}) \cite{Witten:1988ze, Witten:1988hf} approach of topological phases. 
But what is the physical essence of topological orders?
After all, Laughlin's approach focus mainly on the wavefunction,
and TQFT only capture the low energy long-wavelength physics and TQFT may not full classify or describe topological phases of matter in any dimension.
The essence of topological orders is \emph{not} the quantized Hall conductance (which will be broken down to non-quantized if 
the particle number conservation is broken).
The essence of topological orders should not depend on the notion of symmetry.
The essence of topological orders is actually the topological GSD, the non-Abelian geometric phases and the long-range entanglement.
For topological orders, there are degenerate ground states depending on the spatial-topology,
and we can characterize the topological order by adiabatically transporting the ground state sectors.
However, for SPTs, unfortunately we do not have the concept of non-Abelian geometric phases.
How do we capture the essence of SPTs? There are indeed such a tool we can develop, named \emph{symmetry-twist} \cite{WenSPTinv,HW1339,JuvenSPT1}.
The essence of SPTs can be captured by \emph{twisting the symmetry}, namely we can modify the boundary conditions on some branch cut
acting on the Hamiltonian by modifying the Hamiltonian along the cut. This will transport the original state to another unique ground state different by a U(1) phase.
We can obtain the U(1) phase by overlapping the the two states.
Importantly, this U(1) phase will be a universal SPT invariant only if we close the orbit in
the {symmetry-twist} phase space by transporting the states to the original state.
Moreover, regarding the low energy field theory of SPT, 
the quantum field theory (QFT) formulation of SPT is not transparent as the usual TQFT for topological orders without symmetry (only gauge redundancy).
We will present these issues 
in Chap.\ref{aofSymmetry}.

Further illumination on SPTs and topological orders can be found in review articles  listed in Sec.\ref{outline} under \cite{2010RMP_HasanKane,2011_RMP_Qi_Zhang,Turner-Vish,Senthil1405,Wen1210.1281}.

\newpage

\subsection{Evidence of SPTs and topological orders: Experimental progress}
Other than the previous mentioned IQHs and FQHs, there are further rapidly-developing experiments realizing both TOs and SPTs experimentally (exp.) and theoretically (theo.). 
Examples of SPTs are Haldane spin-1 chain protected by
spin rotational symmetry\cite{H8364,{AKL8877}} and the topological
insulators \cite{2010RMP_HasanKane,2011_RMP_Qi_Zhang, MooreBalents, FuKaneMele, Roy} protected by fermion number conservation U(1) and
time reversal symmetry $Z_2^T$. 
See Table \ref{table:exp_progress} for a short summary. The full review on the theory or experimental progress is beyond the focus and the scope of my intention.
The readers can look for the cited references for more details. 
%
\begin{table}[!h]
\begin{tabular}{l}
\hline
\hline
$\bullet$ Symmetry breaking phases:\\
\hline
\hline
-500 (bc) Ferromagnet (exp.)\\ \; \\
\hline %
\hline %
$\bullet$ Topologically ordered states (TOs)\\
\hline %
\hline
1904 Superconductor (exp.) [Onnes 04] (Z2 topo. order)\\
1980 IQH states (exp.) [von Klitzing 80 \cite{Klitzing}] (with no topo. excitations, free fermion)\\
1982 FQH states (exp. theo.) [Tsui-Stormer-Gossard 82, Laughlin 83 \cite{TsuiSG8259,Laughlin8395}] \;\\
1987 Chiral spin liquids (theo.) [Kalmeyer-Laughlin 87, Wen-Wilczek-Zee 89] \;\\
1991 Non-Abelian FQH states, (theo.) [Moore-Read 91, Wen 91] (CFT, slave particle)\\
1991 Z2-spin liquids (theo.) [Read-Sachdev 91, Wen 91, Kitaev 97] \;\\
1992 All Abelian FQH states (theo.) [Wen-Zee 92] (K-matrix)\\
2000 $p_x + ip_y$-superconductor (theo.) [Read-Green 00]  \; \\ 
2002 Hundreds symmetry enriched topological orders (theo.) [Wen 02] (PSG)\\
2005 All 2+1D topo. orders with gapped edge (theo.) [Levin-Wen 05] (UFC)\\
2009 $\nu=5/2$ non-Abelian FQH states (exp. ?) [Willett et al 09] \;\\
\;\\
\hline
\hline
 \begin{minipage}[!t]{6in} $\bullet$ SPTs (no topological order and no symmetry-breaking,
also called topological states despite having no topological order) 
\end{minipage}\\
\hline
\hline
1983 Haldane phase (theo.) [Haldane 83] \;\\
1988 Haldane phase (exp. CsNiCl3) [Morra-Buyers-Armstrong-Hirakawa 88] \;\\
2005 Topological insulators  (TI)  \;\\
2005 TI 2D (theo.) [Kane-Mele 05 \cite{KaneMele2}, Bernevig-Hughes-Zhang 06 \cite{BernevigHughesZhang}] \;\\
2006 TI 3D (theo.) 
[Moore-Balents, Fu-Kane-Mele, Roy \cite{MooreBalents, FuKaneMele, Roy}]  \; \\ 
2007 Topological insulators (TI: exp.) [Molenkamp etal 07 \cite{Molenkamp}] \;\\
2010 Topological crystalline insulators (TCI: theo.~Fu et al 10, 12 \cite{Fu_TCI, 2010THsieh})\;\\ 
2012 Topological crystalline insulators (TCI: exp. 12 \cite{Ando,Story,HasanTCI})\;\\
2011 SPT states in any dim. for any symm. (theo.) [Chen-Gu-Liu-Wen 11 \cite{XieSPT4,XieSPT5}] \;\\ 
\dots \;\;\; \; \dots\\
\hline
\hline
\end{tabular}
\caption{ Theory and experiment progress for TOs and SPTs in a simplified timeline.
Here topological insulator in 2D means the Quantum Spin Hall effect (QSH). 
Here ``exp.'' abbreviates the experiment and ``theo.''  abbreviates the theory.
 }
\label{table:exp_progress}
\end{table}


\section{Motivations and Problems} 
\subsection{Symmetry, Topology and Anomalies of Quantum Matter} \label{sec:STAoverview}

With the background knowledge on quantum matter, let us now motivate in a colloquial style of colloquium 
on how the symmetry, topology and anomalies can be involved in quantum matter. This overview can guide us to pose new questions
and the {statement of the problems} in the next in Sec.\ref{sec:problem}.

{\bf Symmetry}, in everyday terms, 
means the system stays invariant under certain transformation.
To describes the states of matter governed by symmetry,
 Ginzburg-Landau (G-L) theory \cite{Landau1} 
 semi-classically dictates the global symmetry realized onsite and locally. 
However, quantum wavefunctions become fuzzy due to Heisenberg's uncertainty principle and spread non-onsite. 
The symmetry operation can also act non-onsite 
 --- the symmetry concept is enriched when understood at a fully quantum level.
This new concept of non-onsite symmetry 
can be realized on the 
boundary of some bulk gapped insulating phases, it
unearths many missing states 
buried beneath G-L theory. 
States are identified via  
local-unitary transformations, distinct new states are 
termed SPTs \cite{XieSPT4,XieSPT5}. 

{\bf Anomalies} are phenomena that cannot be realized in their own spacetime dimensions.
A classical analogy is that two-dimensional (2D) waves propagate on the surface of the ocean require some extended dimension, the 3D 
volume of bulk water.
Similarly, quantum anomalies describe the anomalous boundary physics at the quantum level \cite{Treiman:1986ep} --- the obstruction to regularizing classical symmetries on the boundary quantized lattice
without an extended bulk.
One of the earlier attempts on connecting quantum anomalies and topological defects are done by Jackiw \cite{Jackiw:1995be} and Callan-Harvey \cite{Callan:1984sa}.
In their work, the use of field theory is implicitly assumed to represent many-body quantum system.
In my work, I will directly establish the quantum anomalies realized on a \emph{discretized regularized lattice} of many-body quantum system.

The 
field theory regularization at high energy in HEP corresponds to the short distance lattice cutoff in CMP.
Using the lattice cutoff as a mean of regularization, 
we have the advantage of distinguishing different types of global symmetry operations, namely onsite and non-onsite. 
We learn that the quantum variables of onsite symmetry can be promoted to dynamical ones and thus can be easily ``gauged.'' 
In contrast,  non-onsite symmetry manifests ``short-range or long-range entangle'' properties, 
hence hard to be gauged: it is an anomalous symmetry. 
By realizing that such an obstruction 
to gauging a global symmetry coincides with the 
't Hooft anomalies \cite{'tHooft:1979bh}, we are led to the 
{\bf first lesson}:
\be
\text{\colorbox{lgray}{
\text{``The correspondence between non-onsite global symmetries and gauge anomalies.''} 
}}\nonumber 
\ee
The correspondence is explicit at the weak gauge coupling. 
Ironically, we find that gauge anomalies need only to be global symmetry anomalies.
Gauge symmetries are not symmetries but redundancies; only global symmetries are real symmetries. 
Meanwhile, the non-onsite symmetry is rooted in the SPT boundary property. Thus we realize the 
{\bf second lesson}:
\be
\text{\colorbox{lgray}{
``The correspondence between gauge anomalies and SPT boundary modes \cite{{Wen:2013oza},{Wang:2013yta}}.''
}} \nonumber
\ee

 {\bf Topology}, 
in colloquial terms, people may 
mistakenly associate the use of topology with the twisting or the winding of electronic bands. 
More accurately, the topology should be defined as a global property instead of local geometry, \emph{robust against any local perturbations even those breaking all symmetries}. 
Thus topological insulators and SPTs are not really topological, due to their lack of robustness 
against short-range perturbations 
breaking their symmetry (see also Table \ref{table:topo}). Our key observation is that since the boundary gapless modes and anomalous global symmetries of SPTs 
are tied to gauge anomalies, the 
further robust boundary gapless modes of \emph{intrinsic topological orders} 
must be associated to some anomalies requiring no global symmetry. 
We realize these anomalies violate spacetime diffeomorphism covariance on their own dimensions. 
This hints at our 
{\bf third lesson}: 
\colorbox{lgray}{\parbox[t]{\linegoal}{
``The correspondence between 
gravitational anomalies and TO's boundary modes \cite{{Wen:2013oza},{Kong:2014qka}}.''}}

Prior to our recent work \cite{Wang:2014oya}, the previous two-decades-long study 
of topological orders in the CMP community primarily focuses on 2D topological orders using modular SL(2,$\mathbb{Z}$) data \cite{Wen1210.1281}. 
Imagine a bulk topological phase of matter placed on a donut as a 2-torus; we deform its 
space and then reglue it back to maintain the same topology. 
This procedure derives 
the mapping class group MCG of a 2-torus $T^2$, which is 
the modular group MCG($T^2$)=SL(2,$\mathbb{Z}$) 
generated by an $\mathcal{S}$ matrix via 
$90^{\circ}$ rotation
and a $\mathcal{T}$ matrix via 
the Dehn twist. 
Modular SL(2,$\mathbb{Z}$) data capture the non-Abeliang geometric phases of ground states \cite{Wilczek:1984dh}
and describe the braiding statistics of quasiparticle excitations.
Clearly topological orders can exist in higher dimensional spacetime, such as 3+1D, what are their excitations 
and how to characterize  their braiding statistics? This is an ongoing open research direction.

Regularizing chiral fermion or chiral gauge theory on the lattice non-perturbatively is a long-standing challenge, 
due to Nielsen-Ninomiya's no-go theorem on the fermion-doubling problem \cite{Nielsen:1980rz}. 
Mysteriously our particle physics Standard model is a chiral gauge theory, thus the no-go theorem is a big challenge for us to bypass for understanding 
non-perturbative strong interacting regime of particle physics.
Fermion-doubling problem in the free fermion language is basically saying that the energy band cross the zero energy even times
in the momentum $k$-space of Brillouin zone due to topological reason, thus with equal number of left-right moving chiral modes --- the fermions are doubled.
It suggests that the HEP no-go theorem is rooted in the CMP thinking.
Providing that our enhanced understanding through topological states of matter, can we tackle this challenge?

Moreover, the nontrivial bulk braiding statistics of excitations and the boundary quantum anomalies have certain correspondence.
Will the study of the bulk-edge correspondence of TOs/SPTs not only guide us to understand exotic phases in CMP, but also resolve
the non-perturbative understanding of particle physics contents, the Standard Model and beyond in HEP problems? 


\subsection{Statement of the problems} \label{sec:problem}

The above discussion in Sec.\ref{sec:STAoverview} had outlined our thinking and the strategy to solve certain physics issues.
But what exactly are the physics issues and problems? Here let us be more specific and pin down them straightforwardly and clearly. In my thesis,
I attempt to address the six questions {\bf Q.I-VI} below and analytically formulate an answer to them:

\noindent
({\bf Q.I)}. {\bf SPT invariant and its field-theory representation} \cite{JuvenSPT1, Gu:2015lfa}: Topological orders (TOs) and SPTs are very different. For TOs, there are topology-dependent degenerate ground states on a topology-nontrivial manifold (such as $d$-torus $T^d$). We can transport the ground states and determine the non-Abelian geometric phases generated in the coupling constant moduli-space. More conveniently, we
 can overlap the wavefunctions to obtain the amplitude data. These are
 the universal topological invariant of TOs. How about SPTs? The challenge is that, with symmetry-protection and without
symmetry-breaking,  there is only a unique ground state.
 Can we obtain the universal SPT invariants? If it is obtainable 
 from a lattice model, then, is there a field-theory representation of SPT invariants?
 Can we recover the group-cohomology classification of SPT and more beyond than that using continuous field theory approach?\\

\noindent
({\bf Q.II)}. {\bf Bosonic anomalies} \cite{Wang:2014tia, Santos:2013uda}: Quantum anomalies occur in our real-world physics, such as pion decaying to two photons via Adler-Bell-Jackiw chiral anomaly \cite{{Adler:1969gk},{Bell:1969ts}}. Anomalies also constrain beautifully on the Standard Model of particle physics, in particular to the Glashow-Weinberg-Salam theory, via anomaly-cancellations 
of gauge and gravitational couplings. 
The above two familiar examples of anomalies concern chiral fermions and continuous symmetry 
(e.g. U(1), SU(2), SU(3) in the weak coupling limit).
Out of curiosity, we ask: 
``Are there concrete examples of quantum anomalies for bosons instead? And anomalies for discrete symmetries? 
Can they be formulated by a continuous field theory and a regularized lattice model?
Are they potentially testable experimentally in the lab in the near future?'' 

\noindent
({\bf Q.III)}. {\bf Topological gapping criteria. Topological degeneracy on a manifold with gapped domain walls and boundaries} \cite{Wang:2012am, Lan:2014uaa}:
By now 2D topological orders are well-studied. 
We understand the proper label of a single 2D
topological order
by a set of ``\emph{topological invariants}'' or ``\emph{topological order parameters}''---
the aforementioned modular SL$(2,\Z)$ $\cS,\cT$ matrices and the chiral central charge $c_-$. 
The $\cS,\cT$ matrices can be derived from geometric phases and encode the quasiparticle (or anyon) statistics. 
Non-zero chiral central charge $c_-$ implies the topological gapless edge modes.
However, it is less known how separate topological orders are related. 
To this end, it is essential to study the following circumstance: 
there are several domains in the system and each domain contains
a topological order, while the whole system is gapped.
In this case, different topological orders are connected by
gapped domain walls. Under what criteria can two topological orders be
connected by a gapped domain wall, and how many different types of gapped domain walls are there? 
Since a gapped boundary is a gapped domain wall between a nontrivial topological order and the vacuum, we can meanwhile
address that under what criteria can topological orders allow gapped boundaries?
When a topologically ordered system has a gapped
bulk, gapped domain walls and gapped boundaries, how to
calculate its GSD on any orientable manifold?

\noindent
({\bf Q.IV)}. {\bf Define lattice chiral fermion/gauge theory non-perturbatively} \cite{Wang:2013yta}: The Standard Model is a chiral gauge theory with chiral fermions ---
where the weak-interaction gauge fields couple to the right-hand and
the left-hand fermions differently. We know the perturbative Lagrangian definition of
the Standard Model since 1970. However, we do not know whether the Standard Model can be regularized
on the lattice, due to the Nielsen-Ninomiya fermion-doubling no go theorem. And we do not have a non-perturbative definition of Standard Model 
as a Hamiltonian quantum mechanical theory. 
We ask: ``Whether there is a \emph{local short-range finite quantum Hamiltonian system} realizing \emph{onsite symmetry} G defined on a spatial lattice 
with a continuous time, such that its low energy physics produces a anomaly-free chiral matter theory of symmetry G?'' 

\noindent
({\bf Q.V)}. {\bf String and particle exotic braiding statistics} \cite{Wang:2014oya,Wan:2014woa}:
Higher dimensional topological orders are the new research frontier and are mostly not yet systematically explored.
In 3+1D, the excitations can involve not only \emph{particle} excitations but also \emph{string} excitations.
We can ask: How to (at least partially) classify and characterize 3+1D topological orders? How to characterize the braiding statistics of
strings and particles? How to formulate or construct certain 3+1D topological orders on the lattice?
What is the physical interpretation of braiding statistics data?

\noindent
({\bf Q.VI)}. {\bf Topological invariants, as quantum statistics derived from spacetime surgery} [Chap.\ref{QS_stSurgery}]:
We have mentioned that the mapping class group data
from overlapping the wavefunction from a spatial manifold mapped back to another wavefunction on the same spatial manifold, 
such as the modular SL$(2,\Z)$ data $\cS$ and $\cT$ matrices on $T^2$-torus, provide the universal topological invariants for topological orders (TOs).
Now, we can digest that, 
the (projective) representations of these modular transformation and the mapping class group encode the
information of TOs, thus encode the information of \emph{quantum topology}.
However, it seems that the mapping class group, or more generally the \emph{spatial topology} dictates certain hidden rules governing the \emph{quantum topology}.
We can ask: How the \emph{spacetime topology} and the \emph{quantum topology} are related or associated with each other? Can the spacetime topology constrains
the existence of certain quantum phases? Reversely, or more profoundly and philosophically, 
can the quantum topology constrains the existence of specific spacetime topology?


\section{Summary of the key results} \label{sec:key_result}

Now we shall summarize the key results and answers to the above questions, in a less-formal but physical intuitive manner.
The general perspective on topological states in terms of 
symmetry, topology and anomalies is simple, organized in Table \ref{table:STA_SPTTO}.

\begin{center}
\begin{table}[!h]
\begin{tabular}{|c||c| }
\hline   
Aspects of: &  Realization in the topological states\\  
\hline
\hline 
{Symmetry} 
& \begin{minipage}[t]{4.6in}  SPTs with global symmetry, classical topology and gauge or mixed gauge-gravity anomalies \end{minipage}\\  
\hline
{Topology} 
&  \begin{minipage}[t]{4.6in}  Topological orders with quantum topology and gravitational anomalies
\end{minipage}\\ 
\hline
{Anomalies} 
& \begin{minipage}[t]{4.6in}  Phenomena happens on the boundary of topological states. 
The properties are connected to the quantum nature
of the bulk, such as the symmetry-protection in SPTs or exotic braiding statistics in TOs.\\
\end{minipage}\\ 
\hline
\end{tabular}
\caption{Perspective on topological states in terms of 
symmetry, topology and anomalies.} 
\label{table:STA_SPTTO}
\end{table}
%
\begin{table}[!h]
 $$\text{ Topology } \to
 \left\{ 
 \begin{array}{lll} 
 \text{ Classical topology: homotopy, mapping and winding numbers, K-theory. } \\
 \text{ Quantum topology: algebraic topology, (co-)homology, tensor category.}  \\ 
 \text{ Spacetime topology: fiber bundles, geometric-topology, surgery theory.} 
  \end{array}  \right.
 $$
 \caption{The interplay of classical, quantum and spacetime topology.}
\label{table:topo}
\end{table}
\end{center}
%
%
%
%

The topology issues studied in SPTs and topological orders are rather different. For SPTs, it is kind of 
classical topology, concerning the continuous mapping,
in terms of homotopy, mapping and winding numbers, or K-theory. 
Classical topology is less robust, and SPTs are not stable against local perturbation which breaks the symmetry.
For topological orders, it concerns quantum topology, which is more algebraic and more robust.
Topological order is robust against any local perturbation.
See Table \ref{table:topo} for a summary.

We summarize the answers in {\bf A.I-VI} corresponding to the previous questions {\bf Q.I-VI}. 

\noindent
({\bf A.I)}. {\bf SPT invariant and its field-theory representation} \cite{JuvenSPT1, Gu:2015lfa}:
Even though SPT has a unique ground state on a closed manifold without symmetry-braking,
we achieve to simulate analogous \emph{geometric phase} of SPTs. The key idea is to do the \emph{symmetry twist},
due to the existence of a global symmetry group $G$. 
To define the symmetry twist, we note that the Hamiltonian
$H=\sum_x H_x$ is invariant under the global symmetry
transformation $U=\prod_\text{all sites} U_x$, namely $H=U H U^{-1}$.
If we perform the symmetry
transformation $U'=\prod_{x\in \prt R} U_x$
only near the boundary of a region $R$ (say on one side of ${\prt R}$),
the local term $H_x$ of $H$ in the Hamiltonian near the boundary of $R$
will be modified: $H_x\to H_x'|_{x\text{ near } \prt R}$.
It is important to remark that the original symmetry $U(g)$ is \emph{local, unitary and onsite} and has a tensor product structure, 
but the symmetry-twist transformation is \emph{not unitary and not a symmetry} transformation.
Instead, the symmetry-twist operation is a \emph{modification to the original Hamiltonian}. In short,
\bea
&& U(g)=\otimes_j U_j(g), \;\; U(g) H U(g)^{-1}=H, \;\;g \in G. \nonumber \\
&& H=  \sum_x H_x \overset{\text{sym.twst along $\prt R$}}{\longrightarrow}\sum_{x \not\in \prt R} H_x+  \sum_{x \in \prt R} H_x'|_{x\text{ near } \prt R}
\eea
Suppose the branch cut $\prt R$ is between the sites indices $x_{0,j}$ and $x_{1,j}$ varying $j$ while moving along the $\prt R$, 
and suppose the interacting Hamiltonian are nearest neighbored interacting with the local term $H_{x_{i},x_{j}}$.
We stress that the symmetry twist can be performed as part of symmetry transformation to modify $H_x'$ as 
$H_{x_{0,j},x_{1,j}} \to H_{x_{0,j},x_{1,j}}' \equiv U_{x_{1,j}}(g) H_{x_{0,j},x_{1,j}} U_{x_{1,j}}(g)^{-1}$,
however the symmetry twist maintains $H_{x_{1,j},x_{2,j}} \to H_{x_{1,j},x_{2,j}}$. Thus overall the symmetry twist is \emph{not a symmetry transformation}
but a modification on $H$ to $H'$.

The above is a lattice Hamiltonian approach. Can we interpret in terms of a continuous field theory perspective?
For systems that realize topological orders, we can adiabatically deform the ground state $| \Psi_{g.s.} (g) \rangle$ of parameters $g$ via:
\bea
&&\langle \Psi_{g.s.} (g+\delta g) | \Psi_{g.s.} (g) \rangle \simeq \dots    \mathbf{Z}_0 \dots
\eea
to detect the volume-independent universal piece of
spacetime partition function, $\mathbf{Z}_0$, which
reveals non-Abelian geometric phase of ground states.
We can use $\mathbf{Z}_0$
with the \emph{symmetry twist} to probe the SPTs.
A symmetry twist implies a change along a codimension-1 surface, which modifies the SPT partition function from $\mathbf{Z}_0$ %
to $\mathbf{Z}_0(\text{sym.twist})$.
{Just like the geometric phases of the degenerate ground states characterize topological orders \cite{Kong:2014qka}, 
we believe that $\mathbf{Z}_0(\text{sym.twist})$, on different spacetime manifolds and for different symmetry twists, fully characterizes
SPTs.
The symmetry twist is similar to gauging the on-site
symmetry except that the symmetry twist is non-dynamical. We can use the gauge connection 1-form $A$ to
describe the corresponding symmetry twists, with probe-fields $A$ coupling to
the matter fields of the system.  So we can write  
\begin{align} 
\mathbf{Z}_0(\text{sym.twist})
=\ep^{\ti \mathbf{S}_0(\text{sym.twist})}=\ep^{\ti \mathbf{S}_0(A)}.
\end{align}
Here $\mathbf{S}_0(A)$ is the SPT invariant that we search for. 
This is a partition function of classical probe fields, or a topological response theory, obtained by integrating out the matter fields of SPTs path integral 
\cite{JuvenSPT1, Gu:2015lfa}.

\noindent
({\bf A.II)}. {\bf Bosonic anomalies}  \cite{Wang:2014tia, Santos:2013uda}: 
We classify and characterize several types of bosonic anomalies found on the boundary of bosonic SPTs, as an example on 1+1D edge. 
One has induced fractional quantum numbers via symmetry-breaking (similar to Jackiw-Rebbi \cite{Jackiw:1975fn} and Goldstone-Wilczek \cite{Goldstone:1981kk} effects); 
one has degenerate zero modes (carrying the projective representation protected by the unbroken part of the symmetry)
either near the 0D kink of a symmetry-breaking domain wall, or on a symmetry-preserving 1D system dimensionally reduced from a thin 2D tube 
with a monodromy defect 1D line embedded. 
More generally, the energy spectrum and conformal dimensions of gapless edge modes under an external gauge flux insertion 
(or twisted by a branch cut, i.e., a monodromy defect line) through the 1D ring
can distinguish many SPT classes. 
The last one exhibits the many-body Aharonov-Bohm (A-B) effect \cite{Santos:2013uda}. 
%
The aforementioned edge properties are explicitly formulated 
in terms of (i) a long wavelength
continuum field theory involving scalar chiral bosons \cite{SPTCS1,{Lu:2012dt},{Ye:2013upa},{Wang:2014tia}}, 
for a generic finite Abelian symmetry group $G=\prod_u Z_{N_u}$. 
We can express the multiplet (non-)chiral boson action
$\mathbf{S}_{edge}=\int \frac{K_{IJ}}{4\pi}  dt dx\,\partial_t \phi_I  \partial_x \phi_J$ with a $K$-matrix:
$
K_{} =\bigl( {\begin{smallmatrix}
0 &1 \\
1 & 0
\end{smallmatrix}}  \bigl) \oplus \bigl( {\begin{smallmatrix}
0 &1 \\
1 & 0
\end{smallmatrix}}  \bigl)  \oplus \bigl( {\begin{smallmatrix}
0 &1 \\
1 & 0
\end{smallmatrix}}  \bigl)  \oplus \dots
$, and a multiplet of scalar bosons: $\phi_I =(\phi_1, \phi_1', \phi_2, \phi_2', \dots )$ with chiral-anti-chiral pair of scalar modes $(\phi_u, \phi_u')$.
One of our key formulas is the symmetry operator $S$ completing the group cohomology classification, 
\bea 
 S&=&\prod_{u,v,w \in \{1,2,3\}} \exp\big[ \frac{\ti \, 2\pi}{N_u}\,
\int^{L}_{0}\,dx\,( \frac{1}{2\pi} \partial_{x}\phi_{u}' + \frac{ p_{\tI}  }{ 2\pi} \partial_{x}\phi_{u} \nonumber\\ 
&+&\epsilon^{u v}  \frac{ p_{\tII} }{ 2\pi}  \partial_{x}\phi_{v} +  { \frac{{p_{\tIII}} N_1N_2N_3}{(2\pi)^2 N_{123}}}  \epsilon^{u v w} \phi^{}_{w}(x)\cdot \partial_x \phi^{}_{v}(x)  ) \big],
\eea
where $p_{\tI}, p_{\tII}, p_{\tIII}$ are nontrivial SPT class indices. 
Here our discrete formulations: (ii) Matrix Product Operators and (iii) discrete quantum lattice models provide crucial insights to derive (i) the field theory formulation.
The symmetry transformation can be readily checked by calculating $S \phi_I S^{-1}$ with the commutation relations given above.
Our \emph{lattice} approach yields a \emph{regularization with anomalous non-onsite symmetry for the field theory} description.

\noindent
({\bf A.III)}. {\bf Topological gapping criteria. Topological degeneracy on a manifold with gapped domain walls and boundaries} \cite{Wang:2012am, Lan:2014uaa}.
The observation in the third lesson guides us to study, under what mechanisms, 
would edge modes  
protected to be gapless: (i) the chirality, associated with perturbative gravitational anomalies and gauge anomalies; 
(ii) the global symmetry protection; 
(iii) the bulk nontrivial statistics \cite{{Wang:2012am},{Levin:2013gaa}}:
even for non-chiral states, global gravitational anomalies can protect gapless edge modes \cite{Lan:2014uaa}.
Conversely, to determine the edge mode gapping criteria of TOs, we use both TQFT 
\cite{Wang:2012am} as well as using SL$(2,\Z)$ modular data and
category theory \cite{Lan:2014uaa}. 
Once the boundary modes are gapped, we can introduce the new concept of boundary degeneracy 
for ground states 
on a generic (open or closed) manifold of gapped boundaries and domain walls \cite{{Wang:2012am},{Levin:2013gaa}}, 
which provides richer information than the old concept of bulk degeneracy on a closed manifold.
Intuitively a gapped boundary is labeled by a set of anyons where they share \emph{trivial self and mutual braiding statistics}. %
We can call this set of particles as \emph{condensible} particles. 
In the Abelian TOs, this set of anyons, with the fusion as operator, 
form a mathematical group structure like a lattice.\footnote{For experts in Chern-Simons TQFT, it is the Chern-Simons quantized lattice Hilbert space.}
We derive the GSD for TOs on a manifold with gapped boundaries, here in an intuitive simplified level, 
as the order of a quotient group between two lattices of Hilbert space,
\be
\GSD=\left|  \frac{\text{lattice of condensible anyons}}{\text{lattice of condensible non-fractionalized particles}} \right|, 
\ee
subject to the implicit constraint on neutrality condition of anyon transporting between all gapped boundaries.
Our generic result of boundary GSD recovers the known result of bulk GSD for a level $k$ or $K$-matrix Chern-Simons theory 
($k=3$ for a filling-fraction $\frac{1}{3}$-FQH state discussed in Sec.\ref{Sec:ClassChara}) with
$$ \GSD=k^g, \text{ or } |\det K|^g. $$
on a genus $g$-Riemman surface.
We predict that the $Z_2$ toric code \cite{Kitaev:1997wr} and $Z_2$ double-semion model (more generally, 
the $Z_k$ gauge theory and the $U(1)_k\times U(1)_{-k}$ non-chiral fractional quantum Hall state at even integer $k$) 
can be numerically and experimentally distinguished, by measuring their boundary degeneracy on an annulus or a cylinder.
{For a more generic non-Abelian TOs, we formulate gapped domain walls and GSD in terms of modular data $\cS$ and $\cT$
in Chap.\ref{aofTopology}.} 
Since gapped domain walls talk to each other through
long-range entanglement, the GSD with domain walls reveals 
more physics than that without domain walls. We
foresee its practicality in experiments, since we can read
even more physics by putting the system on open surfaces
with gapped domain walls.

\noindent
({\bf A.IV)}. {\bf Define lattice chiral fermion/gauge theory non-perturbatively} 
  is a long-standing challenge, 
due to Nielsen-Ninomiya's no-go theorem on the fermion-doubling problem \cite{Nielsen:1980rz}.  
However, based on our framework, we carefully examine this theorem to discover 
at least two approaches 
to bypass the challenge (see Fig.\ref{G-WandOurs} (a)): 
One approach is known to be 
Ginsparg-Wilson (G-W) fermions \cite{Ginsparg:1981bj, {Wilson:1974sk}, {Neuberger:1997fp}, {Neuberger:1998wv}, {Hernandez:1998et}} fulfilling the chiral symmetry non-onsite, where
we find G-W fermions are the boundary modes of some 
SPTs.
The second approach \cite{Wang:2013yta} is 
a bulk trivial insulator placed on a cylinder with gappable boundary modes of onsite symmetry. 
We introduce proper interactions 
within the mirror sector to have one gapped boundary 
and leave the light sector gapless  with chirality on the other boundary --- this approach 
belongs to the mirror-decoupling framework independently studied since Eichten-Preskill \cite{Eichten:1985ft}. 
\begin{figure}[!h]
{\includegraphics[width=0.98\textwidth]{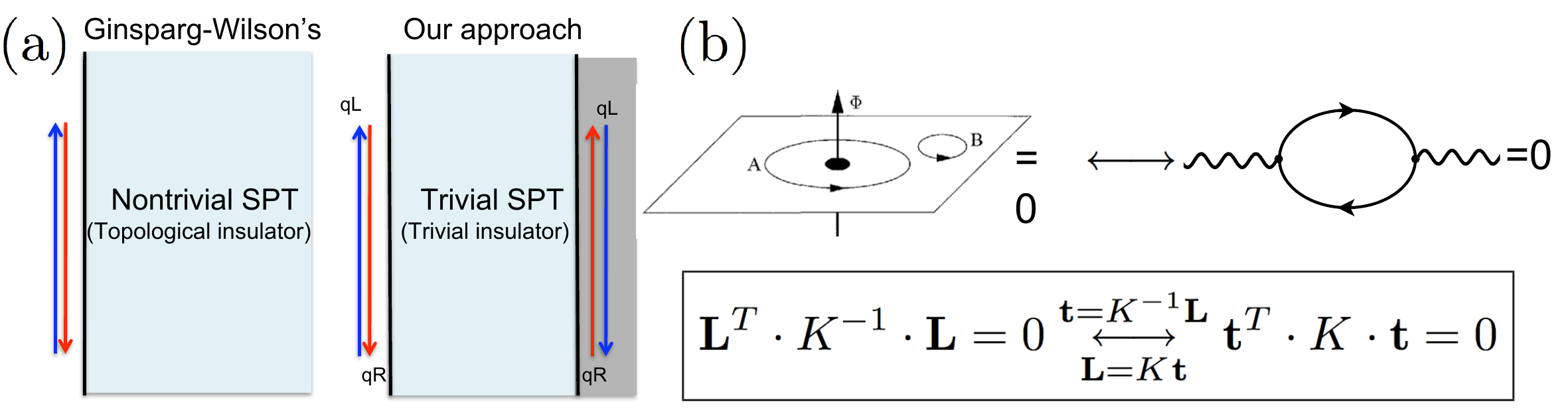}  }  
\caption{(a) Gilzparg-Wilson fermions can be viewed as putting gapless states on the edge of a nontrivial SPT state (e.g. topological insulator).
Our approach can be viewed as putting gapless states on the edge of a trivial SPT state (trivial insulator) and introduce proper strong interactions to gapped out
the mirror sector (in the shaded region). (b) The equivalence of the boundary gapping criteria and the 't Hooft anomaly matching conditions.
Our proof is based on a bulk theory of Abelian SPT described by a $K$-matrix Chern-Simons action 
$\frac{K_{IJ}}{4\pi} \int   a_{I} \wedge \dd  a_{J}$.
A set of anyons, labeled by a matrix $\mathbf{L}$, with trivial mutual and self statistics is formulated as $\mathbf{L}^T \cdot K^{-1} \cdot \mathbf{L}=0$. 
This condition is equivalent to a 1-loop anomaly-matching condition for fermions, or more generally as ${{\mathbf{t}^T} K {\mathbf{t}}=0}$
for both bosons and fermions, where $\mathbf{t}$ is a matrix formed by the charger coupling between matter fields and external gauge fields
(as solid lines and wavy lines respectively in the Feynman diagram).
}
\label{G-WandOurs}
\end{figure}
This observation guides us to our topological non-perturbative proof \cite{Wang:2013yta},  the {\bf fourth
lesson} on anomalies, on \\
\colorbox{lgray}{\parbox[t]{\linegoal}{
``the equivalence of 't Hooft anomaly matching conditions and the boundary gapping criteria. ''
}}\\
See Fig.\ref{G-WandOurs} (b) for the equivalence relation in a picture.
What we discover reinforces Niels Bohr's insight: ``the hallmark of a deep truth [here the fermion-doubling theorem] is that its negation is a further deep truth.''

\noindent
({\bf A.V)}. {\bf String and particle exotic braiding statistics and higher dimensional TO lattice models} \cite{Wang:2014oya,Wan:2014woa}:
We initiate the characterization of higher dimensional topological orders and their braiding statistics of 
string/particle excitations \cite{{Wang:2014xba},{Jiang:2014ksa},Wang:2014oya}. 
For example, 
we place a bulk topological phase in a 3D Rubik's cube with parallel faces identified as a 3-torus,
and deform the space but end up maintaining the same topology -- 
this yields SL(3,$\mathbb{Z}$) data.
The SL($d$,$\mathbb{Z}$) matrix data can be viewed from two perspectives: 
first, using the \emph{spacetime path integral}, the initial and final wavefunction-overlapping yields those data
by the above deformation process on $d$-torus. 
Second, using the mathematical Representation Theory, the modular data {are encoded by} gauge groups and cohomology twist inputs. 
We have derived both approaches and depicted vivid physics connections to multi-string braiding. 
In addition, we systematically construct a lattice Hamiltonian realization of Dijkgraaf-Witten twisted gauge theory \cite{Dijkgraaf:1989pz}
for 3+1D \cite{Wan:2014woa}, which achieves an extension of Kitaev quantum double models in 2+1D \cite{Kitaev:1997wr} to 
the generalized twisted cases and to higher dimensions.

\noindent
({\bf A.VI)}. {\bf Topological invariants, as quantum statistics derived from spacetime surgery} [Chap.\ref{QS_stSurgery}]:
The interplay between quantum topology and spacetime topology is examined, in a few simple examples.
By performing the surgery theory of geometric-topology on the spacetime,  
we show that the quantum fusion rule and quantum statistics are constrained by the intrinsic properties of spacetime topology.
The exotic quantum statistics is defined in the adiabatic braiding process in the gapped phases of matter with topological orders, therefore 
the spacetime topology strictly constrains the quantum topology thus dictates the possible gapped phases of matter.

\section{Outline of thesis and a list of journal publications} \label{outline}

The thesis is organized to address the questions in Sec.\ref{sec:problem} and illuminate more in depth in the Sec.\ref{sec:key_result}.
 
In Chapter \ref{gphase}, we warm up by discussing the
concepts of {geometric phase, wavefunction overlapping and topological invariants}. 

In Chapter \ref{aofSymmetry}, on Aspects of Symmetry, we address the issue in {\bf Q.I} and {\bf A.I} on the procedure to do the symmetry-twist, on the SPT invariant 
derived from the lattice model or and its field-theory representation.
We also address part of the issues in {\bf Q.II} and {\bf A.II} on bosonic anomalies and their SPT invariants and SPT observables.
The reason we discuss part of anomaly issue here is that remarkably the anomalous edge global symmetry corresponds to the 
(gauge) anomalies. The global symmetry can be coupled to a weakly-coupling gauge fields or external probe fields.
So anomalous symmetries manifest quantum anomalies. The issues of anomalous symmetry and anomalies are intertwined.

In Chapter \ref{aofTopology}, on Aspects of Topology,
we work out {\bf Q.III} and {\bf A.III}, topological gapping criteria, topological degeneracy on a manifold with gapped domain walls and boundaries.
We tackle the challenge of {Q.V} and {A.V} on string and particle exotic braiding statistics and TO lattice models in 3+1D. 

In Chapter \ref{aofAnomalies}, on Aspects of Anomalies, we address {\bf Q.IV} and {\bf A.IV} on a non-perturbative
 definition of lattice chiral fermion/gauge theory. 
Other part of discussions are extension of previous topics from {\bf Q.I} and {\bf A.I} on SPT field-theory representation with mixed gauge-gravity anomalies, 
and {\bf Q.II} and {\bf A.II} on bosonic anomalies.     

In Chapter \ref{QS_stSurgery}, we address the issue in {\bf Q.VI} and {\bf A.VI} on 
quantum statistics data as topological invariants derived from spacetime surgery.
We formulate
the constraints of braiding statistics and fusion analogous to Verlinde's formula in 2+1D and 3+1D. 
This approach should be applicable to any spacetime dimension.
A more complete study will be reported elsewhere in the future publication.

 {\bf Review articles}: For a colloquium overview of topological insulators and superconductors can be found in \cite{2010RMP_HasanKane,2011_RMP_Qi_Zhang}.
An earlier version of overview on topological phases is in \cite{Turner-Vish}.
An intuitive but less-formal guide to topological insulators and SPTs can be found in \cite{Senthil1405}.
A more recent review on topological order is in \cite{Wen1210.1281}.
Obviously I must thank S. Coleman's wonderful book \cite{ColemanA} and its inspiration on my thesis and its title.
Two important reviews on quantum field theory and anomalies: Treiman-Witten-Jackiw-Zumino\cite{Treiman:1986ep} and Farhi-Jackiw\cite{Farhi:1982vt}.

Part of the thesis is the overview and the summary for part of the work published elsewhere.
%
\begin{enumerate}
\item Juven Wang and Xiao-Gang Wen. Boundary Degeneracy of Topological Order.
Phys.Rev., B91(12):125124, 2015. {\em ArXiv e-prints} 1212.4863. \cite{Wang:2012am}
\item Peng Ye and Juven Wang. Symmetry-protected topological phases with charge 
and spin symmetries: Response theory and dynamical gauge theory in two and 
three dimensions. Phys.Rev., B88(23):235109, 2013. {\em ArXiv e-prints} 1306.3695. \cite{Ye:2013upa}

\item Juven Wang and Xiao-Gang Wen. A Lattice Non-Perturbative Hamiltonian Con- 
struction of 1+1D Anomaly-Free Chiral Fermions and Bosons - on the equiva- 
lence of the anomaly matching conditions and the boundary fully gapping rules. 
2013. {\em ArXiv e-prints} 1307.7480. \cite{Wang:2013yta}

\item Luiz H. Santos and Juven Wang. Symmetry-protected many-body Aharonov- 
Bohm effect. Phys.Rev., B89(19):195122, 2014. {\em ArXiv e-prints} 1310.8291\cite{Santos:2013uda}

\item Juven Wang, Luiz H. Santos, and Xiao-Gang Wen. Bosonic Anomalies, Induced 
Fractional Quantum Numbers and Degenerate Zero Modes: the anomalous edge 
physics of Symmetry-Protected Topological States. 2014. {\em ArXiv e-prints} 1403.5256 \cite{Wang:2014tia}

\item Juven Wang and Xiao-Gang Wen. Non-Abelian string and particle braiding 
in topological order: Modular SL(3,Z) representation and (3+1) -dimensional 
twisted gauge theory. Phys.Rev., B91(3):035134, 2015. {\em ArXiv e-prints} 1404.7854 \cite{Wang:2014oya}

\item Juven C. Wang, Zheng-Cheng Gu, and Xiao-Gang Wen. Field theory represen- 
tation of gauge-gravity symmetry-protected topological invariants, group coho- 
mology and beyond. Phys.Rev.Lett., 114(3):031601, 2015. {\em ArXiv e-prints} 1405.7689 \cite{Wang:2014pma}

\item Tian Lan, Juven C. Wang, and Xiao-Gang Wen. Gapped Domain Walls, Gapped 
Boundaries and Topological Degeneracy. Phys.Rev.Lett., 114(7):076402, 2015. 
{\em ArXiv e-prints} 1408.6514 \cite{Lan:2014uaa} 

\item Yidun Wan, Juven C. Wang, and Huan He. Twisted Gauge Theory Model of 
Topological Phases in Three Dimensions. 2014. {\em ArXiv e-prints} 1409.3216 \cite{Wan:2014woa}

\item Zheng-Cheng Gu, Juven C. Wang, and Xiao-Gang Wen. Multi-kink topolog- 
ical terms and charge-binding domain-wall condensation induced symmetry- 
protected topological states: Beyond Chern-Simons/BF theory. 2015
{\em ArXiv e-prints} 1503.01768 \cite{Gu:2015lfa}

\item Juven C. Wang,  Clifford H. Taubes, Xiao-Gang Wen and Shing-Tung Yau.    
Quantum statistics and spacetime surgery: Fusion algebra, braiding statistics, and mapping class group (to appear)
\end{enumerate}

and other work to appear in near-future publications.

%% file: G-phase.tex
\chapter{Geometric phase, wavefunction overlap, 
spacetime path integral and topological invariants} \label{gphase}

\section{Overview}
In this chapter, we work through the familiar concept of geometric phase, firstly emphasized by Berry. 
The hint and the similar idea have been noticed by Pancharatnam, Aharonov-Bohm and others that there is an extra geometric phase
in addition to the familiar dynamical phase when we perform the adiabatic evolution on the physical system governed by Hamiltonian. 
What Berry emphasized is that when the adiabatic evolution trajectory in the Hamiltonian coupling constant space
is \emph{closed}, then this \emph{trajectory-dependent geometric phase is a physical measurable invariant quantity}---
this particular geometric phase is invariant in a sense of gauge invariance. 
%
Wilczek-Zee noticed the non-Abelian geometric matrix 
for adiabatically evolving degenerate states.
Wen had the insights to discover 
topological orders (TOs) and its GSD for quantum Hall fluids and
apply the non-Abelian geometric matrix together with Chern-Simons theory of Witten's TQFT to characterize and classify TOs.
Here we digest the development of these ideas in a coherent physical way, and will develop this approach further to study TOs and SPTs in any dimensions in Chap.\ref{aofSymmetry}, \ref{aofTopology} and \ref{QS_stSurgery}. 

\section{Geometric (Berry's) phase and the non-Abelian structure}  \label{sec:Berry}

\subsection{Geometric (Berry's) phase}
Start with the study of wavefunction $\Psi(\tau(t))$ evolving under the time-dependent Hamiltonian $H=H(\tau(t))$
of some time-dependent coupling constant $\tau$.
In general, for a series of $\tau^\alpha$ parameters, we will simply denote them as the parameter $\tau$. 
Our goal is determine $\Psi(\tau(t))$ by solving $$H | \Psi(\tau(t)) \rangle= \ii \hbar \;\partial_t | \Psi(\tau(t)) \rangle.$$
Let us assume $\Psi(\tau(t))$ starts at $t=0$ from an energy-eigenstate $\Psi(\tau(0))=\phi(\tau(0)) \equiv \phi_{E_0}(\tau(t))$ with an eigen-energy $E_0$.
At every moment we can still find a set of eigenstate $\phi(\tau(t))$ as bases following $H(\tau(t)) \phi(\tau(t))=E(\tau(t)) \phi(\tau(t))$.

We consider the \emph{adiabatic} evolution (without sudden exchange energy with external environment), here in the sense that the time scale $dt$ of changing
the energy $d E$ are bounded between two other scales. 
One scale is the energy gap of higher/lower excitations, defined as $\Delta \simeq \Delta_j \equiv |E_j-E_0|$. 
The other scale is the energy splitting $\delta$ between nearly degenerate ground states around the energy scale $E_0$.\footnote{
A deep side remark:
For some miraculous situation, say, ground states of certain many-body systems such as topological orders, 
the degenerate states are \emph{topologically robust to stay together as nearly degenerate} in the energy spectrum.
However, to hold the nearly degenerate into an exact degeneracy requires some extra symmetry. This extra symmetry may not be robust against
local perturbation, and this symmetry is not required for topological orders.
In other words, the \emph{topological ground state degeneracy is not an exact degeneracy, but only an approximate degeneracy}!
}
The {adiabatic} evolution requires that the
energy changes at the unit time: $\frac{\dd E}{ \dd t} \times \text{unit time} \equiv \epsilon$
and the rate of change for a unit energy: $ \frac{\dd E}{ \dd t \cdot \text{unit energy}}   \equiv 1/T$ are bounded by:
\be \label{adibatic1}
\Delta \gg {T}^{-1} \simeq \epsilon, \;\;\;\;\; \text{     or equivalently    }  \;\;\;\;\;{\Delta}^{-1} \ll T \simeq {\epsilon}^{-1}.
\ee
How the scale $\delta$ is set in depends on the physics we look for.
If we like to focus on a single eigenstate without being interfered by other nearly degenerate states, then we will have to set a finer condition:
$\Delta > \delta \gg {T}^{-1} \simeq \epsilon$, or equivalently,  ${\Delta}^{-1} < {\delta}^{-1}  \ll T \simeq {\epsilon}^{-1}$.
Indeed, this is difficult, and it will be easier if we start with an isolated eigenstate instead of the troublesome nearly degenerate states!
However, one interesting piece of physics emerges when we consider the nearly degenerate states together. We will explore in Sec.\ref{sec:nAbBerry},
the geometric phase becomes non-Abelian if we are in the time scale where all nearly degenerate states are important: 
\be \label{adibatic3}
\Delta \gg {T}^{-1} \simeq \epsilon \gg \delta, \;\;\;\;\; \text{     or equivalently    }  \;\;\;\;\;{\Delta}^{-1} \ll T \simeq {\epsilon}^{-1} \ll {\delta}^{-1}.
\ee

Now let us stick to the simplest condition Eq.(\ref{adibatic1}) for a moment with an isolated eigenstate. We will generalize it to degenerate states in Sec.\ref{sec:nAbBerry}.
In this adiabatic evolution, it is reasonable to write a generic wavefunction 
$\Psi(\tau(t))=\e^{-\frac{\ii}{\hbar} \int_0^t \dd s E_0(s)}  \cdot \e^{\ii \gamma} \cdot \phi(\tau(t))$.
The first piece is dynamical phase which exists even for a time-independent Hamiltonian.
The second piece $\e^{\ii \gamma}$ can be solved, and one finds the geometric phase:
\be
 \gamma = \int_{\tau(0)}^{\tau(t)}  {\dd \tau^\alpha} \langle \phi(\tau(t)) | \ii\frac{\prt}{\prt\tau^\alpha}| \phi(\tau(t))\rangle 
\ee
Several properties of $\gamma$ are derived:\\
(i) $\gamma$ is a pure phase, 
 $\gamma \in \R$ because $\langle \phi | \frac{\dd}{\dd \tau}| \phi\rangle$ is an imaginary number. The wavefunction maintains unitary in the same eigenstate.\\
(ii) It has no explicit $\hbar$-dependence. So we may say the geometric phase remains even at the classical limit $\hbar \to 0$.\\
(iii) It is \emph{geometric and trajectory-dependent}. But it has no explicit time-dependent and it is parametrization independent to $\tau$, 
thus $\gamma$ does not change no matter how fast or slow the process is as long as the process stays adiabatic. 
In contrast, the dynamical phase is explicitly time-dependent.\\
(iv) It can be written in terms of gauge potential as $A \equiv A_\alpha {\dd \tau^\alpha} =\langle \phi(\tau(t)) | \ii\frac{\prt}{\prt \tau^\alpha}| \phi(\tau(t))\rangle{\dd \tau^\alpha}$,
so
$
 \gamma = \int_{\tau(0)}^{\tau(t)}  A= \int_{\tau(0)}^{\tau(t)} {\dd \tau^\alpha} \; A(\tau)_\alpha 
$.
In the most general case, $A$ is a connection on a U(1)-bundle with a base-manifold in the $\tau$-parameter space. If the U(1)-bundle is trivial,
the connection $A$ becomes a Lie-algebra valued 1-form, more specifically a 1-form gauge field.\\ 
(v) A choice of eigenbasis up to U(1) phase becomes the \emph{gauge transformation}, if we change the eigenstate by a unitary transformation
$
\phi'(\tau)=\Omega(\tau) \phi(\tau)
$ where $\Omega^{-1}\Omega=\Omega^{\dagger}\Omega=1$,
then the gauge field $A'=A+\ii  ({\dd}\Omega) \Omega^{-1} \Rightarrow 
A'_\alpha=A_\alpha+\ii  (\frac{\prt}{\prt \tau^\alpha}\Omega) \Omega^{-1}$.
For a single energy level, $\Omega(\tau)=\e^{\ii f(\tau)}$, so the gauge field is transformed as $A'=A- \dd f$.\\
(vi) A 
universal feature of the geometric phase $\gamma$ arises, if the trajectory is \emph{closed} with an enclosed region $R$. 
Namely, because $\oint \dd f=0$, the $\gamma$ does not depend on the choices of eigenbasis and the basis (``gauge'') transformation.
If there are multiple parameters of $\tau^\alpha=(\tau^1, \tau^2, \dots)$, we further define the field strength as:
$F_{\alpha\beta}=\prt_\alpha A_\beta-\prt_\beta A_\alpha=
\ii (\langle \frac{\prt \phi}{\prt \tau^\alpha}   | \frac{\prt \phi}{\prt \tau^\beta} \rangle-
\langle \frac{\prt \phi}{\prt \tau^\beta}   | \frac{\prt \phi}{\prt \tau^\alpha} \rangle)$, so
\be \label{eq:geophase1}
\gamma_{\text{enclosed}}=\oint_{\prt R}  A = \oint_{\prt R}  A(\tau)_\alpha {\dd \tau^\alpha}  = \int_R F= \int_R F_{\alpha\beta} {\dd \tau^\alpha} \wedge {\dd \tau^\beta}
\ee
However, we stress and clarify that the 1-form ``gauge field'' $A$ and
the 2-form ``field-strength curvature'' $F$  
here are \emph{not} those in the usual dynamical gauge theory, since it lives in the probed coupling constant $\tau$ with values determined by semi-classical external probes. \\
(vii) We can rewrite the geometric phase $\gamma$ in terms of Kubo formula or the linear response theory form.
Insert a complete eigenbasis identity matrix $I=\sum_{E_j}| \phi_{E_j}\rangle \langle\phi_{E_j}  |$ into the
$F_{\alpha\beta}$-term (be aware, not just for the single eigenstate or nearly degenerate states with energy $E_0(\tau(t))$, but the 
whole energy spectrum $E_j$) and then use
the analogous 
Hellmann-Feynman relation $\langle \phi_{E_j}  | \frac{\prt H}{\prt \tau^\alpha} |\phi_{E_0} \rangle+(E_j-E_0)\langle \phi_{E_j}  | \frac{\prt \phi_{E_0}}{\prt \tau^\alpha} \rangle$, plug in Eq.(\ref{eq:geophase1}), we obtain:
\bea \label{eq:geophase2}
&&\gamma_{\text{enclosed}}=\ii\int_R  (\langle \frac{\prt \phi_{E_0}}{\prt \tau^\alpha}   | \frac{\prt \phi_{E_0}}{\prt \tau^\beta} \rangle-
\langle \frac{\prt \phi_{E_0}}{\prt \tau^\beta}   | \frac{\prt \phi_{E_0}}{\prt \tau^\alpha} \rangle){\dd \tau^\alpha} \wedge {\dd \tau^\beta}\\
&&= \label{eq:geophase3}
\ii\int_R \sum_j 
\frac{ 
\langle \phi_{E_0}  | \frac{\prt H}{\prt \tau^\alpha} |\phi_{E_j} \rangle
\langle \phi_{E_j}  | \frac{\prt H}{\prt \tau^\beta} |\phi_{E_0} \rangle -
\langle \phi_{E_0}  | \frac{\prt H}{\prt \tau^\beta} |\phi_{E_j} \rangle
\langle \phi_{E_j}  | \frac{\prt H}{\prt \tau^\alpha} |\phi_{E_0} \rangle  }{(E_j-E_0)^2}  {\dd \tau^\alpha} \wedge {\dd \tau^\beta}.\;\;\;\;\;\;\;\;
\eea
We will comment more on the deeper meanings of Eq.(\ref{eq:geophase2}) and  Eq.(\ref{eq:geophase3}) after introducing
the more general non-Abelian geometric matrix structure (also called non-Abelian Berry phase or non-Abelian gauge structure) emerged in degenerate energy states.
\subsection{Non-Abelian Geometric (Berry's) Structure} \label{sec:nAbBerry}
%
We can generalize the approach above to the case with 
$N$-fold degenerate energy states of energy $E_0$, firstly performed by Wilczek and Zee \cite{Wilczek:1984dh}.
For degenerate states $|\phi_{E_{0,a}} \rangle \equiv |\phi_{a} \rangle$  with $a=1,\dots,N$, we define the 1-form gauge field  as
\be
A=A_{ba,\alpha} \dd \tau^\alpha =\langle \phi_a(\tau) | \ii\frac{\prt}{\prt \tau^\alpha}| \phi_b(\tau)\rangle \dd\tau^\alpha
\ee
We can prove that
$A_{ab}$ is Hermitian: $A_{ab}=A_{ba}^*$. If the coupling $\tau$ is defined to be real, we have $A=A^\dagger$.
The gauge transformation becomes an $\Omega(\tau)$ transformation between degenerate states:
$
|\phi'_{a} \rangle=\Omega_{ab} | \phi_b\rangle
$.
We find
$A_{ba,\alpha}'= \Omega_{bd} A_{dc}(\Omega^{-1})_{ca}+\ii(\frac{\prt}{\prt \tau^\alpha} \Omega_{bc}) (\Omega^{-1})_{ca}$,
namely,
\be
A'= \Omega_{} A_{}\Omega^{-1}_{}+\ii({\dd} \Omega_{}) (\Omega^{-1})_{}.
\ee
(viii) $\gamma$ as a rank-$N$ \emph{matrix}: The geometric meaning of Eq.(\ref{eq:geophase1}) still holds for non-Abelian geometric structure, 
where $\gamma$ now becomes a rank-$N$ \emph{matrix} $\gamma_{ba}$ for $N$-degenerate states at $E_0$.
The geometric phase factor $\e^{\ii \gamma}$ for Schr\"odinger equation 
becomes a non-Abelian rank-$N$ matrix:
\be
[\e^{\ii \gamma}]_{ba}
=\text{P} [\e^{\ii \int A_{\alpha} \dd \tau^\alpha }]_{ba}
\ee
with the path ordering P.
For $N$-degenerate states, we can generalize 
Eq.(\ref{eq:geophase2}) and  Eq.(\ref{eq:geophase3})
by replacing $|\phi_{E_0} \rangle$ by $|\phi_{E_{0,a}} \rangle=|\phi_{E_{0,a}} \rangle$.
The modification of Eq.(\ref{eq:geophase3}) requires a higher order perturbation theory to fix the singular $O((E_j-E_0)^{-2})$ piece, but the term written in
Eq.(\ref{eq:geophase3}) still exists for $E_j \neq E_0$.

\noindent
(ix) We remark that the meanings of Eq.(\ref{eq:geophase2}) and  Eq.(\ref{eq:geophase3}) are rather \emph{different}.
First, Eq.(\ref{eq:geophase2}) only evaluates the single eigenstate $ |\phi_{E_0} \rangle$ 
or nearly degenerate states  $|\phi_{E_{0,i}} \rangle$ with the same energy $E_0$, and study its-trajectory enclosed region $R$'s local curvature  $F$ 
in the Hamiltonian coupling constant space.
Second, the form in Eq.(\ref{eq:geophase3}) with a second-order perturbation form $O((E_j-E_0)^{-2})$ is associated to the
fact that the curvature $F$ has a second-order derivative. 
However, Eq.(\ref{eq:geophase3}) also gather extra information about the nearby \emph{higher/lower energy excitations} $|\phi_{E_j} \rangle$,
the variation of the full Hamiltonian $\frac{\prt H}{\prt \tau^\alpha}$,
and how \emph{dense} these excitations with energy $E_j$ are around $E_0$ in the spectrum.

We stress that, from Eq.(\ref{eq:geophase2}) and  Eq.(\ref{eq:geophase3}) and the remark (ix), 

\colorbox{lgray}{\parbox[t]{\linegoal}{
{Geometric phase captures important properties encoded in the target eigenstates,  }
{(such as degenerate ground states, say $|\phi_{E_{0,i}} \rangle$) as in Eq.(\ref{eq:geophase1}) and (\ref{eq:geophase2}). Moreover,}
{geometric phase also encodes information about the nearby excitation states, as in}
{Eq.(\ref{eq:geophase3}). In Chap.\ref{aofTopology} we will utilize this fact to compute the Abelian or non-Abelian}
{\emph{braiding statistics of energetic excitations of strings and particles} by studying the}
{\emph{geometric phase or non-Abelian geometric structure of degenerate ground states}.}
{Non-Abelian geometric structure only means that $\gamma$ is a matrix, the braiding statistics}
{can still be Abelian or non-Abelian statistics.}
}}\\

%
We remark on the parametrization of the parameter space: 

\colorbox{lgray}{\parbox[t]{\linegoal}{
{The usual gauge theory has a form $A = A_\mu(x) \dd x^\mu$ with explicit spacetime}
{dependence $x^\mu$, for a gauge group $G$, then the $A$ is a connection of a $G$-bundle.}
{The ``gauge field'' $A=A(\tau)_\alpha {\dd \tau^\alpha}$ we study depends on the Hamiltonian coupling constant}
{$\tau$ space. The ``gauge group' depends on the \emph{accidental global symmetry} of degenerate ground}
{states. such as $\U(N)=\U(1) \times \SU(N)$ for $N$-fold degeneracy. The state $| \Psi\rangle$ lives in the}
{Hilbert space.}
}}

\section{Quantum Hall Liquids: From one electron to many electrons on the torus to the effective Chern-Simons theory}


Here we like to introduce an additional idea, to use the Berry phase and geometric matrix structure to characterize and classify 2D quantum Hall liquids.
For an interacting system with $N$ electrons described by the toy-model Hamiltonian \cite{W1221}:
\be
H=-\frac{1}{2}\frac{\hbar^2}{m_e}\sum_i^N (\prt_{\vec{r}_i} - \ii \vec{A}(\vec{r}_i) )^2+ \sum_{i <j} V_K (\vec{r}_i-\vec{r}_j),
\ee
with a higher order derivative interacting potential $V_K(\vec{r}) \equiv \sum_{n=1}^K v_n (-1)^n \prt^n_{\bar{z}} \delta(z) \prt^n_{{z}} $.
The coordinates of $i$-th electron is $z_i \equiv x_i + \ii y_i$. 
On the other hand, Laughlin wavefunction was proposed to be an ansatz for the FQHs with a filling fraction $\nu=1/K$:
\begin{equation}
\label{LaughlinWF}
\Psi_K(\{z_i\})= \left[ \prod (z_i-z_j)^{K} \right] \e^{-\frac{1}{ 4 \ell_B^2} \sum |z_i|^2},
\end{equation}
with $\ell_B=\sqrt{\hbar c/eB}$ and $2\pi \ell_B^2 B =hc/e=\Phi_0$ as a unit flux.
It is an only approximate ansatz for the real system, because it does not take into account a finite size system with a finite radius confining potential
and the Coulomb interactions.
However, the Laughlin wavefunction turns out to be an exact ground state of the above Hamiltonian on a 2D plane 
for an appropriate vector potential 
$\vec{A}=\frac{1}{2}(-By,Bx)$ in a symmetric gauge and
the background magnetic field $B$.
More generally, we can consider a single or multilayer many-body electronic systems as a FQHs described by a wavefunction with a $K_{IJ}$ $K$-matrix data,
\begin{equation}
\label{KWF}
\Psi_{K_{IJ}}(\{z_i\})= \left[ \prod (z_i^{(I)}-z_j^{(J)})^{K_{IJ}} \right] \e^{-\frac{1}{ 4 \ell_B^2} \sum_{{(I)},i} |z_i^{(I)}|^2}.
\end{equation}
It is believed 
that the generic system with this type of wavefunction and their phases at long wavelength / low energy can be encoded into an effective action of Abelian
Chern-Simons theory:
\be \label{eq:KCStheory}
S=\frac{K_{IJ}}{4\pi} \int   a_{I} \wedge \dd  a_{J}. 
\ee 
For example, one can check 
physics observables such as the Hall conductance: 
\be
\sigma_{xy}= \nu \frac{e^2}{h} =(\frac{n h }{e B}) \frac{e^2}{h}=\frac{n e}{B}= \frac{q \cdot K^{-1} q }{2\pi} \frac{e^2}{\hbar}.
\ee
Here $q$ is the charge coupling to the external electromagnetic field.
Below we would like to study the Berry's geometric phase and matrix by putting the TO systems on a 2-torus $T^2$.
We will discuss both the many electron Hamiltonian pictures and the effective Chern-Simons theory picture to capture the 
geometric matrix. One crucial remark we will come back to justify is that, in order to characterize and classify topological states of matter:

\colorbox{lgray}{\parbox[t]{\linegoal}{
It is important to study the geometric matrix $\gamma$ of a wavefunction on a spatial manifold 
with nontrivial topology, such as a $T^d$ torus. We will see that while the U(1) geometric 
phase arises for a contractible closed trajectory of $\tau(t)$, the geometric matrix may only arise 
 for a non-contractible closed trajectory. A non-contractible closed trajectory
 in the Hamiltonian  coupling constant  $\tau$ space occurs when the different coupling constants are identified
 as the  same family of Hamiltonian, due to the spatial nontrivial-topology manifold.
}}

\subsection{One electron to many electrons of FQHs on a 2-torus} \label{Haldane-Rezayi}


Haldane-Rezayi \cite{HR8529} gave an explicit one-electron wavefunction under magnetic field on $T^2$, which can be generalized to many electrons wavefunction. Let us say the $T^2$ identifies the coordinate $z=x+\ii y$ to $z \sim z+1$ and  $z \sim z+\tau$.
Consider 
$
H=-\frac{1}{2}\frac{\hbar^2}{m_e}  (\prt_{\vec{r}} - \ii \vec{A}(\vec{r}) )^2,
$
with the uniform external magnetic field in the Landau gauge $\vec{A}=(A_x,A_y)=(-By,0)$.
At the lowest Landau level, the degenerate ground states have the following form: 
\begin{equation}
\label{psifz}
\Psi(z) =\Psi(x,y)= f( z) \e^{-\frac{B}{2} y^2}, 
\end{equation}
where $f( z)$ includes an odd elliptic theta function $\theta_\alpha (z \mid \tau)=\sum_{n\in \Z} \exp [ \ii\pi \tau (n + \al)^2 + \ii 2 \pi (n + \al) z ]$ on a 2-torus geometry.
The wavefunction boundary conditions constrained by $z \sim z+1$ and  $z \sim z+\tau$ gives rise to a relation
$B=2\pi \frac{N_\phi}{\tau_y}$ between the external $B$ field and an integer ${N_\phi}$ which counts the total
 number of flux quanta
penetrating 
through the torus.  
For many electrons on a 2-torus, we can generalize the wavefunction to:
\be
\Psi(\{ z_{\ii} \}) = \Psi( \{x_i, y_i\})=f( \{z_i\}) \e^{-\frac{\pi N_\phi}{\tau_y} \sum y_i^2}
\ee
The detailed studies of wavefunction of many electrons can be found in 
in Ref.\cite{W1221}.
To induce non-contractible loop in the Hamiltonian coupling constant space,
we can imagine the periodicity of the spatial $T^2$ torus also gives the identification of different Hamiltonian coupling constants, 
for example, by threading the $B$ flux with some periodic values through the torus.  
To study the Berry phase of non-Abelian geometric matrix for many-body wavefunction seems more difficult \cite{W1221}.
Instead, we can define the modified translation operator, called magnetic translation operator, to incorporate the flux effect into translation.
The non-commutative features of magnetic translation operators capture similar physics like Berry phase.
However, in the next we will study
the geometric matrix directly, by implementing an effective low energy field theory to capture the essential degrees of freedom at ground states.


\subsection{The effective Chern-Simons theory and its geometric matrix}

In the beginning of this section, we mention that TOs and topological states of matter of many electrons can be captured by a Chern-Simons (CS) theory Eq.(\ref{eq:KCStheory}),
one can add an additional kinetic Maxwell term $\frac{1}{4} f_I \wedge * f_I$ where $f_I=\dd a_I$ to introduce dynamics.
Such an approach is firstly used in \cite{W9039} for a level-$k$ CS theory to study chiral spin liquids,
then later it is generalized to study a generic Abelian FQHs by a $K_{IJ}$-matrix CS theory \cite{KW9327}.
For a ${T^2}$ torus of the size $L_1 \times L_2$,
we can express $a_{Ii}(x)=\frac{\theta_{Ii}(x)}{L_i}+\tilde{a}_{Ii}(x)$ with global and local degrees of freedom respectively. 
Here $x \equiv(x_0, x_1, x_2)$, $x_0=t$ and $i=0,1,2$ for spacetime coordinates.
The gauge invariant physical observable
$\e^{\ii \oint a_{Ii} dx}=\e^{\ii \theta_{Ii}}$ is from the global part, and $\theta_{Ii}$ is compact. 
 We can choose a temporal gauge
$a_{I0}=0$, 
then the action becomes:
$S=\int \dd t  \frac{K_{IJ}}{4\pi} ( \dot{\theta}_{I1} {\theta}_{J2}-\dot{\theta}_{I2} {\theta}_{J1} )+\frac{1}{2} m^{ij} \dot{\theta}_{Ii} \dot{\theta}_{J2}+S(\tilde{a})$,
where the first part concerns the global $\theta_{Ii}$, the $S(\tilde{a})$ concerns the local $\tilde{a}_{Ii}$.
The kinetic term  with $m^{ij}=m^{ij} (\tau) \equiv g^{00}(\tau)g^{ij}(\tau)$ depends on the coupling constant $\tau$ from microscopic interactions of many-body systems or lattice models. The Hamiltonian
in the effective $\theta_{Ii}$ degrees of freedom is:
\be \label{eq:CS-Ham}
H=\frac{-1}{2}(m(\tau))^{-1}_{ij} \sum_I (\prt_{{\theta}_{Ii}}-\ii \tA_{Ii}^\theta)(\prt_{{\theta}_{Ij}}-\ii \tA_{Ij}^\theta),
\ee
where $\tA_{Ii}^\theta$ depends on $\theta$, $\tau$ and the $K$ matrix, and we intentionally omit the local excitation $H(\tilde{a})$.
We can do a coordinate transformation $({\theta}_{I1}, {\theta}_{I2}) \to (X_I, Y_I)$, so that the Hamiltonian can be rewritten as:
$
H=\frac{-1}{2} \sum_I (\prt_{{X}_{I}}-\ii \tA_{IX})^2+(\prt_{{Y}_{I}}-\ii \tA_{IY})^2.
$
%
We can view $\tau =\Re \tau+ \ii \Im \tau \equiv \tau_X + \ii \tau_Y$, the compact periodicity is adjusted and identified as
\bea
&&({\theta}_{I1}, {\theta}_{I2}) \sim ({\theta}_{I1} + 2\pi, {\theta}_{I2}) \sim ({\theta}_{I1}, {\theta}_{I2} + 2\pi)\sim ({\theta}_{I1}+ 2n\pi, {\theta}_{I2} + 2m\pi);\;\;\;\;\; n,m \in \Z.  \nonumber\\
&& (X_I, Y_I) \sim (X_I+1, Y_I) \sim (X_I+\tau_X, Y_I+\tau_Y) \sim (X_I+\tau_X, Y_I+\tau_Y).
\eea
And we know the ground state wavefunction for this field theory Hamiltonian is in the same form as Haldane-Rezayi's one-electron Hamiltonian in Sec.\ref{Haldane-Rezayi},
we learn:

\colorbox{lgray}{\parbox[t]{\linegoal}{ One-electron ground state wavefunction under an external magnetic field $B$ on $T^2$ has the  
equivalent form as the ground state wavefunction of the effective level-$K$ Chern-Simons field 
theory by adding the kinetic Maxwell term, where $K=\frac{B \; \Im \tau}{2\pi}$, or more precisely for the general $K$-matrix: 
$(A_{IX},A_{IY})=\frac{2\pi}{\tau_Y} K_{IJ}(-Y_J,0)$, so $B_I=(\nabla \times A)_I=\frac{2\pi}{\tau_Y} K_{II}$.
The kinetic Maxwell term 
gives dynamics to the Hamiltonian field theory, while its mass matrix $m(\tau)$
depends on the coupling constant $\tau$.}} 
%

Because of the periodicity of $\theta_{Ii}$, we can identify a set of Hamiltonian of the different coupling constants $(\tau_X,\tau_Y)$ as the same family.
Namely,
\bea
&&(\theta_1, \theta_2) \to (\theta_1', \theta_2')=(\theta_1-\theta_2, \theta_2), \text{ then } (m'(\tau))^{-1} = (m(\tau+1))^{-1}\\
&&(\theta_1, \theta_2) \to (\theta_1', \theta_2')=(\theta_2, -\theta_1), \text{ then } (m'(\tau))^{-1} = (m(1/\tau))^{-1} 
\eea
More generally $\tau \to \frac{a \tau +b}{c \tau +d}$ with $a,b,c,d\in \Z$ and $\det(\begin{smallmatrix} a& b\\ c& d\end{smallmatrix})=1$,
so $\tau$ is generated by the SL$(2,\Z)$ group. 
%
The full ground state $| \Phi_n(a \mid {\tau}) \rangle$ can be solved from 
Eq.(\ref{eq:CS-Ham}) $H+H(\tilde{a})$ in a good basis with a $\GSD=|\det K|$-number of independent states 
($n=1,\dots,|\det K|$),
so $| \Phi_n(a \mid {\tau}) \rangle$ can be expressed in terms of global part $| \psi_n(\theta \mid {\tau}) \rangle$
and local part $| \tilde{\Phi}_n(\tilde{a} \mid {\tau}) \rangle$, 
so $| \Phi_n(a \mid {\tau}) \rangle=| \psi_n(\theta \mid {\tau}) \rangle | \tilde{\Phi}_n(\tilde{a} \mid {\tau}) \rangle$.
One can compute the geometric matrix $\gamma_{nl}$ and obtain:
\bea
\e^{\ii \gamma_{nl}(\tau',\tau)} &\equiv&
 \langle  \Phi_n({\tau}) |  \Phi_l({\tau'}) \rangle \cdot
  \text{P}[ \exp[\ii \int^{\tilde{\tau}(t)=\tau'}_{\tilde{\tau}(0)=\tau}  \dd \tilde{\tau} \langle  \Phi_n(\tilde{\tau}) | \frac{\prt}{\prt \tilde{\tau}} | \Phi_l(\tilde{\tau}) \rangle]] \nonumber\\
  &=& \langle  \psi_n({\tau}) |  \psi_l({\tau'}) \rangle \cdot
  \text{P}[\exp[\ii \int^{\tilde{\tau}(t)=\tau'}_{\tilde{\tau}(0)=\tau}  \dd \tilde{\tau} \langle  \psi_n(\tilde{\tau}) | \frac{\prt}{\prt \tilde{\tau}} | \psi_l(\tilde{\tau}) \rangle]]
 \e^{\ii \Theta_{\tilde{\Phi}}^{\text{global}}} \nonumber\\
  &=&\langle  \psi_n({\tau}) |  \psi_l({\tau'}) \rangle \cdot \e^{\ii \Theta_0}. 
\eea
The first line is generic true by definition. 
Remarkably the second line is true if $\tau$ and $\tau'$ couplings are identified as the same family of Hamiltonian,\footnote{And if the parallel transport of 
$\langle  \Phi_n(\tilde{\tau}) | \frac{\prt}{\prt \tilde{\tau}} | \Phi_l(\tilde{\tau}) \rangle$ adjusts its basis to absorb the non-Abelian matrix.} 
namely,
identified by the SL$(2,\Z)$ group; the end of the computation shows that,

\colorbox{lgray}{\parbox[t]{\linegoal}{
When $\tau \to \tau'$ is generated and identified by SL$(2,\Z)$ group, then the geometric matrix $\e^{\ii \gamma_{nl}(\tau',\tau)}$ 
contains the universal piece contributed by a matrix $\langle  \psi_n({\tau}) |  \psi_l({\tau'}) \rangle$ (the non-Abelian geometric matrix).
On the other hand, the remained contribution is simply a path-dependent non-universal U(1) factor: $\e^{\ii \Theta_0}$.
}}
Moreover,

\colorbox{lgray}{\parbox[t]{\linegoal}{
We can use the generators of SL$(2,\Z)$ group to extract the geometric matrix data:
$\cT_{nl}=\langle  \psi_n({\tau}) |  \psi_l({\tau+1}) \rangle$ and
$\cS_{nl}=\langle  \psi_n({\tau}) |  \psi_l({-{\tau}^{-1}}) \rangle$, which takes the
inner product of two wavefunctions from two different Hamiltonian identified in the same family
and in the same Hilbert space.
}}
In \cite{KW9327}, it is checked that \\
\colorbox{lgray}{\parbox[t]{\linegoal}{``The braiding statistics of anyon excitations (computed by using Chern-Simons theory by adding source terms: $a_\mu j^\mu$ with $j^\mu$ contains 
the anyonic charge vector $q$)'' 
coincides with
``the non-Abelian Berry's geometric phase / matrix calculation for the degenerate ground states using a non-contractible loop trajectory in the coupling constant space $\tau$. At least for Abelian CS theory, both results agrees on $\cS_{\vec{n} \vec{l}}=\frac{1}{\sqrt{|\det K|}} \exp[\ii 2 \pi\vec{n} K^{-1}  \vec{l}]$ up to a total U(1) phase.''}}\\
This result shows that \emph{ground states} indeed encode information of higher energy quasi-excitations such as their braiding statistics,
and this also agrees with the observation made in the remark after (ix) in Sec.\ref{sec:nAbBerry}.


\section{Intermission: Summary of different related approaches, mapping class group and modular SL($d,\Z$) representation} \label{sec:3approaches}

From the study of {\bf geometric phase} insights above, we are able to take several routes to extend this geometric phase/matrix idea further.
First route, is to, instead of taking (i) \emph{inner product of two wavefunctions of two different Hamiltonians}, 
we can directly take (ii) \emph{wavefunction overlap} between two different ground states of the \emph{same Hamiltonians}.
This is proposed as the \emph{wavefunction overlap} approach \cite{Kong:2014qka,MW1418}.\\
{\bf Wavefunction overlap}:
Given two independent states of degenerate ground states $| \psi_\alpha \rangle$ and $| \psi_\beta \rangle$, 
with any element $\hat{O}$ as the transformation of the wavefunction which is induced by 
the automorphism group AMG of the spatial manifold $\cM$ (a way of mapping the manifold to itself while preserving all of its structure): 
$\AMG(\cM)$.
We can compute the projective representation of $O_{\alpha,\beta}$ by:
\be
\langle \psi_\alpha | \hat{O} | \psi_\beta \rangle = \e^{-\#V} \cdot O_{\alpha,\beta } \cdot \dots
\ee
The first term $\e^{-\#V}$ is the volume-dependent term due to the overlapping factor depends on the number of lattice sites $N_{\text{lattice}}$ by:
$\#^{N_{\text{lattice}}} \propto \#^V$ where all these $\#$ are non-universal numbers. 
The $\dots$ terms are non-universal subleading terms $\e^{O(1/V)}$ approaches to 1 as $V \to \infty$. 
More specifically, for topological orders with gappable boundaries, it is possible to use the  
0-th homotopy group of the $\AMG(\cM)$: the mapping class group $\MCG(\cM)=\pi_0[\AMG(\cM)]$ to fully characterize topological orders \cite{Kong:2014qka,MW1418}. We will give one explicit analytic example using toric code, in Sec.\ref{sec:overlap_kitaev}.

{\bf Spacetime path integral or partition function}: Second route, instead of taking an adiabatic evolution by tuning the Hamiltonian coupling constants,
we can compute the spacetime path integral between two wavefunctions under time evolution. In this
case, the Hamiltonian coupling constants need not to be tuned. Actually we need not to know the Hamiltonian,
but just know the spacetime path integral. Indeed this will be the main approach of my thesis, outlined in Sec.\ref{sec:PT}.

{\bf Symmetry-twist and wavefunction overlap for SPT}: Just like Berry phase approach which tunes and modifies the Hamiltonian,
the modification of Hamiltonian is also a useful tool if there is only a unique ground state, such as SPTs. We will combine the 
wavefunction overlap for different SPT Hamiltonians related by symmetry-twist in Sec.\ref{sym-twist_ST}.
The discussion here follows Ref.\cite{HW1339, JuvenSPT1, Gu:2015lfa}.

{\bf MCG}: Let us summarize the particular MCG of $T^d$ torus which we will focus on extensively \cite{Wang:2014oya}.
{
\bea
\text{MCG}(\mathbb{T}^d)= \text{SL}(d,\Z). 
\eea 
For
3D, the mapping class group SL$(3,\Z)$ is generated by the modular transformation $ \hat{\mathsf{S}}^{xyz}$ and $\hat{\mathsf{T}}^{xy}$: 
\begin{align} \label{eq:ST3D}
 \hat{\mathsf{S}}^{xyz} = \begin{pmatrix}
0 & 0 & 1 \\
1 & 0 & 0 \\
0& 1 & 0
\end{pmatrix},
\qquad
 \hat{\mathsf{T}}^{xy}= \begin{pmatrix}
1 & 1 & 0 \\
0 & 1 & 0 \\
0 & 0 & 1
\end{pmatrix}.
\end{align}
}

For
2D, the mapping class group SL$(2,\Z)$ is generated by the modular transformation $ \hat{\mathsf{S}}^{xy}$ and $\hat{\mathsf{T}}^{xy}$: 
\begin{align} \label{eq:Sxy}
 \hat{\mathsf{S}}^{xyz} = \begin{pmatrix}
0 & -1  \\
1 & 0 
\end{pmatrix},
\qquad
 \hat{\mathsf{T}}^{xy}= \begin{pmatrix}
1 & 1  \\
0 & 1  
\end{pmatrix}.
\end{align}

In the case of the unimodular group, 
there are the unimodular matrices of rank $N$ forms GL$(N,\Z)$.
 $\sfS_\sfU$ and $\sfT_\sfU$ have determinant $\det(\sfS_\sfU)=-1$ and determinant $\det(\sfT_\sfU)=1$ for
any general $N$:
\bea
\sfS_\sfU=\begin{pmatrix}
0 & 0 & 0 & \dots & (-1)^{N}\\
1 & 0 & 0 & \dots & 0 \\
0 & 1 & 0 & \dots & 0\\
\vdots & \vdots & \ddots & \dots & \vdots\\
0 & 0 & 0 & \dots & 0
\end{pmatrix},\;\; 
\sfT_\sfU=\begin{pmatrix}
1 & 1 & 0 & \dots & 0\\
0 & 1 & 0 & \dots & 0\\
0 & 0 & 1 & \dots & 0\\
\vdots & \vdots & \vdots  & \ddots & \vdots\\
0 & 0 & 0  & \dots & 1
\end{pmatrix}.
\eea
Note that $\det(\sfS_\sfU)=-1$ in order to generate both determinant 1 and $-1$ matrices.
For the SL$(N,\Z)$ modular transformation, we denote their generators as $\sfS$ and $\sfT$ for a general $N$
with $\det(\sfS)=\det(\sfT)=1$:
\bea
\sfS=\begin{pmatrix}
0 & 0 & 0 & \dots & (-1)^{N-1}\\
1 & 0 & 0 & \dots & 0 \\
0 & 1 & 0 & \dots & 0\\
\vdots & \vdots & \ddots & \dots & \vdots\\
0 & 0 & 0 & \dots & 0
\end{pmatrix},\;\; 
\sfT=\sfT_\sfU.
\eea
Here for simplicity, let us denote $\sfS^{xyz}$ as $\sfS_{3\tD}$, $\sfS^{xy}$ as $\sfS_{2\tD}$, $\sfT^{xy} = \sfT_{3\tD}= \sfT_{2\tD}$. Recall that
SL$({3},\mathbb{Z})$ is fully generated by generators $\sfS_{3\tD}$ and $\sfT_{3\tD}$. Some relations of $\sfS$ and $\sfT$ are:
\bea
\sfS_{2\tD}=(\sfT_{3\tD}^{-1} \sfS_{3\tD})^3 (\sfS_{3\tD}\sfT_{3\tD})^2  \sfS_{3\tD} \sfT_{3\tD}^{-1}.
\eea
By dimensional reduction (note $\sfT_{2\tD}=\sfT_{3\tD}$), we expect that,
\bea
&&\sfS_{2\tD}^4=(\sfS_{2\tD}\sfT_{3\tD})^6=1,
\;\;\;\;\;\;\;\;
(\sfS_{2\tD}\sfT_{3\tD})^3=e^{\frac{2\pi \ti}{8} c_-}\sfS_{2\tD}^2=e^{\frac{2\pi \ti}{8} c_-} C.
\eea
$c_-$ carries the information of central charges. 
The complex U(1) factor $e^{\frac{2\pi \ti}{8} c_-}$ implies that the representation is projective.
We can express
\bea
\sfR\equiv \begin{pmatrix} 0& 1& 0\\ -1& 1& 0\\ 0& 0& 1 \end{pmatrix}=(\sfT_{3\tD} \sfS_{3\tD})^2  \sfT_{3\tD}^{-1}  \sfS_{3\tD}^2 \sfT_{3\tD}^{-1} \sfS_{3\tD} \sfT_{3\tD} \sfS_{3\tD}.
\;\;\;\;\;\;\;\;
\eea
One can check that 
\bea
&& \sfS_{3\tD} \sfS_{3\tD}^\dagger=\sfS_{3\tD}^3=\sfR^6=(\sfS_{3\tD} \sfR)^4=(\sfR \sfS_{3\tD})^4=1,\\
&& (\sfS_{3\tD}\sfR^2 )^4=(\sfR^2  \sfS_{3\tD})^4=(\sfS_{3\tD} \sfR^3)^3=(\sfR^3 \sfS_{3\tD})^3=1,\;\;\;\;\;\;\;\;\;\; \\ 
&& (\sfS_{3\tD}\sfR^2 \sfS_{3\tD})^2 \sfR^2 = \sfR^2(\sfS_{3\tD}\sfR^2 \sfS_{3\tD})^2 \text{ (mod 3)}.
\eea
Such expressions are known in the mathematic literature, part of them are listed in Ref.\cite{Coxeter}.

\cred{}
For the sake of clarity on the notation, we will use $\hat{O}$ ($\hat{S}$, $\hat{\sfS}$, $\hat{T}$ or $\hat{\sfT}$, etc) for the real-space operation on the wavefunction.
We will use the mathcal notation $O$ ($\cS$, $\cT$, etc) for its projective representation in the ground state basis. 




\section{Wavefunction overlap on the Kitaev's toric code lattice model} \label{sec:overlap_kitaev}

Now I follow the \emph{wavefunction overlap} statement described in Sec.\ref{sec:3approaches} to \emph{exact analytically extract the geometric matrix}. 
We will consider 
an exact solvable model: Kitaev's toric code in 2D, which is a $Z_2$ gauge theory.
Consider this toric code on a 2-torus $T^2$ system with a Hamiltonian composed by Pauli matrices:
\bea
&H= - \sum_v A_v - \sum_p B_p\\
&A_v=\prod_v \sigma^x = \sigma^x_{v,1} \sigma^x_{v,2} \sigma^x_{v,3} \sigma^x_{v,4}, \;\;\;\;
B_p=\prod_p \sigma^z = \sigma^z_{p,1} \sigma^z_{p,2} \sigma^z_{p,3} \sigma^z_{p,4}
\eea
\begin{figure}[!h] 
\centering
(a)\includegraphics[scale=0.5]{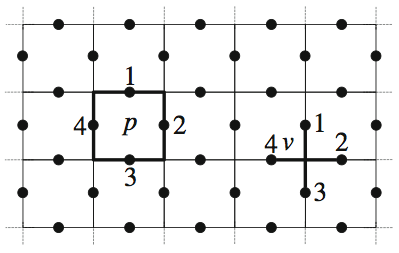}(b)\includegraphics[scale=0.5]{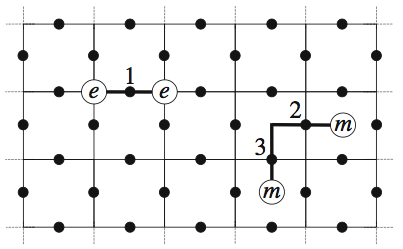} 
\caption{
(a) The square lattice toric code model with $A_v$ and $B_p$ operators. 
(b) The $e$-string operator with end point $e$-charge ($Z_2$ charge) excitations on the vertices, created by a product of $\prod \sigma_z$.
The $m$-string operator with end point $m$-charge ($Z_2$ flux) excitations in the plaquette, created by a product of $\prod \sigma_x$.
See an introduction to toric code in \cite{Kitaev:1997wr,pachos2012introduction}.
} 
\label{torus2D3D} 
\end{figure}
$A_v$ is the vertex operator, $B_p$ is a plaquette operator; both operators act on the nearest four neighbored links. Note that $A_v^2=B_p^2=1$, so the eigenvalues of $A_v, B_p$ are $\pm 1$.
Notice $[A_v,A_v']=[A_v,B_p]=[B_p,B_{p'}]=0$ for all choices of vertices and plaquette $v,v',p,p'$.

Let us denote $ | 0\rangle = | \uparrow \rangle$ and $ | 1\rangle = | \downarrow \rangle$, which satisfies
$
\sigma_z  | 0\rangle = \sigma_z  | \uparrow \rangle =+ | \uparrow \rangle
$.
There are many ground states and many possible bases. The candidate ground states we will start with are the equal-weight superpositions of $m$-loop (corresponds to the $| \xi  \rangle$ state)
or $e$-loop (corresponds to the $|\psi \rangle$ state) configurations:
\bea
&& | \xi  \rangle = \prod_v (\frac{1}{\sqrt{2}}(1+A_v) ) | 0000 \dots \rangle = \prod_v (\frac{1}{\sqrt{2}}(1+A_v) ) | \uparrow \uparrow  \dots \rangle\\
&& |  \psi \rangle =
\prod_p (\frac{1}{\sqrt{2}}(1+B_p) ) | \rightarrow \rightarrow \dots \rangle 
\eea

Note that:
$
B_p  | \xi  \rangle =+ | \xi  \rangle, \;\;\;\; A_{v'}  | \xi  \rangle = | \xi  \rangle 
$.
Due to that $A_{v'}^2=1$, we have
\be
 A_{v'} \prod_v (1+A_v)  | 0000 \dots \rangle =  \prod_v (1+A_v)    | 0000 \dots \rangle,
\ee
Since the right hand side $A_{v'}$ will simply shift those operators without $A_{v'}$ to with one $A_{v'}$ (another operator $\prod_v (1+A_v)$); and shift those operators 
with $A_{v'}$ to the operator without $A_{v'}$ (another operator $\prod_v (1+A_v)$). So
\be
\sum_{v'} A_{v'} \prod_v (1+A_v)  | 0000 \dots \rangle =  N_v \prod_v (1+A_v)    | 0000 \dots \rangle
\ee
\be
\sum_{p} B_{p} \prod_v (1+A_v)  | 0000 \dots \rangle =  \prod_v (1+A_v)   \sum_{p} B_{p} | 0000 \dots \rangle
=N_p \prod_v (1+A_v)   | 0000 \dots \rangle
\ee
here $N_p$ is the number of plaquette. On the torus, we have $N_p=N_v$, hence $
H  | \xi  \rangle =- N_v  | \xi  \rangle
$. So
$| \xi  \rangle$ is one of the lowest energy states, the ground states.
The initial state we like to consider can be a fluctuation of the $m$ loops ($| \xi  \rangle$ state) or $e$ loops ($|  \psi \rangle$ state).
We will focus on $| \xi  \rangle$ creates m string loop, where $m$ is the flux on the plaquette.

Now we like to generate other linear independent states from $| \xi  \rangle$. 
We define $T^2$ torus with two non-contractible directions $X$ and $Y$.
The definitions are the following: \\
(1) The operator $W^X_e$  (or denoted as $W_X$, see also Chap.\ref{QS_stSurgery}) connecting the non-contractible directions $X$ with a series of $\sigma_z$: $\prod \sigma_z$, the $e$ string operator (the end points create two $e$),
which flips the operator $A_v= \prod_v \sigma_x$. 
The $W^X_e$ is the $e$-string along the $X$ direction.\\ 
(2) The operator $W^X_m$ (or denoted as $\Gamma_X$, see also Chap.\ref{QS_stSurgery}) connecting the non-contractible directions $X$ with a series of $\sigma_x$: $\prod \sigma_x$, the $m$ string operator (the end points create two $m$),
which flips the operator $B_p= \prod_p \sigma_z$. 
The $W^X_m$ is the $m$-string along the $X$ direction.\\
(3) The operator $W^Y_e$ (or denoted as $W_Y$, see also Chap.\ref{QS_stSurgery}) connecting the non-contractible directions $Y$ with a series of $\sigma_z$: $\prod \sigma_z$, the $e$ string operator (the end points create two $e$),
which flips the operator $A_v= \prod_v \sigma_x$. 
The $W^Y_e$ is the $e$-string along the $Y$ direction.\\ 
(4) The operator $W^Y_m$ (or denoted as $\Gamma_Y$ in Chap.\ref{QS_stSurgery}) connecting the non-contractible directions $Y$ with a series of $\sigma_x$: $\prod \sigma_x$, the $m$ string operator (the end points create two $m$),
which flips the operator $B_p= \prod_p \sigma_z$. 
The $W^Y_m$ is the $m$-string along the $Y$ direction.

Note that $[W^X_m,W^Y_m]=[W^X_e,W^Y_e]=0$.  $W^X_m W^Y_e=- W^Y_e W^X_m$, and $W^Y_m W^X_e=- W^X_e W^Y_m$.
%
%
Let us start from the superposition of fluctuating $m$-loop state of $| \xi  \rangle = \prod_v (\frac{1}{\sqrt{2}}(1+A_v) ) | 0000 \dots \rangle$.
It is important to note that
\be
W^X_e | \xi  \rangle = | \xi  \rangle, \;\;\;\; W^Y_e | \xi  \rangle = | \xi  \rangle  
\ee
adding a non-contractible $e$-loop $W^X_e$ along $x$ direction on superposed fluctuating $m$-loop state gives the same state $ | \xi  \rangle$.
However, adding a non-contractible $m$-loop $W^X_m$ on  $ | \xi  \rangle$ gives different state.
We require an even number of $\sigma_z$, $\sigma_x$ overlapping on the 
$W^X_e | \xi  \rangle = \prod \sigma_z  \prod_v (1+\prod_v \sigma_x)  | 0000 \dots \rangle
=  \prod_v (1+\prod_v \sigma_x) \prod \sigma_z  | 0000 \dots \rangle= | \xi  \rangle$.\\

Let us choose $W^X_e, W^X_m$ operator as the chosen measurements from all the Hilbert space operators; 
we wish to simultaneously diagonalize the two operators in the eigenstate basis.
Meanwhile we wish to define the trivial vacuum ground state $ | \mathbb{I} \rangle$ as the state where there is only trivial measurement observed by the $W^X_e$'s $e$ loop and 
$W^X_m$'s $m$ loop along the $x$ direction. Namely, there is no $e$ or $m$-string non-contractible loop along the $y$ direction. The goal is to find a relation between
$| \mathbb{I} \rangle$ and $| \xi  \rangle$:
$| \mathbb{I} \rangle =\sum c(n_1,n_2,n_3,n_4)$ $(W^X_m){}^{n_1}$ $(W^Y_m){}^{n_2}$$(W^X_e){}^{n_3}$ $(W^Y_e){}^{n_4}$$| \xi  \rangle$
which satisfies $W^X_e  | \mathbb{I} \rangle = | \mathbb{I} \rangle$ and $W^X_m | \mathbb{I} \rangle= | \mathbb{I} \rangle$.
%
%
It turns out that our $| \xi  \rangle$ as the superposition of the fluctuating $m$ loop states is important to determine $| \mathbb{I} \rangle$.
We will find that {\bf $| \mathbb{I} \rangle$  simply be the superposition of the $| \xi  \rangle$ and the state with a non-contractible $m$-loop winding around the $x$-direction}:
\be
| \mathbb{I} \rangle = \frac{1}{\sqrt{2}}(| \xi  \rangle +W^X_m  | \xi  \rangle  ).
\ee
To prove this, define $| (a, b) \rangle = (W^X_m){}^a (W^Y_m){}^b | \xi \rangle$:
\bea
&& W^X_m | (a, b) \rangle   = (W^X_m){}^{a+1} (W^Y_m){}^b | \xi \rangle =  | (a+1, b) \rangle\\
&& W^Y_m | (a, b) \rangle  =  | (a, b+1) \rangle\\
&& W^X_e | (a, b) \rangle   = (-1)^b (W^X_m){}^{a+1} (W^Y_m){}^b (W^X_e) | \xi \rangle= (-1)^b | (a, b) \rangle\\
&& W^Y_e | (a, b) \rangle   =  (-1)^a | (a, b) \rangle
\eea
Here $W^X_e | \xi \rangle =  | \xi \rangle$ is a nontrivial step. 
We have contractible loop states $| \xi \rangle=\Gamma(\sigma_x) | 0 0 \dots \rangle$, 
and $W^X_e(\sigma_z) \Gamma(\sigma_x) | 0 0 \dots \rangle \propto \Gamma(\sigma_x) | 0 0 \dots \rangle$.
This should imply that
$W^X_e | \xi \rangle =W^Y_e | \xi \rangle =  | \xi \rangle$ if we have the $m$-loop superposed state being acted by $e$-non-contractible loop along any direction on an even lattice site system.

%
Define:
$| (\pm, b) \rangle \equiv \frac{1}{\sqrt{2}}( | (0, b) \rangle \pm |(1 , b) \rangle)$
we see that
$W^X_e | (\pm, b) \rangle =(-1)^b  | (\pm, b) \rangle$, 
$W^X_m | (\pm, b) \rangle$ $=\pm  | (\pm, b) \rangle$
so $W^X_e | (+, 0) \rangle =W^X_m | (+, 0) \rangle= | (+, 0) \rangle$.
\be
| \mathbb{I} \rangle \equiv | (+, 0) \rangle  = \frac{1}{\sqrt{2}}(| \xi  \rangle +W^X_m  | \xi  \rangle  )
\ee
Now, $| \mathbb{I} \rangle$ has no non-contractible $e,m$ along $y$ direction detectible by $W^X_e, W^X_m$.
So all we need to do is creating $e$ and $m$ along $y$ direction by $W^Y_e$ for $e$ and $W^Y_m$ for $m$.
\bea
&&| \mathbb{I} \rangle  =\frac{1}{\sqrt{2}}( | (0, 0) \rangle + |(1, 0) \rangle)  = \frac{1}{\sqrt{2}}(| \xi  \rangle +W^X_m  | \xi  \rangle  )\\
&&| e \rangle  =W^Y_e  | \mathbb{I} \rangle= \frac{1}{\sqrt{2}}( | (0, 0) \rangle - |(1, 0) \rangle)  = \frac{1}{\sqrt{2}}(| \xi  \rangle -W^X_m  | \xi  \rangle  )\\
&&| m \rangle  =W^Y_m  | \mathbb{I} \rangle= \frac{1}{\sqrt{2}}( | (0, 1) \rangle + |(1, 1) \rangle)  = \frac{1}{\sqrt{2}}(W^Y_m | \xi  \rangle + W^X_m W^Y_m  | \xi  \rangle  )\\
&&| em \rangle  =W^Y_e W^Y_m  | \mathbb{I} \rangle= \frac{1}{\sqrt{2}}( | (0, 1) \rangle - |(1, 1) \rangle)  
= \frac{1}{\sqrt{2}}(W^Y_m | \xi  \rangle - W^X_m W^Y_m  | \xi  \rangle  ).
\eea
Now we can do the modular SL$(2,\Z)$ transformation $\cS$ which sends $(x,y) \to (-y,x)$.
\bea
&&\cS | \mathbb{I} \rangle  =\frac{1}{\sqrt{2}}( | (0, 0) \rangle + |(0, 1) \rangle) =\frac{1}{{2}}(| \mathbb{I} \rangle+
| e \rangle+ | m \rangle+ | e m \rangle) \\
&&\cS | e \rangle  = \frac{1}{\sqrt{2}}( | (0, 0) \rangle - |(0, 1) \rangle)
=\frac{1}{{2}}(| \mathbb{I} \rangle+ | e \rangle - | m \rangle - | e m \rangle)  \\
&&\cS | m \rangle  = \frac{1}{\sqrt{2}}( | (-1, 0) \rangle + |(-1, 1) \rangle)
=\frac{1}{{2}}(| \mathbb{I} \rangle - | e \rangle+ | m \rangle - | e m \rangle)\\
&&\cS | em \rangle  = \frac{1}{\sqrt{2}}( | (-1, 0) \rangle - |(-1, 1) \rangle) 
=\frac{1}{{2}}(| \mathbb{I} \rangle - | e \rangle - | m \rangle+ | e m \rangle).
\eea
Clearly, we have now obtained the ideal $\cS$ matrix in the ideal quasi-particle basis using the \emph{wave function overlap approach}:
\be
\cS=\begin{pmatrix}  
\langle \mathbb{I} |\cS | \mathbb{I}\rangle &\langle \mathbb{I} |\cS | e\rangle &\langle \mathbb{I} |\cS |m \rangle &\langle \mathbb{I} |\cS | em\rangle\\
   \langle e |\cS | \mathbb{I}\rangle&\langle e |\cS | e\rangle & \langle e |\cS | m\rangle & \langle e |\cS | em\rangle\\
   \langle m |\cS | \mathbb{I}\rangle& \langle m |\cS | e\rangle &\langle m |\cS | m\rangle& \langle m |\cS | em\rangle\\
    \langle em |\cS | \mathbb{I}\rangle&  \langle em |\cS | e\rangle & \langle em |\cS | m\rangle &\langle em |\cS | em\rangle\end{pmatrix}=
\frac{1}{2}\begin{pmatrix}
    1&1&1&1\\
    1&1&-1&-1\\
    1&-1&1&-1\\
    1&-1&-1&1
  \end{pmatrix}.
\ee
This can also be obtained as minimal entangled states (MES), through another approach by entanglement entropy \cite{ZhangGroverAshvin}.

{\bf Comparison:} The final comment is that the
{\bf adiabatic Berry phase / geometric matrix} calculation has the drawbacks of requiring tuning Hamiltonian coupling constants thus demanding to access a large class of systems; but it has the advantages of dealing with non-translational-symmetry, non-periodic and non-lattice system.
On the other hand, the {\bf wavefunction overlap} has advantages of fixing a single Hamiltonian with different degenerate ground states; 
but it has the drawbacks of restricting to translational-symmetry, periodic and equal-size-lattice system and to the given symmetry of lattice systems 
(usually easier to extract $\cS$ but harder to extract $\cT$) \cite{ZhangGroverAshvin}.
There is also a drawback that in general there is a volume-dependent factor 
$\langle \psi_\alpha | \hat{O} | \psi_\beta \rangle = \e^{-\#V} \cdot O_{\alpha,\beta } \cdot \dots$, although in
our square-lattice toric code example, we did not observe the volume-dependent term in $\cS$, but it indeed occurs
in $\cT$, at least for square and triangle lattice studied in \cite{MW1418}.

\section{Spacetime path integral approach for the modular $\cS$ and $\cT$ in 2+1D and 3+1D: group cohomology cocycle} \label{sec:PT}

Below we will describe the spacetime path integral approach with discretized lattice triangulation of spacetime.
There are two versions, one is for topological order where degrees of freedom live on the links (Sec.\ref{sec:PT_TO}) while gauge theory summing over 
all possible gauge configurations, the other is SPTs where degrees of freedom live on the sites (Sec.\ref{sec:PT_SPT}).
In short, on any closed manifold, the former has $|\mathbf{Z}_{\text{TO}}| \geq 1$, but
the former is restricted to $|\mathbf{Z}_{\text{SPT}}| = 1$.

\subsection{For topological orders} \label{sec:PT_TO}

Some kinds of (but not all of) topological orders can be described by twisted gauge theory, those are gauge theories with cocycle topological terms.
See Appendix of Ref.\cite{Wang:2014oya} for a review of cocycles and group cohomology.
For a pretty generic twisted gauge theory, there is indeed another 
way using the spacetime lattice formalism to construct them 
by the Dijkgraaf-Witten topological gauge theory.\cite{Dijkgraaf:1989pz}  We can formulate the path integral $\mathbf{Z}$ (or partition function) 
of a $(d+1)$D gauge theory ($d$D space, 1D time) of a gauge group $G$ as, 
\be \label{eq:path integral}
\mathbf{Z} =\sum_\gamma e^{\ti S[\gamma]}=\sum_\gamma e^{\ti 2\pi \langle \omega_{d+1}, \gamma(\cM_{\text{tri}})\rangle(\text{mod}{2\pi})} 
 =\frac{|G|}{|G|^{N_v}} \frac{1}{|G|} \sum_{\{ g_{ab}\}} \prod_i (\omega_{d+1}{}^{\epsilon_i}(\{ g_{ab}\})) \mid_{v_{c,d} \in T_i}
\ee
where we sum over all mappings $\gamma: \cM \to BG$, from the spacetime manifold $\cM$ to $BG$, the classifying space of $G$.
In the second equality, we triangulate $\cM$ to $\cM_{\text{tri}}$ with the edge $[v_a v_b]$ connecting the vertex $v_a$ to the vertex $v_b$. 
The action $ \langle \omega_{d+1}, \gamma(\cM_{\text{tri}})\rangle$ evaluates the cocycles $\omega_{d+1}$ on the spacetime $(d+1)$-complex $\cM_{\text{tri}}$. 
By the relation between the topological cohomology class of $BG$ and the cohomology group of $G$: 
$H^{d+2}(BG,\Z)=\cH^{d+1}(G,\R/\Z)$,\cite{Dijkgraaf:1989pz} 
we can simply regard $\omega_{d+1}$ as the ${d+1}$-cocycles of the cohomology group $\cH^{d+1}(G,\R/\Z)$. 
The group elements $g_{ab}$ are assigned at the edge $[v_a v_b]$. 
The $|G|/|G|^{N_v}$ factor is to mod out the redundant gauge equivalence configuration, with the number of vertices $N_v$.
Another extra $|G|^{-1}$ factor mods out the group elements evolving in the time dimension.
The cocycle $\omega_{d+1}$ is evaluated on all the $d+1$-simplex $T_i$ (namely a $d+2$-cell) triangulation of the spacetime complex.
In the case of our 3+1D, we have the 4-cocycle $\omega_{4}$ evaluated at the 4-simplex (or 5-cell) as 
\bea
{ \foursimplex{0}{1}{2}{3}{4}}  \label{eq:4simplex} 
=\omega_4{}^{\epsilon}(g_{01},g_{12}, g_{23}, g_{34}). 
\eea
Here the cocycle $\omega_{4}$ satisfies cocycle condition: $\delta \omega_{4}=1$, which ensures the
path integral $\mathbf{Z}$ on the 4-sphere ${S}^4$ (the surface of the 5-ball) will be trivial as 1. This is a feature of topological gauge theory.
The $\epsilon$ is the $\pm$ sign of the orientation of the 4-simplex,
which is determined by the sign of the volume determinant of the 4-simplex evaluated by $\epsilon=\sgn(\det(\vec{01}, \vec{02}, \vec{03}, \vec{04}))$. 

\begin{figure}[!h] 
\centering
\includegraphics[scale=0.36]{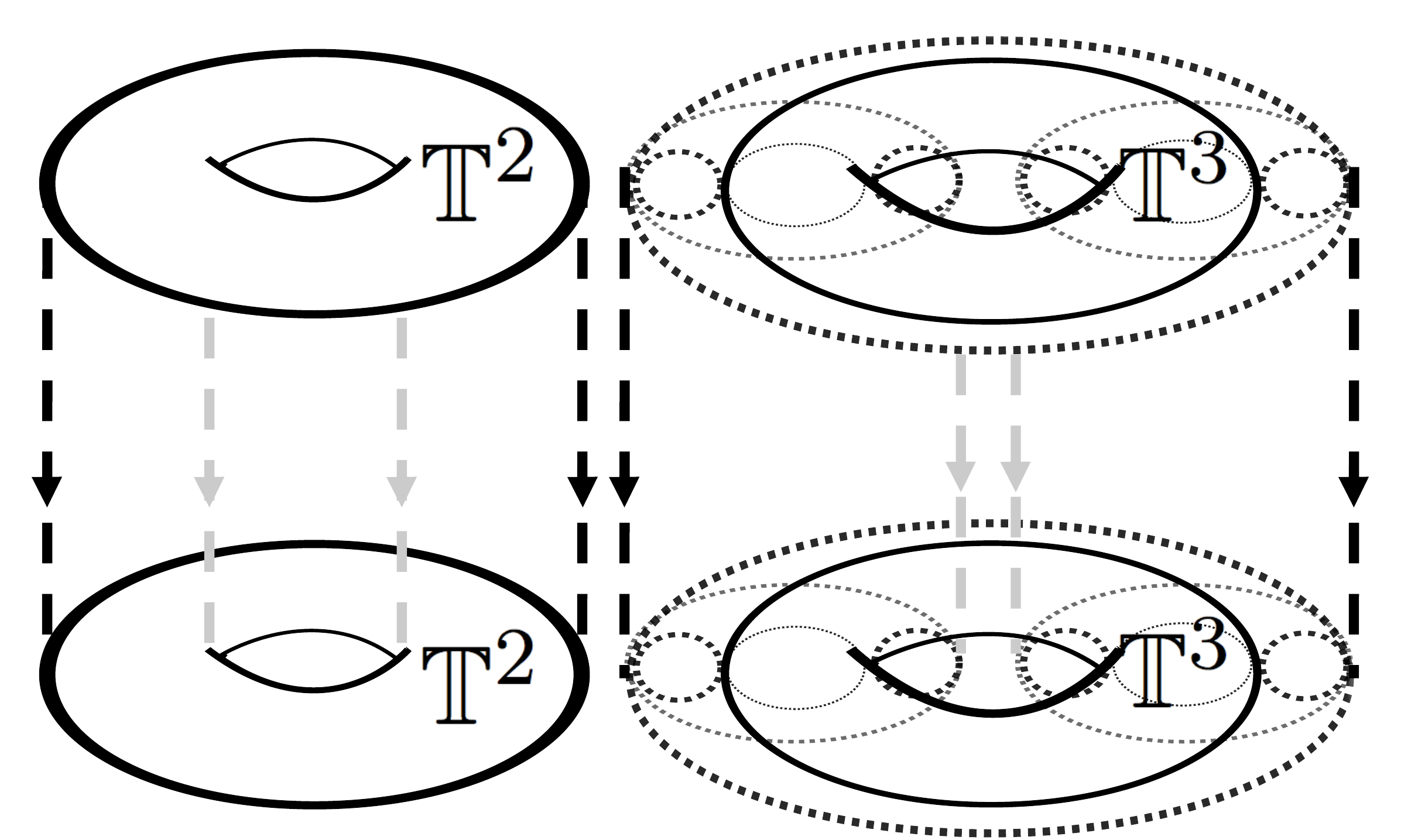}
\caption{
The illustration for ${\sfO}_{\text{(A)(B)}}= \langle  \Psi_{\text{A}}  |  \hat{\sfO}  | \Psi_{\text{B}}\rangle$.
Evolution from an initial state configuration $|\Psi_{in} \rangle$ on the spatial manifold (from the top) 
along the time direction (the dashed line - - -)  to the final state $|\Psi_{out} \rangle$ (at the bottom). 
For the spatial $\mathbb{T}^{d}$ torus, the mapping class group MCG$(\mathbb{T}^{d})$ is the modular SL$(d,\Z)$ transformation.
We show schematically the time evolution on the spatial $\mathbb{T}^{2}$, and $\mathbb{T}^{3}$. The $\mathbb{T}^{3}$ is shown as a $\mathbb{T}^{2}$ attached an $S^1$ circle at each point.
} 
\label{torus2D3D} 
\end{figure}

{We utilize Eq.(\ref{eq:path integral}) to calculate the path integral amplitude from an initial state configuration $|\Psi_{in} \rangle$ on the spatial manifold evolving
along the time direction  to the final state $|\Psi_{out} \rangle$, see Fig.\ref{torus2D3D}.
In general, the calcuation can be done for the mapping class group MCG on any spatial manifold $\cM_{space}$ as MCG$(\cM_{space})$. 
Here we focus on $\cM_{space}=\mathbb{T}^3$ and MCG$(\mathbb{T}^3)=$ SL$(3,\Z)$, as the modular transformation.
We first note that $|\Psi_{in} \rangle= \hat{\mathsf{O}} | \Psi_{\text{B}} \rangle$, such a generic SL$(3,\Z)$ transformation $\hat{\mathsf{O}}$ 
under SL$(3,\Z)$ representation can be absolutely generated by $\hat{\sfS}^{xyz}$ and $\hat{\sfT}^{xy}$ of Eq.(\ref{eq:ST3D}),\cite{Coxeter}
thus $\hat{\mathsf{O}}=\hat{\mathsf{O}}(\hat{\sfS}^{xyz},\hat{\sfT}^{xy})$ as a function of $\hat{\sfS}^{xyz},\hat{\sfT}^{xy}$.
The calculation of the modular SL$(3,\Z)$ transformation from $| \Psi_{in}\rangle$ to $| \Psi_{out}\rangle=| \Psi_{\text{A}} \rangle$ 
by filling the 4-cocycles $\omega_4$ into the spacetime-complex-triangulation renders the amplitude of the matrix element ${\sfO}_{\text{(A)(B)}}$:  
\bea \label{eq:Oamp}
&&\boxed{ {\sfO}({\sfS}^{xyz},{\sfT}^{xy})_{\text{(A)(B)}}= \langle  \Psi_{\text{A}}  | \hat{\sfO}(\hat{\sfS}^{xyz},\hat{\sfT}^{xy})  | \Psi_{\text{B}}\rangle},
\eea
both space and time are discretely triangulated, so this is a spacetime lattice formalism.}

The modular transformations $\hat{\sfS}^{xy}$, $\hat{\sfT}^{xy}$, $\hat{\sfS}^{xyz}$ of Eq.(\ref{eq:ST3D}),(\ref{eq:Sxy}) act on the 3D real space 
as
\bea
&&\hat{\mathsf{S}}^{xy}\cdot (x,y,z)=(-y,x,z),\\  
&&\hat{\mathsf{T}}^{xy}\cdot (x,y,z)=(x+y,y,z),\\
&&\hat{\mathsf{S}}^{xyz}\cdot (x,y,z)=(z,x,y).
\eea
More explicitly, we present triangulations of them:
\begin{align} \label{TriSxy}
  &\hat{\sfS}^{xy}: \begin{matrix}\SLTWOtransformSLeft\end{matrix}
  \underset{\DashedArrow}{t} 
  \begin{matrix}\SLTWOtransformSRight\end{matrix},\\
\label{TriTxy} 
   &\hat{\sfT}^{xy}: \begin{matrix}\SLTWOtransformTL\end{matrix}
  \;\;\underset{\DashedArrow}{t} 
  \begin{matrix}\SLTWOtransformTR\end{matrix},\\
\label{TriSxyz}
    &\hat{\sfS}^{xyz}: \begin{matrix}\SLTHREEtransformSLeft\end{matrix}
    \underset{\DashedArrow}{t} 
      \begin{matrix}\SLTHREEtransformSR\end{matrix}. 
\end{align}
{The modular transformation SL$(2,\Z)$ is generated by $\hat{\sfS}^{xy}$ and $\hat{\sfT}^{xy}$,
while the SL$(3,\Z)$ is generated by $\hat{\sfS}^{xyz}$ and $\hat{\sfT}^{xy}$.
The dashed arrow $\DashedArrow$ 
represents the time evolution (as in Fig.\ref{torus2D3D}) from $|\Psi_{in} \rangle$ to $|\Psi_{out} \rangle$ under $\hat{\sfS}^{xy}$, $\hat{\sfT}^{xy}$, $\hat{\sfS}^{xyz}$ respectively.
The $\hat{\sfS}^{xy}$ and $\hat{\sfT}^{xy}$ transformations on
a $\mathbb{T}^3$ torus's $x$-$y$ plane with the $z$ direction untouched are equivalent to its transformations on a $\mathbb{T}^2$ torus.}\\

\subsection{For SPTs} \label{sec:PT_SPT}

For convenience we can interchange the non-homogeneous cocycles  (the lattice gauge theory cocycles) and 
the homogeneous cocycles (SPT cocycles).
The definition of the lattice gauge theory $n$-cocycles are indeed related to SPT $n$-cocycles.
\be
\omega_n(A_1,A_2, \dots,A_n)=\nu_n(A_1A_2 \dots A_n, A_2 \dots A_n, \dots, A_n, 1) 
=\nu_n(\tilde{A}_1, \tilde{A}_2, \dots, \tilde{A}_n, 1). 
\ee
here $\tilde{A}_j\equiv A_j A_{j+1} \dots A_n$. Let us focus on 2+1D SPTs with 3-cocycles,
\bea
&&\omega_3(A,B,C)=\nu_3(ABC,BC,C,1)
\Rightarrow \omega_3(g_{01},g_{12},g_{23})\\
&&=\omega_3(g_{0} g_{1}^{-1}, g_{1} g_{2}^{-1} ,g_{2} g_{3}^{-1}) 
=\nu_3(g_{0}g_{3}^{-1} ,g_{1}g_{3}^{-1},g_{2}g_{3}^{-1},1)= \nu_3(g_{0},g_{1},g_{2},g_{3}). \nonumber
\eea
Here $A=g_{01}$, $B=g_{12}$, $C=g_{23}$, with $g_{ab} \equiv g_a g_b^{-1}$. We use the fact that SPT $n$-cocycle $\nu_n$ belongs to the $G$-module, such that 
for $r$ are group elements of $G$, it obeys $r \cdot \nu_n({r}_0, {r}_1, \dots, {r}_{n-1}, 1)=\nu( r {r}_0, r {r}_1, \dots, r {r}_{n-1}, r)$ 
 (here we consider only Abelian group $G=\prod_i Z_{N_i}$).
In the case without time reversal symmetry, so group action $g$ on the $G$-module is trivial.
%
%

In short, there is no obstacle so that we can simply use the lattice gauge theory 3-cocycle $\omega(A,B,C)$  to study the SPT 3-cocycle $\nu(ABC,BC,C,1)$.
All we need to do 
is computing the 2+1D SPT path integral $\textbf{Z}_{\text{SPT}}$ (i.e. partition function) using 3-cocycles $\omega_3$,\cite{Wang:2014tia} 
\be
\textbf{Z}_{\text{SPT}}=|G|^{-N_v} \sum_{\{ g_{v}\}} \prod_i (\omega_3{}^{s_i}(\{ g_{v_a} g_{v_b}^{-1} \}))
\ee
Here $|G|$ is the order of the symmetry group, $N_v$ is the number of vertices, $\omega_3$ is 3-cocycle, and ${s_i}$ is the exponent 1 or $-1$ (i.e. the complex conjugate $\dagger$) depending on the orientation of each tetrahedron($3$-simplex). The summing over group elements $g_{v}$ on the vertex produces a symmetry-perserving ground state.
We consider a specific $M^3$, a $3$-complex, 
which can be decomposed into tetrahedra (each as a $3$-simplex). 
There the 3-dimensional spacetime manifold is under triangulation 
(or cellularization) into many tetrahedra.

\section{Symmetry-twist, wavefunction overlap and SPT invariants} \label{sym-twist_ST}

\begin{figure}[!h]
\centering
\includegraphics[height=1in]{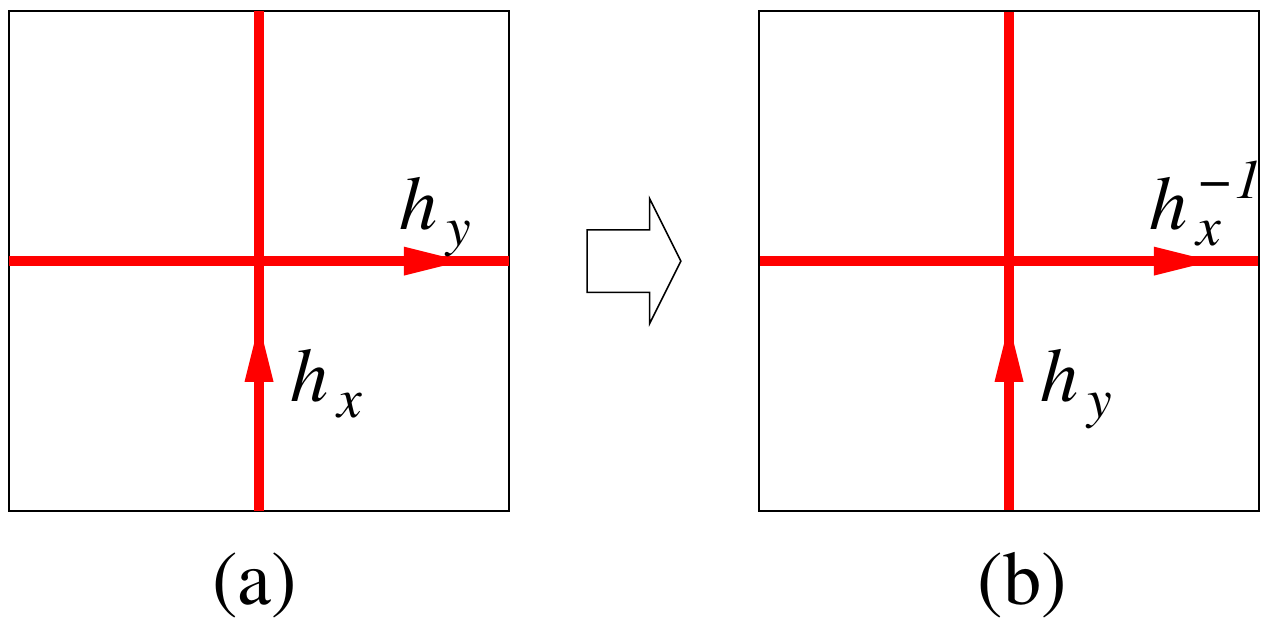}
\caption{ $\hat S$-move is $90^\circ$ rotation.  We apply the symmetry-twist along $x$ and $y$ axis, where $h_x$ and $h_y$ are the twisted boundary condition assigned
respect to its codimension directions. 
(a) A system on $T^2$ with $h_x$ and $h_y$ symmetry twists.
Here $T^2$ has the same size in $x$ and $y$ directions in order to have meaningful wavefunction overlap.
(b) The resulting symmetry twists after the $\hat S$-move. }
\label{Smov}
\end{figure}

Let us consider a 2D many-body lattice system as an example, 
we can write a generic wavefunction as $| \psi_\beta  \rangle = \sum_{ \{  g_{i_x, i_y  } \}   } \psi_\beta( \{  g_{i_x, i_y  } \}  ) | \{  g_{i_x, i_y  } \}\rangle$,
here $| \{  g_{i_x, i_y  } \}\rangle$ is a tensor product ($\otimes$) states for each site $\{ i_x, i_y  \}$ assigned with a group element $g_{i_x, i_y  }$. 
Now we would like to modify the Hamiltonian along a branch cut described in the Chap.1, say along the $x$ and $y$ axes shown in Fig.\ref{Smov}.
$
H=  \sum_x H_x \overset{\text{sym.twst along $\prt R$}}{\longrightarrow}\sum_{x \not\in \prt R} H_x+  \sum_{x \in \prt R} H_x'|_{x\text{ near } \prt R}
$.

\begin{figure}[!h]
\centering
\includegraphics[height=1in]{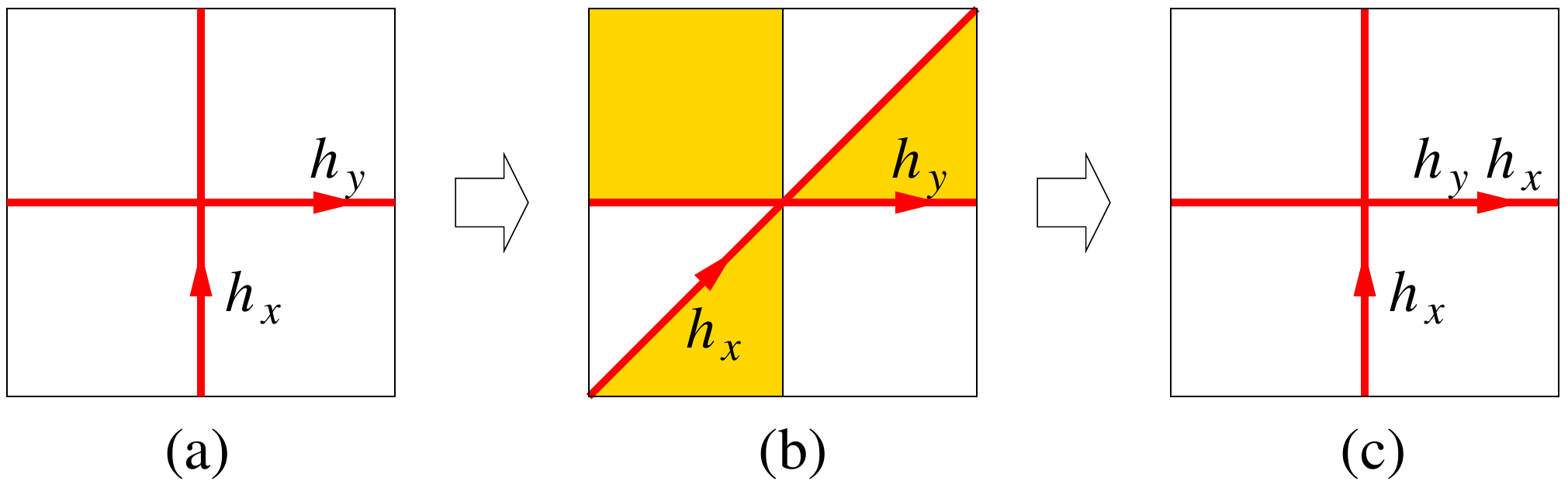}
\caption{ $\hat T$-move is the Dehn twist followed by a symmetry transformation
$h_x$ in the shaded area. (a) A system on $T^2$ with $h_x$ and $h_y$ symmetry twists.
(b) The resulting symmetry twists after the $\hat T$-move. 
(c) After a local symmetry transformation $h_x$ in the shaded region {with a counterclockwise orientation $h_x$ acting on the boundary
of the shaded region}, the symmetry twists previously shown in (b) become the the new symmetry twists in $x$- and $y$-directions.
}
\label{Tmov}
\end{figure}

In order for the ordering sequence of applying $h_x$ and $h_y$ (which is applied first: $h_x$ or $h_y$ ) makes no difference for their energy costs (near the branch cut there is still some tiny energy cost\footnote{
For a $T^2$ torus, we have a non-contractible closed loop.
In general,
if the symmetry twist is on a branch cut of a  non-contractible closed loop, the spectra will be modified. 
if it is on a branch cut of a contractible closed loop, the spectra will not be modified.
}), we have $[h_x, h_y]=0$.

For the sake of simplicity, we will consider a perfect square lattice with equal periodicity $L_x=L_y$.
Write $\Psi_{h_x,h_y}(\{g_{i_x,i_y}\})$ as the wave 
function of $|\Psi_{h_x,h_y}\>$ where ${h_x,h_y}$ are the parameter labels of symmetry-twist and $g_{i_x,i_y}$ are the variables:
\begin{align}
|\Psi_{h_x,h_y}\> = \sum_{\{g_{i_x,i_y}\}}\Psi_{h_x,h_y}(\{g_{i_x,i_y}\})
|\{g_{i_x,i_y}\}\>.
\end{align}
It is conjectured that the two symmetry-twists ${h_x,h_y}$ are good enough, at least for Abelian SPT of group $G$ since there are $|G|^2$-states, 
in a sense that its gauged theory is TOs with exactly the same number of degeneracy on the $T^2$: $|G|^2$.

Naively, we may guess the wavefunction overlap for SL$(2,\Z)$ transformation are:
\be
\langle \psi_\alpha | \hat{\mathsf{S}} | \psi_\beta  \rangle
\overset{?}{=} \sum_{ \{  g_{i_x, i_y  } \}   }  \psi_\alpha^*( \{  g_{i_x, i_y  } \}  ) \psi_\beta( \{  g_{i_y^{-1}, i_x  } \}  ), 
\langle \psi_\alpha | \hat{\mathsf{T}} | \psi_\beta  \rangle
\overset{?}{=} \sum_{ \{  g \}, {i_x, i_y  }  }  \psi_\alpha^*( \{  g_{i_x, i_y  } \}  ) \psi_\beta( \{  g_{i_x+ i_y, i_y  }   \}  ).
\ee
However, this form is close to the answer, but not entirely correct. Here $\alpha$ and $\beta$ should specify the data of symmetry twists.
Moreover, because of the SL$(2,\Z)$ transformation on the spatial $T^2$ torus, the symmetry twists $\alpha$ and $\beta$ should also
be constrained and related. The symmetry twists $\alpha$ and $\beta$ should be the \emph{same symmetry twists} 
after taking into account the SL$(2,\Z)$ transformation. 
{This means that we will overlap two wavefunctions in the same family of Hamiltonian, both twisted by the same symmetry-twist.}

The state $|\Psi_{h_x,h_y}\>$ changes under the modular transformation.
Let us define
\begin{align}
 |\Psi^S_{h_x,h_y}\> &= \sum_{\{g_{i_x,i_y}\}}\Psi_{h_x,h_y}(\{g_{-i_y,i_x}\})
|\{g_{i_x,i_y}\}\>=
\sum_{\{g_{i_x,i_y}\}}\Psi_{h_x,h_y}(\{g_{ \hat{S} \cdot (i_x,i_y)}\})
|\{g_{i_x,i_y}\}\>,
\nonumber\\
|\t \Psi^T_{h_x,h_y}\> &= \sum_{\{g_{i_x,i_y}\}}\Psi_{h_x,h_y}(\{g_{i_x+i_y,i_y}\})
|\{g_{i_x,i_y}\}\>
=
\sum_{\{g_{i_x,i_y}\}}\Psi_{h_x,h_y}(\{g_{ \hat{T} \cdot (i_x,i_y)}\})
|\{g_{i_x,i_y}\}\>.
\end{align}
We note that the state $|\Psi^S_{h_x,h_y}\>$ and the state
$|\Psi_{h'_x,h'_y}\>$ have the same symmetry twists if
$(h'_x,h'_y)=(h_y^{-1},h_x)$.  Thus we can define a matrix
\begin{align}
\label{tS}
 \hat{S}_{(h'_x,h'_y),(h_x,h_y)} =
\del_{h'_x,h_y^{-1}}
\del_{h'_y,h_x} 
\<\Psi_{h'_x,h'_y}|\Psi^S_{h_x,h_y}\>
=
\del_{(h'_x,h'_y),\hat{S} \cdot(h_x,h_y)}
\<\Psi_{h'_x,h'_y}|\Psi^S_{h_x,h_y}\>.
\end{align}
However,  $|\t\Psi^T_{h_x,h_y}\>$ and $|\Psi_{h'_x,h'_y}\>$ always have
different branch cuts of the symmetry twists (see Fig. \ref{Tmov}(b)).  To make their  symmetry
twists comparable, we perform an 
additional local symmetry transformation $h_x$ in
the shaded region 
Fig. \ref{Tmov}(b), which changes $|\t\Psi^T_{h_x,h_y}\>$ to
$|\Psi^T_{h_x,h_y}\>$. 
Now $|\Psi^T_{h_x,h_y}\>$ and $|\Psi_{h'_x,h'_y}\>$ have
the same symmetry twists if 
$(h'_x,h'_y)=(h_x,h_yh_x)$
(see Fig.
\ref{Tmov}(c)).  Thus we define a matrix
\begin{align}
\label{tT}
 \hat{T}_{(h'_x,h'_y),(h_x,h_y)} =
\del_{h'_x,h_x h_y}
\del_{h'_y, h_x} 
\<\Psi_{h'_x,h'_y}|\Psi^T_{h_x,h_y}\>
=
\del_{(h'_x,h'_y),\hat{T} \cdot(h_x,h_y)}
\<\Psi_{h'_x,h'_y}|\Psi^T_{h_x,h_y}\>.
\end{align}

There is an additional operation called \emph{group actions} $\hat{U}(g)$, which sends group elements to other elements in conjugacy class:
$h \to g h g^{-1}$.
In general, $\hat{U}(g) |\Psi_{(h_x,h_y)}\>= U_{(gh_xg^{-1},gh_yg^{-1}), (h_x,h_y)}|\Psi_{gh_xg^{-1},gh_yg^{-1}}\>$.
The factor $U_{h_x,h_y}(g)$ 
is a U(1) phase occurs when evolving $ |\Psi_{(h_x,h_y)}\>$ to $|\Psi_{gh_xg^{-1},gh_yg^{-1}}\>$ state.
The nontrivial geometric matrix element is: 
\bea
\hat{U}(g)_{(h'_x,h'_y),(h_x,h_y)}&=&
\del_{h'_x, g h_x g^{-1} }
\del_{h'_y,  g h_y g^{-1}} 
{\<\Psi_{(g h_x g^{-1},g h_y g^{-1})}|\hat{U}(g) |\Psi_{(h_x,h_y)}\>} \nonumber\\
&=&{\del_{h'_x, g h_x g^{-1} }
\del_{h'_y,  g h_y g^{-1}}  U_{h_x,h_y}(g)}. \;\;\;
\eea 

More generally, for any combination of SL$(2,\Z)$ transformation, we find that \footnote{Here $\sigma_x \cdot\hat{O}^t \cdot \sigma_x \cdot(h_x,h_y)$
is defined as 
$
\big(\begin{smallmatrix}0 & 1 \\   1 &0 \end{smallmatrix} \big)\hat{O}^t \big(\begin{smallmatrix} 0 & 1 \\   1 &0 \end{smallmatrix} \big) \big(\begin{smallmatrix} h_x \\ h_y\end{smallmatrix}\big)
$.
In particular, $\sigma_x \cdot\hat{S}^t \cdot \sigma_x =\hat{S}$
and $\sigma_x \cdot\hat{T}^t \cdot \sigma_x =\hat{T}$.
It is checked by considering the invariant inner product form $(h_y,h_x) \cdot (\hat{x}, \hat{y})=(h_y',h_x') \cdot (\hat{x'}, \hat{y'})$
}
\begin{align}
\label{tO}
\boxed{
 \hat{O}_{(h'_x,h'_y),(h_x,h_y)} 
=
\del_{(h'_x,h'_y),\sigma_x \cdot\hat{O}^t \cdot \sigma_x \cdot(h_x,h_y)}
\<\Psi_{h'_x,h'_y}|\Psi^O_{h_x,h_y}\>
},
\end{align}
here $|\Psi^O_{h_x,h_y}\>$ is obtained from $\sum_{\{g_{i_x,i_y}\}}\Psi_{h_x,h_y}(\{g_{ \hat{O} \cdot (i_x,i_y)}\})
|\{g_{i_x,i_y}\}\>$ subject to an additional change on a shaded area in order to have the same branch-cut configuration as $|\Psi_{h_x,h_y}\>$.

\begin{figure}[!h] 
\centering
\includegraphics[height=1.3in]{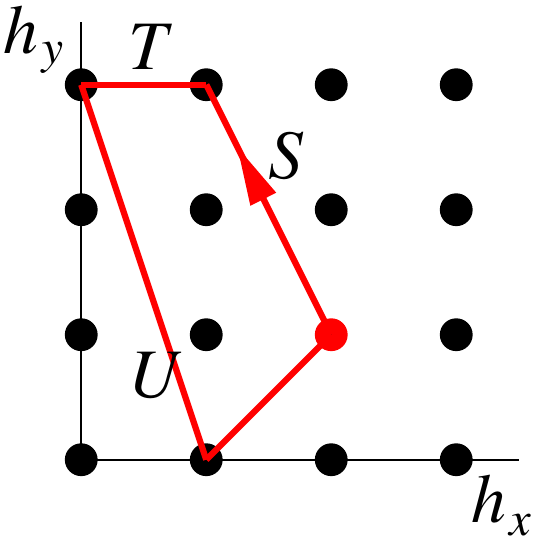}
\caption{The geometric matrix computed from a closed orbit (here in the $(h_x,h_y)$ space) gives rise to a universal SPT invariants.}
\label{closed_orbit}
\end{figure}

In general, there are an area dependent factor with non-universal constants $c_S, c_T$:
\begin{align}
\label{ST}
 \hat{ S}_{(h'_x,h'_y),(h_x,h_y)}&=\e^{-c_S L^2+ o(1/L)} \del_{(h'_x,h'_y),\hat{S} \cdot(h_x,h_y)} \cS_{(h_x,h_y)},
\nonumber\\
 \hat{T}_{(h'_x,h'_y),(h_x,h_y)}&=\e^{-c_T L^2+ o(1/L)} \del_{(h'_x,h'_y),\hat{T} \cdot(h_x,h_y)} \cT_{(h_x,h_y)},\nonumber\\
 \hat{U}(g)_{(h'_x,h'_y),(h_x,h_y)}&= \del_{h'_x, g h_x g^{-1} }
\del_{h'_y,  g h_y g^{-1}}  U_{h_x,h_y}(g).
\end{align}
Importantly the $U$ move has no additional volume-dependent factor because it is a group action which does not deform through any diffeomorphism or MCG elements.
Our goal is to extract the geometric matrix: $\cS_{(h_x,h_y)}, \cT_{(h_x,h_y)}, U_{h_x,h_y}(g)$.
However, such geometric matrices are not yet universal enough.
When $(h_x',h_y')\neq (h_x,h_y)$, the complex phases
$ S_{h_x,h_y},
 T_{h_x,h_y},
 U_{h_x,h_y}(h_t)$ are not well defined, since they depend on the choices of the phases of
$|\Psi_{(h_x,h_y)}\>$ and
$|\Psi_{(h_x',h_y')}\>$.
To obtain the universal geometric matrix, we need to send $|\Psi_{(h_x,h_y)}\>$  back to $|\Psi_{(h_x,h_y)}\>$ 
--- the product of ${S_{h_x,h_y},T_{h_x,h_y},U_{h_x,h_y}(h_t)}$ around a
closed orbit $(h_x,h_y) \to (h_x',h_y') \to \cdots \to (h_x,h_y)$ is universal
(see Fig. \ref{closed_orbit}).  We believe that those products for various
closed orbits
completely characterize the 2+1D SPTs. The same idea applies to SPTs in any dimension.

%% file: chap2.tex
\chapter{Aspects of Symmetry} \label{aofSymmetry}


In Sec.\ref{sec:1405.7689}, we extend the idea of symmetry-twist as a similar way of gauging and a way to couple to external probed field, therefore
we can develop an effective probed field action and partition functions for SPT.
In Sec.\ref{sec:1403.5256}, we express SPT invariants in terms of physical observables such as 
fractional quantum numbers and degenerate zero modes.

\section{Field theory representation of gauge-gravity SPT invariants, group cohomology and beyond: Probed fields} 
\label{sec:1405.7689}


Gapped systems 
without symmetry breaking\cite{GL5064,LanL58} can have intrinsic 
topological order. 
However, even without symmetry breaking and without topological order, gapped
systems can still be nontrivial if there is certain global symmetry
protection,
known as Symmetry-Protected Topological
states (SPTs). 
Their non-trivialness can be found in the gapless / topological
boundary modes protected by a global symmetry, which shows {gauge or
gravitational
anomalies}.\cite{Wen:2013oza,{Qi:2008ew},Wang:2014tia,Kapustin:2014lwa,Kapustin:2014zva,{Freed:2014eja},{Wang:2013yta},K1467,K1459,Kong:2014qka}
More
precisely, they are short-range entangled states which can be deformed to a
trivial product state by local unitary
transformation\cite{VCL0501,V0705} 
if the deformation breaks the
global symmetry.  Examples of SPTs are Haldane spin-1 chain protected by
spin rotational symmetry\cite{H8364,{AKL8877}} and the topological
insulators 
protected by fermion number conservation and
time reversal symmetry.

While some classes of topological orders can be described by topological quantum field theories (TQFT),\cite{Witten:1988hf,{Witten:1988ze}}
it is less clear
{\it how to systematically construct field theory 
with a global symmetry to classify or characterize SPTs for any dimension}.
This challenge originates from the fact that SPTs 
is naturally defined on a discretized spatial lattice
or on a discretized spacetime path integral
by a group cohomology construction\cite{XieSPT4,{Dijkgraaf:1989pz}} 
instead of continuous fields.
Group cohomology construction of SPTs also reveals a duality
between some SPTs and the Dijkgraaf-Witten topological gauge theory.\cite{{Dijkgraaf:1989pz},{LG1209}}

Some important progresses have been recently made to tackle the above question.
For example, there are 2+1D 
Chern-Simons theory, 
non-linear sigma models, 
and an orbifolding approach
implementing modular invariance on 1D edge modes. 
The above approaches have their own benefits, but they may be either limited to certain dimensions, or
be limited to some special cases.
Thus, the previous works may not
fulfill all SPTs predicted from group cohomology classifications.

In this work, we will provide a more systematic way to tackle this problem, 
by constructing topological response field theory and topological invariants for SPTs (SPT invariants) in any dimension protected by a symmetry group $G$.
The new ingredient of our work suggests 
a one-to-one 
correspondence between the continuous semi-classical probe-field partition function and the discretized
cocycle of cohomology group, $\cH^{d+1}(G,\R/\Z)$, predicted to classify $d+1$D SPTs with a symmetry group $G$. 
Moreover, our
formalism can even attain SPTs beyond group cohomology classifications.\cite{K1467,K1459} 

\subsection{Partition function and SPT invariants}
\begin{figure}[!t]
\begin{center}
\includegraphics[scale=0.5]{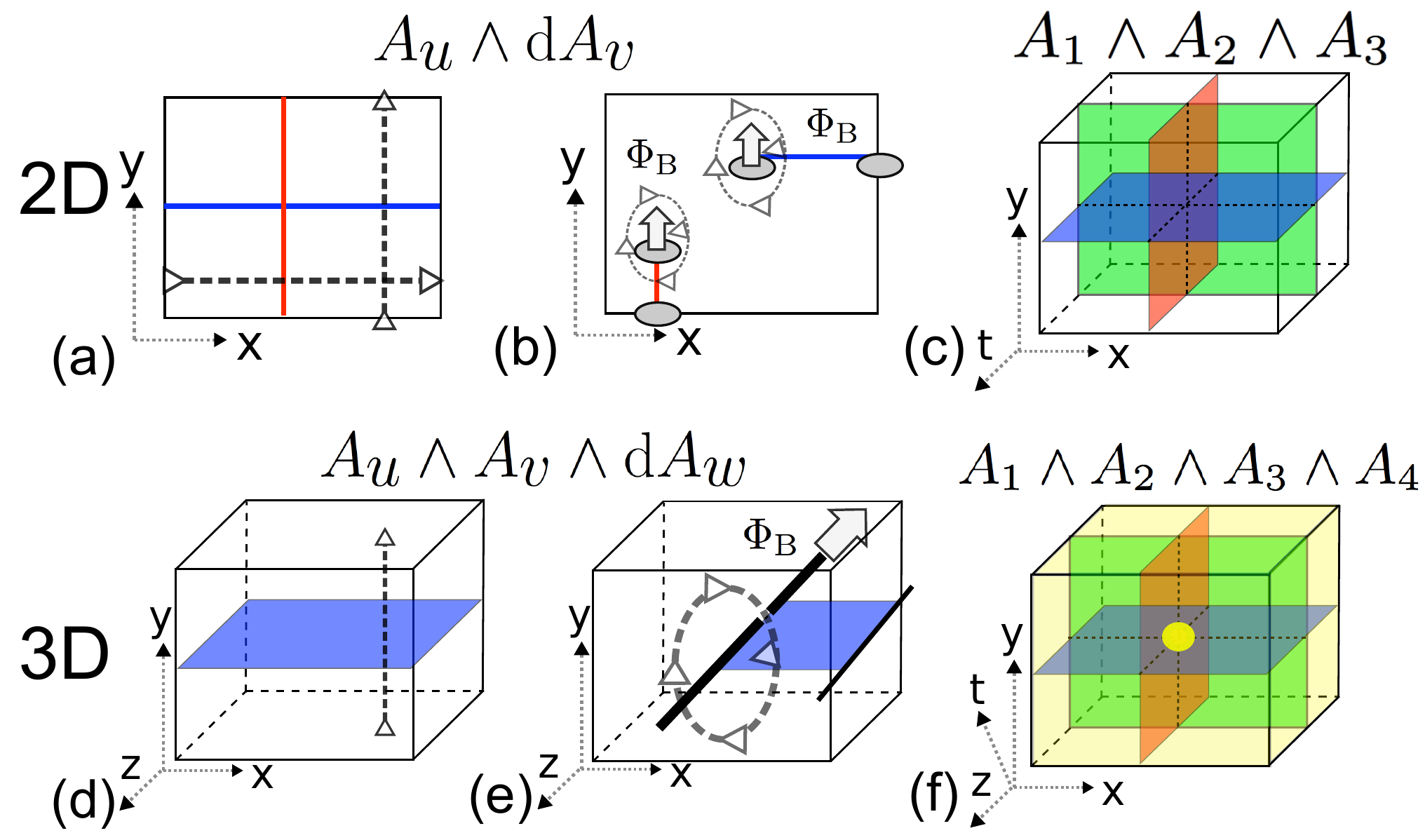} \end{center}
\caption{
On a spacetime manifold, the 1-form probe-field $A$ can be implemented on a
codimension-1 symmetry-twist 
(with flat $\dd A=0$)
modifying the Hamiltonian $H$, but the global symmetry $G$ is preserved as a
whole. The symmetry-twist is analogous to a branch cut, going along the arrow -
- -$\vartriangleright$ would obtain an Aharonov-Bohm phase $\ep^{ig}$ with $g \in
G$ by crossing the branch cut (Fig.(a) for 2D, Fig.(d) for 3D).
{
However if the
symmetry twist ends, 
its ends 
are \emph{monodromy defects}
with $\dd A \neq 0$, effectively with a gauge flux insertion.
Monodromy
defects in Fig.(b) of 2D act like 0D point particles carrying
flux, 
in Fig.(e) of 3D act like 1D line strings carrying flux. 
}
The non-flat monodromy defects with $\dd A\neq 0$ are essential to realize $\int A_u \dd A_v$
and $\int A_u A_v \dd A_w$ for 2D and 3D, while the flat connections ($\dd A=0$) are
enough to realize the {\it top} Type $\int A_1  A_2  \dots A_{d+1}$ whose
partition function on a spacetime $\mathbb{T}^{d+1}$ torus with $(d+1)$
codimension-1 sheets intersection (shown in Fig.(c),(f) in 2+1D, 3+1D) renders a nontrivial element for
Eq.(\ref{eq:SPTZ}).
} \label{fig:1} \end{figure}

For systems that realize topological orders, we can adiabatically deform the ground state $| \Psi_{g.s.} (g) \rangle$ of parameters $g$ via:
\bea
&&\langle \Psi_{g.s.} (g+\delta g) | \Psi_{g.s.} (g) \rangle \simeq \dots    \mathbf{Z}_0 \dots
\eea
to detect the volume-independent universal piece of
partition function, $\mathbf{Z}_0$, which
reveals non-Abelian geometric phase of ground states. 
For systems that realize  SPTs, however, their fixed-point partition
functions $\mathbf{Z}_0$ always equal to 1 due to 
its unique ground state
on any closed topology. We cannot distinguish SPTs via $\mathbf{Z}_0$.
However, due to the existence of a global symmetry, we can use  $\mathbf{Z}_0$
with the {\it symmetry twist} 
To define the symmetry twist, we note that the Hamiltonian
$H=\sum_x H_x$ is invariant under the global symmetry
transformation $U=\prod_\text{all sites} U_x$, namely $H=U H U^{-1}$.
If we perform the symmetry
transformation $U'=\prod_{x\in \prt R} U_x$
only near the boundary of a region $R$ (say on one side of ${\prt R}$),
the local term $H_x$ of $H$ 
will be modified: $H_x\to H_x'|_{x\text{ near } \prt R}$.
Such a change along a codimension-1 surface is called a symmetry twist, {see Fig.\ref{fig:1}(a)(d)},
which modifies $\mathbf{Z}_0$ 
to $\mathbf{Z}_0(\text{sym.twist})$.
\cblue{Just like the geometric phases of the degenerate ground states characterize topological orders,} 
we believe that $\mathbf{Z}_0(\text{sym.twist})$, on different spacetime manifolds and for different symmetry twists, fully characterizes
SPTs. 

The symmetry twist is similar to gauging the on-site
symmetry\cite{LG1209} 
except that the symmetry twist is non-dynamical. We can use the gauge connection 1-form $A$ to
describe the corresponding symmetry twists, with probe-fields $A$ coupling to
the matter fields of the system.  So we can write 
\begin{align} \label{eq:SPTZ}
\mathbf{Z}_0(\text{sym.twist})
=\ep^{\ti \mathbf{S}_0(\text{sym.twist})}=\ep^{\ti \mathbf{S}_0(A)}.
\end{align}
Here $\mathbf{S}_0(A)$ is the SPT invariant that we search for. 
Eq.(\ref{eq:SPTZ}) is a partition function of classical probe fields, or a topological response theory, obtained by integrating out the matter fields of SPTs path integral.
Below we would
like to construct possible forms of $\mathbf{S}_0(A)$ based on the following principles: 
(1) 
$\mathbf{S}_0(A)$ is independent of spacetime metrics ({\it i.e.} topological),
(2) $\mathbf{S}_0(A)$ is gauge invariant (for both large and small gauge transformations), and
(3) ``Almost flat'' connection for probe fields.

\noindent
\underline{{\bf U(1) SPTs}}-- 
Let us start with a simple example of a single global U(1) symmetry.
We can probe the system by coupling the charge fields to an external probe 1-form field $A$ (with a U(1) gauge symmetry), and integrate out the matter fields.
In 1+1D, 
we can write down a partition function by dimensional counting:
$
\mathbf{Z} _0 (\text{sym.twist})
= \exp[{\;\ti\; \frac{\theta}{2\pi} \int F }]
$ with $F \equiv \dd A$,
this is the only term allowed by U(1) gauge symmetry 
$U^{\dagger} (A-\ti \dd) U \simeq A + \dd f$ with $U=\ep^{\ti f}$.
More generally, for an even $(d+1)$D spacetime,
$\mathbf{Z} _0(\text{sym.twist})
= \exp[{\;\ti\;  \frac{\theta}{(\frac{d+1}{2})! (2\pi)^{\frac{d+1}{2}}}   \int F \wedge F \wedge \dots}]
$.
Note that $\theta$ in such an action has no level-quantization ($\theta$ can be
an arbitrary real number).  Thus this theory does {\it not} really correspond
to any nontrivial class, because any $\theta$ is smoothly connected to
$\theta=0$ which represents a trivial SPTs.

In an odd dimensional spacetime, such as 2+1D, we have Chern-Simons coupling for the probe field action
$\mathbf{Z} _0(\text{sym.twist})=$
$\exp[{\;\ti\; \frac{k}{4\pi}  \int A \wedge \dd A}]$.
More generally, for an odd $(d+1)$D,
$
\mathbf{Z} _0(\text{sym.twist})
= \exp[{\;\ti\; \frac{2\pi k}{(\frac{d+2}{2})!(2\pi)^{(d+2)/2}}  \int A \wedge F \wedge \dots}],
$
which is known to have level-quantization $k = 2p$ with $p \in \Z$ for bosons, since U(1) is compact.
We see that only \emph{quantized} topological terms
correspond to non-trivial SPTs, the allowed responses $\mathbf{S}_0(A)$ reproduces
the group cohomology description of the U(1) SPTs: an even dimensional
spacetime has no nontrivial class, while an odd dimension has a $\Z$
class.

\noindent
\underline{{\bf $\prod_u Z_{N_u}$ SPTs}}-- 
Previously the evaluation of  U(1) field on a closed loop (Wilson-loop) $\oint A_u$ can be arbitrary values, whether the loop is contractable or not, since U(1) has continuous value.
{For finite Abelian group symmetry $G=\prod_u Z_{N_u}$ SPTs,
(1) the large gauge transformation $\delta A_u$ is identified by $2\pi$ (this also applies to U(1) SPTs). 
(2) probe fields have discrete $Z_N$ gauge symmetry,}
\bea \label{eq:intA}
\oint \delta A_u= 0{\pmod{2\pi}},\;\;\; \oint A_u =\frac{2\pi n_u}{ N_u} {\pmod{2\pi}}.\;\;\;\;\;
\eea
For a non-contractable loop (such as a $S^1$ circle of a torus), $n_u$ can be a quantized integer which thus allows large gauge transformation.
For a contractable loop, due to the fact that small loop has small $\oint A_u$ but $n_u$ is discrete, $\oint A_u=0$ and $n_u=0$,
which imply the curvature $\dd A=0$, thus $A$ is {\it flat} connection locally.

\noindent
{\bf (i).} For {\bf 1+1D},
the only quantized topological term is:
$
\mathbf{Z} _0(\text{sym.twist})=\exp[{\;\ti\;  k_{\tII}\int A_1 A_2 }].
$
Here and below we omit the wedge product
$\wedge$ between gauge fields as a
conventional notation.
Such a term is {\bf gauge invariant} under transformation if we impose flat connection $\dd A_1=\dd A_2=0$,
since $\delta(A_1 A_2)= (\delta A_1) A_2+ A_1 (\delta A_2)=(\dd f_1) A_2+A_1 (\dd f_2) =-f_1 (\dd A_2)-(\dd A_1) f_2=0$. Here we have abandoned the surface term
by considering a 1+1D
closed bulk spacetime ${\cM^2}$ without boundaries.

\noindent
{\bf $\bullet$ Large gauge transformation}: The  invariance of $\mathbf{Z} _0$ under the allowed large gauge transformation via Eq.(\ref{eq:intA}) implies that
the volume-integration of
$\int \delta(A_1 A_2)$
must be invariant mod $2\pi$, namely
$\frac{(2\pi)^2 k_{\tII}}{N_1}=\frac{(2\pi)^2 k_{\tII}}{N_2} = 0{\pmod{2\pi}}$.
{This rule implies the {\bf level-quantization}.}
\noindent
{\bf $\bullet$ Flux identification}: On the other hand, when the $Z_{N_1}$ flux from $A_1$, $Z_{N_2}$ flux from $A_2$ 
are inserted as $n_1$, $n_2$ multiple units of $2\pi/N_1$, $2\pi/N_2$,
we have $k_{\tII} \int A_1  A_2
=k_{\tII} \frac{(2\pi)^2 }{ N_1 N_2} n_1 n_2$.
We see that $k_{\tII}$ and $k'_{\tII}=k_{\tII}+\frac{N_1N_2}{2\pi}$
give rise to the same partition function $\mathbf{Z} _0$. Thus
they must be identified
$(2\pi) k_{\tII} \simeq (2\pi) k_{\tII} +N_1 N_2$,
as the rule of flux identification. These two rules impose
\bea \label{eq:Z1DAA}
\mathbf{Z} _0(\text{sym.twist})=\exp[{\;\ti\;  p_{\tII}\frac{N_1 N_2 }{(2\pi) N_{12}} \int_{\cM^2} A_1 A_2 }],
\eea
with
$k_{\tII} =p_{\tII}\frac{N_1 N_2 }{(2\pi) N_{12}} $, $p_{\tII} \in \Z_{N_{12}}$.
We abbreviate the greatest common divisor (gcd) $N_{12\dots u} \equiv  \gcd(N_1,N_2, \dots, N_u)$.
Amazingly we have independently recovered the formal group cohomology classification predicted as $\cH^2(\prod_u Z_{N_u},\R/\Z)=\prod_{u<v} \Z_{N_{uv}}$.

\noindent
{\bf (ii).} For {\bf{2+1D}},
we can propose a naive $\mathbf{Z} _0(\text{sym.twist})$ by dimensional counting,
$\exp[{\;\ti\; k_{\tIII}\int A_1 A_2 A_3}]$, which is 
gauge invariant under the flat connection condition.
By the large gauge transformation and the flux identification, we find that the level $k_{\tIII}$ is quantized,
thus
\bea \label{eq:Z2DAAA}
\mathbf{Z} _0(\text{sym.twist})
= \exp[{\;\ti\; p_{\tIII}\frac{N_1 N_2 N_3}{(2\pi)^2 N_{123}} \int_{\cM^3} A_1 A_2 A_3}],
\eea
named as Type III SPTs with a quantized level $p_{\tIII} \in \Z_{N_{123}}$.
The terminology ``Type'' is introduced and used in Ref.\cite{deWildPropitius:1996gt} and \cite{Wang:2014oya}.
As shown in Fig.\ref{fig:1}, the geometric way to understand the 1-form probe
field can be regarded as (the Poincare-dual of) { codimension}-1
sheet assigning a group element $g \in G$ by crossing the sheet
as a branch cut.  These sheets can be regarded as the {\it symmetry
twists}\cite{{Wen:2013ue},Hung:2013cda} in the SPT Hamiltonian formulation.
When three sheets ($yt$, $xt$, $xy$ planes in Fig.\ref{fig:1}(c)) with
nontrivial elements $g_j \in Z_{N_j}$ intersect at a single point of a
spacetime $\mathbb{T}^3$ torus, it produces a nontrivial 
topological invariant 
in Eq.(\ref{eq:SPTZ}) for Type III  SPTs.


There are also other types of partition functions, which require to use
the insert flux $\dd
A\neq 0$ 
only at
the \emph{monodromy defect} (i.e. at the end of branch cut,
see Fig.\ref{fig:1}(b)) to probe
them: 
\bea
\label{eq:Z2DAdA}
&& \mathbf{Z}_0(\text{sym.twist})
= \exp [{\;\ti\; \frac{p }{2\pi} \int_{\cM^3} A_u \dd A_v}],
\eea
{where $u,v$ can be either the same or different gauge fields.
They are Type I, II 
actions: ${p_{\tI,1} } \int A_1 \dd A_1$, 
${p_{\tII,12} } \int A_1 \dd A_2$, etc. 
In order to have $\ep^{\;\ti\; \frac{p_{\tII}
}{2\pi} \int_{\cM^3} A_1 \dd A_2}$ invariant under the large gauge
transformation, 
$p_{\tII}$ must be
integer.  In order to have
$\ep^{\;\ti\; \frac{p_{\tI}
}{2\pi} \int_{\cM^3} A_1 \dd A_1}$
 well-defined, we separate  
$A_1=\bar A_1+A_1^F$ to the non-flat part $A_1$ and the flat part $A_1^F$.
Its partition function becomes $\ep^{{\;\ti\; \frac{p_{\tI} }{2\pi} \int_{\cM^3} A^F_1 \dd \bar A_1}}$. 
The invariance under the large gauge transformation of $A^F_1$
requires
$p_{\tI}$ to be quantized as integers.
We can 
further derive their level-classification via Eq.(\ref{eq:intA}) and two more conditions:
\bea  \label{eq:intdvarA}
\Ointint  \dd A_v =0 {\pmod{2\pi}},\;\;\; \Ointint  \delta \dd  A_v =0.
\eea
The first means that the net sum of all monodromy-defect fluxes on the spacetime manifold must have integer units of $2\pi$.
\cblue{Physically, 
a $2\pi$ flux configuration is trivial for a discrete symmetry group $Z_{N_v}$. Therefore two SPT invariants differ by a $2\pi$ flux configuration on their monodromy-defect should be regarded as the same one.}
The second condition means that the variation of the total flux is zero.
From the above two conditions for flux identification, we find the SPT invariant Eq.(\ref{eq:Z2DAdA}) describes the
$Z_{N_1}$ SPTs $p_\tI \in \Z_{N_1} =\cH^3(Z_{N_1},\R/\Z)$ and
the $Z_{N_1}\times Z_{N_2}$ SPTs $p_\tII \in \Z_{N_{12}}\subset \cH^3(Z_{N_1}\times Z_{N_2},\R/\Z)$.
}

\noindent
{\bf (iii).} For {\bf 3+1D},
we derive the {\it top} Type IV partition function that is independent of spacetime metrics:
\bea \label{eq:Z3DAAAA}
\mathbf{Z} _0(\text{sym.twist})
= \exp[{\ti\frac{p_{\tIV} N_1 N_2 N_3 N_4}{(2\pi)^3 N_{1234}} \int_{\cM^4} A_1 A_2 A_3 A_4}], \;\;\;\;\;\;
\eea
where $\dd A_i=0$ to ensure gauge invariance.
{The large
gauge transformation $\delta A_i$ of Eq.(\ref{eq:intA}),
and flux identification recover
$p_\tIV \in \Z_{N_{1234}}\subset \cH^4(\prod_{i=1}^4
Z_{N_i},\R/\Z)$.
Here the 3D SPT invariant 
is analogous to 2D, when the four
codimension-1 sheets ($yzt$, $xzt$, $yzt$, $xyz$-branes 
in Fig.\ref{fig:1}(f))
with flat $A_j$ of nontrivial element $g_j
\in Z_{N_j}$ intersect at a single point
on spacetime $\mathbb{T}^4$ torus, it renders a nontrivial  
partition function  
for the Type IV SPTs.
}

Another  
response 
is for Type III 3+1D SPTs: 
\begin{align}
\label{eq:Z3DAAdA}
\mathbf{Z} _0(\text{sym.twist})
&=\exp [\ti \int_{\cM^4} \frac{p_{\tIII} N_1N_2}{(2\pi)^2 N_{12}} A_1A_2 \dd A_3]
,
\end{align}
which is gauge invariant only if $\dd A_1=\dd A_2=0$. 
Based on Eq.(\ref{eq:intA}),(\ref{eq:intdvarA}),
the invariance under the large gauge transformations requires $p_\tIII \in \Z_{N_{123}}$. 
Eq.(\ref{eq:Z3DAAdA}) describes
Type III SPTs: $p_\tIII \in \Z_{N_{123}} \subset \cH^4(\prod_{i=1}^3
Z_{N_i},\R/\Z)$. 

Yet another
response
is for Type II 3+1D SPTs: 
\begin{align}
\label{eq:Z3DAAdA3}
\mathbf{Z} _0(\text{sym.twist})
&=\exp [\ti \int_{\cM^4} \frac{p_{\tII} N_1N_2}{(2\pi)^2 N_{12}} A_1A_2 \dd A_2].
\end{align}
The above is gauge invariant only if we choose $A_1$ and $A_2$ such that $\dd
A_1=\dd A_2 \dd A_2=0$.  We denote 
$A_2=\bar A_2+A^F_2$ where
$ \bar A_2  \dd \bar A_2=0$, $\dd A^F_2=0$, $\oint \bar A_2 =0$ mod $2\pi/N_2$,
and $\oint A^F_2 =0$ mod $2\pi/N_2$.  Note that in general $\dd \bar A_2\neq
0$, 
and Eq.(\ref{eq:Z3DAAdA3}) 
becomes $\ep^{\ti \int_{\cM^4} \frac{p_{\tII}
N_1N_2}{(2\pi)^2 N_{12}} A_1A^F_2 \dd \bar A_2}$.
The invariance under the
large gauge transformations of $A_1$ and $A^F_2$ and flux identification requires
$p_\tII \in \Z_{N_{12}} =  \cH^4(\prod_{i=1}^2 Z_{N_i},\R/\Z)$ of Type II SPTs. 
For Eq.(\ref{eq:Z3DAAdA}),(\ref{eq:Z3DAAdA3}), we have assumed the monodromy {\emph{line defect}} at $\dd A \neq 0$ is {\emph{gapped}};
for {\it gapless} defects, one will need to introduce extra anomalous {\it gapless} boundary theories.

\subsection{SPT invariants and physical probes}--\\
{
\underline{{\it Top types:}} 
The SPT invariants can help us to design physical probes for their SPTs, 
as \cblue{observables of numerical simulations or real experiments}.
Let us consider:\\
$\mathbf{Z} _0(\text{sym.twist})$$=\exp[\ti p_{}\frac{ {\prod_{j=1}^{d+1} N_j} }{(2\pi)^d N_{123 \dots (d+1)}}$$\int A_1 A_2 \dots A_{d+1} ]$,
a generic top type $\prod_{j=1}^{d+1} Z_{N_j}$ SPT invariant in $(d+1)$D, and its 
observables.\\
\noindent
$\bullet$ (1). \emph{Induced charges}:
%
If we design the space to have a topology $(S^1)^d$, and add the unit symmetry
twist of the $Z_{N_1}, Z_{N_2}, \dots$, $Z_{N_d}$ to the $S^1$ in 
$d$ directions respectively: $\oint_{S^1} A_j=2\pi  /N_j$.  The SPT invariant
 implies that such a configuration will carry a $Z_{N_{d+1}}$
charge $p_{}\frac{N_{d+1}}{N_{123 \dots (d+1)}}$.}

\noindent
{$\bullet$ (2).\emph{Degenerate zero energy modes}:
We can also apply dimensional reduction to probe SPTs. 
We can design the $d$D space as $(S^1)^{d-1} \times I$, and add
the unit $Z_{N_j}$ symmetry twists along the $j$-th $S^1$
circles for $j=3,\dots,d+1$. 
This induces a 1+1D $Z_{N_1}\times Z_{N_2}$ SPT invariant
$\exp[{\;\ti\; p_{}\frac{N_{12}}{N_{123 \dots (d+1)}}\frac{N_1 N_2}{2\pi N_{12}} \int
A_1 A_2}]$ on the 1D spatial interval $I$.
The 0D boundary of the
reduced 1+1D SPTs has degenerate zero modes that form a projective
representation of $Z_{N_1}\times Z_{N_2}$ symmetry.\cite{Wang:2014tia}
For example, dimensionally reducing 3+1D SPTs Eq.(\ref{eq:Z3DAAAA}) to this 1+1D SPTs, if we break the
$Z_{N_3}$ symmetry on the  $Z_{N_4}$ monodromy defect line,  gapless
excitations on the  defect line will be gapped.  A $Z_{N_3}$ symmetry-breaking
domain wall on the gapped monodromy defect line will carry  degenerate zero modes
that form a projective representation of $Z_{N_1}\times Z_{N_2}$ symmetry.}

\noindent
{$\bullet$ (3).\emph{Gapless boundary excitations}:
For Eq.(\ref{eq:Z3DAAAA}), we design the 3D space as $S^1\times
M^2$, and add the unit $Z_{N_4}$ symmetry twists along the $S^1$ circle. Then
Eq.(\ref{eq:Z3DAAAA}) reduces to
the 2+1D $Z_{N_1}\times Z_{N_2}\times Z_{N_3}$ SPT invariant
$ \exp[{\;\ti\;
p_{\tIV}\frac{N_{123}}{N_{1234}}\frac{N_1 N_2 N_3}{2\pi N_{123}} \int A_1 A_2
A_3}]$ 
labeled by $p_{\tIV}\frac{N_{123}}{N_{1234}} \in \Z_{N_{123}} \subset \cH^3(Z_{N_1}\times
Z_{N_2}\times Z_{N_3},\R/\Z)$.  
Namely, the $Z_{N_4}$
monodromy line defect carries gapless excitations identical to the edge modes
of the 2+1D $Z_{N_1}\times Z_{N_2}\times Z_{N_3}$ SPTs if the symmetry is
not broken. 
}


\noindent
\underline{{\it Lower types:}} 
Take 3+1D  SPTs of Eq.(\ref{eq:Z3DAAdA}) as an example,
there are at least two ways to design physical probes.  First, we can design
the 3D space as $M^2\times I$, where $M^2$ 
is punctured with ${N_3}$ \emph{identical}
monodromy defects each carrying $n_3$ unit $Z_{N_3}$ flux, namely 
%
$\Ointint  \dd A_3 = 2 \pi n_3$ of Eq.(\ref{eq:intdvarA}).
Eq.(\ref{eq:Z3DAAdA}) reduces to $ \exp[{\;\ti\;
p_{\tIII}^{}n_3\frac{N_1 N_2 }{(2\pi) N_{12}} \int A_1 A_2 }] $, which again
describes a 1+1D $Z_{N_1}\times Z_{N_2}$ SPTs, labeled by
$p_{\tIII}^{}n_3$  of
Eq.(\ref{eq:Z1DAA}) in $\cH^2(Z_{N_1}\times Z_{N_2},\R/\Z)=\Z_{N_{12}}$.
{This again has 0D boundary-degenerate-zero-modes.}

Second,
we can design the 3D space as $S^1\times M^2$
and add a symmetry twist of $Z_{N_1}$ along the $S^1$: $\oint_{S^1}
A_1=2\pi n_1 /N_1$,
then the  SPT invariant Eq.(\ref{eq:Z3DAAdA}) reduces to
$\exp[{\;\ti\; \frac{p_\tIII \; n_1 N_2 }{(2\pi)  N_{12}} \int  A_2 \dd A_3}]$,
a 2+1D $Z_{N_2}\times Z_{N_3}$ SPTs labeled by $\frac{p_\tIII \; n_1 N_2
}{N_{12}}$ of Eq.(\ref{eq:Z2DAdA}). \\
\noindent
{$\bullet$ (4).\emph{Defect braiding statistics and fractional charges}:}
These $\int A\dd A$ types in
Eq.(\ref{eq:Z2DAdA}), can be detected by the nontrivial braiding statistics of
monodromy defects, such as the particle/string defects in
2D/3D. 
Moreover,
a $Z_{N_1}$ monodromy defect line carries gapless excitations identical to the
edge of the 2+1D $Z_{N_2}\times Z_{N_3}$ SPTs. If the  gapless excitations
are gapped by $Z_{N_2}$-symmetry-breaking, its domain wall will induce
fractional quantum numbers of $Z_{N_3}$ charge,\cite{Wang:2014tia} 
similar to Jackiw-Rebbi\cite{Jackiw:1975fn} or
Goldstone-Wilczek\cite{Goldstone:1981kk} effect.

\noindent
\underline{\bf U$(1)^m$ SPTs}-- 
It is straightforward to apply the above results to U$(1)^m$ symmetry.
Again, we find only trivial classes for even $(d+1)$D. For odd $(d+1)$D, we can define the lower type action:
$
\mathbf{Z} _0(\text{sym.twist})
= \exp[{\;\ti\; \frac{2\pi k}{(\frac{d+2}{2})!(2\pi)^{(d+2)/2}}  \int A_u \wedge F_v \wedge \dots}].
$
Meanwhile we emphasize that the {\it top} type
action with $k\int A_1 A_2  \dots A_{d+1}$ form will be trivial for U$(1)^m$ case since its coefficient $k$ is no longer well-defined,
at $N \to \infty$ of $(Z_{N})^m$ SPTs states.
For physically relevant $2+1$D,
$k \in 2\Z$ for bosonic  SPTs.
Thus, we will have a $\Z^m \times \Z^{m(m-1)/2}$ classification for U$(1)^m$ symmetry. 


We have formulated the {\bf spacetime partition functions of probe fields} (e.g. $\mathbf{Z} _0(A(x))$, etc), which fields $A(x)$ take values at any coordinates $x$ on a continuous spacetime manifold $\cM$
with no dynamics.
On the other hand, it is known that, $(d+1)$D bosonic SPTs of symmetry group $G$ can be classified by the $(d+1)$-{th} cohomology group $\cH^{d+1}(G,\R/\Z)$
(predicted to be complete at least for finite symmetry group $G$ without time reversal symmetry).
From this prediction that bosonic SPTs can be classified by group cohomology, 
our path integral on the discretized space lattice (or spacetime complex) shall be mapped to the {\bf partition functions of the cohomology group - the cocycles}.
In this section, we ask ``whether we can attain this correspondence from ``partition functions of fields'' to ``cocycles of group cohomology?''
Our answer is ``yes,''
we will bridge this beautiful correspondence between \emph{continuum field theoretic partition functions} and \emph{discrete cocycles} for any $(d+1)$D spacetime dimension for finite Abelian $G=\prod_u Z_{N_u}$.

\begin{center}
\begin{table}[!h]
\noindent
\makebox[\textwidth][c]
{\fontsize{9.5pt}{1em}\selectfont
{
\begin{tabular}{|c||c|c|c|}
\hline
  & partition function $\mathbf{Z}$ &  $(d+1)$-cocycle $\omega_{d+1}$\\
 \hline\hline
0+1&  $\exp(\ti \, p_{\tI} \int A_1) $ &  $ \exp \Big( \frac{2 \pi \ti p_{\tI}  }{N_{1}} \;  a_{1}\Big)$ \\[0mm]  \hline\hline
1+1 &  $\exp(\ti \frac{N_1 N_2 }{(2\pi) N_{12}} p_{\tII} \int A_1 A_2) $ & $ \exp \Big( \frac{2 \pi \ti p_{\tII}  }{N_{12}} \;  a_{1} b_{2}\Big)$  \\[0mm]  \hline\hline
2+1 &  $\exp(\ti \frac{p_{\tI} }{(2\pi) } \int A_1\dd A_1) $  &  $\exp \Big( \frac{2 \pi \ti {p_{\tI} }  }{N_{1}^{2}} \; a_{1}(b_{1} +c_{1} -[b_{1}+c_{1}]) \Big) $\\
            &  $\exp(\ti\, {p_{\tI} } \int C_1) $ 
            &  $\exp \Big( \frac{2 \pi \ti {p_{\tI} }  }{N_{1} } \, a_{1} b_{1} c_{1} \Big) $ \\ \hline
2+1 & $\exp(\ti \frac{p_{\tII} }{(2\pi) } \int A_1\dd A_2) $  & $\exp \Big( \frac{2 \pi \ti {p_{\tII} }  }{N_{1} N_{2}} \; a_{1}(b_{2} +c_{2} -[b_{2} +c_{2}]) \Big) $ \\
            & $\exp(\ti \,  p_{\tII} \frac{N_1 N_2 }{(2\pi) N_{12}}   \int A_1B_2) $ 
            & $ \exp \Big( \frac{2 \pi \ti p_{\tII}   }{N_{12}} \;  a_{1} b_{2} c_{2}\Big)$   \\  \hline
2+1 & $\exp(\ti \, p_{\tIII}\frac{N_1 N_2 N_3}{(2\pi)^2 N_{123}}  \int A_1 A_2 A_3) $ & $ \exp \Big( \frac{2 \pi \ti p_{\tIII}  }{N_{123}} \;  a_{1}b_{2}c_{3} \Big)$ \\  \hline\hline
3+1 & $\exp(\ti \int p_{{ \tII(12)}}^{(1st)}\frac{N_1 N_2 }{(2\pi)^2 N_{12}}  A_1 A_2 \dd A_2) $  & ${\exp \big( \frac{2 \pi \ti p_{{ \tII(12)}}^{(1st)} }{ (N_{12} \cdot N_2  )   }    (a_1 b_2 )( c_2 +d_2 - [c_2+d_2  ]) \big)}$\\
            & $\exp(\ti \, p_{\tII} \frac{N_1 N_2 }{(2\pi) N_{12}}  \int A_1 C_2) $ 
            & ${\exp \big( \frac{2 \pi \ti p_{\tII} }{ N_{12}    }    a_1 b_2 c_2 d_2 \big)}$ \\  \hline
3+1 & $\exp(\ti \int p_{{ \tII(12)}}^{(2nd)}\frac{N_1 N_2 }{(2\pi)^2 N_{12}}  A_2 A_1 \dd A_1) $  & ${\exp \big( \frac{2 \pi \ti p_{{ \tII(12)}}^{(2nd)} }{ (N_{12} \cdot N_1  )   }    (a_2 b_1 )( c_1 +d_1 - [c_1+d_1  ]) \big)}$ \\
            & $\exp(\ti \, p_{\tII} \frac{N_1 N_2 }{(2\pi) N_{12}}  \int A_2 C_1) $ 
            & ${\exp \big( \frac{2 \pi \ti p_{\tII} }{ N_{12}    }    a_2 b_1 c_1 d_1 \big)}$ \\  \hline
3+1 & $\exp(\ti \, p_{{ \tIII(123)}}^{(1st)}\frac{N_1 N_2 }{(2\pi)^2 N_{12}}  \int (A_1 A_2) \dd A_3) $ &  ${\exp \big( \frac{2 \pi \ti p_{{ \tIII(123)}}^{(1st)} }{ (N_{12} \cdot N_3 )  }  (a_1 b_2 )( c_3 +d_3 - [c_3+d_3  ]) \big) }$\\
            & $\exp(\ti \, p_{\tIII}\frac{N_1 N_2 N_3}{(2\pi)^2 N_{123}}  \int A_1 A_2 B_3) $ 
            &  ${\exp \big( \frac{2 \pi \ti p_{\tIII} }{ N_{123}   }  a_1 b_2 c_3 d_3 \big) }$\\  \hline
3+1 &  $\exp(\ti \, p_{{ \tIII(123)}}^{(2nd)}\frac{N_3 N_1 }{(2\pi)^2 N_{31}}  \int (A_3 A_1) \dd A_2 ) $ &  ${\exp \big( \frac{2 \pi \ti p_{{ \tIII(123)}}^{(2nd)} }{ (N_{31} \cdot N_2 )  }  (a_3 b_1 )( c_2 +d_2 - [c_2 +d_2  ]) \big) }$ \\
            & $\exp(\ti \, p_{\tIII}\frac{N_1 N_2 N_3}{(2\pi)^2 N_{123}}  \int A_3 A_1 B_2 ) $ 
             &  ${\exp \big( \frac{2 \pi \ti p_{\tIII} }{ N_{123}   }  a_3 b_1 c_2 d_2 \big) }$\\  \hline
3+1 & $[\exp(\ti \, p_{\tIV}\frac{N_1 N_2 N_3 N_4 }{(2\pi)^3 N_{1234}}  \int A_1 A_2 A_3 A_4) ]$ & $\exp \big( \frac{2 \pi \ti p_{{ \tIV}}^{}}{ N_{1234} }  a_1 b_2 c_3 d_4 \big)$ \\  \hline\hline
4+1 &  $\exp(\ti \, \frac{p_{\tI} }{(2\pi)^2 }  \int A_1\dd A_1 \dd A_1) $  & ${\exp \left( \frac{2 \pi \ti {p_{\tI} } }{ (N_1)^3 } \;a_1( b_1+c_1 - [b_1+c_1 ]) ( d_1 +e_1 - [d_1 +e_1 ]) \right) }$ \\  \hline
4+1 & $\dots$ & $\dots$\\  \hline
4+1 & $\exp(\ti \, p_{\tV}\frac{N_1 N_2 N_3 N_4 N_5}{(2\pi)^4 N_{12345}}  \int A_1 A_2 A_3 A_4A_5) $ & $\exp \big( \frac{2 \pi \ti  p_{\tV} }{ N_{12345} }  a_1 b_2 c_3 d_4 e_5 \big)$\\  \hline
\hline
\end{tabular}
}
}\hspace*{0mm}
\caption{Some derived results on the correspondence between the {\bf spacetime partition function
of probe fields} (the second column) and  the {\bf cocycles of the cohomology group} (the third column) for any finite Abelian group $G=\prod_u Z_{N_u}$.
The first column provides the spacetime dimension: $(d+1)\tD$.
The even/odd effect means that whether their corresponding cocycles are nontrivial or trivial(as coboundary) depends on the level $p$ and $N$ (of the symmetry group $Z_N$) is even/odd.
Details are explained in Sec \ref{subsec:GCcocycle}.
}
\label{table:cocyclefact}
\end{table}
\end{center}


\subsection{Correspondence}

The partition functions
have been treated with careful proper level-quantizations via large gauge transformations and flux identifications.
For $G=\prod_u Z_{N_u}$, the field $A_u, B_u, C_u$, etc, take values in $Z_{N_u}$ variables, thus we can express them as
\be \label{eq:MapABC}
A_u \sim \frac{2\pi g_u}{N_u}, \;\, B_u \sim \frac{2\pi g_u h_u}{N_u}, \;\, C_u \sim \frac{2\pi g_u h_u l_u}{N_u}
\ee
with $g_u, h_u, l_u \in Z_{N_u}$. Here 1-form $A_u$ takes $g_u$ value on one link of a $(d+1)$-simplex, 2-form $B_u$ takes $g_u,h_u$ values on two different links and
3-form $C_u$ takes $g_u,h_u, l_u$ values on three different links of a $(d+1)$-simplex. These correspondence suffices for the flat probe fields.

In other cases, we also need to interpret the non-flat $\dd A\neq 0$ at the monodromy defect as the external inserted fluxes, thus we identify
\be \label{eq:MapdA}
\dd A_u \sim \frac{2\pi (g_u+h_u -[g_u+h_u] )}{N_u},
\ee
here $[g_u+h_u] \equiv g_u+h_u \pmod{N_u}$. Such identification ensures $\dd A_u$ is a multiple of $2\pi$ flux, therefore it is consistent with the constraint 
at the continuum limit.
Based on the Eq.(\ref{eq:MapABC})(\ref{eq:MapdA}), we derive the correspondence in Table \ref{table:cocyclefact},
from the continuum path integral $\mathbf{Z} _0(\text{sym.twist})$ of fields to a U(1) function as the discrete partition function.
In the next subsection, we will verify the U(1) functions in the last column in Table \ref{table:cocyclefact} indeed are the cocycles $\omega_{d+1}$ of cohomology group.
Such a correspondence has been explicitly pointed out in our previous work Ref.\cite{Wang:2014oya} 
and applied to derive the cocycles.


\begin{center}
\begin{table}[!h]
\noindent
\makebox[\textwidth][c]
{\fontsize{9.5pt}{1em}\selectfont
{
\begin{tabular}{|c||c|c|c|}
\hline
& Partition function 
$\mathbf{Z}$ 
& $ \cH^{d+1}$ & K\"unneth formula in $\cH^{d+1}(G,\R/\Z)$  \\
 \hline\hline
0+1 & $\e^{(\ti \,p..\int A_1)}$ & $\Z_{N_1}$ &$\cH^{1}(\Z_{N_1}, \R/\Z) $   \\[0mm]  \hline \hline
1+1 & $\e^{(\ti \,p..\int A_1 A_2)}$ & $\Z_{N_{12}}$ &$ \cH^{1}(\Z_{N_1}, \R/\Z) \boxtimes_\Z \cH^{1}(\Z_{N_2}, \R/\Z) $\\[0mm]  \hline \hline
2+1 & $\e^{(\ti \,p..\int A_1\dd A_1)}$  & $\Z_{N_{1}}$  &$ \cH^{3}(\Z_{N_1}, \R/\Z)$    \\ \hline
2+1 & $\e^{(\ti \,p..\int A_1\dd A_2)}$  & $\Z_{N_{12}}$ &$\cH^{1}(\Z_{N_1}, \R/\Z) \otimes_\Z \cH^{1}(\Z_{N_2}, \R/\Z)$  \\  \hline
2+1 & $\e^{(\ti \,p..\int A_1 A_2 A_3)}$ & $\Z_{N_{123}}$ &$[\cH^1(\Z_{N_1},\R/\Z)\boxtimes_\Z \cH^1(\Z_{N_2},\R/\Z)]\boxtimes_\Z \cH^1(\Z_{N_3}, \R/\Z)$  \\  \hline \hline
3+1 & $\e^{(\ti \,p..\int A_1 A_2 \dd A_2)}$ & $\Z_{N_{12}}$ &$ \cH^{1}(\Z_{N_1}, \R/\Z) \boxtimes_\Z \cH^{3}(\Z_{N_2}, \R/\Z)  $  \\  \hline
3+1 & $\e^{(\ti \,p..\int A_2 A_1 \dd A_1)}$ & $\Z_{N_{12}}$ &$ \cH^{1}(\Z_{N_2}, \R/\Z) \boxtimes_\Z \cH^{3}(\Z_{N_1}, \R/\Z)  $  \\  \hline
3+1 & $\e^{(\ti \,p..\int (A_1 A_2) \dd A_3)}$ & $\Z_{N_{123}}$ &$ {  [ \cH^{1}(\Z_{N_1} , \R/\Z) \boxtimes_\Z \cH^{1}(\Z_{N_2}, \R/\Z) ]     \otimes_\Z  \cH^1(\Z_{N_3}, \R/\Z) }   $  \\  \hline
3+1 & $\e^{(\ti \,p..\int (A_1 \dd A_2) A_3)}$  &  $\Z_{N_{123}}$ &$  [\cH^{1}(\Z_{N_1} , \R/\Z) \otimes_\Z \cH^{1}(\Z_{N_2},\R/\Z) ] \boxtimes_\Z  \cH^1(\Z_{N_3}, \R/\Z) ) $ \\  \hline
3+1 & $\e^{(\ti \,p..\int A_1 A_2 A_3 A_4)}$ &$\Z_{N_{1234}}$ &
$\big[[\cH^1(\Z_{N_1})  \boxtimes_\Z \cH^1(\Z_{N_2})]  \boxtimes_\Z  \cH^1(\Z_{N_3}) \big] \boxtimes_\Z  \cH^1(\Z_{N_4})$  \\  \hline \hline
4+1 & $\e^{(\ti \,p..\int A_1\dd A_1 \dd A_1)}$ &$\Z_{N_{1}}$ &$\cH^{5}(\Z_{N_1},  \R/\Z)$   \\  \hline
4+1 & $\e^{(\ti \,p..\int A_1 \dd A_1 \dd A_2)}$  &$\Z_{N_{12}}$ &$\cH^{3}(\Z_{N_1}  \R/\Z) \otimes_\Z \cH^{1}(\Z_{N_2}  ,\R/\Z)$  \\  \hline
4+1 & $\e^{(\ti \,p..\int A_2 \dd A_2 \dd A_1)}$ &$\Z_{N_{12}}$ &$\cH^{3}(\Z_{N_2}  ,\R/\Z) \otimes_\Z \cH^{1}(\Z_{N_1}  ,\R/\Z)$   \\  \hline
4+1 & $\e^{(\ti \,p..\int A_1 \dd A_1 A_2 A_3 )}$  & $\Z_{N_{123}}$ & ${ \big[  [\cH^{3}(\Z_{N_1}, \R/\Z)  \boxtimes_\Z \cH^{1}(\Z_{N_2}, \R/\Z) ]   \boxtimes_\Z  \cH^1( \Z_{N_3}, \R/\Z) \big]   }$ \\  \hline
4+1 & $\e^{(\ti \,p..\int A_2 \dd A_2 A_1 A_3)}$ & $\Z_{N_{123}}$ & ${ \big[  [\cH^{3}(\Z_{N_2}, \R/\Z)  \boxtimes_\Z \cH^{1}(\Z_{N_1}, \R/\Z) ]   \boxtimes_\Z  \cH^1( \Z_{N_3}, \R/\Z) \big]   }$  \\  \hline
4+1 & $\e^{(\ti \,p..\int A_1 \dd A_2 \dd A_3)}$ & $\Z_{N_{123}}$ & ${ { [\cH^1( \Z_{N_1}, \R/\Z) \otimes_Z \cH^1( \Z_{N_2}, \R/\Z) ]\otimes_Z \cH^1( \Z_{N_3}, \R/\Z)  } }$  \\  \hline
4+1 & $\e^{(\ti \,p..\int A_1 A_2 A_3 \dd A_3)}$ & $\Z_{N_{123}}$ & ${ \big[  [\cH^{1}(\Z_{N_1}, \R/\Z)  \boxtimes_\Z \cH^{1}(\Z_{N_2},\R/\Z) ]   \boxtimes_\Z  \cH^3( \Z_{N_3}, \R/\Z) \big]   }$  \\  \hline
4+1 & $\e^{(\ti \,p..\int  A_1 \dd A_2 A_3 A_4)}$ & $\Z_{N_{1234}}$ & 
${ \Big[ \big[  [\cH^{1}(\Z_{N_1})  \otimes_\Z \cH^{1}(\Z_{N_2}) ]   \boxtimes_\Z  \cH^1( \Z_{N_3}) \big] \boxtimes_\Z  \cH^1( \Z_{N_4}) \Big]   }$  \\  \hline
4+1 & $\e^{(\ti \,p..\int  A_1 A_2 \dd A_3 A_4)}$  & $\Z_{N_{1234}}$ & 
${ \Big[ \big[  [\cH^{1}(\Z_{N_1})  \boxtimes_\Z \cH^{1}(\Z_{N_2}) ]   \otimes_\Z  \cH^1( \Z_{N_3}) \big] \boxtimes_\Z  \cH^1( \Z_{N_4}) \Big]   }$ \\  \hline
4+1 & $\e^{(\ti \,p..\int  A_1 A_2 A_3 \dd A_4)}$  & $\Z_{N_{1234}}$ & 
${ \Big[ \big[  [\cH^{1}(\Z_{N_1})  \boxtimes_\Z \cH^{1}(\Z_{N_2}) ]   \boxtimes_\Z  \cH^1( \Z_{N_3}) \big] \otimes_\Z  \cH^1( \Z_{N_4}) \Big]   }$ \\  \hline
4+1 & $\e^{(\ti \,p.. \int A_1 A_2 A_3 A_4A_5)} $ &$\Z_{N_{12345}}$ &
$ \cH^1(\Z_{N_1})  \boxtimes_\Z \cH^1(\Z_{N_2})  \boxtimes_\Z  \cH^1(\Z_{N_3}) \boxtimes_\Z  \cH^1(\Z_{N_4}) \boxtimes_\Z  \cH^1(\Z_{N_5}) $  
\\  \hline
\end{tabular}
}
}\hspace*{20mm}
\caption{ {\bf From partition functions of fields to K\"unneth formula}. 
Here we consider a finite Abelian group $G=\prod_u Z_{N_u}$.
The field theory result can map to the derived facts about the cohomology group and its cocycles.
The first column provides the spacetime dimension: $(d+1)\tD$.
Here the level-quantization is shown in a shorthand way with only $p ..$ written, the explicit coefficients can be found. 
In some row, we abbreviate $\cH^1(\Z_{n_j},\R/\Z) \equiv \cH^1(\Z_{n_j})$.
The torsion product $\text{Tor}_1^\Z \equiv \boxtimes_\Z $ evokes a wedge product  $\wedge$ structure in the corresponding field theory,
while the tensor product $\otimes_\Z$ evokes appending an extra exterior derivative $\wedge  \dd$ structure in the corresponding field theory.
This simple observation maps the field theoretic path integral to its correspondence in K\"unneth formula.
}
\label{table:cocycleclass}
\end{table}
\end{center}

\begin{center}
\begin{table}[ht]
\noindent
\makebox[\textwidth][c]
{\fontsize{10pt}{1em}\selectfont
{
\begin{tabular}{|c||c|c|c|c|c|c|c|c|}
\hline
      &Type I& Type II & Type III &Type  IV & $\dots$    & $\dots$ & \\[0mm]  \hline
      &$\Z_{N_i}$& $\Z_{N_{ij}}$& $\Z_{N_{ijl}}$ &  $\Z_{N_{ijlm}}$ &  $\dots$
      & $\Z_{{\gcd} \otimes^m_{i}(N_i)}$  & $\Z_{{\gcd} \otimes^d_{i}N^{(i)}}$\\[0mm]  \hline
$\cH^1(G,\R/\Z)$ &$1$& $$& $$ & $$ & $$  & $$  & $$ \\[0mm]  \hline
$\cH^2(G,\R/\Z)$&$0$  &$1$ & $$ & $$ & $$  & $$  & $$\\[0mm]  \hline
$\cH^3(G,\R/\Z)$ &$1$ & $1$ & $1$ & $$ & $$  & $$  & $$\\ \hline
$\cH^4(G,\R/\Z)$ &$0$ & $2$ & $2$ & $1$ & $$  & $$ & $$\\ \hline
$\cH^5(G,\R/\Z)$ & $1$ & $2$ & $4$ & $3$ & $\dots$  & $$  & $$\\ \hline
$\cH^6(G,\R/\Z)$ & $0$ & $3$ & $6$ & $7$ & $\dots$  & $$  & $$\\ \hline
$\cH^d(G,\R/\Z)$ & $\frac{(1-(-1)^d)}{2}$ & $\frac{d}{2}-\frac{(1-(-1)^d)}{4}$ & $\dots$ & $\dots$ & $\dots$ & $\dots$  & $1$\\ \hline
\end{tabular}
}
}\hspace*{20mm}
%
\caption{
The table shows the exponent of the $\Z_{{\gcd} \otimes^m_{i}(N_i)}$ class in the cohomology group
$\cH^d(G,\R/\Z)$ for a finite Abelian group. 
Here we define a shorthand of $\Z_{{\gcd} (N_i,N_j)}\equiv
\Z_{N_{ij}} \equiv  \Z_{{\gcd} \otimes^2_{i}(N_i)}$, etc also for other higher
gcd.
Our definition of the Type $m$  is from its number ($m$) of cyclic gauge
groups in the gcd class $\Z_{{\gcd} \otimes^m_{i}(N_i)}$.  The number of
exponents can be systematically obtained by adding all the numbers of the
previous column from the top row to a row before the wish-to-determine number.
This table in principle can be independently derived by gathering the data of
Table \ref{table:cocycleclass} 
from field theory approach. Thus, we can use field theory to {derive} the group cohomology result. 
}
\label{table:Hgroup}
\end{table}
\end{center}

We remark that the field theoretic path integral's level $p$ quantization and its mod relation also provide an independent way (apart from group cohomology)
to count the number of types of partition functions for a given symmetry group $G$ and a given spacetime dimension.
In addition,
one can further deduce the {\bf K\"unneth formula} 
from a field theoretic partition function viewpoint.
Overall, this correspondence from field theory can be an
independent powerful tool to {\it derive}  the group cohomology and extract the classification data. 

\subsection{Cohomology group and cocycle conditions} \label{subsec:GCcocycle}
To verify that the last column of Table \ref{table:cocyclefact} (bridged from the field theoretic partition function) 
are indeed cocycles of a cohomology group,
here we briefly review the cohomology group $\mathcal{H}^{d+1}(G,\R/\Z)$ (equivalently as $\mathcal{H}^{d+1}(G,\tU(1))$ by $\R/\Z=\tU(1)$),
which is the ${(d+1)}$th-cohomology group of G over G module U(1).
Each class in $\mathcal{H}^{d+1}(G,\R/\Z)$ corresponds to a distinct $(d+1)$-cocycles.
The $n$-cocycles is a $n$-cochain, in addition
they satisfy the $n$-cocycle-conditions $\delta \omega=1$.
The $n$-cochain is a mapping of $\omega_{}^{}(a_1,a_2,\dots,a_n)$:  $G^n \to \tU(1)$ (which inputs
$a_i \in G$, $i=1,\dots, n$, and outputs a $\tU(1)$ value).
The $n$-cochain satisfies the group multiplication rule:
\be
(\omega_{1}\cdot\omega_{2})(a_1,\dots,a_n)= \omega_{1}^{}(a_1,\dots,a_n)\cdot \omega_{2}^{}(a_1,\dots,a_n),
\ee
thus form a group.
The coboundary operator $\delta$
\be \label{eq:delta}
\delta \sfc(g_1, \dots,  g_{n+1}) \equiv
\sfc(g_2, \dots,  g_{n+1})  \sfc(g_1, \dots,  g_{n})^{(-1)^{n+1}}
\cdot \prod_{j=1}^{n} \sfc(g_1, \dots, \; g_j g_{j+1},\; \dots,  g_{n+1})^{(-1)^{j}},
\ee
\noindent
which defines the $n$-cocycle-condition $\delta \omega=1$.
The $n$-cochain forms a group $\text{C}^n$,
while the $n$-cocycle forms its subgroup $\text{Z}^n$.
The distinct $n$-cocycles are not equivalent via $n$-coboundaries,
where Eq.(\ref{eq:delta}) also defines the $n$-coboundary relation:
if n-cocycle $\omega_n$ can be written as
$\omega_n = \delta \Omega_{n-1}$, for any $(n-1)$-cochain $\Omega_{n+1}$, then we
say this $\omega_n$ is a $n$-coboundary.
Due to $ \delta^2=1$, thus we know that the $n$-coboundary further forms a subgroup $\text{B}^n$ . In short,
$
\text{B}^n \subset \text{Z}^n \subset \text{C}^n
$
The $n$-cohomology group is precisely a kernel $\text{Z}^n$ (the group of $n$-cocycles) mod out image $\text{B}^n$ (the group of $n$-coboundary) relation:
\bea
\cH^n(G,\R/\Z)= \text{Z}^n /\text{B}^n.
\eea
For other details about group cohomology (especially Borel group cohomology here), we suggest to read Ref.\cite{{Wang:2014oya},{deWildPropitius:1996gt}} and Reference therein. 

To be more specific cocycle conditions, for finite Abelian group $G$, the 3-cocycle condition for 2+1D is (a pentagon relation),
\be
\delta \omega(a,b,c,d)=\frac{ \omega(b,c,d) \omega(a,bc,d) \omega(a,b,c) }{\omega(ab,c,d)\omega(a,b,cd) }=1.
\ee

The 4-cocycle condition for 3+1D is
\be
\delta \omega(a,b,c,d,e)=\frac{ \omega(b,c,d,e) \omega(a,bc,d,e) \omega(a,b,c,de) }{\omega(ab,c,d,e)\omega(a,b,cd,e) \omega(a,b,c,d) }=1.
\ee
The 5-cocycle condition for 4+1D is
\be
\delta \omega(a,b,c,d,e,f)=\frac{ \omega(b,c,d,e,f) \omega(a,bc,d,e,f) }{\omega(ab,c,d,e,f) } 
\cdot \frac{  \omega(a,b,c,de,f) \omega(a,b,c,d,e)  }{ \omega(a,b,cd,e,f) \omega(a,b,c,d,ef)   }=1.
\ee
We verify that the U(1) functions (mapped from a field theory derivation) in the last column of Table \ref{table:cocyclefact} indeed satisfy cocycle conditions.
Moreover, those  {\bf partition functions purely involve with 1-form $A$ or its field-strength (curvature) $\dd A$ are strictly cocycles but not coboundaries}.
These imply that those terms with only $A$ or $\dd A$ are the precisely nontrivial cocycles in the cohomology group for classification.

However, we find that {\bf partition functions  involve with 2-form $B$, 3-form $C$ or higher forms, although are cocycles but sometimes may also be coboundaries}
at certain quantized level $p$ value. For instance,
for those cocycles correspond to the partition functions of 
${p} \int C_1$,
$p_{} \frac{N_1 N_2 }{(2\pi) N_{12}}   \int A_1B_2$,
$p_{} \frac{N_1 N_2 }{(2\pi) N_{12}}  \int A_1 C_2$,
$p_{} \frac{N_1 N_2 }{(2\pi) N_{12}}  \int A_2 C_1$,
$p_{}\frac{N_1 N_2 N_3}{(2\pi)^2 N_{123}}  \int A_1 A_2 B_3$,
$p_{}\frac{N_1 N_2 N_3}{(2\pi)^2 N_{123}}  \int A_3 A_1 B_2$, etc (which involve with higher forms $B$, $C$),
we find that for $G=(Z_2)^n$ symmetry, $p=1$ are in the nontrivial class (namely not a coboundary),
$G=(Z_4)^n$ symmetry, $p=1, 3$ are in the nontrivial class (namely not a coboundary).
However, for $G=(Z_3)^n$ symmetry of all $p$ and $G=(Z_4)^n$ symmetry at $p=2$,
 are in the trivial class (namely a coboundary), etc.
This indicates an {\bf even-odd effect}, sometimes these cocycles are nontrivial, but sometimes are trivial as coboundary,
depending on the level $p$ is even/odd and the symmetry group $(Z_N)^n$ whether $N$ is even/odd.
{\bf Such an even/odd effect also bring complication into the validity of nontrivial cocycles, thus this is another reason that
we study only field theory involves with only 1-form $A$ or its field strength $\dd A$.
The cocycles composed from $A$ and $\dd A$ in Table \ref{table:cocyclefact} are always nontrivial and are not coboundaries.}

We finally point out that the concept of {\bf boundary term in field theory} (the surface or total derivative term) is connected to the
concept of  {\bf coboundary in the cohomology group}.
For example, $\int (\dd A_1) A_2 A_3$ are identified as the {\bf coboundary} of the linear combination of $\int A_1 A_2 (\dd A_3)$ and $\int A_1 (\dd A_2) A_3$.
Thus, by counting the number of distinct field theoretic actions (not identified by {\bf boundary term})
is precisely counting the number of distinct field theoretic actions (not identified by  {\bf coboundary}).
Such an observation matches the field theory classification to the group cohomology classification. 
Furthermore, we can map the field theory result to the K\"unneth formula listed 
via the correspondence:
\bea
\int A_1   &\sim& \cH^{1}(\Z_{N_1},\R/\Z)\\
\int A_1 \dd A_1 &\sim& \cH^{3}(\Z_{N_1},\R/\Z)\\
\int A_1 \dd A_1 \dd A_1 &\sim& \cH^{5}(\Z_{N_1},\R/\Z)\\
\text{Tor}_1^\Z \equiv \boxtimes_\Z  &\sim& \wedge \\
\otimes_\Z  &\sim& \wedge \dd \\
\int A_1 \wedge A_2  &\sim& \cH^{1}(\Z_{N_1}, \R/\Z) \boxtimes_\Z \cH^{1}(\Z_{N_2}, \R/\Z)\;\;\;\;\; \;\;\;\;\\
\int A_1 \wedge \dd A_2  &\sim& \cH^{1}(\Z_{N_1}, \R/\Z) \otimes_\Z \cH^{1}(\Z_{N_2}, \R/\Z) \;\;\;\;\; \;\;\;\;\\
 &\dots& \nonumber
\eea

To summarize, in this section, we show that, at lease for finite Abelian symmetry group $G=\prod^k_{i=1} Z_{N_i}$,
field theory can be systematically formulated, via the level-quantization developed earlier, 
we can count the number of classes of  SPTs. Explicit examples are organized
in Table \ref{table:cocyclefact} 
where we show that our field theory approach can exhaust all bosonic SPT classes (at least as complete as) in group cohomology:
\bea
&& \cH^2(G,\R/\Z) = \prod_{1 \leq i < j  \leq k}  \Z_{N_{ij}} \label{H2}\\
&& \cH^3(G,\R/\Z) = \prod_{1 \leq i < j < l \leq k}  \Z_{N_i} \times \Z_{N_{ij}} \times  \Z_{N_{ijl}} \label{H3}\\
&&\mathcal{H}^4(G,\R/\Z) =  \prod_{1 \leq i < j < l<m \leq k}
         (\Z_{N_{ij}})^2
         \times (\Z_{N_{ijl}})^2
         \times \Z_{N_{ijlm}}\;\;\; \;\; \;\;\;\;\label{eq:H4}\\
&& \;\; \;\;\;\;\;\; \;\;\;\;\;\; \;\;\;\; \dots   \nonumber
\eea
and we also had addressed the correspondence between field theory and K\"unneth formula.

\subsection{Summary}--
The recently-found SPTs, described by group cohomology, have SPT invariants in terms of
\emph{pure gauge actions} (whose
boundaries have \emph{pure gauge anomalies}\cite{{Wen:2013oza},{Wang:2013yta},{Wang:2014tia},{Kapustin:2014lwa},{Kapustin:2014zva}}).
We have derived the formal group cohomology results from an easily-accessible field theory set-up. 
{For beyond-group-cohomology SPT invariants, while ours of bulk-onsite-unitary symmetry
are \emph{mixed gauge-gravity actions},
those of other symmetries
(e.g. anti-unitary-symmetry time-reversal $\Z_2^T$) may be \emph{pure gravity actions}.\cite{K1459}}
{%
SPT invariants can also be obtained via cobordism
theory,\cite{K1467,K1459,{Freed:2014eja}} or via \emph{gauge-gravity actions} whose
boundaries realizing \emph{gauge-gravitational anomalies}.}
We have incorporated this idea into a field theoretic framework, which should be applicable for both bosonic and fermionic SPTs and for more exotic states awaiting future explorations.

\section{Induced Fractional Quantum Numbers and Degenerate Zero Modes: the anomalous edge physics of Symmetry-Protected Topological States} \label{sec:1403.5256}

\begin{figure}[h!]
\centering
\includegraphics[width=0.2\textwidth]{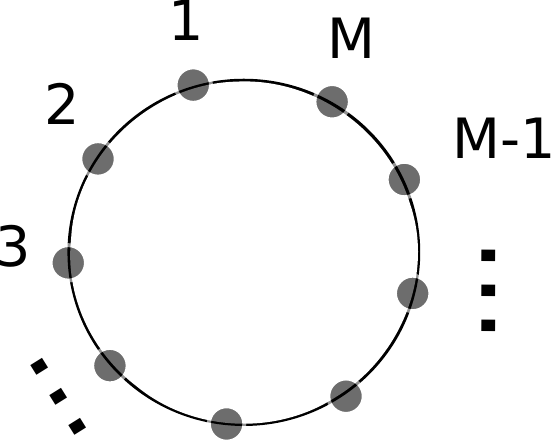}
\caption{ The illustration of 1D lattice model with $M$-sites on a compact ring.
}
\label{fig:2}
\end{figure}

We will now focus on 2+1D bulk/1+1D edge and go further to consider the edge modes of lattice Hamiltonian with $G={Z}_{N_1} \times {Z}_{N_2}  \times {Z}_{N_3}$ symmetry
on a compact ring with $M$ sites (Fig.\ref{fig:2}).
For any 
finite Abelian group $G$, we can derive the distinct 3-cocycles: 
\bea
\omega_{\text{I}}^{(i)}(A,B,C)    &=& 
\exp \Big( \frac{2 \pi \ti p_{i}  }{N_{i}^{2}} \; 
a_{i}(b_{i} +c_{i} -[b_{i}+c_{i}]) \Big)  \label{type1} \;\;\;\;\;\\
\omega_{\text{II}}^{(ij)}(A,B,C) &=&             
\exp \Big( 
\frac{2 \pi \ti p_{ij} }{N_{i}N_{j}}  \;
a_{i}(b_{j} +c_{j} - [b_{j}+c_{j}]) \Big)  \label{type2} \;\;\;\;\;\\
\omega_{\text{III}}^{(ijl)} (A,B,C) &=& \exp \Big( \frac{2 \pi \ti
p_{ijl}  }{{\gcd}(N_{i}, N_{j},N_{l})} \;
a_{i}b_{j}c_{l} \Big),            \label{type3}
\eea
so-called Type I, Type II, Type III 3-cocycles\cite{deWildPropitius:1996gt} respectively.
Since there are at most three finite Abelian subgroup indices shown in Eq.(\ref{type1}),(\ref{type2}),(\ref{type3}),
such a finite group with three Abelian discrete subgroups is the minimal example containing necessary and sufficient information to explore finite Abelian SPTs.
Such a symmetry-group $G$ may have nontrivial SPT class of Type I, Type II and Type III SPTs.
Apparently 
the Type I SPTs studied in our previous work happen,\cite{Santos:2013uda} which are the
class of $p_u \in \mathbb{Z}_{N_u}$ 
in $\cH^3({Z}_{N_1} \times {Z}_{N_2} \times {Z}_{N_3},\tU(1))$. 
Here and below we denote $u, v, w \in \{1,2,3 \}$ and $u,v,w$ are distinct. 
We will also introduce is 
the new class where ${Z}_{N_u}$ and ${Z}_{N_v}$ rotor models ``{\it talk to each other}.''
This will be the mixed Type II class $p_{uv} \in \mathbb{Z}_{N_{uv}}$, 
where symmetry transformation of ${Z}_{N_1}$ global symmetry will affect the ${Z}_{N_2}$ rotor models, while 
similarly ${Z}_{N_2}$ global symmetry will affect the ${Z}_{N_1}$ rotor models. 
There is a new class where three ${Z}_{N_1}$, ${Z}_{N_2}$, ${Z}_{N_3}$ rotor models {\it directly talk to each other}.
This will be the exotic Type III class 
$p_{123} \in \mathbb{Z}_{ N_{123}}$,
where the symmetry transformation of ${Z}_{N_u}$ global symmetry will affect the mixed ${Z}_{N_v},{Z}_{N_w}$ rotor models {\it in 
a mutual way}. 

\cblue{
To verify that our model construction corresponding to the Type I, Type II, Type III 3-cocycle in Eq.(\ref{type1}),(\ref{type2}),(\ref{type3}),
we will implement a technique called ``Matrix Product Operators'' in Sec.\ref{sec:MPO-Type I}. 
We would like to realize a discrete lattice model in Sec.\ref{sec: lattice Type I II}  and a continuum field theory in Sec.\ref{sec:Type II field}, 
to capture the essence of these classes of SPTs.
}



\subsection{Matrix Product Operators and Cocycles \label{sec:MPO-Type I}}

\cblue{
There are various advantages to put a quantum system on a discretize lattice, 
better than viewing it as a continuum field theory. For example, one advantage is that
 the symmetry transformation can be regularized so to understand its property such as onsite or non-onsite.
Another advantage is that we can simulate our model by considering a discretized finite system with a finite dimensional Hilbert space.
For our purpose, to regularize a quantum system on a discrete lattice,
we will firstly use the matrix product operators (MPO) formalism (see Ref.\cite{{Chen:2012hc}} and Reference therein) to 
formulate our symmetry transformations 
corresponding to non-trivial 3-cocycles in the third cohomology group in $\cH^3({Z}_{N_1} \times {Z}_{N_2},\tU(1))=\mathbb{Z}_{N_1} \times \mathbb{Z}_{N_2} \times \mathbb{Z}_{N_{12}}$.} 

First we formulate the unitary operator $S$ 
as the 
MPO: 
\be \label{eq:MPO_general}
S =\sum_{ \{j,j'\}} \tr[T^{j_1 j_1'}_{\alpha_1 \alpha_2}  T^{j_2 j_2'}_{\alpha_2 \alpha_3} \dots T^{j_M j_M'}_{\alpha_M \alpha_1}] |   j'_1, \dots, j'_M\rangle \langle j_1, \dots, j_M |.
\ee
with the its coefficient taking the trace (tr) of a series of onsite tensor $T(g)$ on a lattice, and input a state
$| j_1, \dots, j_M \rangle$ and output another state $|   j'_1, \dots, j'_M\rangle$.
\cblue{
$T=T(g)$ is a tensor with multi-indices and with dependency on a group element $g \in G$ for a symmetry group.
This is the operator formalism of matrix product states (MPS).
Here {\it physical indices} $j_1,j_2, \dots, j_M$ and $j'_1,j'_2, \dots, j'_M$ are labeled by input/output physical eigenvalues (here ${Z}_N$ rotor angle), the subindices $1,2,\dots, M$ are the physical site indices.
%
There are also {\it virtual indices} $\alpha_1, \alpha_2, \dots, \alpha_M$ which are traced in the end.
Summing over all the operation from $\{j,j'\}$ indices, we shall reproduce the symmetry transformation operator $S$. 
}%
\cblue{
What MPO really helps us 
is that}\\
\colorbox{lgray}{\parbox[t]{\linegoal}{
{\emph{by contracting MPO tensors $T(g)$ of $G$-symmetry transformation ${S}$ (here $g\in G$) 
in different sequence on the 
effective 1D lattice of SPT edge modes, 
it can reveal the {\bf nontrivial projective phase} corresponds to {\bf the nontrivial 3-cocycles
of the cohomology group}}.}}}\\

To find out the projective phase 
$e^{i \theta(g_a,g_b,g_c)}$, 
below we use the facts of tensors $T(g_a)$, $T(g_b)$, $T(g_c)$ acting on 
the same site 
with group elements ${g_a,g_b,g_c}$. 
We know a generic projective relation:
\be \label{eq:projective}
T(g_a\cdot g_b)=P_{g_a,g_b}^\dagger T(g_a) T(g_b)P_{g_a,g_b}.
\ee
Here $P_{g_a,g_b}$ is the projection operator.
We contract three 
tensors in two different orders,
\be \label{eq:projective-3cocycle}
{(P_{g_a,g_b} }{ \otimes I_3) P_{g_a g_b, g_c}} \simeq e^{\ti \theta(g_a,g_b,g_c)} {( I_1 \otimes P_{g_b,g_c} ) P_{g_a,g_bg_c}}.
\ee
The left-hand-side contracts the 
$a,b$ first then with the 
$c$, while the right-hand-side contracts the 
$b,c$ first then with the 
$a$.
Here $\simeq$ means the equivalence is up to a projection out of un-parallel states. 
We can derive $P_{g_a,g_b}$ by observing 
that $P_{g_a,g_b}$ inputs one state and outputs two states. 

For Type I SPT class, this MPO formalism has been done quite carefully in Ref.\cite{Chen:2012hc},\cite{Santos:2013uda}. 
Here we generalize it to other SPTs, below we input a group element with $g=(k_1,k_2,k_3)$ and $k_1 \in Z_{N_1}, k_2 \in Z_{N_2}, k_3 \in Z_{N_3}$. 
Without losing generality, we focus on the symmetry
Type I index $p_1 \in \Z_{N_1}$, Type II index $p_{12} \in \Z_{N_{12}}$, Type III index $p_{123} \in  \Z_{N_{123}}$.
By index relabeling, we can fulfill all SPT symmetries 
within the classification. 


We propose our $T(g)$ tensor for Type I, \cite{Chen:2012hc,{Santos:2013uda}} II symmetry with $p_{1} \in \Z_{N_{1}}$, $p_{12} \in \Z_{N_{12}}$ as
\bea \label{eq:Type II T 12}
&&(T^{\phi^{(1)}_{in}, \phi^{(1)}_{out},\phi^{(2)}_{in}, \phi^{(2)}_{out}})^{(p_1,p_{12})}_{\varphi_\alpha^{(1)},\varphi_\beta^{(1)},\varphi_\alpha^{(2)},\varphi_\beta^{(2)},N_1}(\frac{2\pi k_1}{N_1} )=\delta(\phi^{(1)}_{out}-\phi^{(1)}_{in}-\frac{2\pi k_1 }{N_1})\delta(\phi^{(2)}_{out}-\phi^{(2)}_{in}) \nonumber\\
&&
\cdot \int \dd\varphi_\alpha^{(1)} d\varphi_\beta^{(1)} | \varphi_\beta^{(1)} \rangle \langle \varphi_\alpha^{(1)} | \delta(\varphi_\beta^{(1)}-\phi_{in}^{(1)}) e^{\ti p_1 k_1(\varphi^{(1)}_\alpha-\phi^{(1)}_{in})_r
/N_1} \nonumber\\
&& \;\;\;\;\;
\cdot \int \dd\tilde{\varphi}_\alpha^{(2)} d\tilde{\varphi}_\beta^{(2)} | \tilde{\varphi}_\beta^{(2)} \rangle \langle \tilde{\varphi}_\alpha^{(2)} | \delta(\tilde{\varphi}_\beta^{(2)}-\tilde{\phi}_{in}^{(2)}) e^{\ti p_{12} k_{1}(\tilde{\varphi}^{(2)}_\alpha-\tilde{\phi}^{(2)}_{in})_r/N_1}.  
\eea
We propose the Type III $T(g)$ tensor with $p_{123} \in  \Z_{N_{123}}$ as
{\fontsize{11pt}{1em}\selectfont
{
\bea  \label{eq:Type III T}
&&(T^{\phi^{(1)}_{in}, \phi^{(1)}_{out},\phi^{(2)}_{in}, \phi^{(2)}_{out},\phi^{(3)}_{in}, \phi^{(3)}_{out}})^{(p_{123})}_{\varphi_\alpha^{(1)},\varphi_\beta^{(1)},\varphi_\alpha^{(2)},\varphi_\beta^{(2)},
\varphi_\alpha^{(3)},\varphi_\beta^{(3)},N_1,N_2,N_3}(\frac{2\pi k_1}{N_1},\frac{2\pi k_2}{N_2},\frac{2\pi k_3}{N_3}) 
\nonumber\\
&&=\prod_{ \underset{\{1,2,3\}}{u,v,w\in}} \int \dd\varphi_\alpha^{(u)}  | \phi_{in}^{(u)} \rangle \langle \varphi_\alpha^{(u)} |  
\exp[{\ti \,p_{123}  \epsilon^{uvw} k_u\frac{(\varphi^{(v)}_\alpha \phi^{(w)}_{in} )_r}{N_u}  {\frac{N_1N_2N_3}{2\pi N_{123}}}}] 
 \cdot |   \phi^{(u)}_{out} \rangle \langle \phi^{(u)}_{in}|. \;\;\;\;\;\;\;\; \label{eq:Type III T_2}
\eea
}}
Here we consider a lattice with both $\phi^{(u)}$, $\varphi^{(u)}$ as  $Z_{N_u}$ rotor angles.
The tilde notation $\tilde{\phi}^{(u)}$, $\tilde{\varphi}^{(u)}$, for example on $\tilde{\phi}^{(2)}$, means that the variables are in units of $\frac{2\pi}{N_{12}}$, but not in $\frac{2\pi}{N_2}$ unit
(The reason will become explicit later when we regularize the Hamiltonian on a lattice in Sec.\ref{sec: lattice Type I II}).

Take Eq.(\ref{eq:Type II T 12}), by computing the projection operator $P_{g_a,g_b}$ via Eq.(\ref{eq:projective}), we derive the projective phase from Eq.(\ref{eq:projective-3cocycle}):
\be \label{eq:proj Type I}
e^{\ti \theta(g_a,g_b,g_c)} =e^{\ti p_1\frac{2\pi}{N} m_c\frac{m_a+m_b-[m_a+m_b]_N}{N}} =
\omega_{\text{I}}^{(i)}(m_c,m_a,m_b) 
\ee
which the complex projective phase 
indeed induces the Type I 3-cocycle $\omega_{\text{I}}^{(i)}(m_c,m_a,m_b)$ of Eq.(\ref{type1}) 
in the third cohomology group $\cH^3({Z}_N,\tU(1))=\mathbb{Z}_N$. 
(Up to the index redefinition $p_{1} \to -p_{1}$.)
We further derive the projective phase as Type II 3-cocycle of 
Eq.(\ref{type2}),
\bea 
&&e^{\ti \theta(g_a,g_b, g_c)}=e^{\ti p_{12} ({ \frac{2\pi {m^{(1)}_c} }{N_1}  })\big( (m^{(2)}_a+{m^{(2)}_b}) - [m^{(2)}_a+{m^{(2)}_b}]_{N_2} \big) /N_2} 
=\omega_{\text{II}}^{(ij)}(m_3,m_1,m_2) \label{eq:proj Type II}
\eea
up to the index redefinition $p_{12} \to -p_{12}$. 
Here $[m_a+m_b]_N$ with subindex $N$ means taking the value module $N$.

Take Eq.(\ref{eq:Type III T}), we can also derive the projective phase
$e^{\ti \theta(g_a, g_b, g_c)}$ 
of Type III $T(g)$ tensor as 
\bea \label{}
&&e^{\ti \theta(g_a,g_b, g_c)}=e^{ \ti 2\pi p_{123} \epsilon^{u  v w}\big( \frac{{m^{(u)}_c} }{N_{u}} \frac{m^{(v)}_a}{N_v} \frac{m^{(w)}_b}{N_w} \big)\frac{N_1N_2N_3}{N_{123} }} 
\simeq \omega_{\text{III}}^{(uvw)}(m_c,m_a,m_b).
\eea
Adjust $p_{123}$ index (i.e. setting only the $p_{123}$ index in ${m^{(1)}_c}{m^{(2)}_a}{m^{(3)}_b}$ to be nonzero, while others $p_{213}=p_{312}=0$), 
and compute Eq.(\ref{eq:projective-3cocycle}) with only $p_{123}$ index,
we can recover the projective phase reveals Type III 3-cocycle in Eq.(\ref{type3}).

By Eq.(\ref{eq:MPO_general}), we verify that $T(g)$ of Type I, II in Eq.(\ref{eq:Type II T 12}) renders the symmetry transformation operator 
${S}^{(p_1,p_{12})}_{N_1}$: 
\bea  \label{eq:MPOType II S 12} 
{S}^{(p_1,p_{12})}_{N_1}=&&\prod_{j=1}^M  e^{\ti 2\pi L^{(1)}_j/N_1} \cdot \exp[ \ti \frac{p_1}{N_1}  ( \phi^{(1)}_{j+1}-\phi^{(1)}_{j} )_r] 
\cdot \exp[ \ti \frac{p_{12}}{N_1}  ( \tilde{\phi}^{(2)}_{j+1}-\tilde{\phi}^{(2)}_{j} )_r].  
\eea
here $j$ are the site indices, from $1$ to $M$ shown in Fig.\ref{fig:2}.


By Eq.(\ref{eq:MPO_general}), we verify that $T(g)$ of Type III in Eq.(\ref{eq:Type III T}) renders the symmetry transformation operator 
${S}^{(p_{123})}_{N_1,N_2,N_3}$:
\bea  \label{eq:Type III_S}
S^{(p_{123})}_{N_1,N_2,N_3}
={\prod_{j=1}^M} ( \prod^M_{\substack{  {u,v,w \in \{1,2,3\}}   } }  e^{\ti 2\pi L^{(u)}_j/N_u} \cdot W^{\text{III}}_{j,j+1} ). 
\eea
with
\be
W^{\text{III}}_{j,j+1} \equiv \prod_{u,v,w \in \{1,2,3\}} e^{\Big( { \ti {\frac{N_1N_2N_3}{2\pi N_{123}}} \epsilon^{u v w} \frac{p_{123}}{N_{u}}  \big(   \phi^{(v)}_{j+1} \phi^{(w)}_{j} \big) } \Big)}. 
\ee
For both Eq.(\ref{eq:MPOType II S 12}) and Eq.(\ref{eq:Type III_S}),
there is an onsite piece $\langle \phi_j^{(u)}  | e^{i 2\pi L^{(u)}_j/N_u} | \phi_j^{(u)} \rangle$ 
and 
also extra non-onsite symmetry transformation parts: namely, 
$\exp[ \ti \frac{p_1}{N_1}  ( \phi^{(1)}_{j+1}-\phi^{(1)}_{j} )_r]$,
$\exp[ \ti \frac{p_{12}}{N_1}  ( \tilde{\phi}^{(2)}_{j+1}-\tilde{\phi}^{(2)}_{j} )_r]$, and $W^{\text{III}}_{j,j+1}$.
We introduce an angular momentum operator $L^{(u)}_j$ conjugate to $\phi^{(u)}_j$,
such that the $e^{i 2\pi L^{(u)}_j/N_u}$ shifts the rotor angle by $\frac{2\pi}{N_{u}}$ unit, from $|\phi^{(u)}_{j} \rangle$ to $|  \phi^{(u)}_{j} +\frac{2\pi}{N_{u}} \rangle$. 
The subindex $r$ means that we 
further {\it regularize the variable to a discrete compact rotor angle}. 

Meanwhile $p_{1}=p_{1} \text{ mod } N_{1}$, $p_{12}=p_{12} \text{ mod } N_{12}$ and $p_{123}=p_{123} \text{ mod } N_{123}$, 
these demonstrate that our MPO construction fulfills all classes. 
So far we have achieved the SPT symmetry transformation operators Eq.(\ref{eq:MPOType II S 12}),(\ref{eq:Type III_S}) via MPO.
Other technical 
derivations on MPO formalism are 
preserved in Supplemental Materials. 

\subsection{Lattice model \label{sec: lattice Type I II}}

To construct a lattice model, we require the minimal ingredients: (i) $Z_{N_u}$ operators (with $Z_{N_u}$ variables).
(ii) Hilbert space (the state-space where $Z_{N_u}$ operators act on) consists with $Z_{N_u}$ variables-state.
Again we denote $u=1,2,3$ for $Z_{N_1}$,$Z_{N_2}$,$Z_{N_3}$ symmetry.
We can naturally choose the $Z_{N_u}$ variable $\omega_u \equiv e^{\ti\,2\pi/N_u}$, such that 
$\omega_u^{N_u}=1$. 
Here and below we will redefine the quantum state and operators from the MPO basis in Sec.\ref{sec:MPO-Type I} to a lattice basis via:
\be
{\phi}^{(u)}_j \to \phi_{u,j},\;\;\; 
{L}^{(u)}_l \to {L}^{}_{u,l}. \label{eq:MPO-to-lattice}
\ee
The natural physical states on a single site are 
the $Z_{N_u}$ rotor angle state
$| \phi_u=0 \rangle, | \phi_u=2\pi/N_u \rangle, \dots, | \phi_u=2\pi(N_u-1)/N_u \rangle$.

One can find a dual state of rotor angle state $|\phi_u \rangle$, the angular momentum $|L_u \rangle$, such that the basis from $|\phi_u \rangle$ can transform to 
$|L_u \rangle$
via the Fourier transformation, $| \phi_u \rangle=\sum^{N_u-1}_{L_u=0} \frac{1}{\sqrt{N_u}} e^{\ti L_u \phi_u} | L_u \rangle$. 
One can find two proper operators $\sigma^{(u)}_{},\tau^{(u)}_{}$ which make $| \phi_u \rangle$ and $| L_u \rangle$ their own eigenstates respectively. 
With a site index $j$ ($j = 1,...,M$),
we can project $\sigma^{(u)}_{j},\tau^{(u)}_{j}$ operators into the rotor angle 
$| \phi_{u,j} \rangle$ basis, so we can derive $\sigma^{(u)}_{j},\tau^{(u)}_{j}$ operators as $N_u \times N_u$ matrices.
Their forms are :
\bea \label{eq: sigma operator}
&&\sigma^{(u)}_{j}   =  {\begin{pmatrix} 
1 & 0 & 0 & 0 \\
0 & \omega_u & 0 & 0\\  
0 & 0 & \ddots  & 0\\  
0 & 0 & 0 & \omega_u^{N_u-1}
\end{pmatrix}}_j = \langle \phi_{u,j} | e^{\ti \hat{\phi}^{(u)}_{j}} | \phi_{u,j} \rangle, \nonumber\\
&&\tau^{(u)}_{j}   =  {\begin{pmatrix} 
0 & 0 & 0 & \dots &0& 1  \\
1 & 0 & 0 & \dots &0& 0 \\
0 & 1 & 0 & \dots &0& 0 \\
0 & 0 & 1 & \dots &0& 0 \\
\vdots &0 & 0 & \dots &1 & 0 
\end{pmatrix}}_j =\langle \phi_{u,j} | e^{\ti2 \pi \hat{L}^{(u)}_j/N} | \phi_{u,j} \rangle .  
\eea
Operators and variables satisfy the analogue property mentioned in Ref.\cite{Santos:2013uda}, 
such as $(\tau^{(u)})^{N_u}_{j} = (\sigma^{(u)})^{N_u}_{j} =\mathbb{I}$, 
$\tau^{(u)\dagger}_{j}\,\sigma^{(u)}_{j}\,\tau^{(u)}_{j} = \omega_u\,\sigma^{(u)}_{j}$. 
It also enforces the canonical conjugation relation on $\hat{\phi}^{(u)}$ and $\hat{L}^{(u)}$ operators, i.e.
$[\hat{\phi}_j^{(u)},\hat{L}_l^{(v)} ]= \ti \,\delta_{(j,l)}\delta_{(u,v)}$ with the symmetry group index $u,v$
and the site indices $j,l$.
Here $| \phi \rangle$ and $| L \rangle$ are eigenstates of $\hat{\phi}$ and $\hat{L}$ operators respectively.

The linear combination of all $| \phi_1 \rangle$ $| \phi_2 \rangle$ $| \phi_3 \rangle$ states form a complete  $N_1 \times N_2 \times N_3$-dimensional Hilbert space on a single site.

\subsubsection{symmetry transformations \label{Sec: type I II sym transf}}

\noindent
{\bf Type I, II $Z_{N_1} \times Z_{N_2}$ symmetry transformations}\\

Firstly we 
warm up with a generic ${Z}_N$ lattice model realizing the SPT edge modes on a 1D ring with $M$ sites (Fig.\ref{fig:2}).
The SPT edge modes have a special non-onsite symmetry transformation, which means that
its symmetry transformation cannot be written as a tensor product form on each site, thus
$
U(g)_{\text{non-onsite}} \neq \otimes _i U_i(g).
$ 
In general, the symmetry transformation contain a onsite part and another non-onsite part.
The trivial class of SPT (trivial bulk insulator) with unprotected gapped edge modes can be achieved by a simple 
Hamiltonian
as
$-\lambda \, \sum^{M}_{j=1}  ( \tau_{j} + \tau^{\dagger}_{j} )$. (Notice that for the simplest $Z_2$ symmetry, the $\tau_j$ operator  reduces to a spin operator $(\sigma_z)_j$.)
The simple way to find an onsite operator which this Hamiltonian respects and which acts at each site is the $\prod^{M}_{j=1} \tau_j$, a series of $\tau_j$.
On the other hand, to capture the \emph{non-onsite} symmetry transformation, we can use a \emph{domain wall} variable 
in Ref.\cite{Santos:2013uda}, where the symmetry transformation
contains information stored non-locally between different sites (here we will use the minimum construction: 
symmetry stored non-locally between {\it two nearest neighbored sites}). 
\cblue{ 
We propose this {non-onsite} symmetry transformation $U_{j,j+1}$ with a domain wall $(N_{\text{dw}})_{j,j+1}$ operator acting non-locally on site $j$ and $j+1$ as},
\be \label{eq:domain-wall}
U_{j,j+1} 
\equiv \exp \big( \ti \frac{p}{N}  \frac{2\pi}{N}(\delta N_{\text{dw}})_{j,j+1} \big) \equiv\exp[ \ti \frac{p}{N}  (\phi_{1,j+1} - \phi_{1,j})_r],   
\ee
\cblue{The justification of non-onsite symmetry operator Eq.(\ref{eq:domain-wall}) realizing SPT edge symmetry is based on
MPO formalism already done in Sec.\ref{sec:MPO-Type I}.  
}
The domain wall operator $(\delta N_{\text{dw}})_{j,j+1}$ counts the number of units of ${Z}_N$ angle between sites $j$ and $j+1$, so
indeed $(2\pi/N)(\delta N_{\text{dw}})_{j,j+1}$$=(\phi_{1,j+1} - \phi_{1,j})_r$. The subindex $r$ means that we need to further {\it regularize the variable to a discrete ${Z}_N$ angle}.
Here we insert a $p$ index, which is just an available free index with $p=p \text{ mod } N$. 
From Sec.\ref{sec:MPO-Type I}, $p$ is indeed the classification index for the $p$-th of $\mathbb{Z}_N$ class in the third cohomology group $\cH^3(Z_N,\tU(1))=\mathbb{Z}_N$.

Now the question is 
how should we fully regularize this $U_{j,j+1}$ operator into terms of $Z_N$ operators $\sigma^{\dagger}_{j}$ and $\sigma_{j+1}$.
We see the fact that the $N$-th power of $U_{j,j+1}$ renders a constraint
\bea  \label{eq:constraint Type I}
U_{j,j+1}^N 
&=&(\exp[ \ti  \phi_{1,j}]^\dagger \exp[ \ti  \phi_{1,j+1}] )^p=(\sigma^{\dagger}_{j}\sigma_{j+1})^p. \;\;\;
\eea
(Since $\exp[\, \ti\,  \phi_{1,j}]_{ab}= \langle \phi_a | e^{\, \ti \, \phi_j} | \phi_b \rangle = \sigma_{ab,j}$.)
More explicitly, we can write it as a polynomial ansatz 
$ 
U_{j,j+1}=
\exp[\frac{\ti}{N} \sum^{N-1}_{a=0}\, q_{a}\,(\sigma^{\dagger}_{j}\sigma_{j+1})^{a}]
$.
The non-onsite symmetry operator $U_{j,j+1}$ reduces to a problem of solving polynomial coefficients $q_a$ by the constraint Eq.(\ref{eq:constraint Type I}).
Indeed we can solve the constraint explicitly, thus the non-onsite symmetry transformation operator 
acting on a $M$-site ring from $j=1, \dots, M$ is derived: 
\bea \label{eq:Type I symmetry explicit}
U_{j,j+1}=e^
{
-\ti \frac{2\pi}{N^2}p\, 
\Big\{
\left(
\frac{N-1}{2}\,
\right)
\mathbb{I}
+
\sum^{N-1}_{a=1}\,
\frac
{
(\sigma^{\dagger}_{j}\sigma_{j+1})^a
}
{
(\omega^a - 1)
}
\Big\}}.
\eea
%


For a lattice SPTs model with $G=Z_{N_1} \times Z_{N_2}$, 
we can convert MPO's symmetry transformation Eq.(\ref{eq:MPOType II S 12})
to a lattice variable via Eq.(\ref{eq:Type I symmetry explicit}).
We obtain the $Z_{N_u}$ symmetry transformation (here and below $u,v \in \{ 1,2\}, u\neq v$):
\bea \label{S1symp12}
&&\bullet \;\; S^{(p_u,p_{uv})}_{N_u}
\equiv \prod^{M}_{j=1}  e^{\ti 2\pi L^{}_{u,j}/N_u} \cdot \exp[ \ti \frac{p_u}{N_u}  ( \phi_{u,j+1}-\phi_{u,j} )_r]  
 \cdot \exp[ \ti \frac{p_{uv}}{N_u}  ( \tilde{\phi}_{v,j+2}-\tilde{\phi}_{v,j} )_r]   \label{eq:symmetry form lattice Type II S1} \nonumber \\
&&= \prod^{M}_{j=1}\tau^{(u)}_{j}\,
\cdot U^{(N_u,p_u)}_{j,j+1}
\cdot U^{(N_u,p_{uv})}_{j,j+2} 
=\prod^{M}_{j=1}\tau^{(u)}_{j}\, 
\cdot e^{(-\ti \frac{2\pi}{N_u^2}p_u\, 
\Big\{ \left( \frac{N_u-1}{2}\,\right) \mathbb{I} + \sum^{N_u-1}_{a=1}\,\frac {(\sigma^{(u)\dagger}_{j}\sigma^{(u)}_{j+1})^a} {((\omega_u)^a - 1)}
\Big\})}  \nonumber \\ 
&&\cdot e^{ ( -\ti\frac{2\pi }{N_{uv} N_{u}} p_{uv} \, \Big\{ (\frac{N_{uv}-1}{2}) \mathbb{I}+ \sum^{N_{uv}-1}_{a=1}\, \frac { \left( \tilde{\sigma}^{(v) \dagger}_{j}\tilde{\sigma}^{(v)}_{j+2} \right)^a}
{\omega_{uv}^a -1}\Big\} ) }.
\eea
%
%
The operator is unitary, i.e. $S^{(p_u,p_{uv})}_{N_u} S^{(p_u,p_{uv})\dagger}_{N_u} =1$. Here $\sigma_{M+j}\equiv\sigma_{j}$.
The intervals of rotor angles are
\bea 
&&\phi_{1,j} \in \{ n \frac{2\pi}{N_1} | n \in \mathbb{Z}\},\;\;\; \phi_{2,j} \in \{ n \frac{2\pi}{N_2} | n \in \mathbb{Z}\}, 
\tilde{\phi}_{1,j}, \tilde{\phi}_{2,j} \in \{ n \frac{2\pi}{ N_{12} } | n \in \mathbb{Z}\}. \label{eq:field regularize}
\eea
where $\phi_{1,j}$ is $Z_{N_1}$ angle, $\phi_{2,j}$ is $Z_{N_2}$ angle, $\tilde{\phi}_{1,j}$ and $\tilde{\phi}_{2,j}$ are $Z_{N_{12}}$ angles (recall ${\gcd{(N_1,N_2)}} \equiv N_{12}$).
There are 
some  remarks on our above formalism:\\
(i) First,  the $Z_{N_1},Z_{N_2}$ symmetry transformation Eq.(\ref{S1symp12}) 
including both the Type I indices $p_1$, $p_2$ and also Type II indices $p_{12}$ and $p_{21}$. Though $p_1$, $p_2$ are distinct indices, but $p_{12}$ and $p_{21}$ indices are the same index, $p_{12}+p_{21} \to p_{12}$. The invariance $p_{12}+p_{21}$ describes the same SPT symmetry class.\\
(ii) The second remark, 
for Type I non-onsite symmetry transformation (with $p_1$ and $p_2$) are chosen to act on the nearest-neighbor sites (NN: site-$j$ and site-$j+1$); but 
the Type II non-onsite symmetry transformation (with $p_{12}$ and $p_{21}$) are  chosen to be the next nearest-neighbor sites (NNN: site-$j$ and site-$j+2$). The reason is that
we have to avoid the nontrivial Type I and Type II symmetry transformations cancel or interfere with each other. Though in the Sec.\ref{sec:Type II field}, 
we will reveal that the low energy field theory description of non-onsite symmetry transformations for both NN and NNN having the same form in the continuum limit.
In the absence of Type I index, we can have Type II non-onsite symmetry transformation act on nearest-neighbor sites.\\ 
(iii) The third remark, the domain wall picture mentioned in Eq.(\ref{eq:domain-wall}) 
for Type II $p_{12}$ class still hold. But here the lattice regularization is different for terms with $p_{12},p_{21}$ indices.
In order to have distinct $Z_{\gcd{({N}_1,{N}_2)}}$ class with the identification $p_{12} =p_{12}$ mod $N_{12}$.
We will expect that, performing the ${N_u}$ times $Z_{N_u}$ symmetry transformation on the Type II $p_{uv}$ 
non-onsite piece, 
renders a constraint 
\be \label{eq: Type II constraint}
(U^{(N_u,p_{uv})}_{j,j+2})^{N_u}=(\tilde{\sigma}^{(v)\dagger}_{j}\tilde{\sigma}^{(v)}_{j+2})^{p_{uv}},  
\ee

To impose the identification $p_{12} =p_{12}$ mod ${N_{12}}$ and $p_{21} =p_{21}$ mod ${N_{12}}$
so that we have distinct $Z_{\gcd{({N}_1,{N}_2)}}$ classes for the Type II symmetry class
(which leads to impose the constraint $(\tilde{\sigma}^{(1)}_{j})^{N_{12}}=(\tilde{\sigma}^{(2)}_{j})^{N_{12}} = \mathbb{I}$),
we can regularize the $\tilde{\sigma}^{(1)}_{j}$, $\tilde{\sigma}^{(2)}_{j}$  operators in terms of $Z_{\gcd{(N_1,N_2)}}$ variables. 
With ${\omega}_{12} \equiv {\omega}_{21}  \equiv e^{i \frac{2 \pi}{N_{12}} }$, we have  
$\omega_{12}^{N_{12}} =1$.
The  $\tilde{\sigma}^{(u)}_{j}$ matrix has $N_u \times N_u$ components, for $u=1,2$. It is block diagonalizable
with $\frac{N_u}{N_{12}} $ subblocks, and each subblock with ${N_{12}} \times {N_{12}}$ components.
Our regularization provides the nice property:
$\tau^{(1)\dagger}_{j}\,\tilde{\sigma}^{(1)}_{j}\,\tau^{(1)}_{j} = \omega_{12}\,\sigma^{(1)}_{j}$
and $\tau^{(2)\dagger}_{j}\,\tilde{\sigma}^{(2)}_{j}\,\tau^{(2)}_{j} = \omega_{12}\,\sigma^{(2)}_{j}$.
Use the above procedure to regularize Eq.(\ref{eq:MPOType II S 12}) 
on a discretized lattice and solve the constraint Eq.(\ref{eq: Type II constraint}), we obtain
an explicit form of lattice-regularized symmetry transformations Eq.(\ref{S1symp12}). 
For more details on our lattice regularization, see Supplemental Materials. 

\noindent
{\bf Type III symmetry transformations}\\

To construct a Type III SPT with a Type III 3-cocycle Eq.(\ref{type3}),
the key observation is that the 3-cocycle inputs, for example, $a_1 \in Z_{N_1}$, $b_2 \in Z_{N_2}$, $c_3 \in Z_{N_3}$ and outputs a U(1) phase.
This implies that the ${Z}_{N_1}$ symmetry transformation will affect the mixed ${Z}_{N_2},{Z}_{N_3}$ rotor models, etc.
This observation guides us to write down the tensor $T(g)$ in Eq.(\ref{eq:Type III T}) 
and we obtain the symmetry transformation $S^{(p)}_N=S^{(p_{123})}_{N_1,N_2,N_3}$ as Eq.(\ref{eq:Type III_S}):
\bea  \label{eq:Type III_S123}
\bullet &&S^{(p_{123})}_{N_1,N_2,N_3}
={\prod_{j=1}^M} ( \prod^M_{\substack{  {u,v,w \in \{1,2,3\}}   } }  \tau_j^{(u)} \cdot W^{\text{III}}_{j,j+1} ). 
\eea
There is an onsite piece $\tau_j \equiv \langle \phi_j | e^{i 2\pi L^{(u)}_j/N} | \phi_j \rangle$ and also an extra non-onsite symmetry transformation part $W^{\text{III}}_{j,j+1}$.
This non-onsite symmetry transformation $W^{\text{III}}_{j,j+1}$, acting on the site $j$ and $j+1$, is defined by the following, and can be further regularized on the lattice:
\be \label{eq:Type III_W}
\bullet \;\;W^{\text{III}}_{j,j+1}
={ \prod_{u,v,w \in \{1,2,3\}}  \Big( \sigma_{j}^{(v)\dagger} \sigma_{j+1}^{(v)}  \Big)^{ \epsilon^{u v w}   p_{123} {  \frac{ {\log(\sigma_{j}^{(w)})} N_v N_w}{2\pi N_{123}}} }}.
\;\;\;\;\; 
\ee
here we separate $Z_{N_1}$,$Z_{N_2}$,$Z_{N_3}$ non-onsite symmetry transformation to $W^{\text{III}}_{j,j+1; N_1}$, $W^{\text{III}}_{j,j+1; N_2}$, $W^{\text{III}}_{j,j+1; N_3}$ respectively.
Eq.(\ref{eq:Type III_S123}),(\ref{eq:Type III_W}) are fully regularized in terms of $Z_N$ variables on a lattice, 
although they contain \emph{anomalous non-onsite symmetry} operators. 

\subsubsection{lattice Hamiltonians \label{sec:lattice Hamiltonians: type I II}}

We had mentioned the trivial class of SPT Hamiltonian (the class of $p=0$) 
for 1D gapped edge:
\be
H^{(0)}_{N}=-\lambda \, \sum^{M}_{j=1}  ( \tau_{j} + \tau^{\dagger}_{j} )
\ee
Apparently, the 
Hamiltonian is symmetry preserving respect to $S^{(0)}_{N} \equiv\prod^{M}_{j=1} \tau_j$, 
i.e. $S^{(0)}_{N} H^{(0)}_{N}$ $(S^{(0)}_{N})^{-1} =H^{(0)}_{N}$. 
In addition, this Hamiltonian has a symmetry-preserving gapped ground state.

To extend our lattice Hamiltonian construction to $p \neq 0$ class, intuitively we can view the nontrivial SPT Hamiltonians as close relatives of the trivial 
Hamiltonian (which preserves the onsite part of the symmetry transformation with $p=0$), which 
satisfies the symmetry-preserving constraint, i.e.
\be
S^{(p)}_{N} H^{(p)}_{N} (S^{(p)}_{N})^{-1} =H^{(p)}_{N},
\ee
More explicitly, to construct a SPT Hamiltonian of $Z_{N_1} \times Z_{N_2}\times Z_{N_3}$ symmetry
obeying translation 
and symmetry transformation 
invariant (here and below $u, v, w \in \{1,2,3 \}$ and $u,v,w$ are distinct): 
\bea \label{principle type II}
&&\bullet\;\; [H^{(p_u,p_{uv},p_{uvw})}_{N_1,N_2,N_3}  ,T]=0, \;\; 
\bullet\;\;  [H^{(p_u,p_{uv},p_{uvw})}_{N_1,N_2,N_3}  ,
S^{(p)}_{N} ]
=0
\eea
Here $T$ is a translation operator by one lattice site,
satisfying
$
T^{\dagger}\,X_{j}\,T = X_{j+1},
\quad
j = 1,...,M,
$
for any operator $X_{j}$ on the ring such that
$
X_{M+1}
\equiv
X_{1}
$.
Also $T$ satisfies
$
T^{M}
=
\mathbb{I}
$.
We can immediately derive the following SPT Hamiltonian satisfying the rules, 
\be \label{eq:Type II Hamiltonian lattice}
\bullet \;\;
H^{(p_u,p_{uv},p_{uvw})}_{N_1,N_2,N_3} 
\equiv
- \lambda
\!
\sum^{M}_{j=1} 
\!\!
\sum^{N-1}_{\ell = 0}
\!\!
\left(
S^{(p)}_{N} 
\right)^{\!\!-\ell}
\!\!\!
( \tau_{j} + \tau^{\dagger}_{j} )  
\!\!
\left(
S^{(p)}_{N} 
\right)^{\ell} +\dots,
\ee
where we define our notations: $S^{(p)}_{N} \equiv \prod_{ u, v, w \in \{1,2,3 \} }S^{(p_u,p_{uv},p_{uvw})}_{N_u} $  
and $\tau_{j} \equiv \tau_{j}^{(1)} \otimes \mathbb{I}_{N_2 \times N_2}  \otimes \mathbb{I}_{N_3 \times N_3} + \mathbb{I}_{N_1 \times N_1} \otimes \tau_{j}^{(2)}  \otimes \mathbb{I}_{N_3 \times N_3} 
+\mathbb{I}_{N_1 \times N_1}   \otimes \mathbb{I}_{N_2 \times N_2}  \otimes \tau_{j}^{(3)} $.
Here $\tau_{j}$ is a matrix of ${(N_1 \times N_2 \times N_3 )}$ ${\times (N_1 \times N_2 \times N_3)}$-components.
The tower series of sum over power of $(S^{(p)}_{N})$ over $(\tau_{j} + \tau^{\dagger}_{j} )$ will be shifted upon $S^{(p)}_{N} H^{(p)}_{N} (S^{(p)}_{N})^{-1}$, but the overall sum of this Hamiltonian 
is a symmetry-preserving invariant.

\subsection{Field Theory \label{sec:Type II field}}

From a full-refualrized lattice model 
in the previous section, 
we attempt to take the low energy 
limit to realize its corresponding field theory, 
by identifying the commutation relation
$[\hat{\phi}^{(u)}_j,\hat{L}^{(v)}_l ]= \ti \,\delta_{(j,l)} \delta_{(u,v)}$ (here $j, l$ are the site indices, $u, v \in \{ 1,2, 3\}$ 
are the $Z_{N_1},Z_{N_2},Z_{N_3}$ rotor model indices) in the continuum as 
\bea 
&&[\phi_u(x_1), \frac{1}{2\pi}\partial_x \phi_{v}'(x_2)]=  \, \ti \,  \delta(x_1-x_2) \delta_{(u,v)} \label{eq:commutation} 
\eea
which means the $Z_{N_1},Z_{N_2},Z_{N_3}$ lattice operators $\hat{\phi}^{(1)}_j, \hat{L}^{(1)}_l$, $\hat{\phi}^{(2)}_j, \hat{L}^{(2)}_l$, $\hat{\phi}^{(3)}_j, \hat{L}^{(3)}_l$ and field operators $\phi_1,\phi_1'$, $\phi_2,\phi_2'$, $\phi_3,\phi_3'$ are identified by 
\be
\hat{\phi}^{(u)}_j \to \phi_u (x_j),\;\;\; 
\hat{L}^{(u)}_l \to \frac{1}{2\pi}\partial_x \phi_{u}'(x_l). \label{eq:discrete-to-cont}
\ee
We view $\phi_u$ and $\phi_u'$ as the dual rotor angles as before, the relation follows as Sec.\ref{sec:Type II field}. 
We have no difficulty to formulate a K matrix multiplet chiral boson field theory 
(non-chiral `doubled' version of Ref.\cite{Floreanini:1987as}'s action) as 
\be \label{eq:Kmatrix_edge}
\text{{\bf S}}_{\text{SPT},\partial \mathcal{M}^2}= 
\frac{1}{4\pi} \int dt\; dx \; \big( K_{IJ} \partial_t \phi_{I} \partial_x \phi_{J} -V_{IJ}\partial_x \phi_{I}   \partial_x \phi_{J} \big) + \dots.
\ee
requiring a rank-6 symmetric $K$-matrix, 
\be
K_{\text{SPT}} =\bigl( {\begin{smallmatrix} 
0 &1 \\
1 & 0
\end{smallmatrix}}  \bigl) \oplus \bigl( {\begin{smallmatrix} 
0 &1 \\
1 & 0 
\end{smallmatrix}}  \bigl)  
\oplus \bigl( {\begin{smallmatrix} 
0 &1 \\
1 & 0 
\end{smallmatrix}}  \bigl).  
\ee
with a chiral boson multiplet $\phi_I(x)=(\phi_1 (x),$ $\phi_1 ' (x),$ $\phi_2 (x),$ $\phi_2' (x),$ $\phi_3 (x),$ $\phi_3' (x) )$.
The commutation relation Eq.(\ref{eq:commutation}) becomes: $ [\phi_I(x_1), K_{I'J} \partial_x \phi_{J}(x_2)]= {2\pi} \ti  \delta_{I I'} \delta(x_1-x_2)$.
The continuum limit of Eq.(\ref{eq:symmetry form lattice Type II S1}) 
becomes 
\be \label{eq:globalS_Type II_1}
\bullet \;\;  S^{(p_u,p_{uv})}_{N_u}
=
\exp[
\frac{\ti}{N_u}\,
(
\int^{L}_{0}\,dx\,\partial_{x}\phi_{u}'
+
p_u\,\int^{L}_{0}\,dx\,\partial_{x}\phi_{u} 
+
0\int^{L}_{0}\,dx\,\partial_{x}\phi_{v}'
+
p_{uv}\,\int^{L}_{0}\,dx\,\partial_{x} \tilde{\phi}_{v}
)
] \;\;\;
\ee
Notice that we carefully input a tilde on some $\tilde{\phi}_v$ fields. We stress the lattice regularization of $\tilde{\phi}_v$
is different from ${\phi}_v$, see Eq.(\ref{eq:field regularize}), which is analogous to $\tilde{\sigma}^{(1)}$, $\tilde{\sigma}^{(2)}$ and $\sigma^{(1)}$, $\sigma^{(2)}$ in Sec.\ref{Sec: type I II sym transf}. 
We should mention two remarks: 
First, there are higher order terms beyond $\text{{\bf S}}_{\text{SPT},\partial \mathcal{M}^2}$'s quadratic terms when taking continuum limit of lattice. 
At the low energy limit, it shall be reasonable to drop higher order terms. 
Second, in the nontrivial SPT class (some topological terms $p_i \neq 0$, $p_{ij} \neq 0$), the $\det(V)\neq 0$ and all eigenvalues are non-zeros, so the edge modes are gapless.
In the trivial insulating class (all topological terms $p = 0$), the $\det(V)= 0$, so the edge modes may be gapped (consistent with Sec.\ref{sec:lattice Hamiltonians: type I II}).
Use Eq.(\ref{eq:commutation}), we derive the 1D edge global symmetry transformation
$S^{(p_u,p_{uv})}_{N_u}$, 
for  example, $S^{(p_1,p_{12})}_{N_1}$ and $S^{(p_2,p_{21})}_{N_2}$, 
\bea
\label{eq:S1}
&&S^{(p_1,p_{12})}_{N_1}
 {\begin{pmatrix} 
\phi_1 (x)   \\
\phi_1 ' (x)  \\
\tilde{\phi}_2 (x)   \\
\tilde{\phi}_2' (x)
\end{pmatrix}} (S^{(p_1,p_{12})}_{N_1})^{-1}
=
 {\begin{pmatrix} 
\phi_1 (x)   \\
\phi_1 ' (x)  \\
\tilde{\phi}_2 (x)   \\
\tilde{\phi}_2' (x)
\end{pmatrix}} 
+\frac{2\pi}{N_1}{\begin{pmatrix} 
1   \\
p_1 \\
0\\
p_{12}  
\end{pmatrix}}.\\  
&&S^{(p_2,p_{21})}_{N_2}
 {\begin{pmatrix} 
\tilde{\phi}_1 (x)   \\
\tilde{\phi}_1 ' (x)  \\
\phi_2 (x)   \\
\phi_2' (x)
\end{pmatrix}} (S^{(p_2,p_{21})}_{N_2})^{-1}
=
 {\begin{pmatrix} 
\tilde{\phi}_1 (x)   \\
\tilde{\phi}_1 ' (x)  \\
\phi_2 (x)   \\
\phi_2' (x)
\end{pmatrix}} 
+\frac{2\pi}{N_2}{\begin{pmatrix} 
0   \\
p_{21} \\
1\\
p_{2}  
\end{pmatrix}}.  \nonumber
\eea
We can see how $p_{12}$, $p_{21}$ identify the same index
by doing a $M$ matrix with $M \in  \text{SL}(4, \mathbb{Z})$ transformation on the K matrix Chern-Simons theory,
which redefines the $\phi$ field, but still describe the same theory. That means: $K \to K'= M^T K M$ and $\phi \to \phi'=M^{-1} \phi$,
and so the symmetry charge vector $q \to q' =M^{-1} q$. By choosing
\bea
&&M=
\bigg( \begin{smallmatrix}
 1 & 0 & 0 & 0 \\
 {p_1} & 1 &  {p_{21}} & 0 \\
 0 & 0 & 1 & 0 \\
  {p_{12}} & 0 & {p_2} & 1
\end{smallmatrix} \bigg),
 \text{   then the basis is changed to  }  \nonumber\\
&&
K'=
\bigg( \begin{smallmatrix}
 2 {p_1} & 1 &  {p_{12}}+ {p_{21}} & 0 \\
 1 & 0 & 0 & 0 \\
  {p_{12}}+ {p_{21}} & 0 & 2 {p_2} & 1 \\
 0 & 0 & 1 & 0
 \end{smallmatrix} \bigg),
q_1'=
\bigg( \begin{smallmatrix}
1\\
0\\
0\\
0
\end{smallmatrix}
\bigg),\;
q_2'=
\bigg( \begin{smallmatrix}
0\\
0\\
1\\
0
\end{smallmatrix}
\bigg). \;\;\;\;\;\;\nonumber
\eea
The theory labeled by $K_{\text{SPT}},q_1,q_2$ is equivalent to the one  labeled by $K',q_1',q_2'$.
Thus we show that $p_{12}+p_{21} \to p_{12}$ identifies the same index.
There are other ways using the gauged or probed-field version of topological gauge theory (either on the edge or in the bulk) 
to identify the gauge theory's symmetry transformation,\cite{Ye:2013upa}
or the bulk braiding statistics 
to determine this Type II classification
$
p_{12}\; \text{mod}(\gcd(N_1,N_2))
$.

The nontrivial fact that when $p_{12}=N_{12}$ is a trivial class, the symmetry transformation in Eq.(\ref{eq:S1}) 
may not go back to 
the trivial symmetry under the condition $\int^{L}_{0}\,dx\,\partial_{x} \tilde{\phi}_{1}$ $=\int^{L}_{0}\,dx\,\partial_{x} \tilde{\phi}_{2}$ $=2\pi$, implying a soliton
can induce fractional charge (for details see Sec.\ref{sec: Type II frac}).

Our next goal is deriving Type III symmetry transformation Eq.(\ref{eq:Type III_S}). 
By taking the continuum limit of
\bea
&& \epsilon^{(u=1,2,3)(v)(w)}  \phi^{{j+1},(v)}_{in} \phi^{j,(w)}_{in} 
\to 
\big(\partial_x \phi^{(v)}_{in}(x) \phi^{(w)}_{in}(x)-\partial_x \phi^{(w)}_{in}(x) \phi^{(v)}_{in}(x) \big)
\eea
we can massage the continuum limit of Type III symmetry transformation Eq.(\ref{eq:Type III_S}) to ($\gcd({N_1,N_2,N_3}) \equiv N_{123}$)
\be \label{eq:globalS_Type III}
{\;S^{(p_{123})}_{N_1,N_2,N_3}=\prod_{ \underset{ \{1,2,3\}}{u,v,w \in}} \exp\big[ \frac{\ti}{N_u}\,
(
\int^{L}_{0}\,dx\,\partial_{x}\phi_{u}' )\big] \cdot
 \exp\big[ \ti {\frac{N_1N_2N_3}{2\pi N_{123}}} \frac{p_{123}}{N_{u}} \int^{L}_{0}\,dx\, \epsilon^{u v w} \partial_x \phi^{}_{v}(x) \phi^{}_{w}(x) \big].\;
} 
\ee
Here $u,v,w \in \{ 1,2,3\}$ are the label of the symmetry group $Z_{N_1}$, $Z_{N_2}$, $Z_{N_3}$'s indices.
Though this Type III class is already known in the group cohomology sense, 
{ this Type III field theory symmetry transformation result is entirely new and not yet been well-explored in the literature,}  
especially not yet studied in the field theory in the SPT context. 
Our result is an extension along the work of Ref.\cite{{Lu:2012dt}},\cite{{Ye:2013upa}}.

The commutation relation leads to 
\bea
&&[\phi_I(x_i), K_{I'J}  \phi_{J}(x_j)]= 
- {2\pi} \ti\, \delta_{I I'}  \tilde{h}(x_i-x_j). \;\;\;\;\;\;
\eea
Here $\tilde{h}(x_i-x_j) \equiv h(x_i-x_j) -1/2$, where
$h(x)$ is the Heaviside step function, with $h(x)=1$ for $x \geq 0$ and $h(x)=0$ for $x < 0$. 
And $\tilde{h}(x)$ is ${h}(x)$ shifted by 1/2, i.e. $\tilde{h}(x)=1/2$ for $x \geq 0$ and $h(x)=-1/2$ for $x < 0$.
The shifted 1/2 value is for consistency condition for the integration-by-part and the commutation relation.
Use these relations, 
we derive the global symmetry transformation $S^{(p_{123})}_{N_1,N_2,N_3}$ acting on the rotor fields $\phi_u(x),\phi_u'(x)$ (here $u \in \{ 1,2,3\}$) on the 1D edge by 

\bea \label{Type III S phi}
&& 
(S^{(p_{123})}_{N_1,N_2,N_3}) \phi_u(x) {(S^{(p_{123})}_{N_1,N_2,N_3})}^{-1}=\phi_u(x)+\frac{2\pi}{N_{u}} \\
\label{Type III S phi'}
&& 
{(S^{(p_{123})}_{N_1,N_2,N_3}) \phi_u'(x) {(S^{(p_{123})}_{N_1,N_2,N_3})}^{-1}=\phi_u'(x) - \epsilon^{uvw} Q \frac{2\pi}{N_{v}} ( 2 \phi^{}_{w}(x)-\frac{(\phi^{}_{w}(L)+\phi^{}_{w}(0))}{2}})\;\;\;\;\;\;\;\;\;\;
\eea
where one can define a Type III symmetry charge $Q \equiv {p_{123}  \frac{N_1N_2N_3}{2\pi N_{123}}}$.
Here the 1D edge is on a compact circle with the length $L$, here $\phi^{}_{w}(L)$ are $\phi^{}_{w}(0)$ taking value at the position $x=0$ (also $x=L$).
(In the case of infinite 1D line, we shall replace $\phi^{}_{w}(L)$ by $\phi^{}_{w}(\infty)$ and replace $\phi^{}_{w}(0)$ by $\phi^{}_{w}(-\infty)$. ) 
But $\phi^{}_{w}(L)$ may differ from $\phi^{}_{w}(0)$ by $2\pi n$ with some 
number $n$ if there is a nontrivial winding, i.e.
\be
\phi^{}_{w}(L)=\phi^{}_{w}(0)+ 2\pi n=2\pi \frac{n_w}{N_w}+ 2\pi n,
\ee
where we apply the fact that $\phi^{}_{w}(0)$ is a $Z_{N_w}$ rotor angle.
So Eq.(\ref{Type III S phi'}) effectively results in a shift
$
+ \epsilon^{uvw} p_{123} {\frac{N_u }{ N_{123}}}    {(2\pi {n_w} + \pi {N_w}{n} )}  
$
and a rotation $\epsilon^{uvw} Q \frac{2\pi}{N_{v}} ( 2 \phi^{}_{w}(x))$.
Since ${\frac{N_u }{ N_{123}}}$ is necessarily an integer, the symmetry transformation $(S^{(p_{123})}_{N_1,N_2,N_3}) \phi_u'(x) {(S^{(p_{123})}_{N_1,N_2,N_3})}^{-1}$ will
shift by a $2 \pi$ multiple if $p_{123} \, {\frac{N_u }{ N_{123} }}  \,{N_w}{n}$ is an even integer.\\

By realizing the field theory symmetry transformation, we have obtained all classes of SPT edge field theory
within the group cohomology $\cH^3(Z_{N_1} \times Z_{N_2} \times Z_{N_3},\tU(1))$ 
with $p_u \in \mathbb{Z}_{N_u}$, $p_{uv} \in \mathbb{Z}_{N_{uv}}$, $p_{123} \in \mathbb{Z}_{ N_{123}}$.

\color{black}

%% file: chap3.tex
\chapter{Aspects of Topology} \label{aofTopology}



In Sec.\ref{sec:TGSD}, we study the topological boundary gapping criteria (or boundary fully gapping rules).
For Abelian TOs, we apply Chern-Simons theory approach shown in Sec.\ref{sec:1212.4863}.
For non-Abelian TOs, we apply modular $\cS,\cT$ data and introduce a new mathematical object called, the tunneling matrix approach shown in Sec.\ref{sec:1408.6514}.
In Sec.\ref{sec:1404.7854}, we study the string and particle braiding statistics in topological order, based on the modular SL(3,Z) Representation
 and 3+1D twisted gauge theory.


\section{Gapped Domain Walls, Gapped Boundaries and Topological (Bulk and Boundary) Degeneracy} \label{sec:TGSD}

\subsection{For Abelian TOs: Chern-Simons theory approach} 

\label{sec:1212.4863}

\subsubsection{Physical Concepts} \label{sec:II}
\begin{figure}[!h]
\centering
{\includegraphics[width=.75\textwidth]{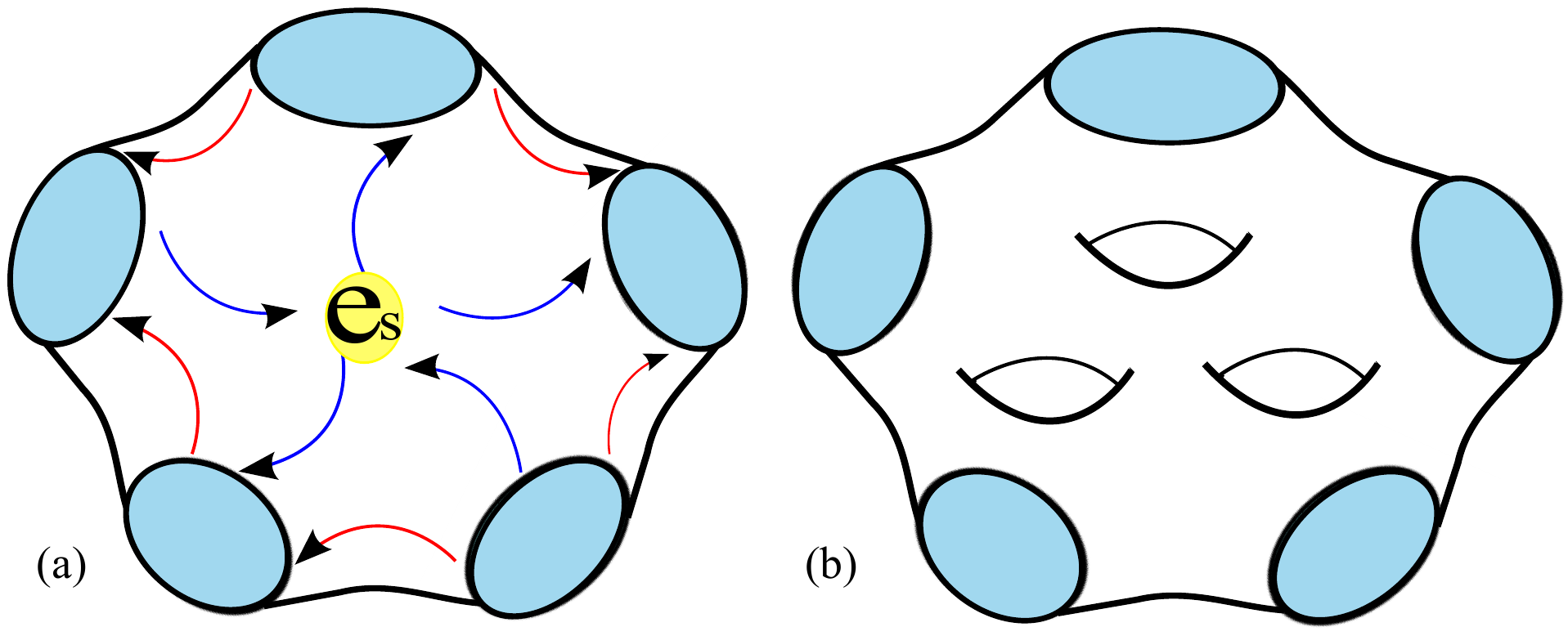}}
\caption{ Topologically ordered states on a 2D manifold with 1D boundaries: (a) 
Illustration of {\it fusion rules} and {\it total neutrality}, 
where anyons are transported from one boundary to another (red arrows), or when they fuse into physical excitations (blue arrows),
on a manifold with five boundaries.
(b) A higher genus compact surface with boundaries (thus with punctures):
a genus-3 manifold with five boundaries.}
\label{hole}
\end{figure}

We start by considering a topologically ordered system on a compact spatial manifold 
with boundaries, 
where each boundary have 
$N$ branches of gapless edge modes. 
Suppose the manifold has total $\eta$ boundaries, we label each boundary as $\partial_\alpha$, with $1 \leq \alpha \leq \eta$. 
Let us focus on the case that the manifold is homeomorphic to a sphere with $\eta$ punctures (\figref{hole}(a)), we will comment on cases with genus or handles (\figref{hole}(b)) later.

If 
particles condense on the boundary due to the interactions of edge modes, it can
introduce mass gap to the edge modes.  
(Note that throughout our study, we regard particles as non-fractionalized particles such as electrons,
and we regard quasiparticles as fractionalized particles such as anyons. From now on, 
we will use \emph{electron} as the synonym of \emph{particle} for the condensed matter systems.)
A set of particles can condense on the
same boundary if they do not have relative quantum fluctuation phases with each
other, thus all {\it condensed particles} are stabilized in the classical
sense.  It requires that {\it condensed particles} have relative zero braiding statistical phase 
(such as Aharonov-Bohm charge-flux braiding phase and flux-flux braiding phase), we call these 
particles with \emph{trivial braiding statistics} 
satisfying Haldane's {\it null} and {\it mutual null} conditions. 
Since electrons or particles have discrete elementary charge unit, we
label them 
as a dimension-$N$ (dim-$N$) lattice $\Gamma_e$ (here the subindex $e$ implies non-fractionalized particles such as electrons), and label 
{\it condensed
particles} as discrete lattice vectors $\ell^{\partial_\alpha}_{}$(with $\ell^{\partial_\alpha}_{} \in
\Gamma_e$) assigned to the boundary $\partial_\alpha$.  We define a {\it complete set} of
{\it condensed particles}, labeled as a lattice $\Gamma^{\partial_\alpha}_{}$, to
include all particles which have null and mutual null statistics to each other: $\Gamma^{\partial_\alpha}_{}=\{ \ell^{\partial_\alpha}_{} \}$.  

Notably 
there 
are different complete sets of condensed particles.
Assigning a {\it complete set} of {\it condensed} ({\it non-fractionalized bosonic}) 
{\it particles} to a boundary corresponds to
assigning certain type of boundary gapping conditions.  
The number of types of {\it
complete sets of condensed particles} constrains 
the number of types of \emph{boundary gapping conditions}, however, the two numbers may differ from each other. 

In principle, each boundary can assign its own boundary condition independently,
this assignment is not determined from the bulk information.
This is why the boundary gapping condition is \emph{beyond the bulk-edge correspondence}.
Below 
we focus on
the non-chiral orders, 
assuming all branches of edge modes can be fully gapped out. 
Later we will derive the criteria when the edge modes can be fully gapped out, at least for Abelian topological orders.

Remarkably there exists a set of {\it compatible anyons} having trivial braiding statistics
respect to the complete set of {\it condensed particles}.  
In other words, {\it compatible anyons} have mutually trivial 
braiding statistics to any elements
in the {\it complete set} of {\it condensed particles}.  For a boundary
$\partial_\alpha$, we label {\it compatible anyons} as discrete lattice vectors
$\ell^{\partial_\alpha}_{qp}$ and find all such anyons to form a {\it
complete set} labeled as $\Gamma^{\partial_\alpha}_{qp}$ with $\Gamma^{\partial_\alpha}_{qp}=\{ \ell^{\partial_\alpha}_{qp} \}$.  
Here $\Gamma^{\partial_\alpha}$ and $\Gamma^{\partial_\alpha}_{qp}$ both have the discrete Hilbert space structure as lattice. 
Note that $\Gamma^{\partial_\alpha} \subseteq \Gamma^{\partial_\alpha}_{qp}$.
And $\Gamma^{\partial_\alpha}$ and $\Gamma^{\partial_\alpha}_{qp}$
have the same dimension of Hilbert space.  
If {\it compatible
anyons} can transport between different boundaries of the compact manifold, 
they must follow {\it total neutrality}: the net transport of {\it
compatible anyons} between boundaries must be balanced by the fusion 
of 
physical particles in the system  (\figref{hole}(a)), so
$\sum_\alpha \ell^{\partial_\alpha}_{qp} \in \Gamma_e$.  Transporting anyons
from boundaries to boundaries in a fractionalized manner (i.e. not in integral
electron or particle units), will result in switching the topological sectors (i.e. switching the ground states) of the
system.  Given data: $\Gamma_e, \Gamma^{\partial_\alpha},
\Gamma^{\partial_\alpha}_{qp}$, we thus derive a generic GSD formula counting the
number 
of elements in a quotient group:
\be
\GSD=\left| \frac{ \{ (\ell^{\partial_1}_{qp},\dots,\ell^{\partial_\eta}_{qp}) \mid \forall \ell^{\partial_\alpha}_{qp} \in \Gamma^{\partial_\alpha}_{qp}, \sum_\alpha \ell^{\partial_\alpha}_{qp} \in \Gamma_e  \} }
{ \{ (\ell^{\partial_1}_{},\dots,\ell^{\partial_\eta}_{}) \mid  \forall \ell^{\partial_\alpha}_{} \in  \Gamma^{\partial_\alpha}_{}  \} } \right|.  \label{gsd_L}
\ee
We derive the form of $\GSD=|L|$ with a group of discrete lattice $L$. Here $|L|$ means the number of elements in $L$, namely the order of $L$.

\subsubsection{Ground state degeneracy of Abelian topological order} \label{sec:III}
To demonstrate our above physical concepts
in a mathematically rigorous setting, let us take Abelian topological order as an example. 
It is believed that Abelian topological order can be fully classified by the $K$ matrix Abelian Chern-Simons theory.
For a system lives on a 2D compact manifold $\mathcal{M}$ with 1D boundaries $\partial \mathcal{M}$, edge modes of each closed boundary (homeomorphic to $S^1$) are described by 
a multiplet-chiral boson theory, 
with the bulk action $S_{bulk}$ and the boundary action $S_{\partial}$:
\bea \label{eq:Sbulk}
S_{bulk} &=& 
\frac{K_{IJ}}{4\pi}\int_\mathcal{M}  dt\; d^2x \; \epsilon^{\mu\nu\rho} a_{I,\mu} \partial_\nu a_{J,\rho}, \\
S_{\partial}&=& \frac{1}{4\pi} \int_{\partial \mathcal{M}} dt \; dx \; K_{IJ} \partial_t \Phi_{I} \partial_x \Phi_{J} -V_{IJ}\partial_x \Phi_{I}   \partial_x \Phi_{J} 
+
\sum_{a} g_{a}  \cos(\ell_{a,I}^{} \cdot\Phi_{I}).  \label{eq:Sedge}
\eea
Here $K_{IJ}$ and $V_{IJ}$ are symmetric integer $N \times N$ matrices, $a_{I,\mu}$ is the 1-form emergent gauge field's $I$-th component in the multiplet.
In terms of edge modes $\Phi_{I}$ with $I=1,2,\dots,N$, this means that there are $N$ branches of edge modes.
The sine-Gordon term $\cos(\ell_{a,I}^{} \cdot\Phi_{I})$ is derived from a local Hermitian gapping term, 
$e^{\ti\ell_{a,I}^{} \cdot\Phi_{I}} + e^{-\ti\ell_{a,I}^{} \cdot\Phi_{I}} \propto \cos(\ell_{a,I}^{} \cdot\Phi_{I})$,
where $\ell_a^{}$ has $N$ components under index $I$ with integer coefficients.

In this work, we investigate the question how generic $g \cos(\ell_{a,I}^{} \cdot\Phi_{I})$ terms
can fully gap edge modes, by turning on large $g$ coupling interactions.
We emphasize that the perturbative relevancy/irrelevancy of $\cos(\ell_{a,I}^{} \cdot\Phi_{I})$ in the renormalization group (RG) language 
is immaterial to our large $g$ coupling limit, since there 
can have energy gap induced by non-perturbative effects at the strong interaction.
Therefore in this work we will include all possible $\ell_{a}$ terms regardless their RG relevancy.

\subsubsection{Canonical quantization of $K$ matrix Abelian Chern-Simons theory edge modes}
In order to understand the energy spectrum or GSD of the edge theory, we study the `quantum' theory, by 
canonical quantizing the boson field $\Phi_{I}$. 
Since $\Phi_{I}$ is the compact phase of a matter 
field, 
its bosonization has zero mode ${\phi_{0}}_{I}$ and winding momentum $P_{\phi_J}$, in addition to non-zero modes: 
\be \label{eq:mode}
\Phi_I(x) ={\phi_{0}}_{I}+K^{-1}_{IJ} P_{\phi_J} \frac{2\pi}{L}x+\ti \sum_{n\neq 0} \frac{1}{n} \alpha_{I,n} e^{-\ti n x \frac{2\pi}{L}}. 
\ee
The periodic boundary size is $L$. The conjugate momentum field of $\Phi_I(x)$ is $\Pi_{I}(x)=\frac{\delta {L}}{\delta (\partial_t \Phi_{I} )}=\frac{1}{2\pi} K_{IJ} \partial_x \Phi_{J}$. 
This yields 
the conjugation relation for zero modes:
$
[{\phi_{0}}_{I},  P_{\phi_J}]=\ti \,\delta_{IJ}
$,
and a generalized Kac-Moody algebra for non-zero modes:
$
[\alpha_{I,n} , \alpha_{J,m} ]= n K^{-1}_{IJ}\delta_{n,-m}
$.
We thus have canonical quantized fields:
$
[\Phi_I(x_1),\Pi_{J}(x_2)]= \ti \,  \delta_{IJ} \delta(x_1-x_2) 
$.\\


\subsubsection{Braiding Statistics and Boundary Fully Gapping Rules} \label{BFGR}
Let us 
first intuitively argue the properties of $\ell_{a}$ as {\it condensed particles} on the edge from $\cos(\ell_{a,I}^{} \cdot\Phi_{I})$ of Eq.(\ref{eq:Sedge}). 
%
%
Let us also determine
the set of lattice spanned by the discrete integer $\ell_{a}$ vectors: $\Gamma^\partial_{}=\{\ell_{a}\}$. 
We shall name $\Gamma^\partial_{}$ as the {\it boundary gapping lattice}.
Here $a$ labels the $a$-th vector in $\Gamma^\partial$. 
From the bulk-edge correspondence, the edge condensed particles labeled by the $\ell_{a}$ vector can be mapped
to some bulk non-fractionalized particle excitations $\ell_{a}$.
It is well-known that the braiding process between two bulk excitations $\ell_{a}$ and $\ell_{b}$ of Eq.(\ref{eq:Sbulk})
causes a mutual-braiding statistical phase term to the whole wavefunction: 
\be \label{eq:Z_ab}
 \exp[\ti \theta_{ab}]= \exp[ \ti \, 2\pi \, \ell_{a,I}^{} K^{-1}_{IJ} \ell_{b,J}^{}].
\ee
We will also denote $\ell_{a,I}^{} K^{-1}_{IJ} \ell_{b,J}^{} \equiv \ell_{a}^T K^{-1} \ell_{b}$.
On the other hand, the self-exchange process between two identical excitations $\ell_{a}$ of Eq.(\ref{eq:Sbulk})
causes a self-braiding statistical phase term to the whole wavefunction: 
\be \label{eq:Z_aa}
 \exp[\ti \theta_{aa}/2]= \exp[ \ti \, \pi \, \ell_{a,I}^{} K^{-1}_{IJ} \ell_{a,J}^{}].
\ee
Without any global symmetry constraint, then any gapping term is allowed. 
Below we argue what are the list of properties that the gapping term satisfies to fully gap the edge modes: 

\noindent
(i) Bosonic self-statistics: 
$\forall \ell_a \in \Gamma^\partial_{}$, $\ell_{a,I}^{} K^{-1}_{IJ} \ell_{a,J}^{} \in 2 \mathbb{Z}$ even integers. This means that the self-statistics of $\ell_{a}$ is bosonic, with a multiple $2\pi$ phase. \\
(ii) Local:  
$\forall \ell_a^{}$, $\ell_b^{} \in \Gamma^\partial_{},\; \ell_{a,I}^{} K^{-1}_{IJ} \ell_{b,J}^{} \in \mathbb{Z}$ integers. 
Winding one $\ell_{a}$ around another $\ell_{b}$ yields a bosonic phase, namely a multiple $2\pi$ statistical phase. 
The bosonic statistics can be viewed as the \emph{local} condition.\\
(iii) Localizing condensate at the classical value without being eliminated by self or mutual quantum fluctuation:
${\forall  \ell_a^{}, \ell_b^{} \in \Gamma^\partial_{},\; \ell_{a,I}^{} K^{-1}_{IJ} \ell_{b,J}^{} =0} $, 
so that $\mathbf{Z}_{\text{statistics}} \sim  \exp[\ti \theta_{ab}] =1$,
the condensation is stabilized and survives in the classical sense.\\
(iv) For the $\cos(\ell_{a,I}^{} \cdot\Phi_{I})$ term, ${\ell}_a^{}$ must be excitations of 
non-fractionalized particle degrees of freedom, since it lives on the `physical' boundary, so ${\ell}_a^{} \in \Gamma_e$ lattice, where 
\be
\Gamma_e=\{ \sum_J c_J K_{IJ} \mid  c_J \in \mathbb{Z} \}.
\ee
This rule imposes an integer charge $ q_I K^{-1}_{IJ} {\ell}^{}_{a,J}$ in the bulk, 
and an integer charge $Q_{I}=\int^L_0 \frac{1}{2\pi} \partial_x \Phi_I dx= K^{-1}_{IJ} P_{\phi_{J}} = K^{-1}_{IJ} {\ell}^{}_{a,J}$ for each branch of edge mode $I$ on the boundary.  Here $q_I$ is the charge vector coupling to an external field $A_\mu$ of gauge or global symmetry, by adding $A_\mu q_I J^\mu_I$ to the ${S}_{bulk}$, 
which corresponds to $q_I A^\mu \partial_\mu \Phi_I$ in the ${S}_{\partial}$.\\
(v) Completeness: we define $\Gamma^\partial_{}$ is a complete set, by including
every possible term $\ell_c$ that has the self null braiding statistics and has the mutually null braiding statistics respect to all the elements $\ell_a \in \Gamma^\partial_{}$.
Namely, mathematically we have $\forall \ell_c^{} \in \Gamma_e$, if $\ell_c^{T} K^{-1} \ell_c^{}=0$ and $\ell_c^{T} K^{-1} \ell_a^{}=0$ for $\forall \ell_a^{} \in \Gamma^\partial_{}$, then $\ell_c^{} \in \Gamma^\partial_{}$ must be true. Otherwise $\Gamma^\partial_{}$ is not complete. \\
(vi) The system is non-chiral. We require the same number of left moving modes and right moving modes to fully gap out the edge modes.

In Sec.\ref{H_and_E} we will use the bulk braiding statistics property of $\ell_{a}$ to determine the gapped edge stability caused by $\cos(\ell_{a,I}^{} \cdot\Phi_{I})$ of Eq.(\ref{eq:Sedge}).
We leave a derivation that these properties above are \emph{sufficient conditions} in Sec.\ref{H_and_E}. 

Indeed the above rules (i)(ii)(iii)(iv)(v)(vi) can be simplified to a set of rules which we call {Boundary Fully Gapping Rules}. 


\paragraph{Boundary Fully Gapping Rules} \label{sec:BFGR}

For an Abelian topological order described by a bulk Chern-Simons theory of Eq.(\ref{eq:Sbulk}) and a boundary theory of Eq.(\ref{eq:Sedge}), we can
add a set of proper interaction terms $\cos(\ell_{a,I}^{} \cdot\Phi_{I})$ on the boundary to gap out the edge modes. 
We will term that the Boundary Fully Gapping Rules, which summarize all the above rules (i)(ii)(iii)(iv)(v)(vi) to determine the gapping term $\ell_a \in \Gamma^\partial_{}$. 
Here $\ell_a$ is some integer vector, namely for every component $\ell_{a,I} \in \mathbb{Z}$. 
The $\Gamma^\partial_{}$ satisfies:\\
 
\noindent
({\bf1}) Null and mutual null conditions: 
${\forall  \ell_a^{}, \ell_b^{} \in \Gamma^\partial_{}}$, ${\ell_{a,I}^{} K^{-1}_{IJ} \ell_{b,J}^{} =0} $. 
This implies self statistics and mutual statistics are bosonic, and the excitation is local. Localized fields are 
not eliminated by self or mutual quantum fluctuations,
so the condensation survives in the classical sense.\\

\noindent
({\bf2}) The dimensions of the lattice $\Gamma^\partial_{}$ is $N/2$, where $N$ must be an even integer. 
Namely, the Chern-Simons lattice $\Gamma^\partial_{}$ assigned to a boundary $\partial$ is spanned by $N/2$ linear independent vectors $\ell_a$.
Mathematically, we write $\Gamma^{\partial}_{}=\{ \underset{a=1, 2, \dots, {N/2}}{\sum}   I^{}_{a} {\ell^{}_{a,I}} \mid\;I^{}_{a} \in \mathbb{Z}\}$. 
\\



\noindent
({\bf 3}) The system is non-chiral.
The signature of $K$ matrix (defined as the number of positive eigenvalues $-$ the number of negative eigenvalues, as $n_L-n_R$) must be zero. 
The non-chiral edge modes implies a measurable observable, the thermal Hall conductance, 
to be zero $\kappa_{xy}= (n_L-n_R)\frac{\pi^2 k_B^2}{3 h}T=0$.
Again, $N=n_L+n_R$ is even.\\

There is an extra rule, which will be important later when we try to reproduce the bulk GSD from the boundary GSD:\\

\noindent
({\bf4}) `Physical' excitation: ${\ell}_a^{} \in \Gamma_e=\{\sum_J c_J K_{IJ} \mid  c_J \in \mathbb{Z} \}$. 
Namely, ${\ell}_a^{}$ is an excitation of non-fractionalized particle 
degree of freedom, 
since it lives on the `physical' boundary.\\


\paragraph{Comments}

Here are some comments for the above rules.
Since any linear combinations of $\ell_a^{} \in \Gamma_e$ still satisfy ({\bf1})({\bf2})({\bf3}), 
we can regard 
$ \Gamma^\partial_{}$ as an {\it infinite discrete lattice group} generated by some basis vectors $\ell_a^{}$.

%
Physically, the rule ({\bf 3}) excludes some violating examples such as odd rank (denoted as rk) $K$ matrix with 
the chiral central charge $c_-=c_L-{c}_R\neq 0$ or the thermal Hall conductance $\kappa_{xy} \neq 0$, which universally has 
gapless chiral edge modes.
For instance, the dim-$1$ boundary gapping lattice: $\{ n(A,B,C) \mid n \in \mathbb{Z} \}$ of $K_{3\times 3}=\diag(1,1,-1)$, with 
$A^2+B^2-C^2=0$, satisfies the rules ({\bf 1})({\bf 2}), but cannot fully gap out chiral edge modes.

\subsubsection{Hamiltonian and Energy Gap} \label{H_and_E}

Here we will justify the Boundary Fully Gapping Rules in Sec.\ref{BFGR} is \emph{sufficient} to fully gap the edge modes. Our approach is to
explicitly calculate the mass gap for the zero energy mode and its higher excitations. We will show that 
\emph{if} the Boundary Fully Gapping Rules hold, there are \emph{stable mass gaps} for all edge modes. 

We consider the even-rank symmetric $K$ matrix, satisfying the rule ({\bf 3}), so the non-chiral system with even number of edge modes can potentially be gappable.

To determine the mass gap of the boundary modes, and to examine the gap in the large system size limit $L\to \infty$, we will take the large $g$ coupling limit of the Hamiltonian: 
$
-g_{a} \int_0^{L} \; dx\; \cos(\ell_{a,I}^{} \cdot\Phi_{I}) \to \frac{1}{2}g_{a}(\ell_{a,I}^{} \cdot\Phi_{I})^2 L
$.
By exactly diagonalizing the quadratic Hamiltonian,
\be
H \simeq (\int^L_0 dx\; V_{IJ}\partial_x \Phi_{I}   \partial_x \Phi_{J} )+\frac{1}{2} \sum_a g_a (\ell_{a,I}^{} \cdot\Phi_{I})^2 L +\dots, 
\ee
with a $\Phi$ mode expansion Eq.(\ref{eq:mode}), we obtain the energy spectra from its eigenvalues.
We realize that:\\
{$\bullet$ \bf Remark 1}: \emph{If} we include all the interaction terms allowed by {\bf Boundary Full Gapping Rules}, we can turn on the energy gap of zero modes ($n=0$) 
as well as the Fourier modes (non-zero modes $n\neq 0$). 
The energy spectrum is in the form of
\be \label{eq:ap-mass-gap-st}
E_n= \big( \sqrt{ \Delta^2 + \# (\frac{2\pi n}{L})^2 } + \dots \big),
\ee
where $\Delta$ is the mass gap. Here $\#$ means some numerical factor.
We emphasize the energy of Fourier modes ($n\neq 0$) behaves towards zero modes 
at long wave-length low energy limit ($L \to \infty$). Such spectra become continuous at $L \to \infty$ limit, which is the expected energy behavior.

\noindent
{$\bullet$ \bf Remark 2}: \emph{If} we include the \emph{incompatible} interaction term, 
e.g. $\ell_a$ and $\ell'$ where $\ell_a^{T} K^{-1} \ell' \neq 0$, 
while the interaction terms contain $\sum_a g_a \cos(\ell_{a} \cdot \Phi) +g' \cos(\ell_{}' \cdot \Phi)$, we obtain the \emph{unstable} energy spectrum:
\be \label{eq:Enunstable}
E_n= \big( \sqrt{ \Delta_m^2 + \# (\frac{2\pi n}{L})^2+ \sum_{a} \# g_{a}  \,g' (\frac{L}{n})^2 +\dots } + \dots \big).
\ee
The energy spectra exhibits an \emph{instability} of the system, because at low energy limit ($L \to \infty$), the spectra become discontinuous (from $n=0$ to $n \neq 0$) and jump to infinity as long as 
there are incompatible cosine terms (i.e. $g_a \cdot g' \neq 0$). 
The dangerous behavior of $(L/n)^2$ implies the quadratic expansion analysis may not describe the full physics.
In that case, the dangerous behavior invalidates localizing of $\Phi$ field at a minimum. This 
invalidates the energy gap, and the \emph{unstable} system potentially seeks to become \emph{gapless phases}.

\noindent
{$\bullet$ \bf Remark 3}: 
%
We provide an alternative way to study the energy gap stability.
We include 
the full cosine interaction term for the lowest energy states, namely the zero and winding modes:
\be \label{eq:cos}
\cos(\ell_{a,I}^{} \cdot\Phi_{I}) \to  
\cos(\ell_{a,I}^{} \cdot ({\phi_{0}}_{I}+K^{-1}_{IJ} P_{\phi_J} \frac{2\pi}{L}x) ).
\ee
The stability of the energy gap can be understood from \emph{under what criteria} we can safely expand the cosine term to extract the leading quadratic terms 
by only keeping the zero modes, namely $\cos(\ell_{a,I}^{} \cdot\Phi_{I}) \simeq 1 - \frac{1}{2}(\ell_{a,I}^{} \cdot\phi_{0I})^2 +\dots$.
The naive reason is the following: if one does not decouple the winding mode $P_{\phi_J}$ term, there is a complicated $x$ dependence in $P_{\phi_J} \frac{2\pi}{L}x$ along the $x$ integration.
The \emph{non-commuting} algebra $[{\phi_{0}}_{I},  P_{\phi_J}]=\ti \delta_{IJ}$ results in the challenge for this cosine expansion.
This challenge can be resolved by requiring $\ell_{a,I}^{}  {\phi_{0}}_{I}$ and $\ell_{a,I'}^{}  K^{-1}_{I'J} P_{\phi_J}$ \emph{commute} 
in Eq.(\ref{eq:cos}),
\bea
[\ell_{a,I}^{}  {\phi_{0}}_{I}, \;\ell_{a,I'}^{}  K^{-1}_{I'J} P_{\phi_J}] &=&\ell_{a,I}^{} K^{-1}_{I'J} \ell_{a,I'}^{}  \; (\ti\delta_{IJ}) 
=(\ell_{a,J}^{} K^{-1}_{I'J} \ell_{a,I'}^{} )(\ti)=0.\;\;\;\;\;\;
\eea  
In fact this is the {Boundary Full Gapping Rule ({\bf 1})} for the self null statistics --- the trivial 
self statistics rule among the interaction gapping terms. 
We can interpret that there is \emph{no quantum fluctuation} destabilize the semi-classical particle condensation.
With this \emph{commuting criterion}, 
we can safely expand Eq.(\ref{eq:cos}) by the trigonometric identity as
\bea
&&\cos(\ell_{a,I}^{} {\phi_{0}}_{I}) \cos(\ell_{a,I}^{} K^{-1}_{IJ} P_{\phi_J} \frac{2\pi}{L}x) 
-\sin(\ell_{a,I}^{} {\phi_{0}}_{I}) \sin(\ell_{a,I}^{} K^{-1}_{IJ} P_{\phi_J} \frac{2\pi}{L}x). 
\eea
Then we integrate over the circumference $L$. 
First, we notice that $\ell_{a,I}^{} K^{-1}_{IJ} P_{\phi_J}$ takes integer values due to $\ell_{a,I}^{} \in \Gamma_e$ and $P_{\phi_J} \in \mathbb{Z}$.
Further we notice that 
due to the periodicity of both $\cos(\dots x)$ and $\sin(\dots x)$ in the region $[0,L)$, so both $x$-integrations over $[0,L)$ vanish. However, 
the exception is $\ell_{a,I}^{} \cdot K^{-1}_{IJ} P_{\phi_J} =0$, then $\cos(\ell_{a,I}^{} K^{-1}_{IJ} P_{\phi_J} \frac{2\pi}{L}x)=1$.
We derive:
\be \label{eq:int_cos}
 g_{a} \int_0^{L}dx\; \text{Eq}.(\ref{eq:cos})=g_{a} L \; \cos(\ell_{a,I}^{} \cdot {\phi_{0}}_{I}) \delta_{(\ell_{a,I}^{} \cdot K^{-1}_{IJ} P_{\phi_J} ,0)}.
\ee
The Kronecker-delta function $\delta_{(\ell_{a,I}^{} \cdot K^{-1}_{IJ} P_{\phi_J} ,0)}=1$ 
indicates that there is a nonzero contribution if and only if $\ell_{a,I}^{} \cdot K^{-1}_{IJ} P_{\phi_J} =0$. 

So far we have shown 
that when the self-null braiding statistics $\ell^T K^{-1} \ell=0$ is true, we have the desired cosine potential expansion via the zero mode quadratic expansion at the large $g_a$ coupling,
$  g_{a} \int_0^{L}dx \cos(\ell_{a,I}^{} \cdot\Phi_{I}) \simeq  - g_{a} L \frac{1}{2}(\ell_{a,I}^{} \cdot\phi_{0I})^2 +\dots$.
If we include not enough gapping terms (less than $N/2$ terms), we cannot fully gap all edge modes. 
On the other hand, if we include more than the {Boundary Full Gapping Rules} (more than $N/2$ terms with incompatible terms), there is a disastrous behavior in the spectrum (see {\bf Remark 2}).
We need to include the mutual-null braiding statistics $\ell^T_a K^{-1} \ell_b=0$ so that the energy gap is stable.

The quadratic Hamiltonian includes both the kinetic and the leading-order of the potential terms:
\be \label{eq:H}
\frac{(2\pi)^2}{4\pi L}  V_{IJ} K^{-1}_{I l_1} K^{-1}_{J l_2} P_{\phi_{l_1}} P_{\phi_{l_2}} +\sum_a g_{a} L  \frac{1}{2}(\ell_{a,I}^{} \cdot\phi_{0I})^2
\ee
By solving the quadratic simple harmonic oscillators, we can show the nonzero energy gaps of zero modes.
The mass matrix can be properly diagonalized, since there are only conjugate variables $\phi_{0I},P_{\phi,J}$ in the quadratic order.
The energy gap is of the order one finite gap, independent of the system size $L$,
\be
\Delta_{} \simeq O(\sqrt{2\pi\, g_a \ell_{a,l_1} \ell_{a,l_2} V_{IJ} K^{-1}_{I l_1} K^{-1}_{J l_2} }). 
\ee
In the diagonalized basis of the Hamiltonian Eq.(\ref{eq:H}), the energy gap $\Delta_{I}$ has the component $I$-dependence. 

More precisely, we find the dimension of independent gapping terms $\Gamma^\partial_{}=\{\ell_a\}$ must be $N/2$, namely satisfying {Boundary Full Gapping Rules ({\bf 2})}.
The number of left and right moving modes must be the same, namely satisfying the non-chiral criterion in {Boundary Full Gapping Rules ({\bf 3})}.
To summarize, by calculating the stability of energy gap, we have thus 
demonstrated that the {Boundary Full Gapping Rules ({\bf 1})({\bf 2})({\bf 3})} are \emph{sufficient} to ensure that the energy gap is stable at large $g$ coupling.

Due to the periodicity of ${\phi_{0}}_{}$, its conjugate variable $P_{\phi}$ forms a discrete quantized lattice. %
This is consistent with the discrete Hilbert space of the ground states, forming the \emph{Chern-Simons quantized lattice} detailed in Sec.\ref{sec:Hilbert}.
We will apply this idea to count the ground state degeneracy of the Chern-Simons theory on a closed manifold or a compact manifold with gapped boundaries. 
The {Boundary Full Gapping Rules ({\bf 4}) will be required for computing the boundary GSD and the bulk GSD. 


\subsubsection{Hilbert Space} \label{sec:Hilbert}

Since ${\phi_{0}}_{}$ is periodic, so 
$P_{\phi}$ forms a discrete lattice.
We now impose 
the rule ({\bf 4}), 
so $\cos(\ell_{a,I}^{} \cdot {\phi_{0}}_{I})$ 
are hopping terms along {\it condensed particle} vector $\ell_{a,I}^{}$ in sublattice of $\Gamma^e$ in the $P_{\phi}$ lattice.    
We can show that rule ({\bf 4}) is essential to derive the bulk GSD by computing the boundary GSD under gluing the boundaries. 

Let $P^{qp}_{\phi}$ represents some {\it compatible anyon} $\ell_{qp}$ which is mutual null to {\it condensed particles} $\ell$ by $\ell_{}^{T} K^{-1}_{} P^{qp}_{\phi}=\ell_{}^{T} K^{-1} \ell_{qp} =0$.
By 
the rule ({\bf1}), thus it means that the compatible anyon $\ell_{qp}$ parallels along 
some $\ell_{}^{}$ vector. However, $\ell_{qp}$  lives on the quasiparticle lattice, i.e. the unit integer lattice of the $P_{\phi}$ lattice. So $\ell_{qp} $ is parametrized by $\frac{1}{|\gcd(\ell_a^{})|} \ell_{a,J}^{}$, with the greatest common divisor defined as $|\gcd(\ell_a)| \equiv  \gcd(|\ell_{a,1}|,|\ell_{a,2}|,\dots,|\ell_{a,N}|)$. 

Now let us consider the Hilbert space of ground states in terms of $P_{\phi}$ lattice.
For the Hilbert space of ground states, we will neglect the kinetic term $H_{kin}=\frac{(2\pi)^2}{4\pi L}  V_{IJ} K^{-1}_{I l1} K^{-1}_{J l2} P_{\phi_{l1}}$ $P_{\phi_{l2}}$ of the order $O(1/L)$ as $L \to \infty$. 
Recall we label the $\alpha$-th boundary of a compact spatial manifold with $\eta$ punctures as $\partial_\alpha$, where $\alpha =1, \dots, \eta$.
Note that $a$ is the index for $a$-th $\ell$ vector: $\ell^{\partial_\alpha}_{a} \in \Gamma^{\partial_\alpha}_{}$. 
If we choose the proper basis $\ell$ vector, based on the rule ({\bf 2}), we have $a=1,\dots,N/2$. 
For the $\alpha$-th boundary $\partial_\alpha$,
a {\it complete set} of {\it condensed particles} forms the {\it boundary gapping lattice}:
\be
\Gamma^{\partial_\alpha}_{}=\{ \sum_{a=1,\dots, N/2}  I^{\partial_\alpha}_{a} {\ell^{\partial_\alpha}_{a,I}} \mid\;I^{\partial_\alpha}_{a} \in \mathbb{Z}\}. 
\ee
Recall $I$ is the $I$-th branch of $K_{N \times N}$ matrix, $I=1,\dots, N$. 

A {\it complete set} of {\it compatible anyon} vectors $\ell_{qp}$ forms the Hilbert space of the winding mode $P_{\phi}$ lattice: 
\bea
\Gamma^{\partial_\alpha}_{qp}=\{\ell^{\partial_\alpha}_{qp,I}  \} =\{ \sum_{a=1,\dots, N/2}   j^{\partial_\alpha}_{a} \frac{ \ell^{\partial_\alpha}_{a,I} }{ |\gcd(\ell^{\partial_\alpha}_{a})|  } \mid \;  j^{\partial_\alpha}_{a} \in \mathbb{Z} \}, \;\;\;\;\;\;\;
\eea
or simply the {\it anyon hopping lattice}. 
 Note $\Gamma^{\partial_\alpha}_{}$, $\Gamma^{\partial_\alpha}_{qp}$ are {\it infinite Abelian discrete lattice group}.
Anyon fusion rules and the {\it total neutrality} condition essentially means the bulk physical charge excitation can {\it fuse} from or {\it split} to multiple anyon charges. 
The rules constrain the set of $j^{\partial_\alpha}_{a}$ values to be limited on the $\Gamma_e$ lattice.

To be more precise mathematically, the anyon fusion rules and the total neutrality condition 
constrain the direct sum 
of the {\it anyon hopping lattice} $\Gamma^{\partial_\alpha}_{qp}$, with $\alpha=1,\dots, \eta$ over all $\eta$ boundaries, must be on the $\Gamma_e$ lattice.
We define such a constrained anyon hopping lattice as $L_{qp \bigcap e}$:
%
\bea
L_{qp \bigcap e}\equiv &\{&\bigoplus_{\alpha=1}^\eta \sum_{a=1}^{N/2}  j^{\partial_\alpha}_{a} \frac{\ell^{\partial_\alpha}_{a,I} }{ |\gcd(\ell^{\partial_\alpha}_{a})|} \; \mid \;  \forall j^{\partial_\alpha}_{a}\in\mathbb{Z},\;   \exists \;c_J\in\mathbb{Z},\nonumber \\
 && \sum_{\alpha=1}^\eta  \sum_{a=1}^{N/2} j^{\partial_\alpha}_{a} \frac{ \ell^{\partial_\alpha}_{a,I}}{ |\gcd(\ell^{\partial_\alpha}_{a})|}=\sum_{J=1}^N c_J K_{IJ}\}.
 \eea  

\paragraph{Number of types of boundary gapping conditions} \label{sec:Ng}

As we exactly solve the number of types of boundary gapping lattices, 
we find that for $\rank(K)=2$, we obtain two boundary gapping lattices.
However, when we consider boundary gapping conditions, we
apply the identification 
and the trivial statistical rules. 
We obtain a list of number of types of boundary gapping conditions $\mathcal{N}^\partial_g$
in Table \ref{table:bGSD_TO}, where $\mathcal{N}^\partial_g \neq 2$ in general.

\begin{center}
\centering
\begin{table}[!h]
\makebox[\textwidth][c]
{\fontsize{10pt}{1em}\selectfont
{
\begin{tabular}{|c||c| c| c| c|}
\hline   
Bosonic TOs & $\mathcal{N}^\partial_g$ & Boundary Conditions & GSD$_{T^2}=|\det K|$ & GSD$_{S^1 \times I^1}$\\   
\hline
\hline 
{
$\begin{array}{ll} K^{}_{Z_2}= 
{\begin{pmatrix} 
0 & 2 \\
2 & 0 
\end{pmatrix}}\\
Z_2 \text{ toric code}
\end{array}
  $}& 2 & $ \left. 
 \begin{array}{ll} 
\{ (1,0),(2,0),\dots \},\\
\{(0,1),(0,2),\dots \}
  \end{array}  \right.$ & 4 & 1, 2 \\ 
\hline
{
$\begin{array}{ll} K^{}_{\diag,2}= 
{\begin{pmatrix} 
2 & 0 \\
0 & -2 
\end{pmatrix}}\\
Z_2 \text{ double-semion}
\end{array}
  $}
& 1 & $ 
\{ (1,1),(2,2),\dots \}
$ & 4 & 2 \\ 
\hline
{
$\begin{array}{ll} K^{}_{Z_3}= 
{\begin{pmatrix} 
0 & 3 \\
3 & 0 
\end{pmatrix}}\\
\text{} Z_3 \text{ gauge theory}
\end{array}
  $}
& 2 & $ \left. 
 \begin{array}{ll} 
\{ (1,0), (2,0), (3,0),\dots \},\\
\{ (0,1),(0,2),(0,3),\dots \}
  \end{array}  \right.$ & 9 & 1, 3  \\ 
  \hline
{
$\begin{array}{ll} K^{}_{Z_4}= 
{\begin{pmatrix} 
0 & 4 \\
4 & 0 
\end{pmatrix}}\\
Z_4 \text{ gauge theory}
\end{array}
  $}& 3 & $ \left. 
 \begin{array}{lll} 
\{ (1,0),(2,0),(3,0),\dots \},\\
\{(0,1),(0,2),(0,3),\dots \},\\
\{ (2,0),(0,2),(2,2),\dots \}
  \end{array}  \right.$ & 16 & 1, 2, 4 \\ 
\hline
{
$\begin{array}{ll} K^{}_{\diag,4}= 
{\begin{pmatrix} 
4 & 0 \\
0 & -4 
\end{pmatrix}}\\
\text{} U(1)_4 \times U(1)_{-4}  \text{ FQH}
\end{array}
  $}& 2 & $ \left. 
 \begin{array}{ll} 
\{ (1,1),(2,2),(3,3),\dots \},\\
\{ (1,3),(2,2),(3,1),\dots \}
  \end{array}  \right.$ & 16 &  2, 4 \\ 
\hline

\hline
Fermionic TOs & $\mathcal{N}^\partial_g$ & Boundary Conditions &  GSD$_{T^2}=|\det K|$ & GSD$_{S^1 \times I^1}$\\ 
\hline
\hline
{
$\begin{array}{ll} K^{}_{\diag,3}= 
{\begin{pmatrix} 
3 & 0 \\
0 & -3 
\end{pmatrix}}\\
\text{} U(1)_3 \times U(1)_{-3}  \text{ FQH}
\end{array}
  $}
& 2 & 
$ \left. 
 \begin{array}{ll} 
\{ (1,1),(2,2),(3,3),\dots \},\\
\{ (1,2),(2,1),(3,3),\dots \}
  \end{array}  \right.$
  & 9 & 1, 3 \\ 
\hline
\end{tabular}
}
}\hspace*{0mm}
\caption{In the first column, we list down some bosonic and fermionic topological orders and their $K$-matrices in Chern-Simons theory. 
Non-fractionalized particles of bosonic topological orders can have only bosonic statistics, but
non-fractionalized particles of fermionic topological orders can have fermionic statistics.
In the second column, we list down their number of types of  boundary gapping conditions $\mathcal{N}^\partial_g$.
In the third column, we list down their boundary gapping conditions in terms of a set of compatible and condensable anyons with trivial braiding statistics.
In the fourth column, we list down their bulk GSD$=|\det K|$ on a closed manifold 2-torus.
In the fifth column, we list down their boundary GSD on an annulus (or a cylinder) with all various types of boundary gapping conditions on two edges.  
The $U(1)_k \times U(1)_{-k}$ FQH means the doubled layer chiral and anti-chiral fractional quantum hall (FQH) states combine to be a non-chiral topological order.}
\label{table:bGSD_TO}
\end{table}
\end{center}

\subsubsection{Examples of boundary GSD: Mutual Chern-Simons theory, ${Z}_k$ topological order, toric code and string-net model} \label{sec:IV}
We now take the ${Z_k}$ gauge theory example with a $K_{Z_k}$-matrix Chern-Simons theory to
demonstrate our understanding of two types of GSD on a cylinder with gapped boundaries in physical pictures. 
By checking all the fusion and braiding properties of quasiparticle excitations, we know that the ${Z_k}$ gauge theory and the $K_{Z_k}=\bigl( {\begin{smallmatrix} 
0 &k \\
k & 0 
\end{smallmatrix}} \bigl)$ Chern-Simons theory are indeed equivalent to the mutual Chern-Simons theory:
$\frac{k}{2\pi}\int  dt\; d^2x \; \epsilon^{\mu\nu\rho} a_{1,\mu} \partial_\nu a_{2,\rho}$.  
All these describe the so-called ${Z}_k$ topological order.
%
\begin{figure}[!h]
\centering
{\includegraphics[width=.45\textwidth]{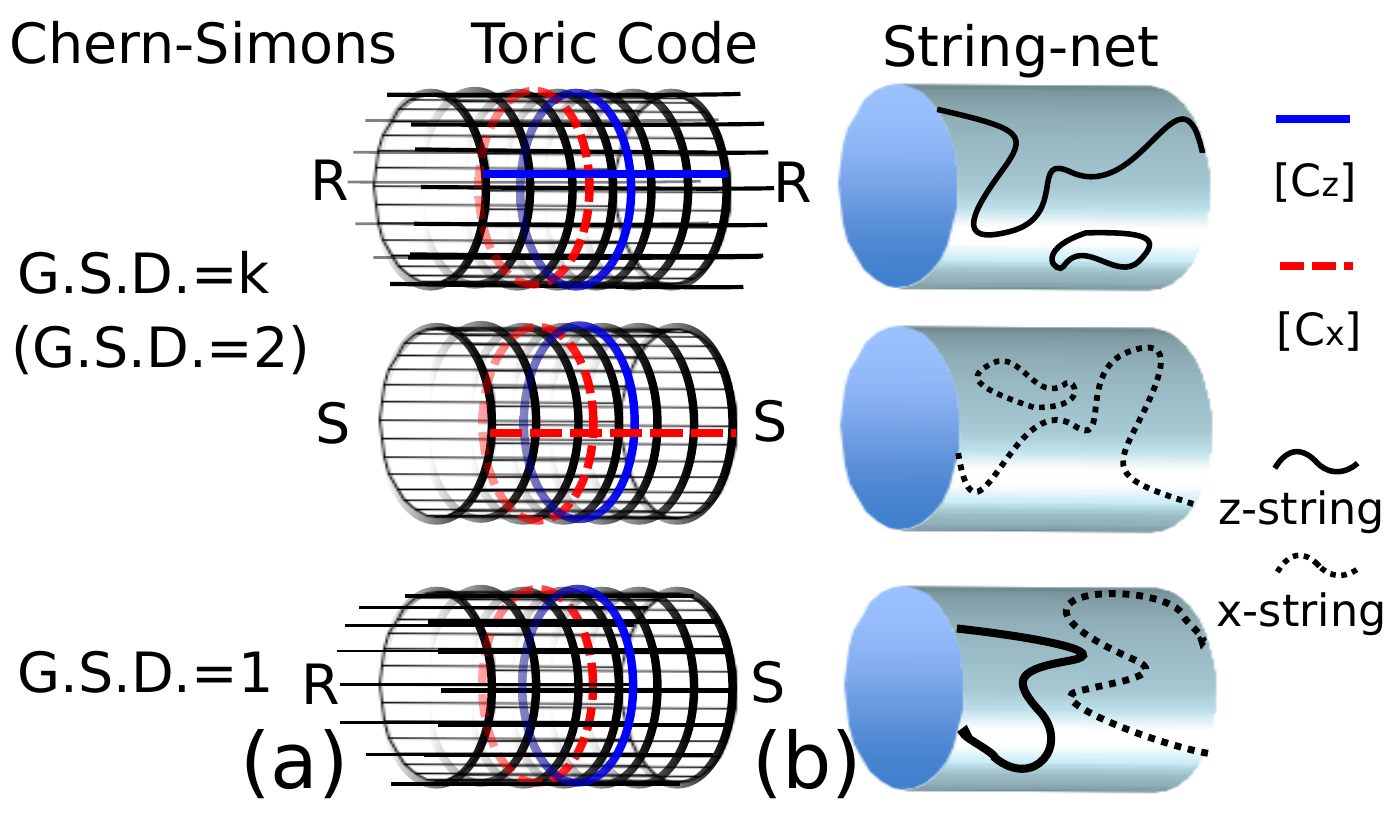}}
\caption{(a) The same boundary conditions on two ends of a cylinder allow a pair of 
cycles $[c_{x}],[c_{z}]$ of a qubit, 
thus $\GSD=2$. Different boundary conditions do not, thus $\GSD=1$.
(b) The same boundary conditions allow z- or x-strings connect two boundaries. Different boundary conditions do not.}
\label{toric_stringnet}
\end{figure}

When $k=2$, it realizes $Z_2$ toric code 
with a Hamiltonian $H_0=-\sum_v A_v - \sum_p B_p$ on a square lattice.\cite{{Kitaev:1997wr}} 
Here the convention is that the vertex operator $A_v=\prod \sigma^x$ goes around four neighbor links of a vertex 
and the plaquette operator $B_p=\prod \sigma^z$ goes around four neighbor links of a plaquette,
with Pauli matrices $\sigma^x$ and $\sigma^z$.
Since the Kitaev's toric code is well-known, the reader can consult other details defined in Ref.\cite{Kitaev:1997wr}.
There are two types of gapped boundaries 
on a cylinder (\figref{toric_stringnet}(a)): First, the $x$ boundary 
(or the rough boundary, denoted as R in FIG.\ref{toric_stringnet}) where $z$-string charge $e$-charge condenses. Second, 
the $z$ boundary (or the smooth boundary, denoted as S in FIG.\ref{toric_stringnet}) where $x$-string ``charge'' $m$-flux condenses.\cite{Kitaev:1997wr} 
We can determine 
the GSD by counting the degree of freedom of the code subspace: 
the number of the qubits --- the number of the independent stabilizers.
For $\Gamma^{\partial_1}_{} = \Gamma^{\partial_2}_{}$, we have the same number of qubits and stabilizers, with one extra constraint $\prod_{\text{all sites}} B_p=1$ for two $x$-boundaries 
(similarly, $\prod_{\text{all sites}} A_v=1$ for two $z$-boundaries). 
This leaves $1$ free qubit, thus 
$\GSD=2^1=2$.
 For $\Gamma^{\partial_1}_{} \neq \Gamma^{\partial_2}_{}$, still the same number of qubits and stabilizers, but has no extra constraint. This leaves no free qubits, thus
 $\GSD=2^0=1$.

We can also count the number of independent logical operators (\figref{toric_stringnet}(a)) in the homology class, with the string-net picture (\figref{toric_stringnet}(b))  in mind. There are two cycles $[c_{x_1}],[c_{z_1}]$ winding around the compact direction of a cylinder. 
If both gapped boundaries of a cylinder are $x$-boundaries, we only have $z$-string connecting two edges: the cycle $[c_{z_2}]$. 
If both gapped boundaries of a cylinder are $z$-boundaries, we only have $x$-string (dual string) connecting two edges: the cycle $[c_{x_2}]$. 
We can define the qubit algebra 
by using the generators of $[c_{x_1}],[c_{z_2}]$ in the first case and 
by using the generators of $[c_{x_2}],[c_{z_1}]$ in the second case.
Cycles of either case can define the algebra $\sigma^x,\sigma^y,\sigma^z$ of a qubit, 
so $\GSD=2$. 
If gapped boundaries of a cylinder are different (one is $x$-boundary, the other is $z$-boundary),
we have no string connecting two edges: there is no nontrivial cycle, which yields no nontrivial Lie algebra, and $\GSD=1$.

Let us use the string-net picture to view the ground state sectors and the GSD.
For both $x$-boundaries ($z$-boundaries), one ground state has an even number of strings (dual strings), the other ground state has an odd number of strings (dual strings), connecting two edges;
so again we obtain $\GSD=2$.
On the other hand, if the boundaries are different on two sides of the cylinder, no cycle is allowed in the non-compact direction, no string and no dual string can connect two edges, so $\GSD=1$. 

Generally, for a $Z_k$ gauge theory (as a level $k$ doubled model) on the compact orientable spatial manifold $\mathcal{M}$ without boundaries or with gapped boundaries, 
without symmetry and without symmetry-breaking,
we obtain its GSD is bounded by the order of the first homology group $H_1(\mathcal{M},Z_k)$ of $\mathcal{M}$ with $Z_k$ coefficient,
or equivalently the $k$ to the power of 
the first Betti number ${b_1(\mathcal{M})}$.

\subsection{For (non-)Abelian TOs: Modular $\cS,\cT$ data and the tunneling matrix} 

\label{sec:1408.6514}

By now we understand how to label a 2D
topological order
by a set of ``\emph{topological order parameters}'' ($\cS,\cT,c_-$),
analogous to ``symmetry-breaking {order parameters}'' for spontaneous symmetry breaking systems. 
However, it is less known how different topological orders are related.
To this end, it is important to investigate the following circumstance: there are several
domains in the system and each domain contains a topological order, while
the whole system is gapped. In this case, different topological orders are
connected by \emph{gapped domain walls}. We now addresses two primary questions: 
\noindent
{\bf (Q1)} ``\emph{Under what criteria can two topological orders be
connected by a gapped domain wall, and how many different types of gapped domain walls are there}?''
Since a gapped boundary is a
gapped domain wall between a nontrivial topological order and the vacuum,
we also address that
``\emph{under what criteria can topological orders allow gapped boundaries}?'' 

\noindent 
{\bf (Q2)} ``\emph{When a topologically ordered system has a gapped bulk, gapped domain walls and gapped boundaries,
how to calculate its ground state degeneracy} (GSD) {on any orientable manifold?}''

We have partially achieved results from Abelian TOs and GSD with gapped boundaries. Here we will take another approach
to generalize our theory to non-Abelian TOs and GSD with gapped domain walls.

\subsubsection{Main result} 
Consider two topological orders, Phases $A$ and $B$, described by $(\cS^A,
\cT^A,c^A_-)$ and $(\cS^B,\cT^B,c^B_-)$.
Suppose there are $N$ and $M$ types of anyons in Phase $A$ and Phase $B$,
then the ranks of their modular matrices are $N$ and $M$ respectively.
If $A$ and $B$ are connected by a gapped domain wall, firstly their central
charges must be the same $c^A_-=c^B_-$.
Next we find that the domain wall can be labeled by
a $M\times N$ \emph{tunneling matrix} $\cW$ whose entries are \emph{fusion-space dimensions} $\cW_{ia}$
satisfying the \emph{commuting condition} \eqref{commute},
and the \emph{stable condition}~\eqref{stable}:
\begin{gather}
\cW_{ia}\in\bbn,
 \label{Winteger}\\
  \cS^B \cW=\cW \cS^A,\quad \cT^B \cW= \cW \cT^A,
  \label{commute}\\
  \cW_{ia}\cW_{jb}\leq\sum_{kc} (\cN^B)_{ij}^k \cW_{kc} (\cN^A)_{ab}^c\,.
  \label{stable}
\end{gather}
$\bbn$ denotes the set of non-negative integers. 
$a,b,c,\dots$ and $i,j,k,\dots$ are anyon indices for Phases $A,B$.
$(\cN^A)_{ab}^c$ and $(\cN^B)_{ij}^k$ are fusion tensors 
of Phases $A,B$.

\eqref{Winteger}\eqref{commute}\eqref{stable} is a set of necessary conditions a gapped domain wall must satisfy,
i.e., \emph{if there is no non-zero solution of $\cW$, the domain wall must be gapless.}
We conjecture that they are also sufficient \add{for a gapped domain wall to exist.}
In the examples studied
, $\cW$ are in one-to-one correspondence with gapped domain walls.
However, for some complicated examples, 
a $\cW$ matrix may correspond to more than one type of gapped domain wall.
This indicates that some additional data are needed to completely classify gapped domain walls.

As a first application of our result, we give a general method to compute the
GSD in the presence of gapped domain walls on any orientable 2D surface.  A
simple case is the GSD on a disk drilled with two holes (equivalently a sphere
with 3 circular boundaries, see Fig.~\ref{fig:gdwfigure}(c)). The gapped
boundaries are labeled by three \emph{vectors}
(one-row or one-column matrices) $\cW^{(1)},\cW^{(2)},\cW^{(3)}$. The GSD is
${\sum}_{ijk} \cW^{(1)}_{i1} \cW^{(2)}_{j1} \cN^k_{ij} \cW^{(3)}_{1k}$.

For gapped boundaries,
our criteria can be understood via \emph{dimension reduction},
i.e., shrinking a 1D gapped boundary $\cW$ to a (composite \footnote{The concepts of trapping anyons,
composite anyon types and fusion spaces are discussed in~\cite{LW14}}) anyon $\bm
q_\cW = \oplus_a\cW_{1a}{a}$.
If the system is on a 2D surface $M^2$ drilled with $n$ gapped boundaries ${\cW^{(1)}}, \dots, {\cW^{(n)}}$, then the GSD is the dimension of the fusion space 
with anyons $\bm{q}_{\cW^{(1)}},  \dots, \bm{q}_{\cW^{(n)}}$,
$\GSD=\dim[\mathcal V(M^2,\bm{q}_{\cW^{(1)}},\dots, \bm{q}_{\cW^{(n)}}  ) ].  $

Since gapped domain walls \emph{talk to each other} through long-range
entanglement, the GSD with domain walls reveals more physics than that without
domain walls.  We foresee its practicality in experiments, since we can read
even more physics by putting the system on open surfaces with gapped domain
walls.
Below we shall properly introduce $\cS,\cT$ and $\cW$ matrices.

\subsubsection{Modular $\cS,\cT$ matrices} 
$\cS$ and $\cT$ are unitary matrices indexed by anyon types $\{1,a,b,c,\dots\}$. 1 labels the trivial anyon type.
The anti-quasiparticle of $a$ is denoted by $a^*$.

$\cT$ describes the self statistics.
It is diagonal $\cT_{ab}=\re^{\ii\theta_a}\delta_{ab}$, where $\re^{\ii\theta_a}$ is the phase factor when
exchanging two anyons $a$.
For the trivial type, 
$\cT_{11}=\re^{\ii\theta_1}=1$.
$\cS$ describes the mutual statistics.
$\cS_{ab}$ is the amplitude of the following process with proper normalization factors:
first create a pair of $aa^*$ and a pair of $bb^*$, then braid $a$ around $b$,
and finally annihilate the two pairs.
$\cS$ is symmetric, $\cS_{ab}=\cS_{ba}$.
If $b=1$, the process is just creation and annihilation, and $\cS_{a1}>0$.
$\cS$ and $\cT$ form a projective representation of
the modular group:
  $\cS^4=\iden,
  (\cS \cT)^3=\re^{2\pi\ii c_-/8} \cS^2$, where $\iden$ denotes the identity matrix.

The anti-quasiparticle can be read from $\cS^2$, ${(\cS^2)_{ab}=\delta_{a^*b}}$.
The fusion tensor $\cN_{ab}^c$ can be calculated via the \emph{Verlinde formula}: 
    \begin{align}
      \cN_{ab}^c=\sum_{m} \frac{\cS_{am} \cS_{bm}\ov{\cS_{cm}}}{\cS_{1m}}\in
      \bbn.
      \label{verlinde}
    \end{align}

{Gapped domain walls}--
Below we demonstrate the physical meanings of the gapped domain wall conditions \eqref{Winteger}\eqref{commute}\eqref{stable}.
First we put Phase $A$ and Phase $B$  on a sphere $S^2$, separated by a gapped
domain wall. Note that there can be many types of domain walls
separating the same pair of phases $A$ and $B$.  What data characterize those different types of domain walls?
We fix the domain wall type, labeled by $W$, and trap 
an anyon $a^*$ in
Phase $A$, an anyon $i$ in Phase $B$ and. This configuration is denoted by
$(S^2,i,W,a^*)$.  The states with such a configuration may be degenerate
and the degenerate subspace is the fusion space $\mathcal V(S^2,i,W,a^*)$.
Here we propose using the
\emph{fusion-space dimensions
  $\cW_{ia} \equiv {\dim}[\mathcal V(S^2,i,W,a^*)]\in\bbn$} to characterize the
gapped domain wall $W$.
{Below we will replace the abstract label $W$ by the concrete data $\cW$.}

\add{There are non-local operators $O_{W,ia^*}$ that create a
pair $aa^*$ in Phase $A$, and then tunnel $a$ through the domain wall to an anyon $i$ in Phase $B$,
$O_{W,ia^*}|\psi_{S^2,W}\rangle\in\mathcal V(S^2,i,W,a^*)$,
where $|\psi_{S^2,W}\rangle$ is the ground state.
Since we care about the fusion states rather than the operators themselves,
we would take the equivalent class $[O_{W,ia^*}]=\{U_{W,ia^*} |(O_{W,ia^*}-U_{W,ia^*})|\psi_{S^2,W}\rangle$ $=0\}$.
We call $[O_{W,ia^*}]$ as \emph{tunneling channels},
which correspond to fusion states in $\mathcal V(S^2,i,W,a^*)$.
Therefore, the fusion space dimension $\cW_{ia}$ is the number of linearly independent tunneling channels.
So we also refer to $\cW$ as the ``tunneling matrix.''
}

{We can compute the dimension of the  fusion space $\mathcal V(S^2,i,W,a^*)$ by first creating a
pair $aa^*$ in Phase $A$, then tunneling $a$ through the domain wall.
In the channel where the tunneling does not leave any topological quasiparticle
on the domain wall, $a$ in Phase $A$ will become
a composite anyon $\bm q_{\cW,a}=\oplus_i \cW_{ia} i$ in Phase $B$.
Thus the fusion-space dimension $\cW_{ia}$ is also the \emph{number} of tunneling channels from,
$a$ of Phase A, to, $i$ of Phase $B$. So we also refer to $\cW$ as
the ``tunneling matrix.''}

The \emph{commuting condition} \eqref{commute} dictates the consistency of anyon statistics in presence of gapped domain walls.
Since modular $\cS,\cT$ matrices encode the anyon statistics,
we require that $\cW$ should commute with them as \eqref{commute}:
$\cS^B \cW=\cW \cS^A$, $\cT^B \cW= \cW \cT^A$.

We may as well create a pair $ii^*$ in Phase $B$ and tunnel $i^*$ to $a^*$.
$\cW^\dag$ describes such tunneling in the opposite direction (i.e.,
$\cW:A\to B,~ \cW^\dag: B\to A$). $\cW^\dag$ and $\cW$ contains the same physical
data. To be consistent, tunneling $i^*$ to $a^*$ should give the same
fusion-space dimension, $(\cW^\dag)_{a^*i^*}=\cW_{i^*a^*}=\cW_{ia}$.
\add{This is guaranteed by $\cW (\cS^A)^2=(\cS^B)^2\cW$ and $(\cS^2)_{ab}=\delta_{a^*b}$.}

\add{
The fusion spaces with four anyons further provide us consistence conditions of $\cW$.
To see this, first notice that there are \emph{generalised} tunneling channels,
$[O_{W,ia^*,x}]$,
which, in addition to tunneling $a$ to $i$, also create quasiparticle $x$ on the domain wall.
If we combine the tunneling channels $[O_{W,ia^*,x}]$ and $[O_{W,jb^*,x^*}]$,
we can create fusion states with a domain wall $W$ and four anyons $i,j,a^*,b^*$,  as Fig.~\ref{fig:stable}(a).
In other words, $[O_{W,ia^*,x}O_{W,jb^*,x^*}]$ form a basis of the fusion space $\mathcal V(S^2,i,j,W,a^*,b^*)$.
Let $\cK_{ia}^x$ denote the number of tunneling channels $[O_{W,ia^*,x}]$,
and we know that $\dim\mathcal V(S^2,i,j,W,a^*,b^*)=\sum_x \cK_{ia}^x \cK_{jb}^{x^*}$.
However, the tunneling process as Fig.~\ref{fig:stable}(b),
i.e., fusing $a,b$ to $c$, using $[O_{W,kc^*}]$ to tunnel $c$ to $k$ and splitting $k$ to $i,j$, forms another basis of the fusion space.
The number of such fusion/tunneling/splitting channels is $\sum_{kc} (\cN^B)_{ij}^k \cW_{kc} (\cN^A)_{ab}^c$.
Therefore, we must have 
\begin{equation}
  \sum_x \cK_{ia}^x \cK_{jb}^{x^*}= \sum_{kc} (\cN^B)_{ij}^k \cW_{kc} (\cN^A)_{ab}^c.
  \label{estable}
\end{equation}
}

We are interested in classifying \emph{stable} gapped domain walls,
i.e., the GSD cannot be
reduced no matter what small perturbations are added near the domain wall.
\add{
For stable gapped domain walls we have $\cW_{ia}=\cK_{ia}^{1}$.}
Unstable gapped domain walls $\mathcal{U}$ split as the sum of stable ones $\cW^{(1)},\cW^{(2)},\dots,\cW^{(N)}$, and $ \mathcal{U}_{ia}=\sum_{n=1}^N\cW^{(n)}_{ia}$, for $N\geq 2$.

{We find that a
gapped domain wall is stable \emph{if and only if} (iff) the tunneling matrix $\cW$ satisfies
the \emph{stable condition} \eqref{stable}:
$\cW_{ia}\cW_{jb}\leq\sum_{kc} (\cN^B)_{ij}^k \cW_{kc} (\cN^A)_{ab}^c$.
It can be understood in the following way.
Consider the number of channels tunneling $a,b$ to $i,j$ through the domain wall.
We may tunnel $a$ to $i$ and
$b$ to $j$ separately. The number of channels is $\cW_{ia}\cW_{jb}$. 
But this way,
we may miss some nontrivial exchanging channels $x$ along the domain wall as
Fig.~\ref{fig:stable}(a). 
If we first fuse
$a,b$ to $c$, tunnel $c$ to $k$ and then split $k$ to $i,j$,
instead we will obtain the
total number of channels $\sum_{kc} (\cN^B)_{ij}^k \cW_{kc} (\cN^A)_{ab}^c$,
as Fig.~\ref{fig:stable}(b).
The missing of exchanging channels leads to the inequality \eqref{stable}.
Such channel counting works only when the gapped domain wall is stable, so
\eqref{stable} is a necessary condition.
But \eqref{commute}\eqref{stable}
together imply that $\cW_{11}=1$, thus $\cW$ cannot be the sum of more than one stable tunneling matrix; it must be stable itself.
Therefore \eqref{stable} with \eqref{commute} is also sufficient
for a gapped domain wall to be stable.}
\add{Now if a gapped domain wall $\cW$ is stable, \eqref{estable} becomes $  \sum_{kc} (\cN^B)_{ij}^k \cW_{kc} (\cN^A)_{ab}^c=\cW_{ia}\cW_{jb}+\sum_{x\neq 1} \cK_{ia}^x \cK_{jb}^{x^*}\geq \cW_{ia}\cW_{jb}$.
We know that \eqref{stable} is necessary for a gapped domain wall to be stable.
Furthermore, setting $i=j=a=b=1$ we know that $\cW_{11}\geq \cW_{11}^2$ and \eqref{commute} requires that $\cW_{11}>0$, thus $\cW_{11}=1$ and $\cW$ cannot be the sum of more than one stable tunneling matrix; it must be stable itself.
Therefore \eqref{stable} with \eqref{commute} is also sufficient
for a gapped domain wall to be stable.
}

\begin{figure}
  \centering
    \includegraphics{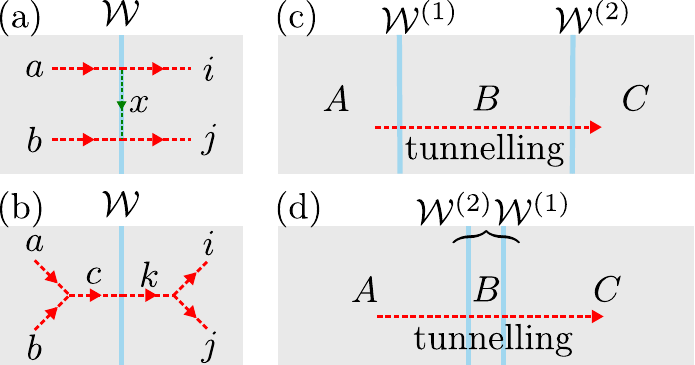}
  \caption{(a)(b) Tunneling channels. 
 (c) Separated domain walls
  $\cW^{(1)}$ and $\cW^{(2)}$. (d) Composite domain wall $\cW^{(2)}\cW^{(1)}$.
}\label{fig:stable}
\end{figure}

\subsubsection{Stability of composite domain walls} 
Let us consider two stable domain walls, $\cW^{(1)}$ between Phases $A$ and $B$, and
$\cW^{(2)}$ between Phases $B$ and $C$, as in Fig.~\ref{fig:stable}(c).
When the two domain walls are far separated, they are both stable.
Any small perturbations added near $\cW^{(1)}$, or near $\cW^{(2)}$, cannot reduce the GSD.

We then shrink the size of the middle Phase $B$, such that the two
domain walls are near enough to be regarded as a single domain wall.
This way we obtain a composite domain wall, whose tunneling matrix is the
composition $\cW^{(2)}\cW^{(1)}$, as Fig.~\ref{fig:stable}(d).
However, this composite domain wall $\cW^{(2)}\cW^{(1)}$ may no longer be stable.
\add{Unless Phase $B$ is vacuum,} we allow more perturbations to $\cW^{(2)}\cW^{(1)}$ than when 
$\cW^{(1)}$ and $\cW^{(2)}$ are far separated. Some operators simultaneously
acting on both $\cW^{(1)}$ and $\cW^{(2)}$ may reduce the GSD,
in which case, the composite domain wall $\cW^{(2)}\cW^{(1)}$ is not stable.

\add{In the special case when Phase $B$ is vacuum, the composite $\cW^{(2)}\cW^{(1)}$ remains stable.}
One can explicitly check this with \eqref{stable}. 

\subsubsection{GSD in the presence of gapped domain walls} 
\begin{figure}
  \centering
  \includegraphics[scale=1]{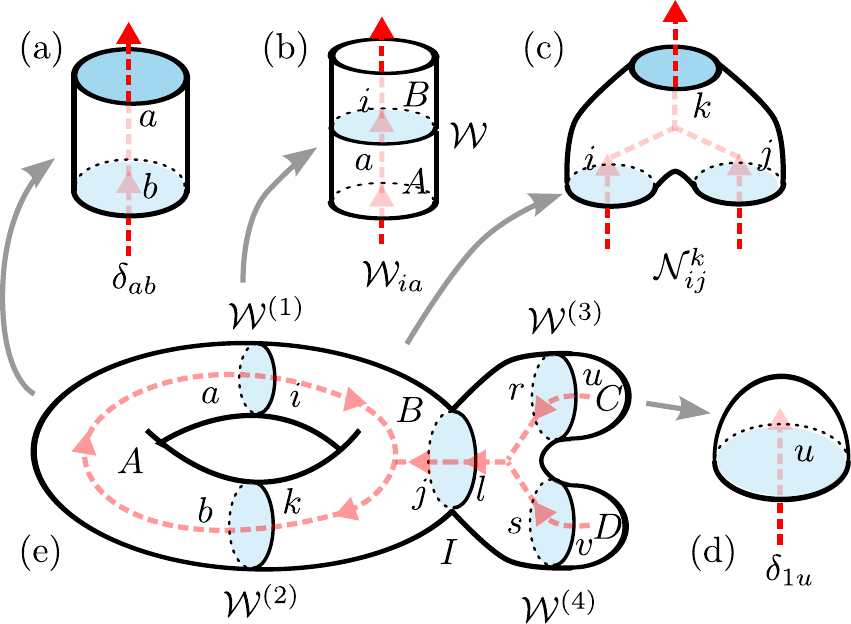}
  \caption{Computing GSD by tensor contraction: Cut a complicated manifold (e)
    into simple segments, add oriented skeletons and anyon indices.
    Associate the segments
    with:  (a) a cylinder with $\delta_{ab}$, 
    (b) a domain wall with its tunneling matrix $\cW_{ia}$, 
    (c) a pair of pants with the fusion tensor $\cN_{ij}^k$
    and (d) a cap with $\delta_{1u}$. Finally, contract all the tensors.
  }
  \label{fig:gdwfigure2}
\end{figure}
Below we derive the GSD, for a 2D system containing several
topological orders separated by loop-like gapped domain walls. 
Domain walls cut a whole 2D system into several segments. 
Without losing generality, let us consider an example in Fig.~\ref{fig:gdwfigure2} 
with topological orders, Phases $A,B,C,D$, and four nontrivial domain walls,
$\cW^{(1)},\cW^{(2)},\cW^{(3)},\cW^{(4)}$, on a manifold Fig.~\ref{fig:gdwfigure2}(e).
We first add extra trivial domain walls $\cW=\iden$, so that all segments
between domain walls are reduced to simpler topologies: caps, cylinders or
pants.
We also add oriented skeletons to the manifold, and put anyon indices
on both sides of the domain walls, 
shown in Fig.~\ref{fig:gdwfigure2}(e).
Next, see Fig.~\ref{fig:gdwfigure2}(a)(b)(c)(d), for the segments with oriented skeletons and anyon indices,
we associate certain tensors: caps with $\delta_{1u}$, cylinders with
$\delta_{ab}$, pants with $\cN_{ij}^k$ in the corresponding topological order,
and domain walls with their tunneling matrices $\cW_{ia}$. 
We may reverse the orientation and at the same time replace the
index $a$ with $a^*$.
Finally we multiply these tensors together and contract all the anyon
indices. 
Physically such tensor contraction computes the total number of winding channels
of anyons, which exactly counts the number of ground states, thus the GSD.

Systems with \emph{gapped boundaries} are
included in our method; just imagine that there are vacuum on
caps connected to the boundaries, e.g., Phases $C,D$ in Fig.~\ref{fig:gdwfigure2}(e)
can be vacuum. Dimensions of generic fusion spaces can also be
calculated, by putting the anyon $a$ on the cap and associating the tensor
$\delta_{au}$ instead of $\delta_{1u}$.

We derive GSD
on exemplary manifolds: 
\begin{enumerate}
  \item A stable domain wall $\cW$ on the sphere: $\GSD=\cW_{11}=1$.
  
  \item A domain wall $\cW$ on the torus: $\GSD=\Tr(\cW)$. 
  Several domain walls $\cW^{(1)},\dots,\cW^{(n)}$ on the torus,
  in Fig.~\ref{fig:gdwfigure}(a): $\GSD=\Tr(\cW^{(1)}\cdots \cW^{(n)})$.
  In particular, $\Tr[\cW^{(1)}(\cW^{(2)})^\dag]$ counts the types of \emph{0D defects} between 1D gapped domain walls $\cW^{(1)},\cW^{(2)}$.
  
%

  \item A sphere with punctures: A cylinder with two gapped boundaries $\cW^L$ and $\cW^R$, in Fig.~\ref{fig:gdwfigure}(b): $\GSD=\sum_a \cW^L_{a1} \cW^R_{1a}$.
  A pair of pants with three gapped boundaries $\cW^{(1)}$, $\cW^{(2)}$ and $\cW^{(3)}$, in Fig.~\ref{fig:gdwfigure}(c): $\GSD={\sum}_{ijk} \cW^{(1)}_{i1} \cW^{(2)}_{j1} \cN^k_{ij} \cW^{(3)}_{1k}$.
  
  \item The rocket graph in Fig.~\ref{fig:gdwfigure2}(e): $\GSD
  =\underset{{a,i,j,k,r,s}}{\sum} \cW^{(1)}_{ia} \cW^{(2)}_{ak}  (\cN^{B})^k_{ij} (\cN^{B})^j_{rs} \cW^{(3)}_{r1} \cW^{(4)}_{s1}$.
\end{enumerate}

\begin{figure}[!h]
\centering
  \includegraphics[scale=0.75]{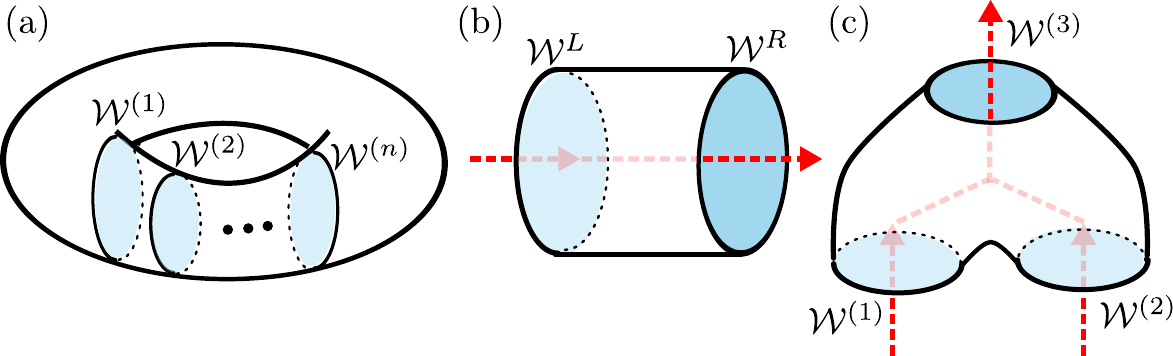}
  \caption{Some 2-manifolds with gapped domain walls. 
  }
  \label{fig:gdwfigure}
\end{figure}

\subsubsection{1+1D gravitational anomaly, topological phase transitions and future directions}
We know that the effective 1+1D edge theory of a 2+1D topological order has
a gravitational anomaly. The gravitational anomalies are classified by the bulk topological order
$(\cS,\cT,c_-)$~\cite{Kong:2014qka}. 
 When $c_-\neq 0$, the edge effective theory
has a perturbative  gravitational anomaly which leads to topological gapless
edge (i.e., the gaplessness of the edge is robust against any change of
the edge Hamiltonian).  Even in the absence of perturbative  gravitational
anomaly, $c_-=0$, certain global gravitational anomalies~\cite{Witten:1985xe} (characterized by
$(\cS,\cT,0)$) can also lead to topological gapless edge~\cite{Wang:2012am, Levin:2013gaa}.   
Our work points out that such global gravitational anomalies are described by
$\cS,\cT$ which do not allow any non-zero solution $\cW$ of
\eqref{Winteger}\eqref{commute}\eqref{stable}. The corresponding 2D
topological order $(\cS,\cT,0)$ will have topological gapless edge.

Since a domain wall sits on the border between two topological orders, 
our study on domain walls can also guide us to better understand the \emph{phase transitions} of topological orders.

\section{Non-Abelian String and Particle Braiding in Topological Order: Modular SL(3,Z) Representation and 3+1D Twisted Gauge Theory} 

\label{sec:1404.7854}

In the 1986 Dirac Memorial Lectures, Feynman explained the braiding statistics of fermions by demonstrating the plate trick and the belt trick. 
Feynman showed that the wavefunction of a quantum system obtains a mysterious $(-1)$ sign by exchanging two fermions, which is associated with 
the fact that an extra $2\pi$ twist or rotation is required to go back to the original state. 
However, it is known that there is richer physics in deconfined topological phases of 2+1D and 3+1D spacetime. 
(Here $d+1$D is $d$-dimensional space and $1$-dimensional time, while $d$D is $d$-dimensional space.)
In 2+1D, there are ``anyons'' with exotic braiding statistics for point particles \cite{wilczek1990fractional}. 
In 3+1D, Feynman only had to consider bosonic or fermionic statistics for point particles, without worrying about anyonic statistics.
Nonetheless, there are string-like excitations, whose braiding process in 3+1D can enrich the statistics of deconfined topological phases.
In this work, we aim to systematically address the string and particle braiding statistics in deconfined gapped phases of 3+1D \emph{topological orders}.
\cblue{ 
Namely, we aim to determine what statistical phase the wavefunction of the whole system gains under the string and particle braiding process.}

Since the discovery of 2+1D topological orders, 
we have now gained quite systematic ways to classify and characterize them, 
by using the induced representations of the mapping class group of the $\mathbb{T}^2$ torus 
(the modular group SL$(2,\Z)$ and the gauge/Berry phase structure of ground states 
and the topology-dependent ground state degeneracy, 
using the unitary fusion
categories, 
and using
simple current algebra, 
a pattern of zeros,
and field theories. 
Our better understanding of topologically ordered states also holds the promises of applying their rich quantum phenomena,
including fractional statistics 
and non-Abelian anyons, to topological quantum computation. 

However, our understanding of 3+1D topological orders is in its infancy and far from systematic. 
This motivates our work attempting to address:\\

\noindent
{\bf Q1}: ``\emph{How do we} (\emph{at least partially}) \emph{classify and characterize 3D topological orders}?''\\

\noindent
By {\it classification}, we mean counting the number of distinct phases of topological orders and giving them a proper label.
By {\it characterization}, we mean to describe their properties in terms of physical observables.
Here our approach to studying $d$D topological orders is to simply generalize the above 2D approach,
to use the ground state degeneracy (GSD) on $d$-torus $\mathbb{T}^d=(S^1)^d$, 
and the associated representations of the mapping class group (MCG) of $\mathbb{T}^d$ (recently proposed in Refs.\cite{Kong:2014qka}),  
\bea
\text{MCG}(\mathbb{T}^d)= \text{SL}(d,\Z). 
\eea 
For
3D, the mapping class group SL$(3,\Z)$ is generated by the modular transformation $ \hat{\mathsf{S}}^{xyz}$ and $\hat{\mathsf{T}}^{xy}$ \cite{Coxeter}.

What are examples of 3D topological orders?
One class of them is described by a discrete gauge
theory with a finite gauge group $G$.  Another class is described by the {\it twisted gauge theory},\cite{Dijkgraaf:1989pz} a gauge theory $G$
with a 4-cocycle twist $\om_4 \in \cH^4(G,\R/\Z)$ of $G$'s fourth cohomology group.  But the twisted gauge
theory characterization of 3D topological orders is not one-to-one: different
pairs $(G,\om_4)$ can describe the same 3D topological order.
In this work, we will use $ \hat{\mathsf{S}}^{xyz}$ and $\hat{\mathsf{T}}^{xy}$ of SL$(3,\Z)$ to characterize 
the topological {\it twisted discrete gauge theory} with finite gauge group $G$, which 
has topology-dependent ground state degeneracy. 
The twisted gauge theories describe a large class of 3D gapped quantum liquids in condensed matter. 
Although we will study the SL$(3,\Z)$ modular data of the ground state sectors of gapped phases, these data can capture
the gapped excitations such as particles and strings. (This strategy is widely-used especially in 2D.) 
There are {\bf two main issues} that we will focus on addressing. 
The first is the {\bf dimensional reduction} from 3D to 2D of SL$(3,\Z)$ modular transformation and cocycles to study 3D topological order.
The second is the {\bf non-Abelian three-string braiding statistics} from a twisted discrete gauge theory of an Abelian gauge group.\\

\begin{figure}[!h] 
\begin{center} 
(a)\includegraphics[scale=0.3]{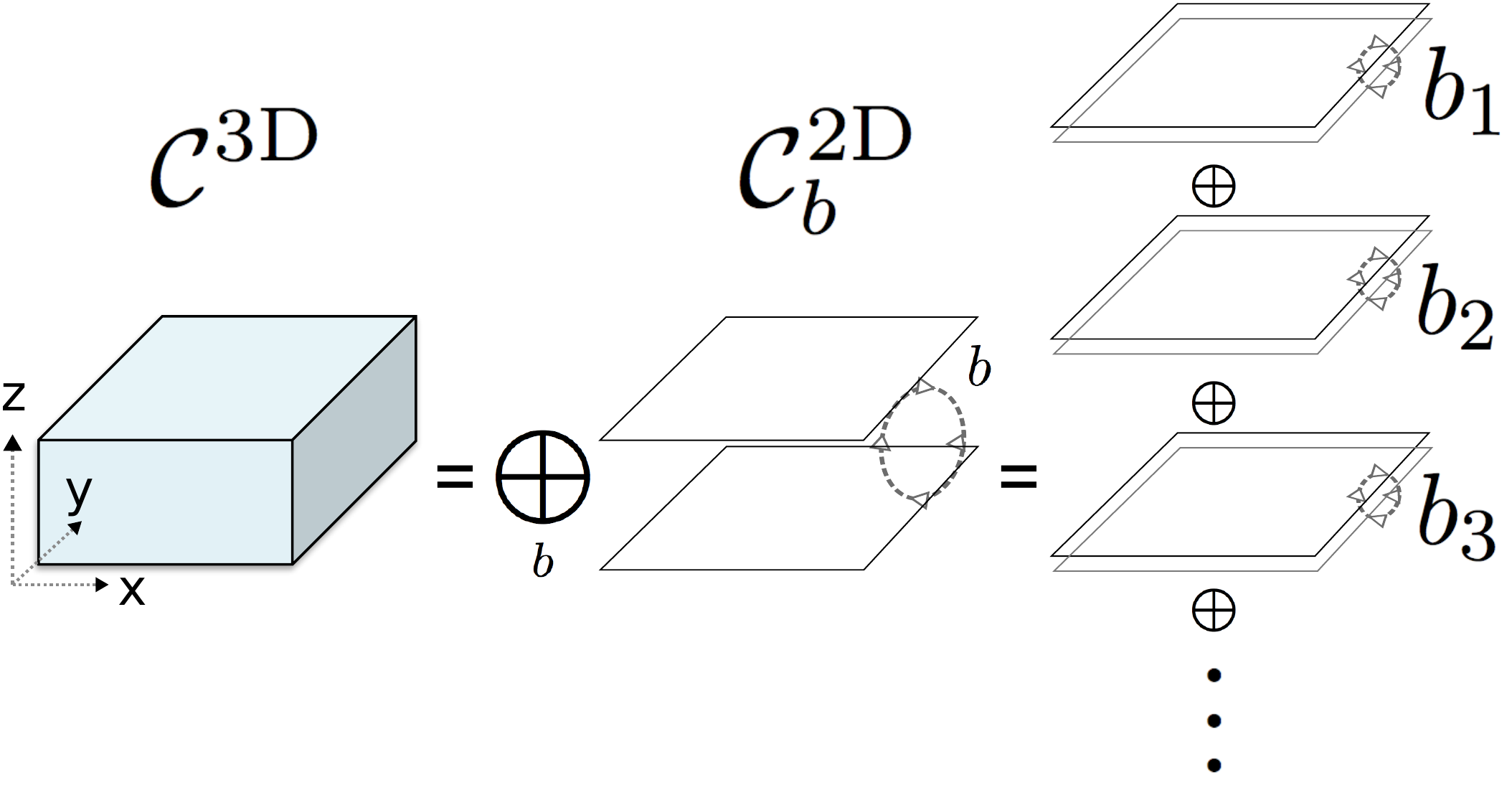}  (b)\includegraphics[scale=0.3]{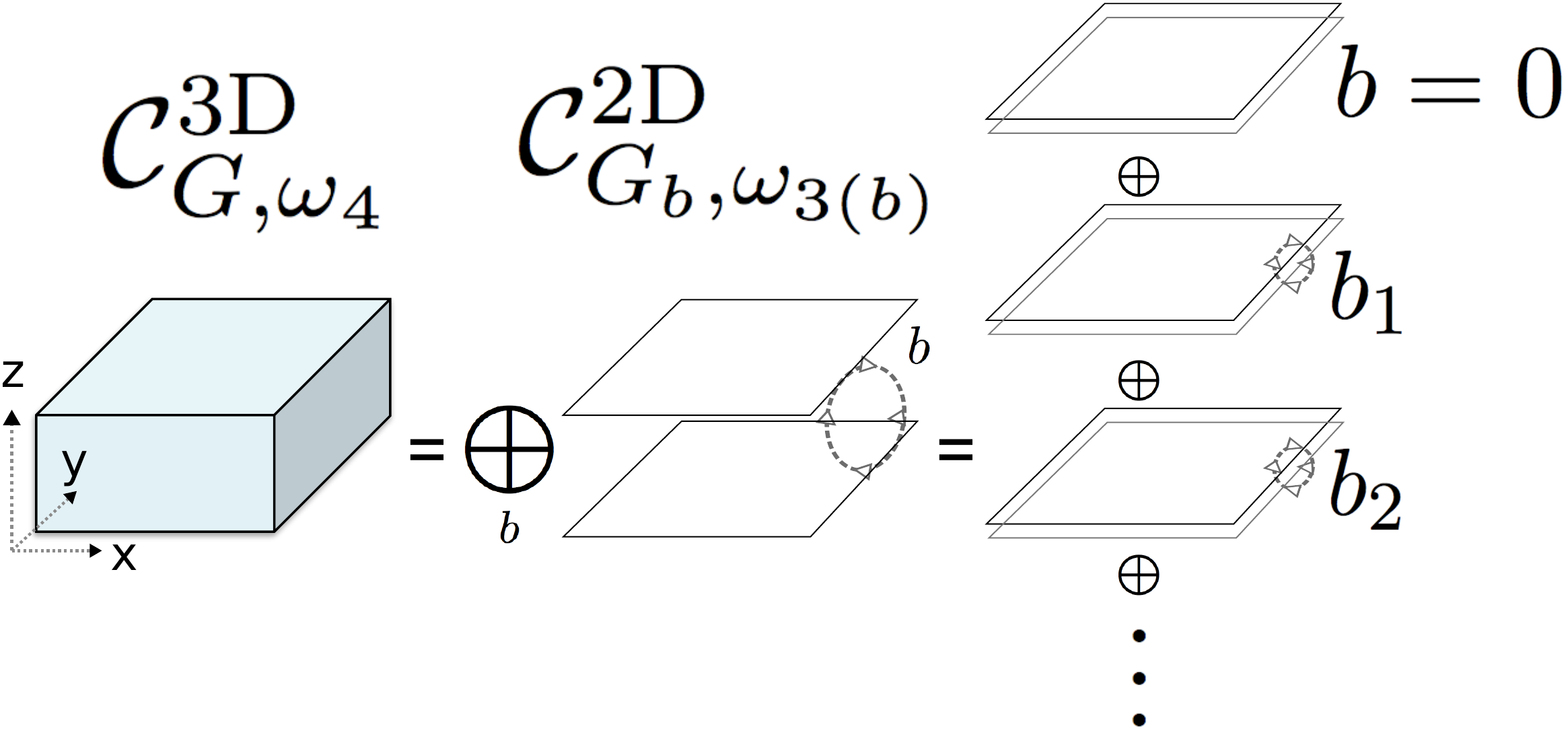}
\end{center}
\caption{(a){
The 3D topological order $\cC^{3\tD}$ can be regarded as  the direct sum of 2D topological orders $\cC^{2\tD}_b$ 
in different sectors $b$, as $\cC^{3\tD} = \oplus_b \cC^{2\tD}_b$, when we compactify a spatial direction $z$ into a circle. 
{\bf This idea is general and applicable to $\cC^{3\tD}$ without a gauge theory description.}
However, when $\cC^{3\tD}$ allows a gauge group $G$ description, the $b$ stands for a group element
(or the conjugacy class for the non-Abelian group) of $G$. 
Thus $b$ acts as a gauge flux 
along the dashed arrow - - -$\vartriangleright$ in the compact direction $z$. 
Thus, $\cC^{3\tD}$ becomes the direct sum of different $\cC^{2\tD}_b$ under distinct gauge fluxes $b$.}
(b){
Combine the reasoning in 
Eq.(\ref{eq:Cb3}) and Fig.\ref{fig:3Dto2D}, we obtain the physical meaning of 
dimensional reduction from a 3+1D twisted gauge theory as a 3D topological order
 to several sectors of 2D topological orders: $\cC^{3\tD}_{G,\omega_4} = \oplus_b \cC_{G,\omega_{3(b)}}^{2\tD}$.
Here $b$ stands for the gauge flux (Wilson line operator) of gauge group $G$. 
Here $\omega_3$ are dimensionally reduced 3-cocycles from 4-cocycles $\omega_4$.
Note that there is a zero flux $b=0$ sector with $\cC_{G,\text{(untwist)}}^{2\tD} =\cC_{G{}}^{2\tD}$. }
} 
\label{fig:3Dto2D} 
\end{figure}

\noindent
($\star$1) {\bf Dimensional Reduction from 3D to 2D: for SL$(3,\Z)$ modular $\mathsf{S}$, $\mathsf{T}$ matrices and cocycles} -
For the first issue, our general philosophy is as follows:\\ 

\noindent
``Since 3D topological orders are foreign and unfamiliar to us, we will \emph{dimensionally reduce 3D topological orders to several sectors of 2D topological orders} in the \emph{Hilbert space
of ground states} (\emph{not in the real space}, see Fig.\ref{fig:3Dto2D}). 
Then we will be able to \emph{borrow the more familiar 2D topological orders to understand 3D topological orders}.''\\

\noindent
We will compute the matrices $\mathsf{S}^{xyz}$ and $\mathsf{T}^{xy}$ that
generate the SL$(3,\Z)$ representation in the quasi-(particle or string)-excitations basis of 3+1D topological order.  
We find an explicit expression of $\mathsf{S}^{xyz}$ and $\mathsf{T}^{xy}$, in terms of the gauge
group $G$ and the 4-cocycle $\om_4$, for both Abelian and non-Abelian gauge
groups.  
We note that SL$(3,\Z)$ contains a subgroup SL$(2,\Z)$, which is
generated by $\hat{\mathsf{S}}^{xy}$ and $\hat{\mathsf{T}}^{xy}$.

In the most generic cases of topological orders (potentially {\it without a gauge group description}), the 
matrices $\mathsf{S}^{xy}$ and $\mathsf{T}^{xy}$ can still be block
diagonalized as the sum of several sectors in the quasi-excitations basis, each sector carrying an index of $b$,
\begin{align} \label{eq:3Dto2DST}
 \mathsf{S}^{xy}=\oplus_b \mathsf{S}^{xy}_b, \;\;\;\; 
 \mathsf{T}^{xy}=\oplus_b \mathsf{T}^{xy}_b,
\end{align}
The pair $(\mathsf{S}^{xy}_b,\mathsf{T}^{xy}_b)$, generating an SL$(2,\Z)$ representation,
describes a 2D topological order $\cC^{2\tD}_b$.
This leads to a dimension reduction of the 3D topological order $\cC^{3\tD}$:
\begin{align} \label{eq:C3DtoC2D}
 \cC^{3\tD} = \oplus_b \cC^{2\tD}_b .
\end{align}

In the more specific case, when the topological order allows a gauge group $G$ description which we focus on here, 
we find that the $b$ stands for a gauge flux for group $G$ (that is, $b$ is a group element for an Abelian $G$, 
while $b$ is a conjugacy class for a non-Abelian $G$). 

The physical picture of the above dimensional reduction is the following (see Fig.\ref{fig:3Dto2D}): 
If we compactify one of the 3D spatial directions (say the $z$ direction) into a small circle,
the 3D topological order $\cC^{3\tD}$ can be viewed as a direct sum of 2D topological orders
$\cC^{2\tD}_b$ with (accidental) degenerate ground states at the lowest energy.

\begin{center} 
\begin{figure}[!ht] 
\includegraphics[scale=0.41]{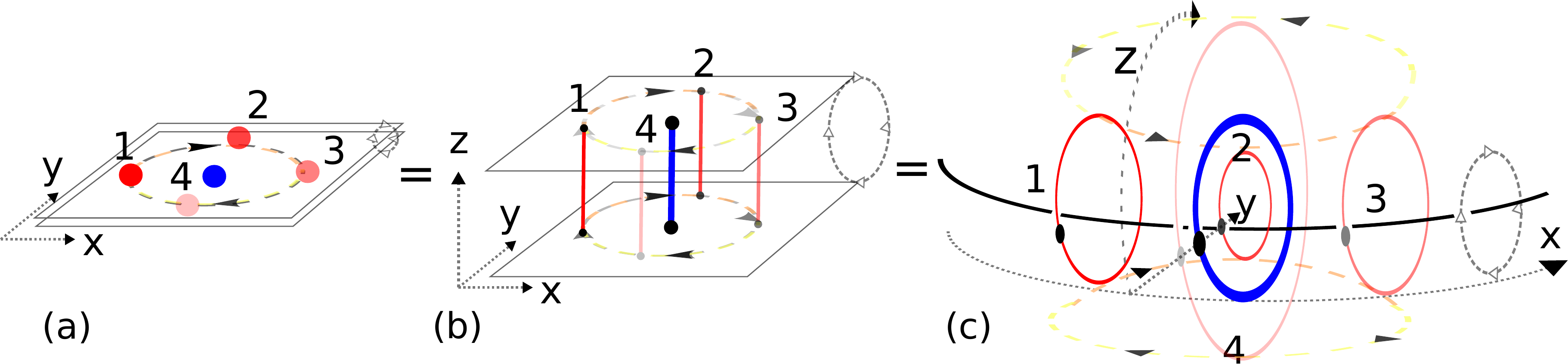} 
\caption{
Mutual braiding statistics following the path $1 \to 2\to 3\to 4 $ along time evolution (see Sec.\ref{Sec:IIIc3string}): 
(a) From a 2D viewpoint of dimensional reduced $\cC^{2\tD}_b$, the 
$2\pi$ braiding 
of two particles is shown.
(b) The compact $z$ direction extends two particles to two closed (red, blue) strings.
(c) An equivalent 3D view, the $b$ flux (along the arrow - - -$\vartriangleright$)  
is regarded as the monodromy caused by a third (black) string.
We identify the coordinates 
$x,y$ and a compact $z$ to see that the full-braiding process
is {\it one (red) string going inside to the loop of another (blue) string, and then going back from the outside.}
{For Abelian topological orders, the mutual braiding process between two excitations (A and B)
in Fig.\ref{fig:3strings_2D_3D_xy}(a) yields a statistical Abelian phase 
${e^{\ti \theta_{\text{(A)(B)} }} \propto {\sfS}^{xy}_{\text{(A)(B)}}}$ proportional to the 2D's $\mathsf{S}^{xy}$ matrix. 
The dimensional-extended equivalent picture Fig.\ref{fig:3strings_2D_3D_xy}(c) implies that the loop-braiding 
yields a phase ${e^{\ti \theta_{\text{(A)(B)}, b}} \propto {\sfS}^{xy}_{b\;\text{(A)(B)}}}$ of Eq.(\ref{eq:Sxyb}) (up to a choice of canonical basis), where $b$ is the flux of the black string.
We clarify that in both (b) and (c) our strings may carry both flux and charge. 
If a string carries only a pure charge, then it is effectively a point particle in 3D.
If a string carries a pure flux, then it is effectively a loop of a pure string in 3D.
If a string carries both charge and flux (as a dyon in 2D), then it is a loop with string fluxes attached with some charged particles in 3D.
Therefore our Fig.\ref{fig:3strings_2D_3D_xy}(c)'s string-string braiding actually represents several braiding processes: the particle-particle, particle-loop and loop-loop braidings, 
all processes are threaded with a background (black) string. 
}
} 
\label{fig:3strings_2D_3D_xy} 
\end{figure}
 \end{center}

In this work, we focus on a generic finite Abelian gauge group $G=\prod_i Z_{N_i} $ (isomorphic to products of cyclic groups) with
generic cocycle twists from the group cohomology.\cite{Dijkgraaf:1989pz} 
We examine the 3+1D twisted gauge theory twisted by 4-cocycle $\omega_4 \in \cH^4(G,\R/\Z)$, and 
reveal that it is a direct sum of 2+1D twisted gauge theories twisted by a dimensionally-reduced 3-cocycle  $\omega_{3(b)} \in \cH^3(G,\R/\Z)$ of
$G$'s third cohomology group, namely
\bea \label{eq:C3DtoC2Dom}
&& \cC^{3\tD}_{G,\omega_4} = \oplus_b \cC_{G_b,\omega_{3(b)}}^{2\tD}.   
\eea
Surprisingly, even for an Abelian group $G$, we find that such a \emph{twisted Abelian gauge theory} can be dual to a twisted or untwisted \emph{non-Abelian gauge theory}. 
We study this fact for 3D as an extension of the 2D examples in Ref.\cite{deWildPropitius:1996gt}. 
By this equivalence, 
we are equipped with (both untwisted and twisted) non-Abelian gauge theory to study its non-Abelian braiding statistics.

($\star$2) {\bf Non-Abelian three-string braiding statistics} -
We are familiar with the 2D braiding statistics: 
there is only \emph{particle-particle braiding}, which yields
bosonic, fermionic or anyonic statistics by braiding a particle around another particle. 
We find that the 3D topological order introduces both particle-like and string-like excitations.  
We aim to address the question:\\

\noindent
{\bf Q2}: ``\emph{How do we characterize the braiding statistics of strings and particles in 3+1D topological orders}?''\\

\noindent
The possible braiding statistics in 3D learned in the past literature are as follows:\\
(i) \emph{Particle-particle braiding}, which can only be bosonic or fermionic due to the absence of nontrivial braid group in 3D for point particles.\\
(ii) \emph{Particle-string braiding}, which is the Aharonov-Bohm effect of $\Z_N$ gauge theory, where
a particle such as a $\Z_N$ charge braiding around a string (or a vortex line) as $\Z_N$ flux, obtaining a $e^{\ti \frac{2 \pi}{N}}$ phase of statistics.
\\
(iii) \emph{String-string braiding}, where a closed string (a red loop), shown in Fig.\ref{fig:3strings_2D_3D_xy}(c) excluding the background black string, 
wraps around a blue loop. 
The related idea known as loop-loop braiding forming the \emph{loop braid group} has been proposed mathematically. 

However, we will 
address 
some extra \emph{new braiding statistics among three closed strings}:\\
(iv) \emph{{Three-string braiding}}, shown in Fig.\ref{fig:3strings_2D_3D_xy}(c), 
where a closed string (a red loop) wraps around another closed string (a blue loop) but the two loops 
are both threaded by a third loop (the black string).
This braiding configuration is discovered recently by Ref.\cite{Wang:2014xba}, 
also 
a related work in Ref.\cite{Jiang:2014ksa} 
for a twisted Abelian gauge theory. 

The new ingredient of our work on braiding statistics can be summarized as follows:
We consider the string and particle braiding of general twisted gauge theories with the most generic finite Abelian gauge group $G=\prod_u Z_{N_u}$, labeled by the data $({G,\omega_4})$. 
We provide a 3D to 2D dimensional reduction approach to realize the three-string braiding statistics of Fig.\ref{fig:3strings_2D_3D_xy}. 
We first show that the SL$(2,\Z)$ representations
$(\mathsf{S}^{xy}_b,\mathsf{T}^{xy}_b)$ fully encode this particular type of Abelian three-closed-string statistics shown in
Fig.\ref{fig:3strings_2D_3D_xy}. 
We further find that, for a twisted gauge theory with
an Abelian $(Z_N)^4$ group, certain 4-cocycles (called as Type IV 4-cocycles) will make the twisted theory to be a
non-Abelian theory. More precisely, {\bf while the two-string braiding statistics of unlink 
is Abelian,
the three-string braiding statistics of Hopf links, 
obtained from threading the two strings with the third string, will become non-Abelian.}
We also demonstrate that $(\mathsf{S}^{xy}_b$ 
encodes this 
three-string braiding statistics.

\subsection{Twisted Gauge Theory and Cocycles of Group Cohomology} \label{sec:II}

In this section, we aim to address the question:

\noindent
{\bf Q3}: ``\emph{How to formulate or construct certain 3+1D topological orders on the lattice}?''

We will consider 3+1D twisted discrete gauge theories. 
Our motivation to study the discrete gauge theory is that it is 
topological and exhibits Aharonov-Bohm phenomena. 
One approach to formulating a discrete gauge theory is the lattice gauge theory.\cite{Kogut:1979wt}
A famous example in both high energy and condensed matter communities 
is the $Z_2$ discrete gauge theory in 2+1D (
also called the $Z_2$ toric code, $Z_2$ spin liquids, $Z_2$ topological order). 
Kitaev's toric code and quantum double model\cite{Kitaev:1997wr} provides a simple Hamiltonian,
\bea  \label{eq:toric}
{H}=-\sum_v A_v -\sum_{p} B_p,
\eea
where a space lattice formalism is used, and $A_v$ is the vertex operator acting on the vertex $v$, $B_p$ is the plaquette (or face) term to ensure the
zero flux condition on each plaquette. Both $A_v$ and $B_p$ consist of only Pauli spin operators for the $Z_2$ model. 
Such ground states of the Hamiltonian are found to be $Z_2$ gauge theory with $|G|^2=4$-fold topological degeneracy on the $\mathbb{T}^2$ torus.
Its generalization to a {\it twisted $Z_2$ gauge theory} is the $Z_2$ double-semions model, captured by the framework of the Levin-Wen {\it string-net model}
\cite{Levin:2004mi,wen2004quantum}.


\subsection{Canonical basis and the generalized twisted quantum double model $D^{\omega}(G)$ to 3D triple basis} \label{Sec:TQD} 
In Chap.\ref{G-phase}, we have formulated a Dijkgraaf-Witten topological gauge theory as higher dimensional TOs.
Thus we have answered the question \emph{{\bf Q3}} using the \emph{spacetime-lattice path integral}.
Our next goal is to \emph{construct its Hamiltonian on the space lattice},
and to find a good basis representing its quasi-excitations, 
such that we can efficiently read the information of ${\sfO}({\sfS}^{xyz},{\sfT}^{xy})$ in this \emph{canonical basis}.
We will outline the \emph{twisted quantum double model} generalized to 3D as the exactly soluble model in the next subsection, 
where the {canonical basis} can diagonalize its Hamiltonian.\\

\noindent
{\bf Canonical basis} - 
For a gauge theory with the gauge group $G$, one may naively think that a good basis for the amplitude Eq.(\ref{eq:Oamp}) is the group elements 
$| g_x, g_y, g_z \rangle$, with $g_i \in G$
as the flux labeling three directions of $\mathbb{T}^3$. 
However, this flux-only label $| g_x, g_y \rangle$ is known to be improper on the $\mathbb{T}^2$ torus already - the canonical basis labeling particles in 2D
is $| \alpha, a \rangle$, requiring both the charge $\alpha$ (as the representation) and the flux $a$ (the group element or the conjugacy class of $G$).
We propose that the proper way to label excitations for a 3+1D twisted discrete gauge theory for any finite 
group $G$
in the canonical basis requires one charge $\alpha$ and two fluxes, $a$ and $b$:
\bea \label{eq:3Dbasis}
|\alpha, a, b \rangle=\frac{1}{\sqrt{|G|}} \sum_{\substack{ {g_y \in C^a,g_z \in C^b}\\{g_x \in Z_{g_y} \cap Z_{g_z}  }} } \Tr[\widetilde{\rho}_\alpha^{g_y,g_z} (g_x) ] |  g_x, g_y, g_z \rangle.\;\;\;\;\;\;
\eea
which is the finite group discrete Fourier transformation on $|  g_x, g_y, g_z \rangle$. 
This is a 
generalization of the 2D result in \cite{deWildPropitius:1996gt} and a very recent 3D Abelian case in \cite{Jiang:2014ksa}.
Here $\alpha$ is the charge of the representation (Rep) label, which is the $\sfC^{(2)}_{a,b}$ Rep of the centralizers $Z_a$, $Z_b$ of the conjugacy classes $C^a, C^b$.
(For an Abelian $G$, the conjugacy class is the group element, and the centralizer is the full $G$.)
$\sfC^{(2)}_{a,b}$ Rep means an inequivalent unitary irreducible projective representation of $G$.
 $\widetilde{\rho}_{\alpha}^{a,b}(c)$ labels this inequivalent unitary irreducible projective $\sfC^{(2)}_{a,b}$ Rep of $G$.
 $\sfC^{(2)}_{a,b}$ is an induced 2-cocycle, dimensionally-reduced from the 4-cocycle $\omega_4$. We illustrate $\sfC^{(2)}_{a,b}$ 
in terms of geometric pictures in Eqs. (\ref{fig:Cabc}) and (\ref{fig:Cabcd}).
\begin{align}
\sfC^{}_{a} (b,c): \begin{matrix}\CaRep\end{matrix}, \label{fig:Cabc}
\end{align}
\begin{align}
\sfC^{(2)}_{a,b} (c,d): \begin{matrix}\CabRepL  \end{matrix} \underset{\DashedArrow}{t \; (d)} \begin{matrix}  \CabRepR \end{matrix} \label{fig:Cabcd}.
\end{align}
{The reduced 2-cocycle $\sfC^{}_{a} (b,c)$ is from the 3-cocycle $\omega_3$ in Eq.(\ref{fig:Cabc}), 
which triangulates a half of $\mathbb{T}^2$ and with a time interval $I$.
The reduced 2-cocycle $\sfC^{}_{a} (b,c)$ is from 4-cocycle $\omega_4$ in Eq.(\ref{fig:Cabcd}), 
which triangulates a half of $\mathbb{T}^3$ and with a time interval $I$.
The dashed arrow $\DashedArrow$ stands for the time $t$ evolution.}

The $\tilde{\rho}_\alpha^{g_y,g_z} (g_x)$ values are determined by the $\sfC^{(2)}_{a,b}$ projective representation formula:
\bea
\label{eq:CabRep}
\widetilde{\rho}_{\alpha}^{a,b}(c)\widetilde{\rho}_{\alpha}^{a,b}(d)=\sfC^{(2)}_{a,b}(c,d)\widetilde{\rho}_{\alpha}^{a,b}(c  d). 
\eea
The trace term $\Tr[\widetilde{\rho}_\alpha^{g_y,g_z} (g_x) ]$ is called {\it the character} in the math literature.
One can view the charge $\alpha_x$ along $x$ direction, the flux $a,b$ along the $y,z$. 

We first 
recall that, in 2D, a reduced 2-cocycle $\sfC^{}_{a} (b,c)$ comes from a slant product $ i_a \omega(b,c)$ of 3-cocycles, \cite{deWildPropitius:1996gt} %
\cite{deWildPropitius:1996gt} 
which is geometrically equivalent to filling three 3-cocycles in a triangular prism of Eq.(\ref{fig:Cabc}).  This is known to
present the {\bf projective representation} 
$\widetilde{\rho}_{\alpha}^{a}(b)\widetilde{\rho}_{\alpha}^{a}(c)=\sfC^{}_{a}(b,c)\widetilde{\rho}_{\alpha}^{a}(bc)$,
because the induced 2-cocycle belongs to the second cohomology group $\cH^2(G,\R/\Z)$.\cite{Mesaros:2012yd}
(See its explicit triangulation and a novel use of the projective representation in Sec VI.B. of Ref.\cite{Wang:2014tia}.)

Similarly, in 3D, a reduced 2-cocycle $\sfC^{}_{a} (b,c)$ comes from doing {\it twice} of the slant products of 4-cocycles forming the geometry of Eq.(\ref{fig:Cabcd}), and renders
\bea \label{eq:2slant}
\sfC^{(2)}_{a,b}= i_b ( \sfC_a (c,d))=i_b ( i_a \omega(c,d)),
\eea 
presenting the $\sfC^{(2)}_{a,b}$-{projective representation} in Eq.(\ref{eq:CabRep}), 
where $\widetilde{\rho}_{\alpha}^{a,b}(c)$: $(Z_a, Z_b)$ ${\rightarrow}$ $\text{GL}\left(Z_a,Z_b\right)$ can be written
as a matrix in the general linear ($\text{GL}$) group. This 3D generalization for the canonical basis in Eq.(\ref{eq:3Dbasis}) is not only natural, but also
consistent to 2D when we turn off the flux along $z$ direction (e.g. set $b=0$). which reduces 3D's $|\alpha, a, b \rangle$ to $|\alpha, a \rangle$ in the 2D case. \\

\noindent
{\bf Generalizing 2D twisted quantum double model $D^{\omega}(G)$ to 3D: twisted quantum triple model?} -- 
A natural way to combine the Dijkgraaf-Witten theory with Kitaev's quantum double model Hamiltonian approach will enable us to study the
Hamiltonian formalism for the twisted gauge theory, which is achieved 
in Ref.\cite{Hu:2012wx},\cite{Mesaros:2012yd} for 2+1D, 
named as the {\it twisted quantum double model}.
In 2D, the widely-used notation $D^{\omega}(G)$ implies the twisted quantum double model with 
its gauge group $G$ and its cocycle twist $\omega$.
It is straightforward to generalize their results to 3+1D. 

To construct the Hamiltonian on the 3D spatial lattice,
we follow \cite{Hu:2012wx} with the form of the twisted quantum double model Hamiltonian of Eq.(\ref{eq:toric}) and put the system on the $\mathbb{T}^3$ torus. 
However, some modification 
for 3D are adopted: the vertex operator $A_v=|G|^{-1} \sum_{[vv']=g \in G } A_v^g$ acts on the vertices of the lattice 
by lifting the vertex point $v$ to $v'$ living in an extra (fourth) dimension as Eq.(\ref{TQD3D_A}), 
  \begin{align} \label{TQD3D_A}
& \begin{matrix}\TQDthreeD{1}{2}{3}{4}{5'}{5} \end{matrix}
  \end{align}
and one computes the 4-cocycle filling amplitude as $\mathbf{Z}$ in Eq.(\ref{eq:path integral}).
To evaluate Eq.(\ref{TQD3D_A})'s $A_v$ operator acting on the vertex $5$, one effectively lifts $5$ to $5'$, and fill 4-cocycles $\omega$ into this geometry to compute the
amplitude $\mathbf{Z}$ in Eq.(\ref{eq:path integral}).
For this specific 
3D spatial lattice surrounding vertex $5$ with $1,2,3$, and $4$ neighboring vertices, 
there are four 4-cocycles $\omega$ filling
in the amplitude of $A_5^{[55']}$.

A plaquette operator $B_p^{(1)}$ still enforces the zero flux condition on each 2D face (a triangle $p$) spanned by three edges of a triangle.
This will ensure zero flux on each face (along the Wilson loop of a 1-form gauge field).
Moreover, zero flux conditions are required if higher form gauge flux are presented. For example,
for 2-form field, one 
adds an additional $B_p^{(2)}$ to ensure the zero flux on a 3-simplex (a tetrahedron $p$).
Thus, $\sum_p B_p^{}$ in Eq.(\ref{eq:toric}) becomes $\sum_p B_p^{(1)}+ \sum_p B_p^{(2)} +\dots$

Analogous to \cite{Hu:2012wx}, the local operators $A_v$, $B_p$ of the Hamiltonian have nice commuting properties:  
$ [A_v^{g}, A_u^{h}]=0$ { if } $v \neq u$, $ [A_v^{g}, B_p]= [B_p, B_p']=0$,
and also $A_v^{g=[vv']} A_{v'}^h=A_v^{gh}$. 
Notice that $A_g$ defines a ground sate projection operator $\sfP_v=|G|^{-1} \sum_g A^g_v$ if we consider a $\mathbb{T}^3$ torus triangulated in a cube with only a point $v$ 
(all eight points are identified).
It can be shown that both $A_g$ and $\sfP$ as projection operators project other states to the ground state $|\alpha, a, b \rangle$,
and $\sfP |\alpha, a, b \rangle=|\alpha, a, b \rangle$ and 
$A_v |\alpha, a, b \rangle \propto |\alpha, a, b \rangle$. 
Since $[A_v^{g}, B_p]=0$, one can simultaneously diagonalize the Hamiltonian Eq.(\ref{eq:toric}) by this {\it canonical basis} $|\alpha, a, b \rangle$ 
as the {\it ground state basis}.

A similar 3D model has been studied recently in \cite{Jiang:2014ksa}. 
There the zero flux condition is imposed in both the vertex operator as well as the plaquette operator.
Their Hilbert space thus is more constrained than that of \cite{Hu:2012wx} or ours. 
However, in the ground state sector, we expect that the physics is the same.
It is less clear to us whether the name, {\bf twisted quantum double model} and its notation $D^{\omega}(G)$, are still proper usages in 3D or higher dimensions. 
With the quantum double basis $| \alpha, a\rangle$ in 2D generalized to a triple basis $| \alpha, a, b\rangle$ in 3D, 
we are tempted to call it
the {\bf twisted quantum triple model} in 3D. It awaits mathematicians and mathematical physicists to explore more details in the future.
\subsection{Cocycle of $\cH^4(G,\R/\Z)$ and its dimensional reduction} 
To 
study the twisted gauge theory of a finite Abelian group, 
we now provide the explicit data on cohomology group and 4-cocycles. 
Here $\mathcal{H}^{d+1}(G,\R/\Z)=\mathcal{H}^{d+1}(G,\tU(1))$ by $\R/\Z=\tU(1)$, as the ${(d+1)}$th-cohomology group of $G$ over $G$ module U(1).
Each class in $\mathcal{H}^{d+1}(G,\R/\Z)$ corresponds to a distinct $(d+1)$-cocycle.
The different 4-cocycles label the distinct topological terms of 3+1D twisted gauge theories. (However, 
different topological terms may share the same data for topological orders, such as the same modular data $\sfS^{xyz}$ and $\sfT^{xy}$.
Thus different topological terms may describe the same topological order.)  
The 4-cocycles $\omega_4$ are 4-cochains, but additionally satisfy the cocycle condition $\delta \omega=1$. 
The $4$-cochain is a mapping $\omega_{4}^{}(a,b,c,d)$:  $(G)^4 \to \tU(1)$, which inputs
$a,b,c,d \in G$, and outputs a $\tU(1)$ phase.
Furthermore, distinct 4-cocycles are not identified by any 4-coboundary $\delta \Omega_3$.
(Namely, distinct cocycles $\omega_4$ and $\omega'_4$ do {\it not} satisfy $\omega_4/ \omega'_4 = \delta \Omega_3$, for any 3-cochain $\Omega_3$.)
The $4$-cochain satisfies the group multiplication rule:
$
(\omega_{4}\cdot\omega_{4}')(a,b,c,d)= \omega_{4}(a,b,c,d)\cdot \omega_{4}'(a,b,c,d), 
$
thus forms a group $\text{C}^4$,
the 4-cocycle  further forms its subgroup $\text{Z}^4$,
and the 4-coboundary further forms a $\text{Z}^4$'s subgroup $\text{B}^4$ (since $ \delta^2 =1$). In short,
$
\text{B}^4 \subset \text{Z}^4 \subset \text{C}^4.
$
The fourth cohomology group is a kernel $\text{Z}^4$ (the group of 4-cocycle) mod out the image $\text{B}^4$ (the group of 4-coboundaries) relation:
$
\cH^4(G,\R/\Z)= \text{Z}^4 /\text{B}^4.
$
We derive the fourth cohomology group of a generic finite Abelian $G=\prod^k_{i=1} Z_{N_i}$ as
\bea
\label{eq:H4}
\mathcal{H}^4(G,\R/\Z) 
=  \prod_{1 \leq i < j < l<m \leq k} 
         (\Z_{N_{ij}})^2
         \times (\Z_{N_{ijl}})^2 
         \times \Z_{N_{ijlm}}.\;\;\; \;\;\;\;
\eea
We construct generic 4-cocycles (not identified by 4-coboundaries) for each type, 
which had summarized in Table \ref{table:cocyclefact}. 

We name 
the Type II 1st and Type II 2nd 4-cocycles for those with topological term indices:
${p_{{ \tII(ij)}}^{(1st)} } \in \Z_{N_{ij}}$ and ${p_{{ \tII(ij)}}^{(2nd)} } \in \Z_{N_{ij}}$ of Eq.(\ref{eq:H4}).
There are Type III 1st and Type III 2nd 4-cocycles for topological term indices:
${p_{{ \tIII(ijl)}}^{(1st)} } \in \Z_{N_{ijl}}$ and ${p_{{ \tIII(ijl)}}^{(2nd)} } \in \Z_{N_{ijl}}$. 
There is 
also Type IV 4-cocycle topological term 
index: $p_{{ \tIV}{(ijlm)}}\in \Z_{N_{ijlm}}$.

Since we earlier alluded to the relation 
$\cC^{3\tD} = \oplus_b \cC^{2\tD}_b$, 
between 3D topological orders (described by 4-cocycles) as the direct sum of sectors of 2D topological orders (described by 3-cocycles),
we wish to see how the dimensionally-reduced 3-cocycle from 4-cocycles can hint at the $\cC^{2\tD}_b$ theory of 2D.
The geometric interpretation of the induced 3-cocycle $\sfC_b(a,c,d) \equiv i_b \omega_4(a,c,d)$ is derived from the 4-cocycle $\omega_4$:
  \begin{align} \label{eq:Cb3}
&\sfC_b(a,c,d): \begin{matrix}\CbThreeRepL  \end{matrix} \underset{\DashedArrow}{t \; (d)}  \begin{matrix}\CbThreeRepR  \end{matrix} 
  \end{align}
  The combination of Eq.(\ref{eq:Cb3}) (with four 4-cocycles filling) times the contribution of Eq.(\ref{fig:Cabc}) (with three 3-cocycles filling)
will produce Eq.(\ref{fig:Cabcd}) with twelve 4-cocycles filling.
Luckily, the Type II and III $\omega_4$ have a simpler form of $\sfC_b(a,c,d) = \omega_4(a,b,c,d)/\omega_4(b,a,c,d)$, while the reduced form of
Type IV $\omega_4$ is more involved.

This indeed promisingly suggests the relation in Eq.(\ref{eq:C3DtoC2Dom}), $\cC^{3\tD}_{G,\omega_4} = \oplus_b \cC_{G,\omega_{3(b)}}^{2\tD}$ with $G_b=G$ the original group. If we view $b$ as the gauge flux along the $z$ direction, and compactify $z$ into a circle, then a single winding around $z$ acts as a monodromy defect carrying
the gauge flux $b$ (group elements or conjugacy classes).
This implies a geometric picture in Fig.\ref{fig:3Dto2D}.

One can tentatively write down a relation,
\bea \label{eq:C3DtoC2Domb}
\cC^{3\tD}_{G,\omega_4} =  \cC_{G,1\text{(untwist)}}^{2\tD} \oplus _{b \neq 0}  \cC_{G,\omega_{3(b)}}^{2\tD}.
\eea 
There is a zero flux $b=0$ sector $\cC_{G,1\text{(untwist)}}^{2\tD} $ (with $\omega_3=1$) where the 2D gauge theory with $G$ is untwisted. 
There are other direct sums of $\cC_{G,\omega_{3(b)}}^{2\tD}$ with nonzero $b$ flux insertion that have twisted $\omega_{3(b)}$. 

However, 
different cocycles can represent the same topological order with the equivalent modular data,
in the next section, we should examine this Eq.(\ref{eq:C3DtoC2Domb}) more carefully not in terms of cocycles, but in terms of the modular data $\sfS^{xyz}$ and $\sfT^{xy}$. 

\subsection{Representation for $\sfS^{xyz}$ and $\sfT^{xy}$}  \label{sec:RepST}

\noindent
{\bf Q4}: ``\emph{What are the generic expressions of SL(3,$\Z$) modular data}?'' 

First, in Sec.\ref{sec:PT}, we apply the \emph{cocycle approach} using the \emph{spacetime path integral} with 
SL$(3,\Z)$ transformation acting along the time evolution to formulate the SL$(3,\Z)$ modular data,
and then in Sec \ref{Rep} we use the more powerful \emph{Representation} (Rep) \emph{Theory} to determine the general expressions of those data in terms of $(G,\omega_4)$.

\subsubsection{Path Integral and Cocycle approach} \label{STcocycle}

{
The cocycles approach uses the spacetime lattice formalism, where we triangulate the spacetime complex of a 4-manifold $\cM=\mathbb{T}^3 \times I$,
(a $\mathbb{T}^3$ torus times a time interval $I$) of 
Eq.(\ref{TriSxy}),(\ref{TriTxy}),(\ref{TriSxyz}) into 4-simplices.
We then apply the path integral $\mathbf{Z}$ in Eq.(\ref{eq:path integral}) and the amplitude form in Eq.(\ref{eq:Oamp}) to obtain
\bea 
&&{\sfT}^{xy}_{\text{(A)(B)}}= \langle   \Psi_{\text{A}}  |  \hat{\sfT}^{xy} | \Psi_{\text{B}}\rangle,\\
&&{\sfS}^{xy}_{\text{(A)(B)}}= \langle  \Psi_{\text{A}}  |  \hat{\sfS}^{xy} | \Psi_{\text{B}}\rangle,\\
&&{\sfS}^{xyz}_{\text{(A)(B)}}= \langle \Psi_{\text{A}}  |   \hat{\sfS}^{xyz} | \Psi_{\text{B}}\rangle,\\
&&\text{GSD}=\Tr[\sfP] =\sum_A \langle  \Psi_{\text{A}}  |  \sfP | \Psi_{\text{A}}\rangle. \label{eq:GSDpathZ}
\eea
Here $| \Psi_{\text{A}}\rangle$ and $| \Psi_{\text{B}}\rangle$ are ground state bases on the $\mathbb{T}^d$ torus, for example, they are
$|\alpha, a\rangle$ (with $\alpha$ charge and $a$ flux) in 2+1D and $|\alpha, a, b\rangle$ (with $\alpha$ charge and $a, b$ fluxes) in 3+1D.
We also include the data of GSD, where the $\sfP$ is the projection operator to ground states discussed in Sec.\ref{Sec:TQD}.
In the case of d-D GSD on $\mathbb{T}^d$ (e.g. 3D GSD on $\mathbb{T}^3$), we simply compute the $\mathbf{Z}$ amplitude filling in $\mathbb{T}^d \times S^1= \mathbb{T}^{d+1}$.
There is no short cut here except doing explicit calculations. 
}
\subsubsection{Representation Theory approach} \label{Rep}

The cocycle approach in Sec.\ref{STcocycle} provides nice physical intuition about the modular transformation process. 
However, the calculation is tedious. There is a powerful approach simply using Representation Theory, we will present the 
general formula of $\hat{\sfS}^{xys}$, $\hat{\sfT}^{xy}$, $\hat{\sfS}^{xy}$ data in terms of $(G, \omega_4)$ directly.
The three steps are outlined as follows:\\ 
(i) Obtain the Eq.(\ref{eq:2slant})'s $\sfC^{(2)}_{a,b}$ by doing the slant product twice from 4-cocycle $\omega_4$,
or triangulating Eq.\ref{fig:Cabc}. 
(ii) Derive $\widetilde{\rho}_{\alpha}^{a,b}(c)$ of $\sfC^{(2)}_{a,b}$-{projective representation} in Eq.(\ref{eq:CabRep}), 
which $\widetilde{\rho}_{\alpha}^{a,b}(c)$ 
is a general linear matrix.\\
(iii) 
Write the modular data in the canonical basis $|\alpha, a,b\rangle$, $|\beta, c,d\rangle$ of Eq.(\ref{eq:3Dbasis}). 
  
  After some long computations, 
  we find 
the most general formula $\sfS^{xyz}$ for a group $G$ (both Abelian or non-Abelian) with cocycle twist $\omega_4$:
\bea \label{Eq.Sxyz} 
&& {\sfS^{xyz}_{(\alpha, a,b)(\beta, c,d)}=\frac{1}{|G|}  \langle \alpha_x, a_y, b_z  | \sum_w \sfS^{xyz}_w | \beta_{x'}, c_{y'}, d_{z'}\rangle }\\
&&{ =\frac{1}{|G|}\sum_{\substack{{g_y\in C^a \cap Z_{g_z} \cap  Z_{g_x},}\\{g_z\in C^b \cap C^c,}\\{g_x\in Z_{g_y} \cap  Z_{g_z} \cap C^d }   }} 
  \text{Tr}\widetilde{\rho}^{g_y, g_z}_{\alpha_x}(g_x)^{*}  \; \text{Tr}\widetilde{\rho}^{g_z,g_x}_{\beta_{y}}(g_y)^{} \delta_{g_x,h_{z'}  } \delta_{g_y,h_{x'}} \delta_{g_z,h_{y'}}}. \;\;\;\;\;\;\nonumber
\eea
Here $C^a,C^b,C^c,C^d$ are conjugacy classes of the group elements $a,b,c,d \in G$. In the case of a non-Abelian $G$, we should regard $a,b$ as 
its conjugacy class $C^a,C^b$ in $|\alpha, a,b\rangle$. $Z_g$ means the centralizer of the conjugacy class of $g$.
For an Abelian $G$, it simplifies to 
\bea 
&& {\sfS^{xyz}_{(\alpha, a,b)(\beta, c,d)}= 
 { \frac{1}{|G|}  
  \text{Tr}\widetilde{\rho}^{a,b}_{\alpha}(d)^{*}  \; \text{Tr}\widetilde{\rho}^{b,d}_{\beta}(a)^{} \delta_{b,c} }
 \equiv \frac{1}{|G|}  \sfS^{\alpha,\beta}_{d,a,b} \delta_{b,c}} \;\;\;\;\;\;\nonumber \\
&&= { 
 { \frac{1}{|G|}  
  \text{Tr}\widetilde{\rho}^{a_y,b_z}_{\alpha_x}(d_{z'})^{*}   \text{Tr}\widetilde{\rho}^{b_z,d_{z'}}_{\beta_{x'}}(a_y)^{} \delta_{b_z,c_{y'}} }
 \equiv \frac{1}{|G|}  \sfS^{\alpha_x,\beta_{y}}_{d_x,a_y,b_z} \delta_{b_z,c_{y'}}}.
 \nonumber
\eea
We write $\beta_{x'}=\beta_{y}$, $d_{z'}=d_x$ due to the coordinate identification under $\hat{\mathsf{S}}^{xyz}$.
The assignment of the directions of gauge fluxes (group elements) are clearly expressed in the second line. 
We may use the first line expression for simplicity.

We also provide the most general formula of $\sfT^{xy}$ in the $|\alpha, a,b\rangle$ basis:
\bea \label{eq:Txy} 
&& {{ \sfT^{xy}=\sfT^{a_y,b_z}_{\alpha_x} =\frac{\text{Tr} \widetilde{\rho}^{a_y,b_z}_{\alpha_x}(a_y)}{\text{dim}({\alpha})} }} \equiv \exp(\ti \Theta^{a_y,b_z}_{\alpha_x}).
\eea
Here ${\text{dim}({\alpha})}$ means the dimension of the representation, equivalently the rank of the
matrix of $\widetilde{\rho}^{a,b}_{\alpha_x}(c)$.
Since SL$(2,\Z)$ is a subgroup of SL$(3,\Z)$, we can express the SL$(2,\Z)$'s $\sfS^{xy}$ by SL$(3,\Z)$'s $\sfS^{xyz}$ and $\sfT^{xy}$
(an expression for both the real spatial basis and the canonical basis): 
\bea \label{eq:Sxydecompose}
&& \sfS^{xy}=((\sfT^{xy})^{-1} \sfS^{xyz})^3 (\sfS^{xyz}\sfT^{xy})^2  \sfS^{xyz} (\sfT^{xy})^{-1}.
\eea

For an Abelian $G$, and when $\sfC^{(2)}_{a,b}(c,d)$ is a 2-coboundary (cohomologically trivial), the dimensionality of Rep is $\text{dim}(\text{Rep}) \equiv \text{dim}(\alpha)=1$, 
and the $\sfS^{xy}$ is simplified:
\bea \label{Eq:SxyAbSimp}
&& 
{ \sfS^{xy}_{(\alpha,a,b)({\beta,c,d})}=  { \frac{1}{|G|}  } 
{ \frac{ \text{tr}\widetilde{\rho}^{a,b}_{\alpha}(a c^{-1})^{*}   }{\text{tr} \widetilde{\rho}^{a,b}_{\alpha}(a)}}
  \frac{  {  \; {\text{tr} \widetilde{\rho}^{c, d}_{\beta}(a c^{-1})}  }  }{\text{tr}\widetilde{\rho}^{c,d}_{\beta}(c)^{} }
  \delta_{b,d} }. \;\;\;\;\;\;
\eea

We can verify the above results by first computing the cocycle path integral approach in Sec.\ref{STcocycle}, and substituting from the flux basis to the canonical basis
by Eq.(\ref{eq:3Dbasis}). We have made several consistent checks, by comparing our $\hat{\sfS}^{xy}$, $\hat{\sfT}^{xy}$, $\hat{\sfS}^{xyz}$ to: (1) the known 2D case for the untwisted theory of a non-Abelian group, 
(2) the recent 3D case for the untwisted theory of a non-Abelian group, 
(3) the recent 3D case for the twisted theory of an Abelian group.
And our expression works for all cases: the (un)twisted theory of (non-)Abelian group. 


\subsubsection{Physics of $\sfS$ and $\sfT$ in 3D}  \label{Sec:physicsST}
The ${\sfS}^{xy}$ and ${\sfT}^{xy}$ in 2D are known to have precise physical meanings. 
At least for Abelian topological orders, there is no ambiguity that 
${\sfS}^{xy}$ in the quasiparticle basis provides the mutual statistics of two particles (winding one around the other by $2\pi$), while
${\sfT}^{xy}$ in the quasiparticle basis provides the self statistics of two identical particles (winding one around the other by $\pi$).
Moreover, the intimate spin-statistics relation   
shows that the statistical phase $e^{\ti \Theta}$ gained by interchanging 
two identical particles is equal to the spin $s$ by $e^{\ti 2 \pi s}$. 
Fig.\ref{fig:spin_statistics} illustrates the {\it spin-statistics relation}.\cite{Finkelstein:1968hy} Thus, people also call ${\sfT}^{xy}$ in 2D as the {\it topological spin}.
Here we ask:\\ 
\noindent
{\bf Q5}: ``\emph{What is the physical interpretation of} SL(3,$\Z$) \emph{modular data in 3D}?''

Our approach again is by dimensional reduction 
of Fig.\ref{fig:3Dto2D}, via Eq.(\ref{eq:3Dto2DST}) and Eq.(\ref{eq:C3DtoC2D}): 
$\mathsf{S}^{xy}=\oplus_b \mathsf{S}^{xy}_b$, $\mathsf{T}^{xy}=\oplus_b \mathsf{T}^{xy}_b$, $ \cC^{3\tD} = \oplus_b \cC^{2\tD}_b$,
reducing the 3D physics to the direct sum of 2D topological phases in different flux sectors,
so we can retrieve the familiar physics of 2D to interpret 3D.\\

For our case with a gauge group 
description, the $b$ (subindex of $\mathsf{S}^{xy}_b$, $\mathsf{T}^{xy}_b$, $\cC^{2\tD}_b$)
labels the gauge flux (group element or conjugacy class $C^b$) winding around the compact $z$ direction in Fig.\ref{fig:3Dto2D}.
This $b$ flux can be viewed as the by-product of a {\it monodromy defect} causing a {\it branch cut} 
(\emph{a symmetry twist}\cite{WenSPTinv, Wang:2014tia,{Santos:2013uda},{Wang:2014pma}}),  
such that the wavefunction will gain a {\it phase} by winding around the compact $z$ direction.
Now we further regard the $b$ flux as a {\it string threading around} in the background, so that winding around this background string (e.g. the black string threading in
Fig.\ref{fig:3strings_2D_3D_xy}(c),\ref{fig:Txy}(c),\ref{fig:Sxyhalf}(c)) gains the $b$ flux effect if there is a nontrivial winding on the compact direction $z$.
The arrow - - -$\vartriangleright$ along the compact $z$ schematically indicates such a $b$ flux effect from the background string threading.
\begin{figure}[!h] 
\centering
\includegraphics[scale=0.25]{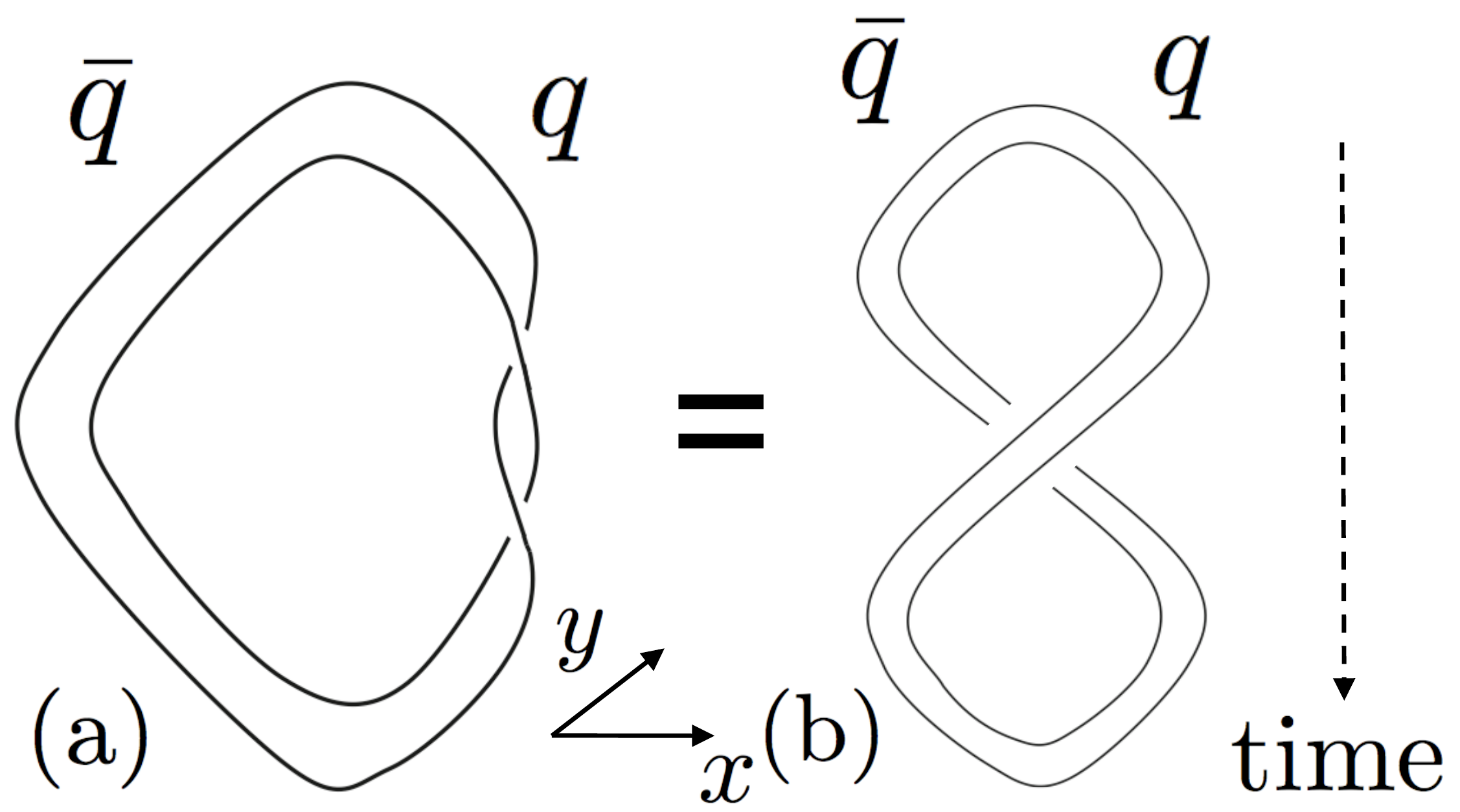}
\caption{
Both process (a) and process (b) start from 
the creation of a pair of particle $q$ and anti-particle $\bar{q}$, but the wordlines evolve along time to the bottom differently.
Process (a) produces a phase $e^{\ti 2 \pi s}$ due to $2\pi$ rotation of q, with spin $s$. 
Process (b) produces a phase $e^{\ti \Theta}$ 
due to the 
exchange statistics.
The homotopic equivalence by deformation implies $e^{\ti 2 \pi s}=e^{\ti \Theta}$.
}
\label{fig:spin_statistics} 
\end{figure}  

\paragraph{${\sfT}^{xy}_b$ and topological spin of a closed string}
We apply the above idea to interpret ${\sfT}^{xy}_b$, shown in Fig.\ref{fig:Txy}. From Eq.(\ref{eq:Txy}), we have
${ {\sfT}^{xy}_b=\sfT^{a_y,b_z}_{\alpha_x}}$ ${\equiv \exp(\ti \Theta^{a_y,b_z}_{\alpha_x})}$ with a fixed $b_z$ label for a given $b_z$ flux sector.
For each $b$, ${\sfT}^{xy}_b$ acts as a familiar 2D ${\sfT}$ matrix $\sfT^{a_y}_{\alpha_x}$, which provides the topological spin of a quasiparticle $(\alpha,a)$
with charge $\alpha$ and flux $a$, in Fig.\ref{fig:Txy}(a).

From the 3D viewpoint, however, this $|\alpha, a\rangle$ particle is actually a closed string 
compactified along the compact $z$ direction.
Thus, in Fig.\ref{fig:Txy}(b), the self-$2\pi$ rotation of the topological spin of a quasiparticle $|\alpha, a\rangle$ is indeed the self-$2\pi$ rotation of a framed closed string.
(Physically we understand that there the string can be {\it framed} with arrows, because the inner texture of the string excitations are allowed in a condensed matter system,
due to defects or the finite size lattice geometry.)
Moreover, from an equivalent 3D view in Fig.\ref{fig:Txy}(c), we can view the self-$2\pi$ rotation of a framed closed string as the {\it self-$2\pi$ flipping} of a framed closed string, which
flips the string inside-out and then outside-in back to its original status. 
This picture works for both 
the $b=0$ zero flux sector and the 
$b \neq 0$ sector under the background string threading. {\bf We thus propose ${ {\sfT}^{xy}_b}$ as the topological spin of a framed closed string,
threaded by a  background string carrying a monodromy $b$ flux.}

\begin{center} 
\begin{figure}[!h] 
\centering
\includegraphics[scale=0.46]{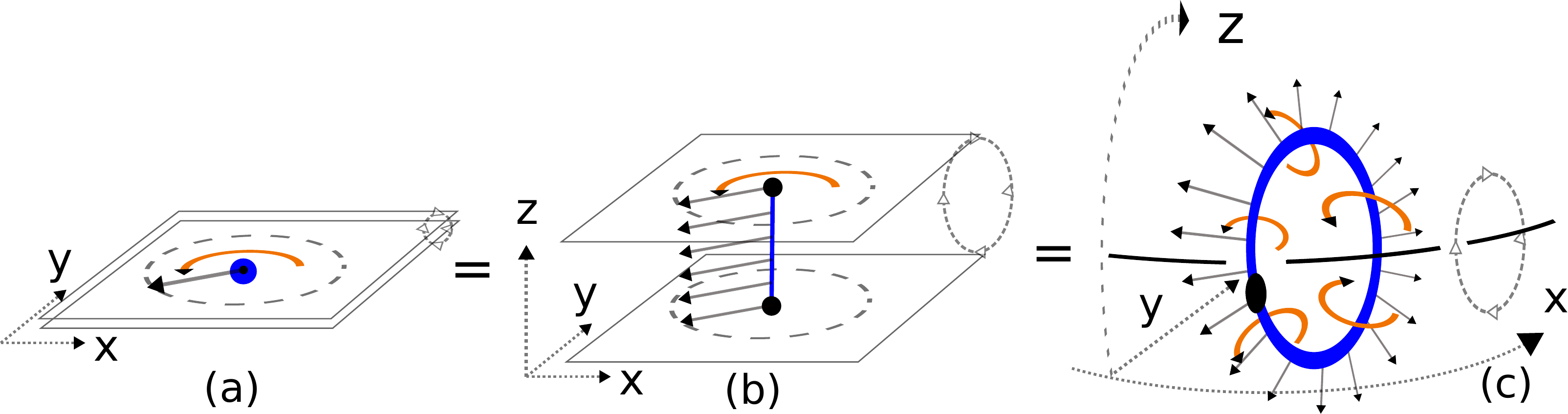}
\caption{
Topological spin of (a) a particle by $2\pi$-self rotation in 2D, (b) a framed closed-string by $2\pi$-self rotation in 3D with a compact $z$,
(c) a closed-string (blue) by $2\pi$-self flipping, threaded by a background (black) string creating monodromy $b$ flux (along the arrow - - -$\vartriangleright$), under a single Hopf link $2^2_1$ configuration.
All above equivalent pictures describe the physics of topological spin in terms of  ${\sfT}^{xy}_b$. 
{For Abelian topological orders, the spin of an excitation (say A) in Fig.\ref{fig:Txy}(a) yields an
Abelian phase 
${e^{\ti \Theta_{\text{(A)} }} = {\sfT}^{xy}_{\text{(A)(A)}}}$ proportional to the diagonal of the 2D's $\mathsf{T}^{xy}$ matrix. 
The dimensional-extended equivalent picture Fig.\ref{fig:Txy}(c) implies that the loop-flipping
yields a phase ${e^{\ti \Theta_{\text{(A)},b }} = {\sfT}^{xy}_{b\;\text{(A)(A)}}}$ of Eq.(\ref{eq:Txy}) (up to a choice of canonical basis), where $b$ is the flux of the black string.}
} 
\label{fig:Txy} 
\end{figure}
\begin{figure}[!h] 
\centering
\includegraphics[scale=0.41]{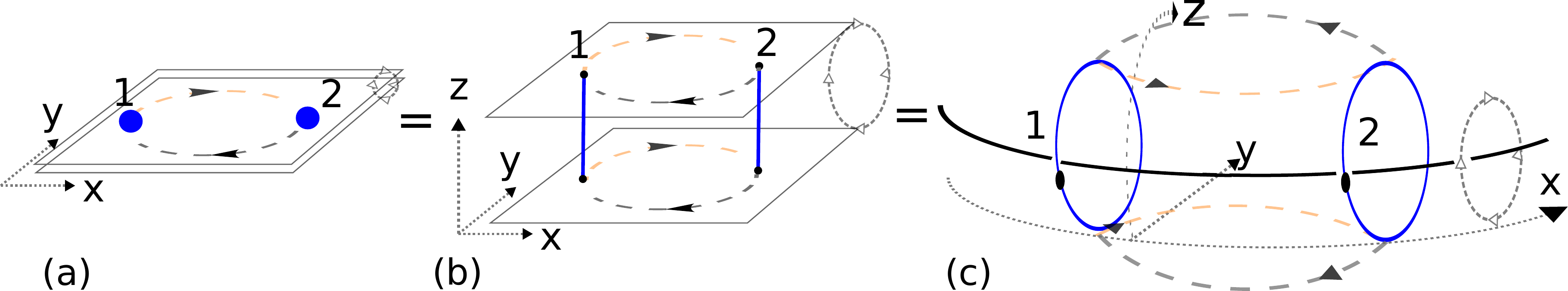}  
\caption{
Exchange statistics of (a) two identical particles at positions 1 and 2 by a $\pi$ winding (half-winding), 
(b) two identical strings by a $\pi$ winding 
in 3D with a compact $z$,
(c) two identical closed-strings (blue) with a $\pi$-winding 
around, both threaded by a background (black) string creating monodromy $b$ flux, under 
the Hopf links $2^2_1 \# 2^2_1$ configuration. Here figures (a)(b)(c) describe the equivalent physics in 3D with a compact $z$ direction.
The physics of exchange statistics of a closed string turns out to be related to the topological spin in Fig.\ref{fig:Txy},
discussed in Sec.\ref{spin-statistics}. 
} 
\label{fig:Sxyhalf} 
\end{figure}
 \end{center}

\paragraph{${\sfS}^{xy}_b$ and three-string braiding statistics} \label{Sec:IIIc3string}

Similarly, we apply the same philosophy to do 3D to 2D reduction for ${\sfS}^{xy}_b$, each effective 2D threading with a distinct gauge flux $b$. 
We can obtain $\mathsf{S}^{xy}_b$ from Eq.(\ref{eq:Sxydecompose}) with SL$(3,\Z)$ modular data.
Here we will focus on interpreting $\mathsf{S}^{xy}_b$ in the Abelian topological order.
Writing $\mathsf{S}^{xy}_b$ in the canonical basis $|\alpha, a,b\rangle$, $|\beta, c,d\rangle$ of Eq.(\ref{eq:3Dbasis}),
we find that, true for Abelian topological order
\bea \label{eq:Sxyb}
\mathsf{S}^{xy}_b 
= \sfS^{xy}_{(\alpha,a,b)({\beta,c,d})} \equiv \frac{1}{|G|} \sfS^{2\tD\;\alpha,\beta}_{a,c\;(b)} \delta_{b,d}.
\eea
As we predict the generality in Eq.(\ref{eq:3Dto2DST}), the $\mathsf{S}^{xy}_b$ here is diagonalized with the $b$ and $d$ identified (as the z-direction flux created by the background string threading).
For a given fixed $b$ flux sector, the only free indices are $|\alpha, a\rangle$ and $|\beta, c\rangle$, all collected in $\sfS^{2\tD\;\alpha,\beta}_{a,c\;(b)}$.
(Explicit data will be presented in Sec.\ref{TypeII,III twist})
Our interpretation is shown in Fig.\ref{fig:3strings_2D_3D_xy}.
From a 2D viewpoint, $\mathsf{S}^{xy}_b$ gives the full $2\pi$ braiding statistics data 
of two quasiparticle $|\alpha, a\rangle$ and $|\beta, c\rangle$  excitations in Fig.\ref{fig:3strings_2D_3D_xy}(a). 
However, from the 3D viewpoint, the two particles are actually two closed strings compactified along the compact $z$ direction.
Thus, the full-$2\pi$ braiding of two particles in Fig.\ref{fig:3strings_2D_3D_xy}(a) becomes that of two closed-strings in Fig.\ref{fig:3strings_2D_3D_xy}(b).
More explicitly, an equivalent 3D view in Fig.\ref{fig:3strings_2D_3D_xy}(c), we identify the coordinates $x,y,z$ carefully to see such a full-braiding process
is that {\it one (red) string going inside to the loop of another (blue) string, and then going back from the outside.}

The above picture  works again for both the $b=0$ zero flux sector as well as the $b \neq 0$ sector under the background string threading.
When $b \neq 0$, the third (black) background string in Fig.\ref{fig:3strings_2D_3D_xy}(c) threading through the two (red, blue) strings. 
The third (black) string creates the monodromy defect/branch cut on the background, 
and carrying $b$ flux along $z$ acting on two (red, blue) strings which have nontrivial winding on the third string.
This three-string braiding was first emphasized in a recent paper,\cite{Wang:2014xba} 
here we make further connection to the data ${ {\sfS}^{xy}_b}$ and understand its physics in a 3D to 2D under $b$ flux sectors.

{\bf We have proposed and shown that ${ {\sfS}^{xy}_b}$ can capture the physics of three-string braiding statistics with
two strings threaded by a third background string causing $b$ flux monodromy, where the three strings have
the linking configuration as the connected sum of two Hopf links $2^2_1\# 2^2_1$.}

\paragraph{Spin-Statistics relation for closed strings} \label{spin-statistics}

Since a spin-statistics relation for 2D particles is shown by Fig.\ref{fig:spin_statistics}, 
we may wonder, by using our 3D to 2D reduction picture, whether a {\it spin-statistics relation for a closed string} holds?

To answer this question, we should compare the {\it topological spin} picture of $\mathsf{T}^{xy}_b=\sfT^{a_y,b_z}_{\alpha_x}$ ${\equiv \exp(\ti \Theta^{a_y,b_z}_{\alpha_x})} $
to the {\it exchange statistic} picture of two closed strings in Fig.\ref{fig:Sxyhalf}. 
Fig.\ref{fig:Sxyhalf} essentially takes a {\it half-braiding} of the $\mathsf{S}^{xy}_b$ process of Fig.\ref{fig:3strings_2D_3D_xy},
and considers doing half-braiding on the same excitations in $|\alpha, a,b\rangle=|\beta, c,d\rangle$.
In principle, one can generalize the framed worldline picture of particles in Fig.\ref{fig:spin_statistics} to the framed worldsheet picture of closed-strings.
(ps. The framed worldline is like a worldsheet, the framed worldsheet is like a worldvolume.)
This interpretation shows that the topological spin of Fig.\ref{fig:Txy} and the exchange statistics of Fig.\ref{fig:Sxyhalf} carry the same data, namely
\bea \label{Spin-Statistics3D}
 {\sfT}^{xy}_b=\sfT^{a_y,b_z}_{\alpha_x} =(\sfS^{2\tD\;\alpha_x,\alpha_x}_{a_y,a_y\;(b_z)})^{\frac{1}{2}} \text{ or } (\sfS^{2\tD\;\alpha_x,\alpha_x}_{a_y,a_y\;(b_z)})^{\frac{1}{2}*} 
\eea
from the data of Eq.(\ref{eq:Txy}),(\ref{eq:Sxyb}). The equivalence holds, up to a (complex conjugate ${}^*$) sign caused by the orientation of the rotation and the exchange.

In Sec.(\ref{TypeII,III twist}), we will show, 
for the twisted gauge theory of Abelian topological orders, such an interpretation Eq.\ref{Spin-Statistics3D} is correct and agrees with our data. 
We term 
this as the {\bf spin-statistics relation for a closed string}.

In this section, we have obtained the explicit formulas of  $\mathsf{S}^{xyz}$,  $\mathsf{T}^{xy}$,  $\mathsf{S}^{xy}$ in Sec.\ref{STcocycle},\ref{Rep}, 
and as well as  captured the physical meanings of $\mathsf{S}^{xy}_b$,  $\mathsf{T}^{xy}_b$ in Sec.\ref{spin-statistics}.
Before concluding, we note that the full understanding of ${\sfS}^{xyz}$ seems to be intriguingly related to the 3D nature. 
It is {\it not} obvious to us that the use of 3D to 2D reduction can capture all physics of ${\sfS}^{xyz}$. 
We will come back to comment this issue in the Sec.\ref{summary}.


\subsection{SL$(3,\mathbb{Z})$ Modular Data and Multi-String Braiding} \label{Sec:SL3ZMulti-String Braiding}

\subsubsection{Ground state degeneracy and Particle, String types} \label{sec:GSD} 

We now proceed to study the topology-dependent ground state degeneracy (GSD), modular data $\mathsf{S}$, $\mathsf{T}$ of 3+1D twisted gauge theory 
with finite group $G=\prod_i Z_{N_i}$. 
We shall comment that the GSD on $\mathbb{T}^2$ of 2D topological order counts the number of quasi-particle excitations, 
which from the Representation (Rep) Theory is simply counting the number of charges $\alpha$ and  fluxes $a$ forming the quasi-particle basis $|\alpha, a \rangle$ spanned the ground state Hilbert space.
{\bf In 2D, GSD counts the number of types of quasi-particles} (or anyons) as well as {\bf the number of basis $|\alpha, a \rangle$.}
For higher dimension,  GSD on $\mathbb{T}^d$ of $d$-D topological order still counts {\bf the number of canonical basis $|\alpha, a, b, \dots \rangle$},
however, may over count the number of types of {\bf particles (with charge), strings (with flux)}, etc excitations. 
From a untwisted $Z_N$ field theory perspective, the fluxed string may be described by a 2-form $B$ field, and the charged particle
may be described by a 1-form $A$ field, with a BF action $\int BdA$.
As we can see the fluxes $a,b$ are over-counted.\\

We suggest that counting the number of types of particles of $d$-dimensions is equivalent to Fig.\ref{fig:qpSphere} process, 
\begin{figure}[!h] 
\centering
\includegraphics[scale=0.35]{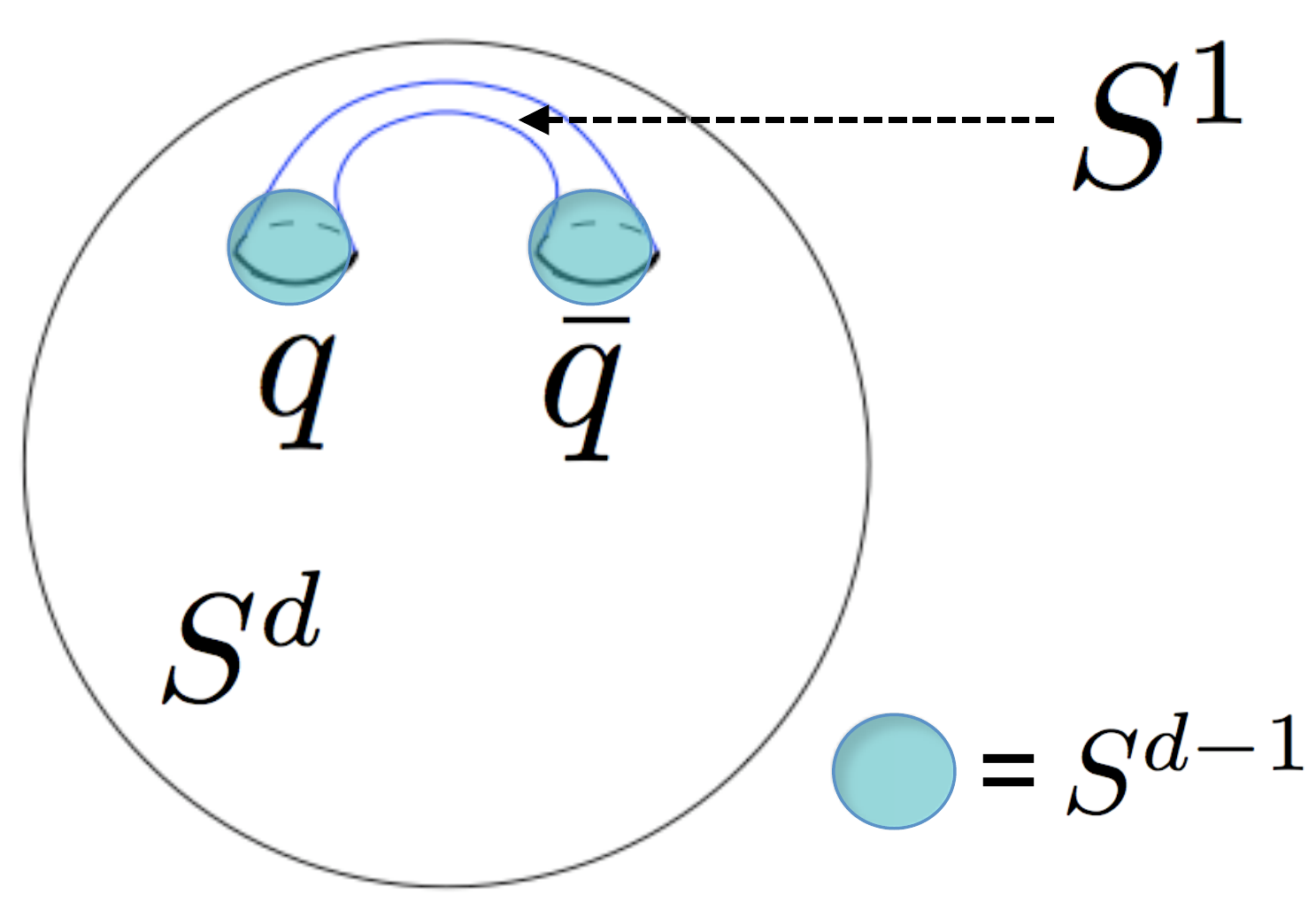}  
\caption{
Number of particle types = GSD on $S^{d-1} \times S^1$.
}
\label{fig:qpSphere} 
\end{figure} 
where we dig a ball $B^d$ with a sphere $S^{d-1}$ around the particle $q$, which resides on $S^d$. And we connect it through a $S^1$ tunnel to its anti-particle $\bar{q}$.
This process causes creation-annihilation from vacuum, and counts how many types of $q$ sectors is equivalent to:
\bea \label{eq:num_part}
\text{the number of particle types}=\text{GSD on } S^{d-1} \times I.\;\;\;
\eea
with $I \simeq S^1$ for this example. For the spacetime integral, one evaluates Eq.(\ref{eq:GSDpathZ}) on $\cM=S^{d-1} \times S^1 \times S^1$.

For counting closed string excitations, one may naively use $\mathbb{T}^2$ to enclose a string, analogously to using $\mathbb{S}^2$ to enclose a particle in 3D.
Then, one may deduce 
$
\text{the number of string types}=\text{GSD on } \mathbb{T}^2 \times S^1\overset{?}{=}\mathbb{T}^3
$, 
and that of spacetime integral on $\mathbb{T}^4$, as we already mentioned earlier which is {\it incorrect} and overcounting.
We suggest, 
\begin{equation} 
\label{eq:num_string}
\text{the number of string types}=\mathsf{S}^{xy}, \mathsf{T}^{xy}\text{'s number of blocks},\;\;\;\;
\end{equation}
whose blocks are labeled by $b$ as the form of Eq.\ref{eq:3Dto2DST}.
We will show the counting by Eq.(\ref{eq:num_part}), (\ref{eq:num_string}) in explicit examples in the next.

\subsubsection{Abelian examples: 3D twisted $Z_{N_1} \times Z_{N_2} \times Z_{N_3}$ gauge theories with Type II, III 4-cocycles}  \label{TypeII,III twist} 
We first study the most generic 3+1D finite Abelian twisted gauge theories with Type II, III 4-cocycle twists. 
It is general enough for us to consider $G=Z_{N_1} \times Z_{N_2} \times Z_{N_3}$ with non-vanished gcd $N_{ij}, N_{ijl}$.
The Type II, III (both their 1st and 2nd kinds) twisted gauge theory have GSD$=|G|^3$ on the spatial $\mathbb{T}^3$ torus. 
As such the canonical basis $|\alpha, a, b \rangle$ of the ground state sector labels the charge ($\alpha$ along $x$) and two fluxes (a, b along $y$, $z$), each of the three has $|G|$ kinds.
Thus, naturally from the Rep Theory viewpoint, we have GSD$=|G|^3$.
However, as mentioned in Sec.\ref{sec:GSD}, the $|G|^3$ overcounts the number of strings and particles. By using Eq.(\ref{eq:num_part}),(\ref{eq:num_string}),
we find there are $|G|$ types of particles and $|G|$ types of strings. The canonical basis $|\alpha, a, b \rangle$ (GSD on $\mathbb{T}^3$) counts twice the flux sectors.
{\bf Several remarks follow}:\\
(1)  For an untwisted gauge theory (topological term $p_{..}=0$), which is the direct product of $Z_N$ gauge theory or $Z_N$ toric code, its statistics has the form 
$\exp \big(  \underset{k}{\sum}  \frac{2 \pi \ti  }{ N_k   }  \; (\beta_k {a}_k-\alpha_k {d}_k )  \big)$ and 
$\exp \big( \underset{k}{\sum}    \frac{2 \pi \ti  }{ N_k   }  \; \alpha_k \cdot a_k  \big)$. This shall be described by the BF theory of $\int B dA$ action. 
With $\alpha, \beta$ as the charge of particles (1-form gauge field $A$), $a,b$ as the flux of string(2-form gauge field $B$). This essentially describes
{\bf the braiding between a pure-particle and a pure-string}.
\\

\noindent
(2)  Both ${\sfS}^{xy}$, ${\sfT}^{xy}$ have block diagonal forms as ${\sfS}^{xy}_b$, ${\sfT}^{xy}_b$ respect to the $b$ flux (along $z$) correctly reflects what 
Eq.(\ref{eq:3Dto2DST}) preludes already.\\

\noindent
(3) ${\sfT}^{xy}$ is in SL$(3,\Z)$ canonical basis automatically and full-diagonal, but ${\sfS}^{xy}$ may not be in the canonical basis 
for each blocks of ${\sfS}^{xy}_b$, due to its SL$(2,\Z)$ nature. 
We can find the proper basis in each $b$ block. 
Nevertheless, the eigenvalues of ${\sfS}^{xy}$ 
are still proper and invariant regardless any basis. \\

\noindent
(4) {\bf Characterization of topological orders}:  We can further compare the 3D ${\sfS}^{xy}_b$ data to SL$(2,\Z)$'s data of 2D ${\sfS}^{xy}$ of 
$\mathcal{H}^3(G,\R/\Z)$. 
All of the dimensional reduction of these data 
(${\sfS}^{xy}_b$ 
and  ${\sfT}^{xy}_b$ 
) 
agree with 3-cocycle (induced from 4-cocycle $\omega_4$). 
Gathering all data, we conclude that Eq.(\ref{eq:C3DtoC2Domb}) becomes explicitly. For example, Type II twists for $G=(Z_2)^2$ as,
\bea
&& {\cC^{3\tD}_{{(Z_{2})^2},1}} =  4 \cC_{{(Z_{2})^2},1}^{2\tD}  \;\;\;\;\; \label{eq.Z2toric_3D}\\
&& {\cC^{3\tD}_{(Z_{2})^2},\omega_{4,\tII}} =  \cC_{(Z_{2})^2}^{2\tD} \oplus   \cC_{{(Z_{2})^2},\omega_{3,\tI}}^{2\tD} 
\oplus 2 \cC_{{(Z_{2})^2},\omega_{3,\tII}}^{2\tD} \;\;\;\;\; \label{eq.Z2twist_3D}
\eea
{Such a Type II $\omega_{4,\tII}$ can produce a $b=0$ sector of ($Z_2$ toric code $\otimes$ $Z_2$ toric code) of 2D as $\cC_{(Z_{2})^2}^{2\tD}$,
some $b \neq 0$ sector of ($Z_2$ double-semions $\otimes$ $Z_2$ toric code) 
as $\cC_{{(Z_{2})^2},\omega_{3,\tI}}^{2\tD}$ and another $b \neq 0$ sector $\cC_{{(Z_{2})^2},\omega_{3,\tII}}^{2\tD}$, for example.}
This procedure can be applied to other types of cocycle twists.\\

\noindent
(5) {\bf Classification of topological orders}: \\
We shall interpret the decomposition in Eq.(\ref{eq:C3DtoC2Domb}) as the implication of classification. 
Let us do the {\it counting of number of phases} in the simplest example of Type II, $G=Z_2\times Z_2$ twisted theory.
There are four types in $(p_{{ \text{II}(12)}}^{(1st)}, p_{{ \text{II}(12)}}^{2nd} ) \in \mathcal{H}^4(G,\R/\Z)=(\Z_2)^2$.
However, we find there are {\bf only two distinct topological orders} out of four. 
One is the trivial $(Z_2)^2$ gauge theory as Eq.(\ref{eq.Z2toric_3D}), the other is the nontrivial type as Eq.(\ref{eq.Z2twist_3D}).
There are two ways to see this, 
(i) from the full ${\sfS}^{xyz}$, ${\sfT}^{xy}$ data. 
(ii) viewing the sector of ${\sfS}^{xy}_b$, ${\sfT}^{xy}_b$ under distinct fluxes $b$, which is from a $\mathcal{H}^3(G,\R/\Z)$ perspective.
We should beware that in principle {\it tagging particles, strings or gauge groups is not allowed}, so one can identify many seemingly-different orders by relabeling their excitations. 

\noindent
(6)  {\bf Spin-statistics relation of closed strings} in Eq.(\ref{Spin-Statistics3D}) is verified 
 to be correct here, while we take the complex conjugate in Eq.(\ref{Spin-Statistics3D}).
 This is why we draw the orientation of Fig.\ref{fig:Txy},\ref{fig:Sxyhalf} oppositely.
Interpreting ${\sfT}^{xy}$ as the {\bf topological spin} also holds.\\

\noindent
(7) {\bf Cyclic relation for $\sfS^{xyz}$ in 3D}:
For all the above data (Type II, Type III), 
there is a special cyclic relation for $\sfS^{\alpha,\beta}_{a,b,d}$
when the charge labels are equal $\alpha=\beta$ 
(e.g. for pure fluxes $\alpha=\beta=0$, namely for pure strings):
\bea \label{eq:cyclic_S_eq}
\sfS^{\alpha,\alpha}_{a,b,d} \cdot \sfS^{\alpha,\alpha}_{b,d,a} \cdot \sfS^{\alpha,\alpha}_{d,a,b}=1.
\eea
However, such a cyclic relation does not hold (even at the zero charge) for $\sfS^{2\tD\;\alpha,\beta}_{a,c\;(b)}$, namely 
$\sfS^{2\tD\;\alpha,\beta}_{a,c\;(b)} \cdot \sfS^{2\tD\;\alpha,\beta}_{c,b\;(a)} \cdot \sfS^{2\tD\;\alpha,\beta}_{b,a\;(c)} \neq 1$ in general.
Some other cyclic relations are studied recently in Ref.\cite{{WL1437},{JMR1462}}, 
for which we have not yet made detailed comparisons but the perspectives may be different.
In Ref.\cite{JMR1462}, their cyclic relation is determined by triple linking numbers associated with the membrane operators.  
In Ref.\cite{WL1437}, their cyclic relation is related to the loop braiding of Fig.\ref{fig:3strings_2D_3D_xy}, which has its relevancy instead 
to $\sfS^{2\tD\;\alpha,\beta}_{a,c\;(b)}$, not our cyclic relation of $\sfS^{\alpha,\beta}_{a,b,d}$ for 3D. 
\subsubsection{Non-Abelian examples: 3D twisted $(Z_n)^4$ gauge theories with Type IV 4-cocycle}  \label{sec:TypeIV4cocycle}

We now study a more interesting example, a generic 3+1D finite Abelian twisted gauge theory with Type IV 4-cocycle twists with $p_{ijlm}\neq 0$. 
For generality, our formula also incorporates  Type IV twists together with the aforementioned Type II, III twists. 
So
all 4-cocycle twists will be discussed in this subsection.
Differ from the previous example of Abelian topological order with Abelian statistics in Sec.\ref{TypeII,III twist},
we will show Type IV 4-cocycle $\omega_{4,\tIV}$  
will cause the {\it gauge theory becomes non-Abelian, having non-Abelian statistics even if the original $G$ is Abelian}.
Our inspiration rooted in a 2D example for Type III 3-cocycle twist 
 will cause a similar effect, discovered 
in Ref.\cite{deWildPropitius:1996gt}.
In general, one can consider $G=Z_{N_1} \times Z_{N_2} \times Z_{N_3} \times Z_{N_4}$ with non-vanished gcd $N_{1234}$;
however, we will focus on $G=(Z_n)^4$ with $N_{1234}=n$, with $n$ is prime for simplicity.
From $\cH^4(G,\R/\Z)=\Z_{n}^{21}$, we have $n^{21}$ types of theories, while $n^{20}$ are Abelian gauge theories, and 
$n^{20} \cdot (n-1)$ types with Type IV $\omega_4$ show 
non-Abelian statistics.
\\

\noindent
{\bf Ground state degeneracy (GSD)-}

We compute the GSD of gauge theories with a Type IV twist on the spatial $\mathbb{T}^3$ torus, truncated from $=|G|^3=|n^4|^3=n^{12}$ to:
\bea
\text{GSD}_{\mathbb{T}^3,\tIV}= \big(n^8+n^9-n^5\big) &+&\big(n^{10} -n^{7} -n^{6} +n^{3}\big) \;\;\;\;\;\;\;\\
 \equiv \text{GSD}^{Abel}_{\mathbb{T}^3,\tIV} &+& \text{GSD}^{nAbel}_{\mathbb{T}^3,\tIV} \label{eq:T3TypeIVdec}
\eea
(We derive the above only for a prime $n$. The GSD truncation is less severe and is in between $\text{GSD}_{\mathbb{T}^3,\tIV}$ and $|G|^3$ for a non-prime $n$.)
As such, the canonical basis $|\alpha, a, b \rangle$ of the ground state sector on $\mathbb{T}^3$ {\it no longer} has $|G|^3$ labels 
with the $|G|$ number charge and two pairs of $|G| \times |G|$ number of fluxes 
as in Sec.\ref{TypeII,III twist}. This truncation is due to the nature of non-Abelian physics of Type IV $\omega_{4,\tIV}$ twisted.
We explain our notation in Eq.(\ref{eq:T3TypeIVdec}); the ($n$)$Abel$ indicates the contribution from (non-)Abelian excitations. 
From the Rep Theory viewpoint, we can recover the truncation back to $|G|^3$ by carefully reconstructing the {\it quantum dimension of excitations}. 
We obtain
\bea \label{eq:GSDtotalT3}
&& |G|^3= { \big(\text{GSD}^{Abel}_{\mathbb{T}^3,\tIV}\big) +\big(\text{GSD}^{nAbel}_{\mathbb{T}^3,\tIV}\big) \cdot n^2}  \\ 
&& ={ \{n^4+n^5-n\} \cdot n^4 \cdot (1)^2 } 
 {+\{(n^{4})^2 -n^{5} -n^{4} +n\} \cdot n^2 \cdot (n)^2} \nonumber \\ 
&& = { \{  \text{Flux}^{Abel}_\tIV  \} \cdot n^4 \cdot (\text{dim}_1)^2+  \{ \text{Flux}^{nAbel}_\tIV  \} \cdot n^2 \cdot (\text{dim}_n)^2} \nonumber
\eea
The $\text{dim}_{m}$ means the dimension of Rep\,as $\text{dim(Rep)}$ is $m$, which is also the {\it quantum dimension} of excitations. Here we have 
a dimension $1$ for Abelian and $n$ for non-Abelian. 
In summary, we understand the decomposition precisely 
in terms of each (non-)Abelian contribution as follows: 
\bea \label{eq:GSDarray}
\left\{
 \begin{array}{l}
\text{flux sectors}=|G|^2=|n^4|^2= \text{Flux}^{Abel}_\tIV  +\text{Flux}^{nAbel}_\tIV \\ 
\text{GSD}_{\mathbb{T}^3,\tIV}=  \text{GSD}^{Abel}_{\mathbb{T}^3,\tIV}+ \text{GSD}^{nAbel}_{\mathbb{T}^3,\tIV} \\
\text{dim(Rep)}^2=1^2, n^2\\
\text{Numbers of charge Rep}= n^4, n^2.
\end{array}
\right.
\eea
 
Actually, the canonical basis $|\alpha, a, b \rangle$ (GSD on $\mathbb{T}^3$) still works,
the sum of Abelian $\text{Flux}^{Abel}_\tIV$ and non-Abelian  $\text{Flux}^{nAbel}_\tIV $ counts the flux number of $a,b$ as the unaltered $|G|^2$.
The charge Rep $\alpha$ is unchanged with a number of $|G|=n^4$ for Abelian sector with a rank-1 matrix (1-dim linear or projective) representation, 
however, the charge Rep $\alpha$ is truncated to a smaller number $n^2$ for non-Abelian sector also with a larger rank-n matrix (n-dim projective) representation.

Another view on $\text{GSD}_{\mathbb{T}^3,\tIV}$ can be inspired by a generic formula like Eq.(\ref{eq:3Dto2DST})
\bea
\text{GSD}_{\cM' \times S^1}=\oplus_b \text{GSD}_{b,{\cM'}}= \sum_b \text{GSD}_{b,\cM'}, 
\eea
where we sum over GSD in all different $b$ flux sectors, with $b$ flux along $S^1$. 
Here we can take $\cM' \times S^1=\mathbb{T}^3$ and $\cM'=\mathbb{T}^2$.
For non-Type IV (untwisted, Type II, III) $\omega_4$ case, 
we have $|G|$ sectors of $b$ flux and each has $\text{GSD}_{b,{\mathbb{T}^2}}=|G|^2$.
For Type IV $\omega_4$ case $G=(Z_n)^4$ with a prime $n$, we have 
\bea
&&\text{GSD}_{\mathbb{T}^3,\tIV} 
=|G|^2+(|G|-1)\cdot |Z_n|^2\cdot (1 \cdot  |Z_n|^3 + (|Z_n|^2-1) \cdot n) \nonumber \\
&&=n^8+(n^4-1)\cdot n^2\cdot (1 \cdot n^3 + (n^3-1) \cdot n).
\eea
As we expect, the first part is from the zero flux $b=0$, which is the normal untwisted 2+1D $(Z_n)^4$ gauge theory (toric code) as $\cC^{2D}_{(Z_n)^4}$
with $|G|^2=n^8$ on 2-torus.
The remaining $(|G|-1)$ copies are inserted with nonzero flux ($b\neq 0$) as $\cC^{2D}_{(Z_n)^4,\omega_3}$ with Type III 3-cocycle twists of Table \ref{tableA1}.
In some case but not all cases, $\cC^{2D}_{(Z_n)^4,\omega_3}$ is $\cC^{2D}_{{(Z_n)}_{\text{untwist}} \times (Z_n)^3_{\text{twist}},\omega_3}$.
In either case, the $\text{GSD}_{b,{\mathbb{T}^2}}$ for $b \neq 0$ has the same decomposition always equivalent to a untwisted $Z_n$ gauge theory with $GSD_{\mathbb{T}^2}=n^2$ direct product with
\bea
&&\text{GSD}_{\mathbb{T}^2,\omega_{3,\tIII}} =(1 \cdot n^3 + (n^3-1) \cdot n) 
 \equiv \text{GSD}^{Abel}_{\mathbb{T}^2,\omega_{3,\tIII}} + \text{GSD}^{nAbel}_{\mathbb{T}^2,\omega_{3,\tIII}}, 
\eea
which we generalize the result derived for 2+1D Type III $\omega_3$ twisted theory with $G=(Z_2)^3$ in Ref.\cite{deWildPropitius:1996gt} to $G=(Z_n)^3$ of a prime $n$.

To summarize, from the GSD counting, we already foresee there exist {\bf non-Abelian strings in 3+1D Type IV twisted gauge theory, 
with a quantum dimension $n$.} Those non-Abelian strings (fluxes) carries $\text{dim(Rep)}=n$ non-Abelian charges. Since charges are sourced by particles,
those {\bf non-Abelian strings are not pure strings but attached with non-Abelian particles.} 
(For a projection perspective from 3D to 2D, a nonAbelain string of $\cC^{3\tD}$ is a non-Abelain dyon with both charge and flux of $\cC^{2\tD}_b$.)

{\bf Some remarks follow}:\\
\noindent
(1) {\bf Dimensional reduction from 3D to 2D sectors with $b$ flux}: From the above $\sfS^{xyz},\sfT^{xy}$, there is no difficulty deriving $\sfS^{xy}$ from Eq.(\ref{eq:Sxydecompose}).
From all these modular data $\sfS^{xy}_b, \sfT^{xy}_b$ data, we find consistency 
with the dimensional reduction of 3D topological order by 
comparison with induced 3-cocycle $\omega_3$ from $\omega_4$. 
Let us consider a single specific example, given the Type IV $p_{1234}=1$ and other zero Type II,III indices $p_{..}=p_{...}=0$,
\bea \label{eq:TypeIV3D2Dbranch}
&& \cC^{3\tD}_{(Z_2)^4, \omega_{4,\tIV}}  = \oplus_b \cC_b^{2\tD}  
 =\cC^{2\tD}_{(Z_2)^4} \oplus  10 \; \cC^{2\tD}_{(Z_2) \times (Z_2)^3_{(ijl)},  \omega_{3,\tIII}^{(ijl)} }  \oplus 5 \cC^{2\tD}_{(Z_2)^4, \omega_{3,\tIII} \times \omega_{3,\tIII} \times \dots} \nonumber\\   
&& =\cC^{2\tD}_{(Z_2)^4} \oplus  10 \; \cC^{2\tD}_{(Z_2) \times (D_4)}  \oplus 5 \cC^{2\tD}_{ (Z_2)^4, \omega_{3,\tIII} \times \omega_{3,\tIII} \times \dots} \nonumber   
\eea
The $\cC^{2\tD}_{(Z_2)^4}$ again is 
the normal ${(Z_2)^4}$ gauge theory at $b=0$.
The 10 copies of $\cC^{2\tD}_{(Z_2) \times (D_4)}$ 
have an untwisted dihedral $D_4$ gauge theory ($|D_4|=8$) product with the normal ${(Z_2)}$ gauge theory.
The duality to $D_4$ theory in 2D can be expected,\cite{deWildPropitius:1995cf} see Table \ref{table:Z2cubeclass}.
(As a byproduct of our work, we go beyond Ref.\cite{deWildPropitius:1995cf}
to give the complete classification of all twisted 2D $\omega_3$ of $G=(Z_2)^3$ and their 
corresponding topological orders and twisted quantum double $D^{\omega}(G)$ in Appendices of \cite{Wang:2014oya}.)  
The remaining 5 copies $\cC^{2\tD}_{ (Z_2)^4, \omega_{3,\tIII} \times \omega_{3,\tIII} \times  \dots} $ must contain the twist on the full group $(Z_2)^4$, not just its subgroup. 
This peculiar feature suggests the following remark.\\

\noindent
(2) 
Sometimes there may exist a duality between a twisted Abelian gauge theory and a untwisted non-Abelian gauge theory,\cite{deWildPropitius:1995cf}
one may wonder whether one can find a dual non-Abelian gauge theory for  $\cC^{3\tD}_{(Z_2)^4, \omega_{4,\tIV}}$?
We find that, however, {\bf $\cC^{3\tD}_{(Z_2)^4, \omega_{4,\tIV}}$ cannot be dual to a normal gauge theory (neither Abelian nor non-Abelian),
but must be a  twisted (Abelian or non-Abelian) gauge theory}. The reason is more involved. 
Let us first recall the more familiar 2D case. One can consider $G=(Z_2)^3$ example with $\cC^{2\tD}_{(Z_2)^3, \omega_{3}}$, 
with $\cH^3(G,\R/\Z)=(\Z_{2})^{7}$. 
There are $2^6$ for non-Abelian types with Type III $\omega_3$ (the other $2^6$ Abelian without with Type III $\omega_3$).
We find the $64$ non-Abelian {\bf types} of 3-cocycles $\omega_3$ go to 5 {\bf classes} labeled $\omega_3[1]$, $\omega_3[{3d}]$,
$\omega_3[{3i}]$, $\omega_3[{5}]$ and $\omega_3[{7}]$, and their twisted quantum double model $D^\omega(G)$ are shown in Table \ref{table:Z2cubeclass}. 
The number in the bracket $[..]$ is related to the number of pairs of $\pm \ti$ in the $\sfT$ matrix and the $d/i$ stand for the linear dependence($d$)/independence($i$)
of fluxes generating cocycles. 
\begin{table}[!h]
\centering
\begin{tabular}{|c||c| c|} 
\hline
Class \; & \;  Twisted quantum double $D^\omega(G)$ & \; Number of Types\\ \hline
$\omega_3[1]$ \; & \; $D^{\omega_3{[1]}}(Z_2{}^3)$, $D(D_4)$ & \;  7\\ \hline
$\omega_3[{3d}]$ \; & \; $D^{\omega_3{[3d]}}(Z_2{}^3)$, $D^{\gamma^4}(Q_8)$ & \;  7\\ \hline
$\omega_3[{3i}]$ \; & \; $D^{\omega_3[{3i}]}(Z_2{}^3)$, $D^{}(Q_8)$, $D^{\alpha_1}(D_4)$, $D^{\alpha_2}(D_4)$ & \;  28\\ \hline
$\omega_3[{5}]$ \; & \; $D^{\omega_3{[5]}}(Z_2{}^3)$, $D^{\alpha_1\alpha_2}(D_4)$ & \; 21 \\ \hline
$\omega_3[{7}]$ \;  & \; $D^{\omega_3{[7]}}(Z_2{}^3)$ & \; 1 \\ \hline
 \end{tabular}
\caption{ $D^\omega(G)$, the twisted quantum double model of $G$ in 2+1D, and their 3-cocycles $\omega_{3}$(involving Type III) types in $\cC^{2\tD}_{(Z_2)^3, \omega_{3}}$. 
We classify the 64 types of 2D non-Abelian twisted gauge theories to 5 classe. 
Each class has distinct non-Abelian statistics. Both dihedral group $D_4$ and quaternion group $Q_8$ are non-Abelian groups of order 8, as $|D_4|=|Q_8|=|(Z_2)^3|=8$.
}
\label{table:Z2cubeclass}
\end{table}
From Table \ref{table:Z2cubeclass}, we show that two classes of 3-cocycles for  $D^{\omega_3}(Z_2)^3$ of 2D can have dual descriptions by gauge theory of non-Abelian dihedral group $D_4$, quaternion 
group $Q_8$. 
However, the other three classes of 3-cocycles  for  $D^{\omega_3}(Z_2)^3$ do not have a dual (untwisted) non-Abelian gauge theory.

Now let us go back to consider 3D $\cC_{G,\omega_{4,\tIV}}^{3\tD}$, 
with $|Z_2|^4=16$. 
From Ref.\cite{MW14}, we know 3+1D $D_4$ gauge theory has decomposition by its 5 centralizers.
Apply the rule of decomposition 
to other groups, it implies that for untwisted group $G$ in 3D $\cC^{3\tD}_{G}$, we can decompose it  
into sectors 
of $\cC^{2\tD}_{G_b,b}$, here $G_b$ becomes the {\bf centralizer} of the {\bf conjugacy class}(flux) $b$:
$
\cC^{3\tD}_{G} =\oplus_b \cC^{2\tD}_{G_b,b}.
$
Some useful information 
is: 
\bea
&&\cC^{3\tD}_{(Z_2)^4} = 16 \cC^{2\tD}_{(Z_2)^4}  \label{eq:Z2fbranch}\\
&&\cC^{3\tD}_{D_4} = 2\cC^{2\tD}_{D_4} \oplus 2\cC^{2\tD}_{(Z_2)^2} \oplus \cC^{2\tD}_{Z_4},\\ 
&&\cC^{3\tD}_{Z_2 \times D_4} = 4\cC^{2\tD}_{Z_2 \times  D_4} \oplus 4\cC^{2\tD}_{(Z_2)^3} \oplus 2 \cC^{2\tD}_{Z_2 \times Z_4}, \label{eq:D4Z2branch}\\
&&\cC^{3\tD}_{Q_8} = 2\cC^{2\tD}_{Q_8} \oplus 3 \cC^{2\tD}_{Z_4},\\
&&\cC^{3\tD}_{Z_2 \times Q_8} =4\cC^{2\tD}_{Z_2 \times Q_8} \oplus 6 \cC^{2\tD}_{Z_2 \times Z_4}.
\eea
and we find that no such decomposition is possible from $|G|=16$ group to match Eq.(\ref{eq:TypeIV3D2Dbranch})'s. Furthermore,
if there exists a non-Abelian $G_{nAbel}$ to have Eq.(\ref{eq:TypeIV3D2Dbranch}), those $(Z_2)^4$, $(Z_2) \times (D_4)$ or the twisted $(Z_2)^4$ 
must be the centralizers of $G_{nAbel}$. But one of the centralizers (the centralizer of the identity element as a conjugacy class $b=0$) of $G_{nAbel}$ must be $G_{nAbel}$ itself, which has already ruled out 
from Eq.(\ref{eq:Z2fbranch}),(\ref{eq:D4Z2branch}). Thus, we prove that {\bf $\cC^{3\tD}_{(Z_2)^4, \omega_{4,\tIV}}$ is not a normal 3+1D gauge theory 
(not ${Z_2 \times  D_4}$, neither Abelian nor non-Abelian) but must only be a twisted gauge theory.}
 

\begin{figure}[!h] 
\centering
\includegraphics[scale=0.4]{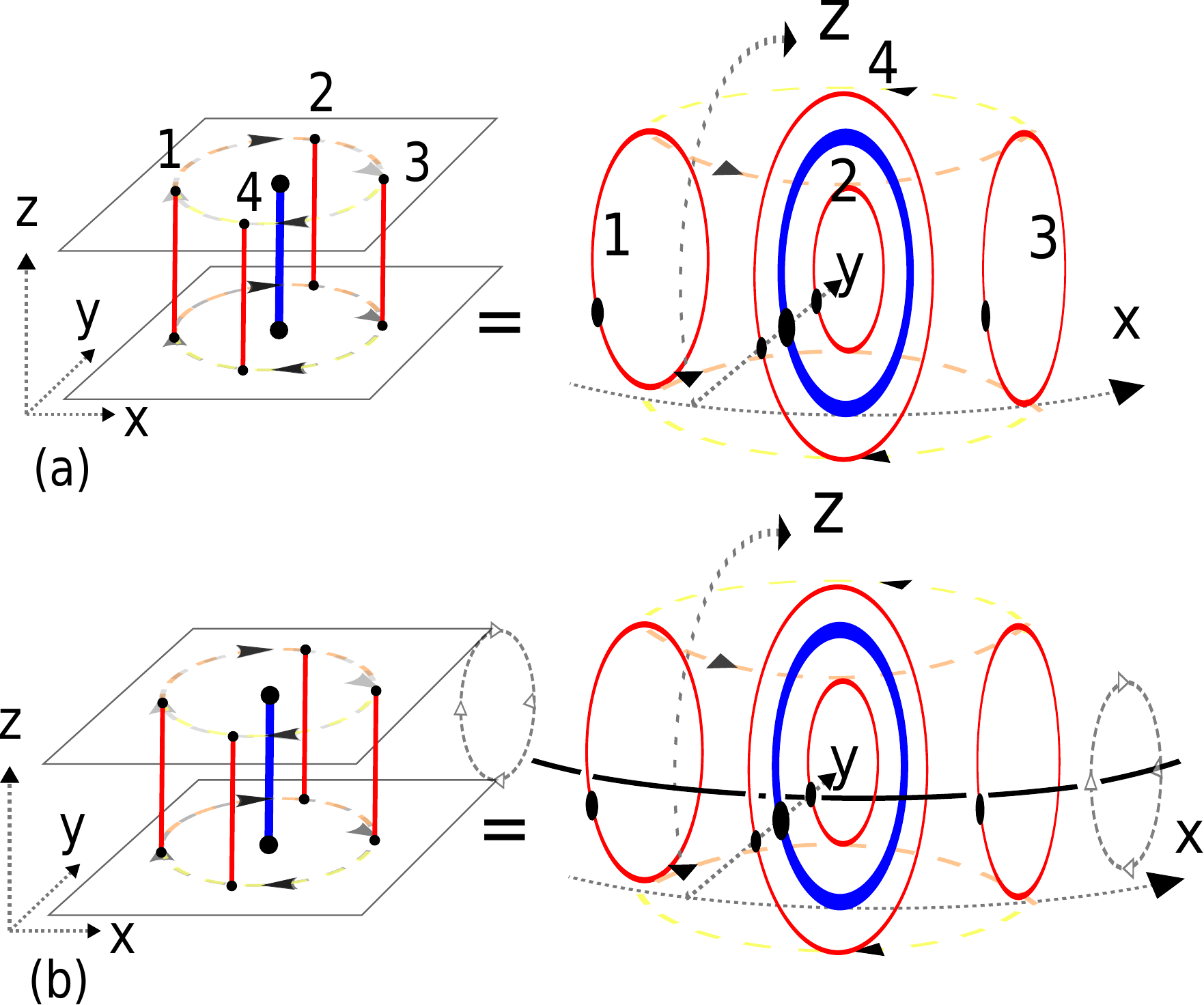}
\caption{
For 3+1D Type IV $\omega_{4,\tIV}$ twisted gauge theory $\cC_{G,\omega_{4,\tIV}}^{3\tD}$:
(a) {\bf Two-string statistics in unlink $0^2_1$ configuration is Abelian}. (The $b=0$ sector as $\cC^{2\tD}_G$.)
(b) {\bf Three-string statistics in two Hopf links $2^2_1 \# 2^2_1$ configuration is non-Abelian}.
(The $b \neq 0$ sector in  $\cC_b^{2\tD}=\cC_{G,\omega_{3,\tIII}}^{2\tD}$.) 
The $b \neq 0$ flux sector creates a monodromy effectively acting as the third (black) string threading the two (red,blue) strings.
}
\label{fig:3strings_2D_3D_Ab_to_nAb} 
\end{figure}

\noindent
(3) We discover that, see Fig.\ref{fig:3strings_2D_3D_Ab_to_nAb}, for any twisted gauge theory $\cC_{G,(\omega_{4,\tIV} \cdot \omega_{4,..})}^{3\tD}$ with Type IV 4-cocycle $\omega_{4,\tIV}$
(whose non-Abelian nature is not affected by adding other Type II,III $\omega_{4,..}$), 
by {\bf threading a third string through two-string unlink $0^2_1$ 
into three-string Hopf links $2^2_1 \# 2^2_1$ configuration, Abelian two-string statistics is promoted to non-Abelian three-string statistics.}
We can see the physics from Eq.(\ref{eq:TypeIV3D2Dbranch}), the $\cC_b^{2\tD}$ is Abelian in $b=0$ sector; but non-Abelian in $b \neq 0$ sector. 
The physics of Fig.\ref{fig:3strings_2D_3D_Ab_to_nAb} is then obvious, 
by applying our discussion in Sec.\ref{Sec:physicsST} about the equivalence between string-threading and the $b\neq 0$ monodromy causes a branch cut.\\

\noindent
(4) {\bf Cyclic relation for non-Abelian $\sfS^{xyz}$ in 3D}: 
Interestingly, for the $(Z_2)^4$ twisted gauge theory with non-Abelian statistics,
we find that a similar cyclic relation Eq.(\ref{eq:cyclic_S_eq}) still holds as long as two conditions are satisfied: (i) the charge labels are equivalent $\alpha=\beta$
and (ii) $\delta_{a\in\{b,d, bd\}} \cdot \delta_{d\in\{a,b, ab\}} \cdot \delta_{b \in\{d, a, da\}} =1$.
However, Eq.(\ref{eq:cyclic_S_eq}) is modified with a factor depending on the dimensionality of Rep $\alpha$:
\bea \label{eq:cyclic_nAb_S_eq} 
\sfS^{\alpha,\alpha}_{a,b,d} \cdot \sfS^{\alpha,\alpha}_{b,d,a} \cdot \sfS^{\alpha,\alpha}_{d,a,b} \cdot |{\dim(\alpha)}|^{-3} =1.
\eea
This identity should hold for any Type IV non-Abelian strings.
This is a cyclic relation of 3D nature, 
 instead of a dimensional-reducing 2D nature of $\sfS^{2\tD\;\alpha,\beta}_{a,c\;(b)}$ in Fig.\ref{fig:3strings_2D_3D_xy}.  

So far we had obtain some string-particle braiding identity via the representation theory and the twisted lattice gauge theory model (of Dijkgraaf-Witten topological gauge theory). In Chap.\ref{QS_stSurgery}, we will explore more possible identities through another more unified approach: geometric-topology surgery theory and quantum partition functions.

%% file: chap4.tex
\chapter{Aspects of Anomalies} \label{aofAnomalies}


We review chiral fermionic Adler-Bell-Jackiw anomalies and quantum Hall states in Sec.\ref{sec:chiralanomaly}.
The we develop and construct the physical systems with bosonic anomalies in Sec.\ref{sec:bAnom}.
We then attempt to construct a non-perturbative lattice chiral fermion/gauge theory by tackling the Nielson-Ninomya fermion-doubling no-go theorem
in Sec.\ref{sec:lattice_chiral}. With the understanding of fermionic and bosonic anomalies
in topological states of matter, we examine some examples of
mixed gauge-gravity anomalies in Sec.\ref{sec:mixedgaugegrav} --- by constructing mixed gauge-gravity actions whose boundaries 
realize mixed gauge-gravity anomalies. Those mixed gauge-gravity actions can be regarded as effective probe field actions for
SPTs beyond-Group-Cohomology classification.

\section{Chiral Fermionic Adler-Bell-Jackiw Anomalies and Topological Phases} \label{sec:chiralanomaly}

First we present a \emph{chiral fermionic anomaly} (ABJ anomalies\cite{{Adler:1969gk},{Bell:1969ts}}) of \emph{a continuous} U(1) \emph{symmetry} realized in topological phases 
in condensed matter,
in contrast to the \emph{bosonic anomalies of discrete symmetries} studied in the next. 

\begin{figure}[h!] 
(a){\includegraphics[width=.4\textwidth]{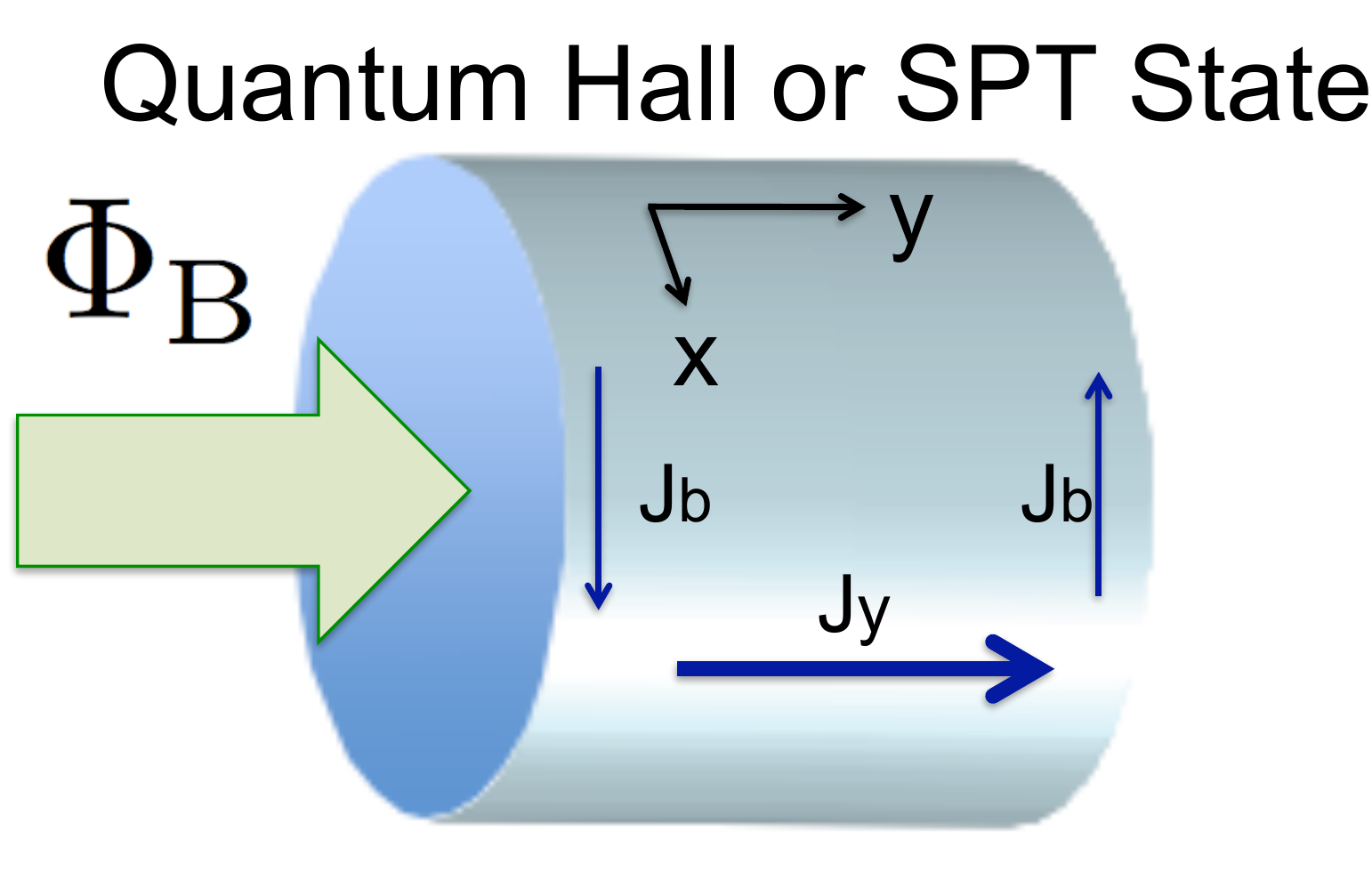}} (b){\includegraphics[width=.56\textwidth]{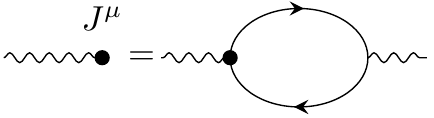}}

\caption{(a) For topological phases, the anomalous current $J_b$ of the boundary theory along $x$ direction leaks to $J_y$ along $y$ direction in the extended bulk system. 
$\Phi_B$-flux insertion $d\Phi_B/dt=-\oint E \cdot d L$ induces the electric $E_x$ field along the $x$ direction.
The effective Hall effect dictates that $J_y=\sigma_{xy}E_x=\sigma_{xy}\varepsilon^{\mu\nu}\,\partial_{\mu} A_{\nu}$, with the effective Hall conductance $\sigma_{xy}$
probed by an external U(1) gauge field $A$. 
(b) {In the fermionic language, the 1+1D chiral fermions (represented by the solid line) and the external U(1) gauge field (represented by the wavy curve) 
contribute to a 1-loop Feynman diagram correction to the axial current $j^\mu_A$. This leads to the non-conservation of $j^\mu_A$ as the anomalous current
$\partial_{\mu}\,  j_{\textrm{A}}^{\mu}= \varepsilon^{\mu\nu} (qK^{-1}q/2\pi)\,F_{\mu\nu}.$ }
}
\label{cylinder}
\end{figure}

Specifically we consider an 1+1D U(1) quantum anomaly realization through 1D edge of U(1) quantum Hall state, such as in Fig.\ref{cylinder}.
We can formulate a Chern-Simons action $S=\int\big(\frac{K}{4\pi} \;a\wedge d a+ \frac{q}{2\pi}A \wedge d a\big)$ with an internal statistical gauge field $a$ and an external U(1) electromagnetic gauge field $A$. 
Its 1+1D boundary is described by a (singlet or multiplet-)chiral boson theory of a field $\Phi$ ($\Phi_L$ on the left edge, $\Phi_R$ on the right edge). 
Here the field strength $F=dA$ is equivalent to the external U(1) flux in the 
flux-insertion thought experiment 
threading through the cylinder
Without losing generality, 
let us first focus on 
the boundary with only one edge mode. We derive its equations of motion as 
\bea 
\partial_{\mu}\,j_{\textrm{b} }^{\mu}
&=&
\frac{\sigma_{xy}}{2}\,
\varepsilon^{\mu\nu}\,F_{\mu\nu}
={\sigma_{xy}}\,
\varepsilon^{\mu\nu}\,\partial_{\mu} A_{\nu}
=
J_{y}, \label{eq:J=sy}\\
\partial_{\mu}\,  j_{\textrm{L}}&=&\partial_{\mu} (\frac{q}{2\pi}\epsilon^{\mu\nu} \partial_\nu \Phi_L)=\partial_{\mu} (q\bar{\psi} \gamma^\mu  P_L  \psi)=+J_{y},\;\;\;\;\; \\
\partial_{\mu}\,  j_{\textrm{R}}&=&-\partial_{\mu} (\frac{q}{2\pi}\epsilon^{\mu\nu} \partial_\nu \Phi_R)=\partial_{\mu} (q\bar{\psi}  \gamma^\mu P_R \psi)=-J_{y}.\;\;\;\;\;\;\;
\eea
We show the Hall conductance
from its definition $J_y ={\sigma_{xy}} E_x $ in Eq.(\ref{eq:J=sy}), as 
$\sigma_{xy}=qK^{-1}q/(2\pi)$.

Here $j_{\textrm{b}}$ stands for the edge current. 
A left-moving current $j_L=j_{\textrm{b}}$ is on one edge, and a right-moving current $j_R=-j_{\textrm{b}}$ is on the other edge, shown in Fig.\ref{cylinder}. 
By bosonization, we convert a compact bosonic phase $\Phi$ to the fermion field $\psi$.
The vector current is $ j_{\textrm{L}}+j_{\textrm{R}} \equiv j_{\textrm{V}}$, and the U(1)$_V$ current is conserved.
The axial current is $ j_{\textrm{L}}-j_{\textrm{R}} \equiv j_{\textrm{A}}$, and we derive the famous ABJ U(1)$_A$ anomalous current in 1+1D 
(or Schwinger's 1+1D quantum electrodynamic [QED] anomaly). 
\bea
\partial_{\mu}\,  j_{\textrm{V}}^{\mu}&=&\partial_{\mu}\, ( j_{\textrm{L}}^{\mu}+j_{\textrm{R}}^{\mu} )= 0,\\ 
\partial_{\mu}\,  j_{\textrm{A}}^{\mu}&=&\partial_{\mu}\, ( j_{\textrm{L}}^{\mu}-j_{\textrm{R}}^{\mu} )=\sigma_{xy} \varepsilon^{\mu\nu}\,F_{\mu\nu}. 
\eea
This simple bulk-edge derivation is consistent with field theory 1-loop calculation through Fig.\ref{cylinder}. 
It shows that the combined boundary theory on the left and right edges (living on the edges of a 2+1D U(1) Chern-Simons theory) can
be viewed as an 1+1D anomalous world of Schwinger's 1+1D QED. 
This is an example of chiral fermionic anomaly of a continuous U(1) symmetry when $K$ is an odd integer.
(When $K$ is an even integer, it becomes a chiral bosonic anomaly of a continuous U(1) symmetry.)

\section{Bosonic Anomalies} 
\label{sec:bAnom}

Now we focus on \emph{characterizing the bosonic anomalies as precisely as possible}, and attempt 
to connect our \emph{bosonic anomalies to the notion defined in the high energy physics context.}
In short, we aim to \emph{make connections between the meanings of boundary bosonic anomalies studied in both high energy physics and condensed matter theory}.

We 
specifically highlight 
three learned aspects about 
SPTs-\\%
\underline{$[1]$. \emph{Non-onsite symmetry on the edge}}: 
An important feature of SPT is that the \emph{global symmetry}  
\cblue{acting on a local Hamiltonian of edge modes}  
is realized 
\emph{non-onsite}. 
For a given symmetry group $G$, the non-onsite symmetry means that its symmetry transformation \emph{cannot} be written as a tensor product form on each site,
\be
U(g)_{\text{non-onsite}} \neq \otimes _i U_i(g),
\ee 
for $g \in G$  of the symmetry group. On the other hand, the onsite symmetry transformation $U(g)$ can be written in a tensor product form acting on each site $i$,
i.e.
$U(g)_{\text{onsite}}= \otimes _i U_i(g)$, for $g \in G$. 
(The symmetry transformation acts as an operator $U(g)$ with $g \in G$, transforming the state $| v \rangle$ globally by  $U(g) | v \rangle$.)
Therefore, to study the SPT edge mode, one should realize how the non-onsite symmetry acts on the 
boundary as in Fig.\ref{fig:1}.
\\
\underline{$[2]$. \emph{Group cohomology construction}}: 
It has been proposed that $d+1$ dimensional($d+1$D) SPTs of symmetry-group-$G$
interacting boson system can be constructed by the 
number of distinct cocycles in the $d+1$-th cohomology group, $\cH^{d+1}(G,\tU(1))$, with $\tU(1)$ coefficient. 
(See also the first use of cocycle in the high energy context by Jackiw in Ref.\cite{{Treiman:1986ep}}) 
While another general framework of cobordism theory is subsequently proposed 
to account for subtleties when symmetry $G$ involves time-reversal, 
in our work we will focus on a 
finite Abelian symmetry group $G=\prod_i Z_{N_i}$,
where the group cohomology is a complete classification.\\
\underline{$[3]$. \emph{Surface anomalies}}:
It has been proposed that the edge modes of SPTs are the source of gauge anomalies, while that of intrinsic topological orders are the source of gravitational anomalies.\cite{Wen:2013oza}
SPT boundary states are known to show at least one of three properties:\\
$\bullet$(1) symmetry-preserving gapless edge modes,\\
$\bullet$(2) symmetry-breaking gapped edge modes, \\
$\bullet$(3)\;symmetry-preserving gapped edge modes with surface topological order. 

We shall now define the meaning of quantum anomaly in a language appreciable by both high energy physics and condensed matter communities -\\
%
\colorbox{lgray}{\parbox[t]{\linegoal}{
The {quantum anomaly} is \emph{an obstruction of a symmetry of a theory to be fully-regularized for a full quantum theory as an onsite symmetry on the UV-cutoff lattice 
in the same spacetime dimension}.}}\\ 
According to this definition, 
to characterize our bosonic anomalies, we will find several possible obstructions to regulate the symmetry at the quantum level:

\noindent
$\star$ \underline{Obstruction of onsite symmetries}:
Consistently we will find throughout our examples to fully-regularize our SPTs 1D edge theory on the 1D lattice Hamiltonian requires the
\emph{non-onsite symmetry}, namely, \emph{realizing the symmetry anomalously}. 
\cblue{The {non-onsite symmetry} on the edge cannot be ``dynamically gauged'' on its own spacetime dimension,
thus this also implies the following obstruction.}

\begin{figure}[h!]
\centering
 \includegraphics[width=0.45\textwidth]{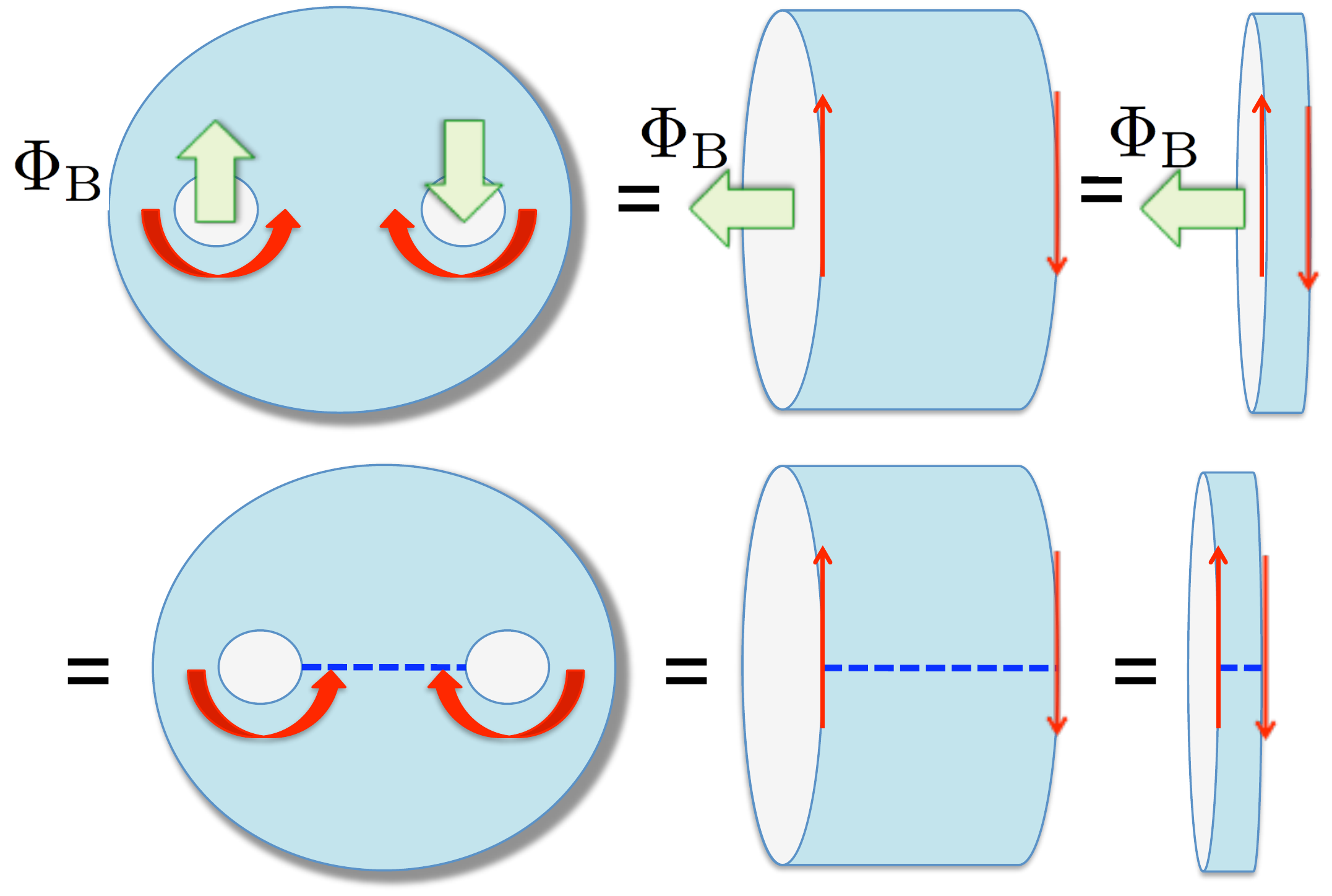}
\caption{ The intuitive way to view the bulk-boundary correspondence for edge modes of SPTs (or intrinsic topological order) under
the flux insertion, or equivalently the monodromy defect / branch cut (blue dashed line) modifying the bulk and the edge Hamiltonians.
SPTs locate on a large sphere with two holes with flux-in and flux-out, is analogous to, a Laughlin type flux insertion through a cylinder,
inducing 
anomalous edge modes(red arrows) moving along the opposite directions on two edges.
}
\label{fig:flux_cut_analogy}
\end{figure}

\noindent
$\star$ \underline{Obstruction of the same spacetime dimension}:
\cblue{We will show 
that the physical observables 
for gapless edge modes (the case $\bullet$(1)) 
are their energy spectral shifts\cite{Santos:2013uda}  
under symmetry-preserving external flux insertion through 
a compact 1D ring.  
The energy spectral shift is caused by the Laughlin-type flux insertion of Fig.\ref{fig:flux_cut_analogy}.
The \emph{flux insertion} can be equivalently regarded as an effective \emph{branch cut} 
modifying the Hamiltonian (blue dashed line in Fig.\ref{fig:flux_cut_analogy}) connecting from the edge to an extra dimensional bulk.
Thus the spectral shifts also indicate the transportation of quantum numbers from one edge to the other edge.
This can be regarded as the anomaly requiring an \emph{extra dimensional bulk}. 
}

\noindent
$\star$ \underline{Non-perturbative effects}:
We know that the familiar 
Adler-Bell-Jackiw anomaly of \emph{chiral fermions},\cite{{Adler:1969gk},{Bell:1969ts}} observed 
in the pion decay in particle-physics can be captured
by the perturbative 1-loop Feynman diagram. 
However, importantly, the result is non-perturbative, being exact from low energy IR to high energy UV.
This effect can be further confirmed via Fujikawa's path integral method 
non-perturbatively.
Instead of the well-known \emph{chiral fermionic anomalies}, do we have \emph{bosonic anomalies} with these non-perturbative effects?

Indeed, yes, we will show two other kinds of bosonic anomalies with non-perturbative effects 
with symmetry-breaking gapped edges (\cblue{the case $\bullet$(2)}): One kind of consequent anomalies for Type II SPTs
under $Z_{N_1}$ symmetry-breaking domain walls 
is 
the {\bf induced fractional $Z_{N_2}$ charge} 
trapped near 0D kink of gapped domain walls.
Amazingly, through a fermionization/bosonization procedure, 
we can apply the field-theoretic Goldstone-Wilczek method to capture the 1-loop Feynman diagram effect \emph{non-perturbatively}, 
as this fractional charge is known to be robust without higher-loop diagrammatic corrections.\cite{Goldstone:1981kk}
We will term this a Type II bosonic anomaly.

The second kind of anomalies for symmetry-breaking gapped edge (\cblue{the case $\bullet$(2)})
is that the edge is gapped under $Z_{N_1}$ symmetry-breaking domain walls, with a consequent 
{\bf degenerate zero energy ground states} due to the projective representation of other symmetries $Z_{N_2} \times Z_{N_3}$.
The zero mode degeneracy is found to be  {$\gcd(N_1,N_2,N_3)$}-fold.
We will term this a Type III bosonic anomaly. 

We will examine a generic finite Abelian 
$G=\prod_i Z_{N_i}$ bosonic SPTs,
and study {\it what is truly anomalous} about the edge under the case of $\bullet$(1) and $\bullet$(2) above. 
(Since it is forbidden to have any intrinsic topological order in a 1D edge, we do not have scenario  $\bullet$(3).)
We focus on addressing the properties of its 1+1D edge modes, their anomalous non-onsite symmetry and bosonic anomalies. 

\begin{center}
\begin{table}[!h] 
\makebox[\textwidth][c]
{\fontsize{10pt}{1em}\selectfont
{
\begin{tabular}{|c||c|c|c|c|}
\hline
$G$-Cohomology &  \multicolumn{4}{c}{Bosonic Anomalies and Physical Observables} \vline \\[0mm]\hline
 3-cocycle & $p \in \cH^3(G,\tU(1))$  & induced frac charge  &  degenerate modes & $\tilde{\Delta}(\tilde{\mathcal{P}})$\\[0mm]  \hline \hline 
Type I $p_1$ & $\color{black}{ \Z_{N_1}}$ & No & No & Yes  \\[0mm]  \hline
Type II $p_{12}$ &   $\color{black}{\Z_{N_{12}}}$ & Yes. {${\frac{p_{12}}{N_{12}}}$ of $Z_{N_2}$ charge}.  &  No &  Yes \\[1mm] \hline 
Type III $p_{123}$ &  $\color{black}{\Z_{N_{123}}}$ & No  & Yes. $N_{123}$-fold. &  Yes   \\[1mm] \hline
\end{tabular}
}
}\hspace*{20mm}
\caption{A summary of bosonic anomalies as 1D edge physical observables to detect the 2+1D SPT of $G=Z_{N_1} \times Z_{N_2} \times Z_{N_3}$ symmetry, here we use $p_i$, $p_{ij}$, $p_{ijk}$ to label the
SPT class index in the third cohomology group $\cH^3(G,\tU(1))$.
For Type II class $p_{12} \in \color{black}{\Z_{N_{12}}}$, we can use a unit of $Z_{N_1}$-symmetry-breaking domain wall to induce a ${\frac{p_{12}}{N_{12}}}$ fractional $Z_{N_2}$ charge, see Sec.\ref{sec: Type II frac}. 
For Type III class $p_{123} \in \color{black}{\Z_{N_{123}}}$, we can either use $Z_{N_1}$-symmetry-breaking domain wall
or use  $Z_{N_1}$-symmetry-preserving flux insertion 
(effectively a monodromy defect) through 1D ring to trap $N_{123}$ multiple degenerate zero energy modes. 
For Type I class $p_{1} \in \color{black}{ \Z_{N_1}}$, our proposed physical observable is the energy spectrum (or conformal dimension $\tilde{\Delta}(\tilde{\mathcal{P}})$
as a function of momentum ${\mathcal{P}}$, see Ref.\cite{Santos:2013uda}) shift under the flux insertion. This energy spectral shift also works for all other (Type II, Type III) classes. 
We denote the fifth column as the energy spectral shift $\tilde{\Delta}(\tilde{\mathcal{P}})$ with the monodromy branch cut or the flux insertion. 
This table serves as topological invariants for Type I, II, III bosonic SPT in the context of Ref.\cite{WenSPTinv}.} 
\label{table3}
\end{table}
\end{center}

\subsection{Type II Bosonic Anomaly: Fractional Quantum Numbers trapped at the Domain Walls
\label{sec: Type II frac}}

We propose the experimental/numerical signatures for certain SPT with Type II class $p_{12} \neq 0$ with (at least) two symmetry group $Z_{N_1} \times Z_{N_2}$,
also as a way to study the physical measurements for Type II bosonic anomaly.
We show that when the $Z_{N_1}$ symmetry is broken by $Z_{N_1}$ domain wall 
created on a ring, there will be some fractionalized $Z_{N_2}$ charges induced near the kink.
We will demonstrate our field theory method can easily capture this effect.

\subsubsection{Field theory approach: fractional $Z_{N}$ charge trapped at the kink of $Z_N$ symmetry-breaking Domain Walls
\label{sec:Type II bnom field}
}

Consider the $Z_{N_1}$ domain wall is created on a ring (the $Z_{N_1}$ symmetry is broken), then the $Z_{N_1}$ domain wall can be captured by 
$\phi_{1}(x)$ for $x \in [0,L)$ takes some constant value $\phi_0$ while $\phi_{1}(L)$ shifted by $2\pi \frac{n_1}{N_1}$ away from $\phi_0$.
This means that $\phi_{1}(x)$ has the fractional winding number:
\be
\int^{L}_{0}\,dx\,\partial_{x} \phi_{1} =\phi_{1}(L) -\phi_{1}(0) = 2\pi \frac{n_1}{N_1},
\ee

\begin{figure}[h!]
\centering
(a) \includegraphics[width=0.28\textwidth]{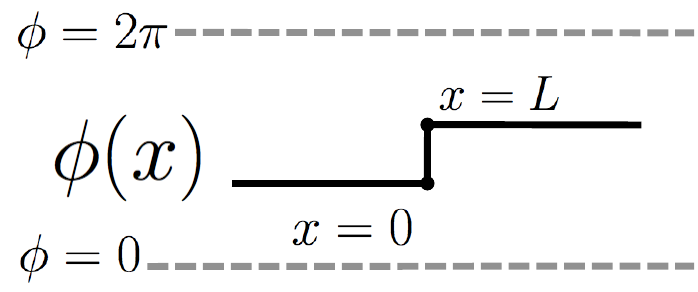} (b) \includegraphics[width=0.28\textwidth]{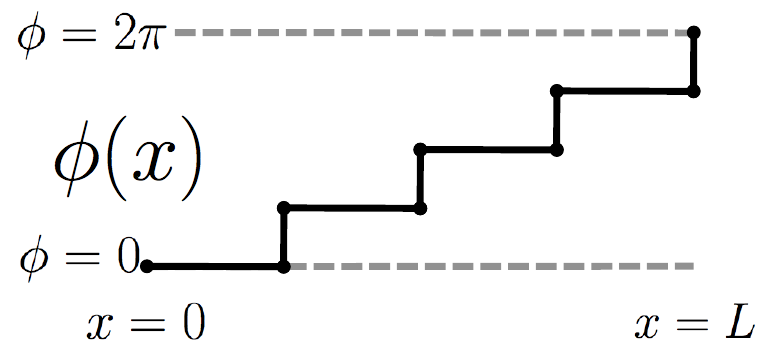} 
(c) \includegraphics[width=0.28\textwidth]{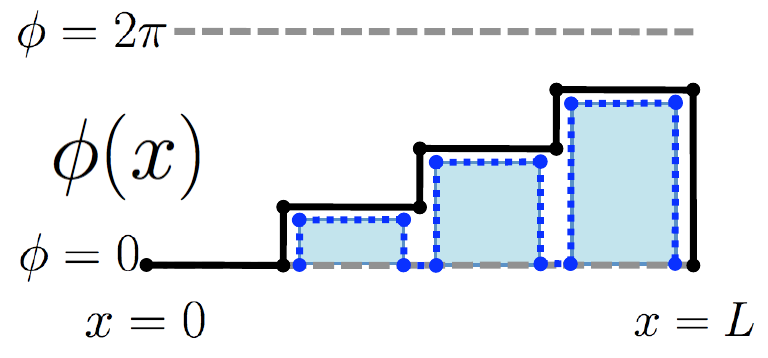} 
\caption{(a) We expect some fractional charge trapped near a single kink around $x=0$ (i.e. $x=0+\epsilon$) and $x=L$ (i.e. $x=0-\epsilon$) in the domain walls.
For $Z_{N_1}$-symmetry breaking domain wall with a kink jump $\Delta \phi_1=2\pi \frac{n_{12}}{N_{12}}$, we predict that the fractionalized $(n_{12}/N_{12})p_{21}$ units of $Z_{N_2}$ charge are induced.
{(b) A nontrivial winding $\int^{L}_{0}\,dx\,\partial_{x} {\phi}(x) = 2\pi$.
This is like a soliton a soliton (or particle) insertion.
For $N_{12}$ number of $Z_{N_1}$-symmetry breaking domain walls, we predict that the integer $p_{21}$ units of total induced $Z_{N_2}$ charge on a 1D ring.
 In average, each kink captures a $p_{21}/N_{12}$ fractional units of $Z_{N_2}$ charge.} 
 (c) {A profile of several domain walls, each with kinks and anti-kinks(in blue color).
For $Z_{N_1}$ symmetry-breaking domain wall, each single kink can trap fractionalized $Z_{N_2}$ charge.
However, overall there is no nontrivial winding, $\int^{L}_{0}\,dx\,\partial_{x} {\phi_1}(x) = 0$ (i.e. no net soliton insertion), so there
is no net induced charge on the whole 1D ring.}}
\label{fig:domain_walls_kink}
\end{figure}

Also recall Eq.(\ref{eq:globalS_Type II_1}) that the Type II $p_{21} \neq 0$ (and $p_{1} = 0, p_{2} = 0$) $Z_{N_2}$ symmetry transformation
\bea
 \;\;  S^{(p_2,p_{21})}_{N_2}
=
\exp[
\frac{ \ti}{N_2}\,
(
p_{21}\,\int^{L}_{0}\,dx\,\partial_{x} \tilde{\phi}_{1}  
+
\int^{L}_{0}\,dx\,\partial_{x}\phi_{2}'
)
], \;\;\;
\eea
can measure the induced $Z_{N_2}$ charge on a state $| \Psi_{\text{domain}} \rangle$ with this domain wall feature as
\be \label{eq:induced charge type II}
S^{(p_2,p_{21})}_{N_2} | \Psi_{\text{domain}} \rangle =\exp[
\frac{ \ti \, p_{21} }{N_2}\,
(
 \tilde{\phi}_{1}(L) -\tilde{\phi}_{1}(0)
)
] | \Psi_{\text{domain}} \rangle 
=
\exp[
(
2\pi \ti \frac{n_{12} \; p_{21} }{N_{12}\, N_2}
)
] | \Psi_{\text{domain}} \rangle.
\ee
We also adopt two facts that:
First, $\int^{L}_{0}\,dx\,\partial_{x} \tilde{\phi}_{1}  = 2\pi \frac{n_{12}}{N_{12}}$ with some integer ${n_{12}}$, where the $\tilde{\phi}_{1}$ is regularized in a unit of $2\pi/{N_{12}}$. 
Second, as $Z_{N_2}$ symmetry is not broken, both $\phi_{2}$ and $\phi_{2}'$ have no domain walls, then the above evaluation takes into account that $\int^{L}_{0}\,dx\,\partial_{x}\phi_{2}'=0$.
This implies that induced charge is fractionalized $(n_{12}/N_{12})p_{21}$ (recall $p_{12}, p_{21}  \in  \mathbb{Z}_{N_{12}}$ ) $Z_{N_2}$ charge. This is the fractional charge trapped at the configuration of a single kink in Fig.\ref{fig:domain_walls_kink}.

On the other hand, one can imagine a series of $N_{12}$ number of 
$Z_{N_1}$-symmetry-breaking domain wall each breaks to different vacuum expectation value(v.e.v.) where the domain wall in the region $[0,x_1)$,$[x_1,x_2)$, $\dots$, $[x_{N_{12}-1},x_{N_{12}}=L)$
with their symmetry-breaking $\phi_1$ value at $0$, 
$2\pi \frac{1}{N_{12}}$, $2\pi \frac{2}{N_{12}}$, $\dots$, $2\pi \frac{N_{12}-1}{N_{12}}$.
This means a nontrivial winding number, like a soliton effect (see Fig.\ref{fig:domain_walls_kink}), 
$
\int^{L}_{0}\,dx\,\partial_{x} \tilde{\phi}_{1} 
= 2\pi 
$ 
and 
$
S^{(p_2,p_{21})}_{N_2} | \Psi_{\text{domain wall}} \rangle$ 
 $=\exp[
(
2\pi \ti \frac{  p_{21} }{\, N_2}
)
] | \Psi_{\text{domain wall}} \rangle
$
capturing
$p_{21}$ integer units of $Z_{N_2}$ charge at $N_{12}$ kinks for totally $N_{12}$ domain walls, in the configuration of Fig.\ref{fig:domain_walls_kink}. In average, each kink captures the $p_{21}/N_{12}$ fractional units of $Z_{N_2}$ charge.

Similarly, we can consider the $Z_{N_2}$ domain wall is created on a ring (the $Z_{N_2}$ symmetry is broken), then the $Z_{N_2}$ domain wall can be captured by 
$\phi_{2}(x)$ soliton profile for $x \in [0,L)$.  
We consider a series of $N_{12}$ number of 
$Z_{N_2}$-symmetry-breaking 
domain walls, each breaks to different v.e.v.
(with an overall profile of 
$
\int^{L}_{0}\,dx\,\partial_{x} \tilde{\phi}_{2} = 2\pi 
$).
By
$
S^{(p_1,p_{12})}_{N_1} | \Psi_{\text{domain wall}} \rangle 
=\exp[
(
2\pi \ti \frac{  p_{12} }{\, N_1}
)
] | \Psi_{\text{domain wall}} \rangle
$, the $N_{12}$ kinks of domain wall captures
$p_{12}$ integer units of $Z_{N_1}$ charge for totally $N_{12}$ domain wall, as in Fig.\ref{fig:domain_walls_kink}. 
In average, each domain wall captures $p_{12}/N_{12}$ fractional units of $Z_{N_1}$ charge.


\subsubsection{Goldstone-Wilczek formula and Fractional Quantum Number \label{sec:Goldstone-Wilczek}}

It is interesting to view our result above in light of the Goldstone-Wilczek (G-W) approach.\cite{Goldstone:1981kk} 
We warm up by computing  $1/2$-fermion charge found by Jackiw-Rebbi\cite{Jackiw:1975fn} using G-W method 
We will then do a more general case for SPT. 
The construction, valid for 1D systems, works as follows.\\

\noindent
{\bf Jackiw-Rebbi model:}  Consider a Lagrangian describing spinless fermions $\psi(x)$ coupled to a classical background profile $\lambda(x)$
via a term $\lambda\,\psi^{\dagger}\sigma_{3} \psi$. In the high temperature phase, the v.e.v. of $\lambda$ is zero and no mass is generated
for the fermions. In the low temperature phase, the $\lambda$ acquires two degenerate vacuum values $\pm  \langle \lambda \rangle$ 
that are related by a ${Z}_2$ symmetry. Generically we have
\be
\langle \lambda \rangle\,\cos\big( \phi(x) - \theta(x) \big),
\ee
where we use the bosonization dictionary 
$
\psi^{\dagger}\sigma_{3} \psi
\rightarrow
\cos(\phi(x))
$
and a phase change $\Delta\theta = \pi$ captures the existence of a domain wall
separating regions with opposite values of the v.e.v. of $\lambda$.
From the fact that the fermion density 
$
\rho(x)
=
\psi^{\dagger}(x)\psi(x) 
= 
\frac{1}{2\pi} \partial_{x} \phi(x)
$ (and the current $J^\mu=\psi^{\dagger}\gamma^\mu\psi =\frac{1}{2\pi}\epsilon^{\mu \nu} \partial_\nu \phi$),
it follows that the induced charge $Q_{\text{dw}}$ on the kink by a domain wall is
\be
Q_{\text{dw}} 
= 
\int^{x_0 + \varepsilon}_{x_0 - \varepsilon}\,dx\,\rho(x)
=
\int^{x_0 + \varepsilon}_{x_0 - \varepsilon}\,dx\,\frac{1}{2\pi} \partial_{x} \phi(x)
=
\frac{1}{2},
\ee
where $x_0$ denotes the center of the domain wall.\\

\noindent
{\bf Type II Bosonic Anomalies:} 
We now consider the case where the ${Z}_{N_1}$ symmetry is spontaneously broken
into different ``{vacuum}'' regions. This can be captured by an effective term in the Hamiltonian 
of the form
\begin{equation}
H_{sb}
=
-\lambda\,\cos \big(\phi_{1}(x) - \theta(x) \big), \quad \lambda > 0,
\end{equation}
and the ground state
is obtained, in the large $\lambda$ limit, by phase locking $\phi_{1} = \theta$, which opens a gap in the spectrum.

Different domain wall regions are described by different choices of the profile $\theta(x)$, as discussed in Sec.\ref{sec:Type II bnom field}. In particular, if we have $\theta(x) = \theta_{k}(x)$
and
$
\theta_{k}(x) = (k-1)\,2\pi/N_{12}, 
$ for
$
x \in [(k-1)L/N_{12}, k L/N_{12} )$, $k = 1,...,N_{12}$.
then we see that that, a domain wall separating regions $k$ and $k+1$ (where the phase difference is $2\pi/N_{12}$)
induces a ${Z}_{N_2}$ charge given by
\bea \label{eq: G-W Q}
&&\delta\,Q_{k,k+1}
=
\int^{k L/N_{12} + \varepsilon}_{k L/N_{12} - \varepsilon}\,dx\,\delta\rho_{2}(x) \nonumber
=
\frac{1}{2\pi} \int^{k L/N_{12} + \varepsilon}_{k L/N_{12} - \varepsilon}\,dx\,
\frac{p_{12}}{N_{2}}\,\partial_{x} \phi^{}_{1}
=
\frac{p_{12}}{N_2 N_{12}}.
\eea
This implies a fractional of ${p_{12}}/{N_{12}}$ induced $Z_{N_2}$ charge on a single kink of $Z_{N_1}$-symmetry breaking domain walls, consistent with Eq.(\ref{eq:induced charge type II}).

\begin{figure}[h!]
\centering
\includegraphics[width=0.4\textwidth]{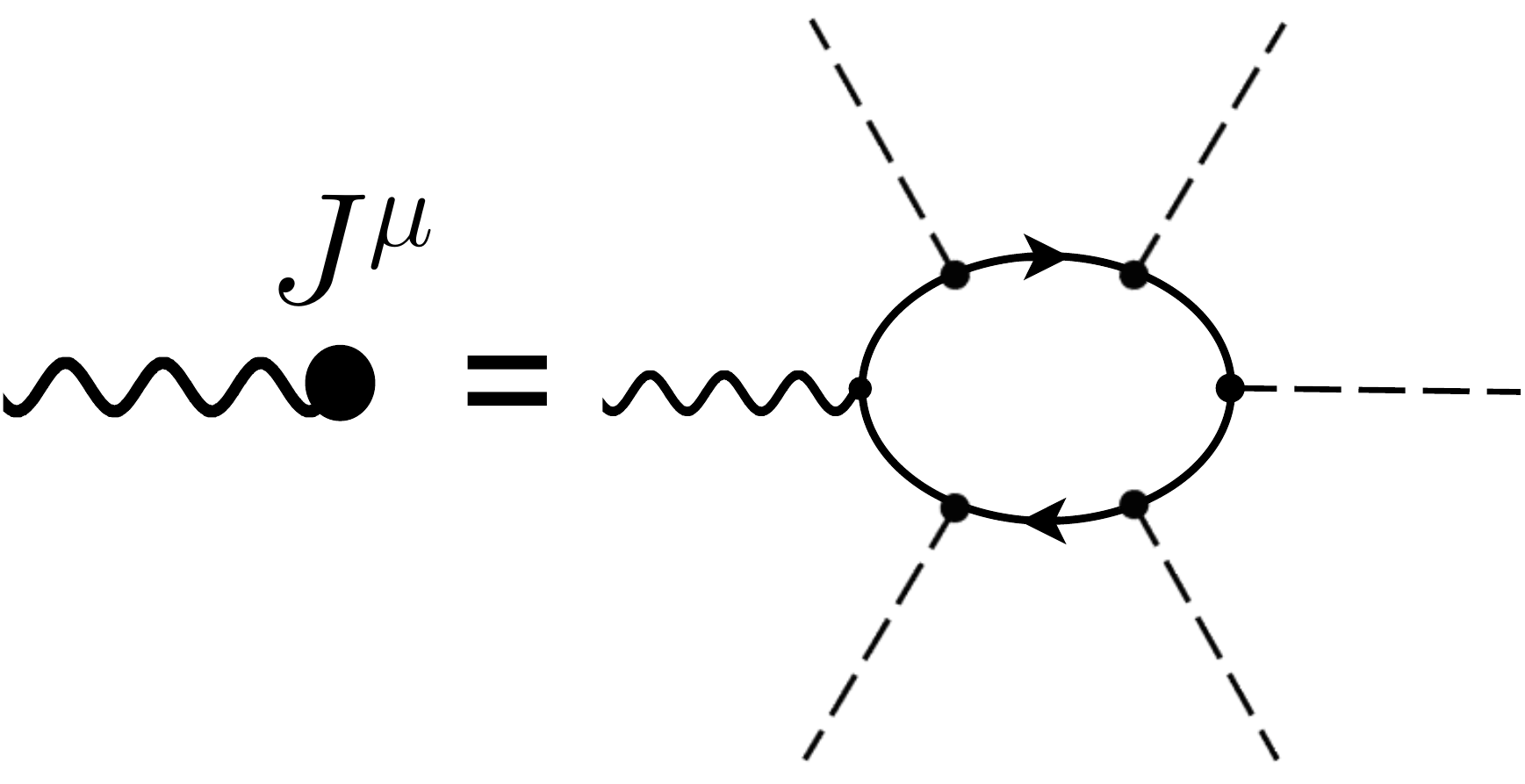} 
\caption{ In the fermionized language, one can capture the anomaly effect on induced (fractional) charge/current under soliton background by the 1-loop diagram.\cite{Goldstone:1981kk}
With the solid line --- represents fermions, 
the wavy line  \protect\middlewave{.42cm}\, 
 represents the external (gauge) field coupling to the induced current $J^\mu$ (or charge $J^0$), and the dashed line - - represents the scalar soliton (domain walls) background.
Here in Sec.\ref{sec:Goldstone-Wilczek}, instead of fermionizing the theory, we directly address in the bosonized language to capture the bosonic anomaly.}
\label{fig:soliton_current}
\end{figure}

Some remarks follow:
If the system is placed on a ring, 
(i) First, with net soliton (or particle) insertions, then the total charge induced is non-zero, see Fig.\ref{fig:domain_walls_kink}.\\
(ii) Second, without net soliton (or particle) insertions, 
then the total charge induced is obviously zero, as domain walls necessarily
come in pairs with opposite charges on the kink and the anti-kink, see Fig.\ref{fig:domain_walls_kink}.\\
(iii) One can also capture this bosonic anomaly in the fermionized language using the 1-loop diagram under soliton background \cite{Goldstone:1981kk},  shown in Fig.\ref{fig:soliton_current}.\\
(iv) A related phenomena has also been examined recently where 
fractionalized boundary excitations cause that 
the symmetry-broken boundary cannot be proliferated to restore the symmetry.

\subsubsection{Lattice approach: Projective phase observed at Domain Walls}

Now we would like to formulate a fully regularized lattice approach to derive the induced fractional charge, 
and compare 
to the complementary field theory done in Sec.\ref{sec:Type II bnom field} and Goldstone-Wilczek approach in Sec.\ref{sec:Goldstone-Wilczek}.
Recall that in the case of a system with onsite symmetry, such as $Z_{N}$ rotor 
model on a 1D ring with a simple Hamiltonian of $\sum_j (\sigma_j + \sigma_j^\dagger)$,
there is an on-site symmetry transformation $S=\prod_j \tau_j $ acting on the full ring. 
We can simply take a segment (from the site $r_1$ to $r_2$) of the symmetry transformation defined as a $D$ operator $D(r_1,r_2) \equiv \prod_{j=r_1}^{r_2} \tau_j $.
The $D$ operator does the job to flip the measurement on $\langle \sigma_\ell \rangle$. What we mean is that $\langle \psi | \sigma_{\ell} |\psi \rangle$
and $\langle \psi' | \sigma_{\ell} |\psi' \rangle \equiv\langle \psi | D^\dagger\sigma_{\ell} D |\psi \rangle=e^{i 2\pi/N}\langle \psi | \sigma_{\ell} |\psi \rangle$ are distinct by a phase $e^{i 2\pi/N}$
as long as $\ell \in [r_1, r_2]$. Thus $D$ operator creates domain wall profile.

For our case of SPT edge modes with non-onsite symmetry studied here, we are readily to generalize the above and take a line segment of non-onsite symmetry transformation with symmetry $Z_{N_u}$ (from the site $r_1$ to $r_2$) and define it as 
a $D_{N_u}$ operator, $D_{N_u}(r_1,r_2)\equiv \prod_{j=r_1}^{r_2} \tau^{(u)}_j \prod_{j=r_1}^{r_2} U_{j,j+1} W^{III}_{j,j+1}$ (from the expression of $S_{N_u}$, with the onsite piece  $\tau^{(u)}_j$ and the non-onsite piece  $U_{j,j+1}$ in Eq.(\ref{S1symp12}) 
and $W_{j,j+1}$ in Eq.(\ref{eq:Type III_W})
).
This $D$ operator effectively creates domain wall on the state with a kink (at the $r_1$) and anti-kink (at the $r_2$) feature, such as in Fig.\ref{fig:domain_walls_kink}.
The total net charge on this type of domain wall (with equal numbers of kink and anti-kinks) is zero, due to no net soliton insertion (i.e. no net winding, so $\int^L_0 \partial_x \phi \, dx =0$).
However, by well-separating kinks and anti-kinks,
we can still compute the phase gained at each single kink. 
We consider the induced charge measurement by $S( D|\psi\rangle) $, which is $(S D S^\dagger) S|\psi\rangle =e^{\ti (\Theta_0+\Theta)} D |\psi\rangle $, where $\Theta_0$ is from the initial charge (i.e. $S|\psi\rangle \equiv e^{\ti \Theta_0} |\psi\rangle$) and $\Theta$ is from the charge gained on the kink. The measurement of symmetry $S$ producing a phase $e^{\ti \Theta}$, implies a nontrivial induced charge trapped at the kink of domain walls. We compute the phase at the left kink on a domain wall for all Type I, II, III SPT classes, and summarize them in Table \ref{table2}.

\begin{center}
\begin{table}[!h]
\makebox[\textwidth][c]
{\fontsize{8pt}{1em}\selectfont
{\begin{tabular}{|c||c|c|c|}
\hline
\text{SPT} & $e^{\ti\Theta_L}$ of $D_{N_u}|\psi\rangle$ acted by $Z_{N_v}$ symmetry $S_{v}$ & $e^{\ti \Theta_L}$ 
under a soliton  $\int_0^L dx\, \partial_x \phi_u= 2\pi$ & Frac charge\\[.5mm]  \hline \hline
$p_{1}$ & $S^{(p_1)}_{N_1} D^{(p_1)}_{N_1} S_{N_1}^{(p_1)\dagger}  $ $\to$  $e^{\ti\Theta_L}=e^{ \ti\,\frac{2\pi p_1}{N_1^2}\,}$ &  
$e^{\ti\Theta_L}=e^{ \ti\,\frac{2\pi p_1}{N_1}\,}$  & No \\[1mm]  \hline
$p_{12}$      & 
$S^{(p_{12})}_{N_2} D^{(p_{12})}_{N_1} S_{N_2}^{(p_{12})\dagger} \to e^{\ti \Theta_L}=e^{ \ti\,\frac{2\pi }{ N_2} \mathbf{\frac{p_{12}}{N_{12}}} }$  & 
$e^{\ti\Theta_L}=e^{ \ti\,\frac{2\pi p_{12}}{N_{2} }\,}$ &  Yes  \\[1mm] \hline 
$p_{123}$ &  $S^{(p_{123})}_{N_2}D^{(p_{123})}_{N_1} S_{N_2}^{(p_{123})\dagger} $ $\to$  $e^{\ti\Theta_L}=e^{ \ti \,\frac{2\pi p_{123} n_3}{N_{123} } \,}$  &  
$e^{\ti\Theta_L}=1$ & No \\[1mm] \hline
\end{tabular}
}}\hspace*{20mm}
\caption{  The phase $e^{i\Theta_L}$ on a domain wall $D_u$ acted by $Z_{N_v}$ symmetry $S_{v}$. This phase is computed at the left kink (the site $r_1$).
The first column shows SPT class labels $p$. The second and the third columns show the computation of phases.
The last column interprets whether the phase indicates a nontrivial induced $Z_{N}$ charge.
Only Type II SPT class with $p_{12}\neq 0$ contains nontrivial induced $Z_{N_2}$ charge with a unit of $\mathbf{{p_{12}}/{N_{12}}}$ trapped at the kink of $Z_{N_1}$-symmetry breaking domain walls.
Here $n_3$ is the exponent inside the $W^{\text{III}}_{}$ matrix, 
$n_3=0,1,\dots,N_3-1$ for each subblock within the total $N_3$ subblocks. 
$N_{12}\equiv\gcd(N_1,N_2)$ and $N_{123}\equiv{\gcd(N_1,N_2,N_3)}$. }
\label{table2}
\end{table}
\end{center}

Although we obtain $e^{\ti\Theta_L}$ for each type, but there are some words of caution for interpreting it. \\
\noindent
{\bf (i)} For Type I class,
with the $Z_{N_1}$-symmetry breaking domain wall, there is no notion of induced $Z_{N_1}$ charge since there is no $Z_{N_1}$-symmetry (already broken) to respect.\\
\noindent
{\bf (ii)} 
$(D^{(p)}_{N})^n$ captures $n$ units of $Z_N$-symmetry-breaking domain wall.
The calculation $S^{(p)}_{N} (D^{(p)}_{N})^n S^{(p)\dagger}_{N}$ renders a $e^{\ti\Theta_L}$ phase for the left kink and 
a $e^{\ti\Theta_R}=e^{-\ti\Theta_L}$ phase for the right anti-kink.
We choose the domain operator as a segment of symmetry transformation.
For Type II class, if we have operators 
$(D^{(p_{12})}_{N_1})^0$ acting on the interval $[0,x_1)$,
while $(D^{(p_{12})}_{N_1})^1$ acting on the interval $[x_1,x_2)$, $\dots$, 
and $(D^{(p_{12})}_{N_1})^{N_{12}}$ acting on the interval  $[x_{N_{12}-1},x_{N_{12}}=L)$,
then we create the domain wall profile shown in Fig.\ref{fig:domain_walls_kink}. 
It is easy to see that due to charge cancellation on each kink/anti-kink,
the $S^{(p_{12})}_{N_2} (D^{(p_{12})}_{N_1})^{N_{12}}  S_{N_2}^{(p_{12})\dagger}$ measurement on a left kink  
captures the same amount of charge trapped by a nontrivial soliton: $\int_0^L dx\, \partial_x \phi_u= 2\pi$. 
\\
\noindent
{\bf (iii)} For Type II class, we consider $Z_{N_1}$-symmetry breaking domain wall (broken to a unit of $\Delta\phi_1=2\pi/N_{12}$), and find that there is induced $Z_{N_2}$ charge with a unit of $\mathbf{{p_{12}}/{N_{12}}}$, consistent with field theory approach in Eq.(\ref{eq:induced charge type II}),(\ref{eq: G-W Q}).
For a total winding is $\int_0^L dx\, \partial_x \phi_1= 2\pi$, there is also a nontrivial induced $\mathbf{ p_{12} }$ units of $Z_{N_2}$ charge.
Suppose a soliton generate this winding number $1$ domain wall profile, even if $p_{12}=N_{12}$ is identified as the trivial class as $p_{12}=0$, we can observe 
$\mathbf{N_{12} }$ units of $Z_{N_2}$ charge, which is in general still not $N_2$ units of $Z_{N_2}$ charge. 
This phenomena has no analogs in Type I, 
and can be traced back to the discussion in Sec.\ref{sec:Type II field}.\\ 
\noindent
{\bf (iv)} For Type III class, with a $Z_{N_1}$-symmetry breaking domain wall: 
On one hand, the $\Theta_L$ phase written in terms of $Z_{N_2}$ or $Z_{N_3}$ charge unit is non-fractionalized but integer.
On the other hand, we will find in Sec.\ref{sec: degenerate zero A} that the $Z_{N_2}$, $Z_{N_3}$ symmetry transformation surprisingly no longer commute. 
So there is no proper notion of induced $Z_{N_2}$, $Z_{N_3}$ charge at all in the Type III class.

\subsection{Type III Bosonic Anomaly: Degenerate zero energy modes (projective representation)
\label{sec: Type III zero} }

We apply the tools we develop 
to study the physical measurements for Type III bosonic anomaly.

\subsection{Field theory approach: Degenerate zero energy modes 
trapped at the kink of $Z_N$ symmetry-breaking Domain Walls \label{sec: degenerate zero A} }
We propose the experimental/numerical signature for certain SPT with Type III symmetric class $p_{123} \neq 0$ under the case of (at least) three symmetry group $Z_{N_1} \times Z_{N_2} \times Z_{N_3}$.
Under the presence of a $Z_{N_1}$ symmetry-breaking domain wall (without losing generality, we can also assume it to be any $Z_{N_u}$), we can detect that the
remained unbroken symmetry $Z_{N_2}$, $Z_{N_3}$ carry projective representation. More precisely,
under the $Z_{N_1}$ domain-wall profile,
\be
\int^{L}_{0}\,dx\,\partial_{x} \phi_{1} =\phi_{1}(L) -\phi_{1}(0) = 2\pi \frac{n_1}{N_1},
\ee
we compute the commutator between two unbroken symmetry operators Eq.(\ref{eq:globalS_Type III}): 
\cblue{
\bea \label{eq:commutator_S2_S3}
S^{(p_{231})}_{N_2} S^{(p_{312})}_{N_3} =S^{(p_{312})}_{N_3} S^{(p_{231})}_{N_2} e^{\ti \frac{2\pi\; n_1}{N_{123}} p_{123}}
\Rightarrow [\log S^{(p_{231})}_{N_2}, \log S^{(p_{312})}_{N_3}]=\ti \frac{2\pi\; n_1}{N_{123}} p_{123},
\eea
}where we identify the index $(p_{231}+p_{312} ) \to p_{123}$ as the same one.
This non-commutative relation Eq.(\ref{eq:commutator_S2_S3}) indicates that $S^{(p_{231})}_{N_2}$ and $S^{(p_{312})}_{N_3}$ are not in a linear representation, but in a projective representation of $Z_{N_2}$, $Z_{N_3}$ symmetry.
This is analogous to the commutator $[T_x,T_y]$ of magnetic translations $T_x$, $T_y$ along $x,y$ direction on a $\mathbb{T}^2$ torus for a filling fraction $1/k$ fractional quantum hall state
(described by U$(1)_k$ level-$k$ Chern-Simons theory): 
\cblue{
\bea
e^{\ti T_x} e^{\ti T_y} =e^{\ti T_y} e^{\ti T_x} e^{\ti\, 2\pi /k}
\Rightarrow [T_x,T_y]= -\ti\, 2\pi /k,
\eea}%
where one studies its ground states on a $\mathbb{T}^2$ torus with a compactified $x$ and $y$ direction gives $k$-fold degeneracy.
\cblue{The k degenerate ground states are $| \psi_m \rangle$ with $m=0, 1, \dots, k-1$, while $| \psi_m \rangle = | \psi_{m+k} \rangle$.
The ground states are chosen to satisfy: $e^{\ti T_x} | \psi_m \rangle =e^{\ti \frac{2\pi m}{k}} | \psi_m \rangle $, $e^{\ti T_y} | \psi_m \rangle = | \psi_{m+1} \rangle $}.
Similarly, for Eq.(\ref{eq:commutator_S2_S3}) we have a $\mathbb{T}^2$ torus compactified in $\phi_2$ and $\phi_3$ directions, such that:\\
(i)  There is a ${N_{123}}$-fold degeneracy for zero energy modes at the domain wall. 
\cblue{We can count the degeneracy by constructing the orthogonal ground states:
consider the eigenstate $| \psi_m \rangle $ of unitary operator $S^{(p_{231})}_{N_2}$, it implies that
$S^{(p_{231})}_{N_2} | \psi_m \rangle =e^{\ti \frac{2\pi\; n_1}{N_{123}} p_{123} m} | \psi_m \rangle$.
$S^{(p_{312})}_{N_3} | \psi_m \rangle =| \psi_{m+1} \rangle $. As long as $\gcd(n_1\,p_{123} ,{N_{123}})=1$,
we have ${N_{123}}$-fold degeneracy of $| \psi_m \rangle $ with $m=0,\dots,{N_{123}}-1$.}\\
(ii) 
Eq.(\ref{eq:commutator_S2_S3}) means the symmetry is realized projectively for the trapped zero energy modes at the domain wall.

We observe these are the signatures of Type III bosonic anomaly. 
This Type III anomaly in principle can be also captured by the perspective
of \emph{decorated $\Z_{N_1}$ domain walls} 
with 
projective $\Z_{N_2} \times \Z_{N_3}$-symmetry. 

\subsubsection{Cocycle approach: Degenerate zero energy modes from $Z_N$ symmetry-preserving monodromy defect (branch cut)
- dimensional reduction from 2D to 1D \label{sec: degenerate zero A}}

\begin{figure}[h!]
\centering
\includegraphics[width=0.5\textwidth]{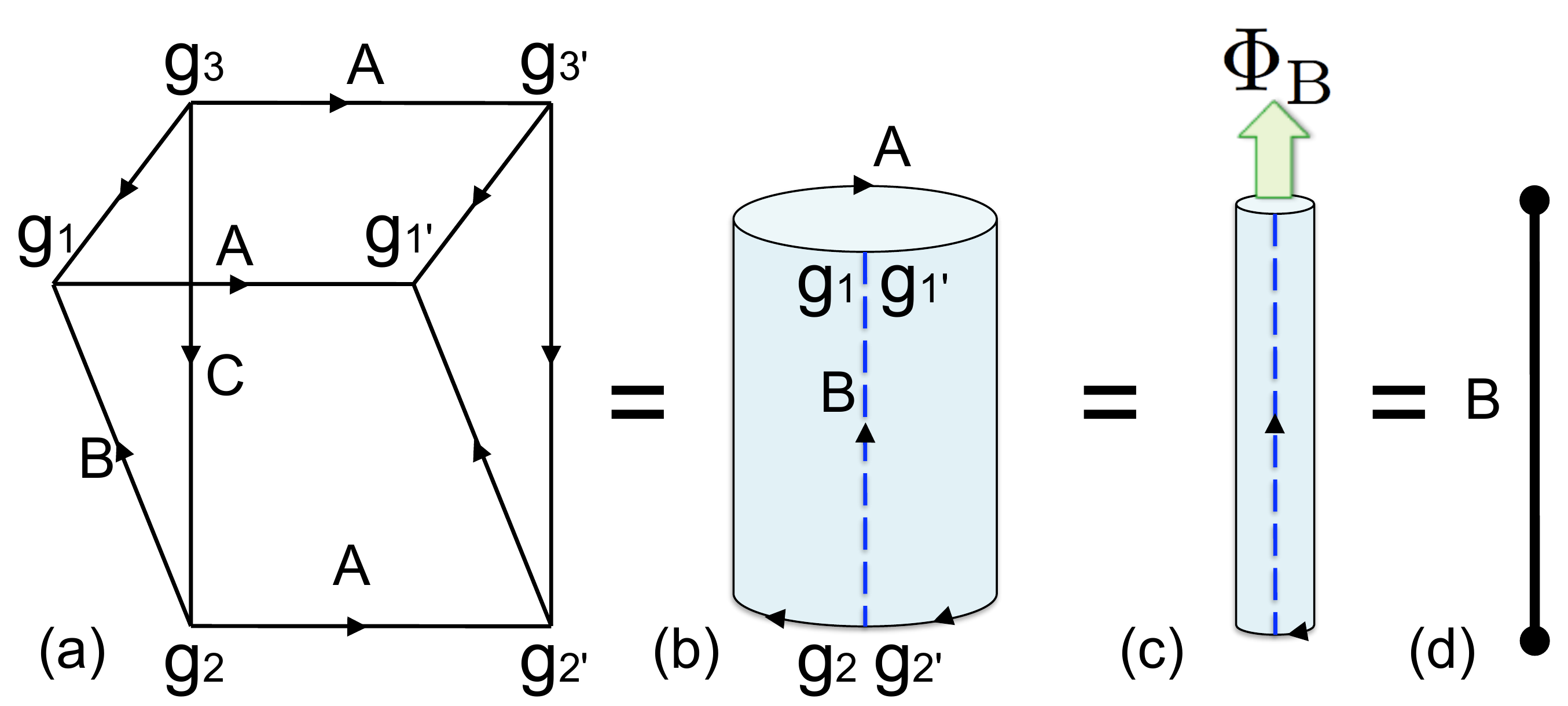} 
\caption{ (a) The induced 2-cocycle from a 2+1D $M^3=M^2 \times  I^1$ topology with a symmetry-preserving $Z_{N_u}$ flux $A$ insertion 
(b) Here $M^2=S^1 \times I^1$ 
is a 2D spatial cylinder, composed by $A$ and $B$, with another extra time dimension $I^1$.
Along the $B$-line we insert a monodromy defect of $Z_{N_1}$, such that $A$ has a nontrivial group element value $A=g_{1'} g_{1}^{-1} =g_{2'} g_{2}^{-1}
=g_{3'} g_{3}^{-1}  \in Z_{N_1}$. The induced 2-cocycle $\beta_A(B,C)$ is a nontrivial element in $\cH^2(Z_{N_v} \times Z_{N_w},\tU(1))$ $=\mathbb{Z}_{N_{vw}}$ (here $u,v,w$ cyclic as $\epsilon^{uvw}=1$), thus which carries a projective representation.
(c) A monodromy defect can viewed as a branch cut induced by a $\Phi_B$ flux insertion (both modifying the Hamiltonians). 
(d) This means that when we do dimensional reduction on the compact ring $S^1$ and view the reduced system as a 1D line segment, there are $N_{123}$ degenerate zero energy modes 
(due to the nontrivial projective representation).}
\label{fig:induced_2-cocycle}
\end{figure}

In Sec.\ref{sec: degenerate zero A}, we had shown the symmetry-breaking domain wall would induce 
degenerate zero energy modes for Type III SPT. In this section,
we will further show that,
a symmetry-preserving $Z_{N_1}$ flux insertion (or a monodromy defect or branch cut modifying the Hamiltonian as in Ref.\cite{Santos:2013uda},\cite{WenSPTinv})
can also have degenerate zero energy modes. 
This is the case that, see Fig.\ref{fig:induced_2-cocycle}, when we put the system on a 2D cylinder and dimensionally reduce it to a 1D line along the monodromy defect.
In this case there is no domain wall, and the $Z_{N_1}$ symmetry is not broken (but only translational symmetry is broken near the monodromy defect / branch cut).

In the below discussion, we will directly use 3-cocycles $\omega_3$ from cohomology group $\cH^3(G,\tU(1))$ to detect the Type III bosonic anomaly.
For convenience we use the non-homogeneous cocycles  (the lattice gauge theory cocycles), though there is no difficulty to convert it to homogeneous cocycles (SPT cocycles).
The definition of the lattice gauge theory $n$-cocycles are indeed related to SPT $n$-cocycles:
\be
\omega_n(A_1,A_2, \dots,A_n)=\nu_n(A_1A_2 \dots A_n, A_2 \dots A_n, \dots, A_n, 1)
=\nu_n(\tilde{A}_1, \tilde{A}_2, \dots, \tilde{A}_n, 1)
\ee
here $\tilde{A}_j\equiv A_j A_{j+1} \dots A_n$. For 3-cocycles
\bea
&&\omega_3(A,B,C)=\nu_3(ABC,BC,C,1) \Rightarrow \omega_3(g_{01},g_{12},g_{23})\\
&&=\omega_3(g_{0} g_{1}^{-1}, g_{1} g_{2}^{-1} ,g_{2} g_{3}^{-1}) 
=\nu_3(g_{0}g_{3}^{-1} ,g_{1}g_{3}^{-1},g_{2}g_{3}^{-1},1)= \nu_3(g_{0},g_{1},g_{2},g_{3}) \nonumber
\eea
Here $A=g_{01}$, $B=g_{12}$, $C=g_{23}$, with $g_{ab} \equiv g_a g_b^{-1}$. We use the fact that SPT $n$-cocycle $\nu_n$ belongs to the $G$-module, such that 
for $r$ are group elements of $G$, it obeys $r \cdot \nu_n({r}_0, {r}_1, \dots, {r}_{n-1}, 1)=\nu( r {r}_0, r {r}_1, \dots, r {r}_{n-1}, r)$ 
 (here we consider only Abelian group $G=\prod_i Z_{N_i}$).
In our case, we do not have time reversal symmetry, so group action $g$ on the $G$-module is trivial.
%
%

\begin{figure}[h!]
\centering
\includegraphics[width=0.65\textwidth]{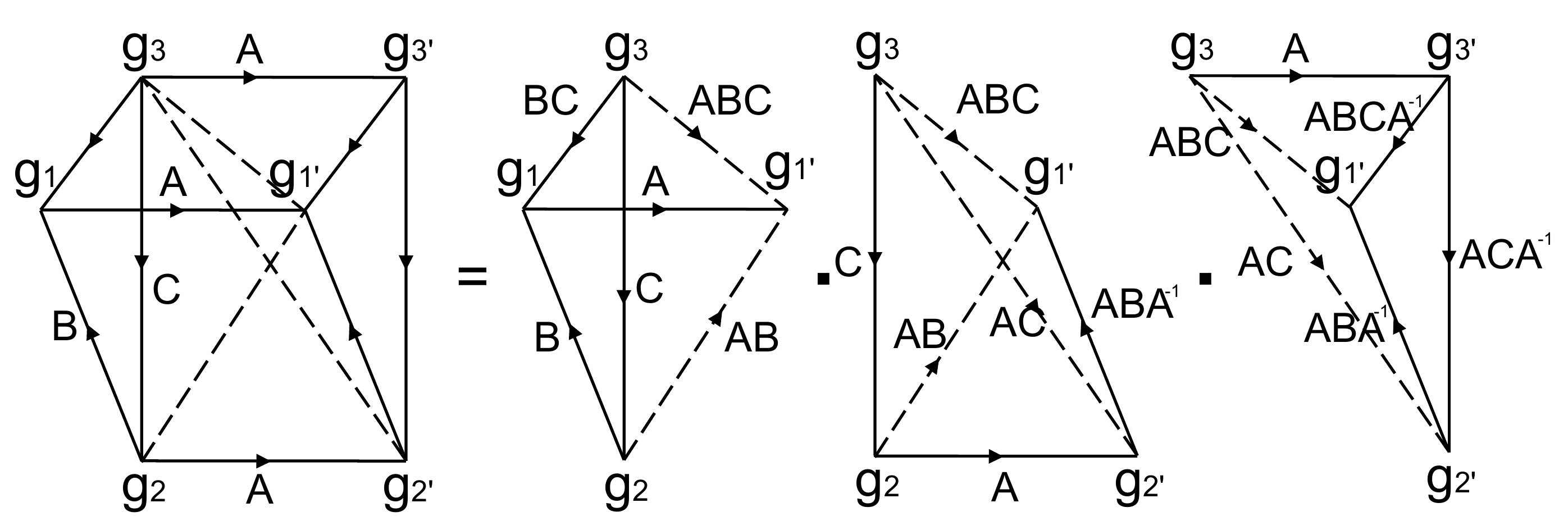} 
\caption{The  
triangulation of a $M^3=M^2 \times  I^1$ topology (here $M^2$ is a spatial cylinder composed by the $A$ and $B$ direction, with a $I^1$ time) into three tetrahedra with branched structures. 
}
\label{fig:decomp_2-cocycle}
\end{figure}

In short, there is no obstacle so that we can simply use the lattice gauge theory 3-cocycle $\omega(A,B,C)$  to study the SPT 3-cocycle $\nu(ABC,BC,C,1)$.
Our goal is to design a geometry of 3-manifold $M^3=M^2 \times  I^1$ with $M^2$ the 2D cylinder with flux insertion (or monodromy defect) and with the $I^1$ time direction
(see Fig.\ref{fig:induced_2-cocycle}(a))
with a sets of 3-cocycles as tetrahedra filling this geometry (Fig.\ref{fig:decomp_2-cocycle}).
All we need to do 
is computing the 2+1D SPT path integral $\textbf{Z}_{\text{SPT}}$ (i.e. partition function) using 3-cocycles $\omega_3$,\cite{{WenSPTinv}} 
\be
\textbf{Z}_{\text{SPT}}=|G|^{-N_v} \sum_{\{ g_{v}\}} \prod_i (\omega_3{}^{s_i}(\{ g_{v_a} g_{v_b}^{-1} \}))
\ee
Here $|G|$ is the order of the symmetry group, $N_v$ is the number of vertices, $\omega_3$ is 3-cocycle, and ${s_i}$ is the exponent 1 or $-1$(i.e. $\dagger$) depending on the orientation of each tetrahedron($3$-simplex). The summing over group elements $g_{v}$ on the vertex produces a symmetry-perserving ground state.
We consider a specific $M^3$, a $3$-complex of Fig.\ref{fig:induced_2-cocycle}(a), which can be decomposed into tetrahedra (each as a $3$-simplex) shown in Fig.\ref{fig:decomp_2-cocycle}.
There the 3-dimensional spacetime manifold is under triangulation 
(or cellularization) into three tetrahedra. 

We now go back to remark that 
the 3-cocycle condition 
indeed means that the path integral $\textbf{Z}_{\text{SPT}}$ on the 3-sphere $S^3$ (as the surface the 4-ball $B^4$) will be trivial as 1.
The 3-coboundary condition 
means to identify the same topological terms (i.e. 3-cocycle) up to total derivative terms.
There is a specific way (called the {\it branching structure}) to determine the orientation of tetrahedron,
thus to determine the sign of $s$ for 3-cocycles $\omega_3{}^{s}$ by the determinant of volume, $s\equiv\det(\vec{v_{32}},\vec{v_{31}},\vec{v_{30}})$. 
Two examples of the orientation with $s=+1, -1$ are:
\bea
&&{\CocycleTriangle{g_0}{g_1}{g_3}{g_2}{1} } = \tetrahedra{g_0}{g_1}{g_2}{g_3}  \label{eq:simplex1} 
=\omega_3(g_0 g_1{}^{-1},g_1 g_2{}^{-1}, g_2 g_3{}^{-1}) 
\eea
\bea
&&{\CocycleTriangleTWO{g_1}{g_0}{g_3}{g_2}{1} } = \tetrahedraTWO{g_1}{g_0}{g_2}{g_3} \label{eq:simplex2} 
=\omega_3{}^{-1}(g_0 g_1{}^{-1},g_1 g_2{}^{-1}, g_2 g_3{}^{-1}). 
\eea
Here we define the numeric ordering $g_{1'}<g_{2'}<g_{3'}<g_{4'}<g_1<g_2<g_3<g_4$, and our arrows connect from the higher to lower ordering.

Now we can 
compute the induced 2-cocycle (the dimensional reduced 1+1D path integral) with a given inserted flux $A$,
determined from three tetrahedra of 3-cocycles, see Fig.\ref{fig:decomp_2-cocycle} and Eq.(\ref{eq:2-cocycle}).
\bea
&&\beta_A(B,C) \equiv {\CocycleTriangle{g_1}{g_2}{g_{1'}}{g_3}{4}  } \centerdot {\CocycleTriangleTWO{g_{2}}{g_{2'}}{g_{1'}}{g_3}{4} } \centerdot {\CocycleTriangle{g_{1'}}{g_{2'}}{g_{3'}}{g_3}{2}  } \label{eq:2-cocycle} \\
&&=\frac{\omega(A,B,C)^{-1} \cdot  \omega(AB A^{-1},A,C)}{  \omega(AB A^{-1},ACA^{-1},A) }
=\frac{\omega(B,A,C)}{\omega(A,B,C) \omega(B,C,A)} \nonumber\\
&&=\frac{\omega(g_1 g_{2}^{-1},  g_{1'} g_1^{-1}, g_2 g_{3}^{-1} )}{\omega(g_{1'} g_1^{-1},g_1 g_{2}^{-1},g_2 g_{3}^{-1}) \omega(g_1 g_{2}^{-1},g_2 g_{3}^{-1},g_{1'} g_1^{-1})} \;\;\;\;\; \nonumber
\eea

We show that 
only when $\omega_3$ is the Type III 3-cocycle $\omega_{\text{III}}$ (of Eq.\ref{type3}), this induced 2-cochain is nontrivial (i.e. a 2-cocycle but not a 2-coboundary). 
In that case,
\be \label{eq:induced 2-cocycle}
\beta_A(B,C)=\exp[\ti \frac{2\pi}{N_{123}}(b_1 a_2 c_3 -a_1 b_2 c_3 -b_1 c_2 a_3)]
\ee 
If we insert $Z_{N_1}$ flux $A=(a_1,0,0)$, then 
we shall compare Eq.(\ref{eq:induced 2-cocycle}) with the nontrivial 2-cocycle $\omega_2(B,C)$ in $\cH^2(Z_{N_2} \times Z_{N_3},\tU(1))$ $=\mathbb{Z}_{N_{23}}$,
\be \label{eq:induced 2-cocycle-2}
\omega_2(B,C)=\exp[\ti \frac{2\pi}{N_{23}}( b_2 c_3 )].
\ee
\cblue{The $\beta_A(B,C)$ is indeed nontrivial 2-cocycle as $\omega_2(B,C)$ 
in the second cohomology group $\cH^2(Z_{N_2} \times Z_{N_3},\tU(1))$. 
Below we like to argue that this Eq.(\ref{eq:induced 2-cocycle-2}) implies the projective representation of the symmetry group $Z_{N_2} \times Z_{N_3}$.
Our argument is based on two facts. 
First, the dimensionally reduced 
SPTs in terms of spacetime partition function Eq.(\ref{eq:induced 2-cocycle-2}) is 
a nontrivial 1+1D SPTs.\cite{{Wang:2014pma}}  
We can physically understand it from the symmetry-twist as a branch-cut modifying the Hamiltonian\cite{Hung:2013cda,{Wang:2014pma}} (see also Sec.\ref{sec:flux}).
Second, 
we know that the 1+1D SPT symmetry transformation $\otimes_x U^x(g)$ along
the 1D's x-site is dictated by 2-cocycle.
The onsite tensor $S(g) \equiv \otimes_x U^x(g)$ acts on a chain 
of 1D SPT renders
\bea
S(g) | \alpha_L, \dots ,\alpha_R\rangle =\frac{\omega_2(\alpha_L^{-1}g^{-1},g)}{\omega_2(\alpha_R^{-1}g^{-1},g)} | g\alpha_L, \dots ,g\alpha_R\rangle,\;\;\;\;\;
\eea
where $\alpha_L$ and $\alpha_R$ are the two ends of the chain, with $g, \alpha_L, \alpha_R, \dots \in G$ all in the symmetry group.
We can derive the effective degree of freedom on the 0D edge $|\alpha_L\rangle$ forms a projective representation of symmetry, we find: 
\be
S(B) S(C) |\alpha_L\rangle 
=\frac{ {\omega_2(\alpha_L^{-1}{C}^{-1} B^{-1},B )}  {\omega_2(\alpha_L^{-1}{C}^{-1},B )}  }{ {\omega_2(\alpha_L^{-1}{C}^{-1} B^{-1},BC)}} S(BC) |\alpha_L\rangle
=\omega_2(B,C) S(BC) |\alpha_L\rangle 
\ee
In the last line, we implement the 2-cocycle condition of $\omega_2$: $\delta \omega_2(a,b,c)=\frac{\omega_2(b,c) {\omega_2(a,bc)} }{ {\omega_2(ab,c)} {\omega_2(a,b)}}=1$.
The projective representation of symmetry transformation $S(B) S(C) =\omega_2(B,C) S(BC)$ is explicitly derived,
and the projective phase is the 2-cocycle $\omega_2(B,C)$ classified by $\cH^2(G,\tU(1))$.
Interestingly, the symmetry transformations on two ends together will form a linear representation,
namely $S(B) S(C) |\alpha_L, \dots ,\alpha_R\rangle = S(BC) |\alpha_L, \dots ,\alpha_R\rangle$. 
}


The same argument holds when $A$ is $Z_{N_2}$ flux or $Z_{N_3}$ flux.
\cblue{From Sec.\ref{sec: Type III zero}}, the projective representation of symmetry implies  
the nontrivial ground state degeneracy
if we view the system as a dimensionally-reduced 1D line segment as in Fig.\ref{fig:induced_2-cocycle}(d).
From the ${N_{123}}$ factor in Eq.(\ref{eq:induced 2-cocycle}), we conclude there is ${N_{123}}$-fold degenerated zero energy modes. 

\begin{figure}[h!]
\centering
\includegraphics[width=0.45\textwidth]{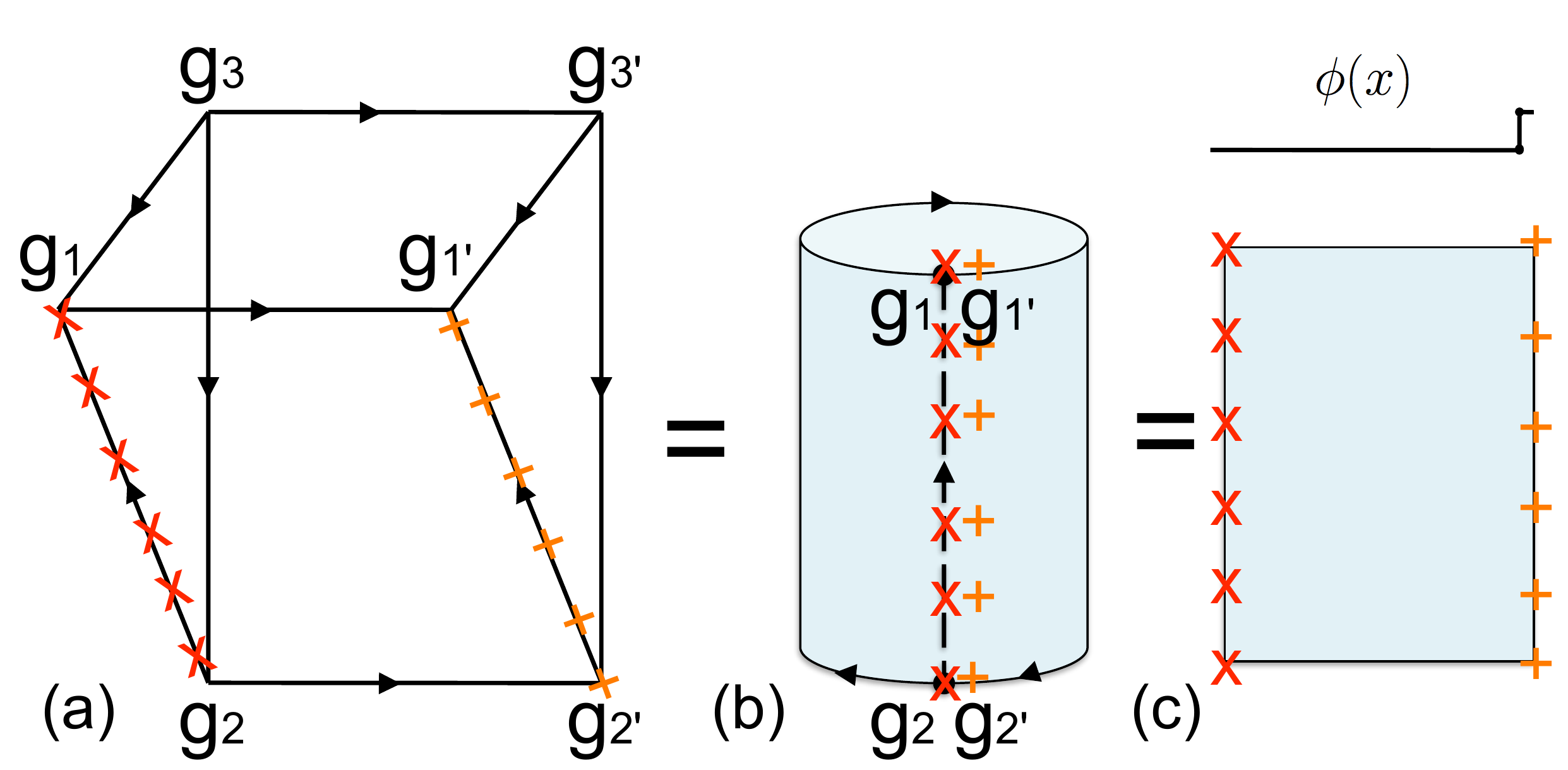} 
\caption{ The $Z_{N_1}$ {\bf symmetry breaking} domain wall along the red $\times$ mark and/or orange + mark, which induces ${N_{123}}$-fold degenerate zero energy modes. 
The situation is very similar to Fig.\ref{fig:induced_2-cocycle} (however, there was $Z_{N_1}$ {\bf symmetry-preserving} flux insertion). 
We show that both cases the induced 2-cochain from calculating path integral $\textbf{Z}_{\text{SPT}}$ renders a nontrivial 2-cocycle of $\cH^2(Z_{N_2} \times Z_{N_3},\tU(1))$ $=\mathbb{Z}_{N_{23}}$,
thus carrying nontrivial projective representation of symmetry.
}
\label{fig:induced_TypeIII_2-cocycle_domain}
\end{figure}

We should make two more remarks: \\
(i) The precise 1+1D path integral is actually summing over $g_{v}$ with a fixed flux $A$ as
$
\textbf{Z}_{\text{SPT}}=|G|^{-N_v} \sum_{\{ g_{v}\}; \text{fixed $A$}}   \beta_A(B,C)
$, but overall our discussion above still holds.\\
(ii) We have used 3-cocycle to construct a symmetry-preserving SPT ground state under $Z_{N_1}$ flux insertion.
We can see that indeed
a $Z_{N_1}$ symmetry-breaking domain wall of Fig.\ref{fig:induced_TypeIII_2-cocycle_domain}
 can be done in almost the same calculation - using 3-cocycles filling a 2+1D spacetime complex(Fig.\ref{fig:induced_TypeIII_2-cocycle_domain}(a)). Although there in Fig.\ref{fig:induced_TypeIII_2-cocycle_domain}(a), we need to fix the group elements $g_1=g_2$ on one side  
 (in the time independent domain wall profile, we need to fix $g_1=g_2=g_3$) and/or fix $g_1'=g_2'$ on the other side.
Remarkably, we conclude that both the {\bf $Z_{N_1}$-symmetry-preserving flux insertion} and {\bf $Z_{N_1}$ symmetry-breaking domain wall} both
provides a ${N_{123}}$-fold degenerate ground states (from the nontrivial projective representation for the $Z_{N_2}$, $Z_{N_3}$ symmetry).
The symmetry-breaking case is consitent with Sec.\ref{sec: degenerate zero A}.


\subsection{Type I, II, III class observables: Flux insertion and non-dynamically ``gauging'' the non-onsite symmetry \label{sec:flux}}

\begin{figure}[h!]
\centering
\includegraphics[width=0.5\textwidth]{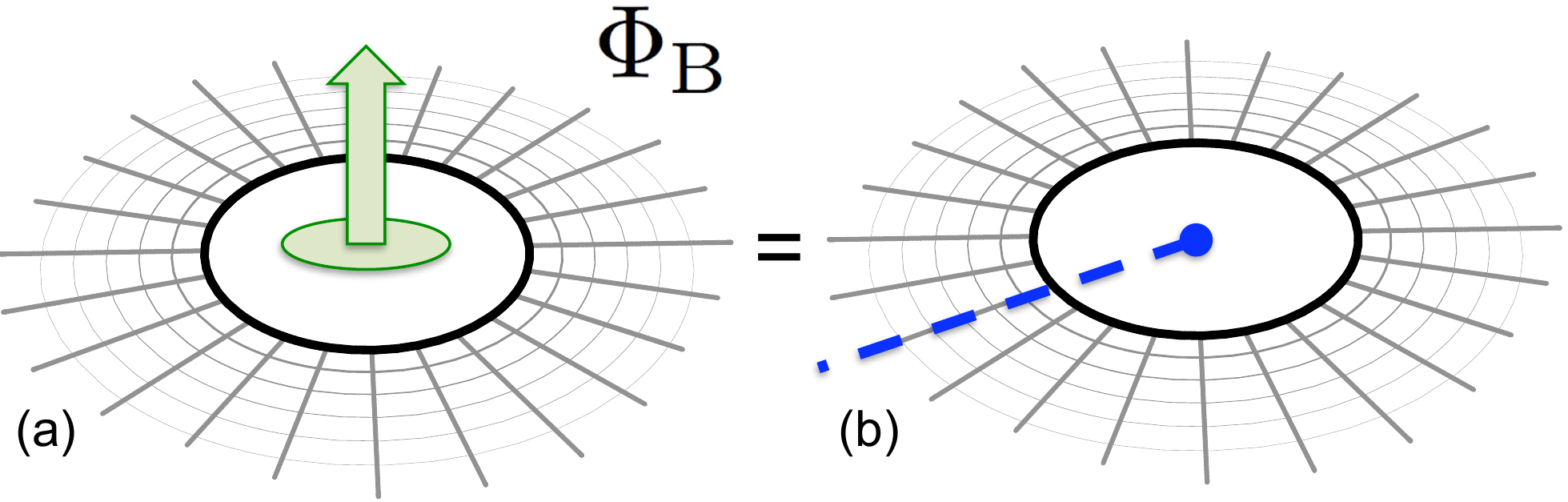}  (c) \includegraphics[width=0.2\textwidth]{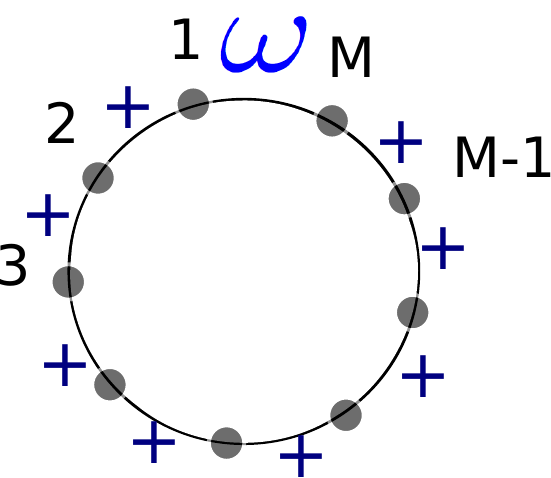}
\caption{
(a) Thread a gauge 
flux $\Phi_B$ through a 1D ring (the boundary of 2D SPT).
(b) The gauge 
flux is effectively captured by 
a branch cut (the dashed line in the blue color). Twisted boundary condition is applied on the branch cut.
The (a) and (b) are equivalent in the sense that both cases capture the equivalent physical observables, such as the energy spectrum.
{
The illustration of an effective 1D lattice model with $M$-sites on a compact ring under a discrete $Z_N$ flux insertion. 
Effectively the gauge 
flux insertion is  captured by a branch cut located between the site-$M$ and the site-$1$.
This results in a $Z_N$ variable $\omega$ insertion as a twist effect modifying the lattice Hamiltonian around the site-$M$ and the site-$1$.
}
}
\label{fig:flux_twist}
\end{figure}

With the Type I, Type II, Type III SPT lattice model built in Chap.\ref{aofSymmetry}, 
in principle we can perform numerical simulations to measure their physical observables, such as (i) the energy spectrum, (ii) the entanglement entropy and (iii) the central charge of the edge modes.
Those are the physical observables for the ``untwisted sectors'', and we would like to further achieve more physical observables on the lattice, by
applying the parallel discussion, 
using $Z_N$ gauge 
flux insertions through the 1D ring. The similar idea can be applied to detect SPTs numerically. 
The gauge flux insertion on the SPT edge modes (lattice Hamiltonian) is like {\it gauging its non-onsite symmetry in a non-dynamical way}. We emphasize that {\it gauging in a non-dynamical way} because the gauge flux is not a local degree of freedom on each site, but a global effect.
The Hamiltonian affected by gauge flux insertions can be realized as the Hamiltonian with twisted boundary conditions, see an analogy made in Fig.\ref{fig:flux_twist}. 
Another way to phrase the flux insertion is that it creates a monodromy defect\cite{WenSPTinv} (or a branch cut) which modify both the bulk and the edge Hamiltonian.
\cblue{Namely, our flux insertion acts effectively as the \emph{symmetry-twist}\cite{{Wang:2014pma}} modifying the Hamiltonian}. 
Here we 
outline the twisted boundary conditions on the Type I, Type II, Type III SPT lattice model of Chap.\ref{aofSymmetry}. 


We firstly review the work done in \cite{Santos:2013uda} of Type I  SPT class and then extends it to Type II, III class. 
We aim to 
build a lattice model with twisted boundary conditions
to capture the edge modes physics in the presence of a unit of 
${Z}_{N}$ flux insertion. 
Since the gauge flux effectively introduces a branch cut
breaking the translational symmetry of $T$ (as shown in Fig.~\ref{fig:flux_twist}),
the gauged (or twisted) Hamiltonian, say $\tilde{H}^{(p)}_{N}$, is {\it not} invariant respect to translational operator $T$, say $[\tilde{H}^{(p)}_{N}, T]\neq 0$.
The challenge of constructing $\tilde{H}^{(p)}_{N}$ is to firstly find a new (so-called {\it magnetic or twisted}) translation operator $\tilde{T}^{(p)}$ incorporating the gauge flux effect at the branch cut, in Fig.~\ref{fig:flux_twist} (b)
and in Fig.\ref{fig:flux_twist}, 
say the branch cut is between the site-$M$ and the site-$1$.
We propose two principles to construct the twisted lattice model.
The first general principle is that
a string of $M$ units of {\it twisted translation operator} $\tilde{T}^{(p)}$ renders 
a 
{\it twisted symmetry transformation} $\tilde{S}^{(p)}_{N}$ incorporating a $Z_N$ unit flux,
\be
\label{Type I principle2-1}
\bullet \;
\tilde{S}^{(p)}_{N} \equiv
( \tilde{T}^{(p)} )^{M}
=\tilde{S}^{(p)}_{N}  \cdot (U^{(N,p)}_{M,1} [ {\sigma}^{\dagger}_{M} {\sigma}^{}_{1} ])^{-1} \cdot U^{(N,p)}_{M,1} [ \omega {\sigma}^{\dagger}_{M} {\sigma}^{}_{1}], \;\, 
\ee
with the unitary operator $(\tilde{T}^{(p)})$, i.e. $(\tilde{T}^{(p)})^\dagger\tilde{T}^{(p)}=\mathbb{I}$.
We clarify that $U^{(N,p)}_{M,1}$ is from Eq.(\ref{eq:Type I symmetry explicit}), 
where $U^{(N,p)}_{M,1} [\dots] \equiv U^{(N,p)}_{M,1} \circ [\dots] $ 
means $U^{(N,p)}_{M,1}$ is a function of 
$\dots$ variables. For example, $ U^{(N,p)}_{M,1} [ \omega {\sigma}^{\dagger}_{M} {\sigma}^{}_{1}]$ means that the variable ${\sigma}^{\dagger}_{M} {\sigma}^{}_{1}$ in Eq.(\ref{eq:Type I symmetry explicit}) is replaced by $\omega {\sigma}^{\dagger}_{M} {\sigma}^{}_{1}$ with an extra $\omega$ insertion.
The second principle is that the twisted Hamiltonian is invariant in respect of 
the twisted translation operator, thus also invariant in respect of twisted symmetry transformation, i.e.
\be \label{Type I principle2-2}
\bullet\;\;\; 
[ \tilde{H}^{(p)}_{N}, \tilde{T}^{(p)} ] =0,\;\;\; 
[\tilde{H}^{(p)}_{N}, \tilde{S}^{(p)}_{N} ]=0.
\ee
%
%
We solve Eq.(\ref{Type I principle2-1}) by finding the twisted lattice translation operator 
\begin{equation}
\tilde{T}^{(p)}
= T\, (U^{(N,p)}_{M,1} ( {\sigma}^{\dagger}_{M} {\sigma}^{}_{1}) )
\tau_{1},
\end{equation}  
for each 
$p \in \mathbb{Z}_N$
classes. 
For the $s$ units of $Z_N$ flux, we have the generalization of $\tilde{T}^{(p)}$ from a unit $Z_N$ flux as, 
\be
\tilde{T}^{(p)}|_{s}
=
T\,
(U^{(N , p)}_{M,1} [ {\sigma}^{\dagger}_{M} {\sigma}^{}_{1}] )^s \,\tau^{s}_{1}.
\ee 

Indeed, there is no difficulty to extend this construction to Type II, III classes.
For Type II SPT classes (with nonzero indices $p_{12}$ and $p_{21}$ of Eq.(\ref{S1symp12}), 
while $p_1=p_2=0$) 
the non-onsite symmetry transformation can be reduced from NNN to NN coupling term 
$U^{(N_1,p_{12})}_{j,j+2} \to U^{(N_1,p_{12})}_{j,j+1}$ , also from   
$U^{(N_2,p_{21})}_{j,j+2} \to U^{(N_2,p_{21})}_{j,j+1}$.
The Type II twisted symmetry transformation has exactly the same form as Eq.(\ref{Type I principle2-1}) except replacing the $U$.
For Type III SPT classes, the Type III twisted symmetry transformation also has the same form as Eq.(\ref{Type I principle2-1})
 except replacing the $U$ to $W$ in Eq.(\ref{eq:Type III_W}).
The second principle in Eq.(\ref{Type I principle2-2}) also follows. 
\\




\noindent
{\bf Twisted Hamiltonian}

The twisted Hamiltonian 
$
\tilde{H}^{(p_1,p_2,p_{12})}_{N_1,N_2}
$
can be readily constructed from
$
{H}^{(p_1,p_2,p_{12})}_{N_1,N_2}
$
of Eq.~(\ref{eq:Type II Hamiltonian lattice}),
with the condition Eq.(\ref{Type I principle2-2}).
(An explicit example for Type I SPT 1D lattice Hamiltonian with a gauge flux insertion has been derived in Ref. \cite{Santos:2013uda}, 
which we shall not repeat here.)

Notice that 
the twisted non-trivial
Hamiltonian breaks the SPT global symmetry (i.e. 
if
$p
\neq
0
$
mod($N$), then 
$
[\tilde{H}^{(p)}_{N}, S^{(p)}_{N}  ]
\neq
0
$),
which can be regarded as the sign of $Z_N$ anomaly.\cite{Wen:2013oza}
On the other hand, in the trivial state $p=0$, 
Eq.~(\ref{Type I principle2-1}) yields
$
\tilde{S}^{(p=0)}_{N} 
=
S^{(p=0)}_{N}
=
\prod^{M}_{j=1}\tau_{j}
$,
where the twisted trivial Hamiltonian still \textit{commutes}
with the global ${Z}_{N}$ onsite symmetry,
and the twisted boundary effect is nothing but the usual toroidal 
boundary conditions.~\cite{philippe1997conformal} 
(See also a discussion along the
context of SPT and the orbifolds \cite{Ryu2013orbifolds}. \\ 

The twisted Hamiltonian 
provides distinct low energy spectrum due to the gauge flux insertion (or the symmetry-twist). 
The energy spectrum thus can be
physical observables to distinguish SPTs. Analytically we can use the field theoretic mode expansion for multiplet scalar chiral bosons
$
\Phi_I(x) ={\phi_{0}}_{I}+K^{-1}_{IJ} P_{\phi_J} \frac{2\pi}{L}x+\ti \sum_{n\neq 0} \frac{1}{n} \alpha_{I,n} e^{-in x \frac{2\pi}{L}},
$
with zero modes ${\phi_{0}}_{I}$ and winding modes $P_{\phi_J}$ satisfying the commutator $[{\phi_{0}}_{I},  P_{\phi_J}]=\ti\delta_{IJ}$. The Fourier modes satisfies a generalized Kac-Moody algebra: 
$[\alpha_{I,n} , \alpha_{J,m} ]= n K^{-1}_{IJ}\delta_{n,-m}$.
The low energy Hamiltonian, in terms of various quadratic mode expansions, becomes
\be
H= \frac{(2\pi)^2}{{4\pi} L} [ V_{IJ} K^{-1}_{I l1} K^{-1}_{J l2} P_{\phi_{l1}} P_{\phi_{l2}}+\sum_{n\neq0} V_{IJ} \alpha_{I,n} \alpha_{J,-n}]+\dots
\ee
Following the procedure outlined in Ref.\cite{Santos:2013uda} 
with gauge flux 
taking into account the twisted boundary conditions, we expect the conformal dimension of gapless edge modes of central charge $c=1$ free bosons
labeled by the primary states $| n_1,m_1,n_2,m_2 \rangle$ (all parameters are integers) with the same compactification radius $R$ for Type I and Type II SPTs
(for simplicity, we assume $N_1 =N_2  \equiv N$):
\bea
 \label{}
\tilde{\Delta}^{(p_1,p_2,p_{12})}_{N} 
&=&
\frac{1}{R^2}\left(n_1+\frac{p_1}{N} +\frac{p_{12}}{N_{}}\right)^{2}
+
\frac{R^2}{4}\left(m_1 + \frac{1}{N}\right)^{2}  \nonumber\\
&+&
\frac{1}{R^2}\left(n_2+\frac{p_2}{N}+\frac{p_{21}}{N_{}}\right)^{2}
+
\frac{R^2}{4}\left(m_2 + \frac{1}{N}\right)^{2} \;\;\;\;\;\;\;\;\;\;
\eea
which is directly proportional to the energy of twisted Hamiltonian. ($p_{12}$ or $p_{21}$ can be used interchangeably.)
The conformal dimension $\tilde{\Delta}^{(p_1,p_2,p_{12})}_{N}({\mathcal{P}_u},{\mathcal{P}_{uv}})$ is intrinsically related to the SPT class labels: $p_1,p_2,p_{12}$,
and is a function of momentum ${\mathcal{P}_u} \equiv (n_u+\frac{p_u}{N} +\frac{p_{uv}}{N_{}})(m_u + \frac{1}{N})$
and ${\mathcal{P}_{uv}} \equiv (n_u+\frac{p_u}{N} +\frac{p_{uv}}{N_{}})(m_v + \frac{1}{N})$.
Remarkably, for Type III SPTs, the nature of \emph{non-commutative symmetry} generators will play the key rule, as if the gauged conformal field theory (CFT)
and its correspoinding gauged dynamical bulk theory has \emph{non-Abelian} features,
we will leave this survey for future works. The bottom line is that different classes of SPT's CFT spectra respond to the flux insertion distinctly, thus we can in principle 
distinguish Type I, II and III SPTs.



\section{
Lattice Non-Perturbative Hamiltonian Construction of Anomaly-Free Chiral Fermions and Bosons}  \label{sec:lattice_chiral}


\subsection{Introduction}

Regulating and defining chiral fermion field theory is a very important
problem, since the standard model is one such
theory.\cite{Lee:1956qn} 
However, the fermion-doubling
problem\cite{Nielsen:1980rz,Nielsen:1981xu,Nielsen:1981hk} 
makes it very difficult to define chiral fermions (in an even dimensional
spacetime) on the lattice.  There is much previous research that tries to solve
this famous problem.  One approach is the lattice gauge theory,\cite{K7959}
which is unsuccessful since it cannot reproduce chiral couplings between the
gauge fields and the fermions.  Another approach is the domain-wall
fermion.\cite{K9242,S9390} However, the gauge field in the domain-wall fermion
approach propagates in one-higher dimension.  Another approach is the
overlap-fermion,\cite{L9995,S9947} 
while the path-integral
in the overlap-fermion approach may not describe a finite quantum theory with a
finite Hilbert space for a finite space-lattice.  There is also the mirror
fermion approach\cite{EP8679,M9259,BMP0628,GP0776} which starts with a lattice
model containing chiral fermions in one original \emph{light sector} coupled to gauge theory, \emph{and} its chiral
conjugated as the \emph{mirror sector}.  Then, one tries to include direct interactions or
boson mediated interactions\cite{S8631} 
between fermions to gap out the
mirror sector only.  However, the later works either fail to demonstrate
\cite{GPR9396,L9418,CGP1247} or argue that it is almost impossible to gap out
(i.e. fully open the mass gaps of) the mirror sector without  breaking the
gauge symmetry in some mirror fermion models.\cite{BD9216}

We realized that the previous failed lattice-gauge approaches always assume
non-interacting lattice fermions (apart from the interaction to the lattice
gauge field).  In this work, we show that lattice 
approach actually works if we include direct fermion-fermion interaction with appropriate strength 
({\it i.e.} the dimensionaless coupling constants are of order
1).\cite{Wen:2013oza,Wen:2013ppa} In other words, a general framework of the
mirror fermion approach actually works for constructing a lattice chiral
fermion theory, at least in 1+1D.  Specifically, any anomaly-free chiral
fermion/boson field theory can be defined as a finite quantum system on a 1D
lattice where the (gauge or global) symmetry is realized as an onsite symmetry,
provided that we allow lattice fermion/boson to have {\it interactions}, {\it
instead of being free}.  (Here, the ``chiral'' theory here means that it
``breaks parity $P$ symmetry.'' Our 1+1D chiral fermion theory breaks parity
$P$ and time reversal $T$ symmetry. 
Our insight comes from
Ref.\,\cite{Wen:2013oza,Wen:2013ppa}, where the connection between gauge
anomalies and symmetry-protected topological (SPT) states (in
one-higher dimension) is found. 



To make our readers fully appreciate our thinking, we shall firstly define our important basic notions clearly:\\

\noindent
$(\diamond 1)$ \emph{Onsite symmetry} 
means that the overall symmetry transformation $U(g)$ of symmetry group $G$ can be defined as the tensor
product of each single site's symmetry transformation $U_i(g)$, via $U(g)= \otimes _i U_i(g)$ with $g \in G$.
\emph{Nonsite symmetry}: means $U(g)_{\text{non-onsite}} \neq \otimes _i U_i(g)$.\\ 
\noindent
$(\diamond 2)$ \emph{Local Hamiltonian with short-range interactions} 
means that the non-zero amplitude of matter(fermion/boson) hopping/interactions in finite time 
has a \emph{finite} range propagation, and cannot be an \emph{infinite} range. 
Strictly speaking, the quasi-local \emph{exponential decay} (of kinetic hopping/interactions) is \emph{non-local} and \emph{not short-ranged}.\\
\noindent
$(\diamond 3)$ \emph{finite(-Hilbert-space) system} means that the dimension of Hilbert space is finite 
if the system has finite lattice sites (e.g. on a cylinder).

Nielsen-Ninomiya theorem\cite{Nielsen:1980rz,Nielsen:1981xu,Nielsen:1981hk}
states that the attempt to regularize chiral fermion on a lattice
as a local \emph{free non-interacting} fermion model with fermion number conservation
({\it i.e.} with U(1) symmetry) 
has fermion-doubling problem\cite{Nielsen:1980rz,Nielsen:1981xu,Nielsen:1981hk} 
in an even dimensional spacetime.
To apply this no-go theorem, however, the symmetry 
is assumed to be an onsite symmetry. 

Ginsparg-Wilson
fermion approach 
copes with this
no-go theorem by solving
Ginsparg-Wilson(GW) relation\cite{Ginsparg:1981bj, {Wilson:1974sk}} based on the quasi-local Neuberger-Dirac 
operator,\cite{Neuberger:1997fp,Neuberger:1998wv,Hernandez:1998et} 
where \emph{quasi-local is strictly non-local}.
In this work, we show 
that the quasi-localness of Neuberger-Dirac 
operator in the GW fermion approach imposing a \emph{non-onsite}\cite{Chen:2012hc,Santos:2013uda} 
U(1) symmetry, instead of an onsite 
symmetry. 
(While here we simply summarize the result, one can read the details of onsite and non-onsite symmetry, and its relation to GW fermion in \cite{Wang:2013yta}.)
For our specific approach for the mirror-fermion decoupling, we
\emph{will not} implement the GW fermions (of non-onsite symmetry) construction,
instead, we will use a lattice fermions with onsite symmetry but with particular properly-designed interactions. 
Comparing GW fermion to our approach, we see that
\begin{itemize}
\item {\bf Ginsparg-Wilson(GW) fermion approach} obtains ``{\it chiral
fermions from a local free fermion lattice model with non-onsite $\U(1)$ symmetry} ({\it without fermion doublers}).''
(Here one regards Ginsparg-Wilson fermion applying the Neuberger-Dirac operator, which is strictly non-onsite and non-local.)

\item {\bf Our approach} obtains ``{\it chiral fermions from  local interacting
fermion lattice model with onsite $U(1)$ symmetry} ({\it without fermion doublers}),
{\it if all $\U(1)$ 
anomalies are canncelled}.''
\end{itemize}

Also, the conventional GW fermion approach discretizes the
Lagrangian/the action on the spacetime lattice, while  we use a local short-range quantum
Hamiltonian on 1D spatial lattice with a continuous time. Such a
distinction causes some difference.  For example, it is known that
Ginsparg-Wilson fermion \emph{can} implement a single Weyl fermion for the free case
without gauge field on a 1+1D space-time-lattice due to the works of Neuberger,
L\"uscher, etc.  
Our approach \emph{cannot} implement a single Weyl fermion on a 1D space-lattice within local short-range Hamiltonian. 
(However, such a distinction may not be important if we are allowed to introduce a non-local infinite-range hopping.)\\ 

\color{black}

\noindent
{{\bf Comparison to Eichten-Preskill and Chen-Giedt-Poppitz models}}:
Due to the past 
investigations, a majority of the high-energy lattice
community believes that the mirror-fermion decoupling (or lattice gauge
approach) fails to realize chiral fermion or chiral gauge theory.  
Thus one may challenge us by asking ``how our mirror-fermion
decoupling model 
is different from Eichten-Preskill
and Chen-Giedt-Poppitz models?''  And ``why the recent numerical attempt of
Chen-Giedt-Poppitz fails?\cite{CGP1247}''
We now stress that, 
our approach provides
\emph{properly designed fermion interaction terms} to make things work, due to the recent understanding to {\bf topological gapped boundary conditions}\cite{{Wang:2012am},{Levin:2013gaa}}: 
\begin{itemize}
\item {\bf Eichten-Preskill(EP)}\cite{EP8679} propose a generic idea of the
mirror-fermion approach for the chiral gauge theory.  There the
\emph{perturbative} analysis on the \emph{weak-coupling and strong-coupling}
expansions are used to demonstrate possible mirror-fermion decoupling phase can
exist in the phase diagram.  The action is discretized on the spacetime
lattice.  In EP 
approach, one tries to gap out the
mirror-fermions via the mass term of composite fermions that do not break the
(gauge) symmetry on lattice.  The  mass term of composite fermions are actually
fermion interacting terms.  So in EP 
approach, one tries to gap
out the mirror-fermions via the direct fermion interaction that do not  break
the (gauge) symmetry on lattice.  However, considering only the symmetry of the
interaction is not enough.  Even when the mirror sector is anomalous, one can
still add the direct fermion interaction that do not  break the (gauge)
symmetry.  So the presence of symmetric direct fermion interaction may or may
not be able to gap out the  mirror sector.  When the mirror sector is
anomaly-free, we will show in this paper, some symmetric interactions are
\emph{helpful} for gapping out the mirror sectors, while other symmetric interactions
are \emph{harmful}. The key issue is to design the proper interaction to gap out
the mirror section, and considering only symmetry is not enough.

\item {\bf Chen-Giedt-Poppitz(CGP)}\cite{CGP1247} follows the EP
general framework to deal with a 3-4-5 anomaly-free model with a single U(1)
symmetry.  All the U(1) symmetry-allowed Yukawa-Higgs terms are introduced to
mediate multi-fermion interactions.  The Ginsparg-Wilson fermion and the
Neuberger's overlap Dirac operator are implemented, the fermion actions are
discretized on the spacetime lattice.  Again, the interaction terms are
designed only based on symmetry, which contain both helpful and harmful
terms, as we will show.

\item {\bf Our model} in general 
belongs to the
mirror-fermion-decoupling idea.  The anomaly-free model we proposed is named as
the 3$_L$-5$_R$-4$_L$-0$_R$ model.  Our 3$_L$-5$_R$-4$_L$-0$_R$ is in-reality
different from Chen-Giedt-Poppitz's 3-4-5 model, since we impliment:\\
\underline{(i) {\bf an onsite-symmetry local lattice model}}: Our lattice Hamiltonian
is built on 1D spatial lattice with \emph{on-site} U(1) symmetry. 
We \emph{neither} implement the GW 
fermion \emph{nor} the
Neuberger-Dirac operator (both have non-onsite symmetry).
\\
\underline{(ii) {\bf a particular set of interaction terms}}  \underline{ \bf{with proper strength}}:\\
Our multi fermion
interaction terms are particularly-designed gapping terms
which obey not only the symmetry but also certain Lagrangian subgroup algebra. 
Those interaction terms are called \emph{helpful} gapping terms, satisfying {\bf Boundary Fully Gapping Rules}.
We will show that the Chen-Giedt-Poppitz's Yukawa-Higgs terms induce extra 
multi-fermion interaction terms which \emph{do not} satsify
{\bf Boundary Fully Gapping Rules}. 
Those extra terms are
incompatible \emph{harmful} terms,
competing with the \emph{helpful} gapping terms and 
causing the preformed mass gap unstable so preventing the mirror sector from being gapped out.
(This can be one of the reasons for the failure of mirror-decoupling in Ref.\cite{CGP1247}.)
We stress that, due to a \emph{topological non-perturbative} reason, 
only a particular set of ideal interaction terms are helpful
to fully gap the mirror sector. Adding more or removing interactions can cause
the mass gap unstable thus the phase flowing to gapless states.
In addition, we stress that only when the helpful interaction terms are in a proper range, \emph{intermediate strength} for dimensionless coupling of order 1,
can they fully gap the mirror sector, and yet not gap the original sector (details in Sec.\ref{sec:mapping-strong-gapping}).
Throughout our work, when we say strong coupling for our model, we really mean intermediate(-strong) coupling in an appropriate range.
In CGP model, however, their strong coupling may be \emph{too strong} (with their kinetic term neglected);
which can be another reason for the failure of mirror-decoupling.\cite{CGP1247}

\underline{(iii) {\bf extra symmetries}}: For our model, a total even number $N$
of left/right moving Weyl fermions ($N_L=N_R=N/2$), we will add only $N/2$
helpful gapping terms under the constraint of the Lagrangian subgroup algebra and {\bf
Boundary Fully Gapping Rules}.  As a result, the full symmetry of our lattice model is
U(1)$^{N/2}$ (where the gapping terms break U(1)$^{N}$ down to U(1)$^{N/2}$).
For the case of our 3$_L$-5$_R$-4$_L$-0$_R$ model, the full U(1)$^2$ symmetry
has two sets of U(1) charges, $\U(1)_{\text{1st}}$ 3-5-4-0 and $\U(1)_{\text{2nd}}$ 0-4-5-3, both are anomaly-free and 
mixed-anomaly-free.  Although the physical consideration
only requires the interaction terms to have
on-site $\U(1)_{\text{1st}}$ symmetry, 
looking for interaction terms with extra U(1) symmetry can
help us to identify the helpful gapping terms and design the proper lattice interactions.
CGP model has only
a single $\U(1)_{\text{1st}}$ symmetry. Here we suggest to improve that model by removing all
the interaction terms that break the $\U(1)_{\text{2nd}}$ symmetry (thus adding all
possible terms that preserve the two U(1) symmetries) with an
intermediate strength.  
\end{itemize}

The plan of our attack is the following. 
In Sec.\ref{sec3-5-4-0} we first consider a 3$_L$-5$_R$-4$_L$-0$_R$
anomaly-free chiral fermion field theory model, with a full $\U(1)^2$ symmetry:
A first 3-5-4-0 $\U(1)_{\text{1st}}$ symmetry
for two left-moving fermions of charge-3 and charge-4, and for
two right-moving fermions of charge-5 and charge-0.
And a second  0-4-5-3 $\U(1)_{\text{2nd}}$ symmetry
for two left-moving fermions of charge-0 and charge-5, and for
two right-moving fermions of charge-4 and charge-3.
If we wish to have a
\emph{single} $\U(1)_{\text{1st}}$ symmetry, we can weakly break the $\U(1)_{\text{2nd}}$ symmetry by adding tiny local $\U(1)_{\text{2nd}}$-symmetry breaking term.

We claim that this model can be put on the lattice with an
onsite $\U(1)$ symmetry, but without fermion-doubling problem.  
We construct a 2+1D lattice model by simply using four layers of the zeroth Landau
levels(or more precisely, four filled bands with Chern numbers \cite{Thouless:1982zz} 
$-1,+1,-1,+1$ on
a lattice) 
right-moving, charge-4 left-moving, charge-0 right-moving, totally four
fermionic modes at low energy on one edge.  Therefore, by putting the 2D bulk
spatial lattice on a cylinder with two edges, one can leave edge states on
one edge untouched so they remain chiral and gapless, while turning on
 interactions to gap out the mirrored edge states on the other edge
with a large mass gap.  

\noindent
[NOTE on usages: 
Here in our work, U(1) symmetry may generically imply 
copies of U(1) symmetry such as U(1)$^M$, with positive integer $M$.  
 (Topological) {\bf Boundary Fully Gapping Rules} are defined as the rules to open the mass gaps of the boundary states.
(Topological) {\bf Gapped Boundary Conditions} are defined to specify 
certain boundary types 
which are gapped (thus topological).
There are two kinds of usages of \emph{lattices} here discussed in our work:
one is the {{\bf Hamiltonian lattice}} model to simulate the chiral fermions/bosons.
The other \emph{lattice} is the {\bf Chern-Simons lattice} structure of 
Hilbert space, which is a quantized lattice due 
to the level/charge quantization of Chern-Simons theory.
]

\noindent
\subsection{3$_L$-5$_R$-4$_L$-0$_R$ Chiral Fermion model  \label{sec3-5-4-0}}

The simplest chiral (Weyl) fermion field theory with U(1) symmetry in $1+1$D is given by the action
\be
\label{WeylS}
S_{\Psi,free}=\int  dtdx \; \ti \psi^\dagger_{L} (\partial_t-\partial_x) \psi_{L}.
\ee
However, Nielsen-Ninomiya theorem claims that
such a theory cannot be put on a lattice with unbroken onsite U(1) symmetry,
due to 
the fermion-doubling problem.\cite{Nielsen:1980rz,Nielsen:1981xu,Nielsen:1981hk} 
While the Ginsparg-Wilson fermion approach can still implement an anomalous single Weyl fermion 
on the lattice,
our approach cannot (unless we modify local Hamiltonian to
infinite-range hopping non-local Hamiltonian). As we will show, our approach is more restricted, only limited to the anomaly-free theory. 
Let us instead consider an anomaly-free
3$_L$-5$_R$-4$_L$-0$_R$ chiral fermion field theory with an action $S_{\Psi_{\A},free}$,
\be
\label{cf}
\int  dtdx \; \Big(
 \ti \psi^\dagger_{L,3} (\partial_t-\partial_x) \psi_{L,3}
+ \ti\psi^\dagger_{R,5} (\partial_t+\partial_x) \psi_{R,5}
+ \ti\psi^\dagger_{L,4} (\partial_t-\partial_x) \psi_{L,4}
+\ti\psi^\dagger_{R,0} (\partial_t+\partial_x) \psi_{R,0}\Big),
\ee
where
$\psi_{L,3}$,
$\psi_{R,5}$,
$\psi_{L,4}$, and
$\psi_{R,0}$ are 1-component Weyl spinor, carrying U(1) charges 3,5,4,0 respectively. The subscript $L$(or $R$)
indicates left(or right) moving along $-\hat{x}$(or $+\hat{x}$).  Although this
theory has equal numbers of left and right moving modes, 
it violates parity 
and time reversal symmetry, 
so it is a chiral theory. 
Such a  chiral fermion field theory
is very special because it is free from U(1) anomaly - it satisfies the anomaly matching
condition\cite{'tHooft:1979bh}  
in $1+1$D, 
which means $\sum_j q_{L,j}^2-q_{R,j}^2=3^2-5^2+4^2-0^2=0$.
We ask:\\
%
\colorbox{lgray}{\parbox[t]{\linegoal}{
{\bf Question 1}:
``Whether there is a \emph{local} \emph{finite} 
Hamiltonian realizing the above U(1) 3-5-4-0 symmetry as an onsite symmetry with \emph{short-range interactions} 
defined on a 1D spatial lattice with a continuous time, such that its low energy physics produces the anomaly-free chiral fermion theory Eq.(\ref{cf})?''
}}\\

\begin{figure}[!h]
\centering
{\includegraphics[width=.48\textwidth]{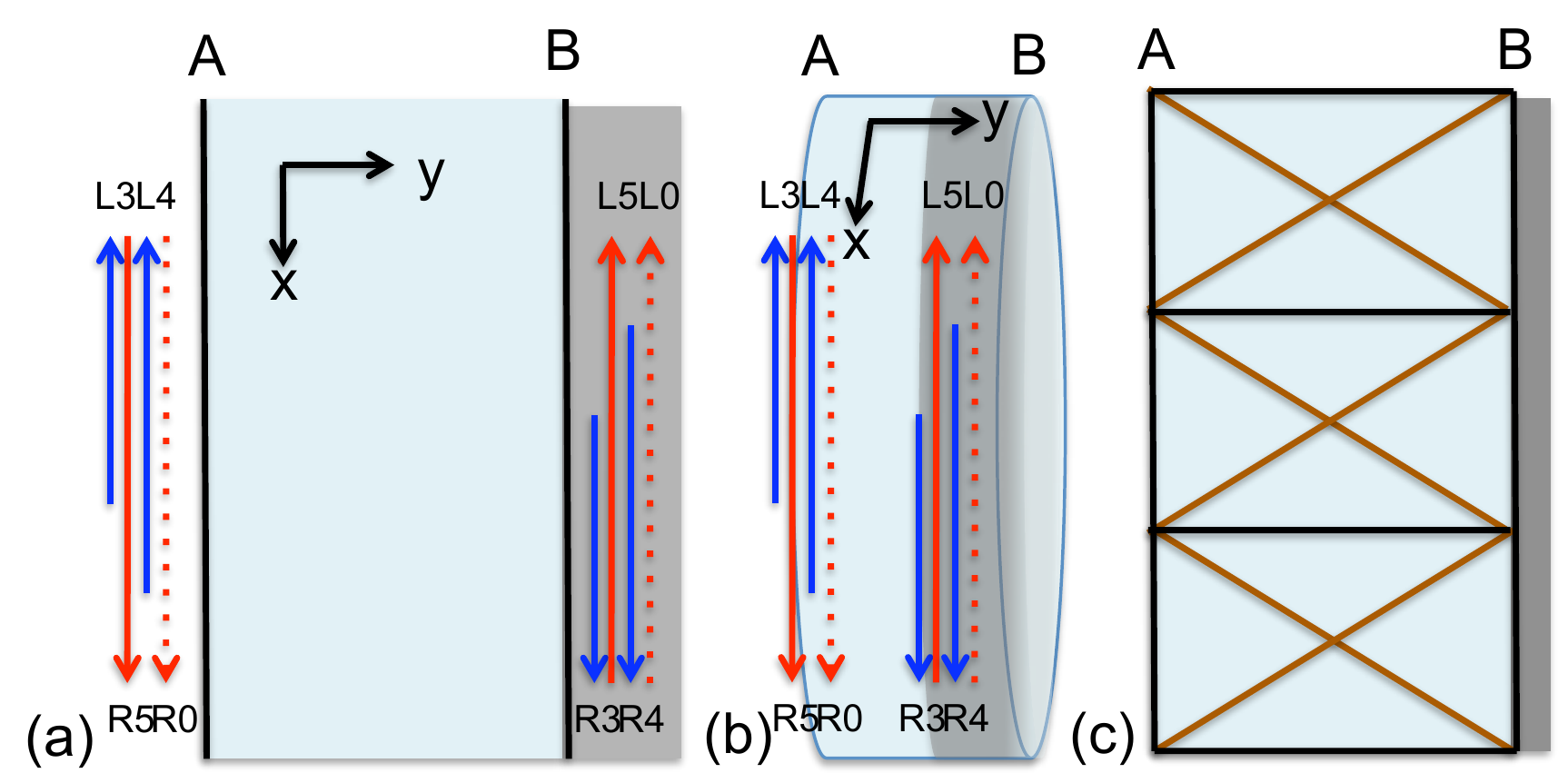}}
\caption{3-5-4-0 chiral fermion model: (a) The fermions carry U(1) charge $3$,$5$,$4$,$0$ for $\psi_{L,3},$$\psi_{R,5},$$\psi_{L,4},$$\psi_{R,0}$ on the edge A, and also for its mirrored partners $\tilde{\psi}_{R,3},$$\tilde{\psi}_{L,5},$$\tilde{\psi}_{R,4},$$\tilde{\psi}_{L,0}$
on the edge B. We focus on the model with a periodic boundary condition along $x$, and a finite-size length along $y$,
effectively as, (b) on a cylinder.
(c) The ladder model on a cylinder 
with the $t$ hopping term along black links, the $t'$ hopping term along brown links. The shadow on the edge B indicates the gapping terms 
with $G_1,G_2$ couplings in Eq.(\ref{H3540}) are imposed.}
\label{3540}
\end{figure}

Yes. We would like to show that the above chiral fermion field theory
can be put
 on a lattice with unbroken onsite U(1) symmetry, 
if we include properly-desgined interactions between fermions.
In fact, we propose that the chiral fermion field theory in Eq.(\ref{cf})
appears as the low energy effective theory of the following 2+1D lattice model
on a cylinder (see Fig.\ref{3540}) with a properly designed Hamiltonian.
To derive such a Hamiltonian, we start from thinking the full two-edges fermion theory with the action $S_\Psi$,
where the particularly chosen multi-fermion interactions $S_{\Psi_{\B},interact}$ will be explained:


\bea 
&S_\Psi&=S_{\Psi_{\A},free}+S_{\Psi_{\B},free}+S_{\Psi_{\B},interact}=\int  dt \; dx \; \bigg( \ti\bar{\Psi}_{\A} \Gamma^\mu  \partial_\mu \Psi_{\A}+ \ti\bar{\Psi}_{\B} \Gamma^\mu  \partial_\mu \Psi_{\B}     \;\;\nonumber\;\;\\
 &\;&+\tilde{g}_{1} \big( (\tilde{\psi}_{R,3} )  (\tilde{\psi}_{L,5} )
 ( \tilde{\psi}^\dagger_{R,4} \nabla_x \tilde{\psi}^\dagger_{R,4}  ) ( \tilde{\psi}_{R,0} \nabla_x \tilde{\psi}_{R,0}  ) +\text{h.c.} \big)  \nonumber\\
&\; &+ \tilde{g}_{2}   \big( (\tilde{\psi}_{L,3} \nabla_x \tilde{\psi}_{L,3} )  (\tilde{\psi}_{R,5}^\dagger  \nabla_x \tilde{\psi}_{R,5}^\dagger  )( \tilde{\psi}_{L,4}   )( \tilde{\psi}_{L,0}   )+\text{h.c.} \big)\bigg),  \label{Lf3-5-4-0}
\eea

The notation for fermion fields on the edge A are $\Psi_{\A}=(\psi_{L,3},\psi_{R,5},\psi_{L,4},\psi_{R,0})$ ,
and fermion fields on the edge B are $\Psi_{\B}=(\tilde{\psi}_{L,5},\tilde{\psi}_{R,3},\tilde{\psi}_{L,0},\tilde{\psi}_{R,4})$.
(Here a left moving mode in $\Psi_{\A}$ corresponds to a right moving mode in $\Psi_{\B}$ because of Landau level/Chern band chirality, the details of lattice model will be explained.)
The gamma matrices in 1+1D are presented in terms of Pauli matrices, with
$\gamma^0=\sigma_x$, $\gamma^1=\ti\sigma_y$, $\gamma^5\equiv\gamma^0\gamma^1=-\sigma_z $,
and $\Gamma^0=\gamma^0\oplus\gamma^0$, $\Gamma^1=\gamma^1\oplus \gamma^1$, $\Gamma^5\equiv\Gamma^0\Gamma^1$ and $\bar{\Psi}_i \equiv \Psi_i \Gamma^0$.

In 1+1D, we can do bosonization, 
where the fermion matter field $\Psi$ turns into bosonic phase field $\Phi$,
more explicitly $\psi_{L,3}\sim e^{\ti \Phi^{\A}_3} $, $\psi_{R,5}\sim e^{\ti \Phi^{\A}_5} $, $\psi_{L,4}\sim e^{\ti \Phi^{\A}_4} $, $\psi_{R,0}\sim e^{\ti \Phi^{\A}_0} $ on A edge,
$\tilde{\psi}_{R,3}\sim  e^{\ti \Phi^{\B}_3} $, $\tilde{\psi}_{L,5} \sim e^{\ti \Phi^{\B}_5} $, $\tilde{\psi}_{R,4} \sim  e^{\ti \Phi^{\B}_4} $, $\tilde{\psi}_{L,0}  \sim  e^{\ti \Phi^{\B}_0}$ on B edge,
up to normal orderings $ :e^{\ti \Phi}:$ and prefactors,\cite{fermionization2} 
but the 
precise factor is not of our interest since our goal is to obtain its non-perturbative lattice realization.
So Eq.(\ref{Lf3-5-4-0}) becomes
\bea 
&&S_{\Phi}=S_{\Phi^{\A}_{free}}+S_{\Phi^{\B}_{free}}+S_{\Phi^{\B}_{interact}}= \nonumber\\
&& 
\frac{1}{4\pi}  \int dt dx  \big(K^{\A}_{IJ}  \partial_t \Phi^{\A}_I   \partial_x \Phi^{\A}_{J} -V_{IJ}  \partial_x \Phi^{\A}_I   \partial_x \Phi^{\A}_{J}\big)+\big(K^{\B}_{IJ}  \partial_t \Phi^{\B}_I   \partial_x \Phi^{\B}_{J} -V_{IJ}  \partial_x \Phi^{\B}_I   \partial_x \Phi^{\B}_{J} \big) \label{b3540}        \;\;\;\nonumber\\
&& +\int dt dx  \bigg( g_{1}  \cos( \Phi^{\B}_{3}+\Phi^{\B}_{5}-2 \Phi^{\B}_{4}+2\Phi^{\B}_{0})+  g_{2}  \cos( 2\Phi^{\B}_{5}-2\Phi^{\B}_{5}+\Phi^{\B}_{4}+\Phi^{\B}_{0}) \bigg). \;\;\;\;\;\;\;
 \eea
 
Here $I,J$ runs over $3,5,4,0$ and $K^{\A}_{IJ}=-K^{\B}_{IJ}=\diag(1,-1,1,-1)$ 
$V_{IJ}=\diag(1,1,1,1)$ are diagonal matrices.
{\bf All we have to prove is that gapping terms, the cosine terms with ${g}_1,{g}_2$ coupling can gap out all states on the edge B.}
First, let us understand more about the full U(1) symmetry.  What are the U(1) symmetries?
They are transformations of 
$$ \text{fermions } \psi \to \psi \cdot e^{\ti q \theta},\;\;\; \text{bosons }\;\;\; \Phi \to \Phi + q \;\theta$$ 
making the full action invariant.
The original \emph{four} Weyl fermions have a full U(1)$^4$ symmetry. Under \emph{two} linear-indepndent interaction terms in $S_{\Psi_{\B},interact}$ (or $S_{\Phi^{\B}_{interact}}$),
U(1)$^4$ is broken down to U(1)$^2$ symmetry.
If we denote these $q$ as a charge vector $\mathbf{t}=(q_3,q_5,q_4,q_0)$,
we find there are such two charge vectors $\mathbf{t}_1=(3,5,4,0)$  and $\mathbf{t}_2=(0,4,5,3)$ for U(1)$_{\text{1st}}$, U(1)$_{\text{2nd}}$ symmetry respectively.

We emphasize 
that finding those gapping terms in this  U(1)$^2$ anomaly-free theory is not accidental. The {\bf anomaly matching condition}\cite{'tHooft:1979bh} %
 here is 
satisfied, for the anomalies $\sum_j q_{L,j}^2-q_{R,j}^2=3^2-5^2+4^2-0^2=0^2-4^2+5^2-3^2=0$, 
and the mixed anomaly: $3 \cdot 0 -5 \cdot 4 + 4 \cdot 5 - 0\cdot 3=0$ which can be formulated as 
\be \label{tKt}
{\boxed{ \mathbf{t}^T_i \cdot (K^{\A}) \cdot \mathbf{t}_j=0}}\;, \;\;\; i,j \in \{1,2\} 
\ee 
with the U(1) charge vector $\mathbf{t}=(3,5,4,0)$, with its transpose $\mathbf{t}^T$.

\;\;\; On the other hand, the {\bf boundary fully gapping rules} 
require two gapping terms, here $g_{1} \cos( \ell_1 \cdot \Phi)+ g_{2}\cos( \ell_2 \cdot \Phi)$,
such that self and mutual statistical angles $\theta_{ij}$ 
defined below among the Wilson-line operators $\ell_i,\ell_j$ are zeros,
\be \label{LKL} 
{\boxed{ \theta_{ij}/(2\pi) \equiv \ell_i^T \cdot (K^{\B})^{-1}  \cdot \ell_j = 0} }\;, \;\;\; i,j \in \{1,2\}    
\ee
Indeed, here we have {$\ell_1=(1,1,-2,2),\ell_2=(2,-2,1,1)$}
satisfying the rules. Thus we prove that the mirrored edge states on the edge B can be fully gapped out.

We will prove the {\bf anomaly matching condition} is equivalent to find a set of gapping terms $g_{a}  \cos(\ell_{a}^{} \cdot\Phi_{}) $, satisfies the {\bf boundary fully gapping rules},
detailed in Sec.\ref{anomaly-hall}, \ref{anomaly-gap}. 
Simply speaking,
the {\bf anomaly matching condition} (Eq.(\ref{tKt})) in 1+1D 
is \emph{equivalent} to (an if and only if relation) 
the {\bf boundary fully gapping rules} (Eq.(\ref{LKL})) in 1+1D boundary/2+1D bulk for an equal number of left-right moving modes($N_L=N_R$, with central charge $c_L=c_R$).
We prove this is true at least for U(1) symmetry case, with the bulk theory is a 2+1D SPT state and the boundary theory is in 1+1D.

{\bf We now propose a lattice Hamiltonian model for this 3$_L$-5$_R$-4$_L$-0$_R$ chiral fermion realizing Eq.(\ref{Lf3-5-4-0})} (thus Eq.(\ref{cf}) at the low energy once the Edge B is gapped out).
Importantly, we \emph{do not} discretize the action Eq.(\ref{Lf3-5-4-0}) on the spacetime lattice.
We \emph{do not} use Ginsparg-Wilson(GW) fermion \emph{nor} the Neuberger-Dirac operator.
GW and Neuberger-Dirac scheme contains \emph{non-onsite symmetry} 
which cause the lattice \emph{difficult to be gauged} to chiral gauge theory.
Instead, the key step is that we 
implement the \emph{on-site symmetry} lattice fermion model. 
The \emph{free kinetic part} is a fermion-hopping model which has a \emph{finite 2D bulk energy gap} 
but with \emph{gapless 1D edge states}. This can be done by using any {\bf lattice Chern insulator}.

We stress that  {\bf any} lattice Chern insulator with onsite-symmetry shall work, and  
we design one as in Fig.\ref{3540}.
Our full Hamiltonian with two interacting $G_1,G_2$ gapping terms is
\bea 
\label{H3540}
H&=&\sum_{q=3,5,4,0} 
\bigg(  \sum_{\langle i, j \rangle}
\big(t_{ij,q}\; \hat{f}^\dagger_{q}(i)
\hat{f}_{q}(j)+h.c.\big) + \sum_{\langle\langle i, j
\rangle\rangle} \big(  t'_{ij,q}
\;\hat{f}^\dagger_{q}(i) \hat{f}_{q}(j)+h.c.\big) \bigg)
\\ 
&+& G_{1} \sum_{j \in \B} \bigg(
\big(\hat{f}_3(j)\big)^1
\big(\hat{f}_5(j)\big)^1
\big(\hat{f}^\dagger_4(j)_{pt.s.}\big)^{2}
\big(\hat{f}_0(j)_{pt.s.}\big)^2
+h.c. \bigg) \nonumber\\
&&+ G_{2} \sum_{j \in \B} \bigg( 
\big(\hat{f}_3(j)_{pt.s.}\big)^2
\big(\hat{f}^\dagger_5(j)_{pt.s.}\big)^2
\big(\hat{f}_4(j)\big)^1
\big(\hat{f}_0(j)\big)^1 
+h.c.
\bigg)
\nonumber 
\eea
\noindent
where $\sum_{j \in \B}$ sums over the lattice points on the right boundary (the edge B in Fig.\ref{3540}),
and the fermion operators $\hat f_{3}$, $\hat f_{5}$, $\hat f_{4}$, $\hat
f_{0}$ carry a U(1)$_{\text{1st}}$ charge 3,5,4,0 
and another U(1)$_{\text{2nd}}$ charge 0,4,5,3 respectively.   
We emphasize that this lattice model has \emph{onsite} U(1)$^2$ symmetry, since this Hamiltonian, including interaction terms, is invariant 
under a global U(1)$_{\text{1st}}$ transformation \emph{on each site} for any $\theta$ angle:
$\hat f_{3} \to \hat f_{3} e^{\ti 3\theta}$, $\hat f_{5} \to \hat f_{5} e^{\ti 5\theta}$, $\hat f_{4} \to \hat f_{4} e^{\ti 4\theta}$, $\hat f_{0} \to \hat f_{0}$,
and invariant under another global U(1)$_{\text{2nd}}$ transformation for any $\theta$ angle:
$\hat f_{3} \to \hat f_{3} $, $\hat f_{5} \to \hat f_{5} e^{\ti 4\theta}$, $\hat f_{4} \to \hat f_{4} e^{\ti 5\theta}$, $\hat f_{0} \to \hat f_{0} e^{\ti 3\theta}$.
The U(1)$_{\text{1st}}$ charge is the reason why it is named as 3$_L$-5$_R$-4$_L$-0$_R$ 
model.

As for notations, 
$\langle i, j \rangle$ stands for nearest-neighbor hopping along black links
and $\langle\langle i, j \rangle\rangle$ stands for next-nearest-neighbor hopping along brown links in Fig.\ref{3540}.
Here $pt.s.$  
stands for point-splitting. 
For example, $(\hat
f_3(j)_{pt.s.})^2\equiv   \hat f_{3}(j) \hat f_{3}(j+\hat
x)$.
We stress that the full Hamiltonian (including interactions) Eq.(\ref{H3540}) is \emph{short-ranged and local}, 
because each term only involves coupling within finite number of neighbor sites.
The hopping amplitudes $t_{ij,3}=t_{ij,4}$ and $t'_{ij,3}=t'_{ij,4}$
produce bands with Chern number $-1$, while the hopping amplitudes
$t_{ij,5}=t_{ij,0}$ and $t'_{ij,5}=t'_{ij,0}$ produce bands with Chern number
$+1$ (see {{Sec.\ref{numeric}}}).\cite{Thouless:1982zz} 
The ground state is obtained by filling the
above four bands.

As Eq.(\ref{H3540}) contains U(1)$_{\text{1st}}$ and an accidental extra U(1)$_{\text{2nd}}$ symmetry, we shall ask:
\frm{ {\bf Question 2}:
``Whether there is a \emph{local}  \emph{finite}  Hamiltonian realizing \emph{only} a U(1) 3-5-4-0
symmetry as an onsite symmetry with \emph{short-range interactions}
defined on a 1D spatial lattice with a continuous time, such that its low energy physics produces the anomaly-free chiral fermion theory Eq.(\ref{cf})?''}

Yes, by adding a small local perturbation to break U(1)$_{\text{2nd}}$ 0-4-5-3 symmetry, we can achieve a faithful
U(1)$_{\text{1st}}$ 3-5-4-0 symmetry chiral fermion theory of Eq.(\ref{cf}).
For example, we can adjust  Eq.(\ref{H3540})'s $H \to H +\delta H $ by adding: 
\bea
&&\delta H = G_{tiny}' \sum_{j \in \B}  
\Big( \big(\hat{f}_3(j)_{pt.s.}\big)^3
\big(\hat{f}^\dagger_5(j)_{pt.s.}\big)^1
\big(\hat{f}^\dagger_4(j)\big)^1 + h.c. \Big) \nonumber\\
&& \Leftrightarrow 
\tilde{g}_{tiny}'   \big( (\tilde{\psi}_{L,3} \nabla_x \tilde{\psi}_{L,3} \nabla_x^2 \tilde{\psi}_{L,3} )  (\tilde{\psi}_{R,5}^\dagger )( \tilde{\psi}_{L,4}^\dagger )+\text{h.c.} \big) \nonumber \\
&& \Leftrightarrow g_{tiny}'  \cos( 3\Phi^{\B}_{3}-\Phi^{\B}_{5}- \Phi^{\B}_{4}) \equiv g_{tiny}'  \cos( \ell_{}' \cdot \Phi^{\B}).
\eea
Here we have $\ell_{}'=(3,-1,-1,0)$. The $g_{tiny}'  \cos( \ell_{}' \cdot \Phi^{\B})$ is not designed to
be a gapping term (its self and mutual statistics 
 happen to be nontrivial:
  ${\ell_{}'^T\cdot (K^{\B})^{-1}  \cdot \ell_{}' \neq 0}$, 
 ${\ell_{}'^T \cdot (K^{\B})^{-1}  \cdot \ell_2 \neq  0}$), but this tiny perturbation term  
 is meant to preserve U(1)$_{\text{1st}}$ 3-5-4-0 symmetry only,
 thus ${\ell_{}'^T \cdot \mathbf{t}_1 } ={\ell_{}'^T \cdot (K^{\B})^{-1}  \cdot \ell_1 =  0}$.
We must set
$(|G_{{tiny}'}|/|G|) \ll 1$ with $|G_1| \sim |G_2| \sim |G|$ about the same magnitude, 
so that
 the tiny local perturbation will not destroy the mass gap.

Without the interaction, i.e. $G_1=G_2=0$, the edge excitations of the above four
bands produce the chiral fermion theory Eq.(\ref{cf}) on the left edge A
and the mirror partners 
on the right edge B. 
So the total low energy effective theory is non-chiral.
In Sec.\ref{numeric}, we will provide an explicit lattice model for this free fermion theory.

However, by turning on the intermediate-strength 
interaction $G_1,G_2\neq 0$, we claim the interaction terms can fully gap out the edge excitations on the right mirrored edge B as in
Fig.\ref{3540}. To find those gapping terms is not accidental - it is guaranteed by our proof 
of equivalence between
{\bf the anomaly matching condition}\cite{'tHooft:1979bh}  
(as ${ \mathbf{t}^T_i \cdot (K)^{-1} \cdot \mathbf{t}_j=0}$ of Eq.(\ref{tKt}) 
) and {\bf the boundary fully gapping rules}\cite{Wang:2012am,Levin:2013gaa} 
(here $G_1,G_2$ terms can gap out the edge) in $1+1$ D.
The low energy effective theory of the interacting lattice
model with only gapless states on the edge A 
is the chiral fermion theory in Eq.(\ref{cf}).  Since the width of the
cylinder is finite, the lattice model Eq.(\ref{H3540}) is actually a 1+1D
lattice model, which gives a non-perturbative lattice definition of the chiral fermion
theory Eq.(\ref{cf}).  Indeed, the Hamiltonian and the lattice need not to be restricted merely to Eq.(\ref{H3540}) and Fig.\ref{3540},
we stress that any on-site symmetry lattice model produces four bands with the desired Chern numbers would work.
We emphasize again that the U(1) symmetry is realized as
an onsite symmetry 
in our lattice model. It is easy to gauge such an onsite
U(1) symmetry 
to obtain a chiral fermion theory coupled to a U(1) gauge field. 


\subsection{From a continuum field theory to a discrete lattice model} 
\label{sec:mapping-field-lattice}

We now comment about the mapping from a continuum field theory of the action Eq.(\ref{cf}) 
to a discretized space Hamiltonian Eq.(\ref{H3540}) with a continuous time.
We \emph{do not} pursue \emph{Ginsparg-Wilson scheme}, and our gapless edge states are \emph{not} derived from the discretization of spacetime action. 
Instead, we will show that the Chern insulator Hamiltonian in Eq.(\ref{H3540}) as we described can provide essential gapless edge states for a free theory (without interactions $G_1,G_2$). 


{\bf Energy and Length Scales}:
We consider a finite 1+1D quantum system with a periodic length scale $L$ for the compact circle of the cylinder in Fig.\ref{3540}.
The finite size width of the cylinder is $w$. The lattice constant is $a$. The mass gap we wish to generate on the mirrored edge is $\Delta_m$, 
which causes a two-point correlator has an exponential decay: 
\be
\langle \psi^\dagger(r) \psi(0) \rangle \sim \langle e^{-\ti \Phi(r)} e^{\ti \Phi(0)}\rangle  \sim \exp(-|r|/\xi)
\ee
with a correlation length scale $\xi$.
The expected length scales follow that
\be \label{eq:lengthscale}
a < \xi \ll  w \ll  L.
\ee
The 1D system size $L$ is larger than the width $w$, the width $w$  is larger than the correlation length $\xi$,
the correlation length $\xi$ is larger than the lattice constant $a$.

\subsubsection{Free kinetic part and the edge states of a Chern insulator}

\subsubsection{Kinetic part mapping and RG analysis} \label{subsubsec:kinetic}

The {\bf kinetic part} of the lattice Hamiltonian
contains the nearest neighbor hopping term  
$\sum_{\langle i, j \rangle}$ 
$\big(t_{ij,q}$  $\hat{f}^\dagger_{q}(i)
\hat{f}_{q}(j)+h.c.\big)
$ together with the
next-nearest neighbor hopping term 
$\sum_{\langle\langle i, j
\rangle\rangle} \big(  t'_{ij,q}$
$hat{f}^\dagger_{q}(i) \hat{f}_{q}(j)$ $+h.c.\big)$, 
which generate the leading order field theory kinetic term 
via
\be \label{subsec:free_map}
t_{ij} \hat{f}^\dagger_{q}(i) \hat{f}_{q}(j) \sim a\; \ti \psi_q^\dagger \partial_x \psi_q + \dots ,
\ee
here hopping constants $t_{ij},t_{ij}'$ with a dimension of energy  $[t_{ij}] = [t_{ij}'] = 1$, 
and $a$ is the lattice spacing with a value $[a]=-1$. Thus,
$[\hat{f}_{q}(j)]=0$ and $[\psi_q] =\frac{1}{2}$.
The map from 
\be \label{eq:subf}
f_q  \to \sqrt{a}\, \psi_q + \dots
\ee
contains subleading terms. 
Subleading terms $\dots$ 
potentially contain higher derivative $\nabla^n_x \psi_q$ 
are only {\bf subleading perturbative effects}
\be 
f_q \to \sqrt{a}\,( \psi_q + \dots + a^n \, \alpha_{\text{small}} \nabla^n_x \psi_q 
+ \dots )\nonumber
\ee
with small coefficients of the polynomial of the small lattice spacing $a$ via $\alpha_{\text{small}}=\alpha_{\text{small}}(a) \lesssim  (a/L)$. 
We comment that only the leading term in the mapping is important,
the full account for the exact mapping from the fermion operator $f_q$ to $\psi_q$ is immaterial
to our model, because of two main reasons:\\

\noindent
$\bullet$(i) Our lattice construction is based on 
several layers of Chern insulators, and the chirality of each layer's edge states are protected by a topological number - the first Chern number $C_1 \in \mathbb{Z}$. 
Such an integer Chern number cannot be deformed by small perturbation,
thus it is {\bf non-perturbative topologically robust}, hence the chirality of edge states will be protected and will not be eliminated by small perturbations.
The origin of our \emph{fermion chirality} (breaking parity and time reversal) is an emergent phenomena
due to the \emph{complex hopping} amplitude of some hopping constant $t_{ij}'$ or $t_{ij}  \in \mathbb{C}$.
Beside, it is well-known that Chern insulator can produce the gapless fermion energy spectrum 
at low energy. 
More details and the energy spectrum are explicitly presented in Sec.\ref{numeric}.
\\

\noindent
$\bullet$(ii) The 
properly-designed interaction 
effect (from boundary fully gapping rules) is a {\bf non-perturbative topological effect} (as we will show in Sec.\ref{anomaly-gap}. 
In addition, we can also do 
the {\bf weak coupling} and the {\bf strong coupling RG}(renormalization group) analysis to show such subleading-perturbation is \emph{irrelevant}.\\

For weak-coupling RG analysis, we can start from the free theory fixed point, and 
evaluate $\alpha_{\text{small}} \psi_q \dots \nabla^n_x \psi_q$ term, which has a higher energy dimension than $\psi_q^\dagger \partial_x \psi_q$, thus irrelevant 
at the infrared low energy, and irrelevant to the ground state of our Hamiltonian.\\

For strong-coupling RG analysis at large $g_1,g_2$ coupling(shown to be the massive phase with mass gap in Sec.\ref{anomaly-gap},
it is convenient to 
use the {\bf bosonized language} to map
the fermion interaction
$U_{\text{interaction}}\big( \tilde{\psi}_{q}, \dots,  \nabla^n_x \tilde{\psi}_{q},\dots  \big) $ of $S_{\Psi_{\B},interact}$
to boson cosine term $g_{a}  \cos(\ell_{a,I}^{} \cdot\Phi_{I})$ of $S_{\Phi^{\B}_{interact}}$.
At the large $g$ coupling fixe point, 
the boson field is pinned down at the minimum of cosine potential,
we thus will consider the dominant term as the discretized spatial lattice (a site index $j$) and only a continuous time:
$
\int dt \, \big( \sum_j  \frac{1}{2}\, g\,  (\ell_{a,I}^{} \cdot\Phi_{I,j})^2 +\dots \big)
$.
Setting this dominant term to be a marginal operator means the scaling dimension of $\Phi_{I,j}$ is 
$
[\Phi_{I,j}] =1/2$ at strong coupling fixed point. 
Since the kinetic term is generated by an operator:
$$
e^{\ti P_\Phi a} \sim e^{\ti a \partial_x \Phi} \sim  e^{\ti (\Phi_{j+1} -  \Phi_j)}
$$
where
$e^{\ti P_\Phi a}$ generates the lattice translation by
$e^{\ti P_\Phi a} \Phi e^{-\ti P_\Phi a} =\Phi+a$,
but $e^{\ti \Phi}$ containing higher powers of irrelevant operators of $(\Phi_{I})^n$ for $n>2$, 
thus the kinetic term is an irrelevant operator at the strong-coupling massive fixed point.

The higher derivative term $\alpha_{\text{small}} \psi_q \dots \nabla^n_x \psi_q$
is generated by the further long range hopping, thus contains higher powers of $:e^{\ti \Phi }:$
thus this subleading terms in Eq(\ref{eq:subf}) 
are {\emph{ further irrelevant perturbation}} at the infrared, comparing to the dominant cosine terms.

\color{black}

\subsubsection{Numerical simulation for the free 
fermion theory with nontrivial Chern number\label{numeric}}

\begin{figure}[tb] 
\centering
{\includegraphics[width=.45\textwidth]{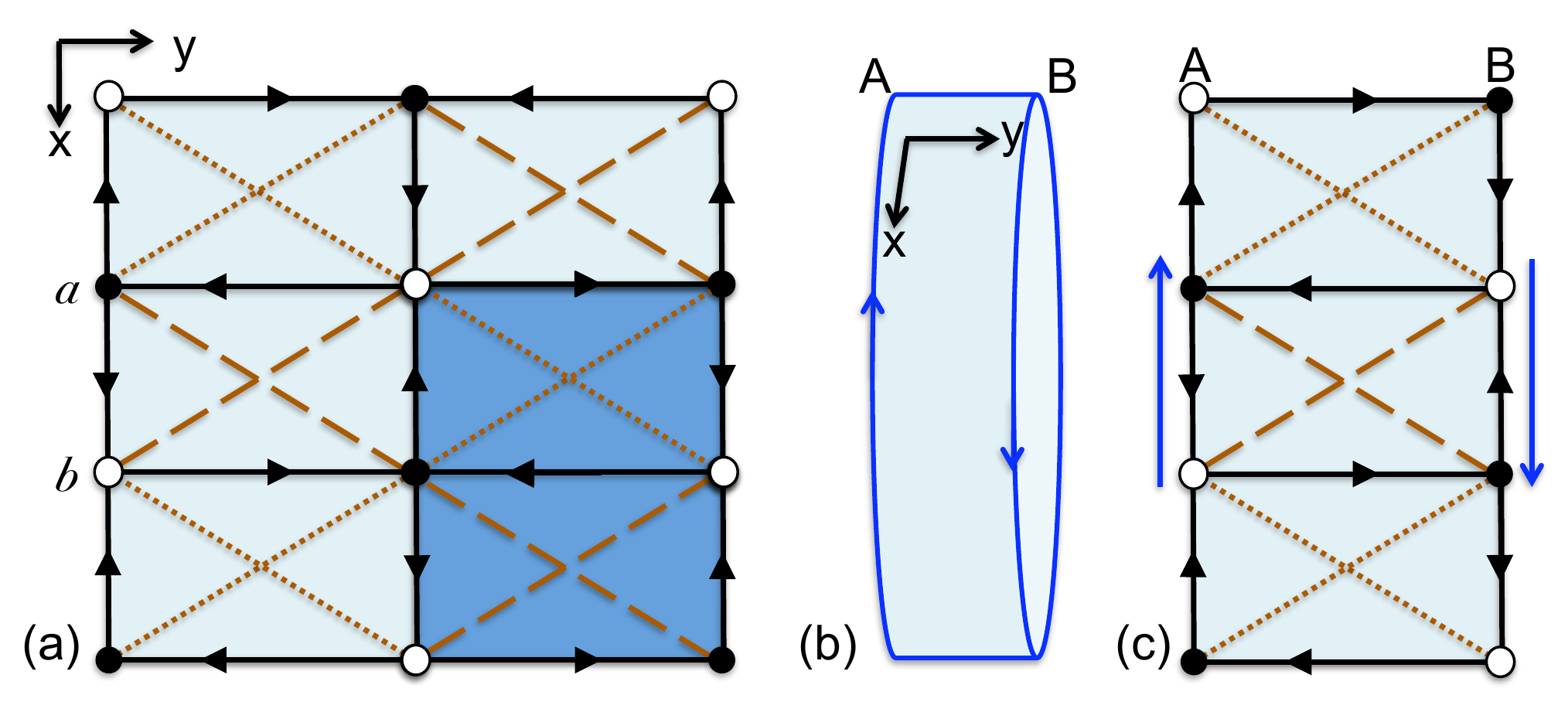}}
\caption{Chiral $\pi$-flux square lattice: (a) A unit cell is indicated as the shaded darker region,
containing two sublattice as a black dot $a$ and a white dot $b$. The lattice Hamiltonian has
hopping constants, $t_1 e^{i\pi/4}$ along the black arrow direction,
 $t_2$ along dashed brown links,
 $-t_2$ along dotted brown links.
(b) Put the lattice on a cylinder.
(c) The ladder: the lattice on a cylinder with a square lattice width.
The chirality of edge state is along the direction of blue arrows.}
\label{chiral-pi-flux}
\end{figure}

 Follow from Sec.\ref{sec3-5-4-0} and \ref{subsubsec:kinetic}, here we provide a concrete lattice realization for free fermions part of Eq.(\ref{H3540}) (with $G_1=G_2=0$),
and show that the Chern insulator provides the desired gapless fermion energy spectrum 
(say, a left-moving Weyl fermion on the edge A and a right-moving Weyl fermion on the edge B, and totally a Dirac fermion for the combined).
We adopt the chiral $\pi$-flux square lattice model 
in Fig.\ref{chiral-pi-flux} as an example.
This lattice model can be regarded as a free theory of 3-5-4-0 fermions of Eq.(\ref{cf}) with its mirrored conjugate.
We will explicitly show filling the first Chern number\cite{Thouless:1982zz} $C_1=-1$ band of the lattice on a cylinder would give the edge states of a free fermion with U(1) charge $3$,
similar four copies of model together render 3-5-4-0 free fermions theory of Eq.(\ref{H3540}).

\begin{figure}[!h] 
\centering
{\includegraphics[width=.35\textwidth]{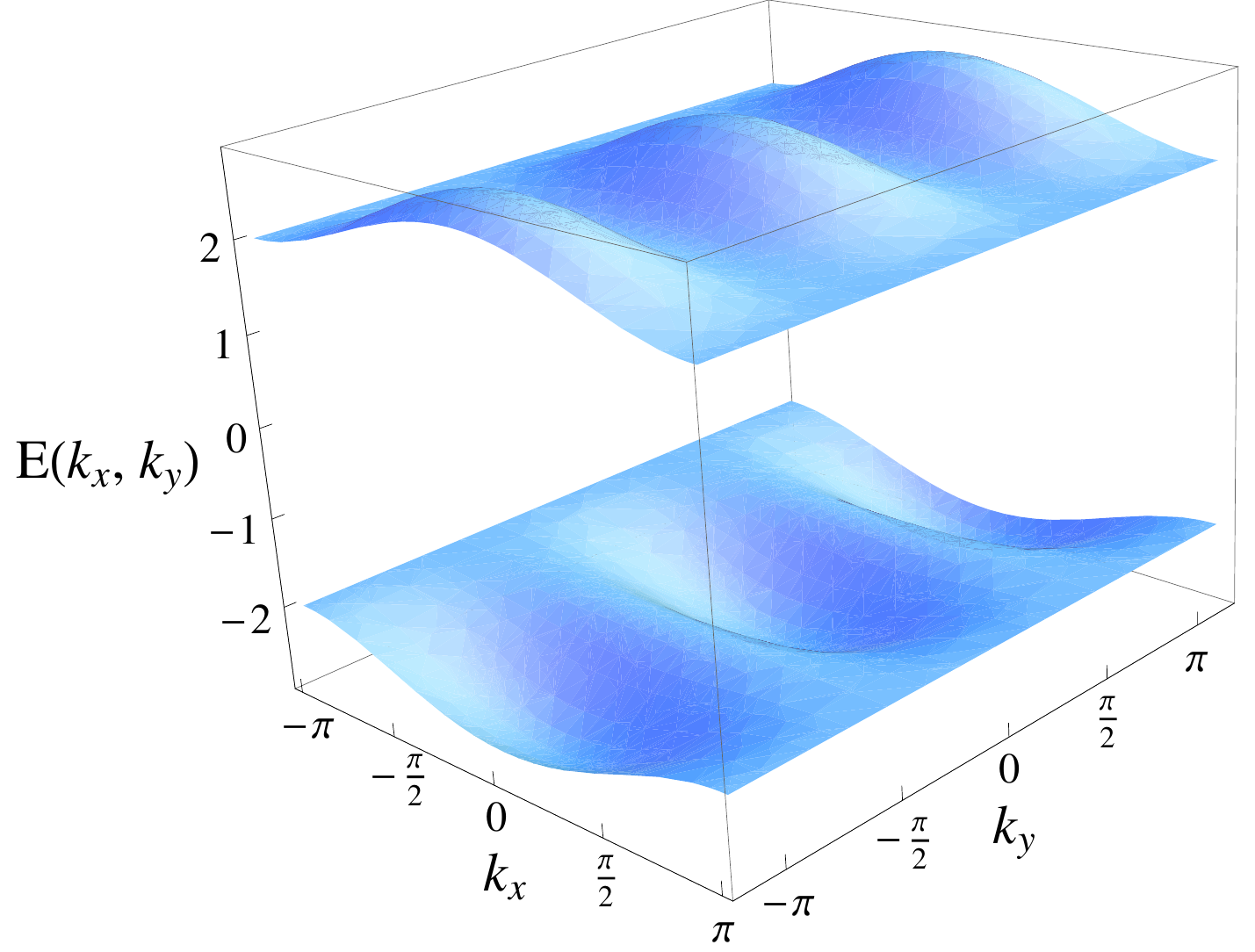}}
\caption{Two nearly-flat energy bands $\E_{\pm}$ in Brillouin zone
for the kinetic hopping terms of our model Eq.(\ref{H3540}).}
\label{BZflatband}
\end{figure}

We design hopping constants $t_{ij,3}=t_1 e^{\ti \pi/4}$ along the black arrow direction in Fig.\ref{chiral-pi-flux}, and its hermitian conjugate determines $t_{ij,3}=t_1 e^{-\ti \pi/4}$ along the opposite hopping direction;
$t'_{ij,3}=t_2$ along dashed brown links, $t'_{ij,3}=-t_2$ along dotted brown links.
The shaded blue region in Fig.\ref{chiral-pi-flux} indicates a unit cell, containing two sublattice as a black dot $a$ and a white dot $b$.
If we put the lattice model on a torus with periodic boundary conditions for both $x,y$ directions,
then we can write the Hamiltonian in $\mathbf{k}=(k_x,k_y)$ space in Brillouin zone (BZ), as
$H=\sum_\mathbf{k} f^\dagger_\mathbf{k} H(\mathbf{k}) f_\mathbf{k}$,
where $f_\mathbf{k}=(f_{a,\mathbf{k}},f_{b,\mathbf{k}})$.
For two sublattice $a,b$, we have a generic pseudospin form of Hamiltonian $H(\mathbf{k})$,
\be
H(\mathbf{k})=B_0(\mathbf{k}) + \vec{B}(\mathbf{k}) \cdot \vec{\sigma}. 
\ee
$\vec{\sigma}$ are Pauli matrices $(\sigma_x,\sigma_y,\sigma_z)$. In this model $B_0(\mathbf{k})=0$ and
$\vec{B}=(B_x(\mathbf{k}),B_y(\mathbf{k}),B_z(\mathbf{k}))$ have three components
in terms of $\mathbf{k}$ and lattice constants $a_x,a_y$. The eigenenergy $\E_{\pm}$ of $H(\mathbf{k})$ provide two nearly-flat energy bands, shown in Fig.\ref{BZflatband}, from
$H(\mathbf{k}) | \psi_{\pm}(\mathbf{k}) \rangle = \E_{\pm} | \psi_{\pm}(\mathbf{k}) \rangle $.

\begin{figure*}[!h] 
{(a)\includegraphics[height=.19\textwidth]{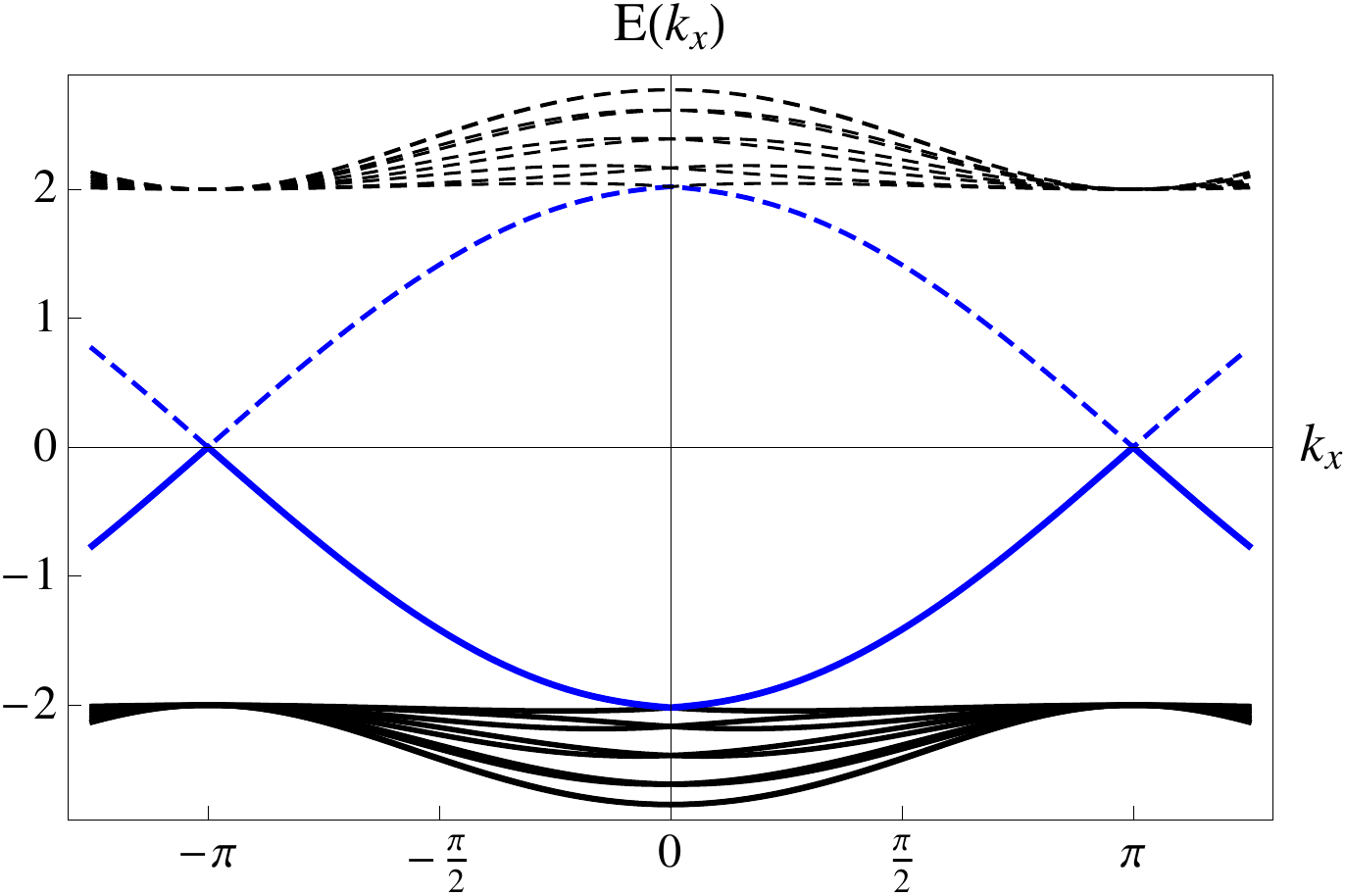} 
~~ (b)\includegraphics[height=.19\textwidth]{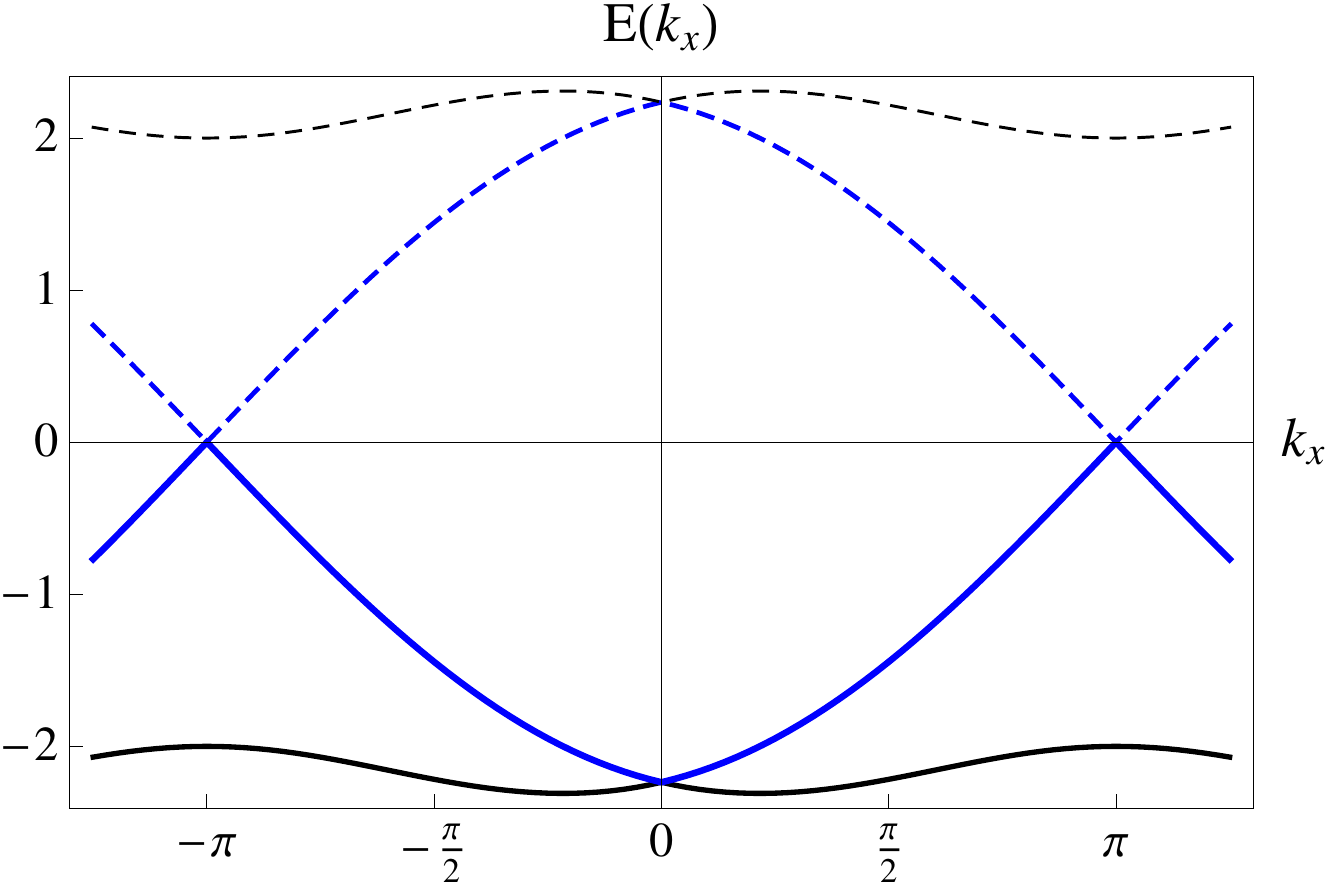}
~~ (c)\includegraphics[height=.19\textwidth]{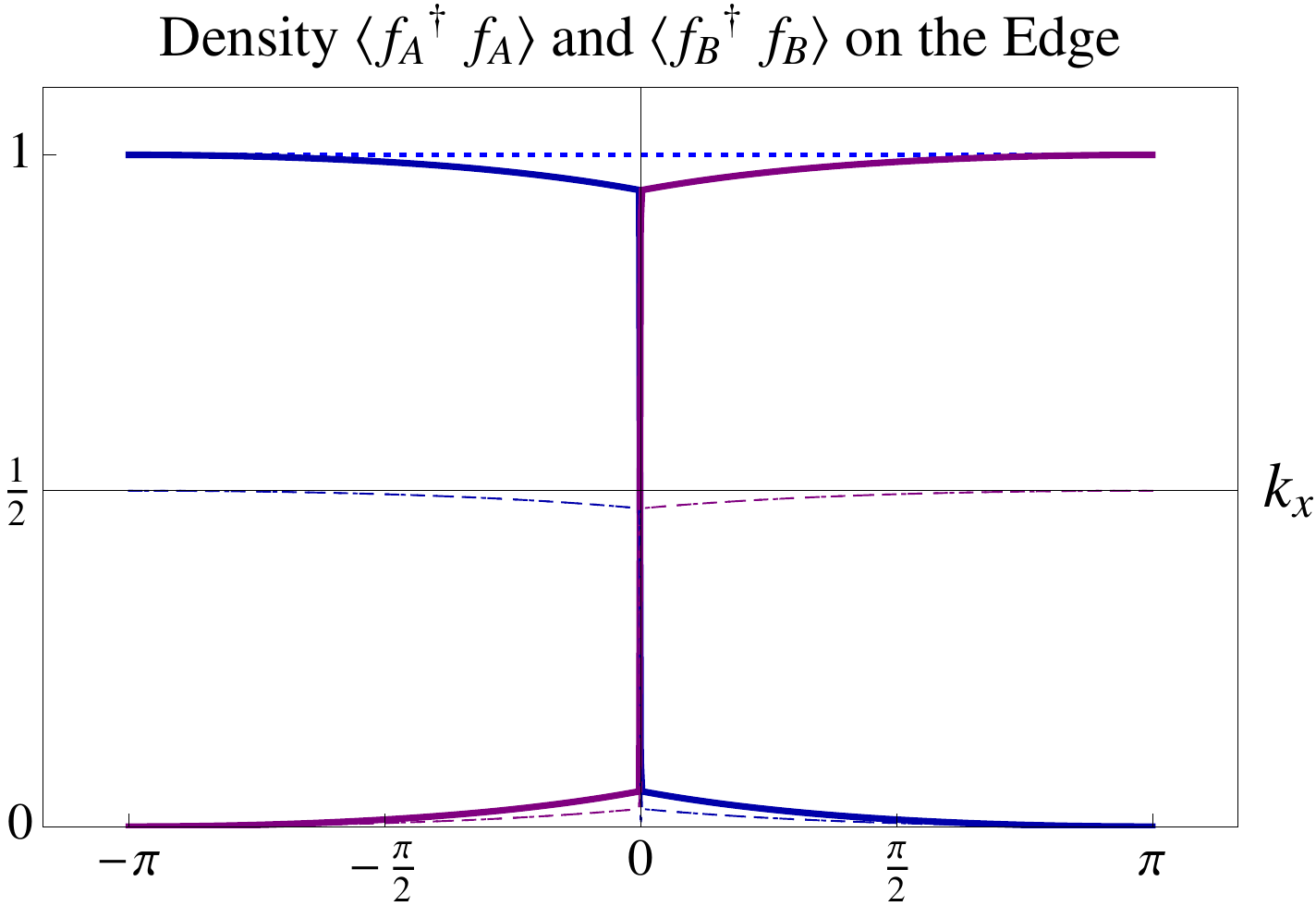}}
\caption{ The energy spectrum $\E(k_x)$ and the density matrix $\langle f^\dagger f \rangle$ of the chiral $\pi$-flux model 
on a cylinder:
{\bf (a)} On a 10-sites width ($9a_y$-width) cylinder: The blue curves are edge states spectrum. The black curves are for states extending in the bulk.
The chemical potential at zero energy fills eigenstates in solid curves, and leaves eigenstates in dashed curves unfilled.
{\bf (b)} On the ladder, a 2-sites width ($1a_y$-width) cylinder: the same as the (a)'s convention.
{\bf (c)} The density $\langle f^\dagger f \rangle$ of the edge eigenstates (the solid blue curve in (b)) on the ladder lattice.
The dotted blue curve shows the total density sums to 1,
the darker purple curve shows $\langle f_{\A}^\dagger f_{\A} \rangle$ on the left edge A, and
the lighter purple curve shows $\langle f_{\B}^\dagger f_{\B} \rangle$ on the right edge B.
The dotted darker(or lighter) purple curve shows density $\langle f_{\A,a}^\dagger f_{\A,a} \rangle$ (or $\langle f_{\B,a}^\dagger f_{\B,a} \rangle$) on sublattice $a$, while
the dashed darker(or lighter) purple curve shows density $\langle f_{\A,b}^\dagger f_{\A,b} \rangle$ (or $\langle f_{\B,b}^\dagger f_{\B,b} \rangle$) on sublattice $b$.
This edge eigenstate has the left edge A density with majority quantum number $k_x<0$, and has the right edge B density with majority quantum number $k_x>0$.
Densities on two sublattice $a,b$ are equally distributed as we desire.
}
\label{kx_chiral}
\end{figure*}

For the later purpose to have the least mixing between edge states on the left edge A and right edge B on a cylinder in Fig.\ref{chiral-pi-flux}(b), here we fine tune 
$t_2/t_1=1/2$.
For convenience, we simply set $t_1=1$ as the order magnitude of $\E_{\pm}$. We set lattice constants $a_x=1/2,a_y=1$ such that BZ has $-\pi \leq k_x <\pi,-\pi \leq k_y <\pi$.
The first Chern number\cite{Thouless:1982zz} of the 
energy band $|\psi_{\pm}(\mathbf{k}) \rangle$ is
\be
C_1=\frac{1}{2\pi}\int_{\mathbf{k} \in \text{BZ}} d^2\mathbf{k}\;  \epsilon^{\mu \nu } \partial_{k_\mu} \langle \psi(\mathbf{k}) | -i \partial_{k_\nu} |  \psi(\mathbf{k}) \rangle.
\ee
We find $C_{1,\pm}=\pm 1$ for two bands.
The $C_{1,-}=-1$ lower energy band indicates the clockwise chirality of edge states when we put the lattice on a cylinder as in Fig.\ref{chiral-pi-flux}(b).
Overall it implies the chirality of the edge state on the left edge A moving along $-\hat{x}$ direction, and on the right edge B moving along $+\hat{x}$ direction 
- the clockwise chirality as in Fig.\ref{chiral-pi-flux}(b), consistent with the earlier result
$C_{1,-}=-1$ of Chern number.
This edge chirality is demonstrated in Fig.\ref{kx_chiral}.

The above construction is for edge states of free fermion with U(1) charge $3$ of 3$_L$-5$_R$-4$_L$-0$_R$ fermion model. Add the same copy with $C_{1,-}=-1$ lower band gives another layer of U(1) charge $4$ free fermion.
For another layers of U(1) charge $5$ and $0$, we simply adjust hopping constant $t_{ij}$ to $t_1 e^{-\ti \pi/4}$ along the black arrow direction and $t_1 e^{\ti\pi/4}$ along the opposite direction in Fig.\ref{chiral-pi-flux}, which makes $C_{1,-}=+1$.
Stack four copies of chiral $\pi$-flux ladders with $C_{1,-}=-1,+1,-1,+1$ provides the lattice model of 3-5-4-0 free fermions with its mirrored conjugate.

The lattice model so far is an effective 1+1D non-chiral theory. 
We claim the interaction terms ($G_1,G_2\neq 0$) can gap out the mirrored edge states on the edge B.
The simulation including interactions can be numerically expansive, even so on a simple ladder model.
Because of higher power interactions, one can no longer diagonalize the model in $\mathbf{k}$ space as the case of the quadratic free-fermion Hamiltonian.
For 
interacting case, one may need to apply exact diagonalization in real space, or density matrix renormalization group (DMRG), 
which is powerful in 1+1D.
We leave this 
interacting numerical study for the 
lattice community or the future work.

\subsection{Interaction gapping terms and the strong coupling scale} \label{sec:mapping-strong-gapping}

Similar to Sec.\ref{subsubsec:kinetic}, for the {\bf interaction gapping terms} of the Hamiltonian, we can do the mapping based on
Eq.(\ref{eq:subf}), where the leading terms on the lattice is 
\bea \label{subsec:gap_map}
&& g_{a}  \cos(\ell_{a,I}^{} \cdot\Phi_{I}) 
= U_{\text{interaction}}\big( \tilde{\psi}_{q}, \dots,  \nabla^n_x \tilde{\psi}_{q},\dots  \big) ) 
\to U_{\text{point.split.}}
\bigg(\hat{f}_{q}(j), \dots  \big( \hat{f}^n_{q}(j)\big)_{pt.s.}, \dots
\bigg) 
+ \alpha_{\text{small}} \dots \;\;\;\;\;\;\;\; \;\;
\eea
Again, potentially there may contain subleading pieces, such as
further higher order derivatives $\alpha_{\text{small}} \nabla^n_x \psi_q$ with a small coefficient $\alpha_{\text{small}}$, 
or tiny mixing of the different U(1)-charge flavors $\alpha_{\text{small}}' {\psi_{q_1}\psi_{q_2}\dots}$. 
However, using the same RG analysis in Sec.\ref{subsubsec:kinetic},
at both the weak coupling and the strong coupling fix points,
we learn that those $\alpha_{\text{small}}$ terms are only {\bf subleading-perturbative effects}
which are {\emph{ further irrelevant perturbation}} at the infrared comparing
to the dominant piece (which is the kinetic term for the weak $g$ coupling, but is replaced by the cosine term for the strong $g$ coupling).


One more question to ask is: what is {\bf the scale of coupling $G$} such that the gapping term becomes dominant and 
the {\bf B edge states form the mass gaps, but 
maintaining} (without interfering with) 
{\bf the gapless A edge states?}

To answer this question, we first know 
the absolute value of energy magnitude 
for each term in the desired Hamiltonian for our chiral fermion model:
\bea  \label{eq:energyscale}
&&|G \text{ gapping term}| \gtrsim |t_{ij},t_{ij}' \text{ kinetic term}| \\
&& \gg  |G \text{ higher order $\nabla^n_x$ and mixing terms}| \gg |t_{ij},t_{ij}' \text{ higher order } \psi_q \dots \nabla^n_x \psi_q|. \nonumber
\eea

For {\bf field theory}, the gapping terms (the cosine potential term or the multi-fermion interactions) are irrelevant for a weak $g$ coupling,
this implies that $g$ needs to be large enough. Here the $g \equiv (g_a)/a^2$ really means the dimensionless quantity $g_a$.

For {\bf lattice model}, however, the dimensional analysis is very different. 
Since the $G$ coupling of gapping terms and the hopping amplitude $t_{ij}$ both have dimension of energy $[G] =[t_{ij}]=1$,
this means that the scale of the dimensionless quantity of $|G|/|t_{ij}|$ is important.  
(The $|t_{ij}|,|t_{ij}'|$ are about the same order of magnitude.)


Presumably we can design the lattice model under Eq.(\ref{eq:lengthscale}), $a < \xi <  w <  L$,
such that their ratios between each length scale are about the same.
We expect the ratio of couplings of ${|G|}$ to ${|t_{ij}|}$ is about the ratio of mass gap ${\Delta_m}$ to kinetic energy fluctuation ${\delta E_k}$ caused by $t_{ij}$ hopping,
thus \emph{very roughly} 
\be
\frac{|G|}{|t_{ij}|} \sim \frac{\Delta_m}{\delta E_k} \sim \frac{(\xi)^{-1}}{(w)^{-1}} \sim \frac{w}{\xi} \sim \frac{L}{w}  \sim \frac{\xi}{a}. 
\ee
We expect that the scales at strong coupling $G$ is about
\be
|G| \gtrsim  |{t_{ij}}|  \cdot  \frac{\xi}{a}  
\ee 
this magnitude can support our lattice chiral fermion model with mirror-fermion decoupling. 
If $G$ is too much smaller than $ |{t_{ij}}|  \cdot  \frac{\xi}{a} $, then 
mirror sector stays gapless.
On the other hand, if $|G|/|{t_{ij}}|$ is too much stronger 
or simply $|G|/|{t_{ij}}| \to \infty $ 
may cause either of two disastrous cases: \\
(i) Both edges would be gapped and the whole 2D plane becomes \emph{dead without kinetic hopping}, if the correlation length reaches the scale of the cylinder width: $\xi \gtrsim w$.\\
(ii) The B edge(say at site $n \hat{y}$) becomes completely gapped, but forms a dead 
overly-high-energy 1D line decoupled from the remain lattice. The neighbored line (along $(n-1) \hat{y}$) next to edge B 
experiences no interaction thus may still form mirror gapless states near B. 
(This may be another reason why CGP fails in Ref.\cite{CGP1247} due to implementing overlarge strong coupling.)\\
So either the two cases caused by too much strong $|G|/|{t_{ij}}|$ is not favorable.
Only $|G| \gtrsim  |{t_{ij}}|  \cdot    \frac{\xi}{a}$, we can have the mirrored sector at edge $B$ gapped,
meanwhile keep the chiral sector at edge $A$ gapless. 
$\frac{|G|}{|{t_{ij}}| }$ is somehow larger than order 1 is what we referred as the {\bf intermediate(-strong) coupling}.
\be
\frac{|G|}{|{t_{ij}}| } \gtrsim   O(1).
\ee
(Our $O(1)$ means some finite values, possibly as large as $10^4,10^6$, etc, but still finite. And the kinetic term is \emph{not} negligible.)
The sign of $G$ coupling shall not matter, since in the cosine potential language, either $g_1,g_2$ greater or smaller than zero
are related by sifting the minimum energy vaccua of the cosine potential.\\

To summarize, the two key messages in Sec.\ref{sec:mapping-field-lattice} are:\\ 
\noindent
$\bullet$ First, the free-kinetic hopping part of lattice model has been simulated and there gapless energy spectra have been computed shown in Figures.
The energy spectra indeed show the gapless Weyl fermions on each edge.
So, 
the continuum field theory to a lattice model mapping is immaterial to the subleading terms of Eq.(\ref{eq:subf}), 
the physics is as good or as exact as we expect for the free kinetic part.
We comment that this lattice realization of quantum hall-like states with chiral edges have been implemented for long in condensed matter, dated back as early such as Haldane's work.\cite{Haldane:1988zza}\\
\noindent
$\bullet$ Second, by adding the interaction gapping terms, the spectra will be modified from the mirror gapless edge to the mirror gapped edge. 
The continuum field theory to a lattice model mapping based on
Eq.(\ref{eq:subf}) for the \emph{gapping terms} in Eq.(\ref{subsec:gap_map}) is as good or as exact as the \emph{free kinetic part} Eq.(\ref{subsec:free_map}),
because the mapping is the same procedure as in Eq.(\ref{eq:subf}). 
Since the subleading correction for the free and for the interacting parts are {\emph{ further irrelevant perturbation}} at the infrared,
the {\bf non-perturbative topological effect} of the gapped edge contributed from the leading terms remains. 

\color{black}

In the next section, 
we will provide a
{\bf topological non-perturbative proof}
to justify that the $G_1,G_2$ interaction terms can gap out mirrored edge states, without employing numerical methods,
but purely based on an analytical derivation.

\subsection{Topological Non-Perturbative Proof of Anomaly Matching Conditions = Boundary Fully Gapping Rules} \label{anomaly-gap-proof}

As Sec.\ref{sec3-5-4-0},\ref{sec:mapping-field-lattice} prelude, we now show that
Eq.(\ref{H3540}) indeed gaps out the mirrored edge states on the edge B in
Fig.\ref{3540}. This proof will support the evidence that Eq.(\ref{H3540})
gives the non-perturbative lattice definition of the 1+1D chiral fermion theory
of Eq.(\ref{cf}).

In Sec.\ref{SPT-CS}, we first provide a generic way to formulate our model, with a insulating bulk but with gapless edge states.
This can be done through so called the {\bf bulk-edge correspondence},
namely the Chern-Simons theory in the bulk and the Wess-Zumino-Witten(WZW) model on the boundary.
More specifically, for our case with U(1) symmetry chiral matter theory, we only needs a U(1)$^N$ rank-$N$ Abelian K matrix Chern-Simons theory in the bulk and the multiplet chiral boson theory on the boundary.
We can further fermionize the multiplet chiral boson theory to the multiplet chiral fermion theory.

In Sec.\ref{anomaly-hall}, we provide a physical understanding between
the anomaly matching conditions and the effective Hall conductance.
This intuition will be helpful to understand the relation between the anomaly matching conditions and Boundary Fully Gapping Rules, to be discussed in Sec.\ref{anomaly-gap}. 

\noindent
\subsection{Bulk-Edge Correspondence - 2+1D Bulk Abelian SPT by Chern-Simons theory \label{SPT-CS}}

With our 3$_L$-5$_R$-4$_L$-0$_R$ chiral fermion model in mind, below we will trace back to fill in the
background how we obtain this model from the understanding of symmetry-protected topological states (SPT).
This understanding in the end leads to a more general construction.

We first notice that the bosonized action of the free part of chiral fermions in Eq.(\ref{b3540}), can be
regarded as the edge states action $S_{\partial}$ of a bulk U(1)$^N$ Abelian K matrix Chern-Simons theory $S_{bulk}$ 
(on a 2+1D manifold ${\cM}$ with the 1+1D boundary ${\partial \cM}$):
\bea
S_{bulk}&=&
\frac{K_{IJ}}{4\pi}\int_{\cM}   a_I \wedge d a_J = \frac{K_{IJ}}{4\pi}\int_{\cM}  dt\, d^2x \varepsilon^{\mu\nu\rho} a^I_{\mu} \partial_\nu a^J_{\rho},\;\;\;\;\;\;
\label{CSbulk} \\
S_{\partial}&=& \frac{1}{4\pi} \int_{\partial \cM} dt \; dx \; K_{IJ} \partial_t \Phi_{I} \partial_x \Phi_{J} -V_{IJ}\partial_x \Phi_{I}   \partial_x \Phi_{J}. \;\;\;\;\;\;\;
\label{CSboundary}
\eea
Here $a_\mu$ is 
intrinsic 1-form gauge field from a low energy 
viewpoint.
Both indices $I,J$ run from $1$ to $N$.
Given $K_{IJ}$ matrix, it is known the ground state degeneracy (GSD) of this theory on the $\mathbb{T}^2$ torus is $\GSD=|\det K|$.
Here $V_{IJ}$ is the symmetric `velocity' matrix, we can simply choose $V_{IJ}=\mathbb{I}$, without losing generality of our argument.
The U(1)$^N$ gauge transformation is $a_I \to a_I + df_I$ and $\Phi_I \to \Phi_I+ f_I$.
The bulk-edge correspondence is meant to have
the gauge non-invariances of the bulk-only and the edge-only cancel with each other, so that the total gauge invariances is achieved 
from the full bulk and edge as a whole.

We will consider only an even integer $N \in 2 \Z^+$. The reason is that only such even number of edge modes, we can potentially gap out the edge states.
(For odd integer $N$, such a set of gapping interaction terms generically \emph{do not} exist, so the mirror edge states remain gapless.)

To formulate 3$_L$-5$_R$-4$_L$-0$_R$ fermion model, as shown in Eq.(\ref{b3540}), we need a rank-4 K matrix $\bigl( {\begin{smallmatrix}
1 &0 \\
0 & -1
\end{smallmatrix}}  \bigl) \oplus \bigl( {\begin{smallmatrix}
1 &0 \\
0 & -1
\end{smallmatrix}}  \bigl)$. Generically, for a general U(1) chiral fermion model, we can use a canonical fermionic matrix
\be \label{eq:Kf}
K^f_{N\times N} =\bigl( {\begin{smallmatrix}
1 &0 \\
0 & -1
\end{smallmatrix}}  \bigl) \oplus \bigl( {\begin{smallmatrix}
1 &0 \\
0 & -1
\end{smallmatrix}}  \bigl) \oplus \bigl( {\begin{smallmatrix}
1 &0 \\
0 & -1
\end{smallmatrix}}  \bigl) \oplus\dots
\ee
Such a matrix is special, because it describes a more-restricted Abelian Chern-Simons theory with GSD$=|\det K^f_{N\times N}|=1$ on the $\mathbb{T}^2$ torus.
In the condensed matter language, the uniques GSD implies it has no long range entanglement, and it has no intrinsic topological order. 
Such a state may be wronged to be only a trivial insulator, but actually this is recently-known to be potentially nontrivial as
the symmetry-protected topological states (SPT).

(This paragraph is for readers with interests in SPT: SPT are short-range entangled states with onsite symmetry in the bulk. 
For SPT, there is no long-range entanglement, no fractionalized quasiparticles (fractional anyons) and no fractional statistics in the bulk. 
The bulk onsite symmetry may be realized as a non-onsite symmetry on the boundary.
If one gauges the non-onsite symmetry of the boundary SPT, the boundary theory becomes
an anomalous gauge theory.\cite{Wen:2013ppa} The anomalous gauge theory is
ill-defined in its own dimension, but can be defined as the boundary of the
bulk SPT.  However, this understanding indicates that if the boundary theory
happens to be anomaly-free, then it can be defined non-perturbatively on the
same dimensional lattice.)

$K^f_{N\times N}$ matrix describe {\bf fermionic SPT states}, which 
is described by  bulk \emph{spin Chern-Simons theory} of $|\det K|=1$.
A spin Chern-Simons theory only exist on the spin manifold, which has spin structure and can further define spinor bundles. 
However, there are another simpler class of SPT states, the {\bf bosonic SPT states}, which
 is described by 
the canonical form  $K^{b\pm}_{N\times N}$ 
with blocks of
$ \bigl( {\begin{smallmatrix}
0 &1 \\
1 & 0
\end{smallmatrix}} \bigl)$ and a set of all positive(or negative) coefficients $\E_8$ lattices $K_{\E_8}$, 
namely,
\bea \label{eq:Kb0}
K^{b0}_{N\times N} &=&\bigl( {\begin{smallmatrix}
0 &1 \\
1 & 0
\end{smallmatrix}}  \bigl) \oplus \bigl( {\begin{smallmatrix}
0 &1 \\
1 & 0
\end{smallmatrix}}  \bigl) \oplus \dots. \\
K^{b \pm}_{N\times N} &=&K^{b0}_{}   \oplus  (\pm K_{\E_8}) \oplus (\pm K_{\E_8}) \oplus  \dots \nonumber
\eea
The $K_{\E_8}$ matrix describe 8-multiplet chiral bosons moving in the same direction, thus it cannot be gapped by adding multi-fermion interaction among themselves.
We will neglect $\E_8$ chiral boson states but only focus on $K^{b0}_{N\times N}$ for the reason to consider \emph{only the gappable states}.
The K-matrix form of Eq.(\ref{eq:Kf}),(\ref{eq:Kb0}) is called
the \emph{unimodular indefinite symmetric integral matrix}.

After fermionizing the boundary action Eq.(\ref{CSboundary}) with $K^f_{N\times N}$ matrix, we obtain multiplet chiral fermions (with several pairs, each pair contain left-right moving Weyl fermions forming a Dirac fermion).
\bea \label{CSferboundary}
&S_\Psi&=\int_{\partial \cM}  dt \; dx \; ( \ti\bar{\Psi}_{\A} \Gamma^\mu  \partial_\mu \Psi_{\A} ).    
\eea
with
$\Gamma^0=\underset{j=1}{\overset{N/2}{\bigoplus}} \gamma^0$, $\Gamma^1 =\underset{j=1}{\overset{N/2}{\bigoplus}}  \gamma^1$, $\Gamma^5\equiv\Gamma^0\Gamma^1$, $\bar{\Psi}_i \equiv \Psi_i \Gamma^0$
and $\gamma^0=\sigma_x$, $\gamma^1=\ti\sigma_y$, $\gamma^5\equiv\gamma^0\gamma^1=-\sigma_z $.\\

\noindent
{\bf Symmetry transformation for the edge states-}

The edge states of $K^f_{N\times N}$ and $K^{b0}_{N\times N}$ Chern-Simons theory are non-chiral in the sense there are equal number of left and right moving modes.
However, we can make them with a charged `chirality' respect to a global(or external probed, or dynamical gauge) symmetry group.
 For the purpose to build up our `chiral fermions and chiral bosons' model with `charge chirality,' we consider the simplest possibility to couple it to a global U(1) symmetry with
 a charge vector $\mathbf{t}$. (This is the same as the symmetry charge vector of SPT states\cite{Lu:2012dt,Ye:2013upa}) 

\noindent
{\bf Chiral Bosons}:
For the case of multiplet chiral boson theory of Eq.(\ref{CSboundary}), the group element $g_{\theta}$ of U(1) symmetry acts on chiral fields as
\bea \label{eq:U(1)}
&& g_{\theta}: W^{\U(1)_{\theta}}=\mathbb{I}_{N \times N}, \;\; \delta \phi^{\U(1)_{\theta}} =\theta \mathbf{t},
\eea
With the following symmetry transformation,
\bea \label{chiralbosonsym}
&&\phi \to   W^{\U(1)_{\theta}}  \phi+\delta \phi^{\U(1)_{\theta}}=\phi+\theta \mathbf{t}
\eea

To derive this boundary symmetry transformation from the bulk Chern-Simons theory via bulk-edge correspondence,
we first write down the charge coupling bulk Lagrangian term, namely $\frac{\mathbf{q}^I}{2\pi} \; \epsilon^{\mu\nu\rho} A_{\mu} \partial_\nu a^I_{\rho}$, 
where the global symmetry current ${\mathbf{q}^I} J^{I \mu}= \frac{\mathbf{q}^I}{2\pi} \; \epsilon^{\mu\nu\rho} \partial_\nu a^I_{\rho}$ is coupled to an external gauge field $A_\mu$.   
The bulk U(1)-symmetry current ${\mathbf{q}^I} J^{I \mu}$ induces a boundary
U(1)-symmetry current $j^{I \mu}=   \frac{\mathbf{q}^I}{2\pi}   \; \epsilon^{\mu\nu} \partial_\nu \phi_{I}$. 
This implies the boundary symmetry operator  
is $S_{sym}=\exp(\ti \,\theta\, \frac{\mathbf{q}^I}{2\pi}   \int  \partial_x \phi_{I})$, 
with an arbitrary U(1) angle $\theta$
The induced symmetry transformation 
on $\phi_I $ is: 
\be \label{chiralbosonsym2}
(S_{sym}) \phi_I (S_{sym})^{-1}=\phi_I- \ti \theta\int dx \frac{\mathbf{q}^l}{2\pi}[\phi_I,\partial_x \phi_l] 
=\phi_I + \theta  (K^{-1})_{I l}  {\mathbf{q}^l}  \equiv \phi_I +\theta \mathbf{t}_I,
\ee
here we have used the canonical commutation relation $[\phi_I,\partial_x \phi_l]=\ti \,(K^{-1})_{I l}  $.
Compare the two Eq.(\ref{chiralbosonsym}),(\ref{chiralbosonsym2}), 
we learn that 
$$\mathbf{t}_I \equiv (K^{-1})_{I l}   {\mathbf{q}^l}.$$
The charge vectors $\mathbf{t}_I$ and ${\mathbf{q}^l}$ are related by an inverse of the $K$ matrix.
The generic interacting or gapping terms\cite{Wang:2012am,Levin:2013gaa,Lu:2012dt} 
for the multiplet chiral boson theory are the sine-Gordon or the cosine term
\be
S_{\partial,\text{gap}}= \int dt \; dx\;  \sum_{a} g_{a}  \cos(\ell_{a,I}^{} \cdot\Phi_{I}).  
\label{eq:Sgap}
\ee
If we insist that $S_{\partial,\text{gap}}$ obeys U(1) symmetry, to make Eq.(\ref{eq:Sgap}) invariant under Eq.(\ref{chiralbosonsym2}), we have to impose 
\bea
&&\ell_{a,I}^{} \cdot\Phi_{I} \to \ell_{a,I}^{} \cdot(\Phi_{I}+ \delta \phi^{\U(1)_{\theta}}) \text{mod}\; 2\pi \nonumber \\
&&\text{so}\;\;\; \boxed{\ell_{a,I}^{} \cdot\mathbf{t}_I =0}  
\Rightarrow \boxed{\ell_{a,I}^{} \cdot (K^{-1})_{I l} \cdot  {\mathbf{q}^l} =0}. \label{eq:t_symmetry}
 \label{eq:gapsym}
\eea

The above generic U(1) symmetry transformation works for bosonic $K^{b0}_{N\times N}$ as well as fermionic $K^f_{N\times N}$. 

\noindent
{\bf Chiral Fermions}:
In the case of fermionic $K^f_{N\times N}$, we will do one more step to fermionize the multiplet chiral boson theory. Fermionize the free kinetic part from Eq.(\ref{CSboundary}) to Eq.(\ref{CSferboundary}),
as well as the interacting cosine term:
\be \label{eq:cosine-multif}
g_{a}  \cos(\ell_{a,I}^{} \cdot\Phi_{I}) 
\to  \prod_{I=1}^{N} \tilde{g}_{a} \big( ({\psi}_{q_I})    (\nabla_x {\psi}_{q_I}) \dots   (\nabla_x^{|\ell_{a,I}|-1} {\psi}_{q_I}) \big)^{\epsilon}
\equiv U_{\text{interaction}}\big( {\psi}_{q}, \dots,  \nabla^n_x  {\psi}_{q},\dots  \big) 
\ee
to multi-fermion interaction. 
The ${\epsilon}$ is defined as the complex conjugation operator which depends on ${\text{sgn}(\ell_{a,I})}$, the sign of $\ell_{a,I}$. 
When ${\text{sgn}(\ell_{a,I})}=-1$, we define ${\psi}^\epsilon \equiv {\psi}^\dagger$ and also for the higher power polynomial terms.
Again, we absorb the normalization factor and the Klein factors through normal ordering of bosonization into the factor $\tilde{g}_{a}$.
The precise factor is not of our concern, 
since our goal is a non-perturbative lattice model. Obviously, the U(1) symmetry transformation for fermions is 
\be \label{chiralfermionsym}
{\psi}_{q_I} \to {\psi}_{q_I} e^{\ti \mathbf{t}_I \theta} ={\psi}_{q_I} e^{\ti (K^{-1})_{I l} \cdot  {\mathbf{q}^l}.  \theta}
\ee
In summary, we have shown a framework to describe U(1) symmetry chiral fermion/boson model using the bulk-edge correspondence, 
the explicit Chern-Siomns/WZW actions are given in Eq.(\ref{CSbulk}), (\ref{CSboundary}), (\ref{CSferboundary}), (\ref{eq:Sgap}), (\ref{eq:cosine-multif}), and their symmetry realization 
Eq.(\ref{chiralbosonsym2}),(\ref{chiralfermionsym}) and constrain are given in Eq.(\ref{eq:t_symmetry}),(\ref{eq:gapsym}). Their physical properties are tightly associated to
the fermionic/bosonic SPT states.

\color{black}

\noindent
\subsection{Anomaly Matching Conditions and Effective Hall Conductance \label{anomaly-hall} }

The bulk-edge correspondence is meant,
not only to achieve the gauge invariance by canceling the non-invariance of bulk-only and boundary-only,
but also to have the boundary anomalous current flow can be transported into the extra dimensional bulk.
This is known as Callan-Harvey effect 
in high energy physics,
 Laughlin thought experiment, 
 or simply the quantum-hall-like state bulk-edge correspondence in condensed matter theory.

The goal of this subsection is to provide a concrete physical understanding of the anomaly matching conditions and effective Hall conductance : \\

\noindent
$\bullet$ (i) The anomalous current inflowing from the boundary is transported into the bulk. We now show that this thinking 
can easily derive the 1+1D U(1) Adler-Bell-Jackiw(ABJ) anomaly,
or Schwinger's 1+1D quantum electrodynamics(QED) anomaly.

\begin{figure}[h!] 
\centering
{(a)\includegraphics[width=.15\textwidth]{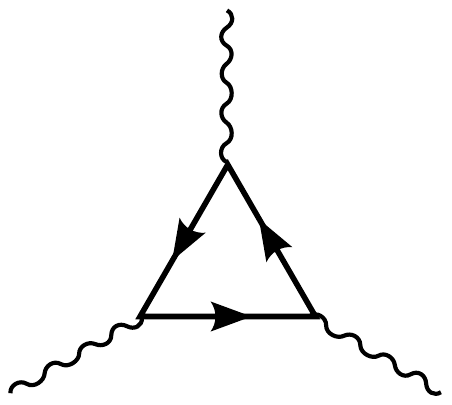}
~~ (b)\includegraphics[width=.20\textwidth]{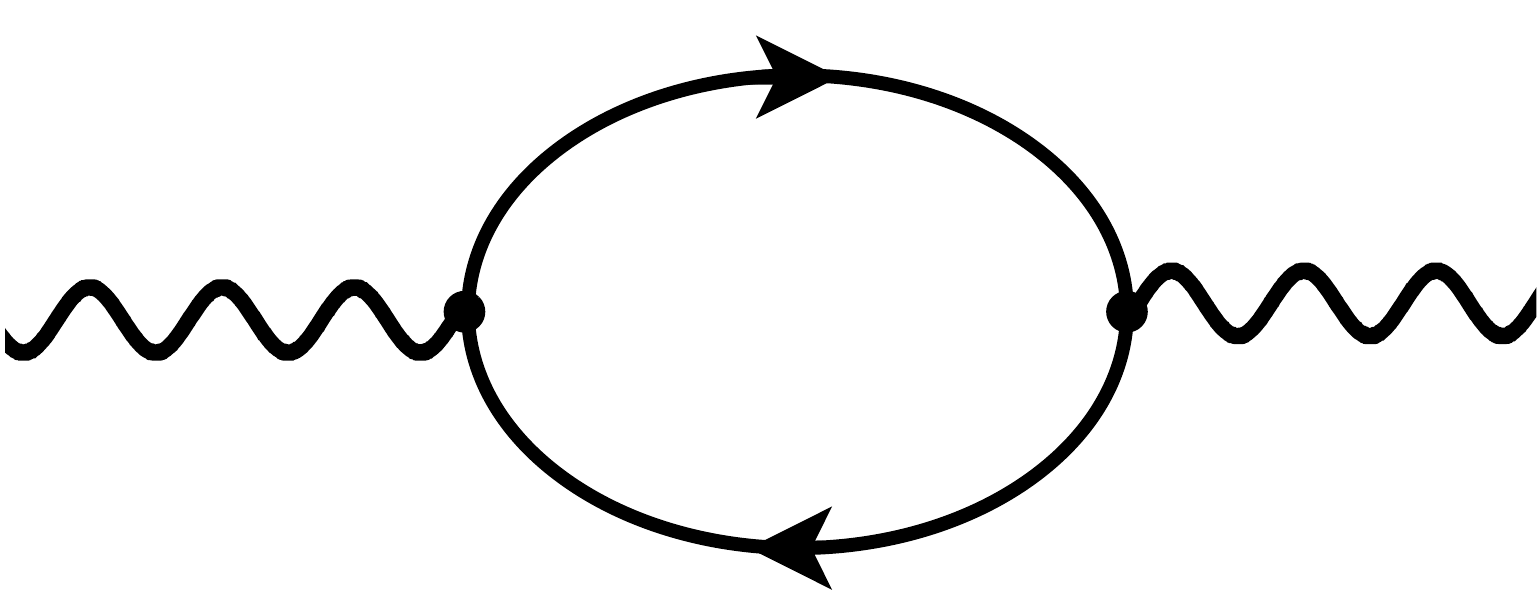}}
\caption{Feynman diagrams with solid lines representing chiral fermions and wavy lines representing U(1) gauge bosons: (a) 3+1D chiral fermionic anomaly shows $\mathcal{A}=\sum_q (q_L^3-q_R^3)$
(b) 1+1D chiral fermionic anomaly shows $\mathcal{A}=\sum_{q} (q_L^2-q_R^2)$ }
\label{anomaly}
\end{figure}

We will focus on the U(1) chiral anomaly, which is ABJ anomaly\cite{Adler:1969gk,Bell:1969ts} type.
It is well-known that ABJ anomaly can be captured by the anomaly factor $\mathcal{A}$ of the 1-loop polygon Feynman diagrams (see Fig.\ref{anomaly}).
The anomaly matching condition requires
\be
\mathcal{A}=\tr[T^aT^bT^c\dots]=0.
\ee
Here $T^a$ is the (fundamental) representation of the global or gauge symmetry algebra, which contributes to the vertices of 1-loop polygon Feynman diagrams. 

For example, the 3+1D chiral anomaly 1-loop triangle diagram of U(1) symmetry in Fig.\ref{anomaly}(a) with chiral fermions on the loop gives $\mathcal{A}=\sum (q_L^3-q_R^3)$.
Similarly, the 1+1D chiral anomaly 1-loop diagram of U(1) symmetry in Fig.\ref{anomaly}(b) with chiral fermions on the loop gives $\mathcal{A}=\sum (q_L^2-q_R^2)$.
Here $L,R$ stand for left-moving and right-moving modes.

How to derive this anomaly matching condition from 
a condensed matter theory viewpoint? 
Conceptually, we understand that\\
\colorbox{lgray}{\parbox[t]{\linegoal}{
A $d$-dimensional anomaly free theory (which satisfies the anomaly matching condition) 
means that there is no anomalous current leaking from its $d$-dimensional spacetime (as the boundary)
to an extended bulk theory of $d+1$-dimension.
}}\\
More precisely, for an 1+1D U(1) anomalous theory realization of the above statement,
we can formulate it as the boundary of a 2+1D bulk as in Fig.\ref{cylinder} with a Chern-Simons action ($S=\int\big(\frac{K}{4\pi} \;a\wedge d a+ \frac{q}{2\pi}A \wedge d a$\big)). 
Here the field strength $F=dA$ is equivalent to the external U(1) flux in the Laughlin's flux-insertion thought experiment 
threading through the cylinder
(see a precise derivation in the Appendix of Ref.\cite{Santos:2013uda}). 
Without losing generality, let us first focus on the boundary action of Eq.(\ref{CSboundary}) as a chiral boson theory with only one edge mode. We derive its equations of motion as shown in Sec.\ref{sec:chiralanomaly} on chiral anomaly and QH states correspondence.
%
Here we derive the Hall conductance, easily obtained from its definitive relation $J_y ={\sigma_{xy}} E_x $ in Eq.(\ref{eq:J=sy}), 
$\sigma_{xy}=qK^{-1}q/(2\pi).$
Here $j_{\textrm{b}}$ stands for the edge current, with a left-moving current $j_L=j_{\textrm{b}}$ on one edge and a right-moving current $j_R=-j_{\textrm{b}}$ on the other edge, as in Fig.\ref{cylinder}. 
We convert a compact bosonic phase $\Phi$ to the fermion field $\psi$ by bosonization.
We can combine currents $ j_{\textrm{L}}+j_{\textrm{R}}$ as the vector current $j_{\textrm{V}}$, then find its U(1)$_V$ current conserved.
We combine currents $ j_{\textrm{L}}-j_{\textrm{R}}$ as the axial current $j_{\textrm{A}}$, then we obtain the famous ABJ U(1)$_A$ anomalous current in 1+1D (or Schwinger 1+1D QED anomaly).
%
%
This simple physical derivation shows that the left and right edges' boundary theories (living on the edge of a 2+1D U(1) Chern-Simons theory) can
combine to be a 1+1D anomalous world of Schwinger's 1+1D QED.

In other words, when the anomaly-matching condition holds ($\mathcal{A}=0$), then
there is no anomalous leaking current into the extended bulk theory,
as in Fig.\ref{cylinder}, so no `effective Hall conductance' for this anomaly-free theory.

It is straightforward to generalize the above discussion to a rank-$N$ K matrix Chern-Simons theory. It is easy to show that
the Hall conductance in a 2+1D system for a generic $K$ matrix is (via $ {\mathbf{q}_l} =K^{}_{I l} \,\mathbf{t}_I$)
\bea
 \boxed{\sigma_{xy}=\frac{1}{2 \pi} \mathbf{q} \cdot {K}^{-1} \cdot \mathbf{q} =\frac{1}{2 \pi} \mathbf{t} \cdot  {K}^{} \cdot  \mathbf{t}.} \;\;\;
\eea
For a 2+1D fermionic system for $K^f$ matrix of Eq.(\ref{eq:Kf}), 
\bea
 \sigma_{xy} =\frac{q^2}{2 \pi} \mathbf{t} {(K^{f}_{N \times N})}^{}\mathbf{t} =\frac{1}{2 \pi}\sum_q (q_{L}^2-q_{R}^2)=\frac{1}{2 \pi}\mathcal{A}.\;\;\;
\eea
%
Remarkably, this physical picture demonstrates that we can reverse the logic, starting from the `{\bf effective Hall conductance} of the bulk system'
to derive the {\bf anomaly factor} from the relation 
\be
{\boxed{ \mathcal{A}\; (\text{anomaly factor})= 2\pi \sigma_{xy} \;(\text{effective Hall conductance})  }  }
\label{eq:A=hall}
\ee
And from the ``no anomalous current in the bulk'' means that ``$\sigma_{xy}=0$'', we can further understand ``the anomaly matching condition $\mathcal{A}=2\pi \sigma_{xy}=0$.'' 

For the U(1) symmetry case, we can explicitly derive the anomaly matching condition for fermions and bosons. 

\noindent
\subsection{Anomaly Matching Conditions and Boundary Fully Gapping Rules  } \label{anomaly-gap}

This subsection is the main emphasis of our work, and we encourage the readers paying extra 
attentions on the result presented here. 
We will first present a heuristic physical argument on the rules that under what situations the boundary states can be gapped,
named as the {\bf Boundary Fully Gapping Rules}. We will then provide a \emph{topological non-perturbative} proof using the notion of Lagrangian subgroup and the exact sequence, following our
previous work Ref.\cite{Wang:2012am} and the work in Ref.\cite{{Levin:2013gaa}}. 
And we will also provide 
 \emph{perturbative} RG analysis, both for strong and weak coupling analysis of cosine potential cases.\\

\subsubsection{{topological non-perturbative} proof}  \label{sec:topo-nonp-proof}

The above physical picture is suggestive, but not yet rigorous enough mathematically.
Here we will formulate some topological non-perturbative proofs for {\bf Boundary Fully Gapping Rules}, and its equivalence to the {\bf anomaly-matching conditions} for the case of
U(1) symmetry. The first approach is using the topological quantum field theory(TQFT) along the logic of Ref.\cite{Kapustin:2010hk}.
The new ingredient for us is to find \emph{the equivalence of the gapped boundary to the anomaly-matching conditions}. 

For a field theory, the boundary condition is defined by a Lagrangian submanifold in the
space of Cauchy boundary condition data on the boundary. 
For a topological gapped boundary condition of a TQFT with a gauge group, 
we must choose a Lagrangian subspace in the Lie algebra of the gauge group. 
A subspace is {\bf Lagrangian} \emph{if and only if} it is both {\bf isotropic and coisotropic}. 

Specifically, for $\mathbf{W}$ be a linear subspace of a finite-dimensional vector space $\mathbf{V}$. Define the symplectic complement of $\bfW$ to be the subspace $\bfW^{\perp}$ as
\be
\bfW^{\perp} = \{v\in \bfV \mid \omega(v,w) = 0, \;\;\; \forall w\in \bfW\}
\ee
Here $\omega$ is the symplectic form, in the matrix form
$
\omega=\begin{pmatrix} 0 & \mathbf{1} \\ -\mathbf{1} & 0\end{pmatrix}
$
with $0$ and $\mathbf{1}$ are the block matrix of the zero and the identity.
The symplectic complement $\bfW^{\perp}$ satisfies:
$(\bfW^{\perp})^{\perp} = \bfW$, $\dim \bfW + \dim \bfW^\perp = \dim \bfV$. 
We have:\\

\noindent
$\bullet$  $\bfW$ is Lagrangian if and only if it is both isotropic and coisotropic, namely, 
 if and only if $\bfW = \bfW_{\perp}$. 
In a finite-dimensional $\bfV$, a Lagrangian subspace $\bfW$ is an isotropic one whose dimension is half that of $\bfV$. 

Now let us focus on the K-matrix $\text{U(1)}^N$ Chern-Simons theory,
the symplectic form $\omega$  is given by (with the restricted $a_{\parallel,I}$ on ${\partial \cM}$ ) 
\be \label{eq:CSsymp}
\omega=\frac{K_{IJ}}{4\pi} \int_{ \cM}  (\delta a_{\parallel,I}) \wedge d (\delta a_{\parallel,J}).
\ee
The bulk gauge group $\text{U(1)}^N  \cong \mathbb{T}_\Lambda$ as the torus, is the quotient space of $N$-dimensional vector space $\bfV$ by a subgroup $\Lambda \cong \Z^N$.
Locally the gauge field $a$ is a 1-form,
which has values in the Lie algebra of $\mathbb{T}_\Lambda$, we can denote this Lie algebra $\mathbf{t}_\Lambda$ as the vector space $\mathbf{t}_\Lambda =\Lambda \otimes \mathbb{R}$.

Importantly, for \emph{topological gapped boundary},
$a_{\parallel,I}$ lies in a Lagrangian subspace of $\mathbf{t}_\Lambda$ implies that the {\bf boundary gauge group} ($ \equiv \mathbb{T}_{\Lambda_0}$)
is a {\bf Lagrangian subgroup}. We can rephrase it in terms of the exact sequence
for the vector space of Abelian group $\Lambda \cong \Z^N$ and its subgroup $\Lambda_0$:
\bea
0 \to \Lambda_0 \overset{\mathbf{h}}{\to} \Lambda \to \Lambda/\Lambda_0 \to 0.
\eea
Here $0$ means the trivial zero-dimensional vector space and 
$\mathbf{h}$ is an injective map from $\Lambda_0$ to $\Lambda$. 
We can also rephrase it in terms of the exact sequence
for the vector space of Lie algebra by $0 \to \mathbf{t}_{(\Lambda/\Lambda_0)}^* \to  \mathbf{t}_{\Lambda}^* \to \mathbf{t}_{\Lambda_0}^* \to  0$.

The generic Lagrangian subgroup condition applies to K-matrix with the above symplectic form Eq.(\ref{eq:CSsymp}) renders three conditions on $\bfW $:\\
\noindent
$\bullet(i)$  The subspace $\bfW$ is isotropic with respect to the symmetric bilinear form $K$. \\
\noindent 
$\bullet(ii)$  The subspace dimension is a half of the dimension of $\mathbf{t}_\Lambda$. \\
\noindent 
$\bullet(iii)$ The signature of $K$ is zero. This means that $K$ has the same number of positive and negative eigenvalues.\\
Now we can examine the if and only if conditions $\bullet(i)$,$\bullet(ii)$,$\bullet(iii)$ listed above. 

For $\bullet(i)$ ``The subspace is isotropic with respect to the symmetric bilinear form $K$'' to be true, we have an extra condition on the injective ${\mathbf{h}}$ matrix (${\mathbf{h}}$ with
$N \times (N/2)$ components) for the $K$ matrix:
\bea \label{eq:TQFThKh}
\boxed{ {\mathbf{h}^T} K {\mathbf{h}}=0}.
\eea
Since $K$ is invertible($\det(K) \neq 0$), by defining a $N \times (N/2)$-component $\mathbf{L} \equiv K {\mathbf{h}}$, we have an equivalent condition:
\be  \label{eq:TQFTLKL}
\boxed{\mathbf{L}^T K^{-1} \mathbf{L}=0}.
\ee

For $\bullet(ii)$, ``the subspace dimension is a half of the dimension of $\mathbf{t}_\Lambda$'' is true if
$\Lambda_0$ is a rank-$N/2$ integer matrix. 

For $\bullet(iii)$, ``the signature of $K$ is zero'' 
is true, because our $K_{b0}$ and fermionic $K_{f}$ matrices implies that we have same number of left moving modes ($N/2$) and right moving modes ($N/2$), with $N\in 2 \Z^+$ an even number. 

Lo and behold, these above conditions $\bullet(i)$,$\bullet(ii)$,$\bullet(iii)$ are equivalent to the {\bf boundary full gapping rules} listed earlier. 
We can interpret $\bullet(i)$ as trivial statistics by
either writing in the column vector of ${\mathbf{h}}$ matrix (${\mathbf{h}} \equiv \Big(\eta_{1}, \eta_{2}, \dots,  \eta_{N/2}  \Big)$ with $N \times (N/2)$-components):
\be 
\boxed{\eta_{a,I'} K_{I'J'} \eta_{b,J'}=0}.
\ee
or writing in the column vector of ${\mathbf{L}}$ matrix ($\mathbf{L} \equiv \Big(\ell_{1}, \ell_{2}, \dots,  \ell_{N/2}  \Big)$ with $N \times (N/2)$-components): 
\be 
\boxed{\ell_{a,I} K^{-1}_{IJ} \ell_{b,J}=0}.
\ee
for any $\ell_a, \ell_b \in \Gamma^\partial \equiv \{ \sum_\alpha c_\alpha \ell_{\alpha,I}^{} |  c_\alpha \in \Z\}$ of boundary gapping lattice(Lagrangian subgroup).
Namely, {the \emph{boundary gapping lattice} $\Gamma^\partial$ is basically the $N/2$-dimensional vector space of a Chern-Simons lattice spanned by the $N/2$-independent 
column vectors of $\mathbf{L}$ matrix ($\mathbf{L} \equiv \Big(\ell_{1}, \ell_{2}, \dots,  \ell_{N/2}  \Big)$).}

Moreover, we can go a step further 
to relate the above rules equivalent to the {\bf anomaly-matching conditions}. 
By adding the corresponding cosine potential $g_a \cos(\ell_a \cdot \Phi)$ 
to the edge states of U(1)$^N$ Chern-Simons theory, we break the symmetry down to
$$
\U(1)^N \to \U(1)^{N/2}.
$$
What are the remained $\U(1)^{N/2}$ symmetry? By Eq.(\ref{eq:t_symmetry}),
this remained $\U(1)^{N/2}$ symmetry is generated by a number of $N/2$ of $\mathbf{t}_{b,I}$ vectors satisfying ${\ell_{a,I}^{} \cdot\mathbf{t}_{b,I} =0}$. 
We can easily construct 
\be \label{eq:t=KinvL}
\mathbf{t}_{b,I} \equiv K^{-1}_{IJ} \ell_{b,J}, \;\;\; \mathbf{t}  \equiv K^{-1} \mathbf{L}
\ee 
with $N/2$ number of them (or define $\mathbf{t}$ as
the linear-combination of $\mathbf{t}_{b,I} \equiv \sum_{I'} c_{I I'} (K^{-1}_{I'J} \ell_{b,J})$). 
It turns out that
$\U(1)^{N/2}$ symmetry is exactly generated by $\mathbf{t}_{b,I}$ with $b=1,\dots, N/2$, and
these remained unbroken symmetry with $N/2$ of U(1) generators are {\bf anomaly-free} and {\bf mixed anomaly-free}, due to
\be
\boxed{\mathbf{t}_{a,I'} K_{I'J'} \mathbf{t}_{b,J'}={\ell_{a,I'} K^{-1}_{I'J'}   \ell_{b,J'}}=0}.
\ee
Indeed, $\mathbf{t}_a$ must be anomaly-free, because it is easily notice that by defining an
$N \times N/2$ matrix
$\mathbf{t} \equiv \Big(\mathbf{t}_{1}, \mathbf{t}_{2}, \dots,  \mathbf{t}_{N/2}  \Big) =\Big(\eta_{1}, \eta_{2}, \dots,  \eta_{N/2}  \Big)$ of
Eq.(\ref{eq:TQFThKh}), thus 
we must have:
\bea \label{eq:TQFTtKt}
\boxed{{\mathbf{t}^T} K {\mathbf{t}}=0}, \;\;\; \text{where }\mathbf{t} = \mathbf{h}. 
\eea
This is exactly the anomaly factor and the effective Hall conductance discussed in Sec.\ref{anomaly-hall}.

In summary of the above, we have provided a topological non-perturbative proof that
the {\bf Boundary Fully Gapping Rules}, 
and its extension
to the equivalence relation to the {\bf anomaly-matching conditions}. 
We emphasize that
{\bf Boundary Fully Gapping Rules} provide a topological statement on the gapped boundary conditions, which is non-perturbative,
while the {\bf anomaly-matching conditions} are also non-perturbative in the sense that the conditions hold at any energy scale,
from low energy IR to high energy UV.
Thus, the equivalence between the twos is remarkable, especially that both are \emph{non-perturbative statements} (namely the proof we provide is
as exact as integer number values without allowing any small perturbative expansion).
Our proof apply to
a bulk U(1)$^N$ K matrix Chern-Simons theory (describing bulk Abelian topological orders or Abelian SPT states)
with boundary multiplet chiral boson/fermion theories. 

\subsubsection{perturbative arguments} \label{sec:III-perturb}

Apart from the non-perturbative proof using TQFT, we can use other well-known techniques to show the boundary is gapped when the {\bf Boundary Fully Gapping Rules} are satisfied.
it is convenient to map the $K_{N\times N}$-matrix multiplet chiral boson theory to $N/2$ copies of non-chiral Luttinger liquids, each copy with an action
\bea \label{eq:non-chiral-Lutt}
&&\int dt \, dx \; \Big( \frac{1}{4\pi}( (\partial_t \bar{\phi}_{a} \partial_x \bar{\theta}_{a} +\partial_x \bar{\phi}_{a} \partial_t \bar{\theta}_{a}) -V_{IJ} \partial_x \Phi_{I}   \partial_x \Phi_{J}) 
+  g_{}  \cos(\beta \; \bar{\theta}_{a}) \Big)
\eea
at large coupling $g$ at the low energy ground state. Notice that the mapping sends 
$\Phi \to \Phi''=( \bar{\phi}_1, \bar{\phi}_2, \dots, \bar{\phi}_{N/2}, \bar{\theta}_1, \bar{\theta}_2, \dots,  \bar{\theta}_{N/2})$ in a new basis, such that
the cosine potential only takes one field $\bar{\theta}_{a}$ decoupled from the full multiplet.
However, this mapping has been shown 
to be possible \emph{if} $\mathbf{L}^T K^{-1} \mathbf{L}=0$ is satisfied.

When the mapping is done, 
we can simply study a single copy of non-chiral Luttinger liquids,
and which, by changing of variables, 
is indeed equivalent to the action of Klein-Gordon fields with a sine-Gordon cosine potential studied by S.\,Coleman. 
We have demonstrated various ways to show the existence of mass gap of this sine-Gordon action 
For example,\\
 
\noindent
$\bullet$ For non-perturbative perspectives, there is a duality between the quantum sine-Gordon action of bosons and the massive Thirring model of fermions in 1+1D.
In the sense, it is an integrable model, and the Zamolodchikov formula is known and Bethe ansatz can be applicable.
The mass gap is known unambiguously at the large $g$.

\noindent
$\bullet$ For perturbative arguments, we can use {\bf RG} to do {\bf weak or strong coupling expansions}.

For \emph{weak coupling} $g$ analysis, it is known that choosing the kinetic term as a marginal term, 
and the scaling dimension of the normal ordered $[\cos(\beta \bar{\theta})]=\frac{\beta^2}{2}$.
In the weak coupling analysis, ${\beta^2} < \beta_c^2 \equiv 4$ will flow to the large $g$ gapped phases (with an exponentially decaying correlator) at low energy,
while ${\beta^2} >\beta_c^2$ will have the low energy flow to the quasi-long-range gapless phases (with an algebraic decaying correlator) at the low energy ground state.
At $\beta=\beta_c$, it is known to have Berezinsky-Kosterlitz-Thouless(BKT) transition.
We find that our model satisfies ${\beta^2} < \beta_c^2$, 
thus necessarily flows to gapped phases, 
because the gapping terms can be written as $g_a \cos(\bar{\theta}_1) +g_b \cos(\bar{\theta}_2)$ in the new basis,
where both ${\beta^2} =1 < \beta_c^2$.
 

%
However, the weak coupling RG may not account the correct physics at large $g$. 

We also perform 
the \emph{strong coupling} $g$ RG analysis, by setting the pin-down fields at large $g$ coupling of $g \cos(\beta \bar{\theta} )$ with the quadratic fluctuations as the marginal operators.
We find the kinetic term changes to an irrelevant operator. And the two-point correlator at large $g$ coupling exponentially decays implies that our starting point is a strong-coupling fixed point of gapped phase.
Such an analysis shows \emph{$\beta$-independence}, where the gapped phase is universal at \emph{strong coupling} $g$ regardless the values of $\beta$
and robust against kinetic perturbation. It implies that there is no instanton connecting different minimum vacua of large-$g$ cosine potential for 1+1D at zero temperature 
for this particular action Eq.(\ref{eq:non-chiral-Lutt}). 

In short, from the mapping to decoupled $N/2$-copies of non-chiral Luttinger liquids with gapped spectra
together with the anomaly-matching conditions proved, 
we obtain the relations:\\
\colorbox{lgray}{\parbox[t]{\linegoal}{
\center{the U(1)$^{N/2}$ anomaly-free theory ($\mathbf{q}^T \cdot {K}^{-1} \cdot \mathbf{q} = \mathbf{t}^T \cdot  {K}^{} \cdot  \mathbf{t}=0$) with gapping terms $\mathbf{L}^T K^{-1} \mathbf{L}=0$ satisfied. \\
$\updownarrow$\\ 
the $K$ matrix multiplet-chirla boson theories with gapping terms $\mathbf{L}^T K^{-1} \mathbf{L}=0$ satisfied.\\
$\downarrow$ \\ 
 $N/2$-decoupled-copies of non-Chiral Luttinger liquid actions with gapped energy spectra.}
 }}

\subsubsection{preserved U(1)$^{N/2}$ symmetry and a unique ground state} 

We would like to discuss the symmetry of the system further. 
As we mention in Sec.\ref{sec:topo-nonp-proof}, the symmetry is broken down from 
$\U(1)^N \to \U(1)^{N/2}$ by adding $N/2$ gapping terms with $N=4$.  
In the case of gapping terms $\ell_1=(1,1,-2,2)$ and $\ell_2=(2,-2,1,1)$,
we can find the unbroken symmetry by Eq.(\ref{eq:t=KinvL}), 
where the symmetry charge vectors are $\mathbf{t}_1=(1,-1,-2,-2)$ and $\mathbf{t}_2=(2,2,1,-1)$.
The symmetry vector can have another familiar linear combination $\mathbf{t}_1=(3,5,4,0)$ and $\mathbf{t}_2=(0, 4, 5, 3)$, which indeed matches to our original 
U(1)$_{\text{1st}}$ 3-5-4-0 and U(1)$_{\text{2nd}}$ 0-4-5-3 symmetries.
Similarly, the two gapping terms can have another linear combinations: $\ell_1=(3,-5,4,0)$ and $\ell_2=(0,4,-5,3)$.
We can freely choose any linear-independent combination set of the following,
%
%
%
\bea \label{eq:Lt3540dual}
&&\mathbf{L}=\left(
\begin{array}{cc}
 3 & 0 \\
 -5 & 4 \\
 4 & -5 \\
 0 & 3
\end{array}
\right),\left(
\begin{array}{cc}
 1 & 2 \\
 1 & -2 \\
 -2 & 1 \\
 2 & 1
\end{array} 
\right) 
\Longleftrightarrow
\mathbf{t}=\left(
\begin{array}{cc}
 3 & 0 \\
 5 & 4 \\
 4 & 5 \\
 0 & 3
\end{array}
\right),\left(
\begin{array}{cc}
 1 & 2 \\
 -1 & 2 \\
 -2 & 1 \\
 -2 & -1
\end{array} 
\right), \dots. \nonumber
\eea
and we emphasize the vector space spanned by the column vectors of $\mathbf{L}$ 
and $\mathbf{t}$ (the complement space of $\mathbf{L}$'s) 
will be the entire 4-dimensional vector space $\mathbb{Z}^4$.

Now we like to answer: \\
(Q4) Whether the $\U(1)^{N/2}$ symmetry stays unbroken when the mirror sector becomes gapped by the strong interactions?\\
(A4) The answer is Yes. We can check: There are two possibilities that $\U(1)^{N/2}$ symmetry is broken. 
\noindent
One is that it is \emph{explicitly broken} by the interaction term.
This is not true. 
\noindent
The second possibility is that the ground state (of our chiral fermions with the gapped mirror sector) \emph{spontaneously or explicitly break}
the $\U(1)^{N/2}$ symmetry. This possibility can be checked by calculating its {\bf ground state degeneracy(GSD) on the cylinder with gapped boundary}.
Using the method developing in our previous work Ref.\cite{Wang:2012am}, also in Ref.\cite{{Kapustin:2013nva},{Wang:2013vna}}, we find GSD=1,
there is only a unique ground state. Because there is only one lowest energy state, it cannot \emph{spontaneously or explicitly break} the remained symmetry.
The GSD is 1 as long as the $\ell_a$ vectors are chosen to be the minimal vector, namely the greatest common divisor(gcd) among each component of any $\ell_a$ is 1,
${|\gcd(\ell_{a,1}, \ell_{a,2}, \dots,  \ell_{a,N/2} \Big) |}=1$, such
that 
$$
\ell_a \equiv \frac{(\ell_{a,1}, \ell_{a,2}, \dots,  \ell_{a,N/2}  )}{|\gcd(\ell_{a,1}, \ell_{a,2}, \dots,  \ell_{a,N/2}) |}.
$$

In addition, 
thanks to Coleman-Mermin-Wagner theorem, there is \emph{no spontaneous symmetry breaking for any continuous symmetry in 1+1D, due to no Goldstone modes 
in 1+1D}, we can safely conclude that $\U(1)^{N/2}$ symmetry stays unbroken. 
\color{black}

 
 To summarize the whole Sec.\ref{anomaly-gap-proof}, we provide both non-perturbative and perturbative analysis on {\bf Boundary Fully Gapping Rules}. 
 This applies to a generic K-matrix U(1)$^{N}$ Abelian Chern-Simons theory with a boundary multiplet chiral boson theory.
 (This generic K matrix theory describes general Abelian topological orders including all Abelian SPT states.)

In addition, in the case when K is \emph{unimodular indefinite symmetric integral matrix}, for both fermions $K=K^{f}$ and bosons $K=K^{b0}$, we have further proved:\\
\colorbox{lgray}{\parbox[t]{\linegoal}{
{\bf Theorem:} The boundary fully gapping rules of 1+1D boundary/2+1D bulk with unbroken U(1)$^{N/2}$ symmetry $\leftrightarrow$ ABJ's U(1)$^{N/2}$ anomaly matching conditions in 1+1D.}}\\
Similar to our non-perturbative algebraic result on topological gapped boundaries, 
the {\bf 't Hooft anomaly matching} here is a non-perturbative statement, being exact from IR to UV, insensitive to the energy scale.

\noindent
\subsection{General Construction of Non-Perturbative Anomaly-Free chiral matter model from SPT \label{model}}

As we already had an explicit example of 3$_L$-5$_R$-4$_L$-0$_R$ 
chiral fermion model introduced in Sec.\ref{sec3-5-4-0},\ref{numeric}, 
and we had paved the way building up tools and notions in Sec.\ref{anomaly-gap-proof}, 
now we are finally here to present our general model construction.
Our construction of non-perturbative anomaly-free chiral fermions and bosons model with onsite U(1) symmetry is the following.\\

\noindent
{\bf Step 1}: We start with a $K$ matrix Chern-Simons theory as in Eq.(\ref{CSbulk}),(\ref{CSboundary})
for \emph{unimodular indefinite symmetric integral $K$ matrices}, both fermions $K=K^{f}$ of Eq.(\ref{eq:Kf})
and bosons $K=K^{b0}$ of Eq.(\ref{eq:Kb0}) (describing generic Abelian SPT states with GSD on torus is $|\det(K)|=1$.)\\

\noindent
{\bf Step 2}: We assign charge vectors $\mathbf{t}_a$ of U(1) symmetry 
as in Eq.(\ref{eq:U(1)}), which satisfies the anomaly matching condition. 
We can assign up to $N/2$ charge vector
$\mathbf{t} \equiv \Big(\mathbf{t}_{1}, \mathbf{t}_{2}, \dots,  \mathbf{t}_{N/2} \Big)$ with a total U(1)$^{N/2}$ symmetry with the matching
$\mathcal{A}= {{\mathbf{t}^T} K {\mathbf{t}}=0}$
such that the model is anomaly and mixed-anomaly free. \\

\noindent
{\bf Step 3}: In order to be a \emph{chiral} theory, it needs to \emph{violate the parity symmetry}.  
In our model construction, assigning $q_{L,j} \neq q_{R,j}$ generally fulfills our aims by breaking both parity and time reversal symmetry.\\ 

\noindent
{\bf Step 4}: By the equivalence of the anomaly matching condition and boundary fully gapping rules(proved in Sec.\ref{sec:topo-nonp-proof}, 
a proper choice of gapping terms of Eq.(\ref{eq:Sgap}) can fully gap out the edge states. 
For $N_L=N_R=N/2$ left/right Weyl fermions, there are $N/2$ gapping terms ($\mathbf{L} \equiv \Big(\ell_{1}, \ell_{2}, \dots,  \ell_{N/2}  \Big)$), and
the U(1) symmetry can be extended to U(1)$^{N/2}$ symmetry by finding the corresponding $N/2$ charge vectors ($\mathbf{t} \equiv \Big(\mathbf{t}_{1}, \mathbf{t}_{2}, \dots,  \mathbf{t}_{N/2}  \Big)$). 
The topological non-perturbative proof found in Sec.\ref{sec:topo-nonp-proof} guarantees the duality relation:
\bea  \label{eq:TQFTLKLTKT}
&&\boxed{\mathbf{L}^T \cdot K^{-1} \cdot \mathbf{L}=0 \underset{ \mathbf{L}=K^{} \mathbf{t}  }{\overset{ \mathbf{t}=K^{-1} \mathbf{L} }{\longleftrightarrow}} \mathbf{t}^T \cdot  {K}^{} \cdot  \mathbf{t}=0}.
\eea
Given $K$ as a $N \times N$-component matrix of $K^{f}$ or $K^{b0}$, we have $\mathbf{L}$ and $\mathbf{t}$ are both $N \times (N/2)$-component matrices.

So our strategy is that constructing the bulk SPT on a 2D spatial lattice with two edges
(for example, 
a cylinder in Fig.\ref{3540},Fig.\ref{cylinder}). 
The low energy edge property of the 2D lattice model has the same continuum field theory\cite{fermionization1} as we had in Eq.(\ref{CSboundary}),
and selectively only fully gapping out states on one mirrored edge with a large mass gap by adding symmetry allowed gapping terms Eq.(\ref{eq:Sgap}),
while leaving the other side gapless edge states untouched.\cite{Wen:2013ppa}

In summary, 
we start with a chiral edge theory of SPT states with $ \cos(\ell_{I}^{} \cdot\Phi^B_{I})$ gapping terms on the edge B, which action is
\bea 
S_{\Phi} 
&=&\frac{1}{4\pi}  \int dt dx  \big(K^{\A}_{IJ}  \partial_t \Phi^{\A}_I   \partial_x \Phi^{\A}_{J} -V_{IJ}  \partial_x \Phi^{\A}_I   \partial_x \Phi^{\A}_{J}\big) \;\;\;\nonumber \\
&+&\frac{1}{4\pi}  \int dt dx \big(K^{\B}_{IJ}  \partial_t \Phi^{\B}_I   \partial_x \Phi^{\B}_{J} -V_{IJ}  \partial_x \Phi^{\B}_I   \partial_x \Phi^{\B}_{J} \big) 
+\int dt dx  \; \sum_{a} g_{a}  \cos(\ell_{a,I}^{} \cdot\Phi_{I}).  \;\;\;\;\;\;\;
 \eea
We fermionize the action to:
\bea
&S_\Psi&=\int  dt \; dx \; ( i\bar{\Psi}_{\A} \Gamma^\mu  \partial_\mu \Psi_{\A}+ i\bar{\Psi}_{\B} \Gamma^\mu  \partial_\mu \Psi_{\B}  
+U_{\text{interaction}}\big( \tilde{\psi}_{q}, \dots,  \nabla^n_x \tilde{\psi}_{q},\dots  \big) ).    
\eea
with
$\Gamma^0$, $\Gamma^1$, $\Gamma^5$ follow the notations of Eq.(\ref{CSferboundary}).

The gapping terms on the field theory side need to be irrelevant operators or marginally irrelevant operators 
with appropriate strength (to be order 1 intermediate-strength for the dimensionless lattice coupling $|G|/|t_{ij}| \gtrsim O(1)$), 
so it can gap the mirror sector, but it is weak enough to keep the original light sector gapless.

Use several copies of Chern bands to simulate the free kinetic part of Weyl fermions, and convert the higher-derivatives fermion interactions $U_{\text{interaction}}$ to the point-splitting
$U_{\text{point.split.}}$ term on the lattice, we propose its corresponding lattice Hamiltonian 
\bea 
\label{}
H&=&\sum_{q} 
\bigg(  \sum_{\langle i, j \rangle}
\big(t_{ij,q}\; \hat{f}^\dagger_{q}(i)
\hat{f}_{q}(j)+h.c.\big) 
+\sum_{\langle\langle i, j
\rangle\rangle} \big(  t'_{ij,q}
\;\hat{f}^\dagger_{q}(i) \hat{f}_{q}(j)+h.c.\big) \bigg)  \nonumber \\ %
&+& \sum_{j \in \B} U_{\text{point.split.}}
\bigg(\hat{f}_{q}(j), \dots  \big( \hat{f}^n_{q}(j)\big)_{pt.s.}, \dots
\bigg).\;\; 
\nonumber 
\eea

Our key to avoid Nielsen-Ninomiya challenge\cite{Nielsen:1980rz,Nielsen:1981xu,Nielsen:1981hk} is that our model has the \emph{properly-desgined} interactions.\\ 
\colorbox{lgray}{\parbox[t]{\linegoal}{
We have obtained a 1+1D non-perturbative lattice Hamiltonian construction (and realization) of anomaly-free massless chiral fermions (and chiral bosons) on one gapless edge.}
}\\

\subsection{Summary \label{summary}} 

We have proposed a 1+1D lattice Hamiltonian definition of non-perturbative anomaly-free chiral matter models with U(1) symmetry. 
Our 3$_L$-5$_R$-4$_L$-0$_R$ fermion model is under the framework of the mirror fermion decoupling approach. 
However, some importance essences make our model distinct from the lattice models of {Eichten-Preskill}\cite{EP8679} and {Chen-Giedt-Poppitz} 3-4-5 model.\cite{CGP1247}
The differences between our and theirs are:\\

\noindent
\underline{\bf Onsite or non-onsite symmetry}.
Our model only implements onsite symmetry, which can be easily to be gauged. 
While {Chen-Giedt-Poppitz} model implements Ginsparg-Wilson(GW) fermion approach with non-onsite symmetry 
To have 
GW relation $\{D,\gamma^5\}=2aD\gamma^5D$ to be true ($a$ is the lattice constant), the
Dirac operator is non-onsite (not strictly local) as $D(x_1,x_2) \sim e^{-|x_1-x_2|/{\xi} }$ but with a distribution range $\xi$. The axial U(1)$_A$ symmetry is modified
$$\delta \psi(y) = \sum_w \ti\, \theta_A \hat{\gamma}_5 (y,w) \psi(w),\;\;\; \delta \bar{\psi}(x) = \ti\, \theta_A \bar{\psi}(x)  {\gamma}_5$$
with the operator $\hat{\gamma}_5(x,y) \equiv \gamma_5 - 2 a \gamma_5 D(x,y)$. 
Since its axial U(1)$_A$ symmetry transformation contains $D$ and the Dirac operator $D$ is non-onsite, the GW approach necessarily implements non-onsite symmetry.
GW fermion has non-onsite symmetry in the way that it cannot be written as the tensor product structure on each site:
$U(\theta_A)_{\text{non-onsite}} \neq \otimes _j U_j(\theta_A)$, for $e^{\ti \theta_A} \in \U(1)_A$. 
The Neuberger-Dirac operator also contains such a non-onsite symmetry feature.
The non-onsite symmetry is the signature property of the boundary theory of SPT states.
The non-onsite symmetry causes GW fermion diffcult to be gauged to a chiral gauge theory,
because the gauge theory is originally defined by gauging the local (on-site) degrees of freedom.\\

\noindent
\underline{\bf Interaction terms}.
Our model has properly chosen a particular set of interactions satisfying the Eq.(\ref{eq:TQFTLKLTKT}), from the Lagrangian subgroup algebra
to define a topological gapped boundary conditions. On the other hand, {Chen-Giedt-Poppitz} model proposed different kinds of interactions - 
all Higgs terms obeying U(1)$_{\text{1st}}$ 3-5-4-0 symmetry (Eq.(2.4) of Ref.\cite{CGP1247}), 
including the 
Yukawa-Dirac term: 
\be \label{eq:yukawa-d}
\int dt dx\Big(\mathrm{g}_{30} \psi_{L,3}^\dagger \psi_{R,0} \phi_h^{-3}+
\mathrm{g}_{40} \psi_{L,4}^\dagger \psi_{R,0}  \phi_h^{-4} 
+\mathrm{g}_{35} \psi_{L,3}^\dagger \psi_{R,5}  \phi_h^{2}+\mathrm{g}_{45} \psi_{L,4}^\dagger \psi_{R,5}  \phi_h^{1}+ h.c.\Big),\;\;\;\;\;\;
\ee
with Higgs field $ \phi_h(x,t)$ carrying charge $(-1)$. There are also Yukawa-Majorana term:
\be \label{eq:yukawa-m}
\int dt dx\Big(\ti \mathrm{g}_{30}^{M} \psi_{L,3} \psi_{R,0} \phi_h^{3}+
\ti \mathrm{g}_{40}^{M} \psi_{L,4} \psi_{R,0}  \phi_h^{4} 
+\ti \mathrm{g}_{35}^{M} \psi_{L,3} \psi_{R,5}  \phi_h^{8}+ \ti \mathrm{g}_{45}^{M} \psi_{L,4} \psi_{R,5}  \phi_h^{9}+ h.c.\Big), \;\;\;\;\;\;
\ee
Notice that the Yukawa-Majorana coupling has an extra imaginary number $\ti$ in the front, and implicitly there is also a
Pauli matrix $\sigma_y$ if we write the Yukawa-Majorana term in the two-component Weyl basis.

The question is: {\bf How can we compare between interactions of ours and Ref.\cite{CGP1247}'s?}
If integrating out the Higgs field $\phi_h$, we find that:\\
$(\star 1)$ Yukawa-Dirac terms of Eq.(\ref{eq:yukawa-d}) 
\emph{cannot} generate any of our multi-fermion interactions of $\mathbf{L}$ in Eq.(\ref{eq:Lt3540dual}) for our 3$_L$-5$_R$-4$_L$-0$_R$ model.\\
$(\star 2)$ Yukawa-Majorana terms of Eq.(\ref{eq:yukawa-m}) 
\emph{cannot} generate any of our multi-fermion interactions of $\mathbf{L}$ in Eq.(\ref{eq:Lt3540dual}) for our 3$_L$-5$_R$-4$_L$-0$_R$ model.\\
$(\star 3)$ Combine Yukawa-Dirac and Yukawa-Majorana terms of Eq.(\ref{eq:yukawa-d}),(\ref{eq:yukawa-m}), one can indeed generate 
the multi-fermion interactions of $\mathbf{L}$ in Eq.(\ref{eq:Lt3540dual}); however, many more multi-fermion interactions 
outside of the Lagrangian subgroup (not being spanned by $\mathbf{L}$) are generated. Those extra unwanted multi-fermion interactions \emph{do not} obey the boundary fully gapping rules.
As we have shown in Sec.\ref{sec:III-perturb}, 
those extra unwanted interactions induced by the Yukawa term will cause the pre-formed mass gap unstable
due to the nontrivial braiding statistics between the interaction terms. {\bf This explains 
why the massless mirror sector is observed in Ref.\cite{CGP1247}.
In short, we know that  Ref.\cite{CGP1247}'s interaction terms are different from us, and know that the properly-designed interactions are crucial, and
our proposal 
will succeed the mirror-sector-decoupling even if Ref.\cite{CGP1247} fails.}\\

\noindent
\underline{\bf{ $\U(1)^{N}\to \U(1)^{N/2} \to \U(1)$}}.
We have shown that for a given $N_L=N_R=N/2$ equal-number-left-right moving mode theory, 
the $N/2$ gapping terms break the symmetry from $\U(1)^{N}\to \U(1)^{N/2}$. 
Its remained $ \U(1)^{N/2}$ symmetry is unbroken and mixed-anomaly free.
{\bf Is it possible to further add interactions to break $ \U(1)^{N/2}$ to a smaller symmetry, such as a single U(1)? }
For example, breaking the U(1)$_{\text{2nd}}$ 0-4-5-3 of 3$_L$-5$_R$-4$_L$-0$_R$ model to only a single U(1)$_{\text{\text{1st}}}$ 3-5-4-0 symmetry remained.
We argue that it is doable. %
Adding any extra explicit-symmetry-breaking term may be incompatible to the original Lagrangian subgroup and thus potentially ruins the stability of the mass gap. 
Nonetheless,
{\bf as long as we add an extra interaction term(breaking the U(1)$_{\text{2nd}}$ symmetry), which is irrelevant operator with 
a tiny coupling},
it can be weak enough not driving the system to gapless states.
Thus, our setting to obtain 3-5-4-0 symmetry is still quite different from {Chen-Giedt-Poppitz} where the {\bf universal strong couplings} are applied.



We show that GW fermion approach implements the {\it non-onsite symmetry} 
thus GW can avoid the fermion-doubling no-go theorem (limited to an {\it onsite symmetry}) to
obtain chiral fermion states. 
%
Remarkably, this also suggests that \\
\colorbox{lgray}{\parbox[t]{\linegoal}{
{The nontrivial edge states of SPT order, 
such as topological insulators 
alike,
can be obtained in its own dimension (without the need of an extra dimension to the bulk)
by implementing the {\it non-onsite symmetry} as Ginsparg-Wilson fermion approach.}}
}\\
To summarize, 
so far we have realized (see Fig.\ref{G-WandOurs}),
\begin{itemize}
\item {\bf Nielsen-Ninomiya theorem} claims that local free chiral fermions on the lattice with onsite (U(1) or chiral) 
symmetry have fermion-doubling problem in even dimensional spacetime.
\item {\bf  Gilzparg-Wilson(G-W) fermions}: quasi-local free chiral fermions on the lattice with non-onsite U(1) symmetry 
have no fermion doublers. 
G-W fermions correspond to gapless edge states of a nontrivial SPT state. 
\item {\bf Our 3-5-4-0 chiral fermion and general model constructions}: local interacting chiral fermions on the lattice with onsite U(1) symmetry\cite{U(1)sym} have no fermion-doublers.
Our model corresponds to unprotected gapless edge states of a trivial SPT state (i.e. a trivial insulator).
\end{itemize}
%

%


We should also clarify that, from SPT classification viewpoint, 
all our chiral fermion models are in the same class of 
$K^f=({\begin{smallmatrix}
1 &0 \\
0 & -1
\end{smallmatrix}} )$ with $\mathbf{t}=(1,-1)$, a {\it trivial class} in the fermionic SPT with U(1) symmetry.\cite{Lu:2012dt,Ye:2013upa} 
All our chiral boson models are in the same class of 
$K^b=({\begin{smallmatrix}
0 &1 \\
1 & 0
\end{smallmatrix}} )$ with $\mathbf{t}=(1,0)$, a {\it trivial class} in the bosonic SPT with U(1) symmetry.\cite{Lu:2012dt,Ye:2013upa} 
In short, we understand that 
{from the 2+1D bulk theory viewpoint, all our chiral matter models are {\it equivalent} to the {\it trivial class} of SPT(trivial bulk insulator)
in 
SPT classification. 
However,
the 1+1D boundary theories with different U(1) charge vectors $\mathbf{t}$ 
can be regarded as {\it different} chiral matter theories on its own 1+1D.}

\noindent
{\bf{Proof of a Special Case and some Conjectures}}

At this stage we already fulfill proposing our models,
on the other hand the outcome of our proposal becomes fruitful with deeper implications.
We prove that,  at least for 1+1D boundary/2+1D bulk SPT states with U(1) symmetry, 
There are equivalence relations between \\
(a) `` 't Hooft anomaly matching conditions satisfied'', \\
(b) ``the boundary fully gapping rules satisfied'',  \\
(c) ``the effective Hall conductance is zero,'' and\\
(d) ``a bulk trivial SPT (i.e.\,trivial insulator), 
with unprotected boundary edge states (realizing an onsite symmetry) which can be decoupled from the bulk.''



Rigorously speaking, what we actually prove in Sec.\ref{sec:topo-nonp-proof} 
is the equivalence of
{\bf Theorem:} ABJ's U(1) anomaly matching condition in 1+1D $\leftrightarrow$ the boundary fully gapping rules of 1+1D boundary/2+1D bulk with unbroken U(1) symmetry
for an equal number of left-right moving Weyl-fermion modes($N_L=N_R$,  $c_L=c_R$) of 1+1D theory.
Note that 
some modifications are needed for more generic cases:\\
(i)  For unbalanced left-right moving modes, the number chirality also implies the additional \emph{gravitational anomaly}.\\
(ii) For a bulk with \emph{topological order} (instead of pure SPT states), even if the boundary is gappable without breaking the symmetry,
there still can be nontrivial signature on the boundary, such as degenerate ground states (with gapped boundaries) or surface topological order.
This modifies the above specific Theorem to a more general Conjecture:
{\bf Conjecture:} The anomaly matching condition in $(d+1)$D $\leftrightarrow$ 
the boundary fully gapping rules of $(d+1)$D boundary/$(d+2)$D bulk with unbroken $G$ symmetry
for an equal number of left-right moving modes($N_L=N_R$) of $(d+1)$D theory, such that the system with arbitrary gapped boundaries has 
\emph{a unique non-degenerate ground state}(GSD=1),\cite{Wang:2012am} 
\emph{no surface topological order},
\emph{no symmetry/quantum number fractionalization}\cite{Wang:2014tia}
and \emph{without any nontrivial(anomalous) boundary signature}.

However, for an arbitrary given theory,
we {\it do not} know ``all kinds of anomalies,''  and thus in principle we {\it do not} know ``all anomaly matching conditions.''  
However, our work reveals some deep connection between the ``anomaly matching conditions'' and the ``boundary fully gapping rules.'' 
Alternatively, if we take the following statement as a definition instead,
{\bf Proposed Definition:}  The {\bf anomaly matching conditions} (all anomalies need to be cancelled) for symmetry $G$ $\leftrightarrow$  
the {\bf boundary fully gapping rules} without breaking symmetry $G$ and without anomalous boundary signatures under gapped boundary.
then the Theorem and the Proposed Definition together reveal that 
The only anomaly type of \emph{a theory with an equal number of left/right-hand Weyl fermion modes}
and only with a U(1) symmetry in 1+1D is ABJ's U(1) anomaly.

Arguably the most interesting future direction is to test our above conjecture for more general cases, 
such as other dimensions or other symmetry groups.
One may test the above statements via the modular invariance\cite{{Levin:2013gaa},Ryu2013orbifolds,Ryu2014orientifolds} of boundary theory.
%
It will also be profound
to address, the boundary fully gapping rules for non-Abelian symmetry,
and the anomaly matching condition for
non-ABJ anomaly\cite{Wen:2013oza,Wen:2013ppa} 
through our proposal.\\

Though being numerically challenging, it will be interesting to test our models on the lattice. 
Our {\bf local spatial-lattice Hamiltonian with a finite Hilbert space, 
onsite symmetry and short-ranged hopping/interaction terms} is exactly a 
{\bf condensed matter system we can realize in the lab}. It may be possible in the future we can simulate the lattice chiral model
in the physical instant time using the condensed matter set-up in the lab (such as in cold atoms system). 
Such a real-quantum-world simulation may be much faster than any classical computer or quantum computer.


\section{Mixed gauge-gravity anomalies: Beyond Group Cohomology and mixed gauge-gravity actions} \label{sec:mixedgaugegrav}

We have discussed the allowed action
$\mathbf{S}_0(\text{sym.twist})$ that is described by pure gauge fields $A_j$.  We
find that its allowed SPTs
coincide with group cohomology
results.  For a curved spacetime, we have more general topological
responses that contain both gauge fields for symmetry twists and gravitational
connections $\Gamma$ for spacetime geometry. 
Such mixed gauge-gravity
topological responses will attain 
SPTs beyond
group cohomology.  The possibility was recently discussed  in Ref.\cite{{K1467,K1459}}.  Here we will propose some additional new examples
for  SPTs with U(1) symmetry.

 In {\bf 4+1D}, the following SPT 
 response exists,
\bea
\mathbf{Z} _0(\text{sym.twist})
&=& \exp[{\ti \frac{k}{3} \int_{\cM^5}  F \wedge  \text{CS}_3(\Gamma) }]
 \nonumber \\
&=& \exp[{\ti \frac{k}{3} \int_{\cN^6}  F \wedge  \tp_1 }],\  k \in \Z \;\;\;\;\;\;\;\;
\eea
where $ \text{CS}_3(\Gamma) $ is the gravitations Chern-Simons 3-form and $\dd(\text{CS}_3)=\tp_1$
is the first Pontryagin class.
This 
SPT response is a Wess-Zumino-Witten form with a surface $\partial \cN^6=\cM^5$.
This renders 
an extra $\Z$-class of 4+1D
U(1) SPTs beyond group cohomology.  They have the following
physical property: If we choose the 4D space to be $S^2 \times M^2$ and
\cblue{put a U(1) monopole at the center of $S^2$}: $\int_{S^2} F=2\pi$, in the large $M^2$
limit, the effective 2+1D theory on $M^2$ space is $k$ copies of E$_8$ bosonic
quantum Hall states.  
A U(1) monopole in 4D space is a 1D loop.
{By cutting $M^2$ into two separated manifolds, each with a 1D-loop boundary,
we see U(1) monopole and anti-monopole 
as these two 1D-loops, each
loop carries $k$ copies of E$_8$ bosonic quantum Hall edge modes.
}
\cblue{Their gravitational response can be detected by thermal transport with a thermal Hall conductance,
$\kappa_{xy}=8 k\frac{\pi^2 k_B^2}{3 h}T$.}
%

%

In {\bf 3+1D}, the following 
SPT response exists
\bea
\mathbf{Z} _0(\text{sym.twist})
= \exp[ \frac{\ti}{ 2 } \int_{\cM^4}  F \wedge w_2],
\eea
where $w_j$ is the $j^\text{th}$ Stiefel-Whitney (SW) class.\footnote{To be more precise, we know that wedge product is only defined
for differential forms. They are not defined for SW classes.}
Let us 
design $\cM^4$ as a complex manifold, thus $w_{2j}=c_j$ mod 2.
The first Chern class
$c_1$ of the  tangent bundle of $\cM^4$ is also the first Chern class of the
determinant line bundle of the tangent bundle of $\cM^4$.
So if we choose the U(1) symmetry twist as 
the determinate line bundle of $\cM^4$,
we can write the above as ($F=2\pi c_1$):
$
\mathbf{Z} _0(\text{sym.twist})
= \exp[ \ti \pi \int_{\cM^4}  c_1 \wedge c_1]$.
On a 4-dimensional complex manifold, we have $\tp_1=c_1^2-2c_2$.
Since the 4-manifold  $\CP^2$ is not a spin manifold, thus $w_2\neq 0$.  From
$\int_{\CP^2} \tp_1=3$, we see that $\int_{\CP^2}  c_1 \wedge c_1 =1$ mod 2.  So
the above topological response is non-trivial, and it 
\cblue{suggests} a $\Z_2$-class
of 3+1D U(1)  SPTs beyond group cohomology.
\cred{Although this topological response is non-trivial, however,
we do not gain extra 3+1D U(1) SPTs beyond group
cohomology, since $\exp[ \frac{\ti}{ 2 } \int_{\cN^4}  F \wedge w_2]
=\exp[ \frac{\ti}{ 4 \pi } \int_{\cN^4}  F \wedge F]$ on any manifold $\cN^4$, and  
since the level of $\int F \wedge F$ of U(1)-symmetry is not quantized on any manifold \cite{Wen:2014zga}.} 
 
 In the above we propose two mixed gauge-gravity actions, and we rule out the second example where the bulk action does not correspond 
 to any nontrivial SPTs. Clearly there are many more types of mixed gauge-gravity anomalies, and there are more examples of
beyond-group-cohomology SPTs one can study by constructing mixed gauge-gravity actions, for example, those found in Ref.\cite{Wen:2014zga}.


%% file: QS_stSurgery.tex

\def \- {\!\smallsetminus\!}
\newcommand{\tL}{\mathrm{L}}
\newcommand{\SL}{\mathrm{SL}}
\chapter{Quantum Statistics and Spacetime Surgery} \label{QS_stSurgery}

\begin{figure}[!h]
\centerline{
(a)\includegraphics[scale=.8]{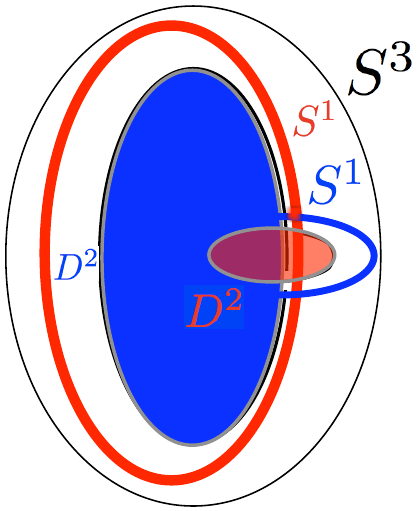} \;\;\;\;\;\; (b)\includegraphics[scale=.8]{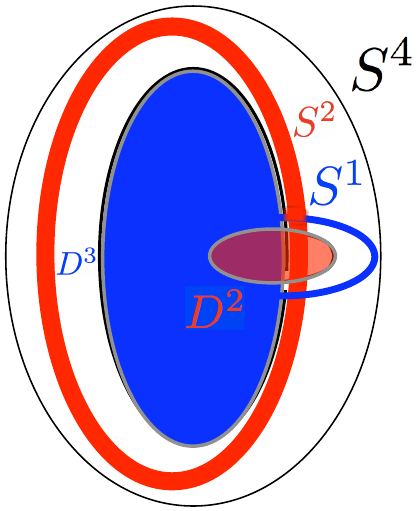} 
} \caption{An illustration of $(D^2 \times S^1) \cup_{S^1 \times S^1} (S^1 \times D^2)=S^3$ and $(D^3 \times S^1) \cup_{S^2 \times S^1} (S^2 \times D^2)=S^4$.
(a) Note that $S^1$ of $(S^1 \times D^2)$ bounds the boundary of $D^2$ within $(D^2 \times S^1)$,
and $S^1$ of $(D^2 \times S^1)$ bounds the boundary of $D^2$ within $(S^1 \times D^2)$.
The blue part illustrates $(D^2 \times S^1)$.
The red part illustrates $(S^1 \times D^2)$.
(b) Note that $S^2$ of $(S^2 \times D^2)$ bounds the boundary of $D^3$ within $(D^3 \times S^1)$,
and $S^1$ of $(D^3 \times S^1)$ bounds the boundary of $D^2$ within $(S^2 \times D^2)$.
The blue part illustrates $(D^3 \times S^1)$.
The red part illustrates $(S^2 \times D^2)$.
} \label{surgery}
\end{figure}

In this chapter, we will apply the geometric-topology techniques, to do the spacetime surgery, and see how
the nontrivial quantum statistics is constrained by the spacetime surgery configuration. An example of surgery on cut and glue is give in 
Fig.\ref{surgery}, explained in the next. 

\section{Some properties of the spacetime surgery from geometric-topology}

We will discuss the surgery on cutting and gluing of $d$-manifold $\cM^d$.
We denote that $S^d$ is a $d$-sphere, $D^d$ is a $d$-ball or called a $d$-disk, $T^d$ is a $d$-torus.
The notation for gluing the boundaries of two manifolds $\cM_1$ and $\cM_2$ is:
\be
\cM_1 \cup_{B;\varphi} \cM_2. 
\ee
It requires that the boundary of $\cM_1$ and $\cM_2$ are the same.
Namely,
\be
\partial \cM_1 =-\partial \cM_2=B.
\ee
We have a extra mapping $\varphi$ allowed by diffeomorphism when gluing two manifolds. In particular, we will focus on
a $\varphi$ of mapping class group (MCG) in our work. Thus, we can apply any element of $\varphi \in \MCG(B)$.
For $\varphi =1$ as a trivial identity map, we may simply denote $\cM_1 \cup_{B} \cM_2=\cM_1 \cup_{B,I} \cM_2.$

The connected sum of two $d$-dimensional manifolds $\cM_1$ an $\cM_2$ is denoted as
\be
\cM_1 \# \cM_2. 
\ee
Say, we cut a ball $D^d$ ($D$ for the disk $D^d$ or the same a ball $B$ for the $B^d$) out of the $\cM_1$ and $\cM_2$.
Each of the complement space defined as $\cM_1 \- D^d$ and $\cM_2\- D^d$ has a boundary of a sphere $S^{d-1}$. We glue the two manifolds $\cM_1$ and $\cM_2$
by a cylinder $S^{d-1} \times I$ where the $I$ is a 1 dimensional interval.

The list of simple (non-exotic) 3-manifolds and 4-manifolds we will focus on those without boundaries and those with boundaries.
Also another list of 2-manifolds (as the boundaries of 3-manifolds) and 3-manifolds (as the boundaries of 4-manifolds), all are summarized in
Table \ref{table:manifolds}.

\begin{center}
\begin{table}[!h]
\begin{tabular}{l}
\hline
\hline
{Manifolds}:\\
 \hline
\hline
3-manifolds without boundaries:\\
$S^3, \;\; S^2 \times S^1,\;\; (S^1)^3=T^3$, etc.\\
3-manifolds with boundaries: \\ 
$D^3, \;\; D^2 \times S^1$, etc.\\
4-manifolds without boundaries:\\ 
$S^4, \;\; S^3 \times S^1, \;\; S^2 \times S^2, \;\; S^2 \times (S^1)^2 =S^2 \times T^2,\;\; (S^1)^4=T^4$,\\ 
$S^3 \times S^1 \# S^2 \times S^2$,  $S^3 \times S^1 \# S^2 \times S^2 \# S^2 \times S^2$. \\
4-manifolds with boundaries: \\
$D^4, \;\; D^3 \times S^1,  \;\;  D^2 \times S^2,  \;\; D^2 \times (S^1)^2 = D^2 \times T^2 \equiv C^4, \;\; S^4 \smallsetminus D^2 \times T^2 $.\\
Certain 2-manifolds as the boundaries of 3-manifolds:\\
$S^2, \;\; (S^1)^2=T^2$, etc.\\
Certain 3-manifolds as the boundaries of 4-manifolds:\\
$S^3, \;\; S^2 \times S^1, \;\; (S^1)^3=T^3$, etc.\\
\hline
\hline
{Surgery}:\\ 
 \hline
\hline
Cutting and gluing 4-manifolds:\\
$S^4=(D^3 \times S^1) \cup_{S^2 \times S^1} (S^2 \times D^2)=(D^2 \times T^2) \cup_{T^3} ({S^4 \smallsetminus D^2 \times T^2})=D^4 \cup D^4$. \\
$S^3 \times S^1= (D^3 \times S^1) \cup_{S^2 \times S^1} (D^3 \times S^1) 
= (D^2 \times T^2) \cup_{T^3; \cS^{xyz}} (D^2 \times T^2)$.\\
 $S^2 \times S^2= (D^2 \times S^2) \cup_{S^2 \times S^1} (D^2 \times S^2)= ({S^4 \smallsetminus D^2 \times T^2}) \cup_{T^3; \cS^{xyz}} ({S^4 \smallsetminus D^2 \times T^2})$.\\
 $S^2 \times S^1  \times S^1= (D^2 \times T^2) \cup_{T^3} (D^2 \times T^2)$.\\
 $S^3 \times S^1  \# S^2 \times S^2 = (S^4 \smallsetminus  D^2 \times T^2) \cup_{T^3; \cS^{xyz}} (D^2 \times T^2)$.\\
  $S^3 \times S^1  \# S^2 \times S^2  \# S^2 \times S^2 = (S^4 \smallsetminus  D^2 \times T^2) \cup_{T^3} (S^4 \smallsetminus  D^2 \times T^2)$.\\
Cutting and gluing 3-manifolds:\\
%
%
$S^3=(D^2 \times S^1) \cup_{T^2;} (S^1 \times D^2)=(D^2 \times S^1) \cup_{T^2;\cS_{xy}} (D^2 \times S^1) =D^3 \cup D^3$.\\
$S^2 \times S^1 =(D^2 \times S^1) \cup_{T^2} (D^2 \times S^1)$.\\
%
%
\hline
\hline
{Mapping Class Group (MCG)}:\\
MCG($T^d$)=SL($d,\Z$), \;\;\;\;\; MCG$({S^2 \times S^1})=\Z_2 \times \Z_2$\\
 \hline
\hline
\end{tabular}
\caption{Manifolds, surgery formula and mapping class group (MCG) that are considered in 
our study.}
\label{table:manifolds}
\end{table}
\end{center}

\section{2+1D quantum statistics and 2- and 3-manifolds}

\subsection{Algebra of world-line operators, fusion, and braiding statistics in 2+1D}
\label{tunalg}

\subsubsection{World-line operators around a torus}

\begin{figure}[t]
\centerline{
\includegraphics[scale=0.5]{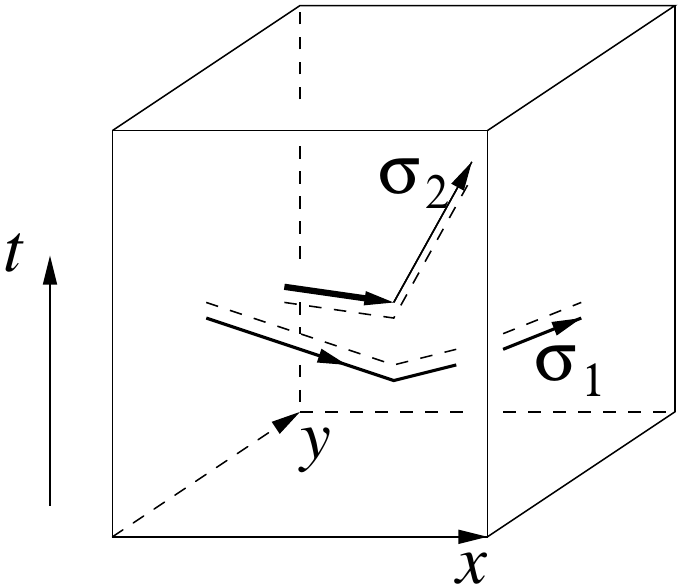}
} \caption{ The tunneling processes $W^x_{\si_1}$ and $W^y_{\si_2}$.
The dash lines are the framing of the world-line of the
 tunneling processes.
} \label{qptun}
\end{figure}

The particle-like topological excitations are created in pairs at the ends of
the corresponding world-line operator (or Wilson loop operator in gauge
theory).  A closed world-line operator is related to a tunneling process of a
topological excitation.  Let us consider the following tunneling process around
a torus: (a) we first create a quasiparticle $\si_1$ and its anti quasiparticle
$\bar\si_1$, then (b) move the quasiparticle around the torus to wrap the torus
$n_1$ times in the $x$ direction and $n_2$ times in the $y$ direction, and last
(c) we annihilate $\si_1$ and $\bar\si_1$.  The whole tunneling process (and
the corresponding closed world-line operator) induces a transformation between
the degenerate ground states on the torus $|\al\>$ labeled by $\al$:
\begin{equation*}
 |\al\>\to W_{\si_1}^{(n_1,n_2)}|\al\> = |\al'\>
(W_{\si_1}^{(n_1,n_2)})_{\al'\al}
\end{equation*}
We like to point out that in general, the quasiparticle is a size-0 point
and is not isotropic. To capture so a non-isotropic property of the
quasiparticle we added a framing to world-line that represent the
tunneling process (see Fig. \ref{qptun}).

Let $ W^x_{\si_1}=W^{(1,0)}_{\si_1}$ and $ W^y_{\si_1}=W^{(0,1)}_{\si_1}$ (see
Fig. \ref{qptun}).  A combination of two tunneling processes in the $x$
direction: $W^x_{\si_1}$ and then $W^x_{\si_2}$, induces a transformation
$W^x_{\si_2}W^x_{\si_1}$ on the degenerate ground states.  A combination of the
same two tunneling processes but with a different time order: $W^x_{\si_2}$ and
then $W^x_{\si_1}$, induces a transformation $W^x_{\si_1}W^x_{\si_2}$ on the
degenerate ground states.  We note that the two tunneling paths with different
time orders can be deformed into each other smoothly. So they only differ by
local perturbations.  Due to the topological stability of the degenerate ground
states, 
local perturbations cannot change the degenerate ground
states. Therefore $W^x_{\si_1}$ and $W^x_{\si_2}$ commute, and similarly
$W^y_{\si_1}$ and $W^y_{\si_2}$ commute too.  We see that $W^x_{\si}$'s can be
simultaneously diagonalized.  Similarly, $W^y_{\si}$'s can also be
simultaneously diagonalized.  Due to the $90^\circ$ rotation symmetry,
$W^x_{\si}$ and $W^y_{\si}$ have the same set of eigenvalues.  But since
$W^x_{\si}$ and $W^y_{\si}$ in general do not commute, we in general cannot
simultaneously diagonalize $W^x_{\si}$ and $W^y_{\si}$.

\begin{figure}[t]
\centerline{
\includegraphics[scale=0.5]{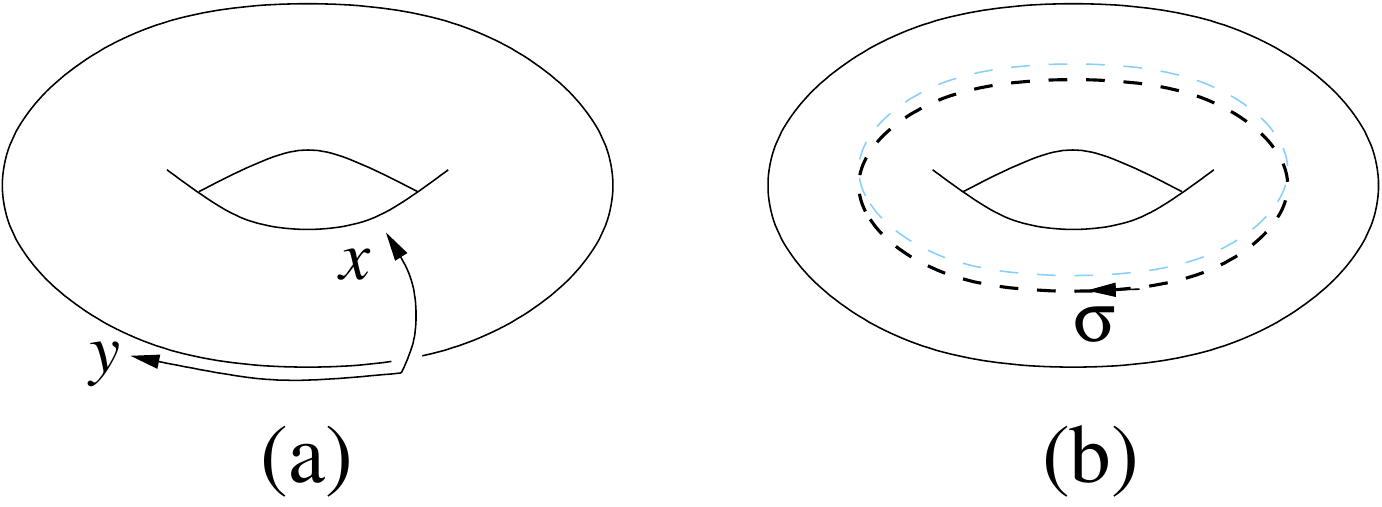}
}
\caption{
(a): The ground state $|\si=0\>$ on a torus that corresponds to the trivial
topological excitation can be represented by an empty solid torus
$S^1_x\times D^2_{yt}$.
(b): The other ground state $\Phi_{\si}$ that corresponds to
a type $\si$ quasiparticle can be represented by
an solid torus with a loop of type $\si$ at the center.
}
\label{sttT}
\end{figure}

We notice that we can view the torus $D^2_{xt} \times S^1_y$ as a surface of
$D^2_{xt} \times S^1_y$, where the disk $D^2_{xt}$ is obtained by shringking the
$S^1_x$ parametried by $x$ to zero.  The path integral on  $D^2_{xt} \times S^1_y$ 
will produce a state on $S^1_x\times S^1_y$, and we will denote such a
state as $|0_{D^2_{xt} \times S^1_y}\>$.  All other degenerate ground states can
be obtained by the action of world-line operators $W^y_{\si}$ (see Fig.
\ref{sttT}) We can define
\begin{equation}
\label{GW}
G^\al_\si= \<\al|W^y_{\si} |0_{D^2_{xt} \times S^1_y}\> .
\end{equation}
We can also define
\begin{equation}
\label{FWW}
F^\al_{\si_1\si_2}= \<\al|W^y_{\si_1} W^y_{\si_2}|0_{D^2_{xt} \times S^1_y}\> .
\end{equation}


%

\subsubsection{World-line amplitudes}

\begin{figure}[t]
\centerline{ 
(a)\includegraphics[scale=0.5]{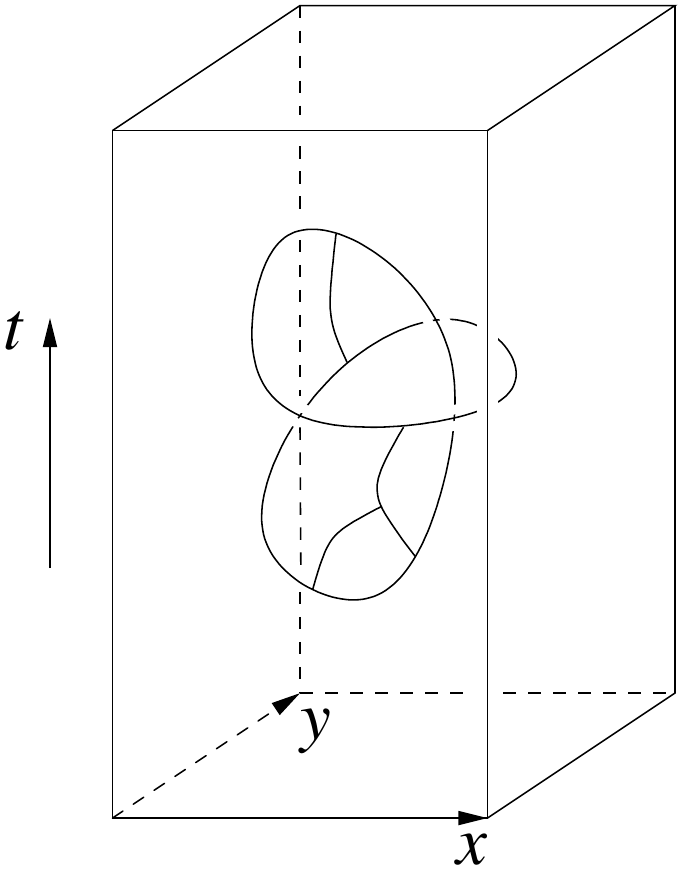} (b)\includegraphics[scale=0.5]{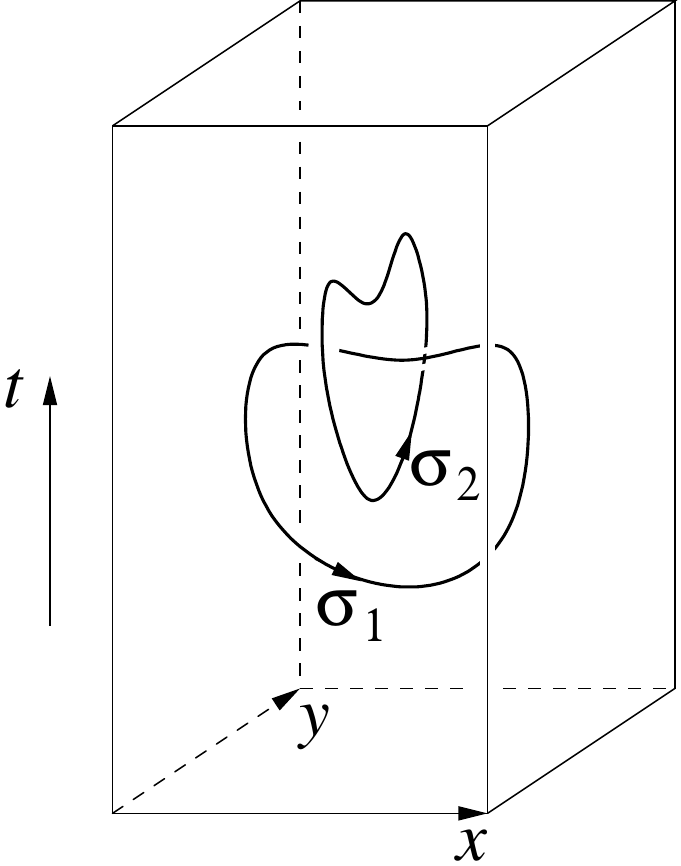}
}
\caption{
(a) A general local  tunneling process. (b)
The amplitude of two linked local loops is a complex number
$S^\text{line}_{\si_1\bar\si_2}$.
}
\label{qptung}
\end{figure}

In general, any local linked world-line operatiors correspond to local
perturbations (see Fig. \ref{qptung}), which are proportional to an identity
operator in the subspace of degenerate ground states.  So, local link
world-line operatiors is simply a complex number.  The two linked closed
world-lines also  correspond to a complex number which is defined as
$\cS^\text{line}_{\si_1\bar\si_2}$ (see Fig.  \ref{qptung}).  Clearly
$\cS^\text{line}_{\si_1\si_2}$ satisfies
\begin{equation}
\label{Sloopsymm}
 \cS^\text{line}_{\si_1\si_2} =\cS^\text{line}_{\si_2\si_1} .
\end{equation}
A single closed world-line of type-$\si$ gives rise to a complex
number as well which is defined as $d_\si$,  and we have
\begin{align}
 \cS^\text{line}_{\si 0}=
 \cS^\text{line}_{0\si }= d_\si = d_{\bar \si}.
\end{align}
For Abelian anyons $\si_1$ and $\si_2$,
$\cS^\text{line}_{\si_1\si_2}$ is closely related to the mutual statistical
angle $\th_{\si_1\si_2}$
between them
\begin{align}
 \e^{\ii \th_{\si_1\si_2}}=\frac{ \cS^\text{line}_{\si_1\si_2} }{d_{\si_1}d_{\si_2} }.
\end{align}
We may refer $\frac{ \cS^\text{line}_{\si_1\si_2} }{d_{\si_1}d_{\si_2} }$ as
the generalized mutual statistics even for non-Abelian  anyons $\si_1$ and
$\si_2$. In general $\frac{ \cS^\text{line}_{\si_1\si_2} }{d_{\si_1}d_{\si_2}
}$ may not be a $U(1)$ phase factor.

%

\subsubsection{Representations of mapping class group}

\begin{figure}[t]
\centerline{
\includegraphics[scale=0.5]{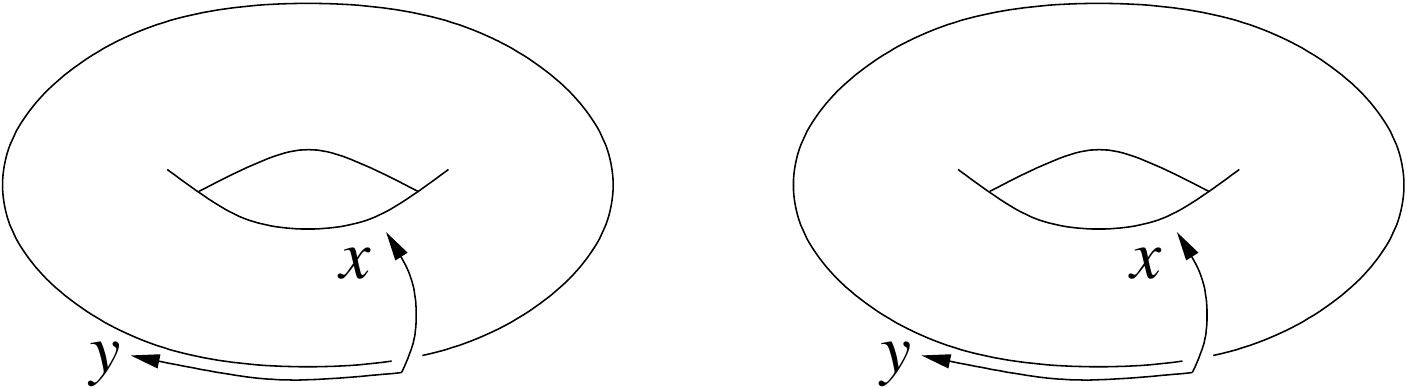}
}
\caption{
Gluing two solid tori $D^2_{xt} \times S^1_y$ without twist forms a $S^2\times
S^1$.  The gluing is done by identifing the $(x,y)$ point on the surface of the
first torus with the $(x,-y)$ point on the surface of the second torus.  If we
add an additional $\cS$ twist, \ie if we identify $(x,y)$ with $(-y,-x)$, the
gluing will produce a $S^3$.  If we add an additional $\cT$ twist, \ie if we
identify $(x,y)$ with $(x+y,-y)$, the gluing will produce a $S^2\times S^1$.
}
\label{glue3d}
\end{figure}


We know that the degenerate ground states on a torus form a representation
mapping class group (MCG) SL$(2,\Z)$ which is generated by $\cS=\bpm 0&-1\\1&0\epm,\ \cT=\bpm 1&1\\0&1\epm$:
\begin{align}
\cS^2=(\cS\cT)^3=\cC,\ \ \ \cC^2=1.
\end{align}
The matrix elements
\begin{align}
&\ \ \ 
\<0_{ D^2_{xt} \times S^1_y }|(W^y_{\si_1})^\dag \hat\cS W^y_{\si_2}|0_{D^2_{xt} \times S^1_y}\> 
= \<0_{D^2_{xt} \times S^1_y}|(W^y_{\si_1})^\dag | \alpha\> \<\alpha | \hat\cS | \beta \> \<\beta | W^y_{\si_2}|   0_{D^2_{xt} \times S^1_y}\> 
\nonumber \\
&=\sum_{\al\bt} (G^\al_{\si_1})^* \cS_{\al\bt} G^\bt_{\si_2}
\end{align}
can be computed via the path integral on two solid tori $D^2_{xt} \times S^1_y$,
one with a closed world-line $W^y_{\si_1}$ at the center and the other with a
closed world-line $W^y_{\si_2}$ at the center.  The two solid tori are glued
along the surface $S^1_x\times S^1_y$ but with a $\cS=\bpm 0&-1\\1&0\epm$ twist
$(x,y)\to (-y,x)$ (see Fig. \ref{glue3d}).  The resulting space-time is
$S^3=D^2_{xt} \times S^1_y \cup_{B; \cS} D^2_{xt} \times S^1_y$ with two linked closed
world-lines $\si_1$ and $\si_2$ (see Fig.  \ref{qptung} which describes two
linked closed world-lines). 
Here $M\cup_B N$ is the union of $M$ and $N$ alone the boundary
$B=\prt M=-\prt N$.

Combined with the above discussion, we find that
\begin{align}
\label{Sgaga}
Z \bpm \includegraphics[scale=0.25]{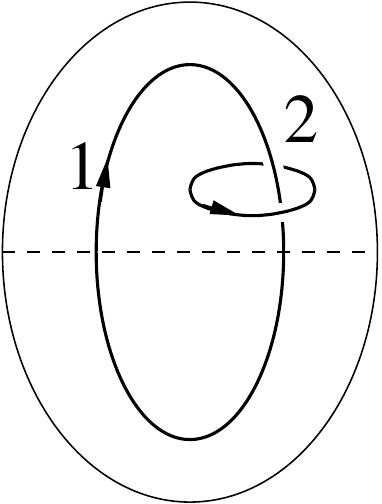}  
 \epm 
&=
\sum_{\al\bt} (G^\al_{\si_1})^* \cS_{\al\bt} G^\bt_{\si_2}
=  
\cS^\text{line}_{\bar\si_1\si_2} Z(S^3)
\nonumber\\
& = 
\cS^\text{line}_{\bar\si_1\si_2}
\sum_{\al\bt} (G^\al_{0})^* \cS_{\al\bt} G^\bt_{0} 
,  
\end{align}
where $Z(S^3)$ is the partition function on $S^3$, which happen to equal to
$\sum_{\al\bt} (G^\al_{0})^*\cS_{\al\bt} G^\bt_{0}$. We obtain an important
relation that connects the representation of MCG and the amplitudes of local
closed world-lines:
\begin{align}
\frac{
\sum_{\al\bt}  (G^\al_{\si_1})^*\cS_{\al\bt} G^\bt_{\si_2}
}{
\sum_{\al\bt} (G^\al_{0})^*\cS_{\al\bt} G^\bt_{0}
} =  \cS^\text{line}_{\bar\si_1\si_2}.
\end{align}
Here $ S^\text{line}_{\bar\si_1\si_2}$ is the amplitude of two linked
world-line loops (see Fig. \ref{qptung}).

Next, let us consider the partition
function on space-time that is formed by disconnected three manifolds
$M$ and $N$, which denoted as $M\sqcup N$.
We have
\begin{align}
Z(M\sqcup N)=Z(M)Z(N) ,
\end{align}
for fixed-point partition function of topologically ordered state.
We divide $M$ into two pieces, $M_U$ and $M_D$: $M=M_D\cup_{\prt M_D} M_U$.
We also divide $N$ into two pieces, $N_U$ and $M_D$: $N=N_D\cup_{\prt N_D} N_U$.
If the boundary of $M_D$ and $N_D$ is the same configuration as the sphere $S^2=\prt M_D=\prt N_D$ with the same defect configurations on $S^2$,
and if the ground state degeneracy of this spatial configuration is unique (namely, the dimension of Hilbert space is $1$),
then we can glue $M_D$ to $N_U$ and $M_U$ to $N_D$ to obtain
the following identity
\begin{align}
\label{ZMN}
& Z \bpm \includegraphics[scale=0.35]{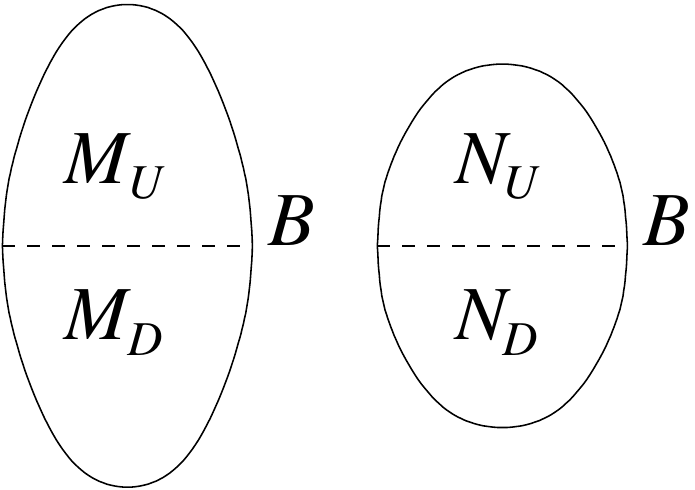} \epm 
=
 Z \bpm \includegraphics[scale=0.35]{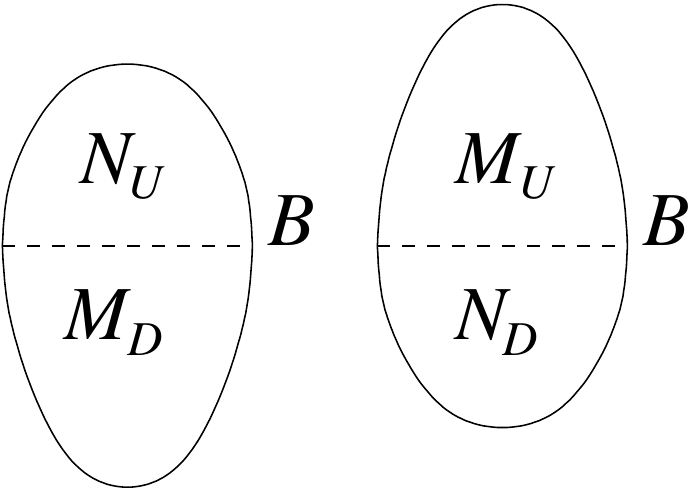} \epm 
\nonumber\\
\text{or }
&\ \ \ \
Z(M_D\cup_B M_U) 
Z(N_D\cup_B N_U) 
= 
Z(M_D\cup_B N_U) 
Z(N_D\cup_B M_U) 
\end{align}
for fixed-point partition function of topologically ordered state.  This is a
very useful relation. In particular, we can use it to show that, for
loops in $S^3$,
\begin{align} \label{Z2Dcut}
&\ \ \ \
 Z \bpm \includegraphics[scale=0.3]{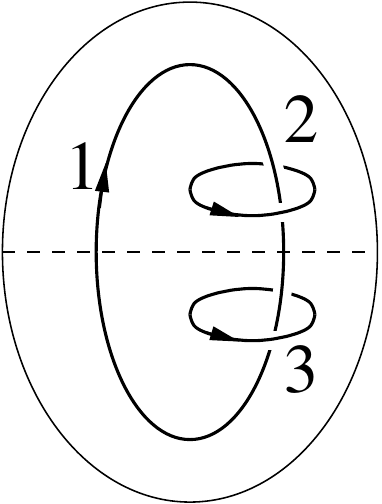} \epm 
 Z \bpm \includegraphics[scale=0.3]{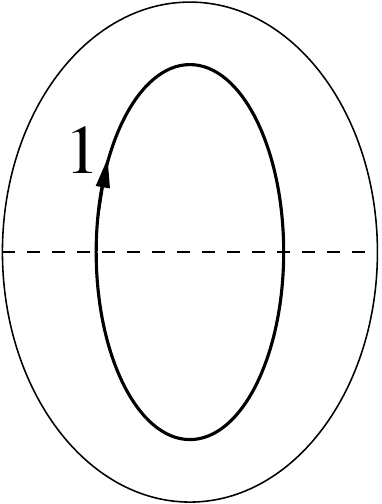} \epm 
=
 Z \bpm \includegraphics[scale=0.3]{4D_3+1D_fig/S3ll12} \epm 
 Z \bpm \includegraphics[scale=0.3]{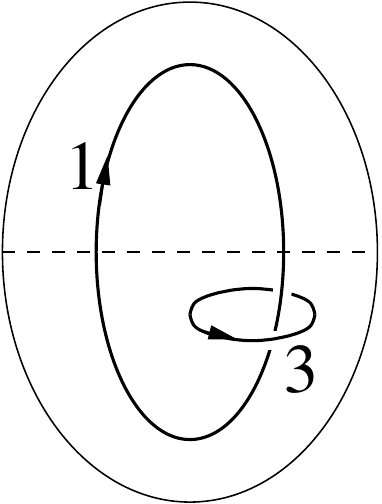} \epm 
\end{align}
Since for two linked loops in $S^3$ we have \eqn{Sgaga}, and
for three linked loops in $S^3$ we have 
\begin{align}
 Z \bpm \includegraphics[scale=0.3]{4D_3+1D_fig/S3lll} \epm 
&= 
 \<0_{D^2_{xt} \times S^1_y}|(W^y_{\si_1})^\dag \cS W^y_{\si_2}W^y_{\si_3}|0_{D^2_{xt} \times S^1_y}\> 
=
\sum_{\al\bt} (G^\al_{\si_1})^*\cS_{\al\bt} F^\bt_{\si_2\si_3}
\end{align}
we find that
\begin{align}
&\ \ \ \
\sum_{\al\bt}(G^\al_{\si_1})^*\cS_{\al\bt} G^\bt_{0}
\sum_{\al\bt}(G^\al_{\si_1})^*\cS_{\al\bt} F^\bt_{\si_2\si_3}
=
\sum_{\al\bt}(G^\al_{\si_1})^*\cS_{\al\bt} G^\bt_{\si_2}
\sum_{\al\bt}(G^\al_{\si_1})^*\cS_{\al\bt} G^\bt_{\si_3}
\end{align}
or
\begin{align}
\cS^\text{line}_{\bar\si_10}\sum_{\si_4} \cS^\text{line}_{\bar\si_1\si_4}  \t N^{\si_4}_{\si_2\si_3} 
=
\cS^\text{line}_{\bar\si_1\si_2}
\cS^\text{line}_{\bar\si_1\si_3} ,
\end{align}
further, by the symmetric $\cS^\text{line}_{\si_4 \bar\si_1}=\cS^\text{line}_{ \bar\si_1 \si_4}$
\bea
&&\sum_{\bar\si_1} \sum_{\si_4} \t N^{\si_4}_{\si_2\si_3} \cS^\text{line}_{\si_4 \bar\si_1} (\cS^\text{line -1})_{\bar\si_1\si_a}
=
\sum_{\bar\si_1} \frac{\cS^\text{line}_{\bar\si_1\si_2}
\cS^\text{line}_{\bar\si_1\si_3}  (\cS^\text{line -1})_{\bar\si_1\si_a}}
{\cS^\text{line}_{\bar\si_10}}, \nonumber \\
&&
\t N^{\si_a}_{\si_2\si_3} 
=
\sum_{\bar\si_1} \frac{\cS^\text{line}_{\bar\si_1\si_2}
\cS^\text{line}_{\bar\si_1\si_3}  (\cS^\text{line -1})_{\bar\si_1\si_a}}
{\cS^\text{line}_{\bar\si_10}},
\eea
where
\begin{align}
\label{tN}
 \t N^{\si_4}_{\si_2\si_3} = \sum_\bt (G^{-1})^{\si_4}_\bt F^\bt_{\si_2\si_3} .
\end{align}
The above is the Verlinde formula. Compare to the usual:
\emph{Verlinde formula} 
    \begin{align}
      \cN_{ab}^c=\sum_{m} \frac{\cS_{am} \cS_{bm}\ov{\cS_{cm}}}{\cS_{1m}}\in
      \mathbb{N}.
      \label{verlinde}
    \end{align}

\subsubsection{Canonical world-line operators and canonical basis}

In general the norm of the state $W^y_{\si} |0_{D^2_{xt} \times S^1_y}\>$ is proportional to the length
of the world-line operator.  We choose a normalization of the local operators
that form the world-line properly, so that
\begin{align}
&\ \ \ 
\<0_{D^2_{xt} \times S^1_y}|(W^y_{\si})^\dag W^y_{\si} |0_{D^2_{xt} \times S^1_y}\>
\nonumber\\
&=
\<0_{D^2_{xt} \times S^1_y}|W^y_{\bar \si}W^y_{\si} |0_{D^2_{xt} \times S^1_y}\>
\end{align}
is independent of the length and the shape of the world-line operator
$W^y_{\si}$.  Here $\bar \si$ is the anti-particle of $\si$ and we have assumed
that $W^y_{\bar \si}=(W^y_{\si})^\dag$.  For such proper normalized world-line
operators, we conjecture that
\begin{align}
 \<0_{D^2_{xt} \times S^1_y}|W^y_{\bar \si}W^y_{\si} |0_{D^2_{xt} \times S^1_y}\>=1,
\end{align}
assuming $\<0_{D^2_{xt} \times S^1_y}|0_{D^2_{xt} \times S^1_y}\>=1$.  In this case, $G^\al_\si=\del_{\al\si}$, and
$F^\al_{\si_1\si_2}=\t N^\al_{\si_1\si_2}=N^\al_{\si_1\si_2}$, where
$N^{\si_3}_{\si_1\si_2}$ are integers which describe the fusion of two
topological excitations via the following fusion algebra
\begin{align}
\label{FApar}
 \si_1\si_2=N_{\si_2\si_1}^{\si_3} \si_3.
\end{align}

\begin{figure}[t]
\centerline{
\includegraphics[scale=0.42]{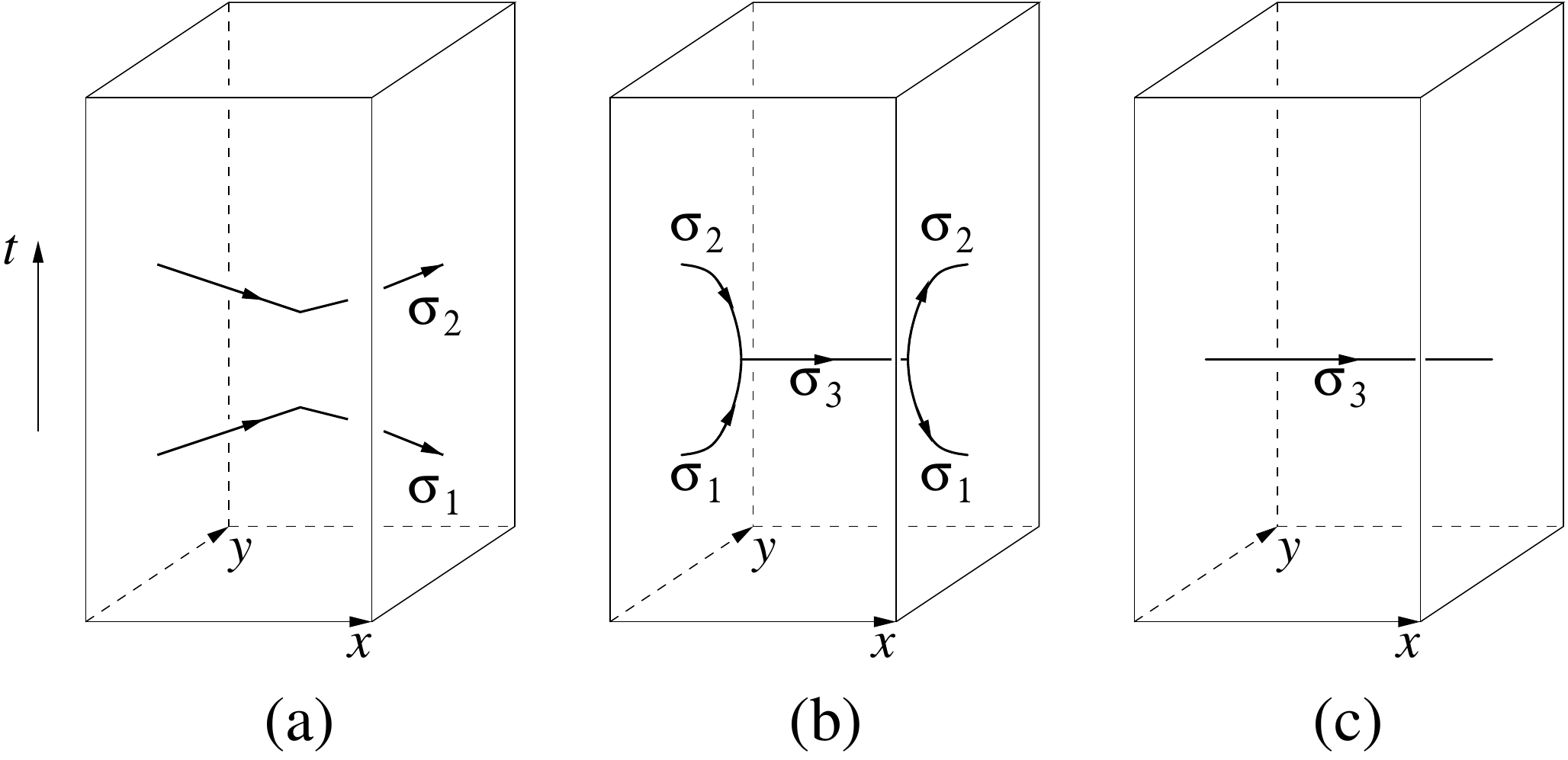}
} \caption{ (a): Two tunneling processes: $W^x_{\si_1}$ and $W^x_{\si_2}$. (b):
The tunneling path of the above two tunneling processes can be deformed using
the fusion of $\si_1 \si_2 \to \si_3$.  (c): The two tunneling processes,
$W^x_{\si_1}$ and $W^x_{\si_2}$, can be represented by a single tunneling
processes, $W^x_{\si_3}$.
}
\label{falgT}
\end{figure}

To see the relation $F^{\si_3}_{\si_1\si_2}=\t
N^{\si_3}_{\si_1\si_2}=N^{\si_3}_{\si_1\si_2}$, we note that
\begin{align}
F^{\si_3}_{\si_1\si_2}= \<0_{D^2_{xt} \times S^1_y}|W^y_{\bar \si_3} W^y_{\si_1} W^y_{\si_2}|0_{D^2_{xt} \times S^1_y}\> .
\end{align}
The above can be viewed as a path integral on $S^2\times S^1$ with three
world-lines $\si_1,\si_2,\bar\si_3$ wrapping around $S^1$.  Now we view $S^1$
as the time direction. Then the  path integral on $S^2\times S^1$ gives rise to
the ground state degeneracy with topological excitations
$\si_1,\si_2,\bar\si_3$ on the $S^2$ space. The  ground state degeneracy is
nothing but $N_{\si_2\si_1}^{\si_3}$ which is always an non-negative integer.
Therefore $F^{\si_3}_{\si_1\si_2}=\t
N^{\si_3}_{\si_1\si_2}=N^{\si_3}_{\si_1\si_2}$.
The fusion  algebra $\si_1\si_2=N_{\si_2\si_1}^{\si_3} \si_3 $ 
gives rise to an operator product algebra for the normalizaed closed
world-line operators (see Fig. \ref{falgT}):
\begin{align}
W_{\si_1} W_{\si_2} =\sum_{\si_3} N^{\si_3}_{\si_1\si_2} W_{\si_3}.
\end{align}

\begin{figure}[t]
\centerline{
\includegraphics[scale=0.5]{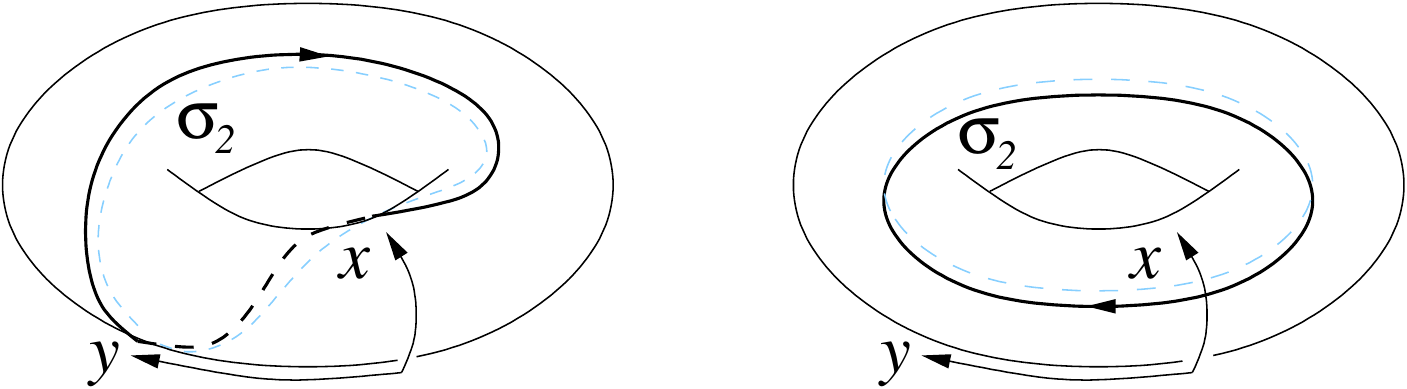}
}
\caption{
Gluing two solid tori ${D^2_{xt} \times S^1_y}$ with an additional $\cT$ twist,
\ie identifying $(x,y)$ with $(x+y,-y)$, will produce a $S^2\times S^1$.  The
loop of $\si_2$ in $y$-direction in the second solid torus at right can be
deformed into a loop of  $\si_2$  in the first solid torus at left. We see that
the loop is twisted by $2\pi$ in the anti-clockwise direction.
}
\label{TTrns}
\end{figure}

\subsection{Relations involving $\cT$}

The matrix elements
\begin{align}
&\ \ \ 
\<0_{D^2_{xt} \times S^1_y}|(W^y_{\si_2})^\dag \hat\cT W^y_{\si_1}|0_{D^2_{xt} \times S^1_y}\> 
\nonumber\\
&= \<0_{D^2_{xt} \times S^1_y}|(W^y_{\si_2})^\dag | \alpha\> \<\alpha | \hat\cT | \beta \> \<\beta | W^y_{\si_1}|   0_{D^2_{xt} \times S^1_y}\> 
\nonumber \\
&=\sum_{\al\bt} (G^\al_{\si_1})^* \cT_{\al\bt} G^\bt_{\si_2}
\end{align}
can be computed via the path integral on two solid tori ${D^2_{xt} \times S^1_y}$, one
with a closed world-line $W^y_{\si_2}$ at the center and the other with a
closed world-line $W^y_{\si_1}$ at the center.  The two solid tori are glued
along the surface $S^1_x\times S^1_y$ but with a $\cT=\bpm 1&1\\0&1\epm$ twist
$(x,y)\to (x+y,y)$ (see Fig.  \ref{glue3d}.  The resulting space-time is 
$S^1\times S^2={D^2_{xt} \times S^1_y} \cup_{S^1\times S^1,\cT}
{D^2_{xt} \times S^1_y}$ with two un-linked closed world-lines $\si_1$ and $\si_2$.
But the  world-line $\si_2$ is twisted by $2\pi$ (see Fig. \ref{TTrns}).
The amplitude of such twisted  world-line
is given by the amplitude of  un-twisted  world-line plus an additional phase
$\e^{\ii \th_{\si_2}}$, where 
$\th_\si/2\pi$ is the spin of the $\si$ topological excitation.
Note that $\th_\si$  is also the statistical angle for $\si$.
This suggests that
\begin{align}
\sum_{\al\bt} (G^\al_{\si_1})^* \cT_{\al\bt} G^\bt_{\si_2}= 
\del_{\si_1\si_2}\e^{\ii \th_{\si_2}}.
\end{align}
In the canonical basis, the above becomes a well known result
\begin{align}
\cT_{\si_1\si_2} = 
\del_{\si_1\si_2}\e^{\ii \th_{\si_2}}.
\end{align}


\section{3+1D quantum statistics and 3- and 4-manifolds}

{ Importantly the \emph{dashed}-cut in Eq.(\ref{ZMN}) indicates a $S^d$ sphere. 
For example, the \emph{dashed}-cut in Eq.(\ref{Z2Dcut}) indicates a $S^2$ sphere, where the lower bounded volume is a $D^3$ ball
and the upper bounded volume is a $S^3 \- D^3=D^3$ ball.
This surgery formula works 
if the cut bounded volume is a $D^3$ ball with a $S^2$ sphere boundary.}

Here we will extend the previous approach to higher dimensions, in 3+1D.

\subsection{3+1D formula by gluing $(D^3 \times S^1) \cup_{S^2 \times S^1} (S^2 \times D^2)=S^4$}

\subsubsection{$S^2$ world-sheet linked by two $S^1$ world-lines in $S^4$}

Here we give two new 3+1D formulas by the gluing procedures
\begin{eqnarray}
(D^3 \times S^1) \cup_{S^2 \times S^1} (S^2 \times D^2)&=& (D^3_{\theta\phi w} \times S^1_{\varphi}) \cup_{S^2_{\theta\phi} \times S^1_{\varphi}} (S^2_{\theta\phi}  \times D^2_{\varphi w}) 
=S^4
\end{eqnarray}
with the identity map between their ${S^2 \times S^1}$ boundaries.
Below we will sure that we can have a linking between a $S^2$ surface acted by a world-sheet operator
and a $S^1$ circle acted by a world-line operator in $S^4$. So we have:
\begin{align}
\text{$S^1$ and $S^2$ linked in $S^4$.}
\end{align}

We can define the $| 0 \rangle$ state on $D^3 \times S^1$ as $| 0_{D^3 \times S^1} \rangle$.
Since the $S^2=S^2_{\theta\phi}$ sphere (say with ${\theta,\phi}$ coordinates) 
as the boundary of the $D^3$ ball is contractible, we cannot have any non-trivial
non-contractible surface operators $V^{{\theta\phi}}_\mu$ acting on the state $| 0_{D^3 \times S^1} \rangle$.
However, we can create new states (perhaps incomplete basis) by generating world-lines on
the non-contractible $S^1$, namely we apply $W^\varphi_\sigma$ acting on
$| 0_{D^3 \times S^1} \rangle$:
$$
W^\varphi_\sigma | 0_{D^3 \times S^1} \rangle
$$
and
\begin{eqnarray}
&&\langle \alpha |W^\varphi_\sigma | 0_{D^3 \times S^1} \rangle=G^\alpha_{\sigma;D^3 \times S^1}, \\
&&\langle \alpha |W^\varphi_{\sigma_1} W^\varphi_{\sigma_2} | 0_{D^3 \times S^1} \rangle=F^\alpha_{\sigma_1 \sigma_2;D^3 \times S^1}.
\end{eqnarray}

On the other hand, we can define the $| 0 \rangle$ state on $S^2 \times D^2$ as $| 0_{S^2 \times D^2} \rangle$.
Since the $S^1=S^1_{\varphi}$ sphere (say with a ${\varphi}$ coordinate) 
as the boundary of the $D^2$ ball is contractible, we cannot have any non-trivial
 loop operators $W^\varphi_\sigma$ acting on the state $| 0_{S^2 \times D^2} \rangle$.
However, we can create new states (perhaps incomplete basis) by generating world-sheet on
the non-contractible $S^2$, namely we apply $V^{{\theta\phi}}_\mu$ acting on
$| 0_{S^2 \times D^2} \rangle$:
$$
V^{{\theta\phi}}_\mu | 0_{S^2 \times D^2} \rangle
$$
and
\begin{eqnarray}
&&\langle \alpha |V^{{\theta\phi}}_\mu | 0_{S^2 \times D^2} \rangle=G^\alpha_{\mu;S^2 \times D^2}, \\
&&\langle \alpha |V^{{\theta\phi}}_{\mu_1} V^{{\theta\phi}}_{\mu_2}  | 0_{S^2 \times D^2} \rangle=F^\alpha_{\mu_1 \mu_2;S^2 \times D^2}.
\end{eqnarray}

We can design the $S^2$ world-sheet operator (from the $V^{{\theta\phi}}_\mu | 0_{S^2 \times D^2} \rangle$) bounds the 
$D^3$ ball of the $D^3 \times S^1$ spacetime, meanwhile we can also design 
the $S^1$ world-line operator (from the $W^\varphi_\sigma | 0_{D^3 \times S^1} \rangle$) bounds the 
$D^2$ ball of the ${S^2 \times D^2}$ spacetime. By doing so, we have:
\begin{align}
\text{The $S^1$ world-line and the $S^2$ world-sheet linked in $S^4$.}
\end{align}
Since we glue the two copies boundaries of $S^2 \times S^1$ via the 
map $\mathbb{I}$, we have the glued partition function:
\begin{align}
\label{Iglue}
Z \bpm \includegraphics[scale=0.45]{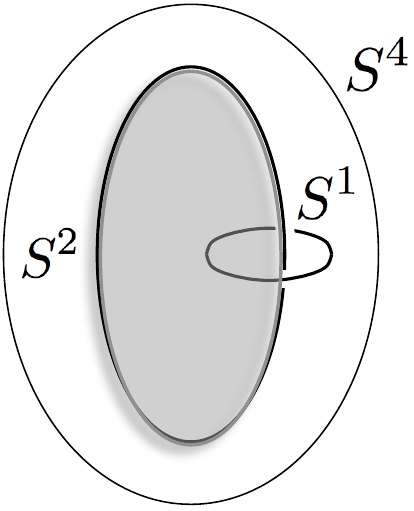} \epm 
&=
\sum_{\al\bt} (G^\alpha_{\mu_1;S^2 \times D^2})^* \mathbb{I}_{\al\bt} G^\bt_{\sigma_2;D^3 \times S^1} \nonumber\\
&=  
\mathbb{I}^\text{Link($S^2$,$S^1$)}_{\mu_1\sigma_2} Z(S^4)  
 = 
\mathbb{I}^\text{Link($S^2$,$S^1$)}_{\mu_1\sigma_2} 
\sum_{\al\bt} (G^\alpha_{0;S^2 \times D^2})^* \mathbb{I}_{\al\bt} G^\bt_{0;D^3 \times S^1} 
,  
\end{align}

\begin{align}
\mathbb{I}^\text{Link($S^2$,$S^1$)}_{\mu_1 \sigma_2} 
&=\frac{\sum_{\al\bt} (G^\alpha_{\mu_1;S^2 \times D^2})^* \mathbb{I}_{\al\bt} G^\bt_{\sigma_2;D^3 \times S^1} }{\sum_{\al\bt} (G^\alpha_{0;S^2 \times D^2})^* \mathbb{I}_{\al\bt} G^\bt_{0;D^3 \times S^1} }
\end{align}
so that $\mathbb{I}^\text{Link($S^2$,$S^1$)}_{00} =1$.

Now we use the gluing to show that, for
the linking of $S^2$ and $S^1$ in $S^4$,
with the gray region below indicates a 2-sphere $S^2$.
\begin{align} \label{eq:S2S1S1inS4}
&\ \ \ \
 Z \bpm \includegraphics[scale=0.45]{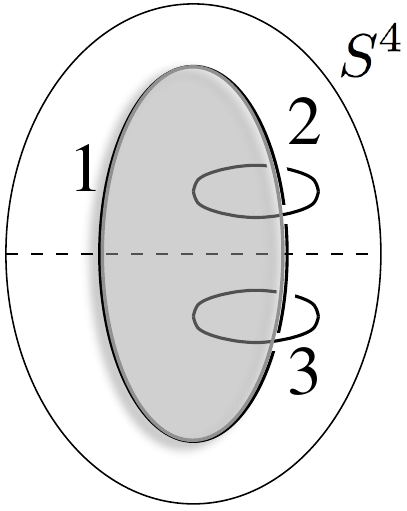} \epm 
 Z \bpm \includegraphics[scale=0.45]{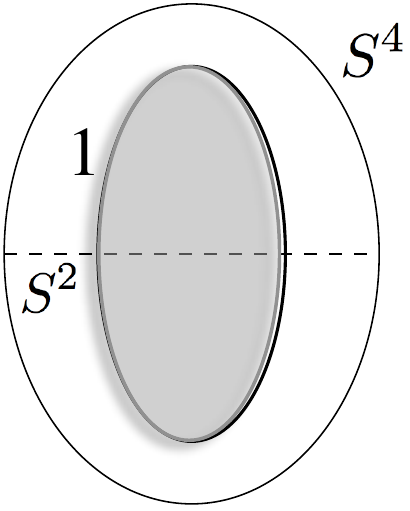} \epm 
=
 Z \bpm \includegraphics[scale=0.45]{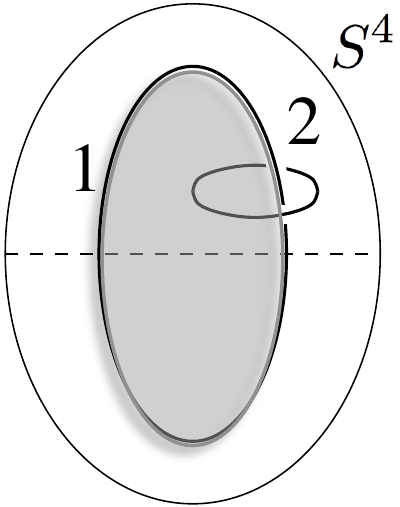} \epm 
 Z \bpm \includegraphics[scale=0.45]{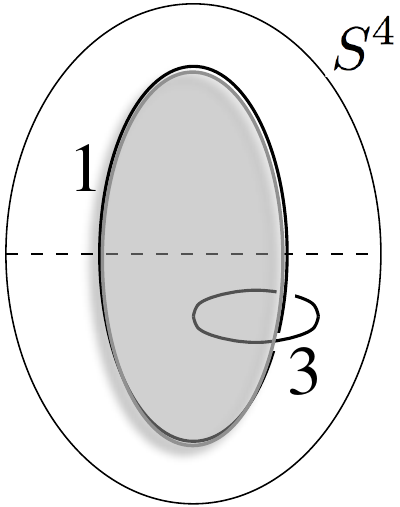} \epm. 
\end{align}
{\bf Importantly the \emph{dashed}-cut indicates a $S^3$ sphere, where the lower bounded volume is a $D^4$ ball
and the upper bounded volume is $S^4 \- D^4= D^4$.
This surgery formula works if the cut bounded volume is a $D^4$ ball with a $S^3$ sphere boundary.}
\color{black}

For the linked configuration in $S^4$ we have 
\begin{align}
 Z \bpm \includegraphics[scale=0.4]{4D_3+1D_fig/Link_S2_S1_S1_in_S4} \epm 
&= 
 \<0_{S^2\times D^2}|(V^{{\theta\phi}}_{\mu_1})^\dag \mathbb{I} W^\varphi_{\sigma_2} W^\varphi_{\sigma_3} | 0_{D^3 \times S^1} \rangle 
=\sum_{\al} (G^\al_{\mu_1;{S^2 \times D^2}})^* \mathbb{I}_{\al\bt} F^\bt_{\si_2 \si_3;{D^3 \times S^1}}.
\end{align}
We find that Eq.(\ref{eq:S2S1S1inS4}) becomes:
%
\begin{align} \label{eq:VF_S2linkS1S1}
{
\mathbb{I}^\text{Link($S^2$,$S^1$)}_{\mu_1 0}
\sum_{\si_4}  (F^{S^1})^{\si_4}_{\si_2\si_3}\mathbb{I}^\text{Link($S^2$,$S^1$)}_{\mu_1\si_4}
=
\mathbb{I}^\text{Link($S^2$,$S^1$)}_{\mu_1\si_2}
\mathbb{I}^\text{Link($S^2$,$S^1$)}_{\mu_1\si_3}},
\end{align}
(here we {\bf cannot} use $\mathbb{I}^\text{Link($S^2$,$S^1$)}_{\mu_1 0}=\mathbb{I}^\text{Link($S^2$,$S^1$)}_{0 \si_1}$.)
where
\begin{align}
\label{tN}
(F^{S^1})^{\si_4}_{\si_2\si_3} \equiv \t N^{\si_4; S^2}_{\si_2\si_3;S^1} \equiv  \sum_\al (G^{-1})^{\si_4}_{\al;D^3 \times S^1}  F^\al_{\si_2 \si_3;D^3 \times S^1}.
\end{align}
Here 
$(F^{S^1})^{\si_4}_{\si_2\si_3}$ means fusing two $S^1$ respectively carrying $\si_2$ and $\si_3$ indices
with an outcome $S^1$ carrying $\si_4$,  
in the spacetime path integral linked by another $S^2$.
The corresponding braiding process for ``two 1D world-lines $S^1$ linked by a 2D world-sheet $S^2$ in a 4D spacetime $S^4$'' 
is that the two 0D particles braid around a 1D string in the spatial slice.
Importantly in the spatial picture the two 0D particles are not threaded by 1D string, so
we remark that {\bf the fusion algebra 
$(F^{S^1})^{\si_4}_{\si_2\si_3}$ 
of fusing two particles $\si_2,\si_3$ does not have a 
based string-loop $\mu_1$ dependence}.
The above is one kind of 3+1D analogy of Verlinde formula.

\subsubsection{$S^2$ world-sheet linked by two $S^1$ world-lines in $S^4$}

We have derived Eq.(\ref{eq:VF_S2linkS1S1}) which has a {$S^1$ world-line linked by two $S^2$ world-sheets in $S^4$}.
Now we can reverse the role of $S^2$ and $S^1$, so that
a $S^2$ world-sheet will be linked by two $S^1$ world-lines in $S^4$.
Notice that
\begin{align}
\label{Iglue}
&Z \bpm \includegraphics[scale=0.45]{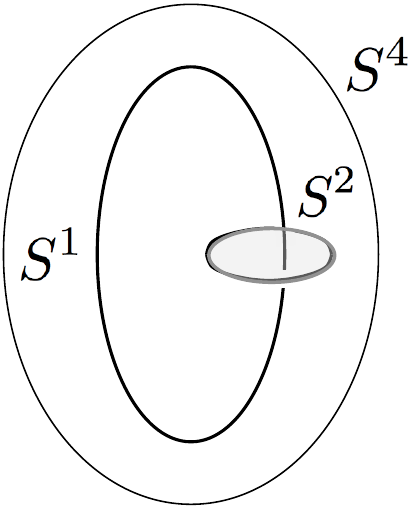} \epm 
=Z \bpm \includegraphics[scale=0.45]{4D_3+1D_fig/Link_S2_S1_in_S4.pdf} \epm \nonumber \\
&=  
\mathbb{I}^\text{Link($S^1$,$S^2$)}_{\sigma_1\mu_2} Z(S^4)  
=
\mathbb{I}^\text{Link($S^2$,$S^1$)}_{\mu_2 \sigma_1} Z(S^4)  
\nonumber\\
& = 
\mathbb{I}^\text{Link($S^1$,$S^2$)}_{\sigma_1\mu_2} 
\sum_{\al} (G^\alpha_{0;D^3 \times S^1})^* G^\al_{0;S^2 \times D^2} 
= 
\mathbb{I}^\text{Link($S^2$,$S^1$)}_{\mu_2\sigma_1} 
\sum_{\al} (G^\alpha_{0;S^2 \times D^2})^* G^\al_{0;D^3 \times S^1},  
\end{align}
We have the orientation dependence (time-orientation and framing) of the braiding process
%
We again use the gluing to obtain that, for
the linking of $S^2$ and $S^1$ in $S^4$,
with the gray region below indicates a 2-sphere $S^2$,
\begin{align} \label{eq:S1S2S2inS4}
&\ \ \ \
 Z \bpm \includegraphics[scale=0.45]{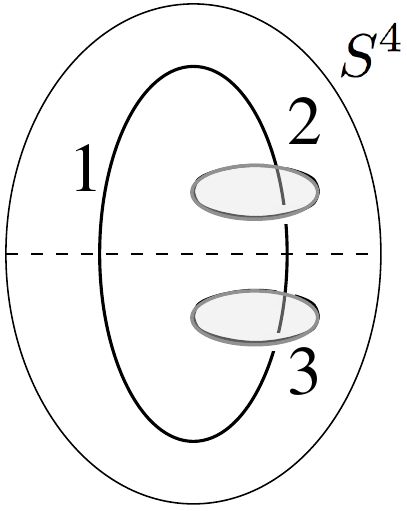} \epm 
 Z \bpm \includegraphics[scale=0.45]{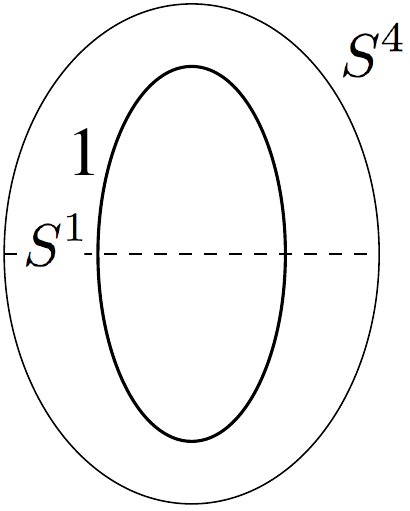} \epm 
=
 Z \bpm \includegraphics[scale=0.45]{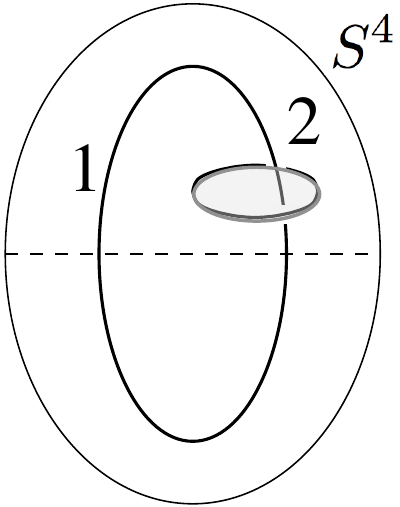} \epm 
 Z \bpm \includegraphics[scale=0.45]{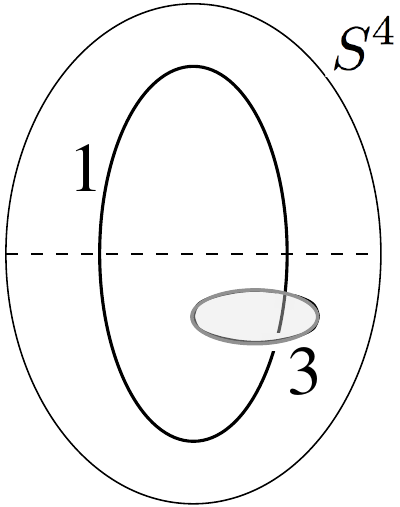} \epm. 
\end{align}
{Importantly the \emph{dashed}-cut indicates a $S^3$ sphere, where the lower bounded volume is a $D^4$ ball
and the upper bounded volume is $S^4 \- D^4$.
This surgery formula works if the cut bounded volume is a $D^4$ ball with a $S^3$ sphere boundary.}
\color{black}

We find that Eq.(\ref{eq:S1S2S2inS4}) implies
\begin{align} \label{eq:VF_S1linkS2S2}
\boxed{
\mathbb{I}^\text{Link($S^1$,$S^2$)}_{\si_10}
\sum_{\mu_4}   (F^{S^2})^{\mu_4}_{\mu_2\mu_3}\mathbb{I}^\text{Link($S^1$,$S^2$)}_{\si_1\mu_4}
=
\mathbb{I}^\text{Link($S^1$,$S^2$)}_{\si_1\mu_2}
\mathbb{I}^\text{Link($S^1$,$S^2$)}_{\si_1\mu_3}},
\end{align}
(here we {\bf cannot} use $\mathbb{I}^\text{Link($S^1$,$S^2$)}_{\si_1 0}=\mathbb{I}^\text{Link($S^1$,$S^2$)}_{0 \mu_1}$.)
where
\begin{align}
\label{tM}
 (F^{S^2})^{\mu_4}_{\mu_2\mu_3}  \equiv \t M^{\mu_4; S^1}_{\mu_2\mu_3;S^2} \equiv \sum_\al (G^{-1})^{\mu_4}_{\al;S^2 \times D^2}  F^\al_{\mu_2\mu_3;S^2 \times D^2}.
\end{align}
Here $(F^{S^2})^{\mu_4}_{\mu_2\mu_3}$ 
means fusing two $S^2$ respectively carrying $\mu_2$ and $\mu_3$ indices
with an outcome $S^2$ carrying $\mu_4$,  
in the spacetime path integral linked by another $S^1$.
The corresponding braiding process for ``two 2D world-sheets $S^2$ linked by a 1D world-line $S^1$ in a 4D spacetime $S^4$'' 
is that the two 1D strings braid around a 0D particle in the spatial slice.
Importantly in the spatial picture the two 1D strings cannot be threaded by a 0D particle, so
we remark that {\bf the fusion algebra $ (F^{S^2})^{\mu_4}_{\mu_2\mu_3} $ 
of fusing two strings $\mu_2,\mu_3$ does not have a 
$\si_1$ base dependence}.
It is possible that the $S^2$ world-sheet in the spacetime picture implies that the string-loop must be shrinkable to a point.
Therefore, the string-loop is \emph{neutral}; in the specialized case such as a gauge theory, it cannot carry \emph{gauge charge}.
The above is another kind of 3+1D analogy of Verlinde formula.

\subsection{3+1D formulas involving $\cS^{xyz}$ and gluing $D^2 \times T^2$ with $S^4 \- D^2 \times T^2$}
In 3+1D, there is 
a three-loop braiding process \cite{Wang:2014xba, Jiang:2014ksa, MW14, Wang:2014oya, JMR1462}  
we discussed before. And we can view the braiding process in the spacetime as the three $T^2$-worldsheets
 \emph{triple-linking} in the $S^4$-spacetime (shown in the first figure of Eq.(\ref{T2T2T2glue})).
We can express that
\bea \label{T2T2T2glue}
&&
{\tL^{\text{Tri}}_{{\mu_3}, {\mu_2}, {\mu_1}}} 
\equiv Z \bpm \includegraphics[scale=0.45]{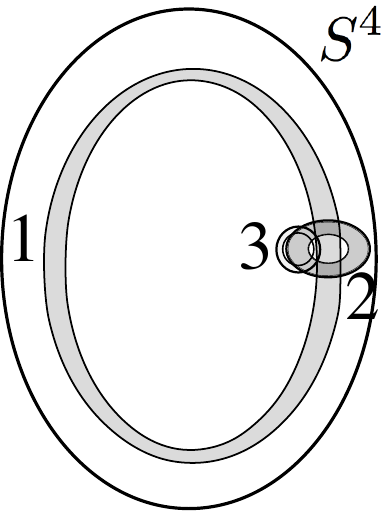} \epm
=Z \bpm \includegraphics[scale=0.45]{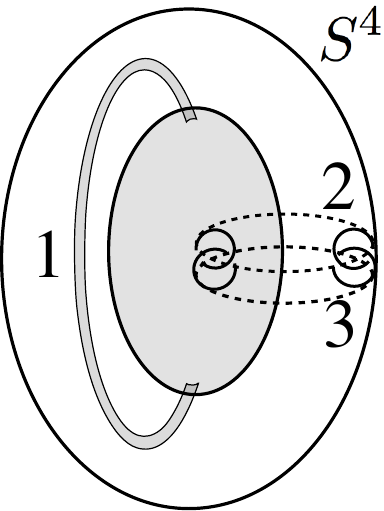} \epm
 \nonumber\\
 && \equiv Z[S^4; \text{Link[Spun[Hopf}[\mu_3,\mu_2]],\mu_1]] 
=\langle 0_{S^4 \- {D^2_{wx} \times T^2_{yz} }}| V^{T^2_{zx} \dagger}_{\mu_3} V^{T^2_{xy} \dagger}_{\mu_2} V^{T^2_{yz}}_{\mu_1} | 0_{{D^2_{wx} \times T^2_{yz} }} \rangle, \;\;\;\;\;\;\;\;
\eea
where we insert the worldsheets $V^{T^2_{yz}}_{\alpha}$ along the generator of the second homology group $H_2(D^2_{wx} \times T^2_{yz},\Z)=\Z$
while we insert worldsheets  $V^{T^2_{xy} \dagger}_{\beta}$ and $V^{T^2_{zx} \dagger}_{\gamma}$  along another two generators of $H_2({S^4\- {D^2_{wx} \times T^2_{yz}}},\Z)=\Z^2$. 
We learn that Eq.(\ref{T2T2T2glue}) can also be viewed as the spinning 
surgery construction of a Hopf link, say $\text{Spun[Hopf}[\mu_3,\mu_2]]$, linked by a third $T^2$-torus of $\mu_1$. 
Overall, we can denote the path integral as $Z[S^4; \text{Link[Spun[Hopf}[\mu_3,\mu_2]],\mu_1]]$, 
displayed in the second figure of Eq.(\ref{T2T2T2glue}).

In general, whenever there is a $T^3$-boundary of 4-submanifolds (such as $D^2 \times T^2$ and ${S^4 \- D^2 \times T^2}$), 
say $\cM_1^4 \cup \cM_2^4= \cM^4$ and $\partial \cM_1^4 =-\partial \cM_2^4=T^3$,
then we can cut out the 4-submanifold $\cM_1^4$ and reglue it back to $\cM_2^4$ via the mapping class group $\MCG(T^3)=\SL(3,\Z)$ 
generated by the two generators:
\bea
\hat{\cS}^{xyz}=\bpm 0& 0&1 \\1& 0&0 \\0& 1&0\epm \text{ and } \;\;\; \hat{\cT}^{xy}=\bpm 1&1 &0\\0&1 & 0 \\0 &0 & 1\epm
\eea
Here in this Chapter, we particularly define their modular $\SL(3,\Z)$-representations in the quasi-excitation bases as  
\bea
&&
{\cS^{xyz}_{\mu_2, \mu_1}} \equiv 
\< 0_{D^2_{xw} \times T^2_{yz}} | V^{T^2_{yz} \dagger}_{\mu_2}  \hat{\cS}^{xyz}  V^{T^2_{yz}}_{\mu_1}  | 0_{D^2_{xw} \times T^2_{yz}} \rangle, \\
&& {\cT^{xy}_{\mu_2, \mu_1}} 
=\< 0_{D^2_{xw} \times T^2_{yz}} | V^{T^2_{yz} \dagger}_{\mu_2}  \hat{\cT}^{xy}  V^{T^2_{yz}}_{\mu_1}  | 0_{D^2_{xw} \times T^2_{yz}} \rangle.
\eea
Thus we can actually do a modular-$\cS^{xyz}$ surgery to simplify the expression ${\tL^{\text{Tri}}_{{\mu_3}, {\mu_2}, {\mu_1}}}$.
We apply the surgery gluing 
\bea
&&(D^2 \times T^2) \cup_{T^3; \cS^{xyz}} (S^4 \- D^2 \times T^2) = S^3 \times S^1 \# S^2 \times S^2.\\
&&({D^2 \times T^2}) \cup_{T^3; \hat{S}^{xyz}} ({D^2 \times T^2})={S^3 \times S^1},
\eea
on the tubular neighborhood ${D^2 \times T^2}$ of the $T^2$-worldsheets, and we also apply their reverse-surgery process. We derive
\be
{\tL^{\text{Tri}}_{{\mu_3}, {\mu_2}, {\mu_1}}}=\sum_{\mu_3', {\Gamma_2}, {\Gamma_2'}, {\Gamma_2''}, {\eta_2}, {\eta_2'}} {\cS^{xyz}_{\mu_3', \mu_3}} (F^{T^2})_{\mu_2 \mu_3'}^{\Gamma_2} {(\cS^{xyz})^{-1}_{\Gamma_2', \Gamma_2}} 
{(\cS^{xyz})^{-1}_{\Gamma_2'', \Gamma_2'}}  (F^{T^2})_{{\mu_1} {\Gamma_2''}}^{\eta_2}  {\cS^{xyz}_{\eta_2', \eta_2}}  \; {\tL^{\text{Tri}}_{0, 0, {\eta_2'}}}. \;\;\; \label{eq:Sxyzsurger}
\ee
There is a Verlinde-like quantum surgery formula for 3+1D involving the  Eq.(\ref{T2T2T2glue}) that we can write down,
by applying Eq.(\ref{eq:Sxyzsurger}) explicitly:
\bea 
\label{eq:Spin[HopfLink]S4}
 &&\ \ \ \
Z \bpm \includegraphics[scale=0.45]{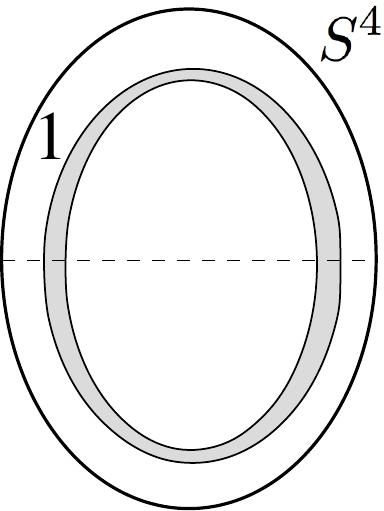} \epm 
 Z \bpm \includegraphics[scale=0.45]{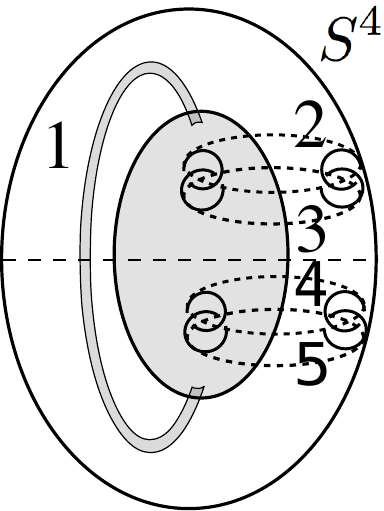} \epm 
_1=
 Z \bpm \includegraphics[scale=0.45]{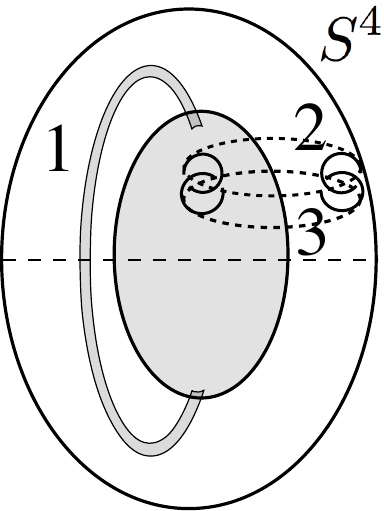} \epm 
 Z \bpm \includegraphics[scale=0.45]{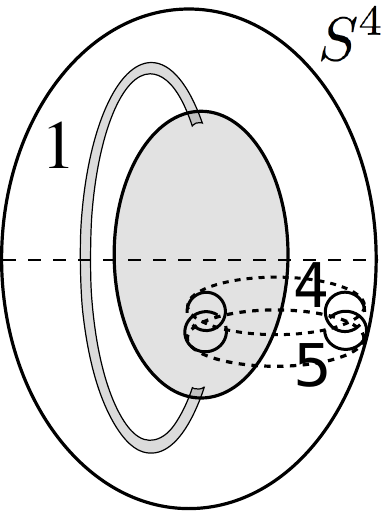} \epm \nonumber \\
 &&\ \ \ \ \Rightarrow 
  Z \bpm \includegraphics[scale=0.45]{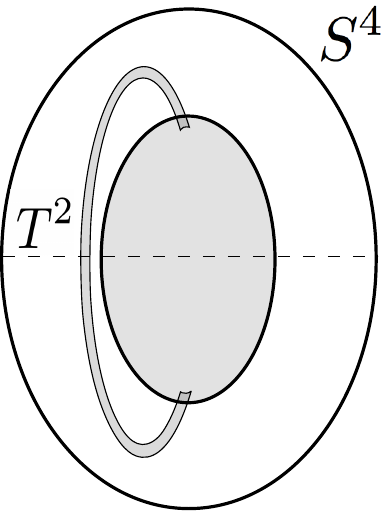} \epm 
 Z \bpm \includegraphics[scale=0.45]{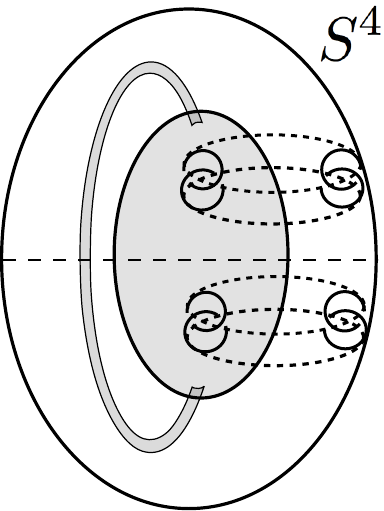} \epm 
=
 Z \bpm \includegraphics[scale=0.45]{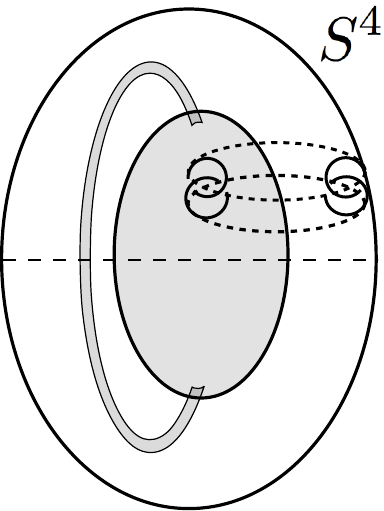} \epm 
 Z \bpm \includegraphics[scale=0.45]{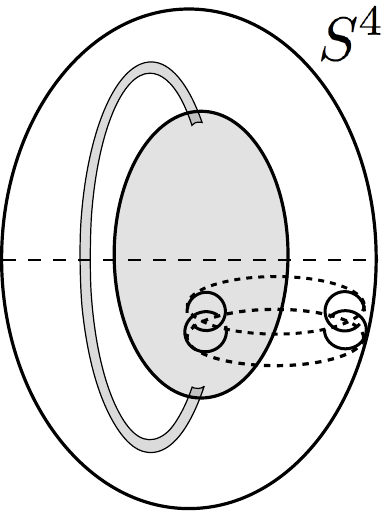} \epm \nonumber\\
&& \Rightarrow \label{eq:VerlindeT3}
 {{\tL^{\text{Tri}}_{0, 0, {\mu_1}}} \cdot
 \sum_{{\Gamma, \Gamma'},{\Gamma_1, \Gamma_1'}}
 (F^{T^2})_{{\zeta_2},{\zeta_4}}^{\Gamma}
{(\cS^{xyz})^{-1}_{\Gamma', \Gamma}}  (F^{T^2})_{{\mu_1} {\Gamma'}}^{\Gamma_1}  {\cS^{xyz}_{\Gamma_1', \Gamma_1}}  \; {\tL^{\text{Tri}}_{0, 0, {\Gamma_1'}}} } \nonumber
\\
&&
{=
{\sum_{{\zeta_2'}, {\eta_2}, {\eta_2'}} 
{(\cS^{xyz})^{-1}_{\zeta_2', {\zeta_2}}}  (F^{T^2})_{{\mu_1} {\zeta_2'}}^{\eta_2}  {\cS^{xyz}_{\eta_2', \eta_2}}  \; {\tL^{\text{Tri}}_{0, 0, {\eta_2'}}} } 
\;
{\sum_{{\zeta_4'}, {\eta_4}, {\eta_4'}}  
{(\cS^{xyz})^{-1}_{\zeta_4', {\zeta_4}}}  (F^{T^2})_{{\mu_1} {\zeta_4'}}^{\eta_4}  {\cS^{xyz}_{\eta_4', \eta_4}}  \; {\tL^{\text{Tri}}_{0, 0, {\eta_4'}}} }}. \;\;\;\;\;\;\;\;\;\;\;
\eea
Here we consider the dashed-cut is a spatial slice of a 3-sphere $S^3$, while the $S^4$ is cut into two pieces of upper and lower $D^4$ balls, namely
$D^4 \cup_{S^3} D^4=S^4$. At the spatial cross-section on the spatial $S^3$, there are only a pair of anyonic loop and anti-loop excitations of closed strings.
The dimension of Hilbert space with a pair of loop and anti-loop excitations on $S^3$ is 1. 
So we can use the identity Eq.(\ref{ZMN}). 
The worldsheet index ${\zeta_2}$ is obtained from fusing ${\mu_2}$ and ${\mu_3}$-worldsheets,
and the worldsheet index ${\zeta_4}$ is obtained from fusing  to fusing ${\mu_4}$ and ${\mu_5}$-worldsheets.
Here $F^{T^2}$ is the fusion algebra defined by fusing parallel $T^2$-worldsheets, as $V^{T^2_{}}_{\mu_1} V^{T^2_{}}_{\mu_2}=  (F^{T^2})_{{\mu_1}{\mu_2}}^{\mu_3} V^{T^2_{}}_{\mu_3}$.
All indices are summed over, except that only the indices ${\mu_1},{\zeta_2}$ and ${\zeta_4}$ are fixed.
Eventually
we derive a new quantum surgery formula Eq.(\ref{eq:VerlindeT3}) analogous to the Verlinde formula for 3+1D topological orders.

We can also consider another type of worldline-worldsheet linking, the $S^1$ and the $T^2$ link in $S^4$. The path integral has the form:
\bea
&&Z(S^4; \text{Link}[T^2,S^1] )=Z \bpm \includegraphics[scale=0.6]{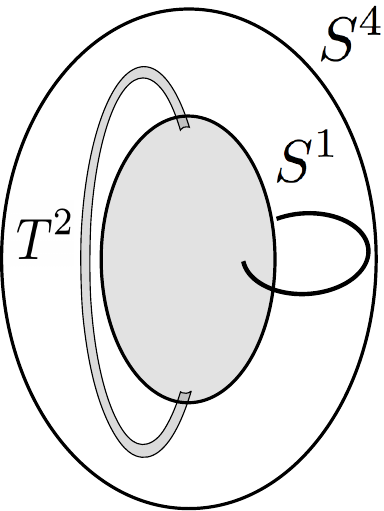} \epm
=Z \bpm \includegraphics[scale=0.6]{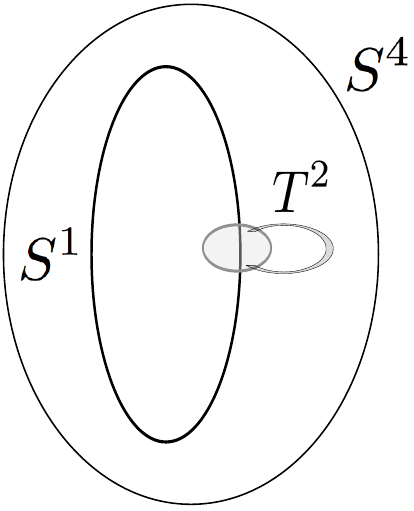} \epm \nonumber \\
&&=\langle  0_{D^2 \times T^2}  | V_{\mu}^{T^2_{}\dagger} W^{S^1}_{\si}   | 0_{S^4 \- D^2 \times T^2} \rangle,
\eea
up to an orientation of the worldline and worldsheet. 
Indeed, $Z(S^4; \text{Link}[T^2,S^1] )$ is a special case of the path integral of $Z[S^4; \text{Link[Spun[Hopf}$ $[\mu_3,\mu_2]],\mu_1]]$ $= {\tL^{\text{Tri}}_{{\mu_3}, {\mu_2}, {\mu_1}}}$ .
We find that the following quantum surgery Verlinde-like formula also hold for 3+1D topological orders,
\bea \label{eq:T2S1}
&&
Z \bpm \includegraphics[scale=0.5]{4D_3+1D_fig/3+1D_T2_S1_0.pdf} \epm 
Z \bpm \includegraphics[scale=0.5]{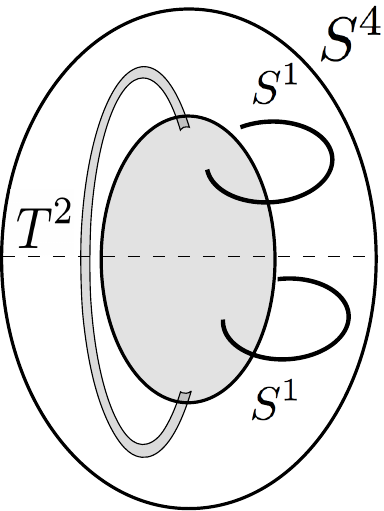} \epm
=Z \bpm \includegraphics[scale=0.5]{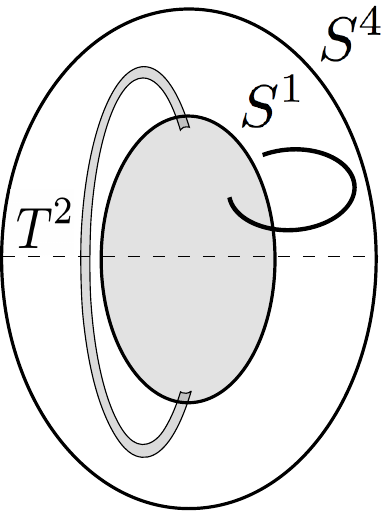} \epm 
Z \bpm \includegraphics[scale=0.5]{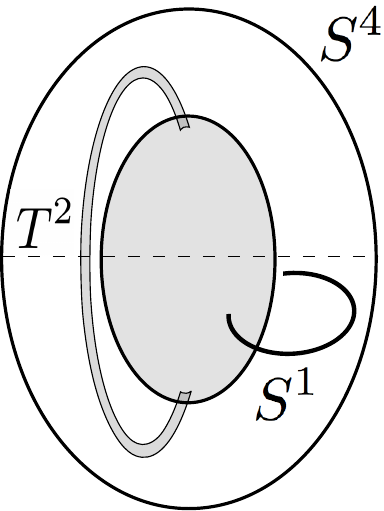} \epm.\;\;\;\;\;\;
\eea
\bea \label{eq:S1T2}
 &&
  Z \bpm \includegraphics[scale=0.5]{4D_3+1D_fig/Link_S1_in_S4.pdf} \epm 
 Z \bpm  \includegraphics[scale=0.5]{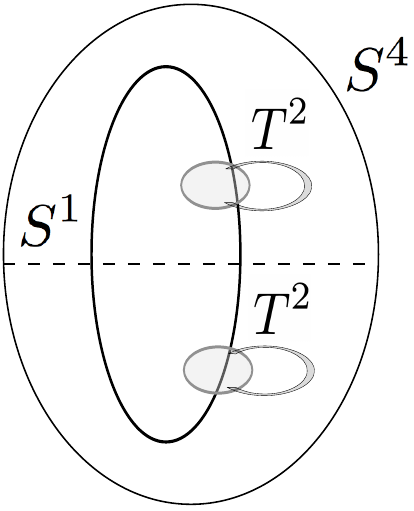} \epm 
=
 Z \bpm  \includegraphics[scale=0.5]{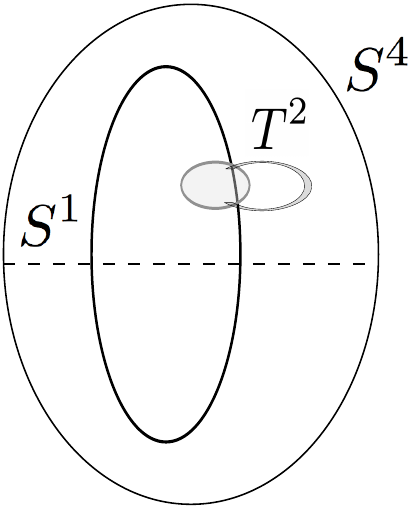} \epm 
 Z \bpm  \includegraphics[scale=0.5]{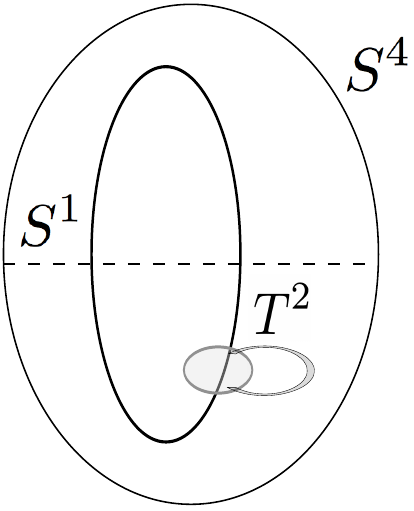} \epm. \;\;\;\;\;\;\;\;\;
\eea
\bea 
\label{eq:T2S1triple}
&&\ \ \ \
  Z \bpm \includegraphics[scale=0.5]{4D_3+1D_fig/3+1D_T2_S1_0.pdf} \epm 
 Z \bpm  \includegraphics[scale=0.5]{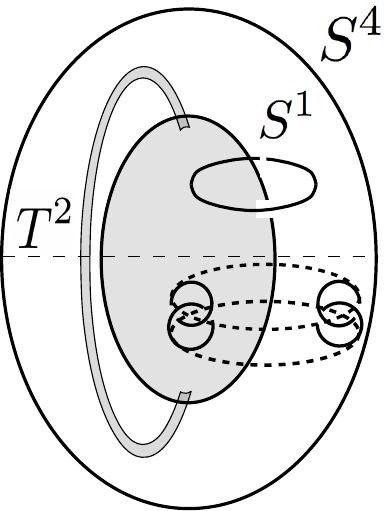} \epm 
  = Z \bpm \includegraphics[scale=0.5]{4D_3+1D_fig/3+1D_T2_S1_up.pdf} \epm Z \bpm \includegraphics[scale=0.55]{4D_3+1D_fig/3+1D_Triple_link_double_SpinHopfLink_S2_Down.pdf} \epm.\;\;\;\;\;\;\;\;
\eea
Here again we consider the dashed-cut is a spatial slice of a 3-sphere $S^3$, while the $S^4$ is cut into two pieces of upper and lower $D^4$ balls, namely
$D^4 \cup_{S^3} D^4=S^4$. At the spatial cross-section on the spatial $S^3$, there are only a pair of anyonic loop and anti-loop excitations of closed strings on
$S^3$ in Eq.(\ref{eq:T2S1})
and also in Eq.(\ref{eq:T2S1triple}),
there are only a pair of anyon and anti-anyon excitations of particles on $S^3$ in Eq.(\ref{eq:S1T2}).
The dimension of Hilbert space for all the above configurations on $S^3$ is 1. So we can apply the identity Eq.(\ref{ZMN}). 
And the formulas for  Eqs.(\ref{eq:T2S1}),(\ref{eq:S1T2}) and (\ref{eq:T2S1triple}) have the same forms as Eq.(\ref{eq:VerlindeT3}), except that
we need to switch some of the fusion algebra of worldsheet $F^{T^2}$ to that of worldline $F^{S^1}$,
also we need to switch some of the worldsheet index $\mu$ to the worldline index $\sigma$.

Nonetheless, if we regard the worldline
fusion algebra $F^{S^1}$ is part of the worldsheet fusion algebra $F^{T^2}$, in short, 
Eq.(\ref{eq:VerlindeT3}) contain the most generic form including Eqs.(\ref{eq:T2S1}),(\ref{eq:S1T2}) and (\ref{eq:T2S1triple}).

\section{Interplay of quantum topology and spacetime topology: Verlinde formula and its generalizations}

The interplay between quantum topology and spacetime topology is examined.
We had derived generalized Verlinde formulas in Eqs.(\ref{eq:VF_S2linkS1S1}) and (\ref{eq:VF_S1linkS2S2}). 
Clearly there are more relations, which we will report elsewhere in the future publications.
By performing the surgery theory of geometric-topology on the spacetime,  
we show that the quantum fusion rule and quantum statistics are constrained by the intrinsic properties of spacetime topology.
The exotic quantum statistics is defined in the adiabatic braiding process in the gapped phases of matter with topological orders, therefore 
the spacetime topology strictly constrains the quantum topology thus dictates the possible gapped phases of matter.



%% file: Conclusion.tex
\chapter{Conclusion: Finale and A New View of Emergence-Reductionism} \label{Conclusion}


Thus far we have explored some physical properties and mathematical structures 
of topological states of matter 
of SPTs and topological orders (TOs).
We have started from the very basic notion of quantum mechanical Berry's geometric phase, 
to reveal its profound connection to many-body topological states of matter.
For example, we use geometric phase and geometric matrix to define modular $\cS$ and $\cT$ matrices which connects to 
the braiding statistics of quasi-excitations of TOs, or 
the SPT invariants such as fractionalized charges and degenerate zero modes from symmetry-twists of SPTs.


One important ingredient of our work is that the constraints of topologies and the meaning of topologies.
We can roughly distinguish them into three types: SPT topology (``classical'' topology),
TO topology (quantum topology) and spacetime topology.
Our understanding now is 
developed much beyond what was previewed in Chap.\ref{chap:Intro}. We summarize in Table \ref{table:topo_conclude}.
%
%
\begin{center}
\begin{table}[!h]
 $$\text{ Topology } \\
 \to
 \left\{ 
 \begin{array}{lll}  
 \text{ Classical: homotopy, mapping and winding numbers, K-theory. } \\ 
 \text{ Quantum: algebraic topology, homology, cohomology, tensor category.}  \\ 
 \text{ Spacetime: fiber bundles, geometric-topology, surgery theory.} 
  \end{array}  \right.$$ \\
 $$ \to
  \left\{ 
 \begin{array}{lll} 
 \text{ ``Classical + Spacetime'' topology: SPT invariants  } \\
 \text{ ``Quantum + Spacetime'' topology: TO invariants } 
 \end{array}  \right.$$\\
  $$\to \left\{ \begin{array}{lll} 
  \to  \text{(Field-theory Rep. {\bf Q.I}. Chap.\ref{aofSymmetry})}\\
  \to \text{(Quantum statistics + spacetime surgery. {\bf Q.VI}. Chap.\ref{QS_stSurgery})}
  \end{array}  \right.
 $$
 \caption{The interplay of classical, quantum and spacetime topology.}
\label{table:topo_conclude}
\end{table}
\end{center}
One important message is that, when we combine the concepts of SPT topology (a sort of ``classical'' topology compared to the TOs) and spacetime topology, we are able to constrain, characterize and classify the possible types of SPT probes as symmetry-twists having branch cuts on the spacetime manifold (Chap.\ref{aofSymmetry}).
This way of thinking leads us to regard 
the SPTs as having a closer tie to the spacetime topology.
SPTs are not merely some quantum matter by itself, 
but quantum matter with symmetry-protection tied to the spacetime topology.
Analogously, another important message is that, when we combine the concepts of TO topology (a sort of ``quantum'' topology compared to the SPTs) 
and spacetime topology, we are able to derive the consistent braiding-fusion formulas such as Verlinde's formula.
These formulas further constrain the possible TOs on a given-dimensional spacetime manifold (Chap.\ref{QS_stSurgery}).
This way of thinking leads us to regard 
the TOs as also having a closer tie to the spacetime topology.
TOs are not merely some quantum matter by itself, 
but further exotic quantum matter tied to the spacetime topology
with robust topological GSD and non-Abelian geometric matrix, even without symmetry-protection.

\begin{figure}[!h]
{\includegraphics[width=0.98\textwidth]{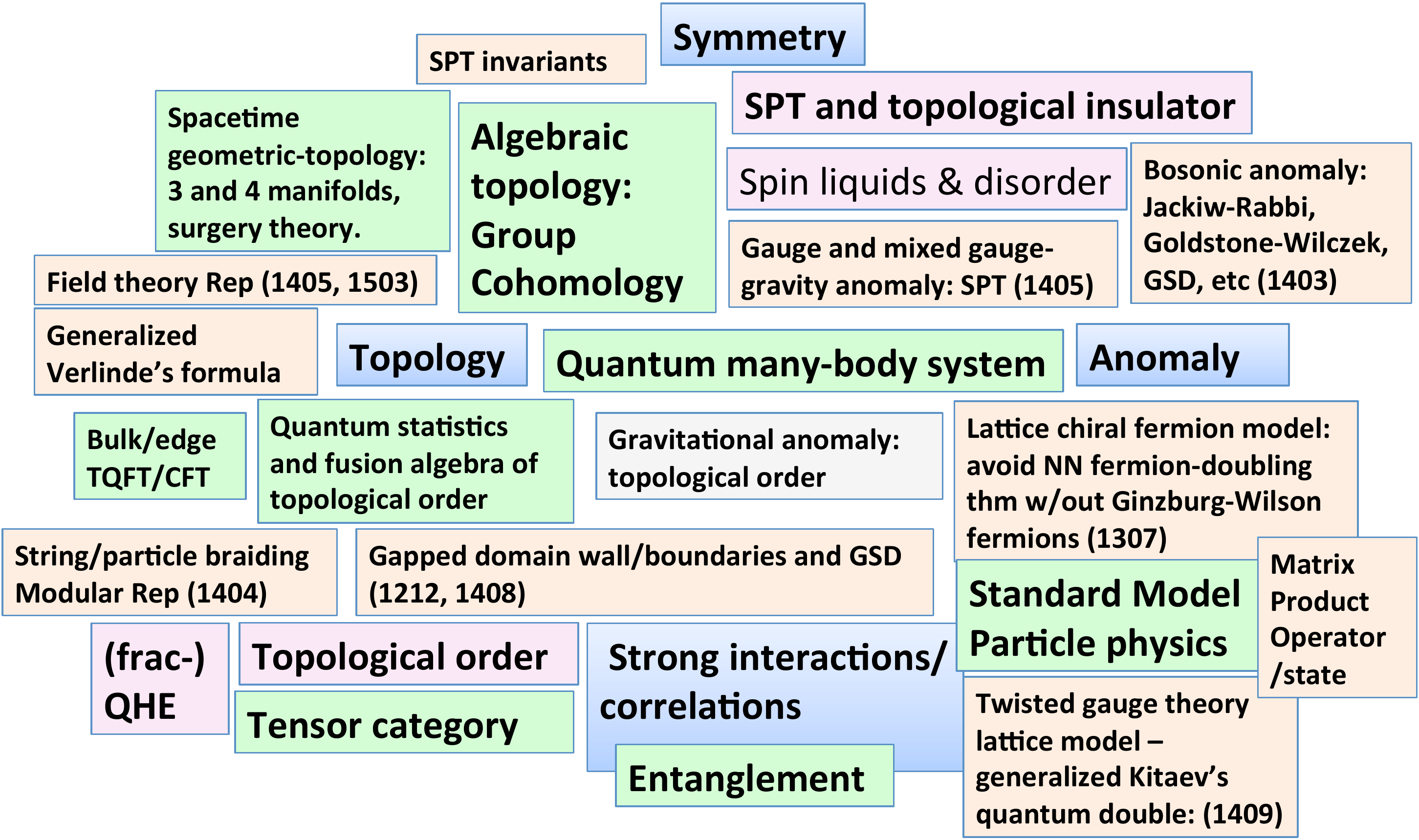}  }  
\caption{The flow chart overviews the main ideas and subjects emerged from the development of the thesis. The numbers shown above
represent the arXiv numbers (year and month) for my journal publication preprints.
}
\label{chart_summary_final_4}
\end{figure}

In Table \ref{table:Summary}, we summarize the alternative views of the reductionism of HEP
and the emergence of CMP on the examples we studied throughout Chap.\ref{gphase} to Chap.\ref{QS_stSurgery}.
The flow chart in Figure \ref{chart_summary_final_4} overviews the main ideas and subjects emerge from the development of the thesis. 

\begin{center}
\begin{table}[!h]
\makebox[\textwidth][c]
{\fontsize{10pt}{1em}\selectfont
{
\begin{tabular}{|c||c| c| }
\hline 
& Reductionism \& HEP & Emergence \& CMP  \\ \hline \hline
\begin{minipage}[t]{1.4in} {$\bullet$ Symmetry-twist}: \cite{Santos:2013uda, Wang:2014tia, Wang:2014pma, Gu:2015lfa} 
\end{minipage}
& \begin{minipage}[t]{2.in} Twisted sector of CFTs, \\
gauging the symmetry, orbifolds\cite{Ryu2013orbifolds} and orientifolds\cite{Ryu2014orientifolds} \end{minipage}  & 
\begin{minipage}[t]{2.in}
Modified Hamiltonian along the branch cut, or twisted boundary conditions of wavefunctions or Hamiltonians\\
 \end{minipage} \\ 
\hline
\begin{minipage}[t]{1.4in} 
$\bullet$ Anomalies: \cite{Wang:2012am,Wang:2013yta,Lan:2014uaa}\\
Bulk-edge correspondence.\\
Fermion-doubling.\\
Bosonic anomalies \cite{Santos:2013uda}.
\end{minipage} & 
\begin{minipage}[t]{2.in} 
t'Hooft anomaly-matching conditions. \\
Lattice chiral fermion/gauge theory.\\
Non-onsite symmetry lattice regularization; 
Jackiw-Rebbi or Goldstone-Wilczek effect.
\end{minipage} 
& 
\begin{minipage}[t]{2.in}
Boundary fully gapping rules;\\
gapping the mirror sector of coupled Chern insulators.\\ 
Induced fractional quantum number and degenerate zero modes
computed on a lattice
\end{minipage} \\
\hline   
\begin{minipage}[t]{1.4in} 
$\bullet$ {Quantum Statistics}: \cite{Wang:2014oya,Wan:2014woa}
String and particle braiding statistics \\
$\bullet$ {Topological order lattice model} \cite{Wang:2014oya,Wan:2014woa}
\end{minipage} 
 & \begin{minipage}[t]{2.in}
 Representation theory, quantum algebra, 
algebraic topology, geometric-topology, surgery theory.\\
Dijkgraaf-Witten lattice spacetime path integral \cite{Dijkgraaf:1989pz}, group cohomology,  
Hopf algebra, TQFT. 
\end{minipage} 
 & \begin{minipage}[t]{2.in}
 Wavefunction overlapping of quantum states\\[2mm]
 Generalized Kitaev's toric code \cite{Kitaev:1997wr}
 to twisted quantum double model \cite{HuWanWuTQD} and twisted gauge theory on the lattice
 \end{minipage} \\
\hline
\end{tabular}
}
}\hspace*{0mm}
\caption{Dictionary between the physics or mathematics used in reductionism and in emergence viewpoints. 
Some of the aspects in HEP are 
done in my work, while some are adopted from the cited references. In any case, my original work connects them to CMP issues.}
\label{table:Summary}
\end{table}
\end{center}
%

There are several {\bf future directions}:

{\bf Gauge-gravity anomalies, SPTs and TOs}: In this thesis, we have tentatively suggested connecting 
gauge anomalies or mixed gauge-gravitational anomalies to the surface anomalies of SPTs,
and
connecting gravitational anomalies to the surface anomalies of TOs. 
The key concept here is viewing the gauge anomalies and external probed gauge fields 
as the phenomena of weakly coupling gauge fields coupled to the global symmetry of SPTs.
There is another (second) kind of anomalies more robust than this.
If we break all the possible global symmetries of topological states, and if there is still any anomalous surface effect,
then these anomalies do not need to be protected by global symmetries.
Thus, they are more robust than gauge anomalies. 
They are associated with spacetime diffeomorphism.
They can be viewed as gravitational anomalies --- including 
perturbative gravitational anomalies (computable from a 1-loop Feynman diagram) or non-perturbative global gravitational anomalies.
We understand that so far we have only studied some examples to support this claim. It will be helpful
to find more new examples to test and clarify this claim.

{\bf Experimental or numerical realization of phases of matter}: It will be important to find the materialization 
through correlated quantum magnets, electronic Mott insulators and ultra-cold atoms. It seems that a complete theory on the
classification of interacting SPTs and topological insulators/superconductors is still missing \cite{Senthil1405}. Several steps of progress have been 
made in specific examples.
What we are searching for a consistent-complete theory is like the Ginzburg-Landau's group theory for symmetry-breaking phases,
and the tensor category theory for 2D TOs. Whether we have the group-cohomology or cobordism theory or something else to classify SPTs, and what theory we should develop to
fully classify higher-dimensional TOs, are the open challenges to be tackled in the future.

{\bf Non-perturbative lattice chiral fermion}:
We have proposed a simple model for a non-perturbative lattice chiral fermion model.
Currently 
three main attempts to regularize the following chiral theory on the lattice are: 
(i) U(1) chiral fermion  \cite{{Wang:2013yta}}, (ii) SO(10) chiral gauge \cite{Wen:2013ppa} and (iii) $\U(1)\times Z_2 \times Z_2^T$ interacting topological superconductor theory \cite{You:2014oaa}. 
Our anomaly-free proof for U(1) chiral fermions is the most rigorous among the three, thanks to the bosonization technique in 1D. 
The possible next step will be concretely proving that an anomaly-free SM-like chiral gauge theory can be constructed using the mirror-decoupling set-up,
by looking into larger extra symmetries as we did in \cite{{Wang:2013yta}} outside a tentative gauge group SO(10) or SU(5).
One can search for 
the potential hidden constraints on the parameters of SM or modified-SM when it can be defined non-perturbatively on the lattice,
such as the Higgs scale and top quark mass.
This may help to
address 
beyond-SM questions, such as hierarchy 
or strong CP problems.
There is also the possibility of synthesizing the idea of mirror-fermion decoupling with other HEP theories
with extra dimensions, such as Kaluza-Klein, Einstein-Bergmann and Randall-Sundrum models.
The second issue 
for this goal is to 
implement the simplest 1+1D 
chiral 
theory on the lattice. 
The first attempt can be programming the tensor network code, using 1D Matrix Product States (MPS) or Density Matrix Renormalization Group (DMRG).
If it is possible, 
the future directions are (1) inventing a new algorithm which can solve a gapless system, going beyond the gapped one of DMRG, 
or (2) seeking 
collaboration with simulation experts in many-body quantum systems or in lattice QCD.  
A new algorithm is desirable and profound because the gapless chiral fermions are highly-entangled states with an entropy $S \propto L^{d-1}  \log (L)$ 
beyond-the-area-law 
(
$L$ is the length dimension in $d$-D), 
which require a new ground state wavefunction ansatz and a new simulation theory, such
as the Multi-scale Entanglement Renormalization Ansatz (MERA). 

{\bf More general statements about anomaly-matching conditions and topological gapping criteria for higher dimensional theory.} 
In Chap.\ref{aofTopology} and Chap.\ref{aofAnomalies}, we discuss the 1+1D statement between anomaly-matching conditions and topological gapping criteria
with U(1)$^n$ symmetry. This concept may hold more generally than the given spacetime and the given symmetry.
It will be remarkable to extend this direction further.
It may be useful to understand the characterization by entanglement entropy (EE), in particular topological entanglement entropy (TEE).
One key feature of topological orders is known to be their long-range entanglement. 
TEE 
shows how quantum wavefunctions can be correlated to each other remotely through the entanglement. 
Up till now, people have been looking at TEE on the system either at a closed manifold or at the interior without physical boundaries. 
Building upon our previous work on gapped boundaries/domain walls \cite{{Wang:2012am},{Lan:2014uaa}},  
we can investigate EE and TEE on the system on generic open or closed manifolds. 
It is an important question, because the gappability of boundary is related to the 
't Hooft anomaly-matching conditions of gauge anomalies or (perturbative or global) gravitational anomalies. 
From the reductionism aspect 
in HEP, the anomaly-matching is the mechanism of the anomaly-inflow.
However,  from the emergence aspect in CMP, the same concept 
is analogous to Laughlin's flux insertion Gedanken-experiment.
One can also check how EE 
fits into the mirror-fermion decoupling set-up, and whether the gapped mirror sector (set at the Planck scale) constrains the light sector with nearly gapless fermions. 
HEP tends to view gauge theory as governed by ``gauge symmetry,'' 
but indeed it has no real symmetries only redundancies.
Ultimately, we can view gauge theory through entanglement. We should understand how the SM structure can be embedded and ask how 
\emph{new entanglement concepts beyond EE or TEE can be introduced
for strongly-interacting gapless phases}.

{\bf Topological phase transitions and gapless phases}: The topological gapping criteria for gapped domain walls located between TOs Phases $A$ and $B$, may be connected to
the topological phase transitions between TOs Phases $A$ and $B$.  Topological phase transitions here may not just include the first order phase transition, but also the second order phase transition. It will be important to see whether extra conditions should be included to derive the criteria for the second order phase transitions.
This information could provide one of crucial hints for studying robust gapless phases of matter. Another crucial hints
can be the surface gapless states due to anomalies protected by the bulk state of matter.

{\bf Further interplay of the spacetime topology and the states of matter topology}:
It is likely the approach on spacetime surgery and the quantum amplitudes can be generalized to study not just TOs but also SPTs, or more generally 
symmetry-enriched topological states (SETs, for topological orders with symmetry-protection). There should be a direct program generalized 
by extending Chap. \ref{aofSymmetry} to \ref{QS_stSurgery}.

{\bf Many-body entanglement structure}: Throughout the thesis, the hidden concept of entanglement is used, but we have not yet explicitly studied it in depth.
Whether the result we obtained so far (such as anomaly-matching and gapping criteria, string braiding statistics and generalized Verlinde formula) 
can be organized in terms of the fundamental principles of quantum entanglement structure will be left for future questions.

If we digest the reductionism and emergence viewpoints further, the two views are rather \emph{different}, but there is \emph{not} actually an obvious 
cut between the two philosophies. 
The principles emerging from a many-body system can be reduced to a few guiding principles and the 
basic ingredients of qubits or spins and their interactions can be regarded as the fundamental blocks of reductionism 
(thus, emergence gives rise to 
reductionism).
On the other hand, it also happens that the fundamental laws of reductionism can be re-arranged and transformed into another set of formulation
where the fundamental ingredients are transformed, too (such as the particle-wave duality; this is sort of reductionism 
gives rise to 
emergence). 
This reminds us of the importance of complementarity of reductionism and emergence.
We should recall that Niels Bohr's coat of arms, ``opposites are complementary,''
and John Wheeler's comment on complementarity: ``Bohr's principle of complementarity is the most revolutionary scientific concept of this century and the heart of his fifty-year search for the full significance of the quantum idea.'' Perhaps to keep the two complementary viewpoints will guide us to digest  
the full 
profundity of the quantum world and the beyond-quantum world.

%% file: biblio.tex
\begin{singlespace}
\bibliography{SPT_new}
\bibliographystyle{unsrt}
\end{singlespace}


%% file: main.bbl
\begin{thebibliography}{100}

\bibitem{Anderson}
P.~W. Anderson.
\newblock More is different.
\newblock {\em Science}, 177(4047):393--396, 1972.

\bibitem{ColemanA}
Sidney Coleman.
\newblock {\em {Aspects of Symmetry}}.
\newblock Cambridge University Press., 1988.

\bibitem{Farhi:1982vt}
E.~Farhi and R.~Jackiw.
\newblock {Dynamical Gauge Symmetry Breaking. A Collection Of Reprints }.
\newblock 1982.

\bibitem{Treiman:1986ep}
S.B. Treiman, Edward Witten, R.~Jackiw, and B.~Zumino.
\newblock {Current Algebra And Anomalies }.

\bibitem{wilczek1990fractional}
F.~Wilczek.
\newblock {\em Fractional Statistics and Anyon Superconductivity}.
\newblock International journal of modern physics. World Scientific, 1990.

\bibitem{Jackiw:1995be}
R.~Jackiw.
\newblock {Diverse topics in theoretical and mathematical physics}.
\newblock 1995.

\bibitem{WilsonRGRMP}
Kenneth~G. Wilson.
\newblock The renormalization group: Critical phenomena and the kondo problem.
\newblock {\em Rev. Mod. Phys.}, 47:773--840, Oct 1975.

\bibitem{Landau1}
L.~D. Landau.
\newblock Theory of phase transformations.
\newblock {\em Phys. Z. Sowjetunion}, 11:26 and 545, 1937.

\bibitem{GL5064}
V.~L. Ginzburg and L.~D. Landau.
\newblock On the theory of superconductivity.
\newblock {\em Zh. Eksp. Teor. Fiz.}, 20:1064--1082, 1950.

\bibitem{LL58}
L.~D. Landau and E.~M. Lifschitz.
\newblock {\em Statistical Physics - Course of Theoretical Physics Vol 5}.
\newblock Pergamon, London, 1958.

\bibitem{wen2004quantum}
X.G. Wen.
\newblock {\em Quantum Field Theory of Many-body Systems: From the Origin of
  Sound to an Origin of Light and Electrons}.
\newblock Oxford graduate texts. Oxford University Press, 2004.

\bibitem{XieSPT4}
Xie Chen, Zheng-Cheng Gu, Zheng-Xin Liu, and Xiao-Gang Wen.
\newblock Symmetry-protected topological orders and the cohomology class of
  their symmetry group.
\newblock {\em Phys. Rev. B}, 87:155114, 2013.

\bibitem{XieSPT5}
Xie Chen, Zheng-Cheng Gu, Zheng-Xin Liu, and Xiao-Gang Wen.
\newblock Symmetry protected topological orders in interacting bosonic systems.
\newblock {\em Science 338, 1604}, 2012.

\bibitem{W9039}
Xiao-Gang Wen.
\newblock Topological orders in rigid states.
\newblock {\em Int. J. Mod. Phys. B}, 4:239, 1990.

\bibitem{Wen1210.1281}
X.-G. {Wen}.
\newblock {Topological order: from long-range entangled quantum matter to an
  unification of light and electrons}.
\newblock {\em ArXiv e-prints}, October 2012.

\bibitem{Klitzing}
K.~v. Klitzing, G.~Dorda, and M.~Pepper.
\newblock New method for high-accuracy determination of the fine-structure
  constant based on quantized hall resistance.
\newblock {\em Phys. Rev. Lett.}, 45:494--497, Aug 1980.

\bibitem{TsuiSG8259}
D.~C. Tsui, H.~L. Stormer, and A.~C. Gossard.
\newblock Two-dimensional magnetotransport in the extreme quantum limit.
\newblock {\em Phys. Rev. Lett.}, 48:1559--1562, 1982.

\bibitem{Laughlin8395}
R.~B. Laughlin.
\newblock Anomalous quantum hall effect: An incompressible quantum fluid with
  fractionally charged excitations.
\newblock {\em Phys. Rev. Lett.}, 50:1395--1398, 1983.

\bibitem{Kitaevhoneycomb}
Alexei Kitaev.
\newblock Anyons in an exactly solved model and beyond.
\newblock {\em Ann. Phys.}, 321:2, 2006.

\bibitem{preskill1998lecture}
John Preskill.
\newblock Lecture notes for physics 229: Quantum information and computation.
\newblock 1998.

\bibitem{pachos2012introduction}
J.K. Pachos.
\newblock {\em Introduction to Topological Quantum Computation}.
\newblock Introduction to Topological Quantum Computation. Cambridge University
  Press, 2012.

\bibitem{KitaevPreskill}
Alexei Kitaev and John Preskill.
\newblock Topological entanglement entropy.
\newblock {\em Phys. Rev. Lett.}, 96:110404, Mar 2006.

\bibitem{LW0605}
Michael Levin and Xiao-Gang Wen.
\newblock Detecting topological order in a ground state wave function.
\newblock {\em Phys. Rev. Lett.}, 96:110405, 2006.

\bibitem{CNYangRMP}
C.~N. Yang.
\newblock Concept of off-diagonal long-range order and the quantum phases of
  liquid he and of superconductors.
\newblock {\em Rev. Mod. Phys.}, 34:694--704, Oct 1962.

\bibitem{Witten:1988ze}
Edward Witten.
\newblock {Topological Quantum Field Theory}.
\newblock {\em Commun.Math.Phys.}, 117:353, 1988.

\bibitem{Witten:1988hf}
Edward Witten.
\newblock {Quantum Field Theory and the Jones Polynomial}.
\newblock {\em Commun.Math.Phys.}, 121:351--399, 1989.

\bibitem{WenSPTinv}
X.-G. {Wen}.
\newblock {Symmetry-protected topological invariants of symmetry-protected
  topological phases of interacting bosons and fermions}.
\newblock {\em Phys. Rev. B}, 89(3):035147, January 2014.

\bibitem{HW1339}
L.-Y. Hung and X.-G. Wen.
\newblock Universal symmetry-protected topological invariants for
  symmetry-protected topological states.
\newblock 2013.

\bibitem{JuvenSPT1}
Juven~C. {Wang}, Z.-C. {Gu}, and X.-G. {Wen}.
\newblock {Field-Theory Representation of Gauge-Gravity Symmetry-Protected
  Topological Invariants, Group Cohomology, and Beyond}.
\newblock {\em Phys. Rev. Lett.}, 114(3):031601, January 2015.

\bibitem{2010RMP_HasanKane}
M.~Z. {Hasan} and C.~L. {Kane}.
\newblock {Colloquium: Topological insulators}.
\newblock {\em Reviews of Modern Physics}, 82:3045--3067, October 2010.

\bibitem{2011_RMP_Qi_Zhang}
X.-L. {Qi} and S.-C. {Zhang}.
\newblock {Topological insulators and superconductors}.
\newblock {\em Reviews of Modern Physics}, 83:1057--1110, October 2011.

\bibitem{Turner-Vish}
A.~M. {Turner} and A.~{Vishwanath}.
\newblock {Beyond Band Insulators: Topology of Semi-metals and Interacting
  Phases}.
\newblock {\em ArXiv e-prints}, January 2013.

\bibitem{Senthil1405}
T.~{Senthil}.
\newblock {Symmetry Protected Topological phases of Quantum Matter}.
\newblock {\em ArXiv e-prints}, May 2014.

\bibitem{H8364}
F.~D.~M. Haldane.
\newblock Continuum dynamics of the 1-{D} heisenberg antiferromagnet:
  Identification with the {O}(3) nonlinear sigma model.
\newblock {\em Physics Letters A}, 93:464, 1983.

\bibitem{AKL8877}
I.~Affleck, T.~Kennedy, E.~H. Lieb, and H.~Tasaki.
\newblock {\em Commun. Math. Phys.}, 115:477, 1988.

\bibitem{MooreBalents}
J.~E. Moore and L.~Balents.
\newblock Topological invariants of time-reversal-invariant band structures.
\newblock {\em Phys. Rev. B}, 75(12):121306, 2007.

\bibitem{FuKaneMele}
Liang Fu, C.~L. Kane, and E.~J. Mele.
\newblock Topological insulators in three dimensions.
\newblock {\em Phys. Rev. Lett.}, 98(10):106803, 2007.

\bibitem{Roy}
Rahul Roy.
\newblock Topological phases and the quantum spin hall effect in three
  dimensions.
\newblock {\em Phys. Rev. B}, 79(19):195322, 2009.

\bibitem{KaneMele2}
C.~L. Kane and E.~J. Mele.
\newblock Z$_2$ topological order and the quantum spin hall effect.
\newblock {\em Phys. Rev. Lett.}, 95:146802, 2005.

\bibitem{BernevigHughesZhang}
B.~Andrei Bernevig, Taylor~L. Hughes, and Shou-Cheng Zhang.
\newblock Quantum spin hall effect and topological phase transition in hgte
  quantum wells.
\newblock {\em Science}, 314(5806):1757--1761, 2006.

\bibitem{Molenkamp}
M.~{K{\"o}nig}, S.~{Wiedmann}, C.~{Br{\"u}ne}, A.~{Roth}, H.~{Buhmann}, L.~W.
  {Molenkamp}, X.-L. {Qi}, and S.-C. {Zhang}.
\newblock {Quantum Spin Hall Insulator State in HgTe Quantum Wells}.
\newblock {\em Science}, 318:766--, November 2007.

\bibitem{Fu_TCI}
L.~{Fu}.
\newblock {Topological Crystalline Insulators}.
\newblock {\em Physical Review Letters}, 106(10):106802, March 2011.

\bibitem{2010THsieh}
T.~H. {Hsieh}, H.~{Lin}, J.~{Liu}, W.~{Duan}, A.~{Bansil}, and L.~{Fu}.
\newblock {Topological crystalline insulators in the SnTe material class}.
\newblock {\em Nature Communications}, 3:982, July 2012.

\bibitem{Ando}
Y.~{Tanaka}, Z.~{Ren}, T.~{Sato}, K.~{Nakayama}, S.~{Souma}, T.~{Takahashi},
  K.~{Segawa}, and Y.~{Ando}.
\newblock {Experimental realization of a topological crystalline insulator in
  SnTe}.
\newblock {\em Nature Physics}, 8:800--803, November 2012.

\bibitem{Story}
P.~{Dziawa}, B.~J. {Kowalski}, K.~{Dybko}, R.~{Buczko}, A.~{Szczerbakow},
  M.~{Szot}, E.~{{\L}usakowska}, T.~{Balasubramanian}, B.~M. {Wojek}, M.~H.
  {Berntsen}, O.~{Tjernberg}, and T.~{Story}.
\newblock {Topological crystalline insulator states in Pb$_{1-x}$Sn$_{x}$Se}.
\newblock {\em Nature Materials}, 11:1023--1027, December 2012.

\bibitem{HasanTCI}
S.-Y. {Xu}, C.~{Liu}, N.~{Alidoust}, M.~{Neupane}, D.~{Qian}, I.~{Belopolski},
  J.~D. {Denlinger}, Y.~J. {Wang}, H.~{Lin}, L.~A. {Wray}, G.~{Landolt},
  B.~{Slomski}, J.~H. {Dil}, A.~{Marcinkova}, E.~{Morosan}, Q.~{Gibson},
  R.~{Sankar}, F.~C. {Chou}, R.~J. {Cava}, A.~{Bansil}, and M.~Z. {Hasan}.
\newblock {Observation of a topological crystalline insulator phase and
  topological phase transition in Pb$_{1-x}$Sn$_{x}$Te}.
\newblock {\em Nature Communications}, 3:1192, November 2012.

\bibitem{Callan:1984sa}
Jr. Callan, Curtis~G. and Jeffrey~A. Harvey.
\newblock {Anomalies and Fermion Zero Modes on Strings and Domain Walls}.
\newblock {\em Nucl.Phys.}, B250:427, 1985.

\bibitem{'tHooft:1979bh}
Gerard 't~Hooft.
\newblock {Naturalness, chiral symmetry, and spontaneous chiral symmetry
  breaking}.
\newblock {\em NATO Sci.Ser.B}, 59:135, 1980.

\bibitem{Wen:2013oza}
Xiao-Gang Wen.
\newblock {Classifying gauge anomalies through symmetry-protected trivial
  orders and classifying gravitational anomalies through topological orders}.
\newblock {\em Phys.Rev.}, D88(4):045013, 2013.

\bibitem{Wang:2013yta}
Juven Wang and Xiao-Gang Wen.
\newblock {A Lattice Non-Perturbative Hamiltonian Construction of 1+1D
  Anomaly-Free Chiral Fermions and Bosons - on the equivalence of the anomaly
  matching conditions and the boundary fully gapping rules}.
\newblock 2013.

\bibitem{Kong:2014qka}
Liang Kong and Xiao-Gang Wen.
\newblock {Braided fusion categories, gravitational anomalies, and the
  mathematical framework for topological orders in any dimensions}.
\newblock 2014.

\bibitem{Wang:2014oya}
Juven Wang and Xiao-Gang Wen.
\newblock {Non-Abelian string and particle braiding in topological order:
  Modular SL(3,$\mathbb{Z}$) representation and (3+1) -dimensional twisted
  gauge theory}.
\newblock {\em Phys.Rev.}, B91(3):035134, 2015.

\bibitem{Wilczek:1984dh}
Frank Wilczek and A.~Zee.
\newblock {Appearance of Gauge Structure in Simple Dynamical Systems}.
\newblock {\em Phys.Rev.Lett.}, 52:2111--2114, 1984.

\bibitem{Nielsen:1980rz}
Holger~Bech Nielsen and M.~Ninomiya.
\newblock {Absence of Neutrinos on a Lattice. 1. Proof by Homotopy Theory}.
\newblock {\em Nucl.Phys.}, B185:20, 1981.

\bibitem{Gu:2015lfa}
Zheng-Cheng Gu, Juven~C. Wang, and Xiao-Gang Wen.
\newblock {Multi-kink topological terms and charge-binding domain-wall
  condensation induced symmetry-protected topological states: Beyond
  Chern-Simons/BF theory}.
\newblock 2015.

\bibitem{Wang:2014tia}
Juven Wang, Luiz~H. Santos, and Xiao-Gang Wen.
\newblock {Bosonic Anomalies, Induced Fractional Quantum Numbers and Degenerate
  Zero Modes: the anomalous edge physics of Symmetry-Protected Topological
  States}.
\newblock 2014.

\bibitem{Santos:2013uda}
Luiz~H. Santos and Juven Wang.
\newblock {Symmetry-protected many-body Aharonov-Bohm effect}.
\newblock {\em Phys.Rev.}, B89(19):195122, 2014.

\bibitem{Adler:1969gk}
Stephen~L. Adler.
\newblock {Axial vector vertex in spinor electrodynamics}.
\newblock {\em Phys.Rev.}, 177:2426--2438, 1969.

\bibitem{Bell:1969ts}
J.S. Bell and R.~Jackiw.
\newblock {A PCAC puzzle: pi0 --> gamma gamma in the sigma model}.
\newblock {\em Nuovo Cim.}, A60:47--61, 1969.

\bibitem{Wang:2012am}
Juven Wang and Xiao-Gang Wen.
\newblock {Boundary Degeneracy of Topological Order}.
\newblock {\em Phys.Rev.}, B91(12):125124, 2015.

\bibitem{Lan:2014uaa}
Tian Lan, Juven~C. Wang, and Xiao-Gang Wen.
\newblock {Gapped Domain Walls, Gapped Boundaries and Topological Degeneracy}.
\newblock {\em Phys.Rev.Lett.}, 114(7):076402, 2015.

\bibitem{Wan:2014woa}
Yidun Wan, Juven~C. Wang, and Huan He.
\newblock {Twisted Gauge Theory Model of Topological Phases in Three
  Dimensions}.
\newblock 2014.

\bibitem{Jackiw:1975fn}
R.~Jackiw and C.~Rebbi.
\newblock {Solitons with Fermion Number 1/2}.
\newblock {\em Phys.Rev.}, D13:3398--3409, 1976.

\bibitem{Goldstone:1981kk}
Jeffrey Goldstone and Frank Wilczek.
\newblock {Fractional Quantum Numbers on Solitons}.
\newblock {\em Phys.Rev.Lett.}, 47:986--989, 1981.

\bibitem{SPTCS1}
Michael Levin and Ady Stern.
\newblock {\em Phys. Rev. B}, 86:115131, 2012.

\bibitem{Lu:2012dt}
Y.-M. {Lu} and A.~{Vishwanath}.
\newblock {Theory and classification of interacting integer topological phases
  in two dimensions: A Chern-Simons approach}.
\newblock {\em Phys. Rev. B}, 86(12):125119, September 2012.

\bibitem{Ye:2013upa}
Peng Ye and Juven Wang.
\newblock {Symmetry-protected topological phases with charge and spin
  symmetries: Response theory and dynamical gauge theory in two and three
  dimensions}.
\newblock {\em Phys.Rev.}, B88(23):235109, 2013.

\bibitem{Levin:2013gaa}
Michael Levin.
\newblock {Protected edge modes without symmetry}.
\newblock {\em Phys.Rev.}, X3(2):021009, 2013.

\bibitem{Kitaev:1997wr}
Alexei~Yu Kitaev.
\newblock Fault-tolerant quantum computation by anyons.
\newblock {\em Annals of Physics}, 303:2--30, 2003.

\bibitem{Ginsparg:1981bj}
Paul~H. Ginsparg and Kenneth~G. Wilson.
\newblock {A Remnant of Chiral Symmetry on the Lattice}.
\newblock {\em Phys.Rev.}, D25:2649, 1982.

\bibitem{Wilson:1974sk}
Kenneth~G. Wilson.
\newblock {Confinement of Quarks}.
\newblock {\em Phys.Rev.}, D10:2445--2459, 1974.

\bibitem{Neuberger:1997fp}
Herbert Neuberger.
\newblock {Exactly massless quarks on the lattice}.
\newblock {\em Phys.Lett.}, B417:141--144, 1998.

\bibitem{Neuberger:1998wv}
Herbert Neuberger.
\newblock {More about exactly massless quarks on the lattice}.
\newblock {\em Phys.Lett.}, B427:353--355, 1998.

\bibitem{Hernandez:1998et}
Pilar Hernandez, Karl Jansen, and Martin Luscher.
\newblock {Locality properties of Neuberger's lattice Dirac operator}.
\newblock {\em Nucl.Phys.}, B552:363--378, 1999.

\bibitem{Eichten:1985ft}
Estia Eichten and John Preskill.
\newblock {Chiral Gauge Theories on the Lattice}.
\newblock {\em Nucl.Phys.}, B268:179, 1986.

\bibitem{Wang:2014xba}
Chenjie Wang and Michael Levin.
\newblock {Braiding statistics of loop excitations in three dimensions}.
\newblock {\em Phys.Rev.Lett.}, 113(8):080403, 2014.

\bibitem{Jiang:2014ksa}
Shenghan Jiang, Andrej Mesaros, and Ying Ran.
\newblock {Generalized Modular Transformations in (3+1)D Topologically Ordered
  Phases and Triple Linking Invariant of Loop Braiding}.
\newblock {\em Phys.Rev.}, X4(3):031048, 2014.

\bibitem{Dijkgraaf:1989pz}
Robbert Dijkgraaf and Edward Witten.
\newblock {Topological Gauge Theories and Group Cohomology}.
\newblock {\em Commun.Math.Phys.}, 129:393, 1990.

\bibitem{Wang:2014pma}
Juven~C. Wang, Zheng-Cheng Gu, and Xiao-Gang Wen.
\newblock {Field theory representation of gauge-gravity symmetry-protected
  topological invariants, group cohomology and beyond}.
\newblock {\em Phys.Rev.Lett.}, 114(3):031601, 2015.

\bibitem{W1221}
X.-G. Wen.
\newblock {Modular transformation and bosonic/fermionic topological orders in
  Abelian fractional quantum Hall states}.
\newblock 2012.

\bibitem{HR8529}
F.~D.~M. Haldane and E.~H. Rezayi.
\newblock Periodic laughlin-jastrow wave functions for the fractional quantized
  hall effect.
\newblock {\em Phys. Rev. B}, 31:2529--2531, 1985.

\bibitem{KW9327}
E.~Keski-Vakkuri and Xiao-Gang Wen.
\newblock Ground state structure of hierarchical {QH} states on torus and
  modular transformation.
\newblock {\em Int. J. Mod. Phys. B}, 7:4227, 1993.

\bibitem{MW1418}
Heidar Moradi and Xiao-Gang Wen.
\newblock Universal wave function overlap and universal topological data from
  generic gapped ground states.
\newblock 2014.

\bibitem{Coxeter}
H.~S.~M. Coxeter and W.~O.~J. Moser.
\newblock {\em Generators and relations for discrete groups}.
\newblock Berlin-Gottingen-Heidelberg, Springer, 1957.

\bibitem{ZhangGroverAshvin}
Y.~{Zhang}, T.~{Grover}, A.~{Turner}, M.~{Oshikawa}, and A.~{Vishwanath}.
\newblock {Quasiparticle statistics and braiding from ground-state
  entanglement}.
\newblock {\em Phys. Rev. B}, 85(23):235151, June 2012.

\bibitem{LanL58}
L.~D. Landau and E.~M. Lifschitz.
\newblock {\em Statistical Physics - Course of Theoretical Physics Vol 5}.
\newblock Pergamon, London, 1958.

\bibitem{Qi:2008ew}
Xiao-Liang Qi, Taylor Hughes, and Shou-Cheng Zhang.
\newblock {Topological Field Theory of Time-Reversal Invariant Insulators}.
\newblock {\em Phys.Rev.}, B78:195424, 2008.

\bibitem{Kapustin:2014lwa}
Anton Kapustin and Ryan Thorngren.
\newblock {Anomalies of discrete symmetries in three dimensions and group
  cohomology}.
\newblock {\em Phys.Rev.Lett.}, 112(23):231602, 2014.

\bibitem{Kapustin:2014zva}
Anton Kapustin and Ryan Thorngren.
\newblock {Anomalies of discrete symmetries in various dimensions and group
  cohomology}.
\newblock 2014.

\bibitem{Freed:2014eja}
Daniel~S. Freed.
\newblock {Short-range entanglement and invertible field theories}.
\newblock 2014.

\bibitem{K1467}
A.~Kapustin.
\newblock {Symmetry Protected Topological Phases, Anomalies, and Cobordisms:
  Beyond Group Cohomology}.
\newblock 2014.

\bibitem{K1459}
A.~Kapustin.
\newblock {Bosonic Topological Insulators and Paramagnets: A view from
  cobordisms}.
\newblock 2014.

\bibitem{VCL0501}
F.~Verstraete, J.~I. Cirac, J.~I. Latorre, E.~Rico, and M.~M. Wolf.
\newblock Renormalization-group transformations on quantum states.
\newblock {\em Phys. Rev. Lett.}, 94:140601, 2005.

\bibitem{V0705}
Guifre Vidal.
\newblock Entanglement renormalization.
\newblock {\em Phys. Rev. Lett.}, 99:220405, 2007.

\bibitem{LG1209}
M.~Levin and Z.-C. Gu.
\newblock {Braiding statistics approach to symmetry-protected topological
  phases}.
\newblock {\em Phys. Rev. B}, 86:115109, 2012.

\bibitem{deWildPropitius:1996gt}
Mark de~Wild~Propitius.
\newblock {Spontaneously broken Abelian Chern-Simons theories}.
\newblock {\em Nucl.Phys.}, B489:297--359, 1997.

\bibitem{Chen:2012hc}
Xie Chen and Xiao-Gang Wen.
\newblock {Chiral symmetry on the edge of 2D symmetry protected topological
  phases}.
\newblock {\em Phys.Rev.}, B86:235135, 2012.

\bibitem{LW14}
Tian Lan and Xiao-Gang Wen.
\newblock {Topological quasiparticles and the holographic bulk-edge relation in
  (2+1) -dimensional string-net models}.
\newblock {\em Phys.Rev.}, B90(11):115119, 2014.

\bibitem{Witten:1985xe}
Edward Witten.
\newblock {GLOBAL GRAVITATIONAL ANOMALIES}.
\newblock {\em Commun.Math.Phys.}, 100:197, 1985.

\bibitem{Kogut:1979wt}
John~B. Kogut.
\newblock {An Introduction to Lattice Gauge Theory and Spin Systems}.
\newblock {\em Rev.Mod.Phys.}, 51:659, 1979.

\bibitem{Levin:2004mi}
Michael~A. Levin and Xiao-Gang Wen.
\newblock {String net condensation: A Physical mechanism for topological
  phases}.
\newblock {\em Phys.Rev.}, B71:045110, 2005.

\bibitem{Mesaros:2012yd}
Andrej Mesaros and Ying Ran.
\newblock {Classification of symmetry enriched topological phases with exactly
  solvable models}.
\newblock {\em Phys.Rev.}, B87(15):155115, 2013.

\bibitem{Hu:2012wx}
Yuting Hu, Yidun Wan, and Yong-Shi Wu.
\newblock {Twisted quantum double model of topological phases in two
  dimensions}.
\newblock {\em Phys.Rev.}, B87(12):125114, 2013.

\bibitem{Finkelstein:1968hy}
D.~Finkelstein and J.~Rubinstein.
\newblock {Connection between spin, statistics, and kinks}.
\newblock {\em J.Math.Phys.}, 9:1762--1779, 1968.

\bibitem{WL1437}
C.~Wang and M.~Levin.
\newblock {Braiding statistics of loop excitations in three dimensions}.
\newblock 2014.

\bibitem{JMR1462}
S.~Jiang, A.~Mesaros, and Y.~Ran.
\newblock {Generalized modular transformations in 3+1D topologically ordered
  phases and triple linking invariant of loop braiding}.
\newblock 2014.

\bibitem{deWildPropitius:1995cf}
Mark Dirk~Frederik de~Wild~Propitius.
\newblock {Topological interactions in broken gauge theories}.
\newblock 1995.

\bibitem{MW14}
Heidar Moradi and Xiao-Gang Wen.
\newblock {Universal Topological Data for Gapped Quantum Liquids in Three
  Dimensions and Fusion Algebra for Non-Abelian String Excitations}.
\newblock {\em Phys.Rev.}, B91(7):075114, 2015.

\bibitem{Ryu2013orbifolds}
O.~M. {Sule}, X.~{Chen}, and S.~{Ryu}.
\newblock {Symmetry-protected topological phases and orbifolds: Generalized
  Laughlin's argument}.
\newblock {\em Phys. Rev. B}, 88(7):075125, August 2013.

\bibitem{Lee:1956qn}
T.D. Lee and Chen-Ning Yang.
\newblock {Question of Parity Conservation in Weak Interactions}.
\newblock {\em Phys.Rev.}, 104:254--258, 1956.

\bibitem{Nielsen:1981xu}
Holger~Bech Nielsen and M.~Ninomiya.
\newblock {Absence of Neutrinos on a Lattice. 2. Intuitive Topological Proof}.
\newblock {\em Nucl.Phys.}, B193:173, 1981.

\bibitem{Nielsen:1981hk}
Holger~Bech Nielsen and M.~Ninomiya.
\newblock {No Go Theorem for Regularizing Chiral Fermions}.
\newblock {\em Phys.Lett.}, B105:219, 1981.

\bibitem{K7959}
J.~B. Kogut.
\newblock {An introduction to lattice gauge theory and spin systems}.
\newblock {\em Rev. Mod. Phys.}, 51:659 -- 713, 1979.

\bibitem{K9242}
D.~B. Kaplan.
\newblock {A Method for Simulating Chiral Fermions on the Lattice}.
\newblock {\em Phys. Lett. B}, 288:342, 1992.

\bibitem{S9390}
Y.~Shamir.
\newblock {Chiral fermions from lattice boundaries}.
\newblock {\em Nucl. Phys. B}, 406:90, 1993.

\bibitem{L9995}
M.~L{\"u}scher.
\newblock {Abelian chiral gauge theories on the lattice with exact gauge
  invariance}.
\newblock {\em Nucl. Phys. B}, 549(99):295 -- 334, 1999.

\bibitem{S9947}
H.~Suzuki.
\newblock {Gauge Invariant Effective Action in Abelian Chiral Gauge Theory on
  the Lattice}.
\newblock {\em Prog. Theor. Phys}, 101:1147 -- 1154, 1999.

\bibitem{EP8679}
E.~Eichten and J.~Preskill.
\newblock {Chiral gauge theories on the lattice}.
\newblock {\em Nucl. Phys. B}, 268:179, 1986.

\bibitem{M9259}
I.~Montvay.
\newblock {Mirror fermions in chiral gauge theories}.
\newblock {\em Nucl. Phys. Proc. Suppl.}, 29BC:159, 1992.

\bibitem{BMP0628}
T.~Bhattacharya, M.~R. Martin, and E.~Poppitz.
\newblock {Chiral Lattice Gauge Theories from Warped Domain Walls and
  Ginsparg-Wilson Fermions}.
\newblock {\em Phys. Rev. D}, 74:085028, 2006.

\bibitem{GP0776}
J.~Giedt and E.~Poppitz.
\newblock {Chiral Lattice Gauge Theories and The Strong Coupling Dynamics of a
  Yukawa-Higgs Model with Ginsparg-Wilson Fermions}.
\newblock {\em Journal of High Energy Physics}, 10:76, 2007.

\bibitem{S8631}
J.~Smit.
\newblock {\em Acta Phys. Pol.}, B17:531, 1986.

\bibitem{GPR9396}
M.~Golterman, D.~Petcher, and E.~Rivas.
\newblock {Absence of Chiral Fermions in the Eichten--Preskill Model}.
\newblock {\em Nucl. Phys. B}, 395:596 -- 622, 1993.

\bibitem{L9418}
L.~Lin.
\newblock {Nondecoupling of Heavy Mirror-Fermion}.
\newblock {\em Phys. Lett. B}, 324:418 -- 424, 1994.

\bibitem{CGP1247}
C.~Chen, J.~Giedt, and E.~Poppitz.
\newblock {On the decoupling of mirror fermions}.
\newblock {\em Journal of High Energy Physics}, 131:1304, 2013.

\bibitem{BD9216}
T.~Banks and A.~Dabholkar.
\newblock {Decoupling a Fermion Whose Mass Comes from a Yukawa Coupling:
  Nonperturbative Considerations}.
\newblock {\em Phys. Rev. D}, 46:4016 -- 4028, 1992.

\bibitem{Wen:2013ppa}
Xiao-Gang Wen.
\newblock {A lattice non-perturbative definition of an SO(10) chiral gauge
  theory and its induced standard model}.
\newblock {\em Chin.Phys.Lett.}, 30:111101, 2013.

\bibitem{Thouless:1982zz}
D.J. Thouless, M.~Kohmoto, M.P. Nightingale, and M.~den Nijs.
\newblock {Quantized Hall Conductance in a Two-Dimensional Periodic Potential}.
\newblock {\em Phys.Rev.Lett.}, 49:405--408, 1982.

\bibitem{Kapustin:2010hk}
Anton Kapustin and Natalia Saulina.
\newblock {Topological boundary conditions in abelian Chern-Simons theory}.
\newblock {\em Nucl.Phys.}, B845:393--435, 2011.

\bibitem{Ryu2014orientifolds}
C.-T. {Hsieh}, O.~M. {Sule}, G.~Y. {Cho}, S.~{Ryu}, and R.~G. {Leigh}.
\newblock {Symmetry-protected topological phases, generalized Laughlin
  argument, and orientifolds}.
\newblock {\em Phys. Rev. B}, 90(16):165134, October 2014.

\bibitem{Wen:2014zga}
Xiao-Gang Wen.
\newblock {Construction of bosonic symmetry-protected-trivial states and their
  topological invariants via $G\times SO(\infty)$ non-linear $\sigma$-models}.
\newblock 2014.

\bibitem{HuWanWuTQD}
Y.~{Hu}, Y.~{Wan}, and Y.-S. {Wu}.
\newblock {Twisted quantum double model of topological phases in two
  dimensions}.
\newblock {\em Phys. Rev. B}, 87(12):125114, March 2013.

\bibitem{You:2014oaa}
Yizhuang You, Yoni BenTov, and Cenke Xu.
\newblock {Interacting Topological Superconductors and possible Origin of $16n$
  Chiral Fermions in the Standard Model}.
\newblock 2014.

\end{thebibliography}
